%% file: main.tex
\documentclass[11pt,oneside]{book}

\input{preamble}

\title{\bfseries High-Dimensional Data Analysis for Elliptically Symmetric Distributions}
\author{Long Feng\\Nankai University\\\texttt{flnankai@nankai.edu.cn}}
\date{}

\begin{document}

\frontmatter
\maketitle
\tableofcontents

\include{chapters/preface}

\mainmatter
\include{chapters/ch1_foundations}

\include{chapters/ch2_location}

\include{chapters/ch3_matrix}

\include{chapters/ch4_other_tests}
\include{chapters/ch5_classification}
\include{chapters/ch6_pca_factor}
\include{chapters/ch7_clustering}

\appendix
\include{chapters/appendix_probability}

\backmatter
\bibliographystyle{apacite}
\bibliography{references}
\clearpage
\printindex

\end{document}

%% file: preamble.tex
\usepackage[a4paper,margin=1in]{geometry}
\usepackage[T1]{fontenc}
\usepackage[utf8]{inputenc}
\usepackage{lmodern}
\usepackage{microtype}
\usepackage{amsmath,amssymb,amsthm,mathtools}
\usepackage{mathrsfs}
\usepackage{bm}
\usepackage{bbm}
\usepackage{array}
\usepackage{booktabs}
\usepackage{enumitem}
\usepackage{graphicx}
\usepackage{xcolor}
\usepackage{makeidx}
\usepackage[natbibapa]{apacite}
\usepackage{hyperref}

\hypersetup{
  colorlinks=true,
  linkcolor=blue!50!black,
  citecolor=blue!50!black,
  urlcolor=blue!50!black,
  pdftitle={High-Dimensional Data Analysis for Elliptically Symmetric Distributions},
  pdfauthor={Long Feng, Nankai University, flnankai@nankai.edu.cn}
}

\makeindex
\mathtoolsset{showonlyrefs}

\numberwithin{equation}{chapter}

\newtheorem{theorem}{Theorem}[chapter]
\newtheorem{proposition}{Proposition}[chapter]
\newtheorem{lemma}{Lemma}[chapter]
\newtheorem{corollary}{Corollary}[chapter]
\theoremstyle{definition}
\newtheorem{definition}{Definition}[chapter]
\newtheorem{assumption}{Assumption}[chapter]
\newtheorem{example}{Example}[chapter]
\theoremstyle{remark}
\newtheorem{remark}{Remark}[chapter]

\newcommand{\R}{\mathbb{R}}
\newcommand{\E}{\mathbb{E}}
\newcommand{\Prob}{\mathbb{P}}
\newcommand{\Var}{\operatorname{Var}}
\newcommand{\Cov}{\operatorname{Cov}}
\newcommand{\Corr}{\operatorname{Corr}}
\newcommand{\tr}{\operatorname{tr}}
\newcommand{\diag}{\operatorname{diag}}
\newcommand{\rank}{\operatorname{rank}}
\newcommand{\sign}{\operatorname{sign}}
\newcommand{\vech}{\operatorname{vech}}
\newcommand{\vecop}{\operatorname{vec}}
\newcommand{\Span}{\operatorname{span}}
\newcommand{\argmin}{\operatorname*{arg\,min}}
\newcommand{\argmax}{\operatorname*{arg\,max}}
\newcommand{\iid}{\stackrel{\mathrm{iid}}{\sim}}
\newcommand{\indep}{\perp\!\!\!\perp}
\newcommand{\op}{\mathrm{op}}
\newcommand{\vct}[1]{\bm{#1}}
\newcommand{\mtr}[1]{\bm{#1}}
\newcommand{\trans}{^{\mathsf T}}

\newcommand{\norm}[1]{\left\lVert #1\right\rVert}
\newcommand{\abs}[1]{\left\lvert #1\right\rvert}
\newcommand{\frobnorm}[1]{\left\lVert #1\right\rVert_{\mathrm F}}
\newcommand{\opnorm}[1]{\left\lVert #1\right\rVert_{\mathrm{op}}}
\newcommand{\maxnorm}[1]{\left\lVert #1\right\rVert_{\max}}
\newcommand{\onenorm}[1]{\left\lVert #1\right\rVert_{1}}
\newcommand{\twonorm}[1]{\left\lVert #1\right\rVert_{2}}
\newcommand{\spn}{\mathbb{S}}

\newcommand{\vx}{\vct{x}}
\newcommand{\vX}{\vct{X}}
\newcommand{\vy}{\vct{y}}
\newcommand{\vY}{\vct{Y}}
\newcommand{\vZ}{\vct{Z}}
\newcommand{\vu}{\vct{u}}
\newcommand{\vv}{\vct{v}}
\newcommand{\vz}{\vct{z}}
\newcommand{\va}{\vct{a}}
\newcommand{\vh}{\vct{h}}
\newcommand{\vt}{\vct{t}}
\newcommand{\vw}{\vct{w}}
\newcommand{\vmu}{\bm{\mu}}

\newcommand{\vtheta}{\bm{\theta}}
\newcommand{\valpha}{\bm{\alpha}}
\newcommand{\vbeta}{\bm{\beta}}
\newcommand{\vr}{\vct{r}}
\newcommand{\vs}{\vct{s}}
\newcommand{\veta}{\bm{\eta}}
\newcommand{\vrho}{\bm{\rho}}

\newcommand{\vb}{\vct{b}}
\newcommand{\vC}{\vct{C}}

\newcommand{\vG}{\vct{G}}
\newcommand{\vR}{\vct{R}}
\newcommand{\vS}{\vct{S}}
\newcommand{\vf}{\vct{f}}
\newcommand{\vdelta}{\bm{\delta}}
\newcommand{\vlambda}{\bm{\lambda}}
\newcommand{\vLambda}{\bm{\Lambda}}
\newcommand{\vxi}{\bm{\xi}}
\newcommand{\vvarepsilon}{\bm{\varepsilon}}
\newcommand{\mA}{\mtr{A}}
\newcommand{\mB}{\mtr{B}}
\newcommand{\mC}{\mtr{C}}
\newcommand{\mD}{\mtr{D}}
\newcommand{\mE}{\mtr{E}}
\newcommand{\mF}{\mtr{F}}
\newcommand{\mG}{\mtr{G}}
\newcommand{\mI}{\mtr{I}}
\newcommand{\mJ}{\mtr{J}}
\newcommand{\mP}{\mtr{P}}
\newcommand{\mS}{\mtr{S}}

\newcommand{\mU}{\mtr{U}}
\newcommand{\mV}{\mtr{V}}
\newcommand{\mX}{\mtr{X}}
\newcommand{\mY}{\mtr{Y}}
\newcommand{\mZ}{\mtr{Z}}
\newcommand{\mSigma}{\bm{\Sigma}}
\newcommand{\mOmega}{\bm{\Omega}}
\newcommand{\mLambda}{\bm{\Lambda}}
\newcommand{\mM}{\mtr{M}}
\newcommand{\mTheta}{\bm{\Theta}}
\newcommand{\mGamma}{\bm{\Gamma}}
\newcommand{\mXi}{\bm{\Xi}}
\newcommand{\mPi}{\bm{\Pi}}

\newcommand{\vV}{\vct{V}}
\newcommand{\vW}{\vct{W}}
\newcommand{\vU}{\vct{U}}
\newcommand{\vDelta}{\bm{\Delta}}
\newcommand{\vgamma}{\bm{\gamma}}
\newcommand{\mQ}{\mtr{Q}}
\newcommand{\mR}{\mtr{R}}
\newcommand{\mW}{\mtr{W}}
\newcommand{\mDelta}{\bm{\Delta}}

\setlist[itemize]{topsep=4pt,itemsep=2pt}
\setlist[enumerate]{topsep=4pt,itemsep=2pt}

\allowdisplaybreaks

\newcommand{\mH}{\mtr{H}}

\newcommand{\mK}{\mtr{K}}
\newcommand{\mO}{\mtr{O}}
\newcommand{\vsigma}{\bm{\sigma}}
\newcommand{\Range}{\operatorname{Range}}

\newcommand{\idx}[1]{\index{#1}}

%% file: chapters/preface.tex
\chapter*{Preface}
\addcontentsline{toc}{chapter}{Preface}

High-dimensional data analysis has become one of the central themes of modern
statistics. In contemporary applications arising from genomics, finance,
macroeconomics, imaging, network analysis, and machine learning, the ambient
dimension is often comparable to or much larger than the available sample size.
Over roughly the last two decades, statistical methodology has therefore moved
far beyond the classical fixed-dimension paradigm. New theories have been
developed for location inference, covariance and precision matrix estimation,
multiple testing, factor modelling, classification, change-point analysis, and
dimension reduction in regimes where conventional multivariate procedures are no
longer applicable. Much of this literature began under Gaussian or light-tailed
assumptions, but practical data sets frequently exhibit heavy tails, outliers,
scale heterogeneity, and dependence structures that are poorly captured by the
multivariate normal model.

The purpose of this monograph is to provide a systematic account of
high-dimensional data analysis under \emph{elliptical symmetry}. The elliptical
family occupies a particularly important position in robust multivariate
statistics: it preserves the geometry of location and scatter while allowing a
much broader range of tail behaviours than the Gaussian model. This makes it a
natural framework for studying spatial signs, spatial ranks, multivariate
Kendall's tau matrices, Tyler-type shape estimators, and other methods that are
stable under heavy tails. At the same time, many classical high-dimensional
questions --- mean testing, covariance comparison, alpha testing, classification,
principal component analysis, and factor modelling --- can be reformulated in a
coherent way once the distinction between covariance, scatter, and shape is made
explicit. The present book is written from precisely this viewpoint.

The organization of the book reflects that objective. Chapter~1 introduces the
basic theory of elliptically symmetric distributions, together with spatial
signs, spatial ranks, multivariate Kendall's tau matrices, Tyler's shape
estimator, Hettmansperger--Randles estimation, and the closure properties of the
elliptical family under linear transformation and conditioning. Chapter~2 turns
to high-dimensional location estimation and testing. It begins with the classical
low-dimensional theory, reviews the main Gaussian and light-tailed benchmark
procedures, and then develops the spatial-sign, spatial-rank, weighted-sign,
max-type, max-sum, and strong-correlation methods that are central to the
elliptical program, including regularized Hotelling procedures that retain power under
pervasive dependence. Chapter~3 studies matrix estimation and testing, including covariance,
shape, inverse shape, proportionality testing, shrinkage methods, high-dimensional
Hettmansperger--Randles estimation, and tensor elliptical graphical models. Chapter~4 covers
other important testing problems, including alpha tests in factor pricing models,
change-point analysis, white-noise testing, independence testing, direct goodness-of-fit tests
for elliptical symmetry, and related adaptive procedures. Chapter~5 is devoted to classification, with particular
emphasis on robust high-dimensional LDA and QDA. Chapter~6 treats principal
component analysis, factor models, elliptical component analysis, generalized
spatial-sign PCA, and Kendall-based robust factor methods. Chapter~7 studies
high-dimensional clustering. It reviews classical and Gaussian/light-tailed
benchmark procedures, then develops robust spatial-median clustering and
semiparametric elliptical mixture clustering under heavy-tailed elliptical
models.

This book is written as a research monograph rather than as a purely pedagogical
textbook. We have tried to make the statistical constructions, assumptions, and
main proof ideas as transparent as possible, but we have not attempted to write
out every technical proof in complete detail. Doing so for every theorem would
make the book excessively long and, in many places, unnecessarily repetitive.
Instead, we emphasize the principal reductions, the main probabilistic tools, and
the structural ideas behind the arguments. Readers interested in the full details
are encouraged to consult the original papers cited in the bibliographic notes.

A computational companion is also planned. In a later stage, the author intends
to prepare an \textsf{R} package, together with reproducible code, so that the
methods in this monograph can be implemented more directly in practice. That
companion will be developed with the aid of modern code-generation tools,
including Codex.

Finally, this monograph is also a personal synthesis of the author's research
program over the past decade, together with the closely related literature to
which it naturally connects. It is offered as a reference for researchers and
students who wish to understand how high-dimensional methodology can be rebuilt
under elliptical symmetry. Comments, corrections, and suggestions are welcome and
may be sent directly to the author. Future revisions of the manuscript will
continue to be updated, including with the assistance of ChatGPT for editorial
refinement.

\chapter*{Notation and conventions}
\addcontentsline{toc}{chapter}{Notation and conventions}

Throughout the book, vectors are denoted by bold lower-case letters, matrices by
bold upper-case letters, and scalars by italic letters. For a symmetric matrix
$\mA$, we write $\lambda_1(\mA)\ge \cdots \ge \lambda_p(\mA)$ for its ordered
eigenvalues, $\tr(\mA)$ for its trace, $\det(\mA)$ for its determinant,
$\opnorm{\mA}$ for its operator norm, and $\frobnorm{\mA}$ for its Frobenius norm.
The Euclidean norm of a vector $\vx$ is written as $\twonorm{\vx}$. The symbol
$a_n \asymp b_n$ means that $a_n/b_n$ is bounded away from zero and infinity, and
$\overset{d}{\to}$ denotes convergence in distribution.

The basic elliptical model is written as
\begin{equation}
  \vX = \vmu + \xi \mA \vu,
\end{equation}
where $\vu \sim \mathrm{Unif}(\spn^{p-1})$, $\xi\ge 0$ is independent of $\vu$,
and $\mS = \mA \mA\trans$ is a scatter matrix. Since $\mS$ is identifiable only
up to a positive scalar factor, the normalized shape matrix is defined by
\begin{equation}
  \mV = \frac{p \mS}{\tr(\mS)}, \qquad \tr(\mV)=p.
\end{equation}
Only when second moments exist do we identify the covariance matrix as
\begin{equation}
  \mSigma = \Cov(\vX) = \frac{\E(\xi^2)}{p}\mS .
\end{equation}

For a nonzero vector $\vx\in \R^p$, the spatial sign is
\begin{equation}
  U(\vx) = \frac{\vx}{\twonorm{\vx}},
\end{equation}
with the convention $U(\vct{0})=\vct{0}$. The population spatial rank at $\vx$
with respect to a distribution $F$ is
\begin{equation}
  R(\vx;F)=\E\bigl\{U(\vx-\vX)\bigr\}, \qquad \vX\sim F.
\end{equation}

\begin{table}[htbp]
\centering
\caption{Core notation used throughout the book.}
\label{tab:preface-core-notation}
\begin{tabular}{@{}p{0.24\textwidth}p{0.66\textwidth}@{}}
\toprule
Symbol & Meaning \\
\midrule
$\vX,\vY$ & Random vectors in $\R^p$ \\
$\vmu,\vtheta,\valpha$ & Location-type parameters \\
$\mS$ & Scatter matrix \\
$\mV$ & Normalized shape matrix, usually with $\tr(\mV)=p$ \\
$\mSigma$ & Covariance matrix when second moments exist \\
$\mOmega$ & Precision matrix or shape-based inverse \\
$U(\vx)$ & Spatial sign of $\vx$ \\
$R(\vx;F)$ & Population spatial rank of $\vx$ under $F$ \\
$\bar{\vX}$ & Sample mean \\
$\widehat{\vtheta}_{\mathrm{SM}}$ & Sample spatial median \\
$\widehat{\mS}_{\mathrm{sign}}$ & Sample spatial sign covariance matrix \\
$T_{\mathrm{sum}}$ & Generic sum-type statistic \\
$T_{\max}$ & Generic max-type statistic \\
$s$ & Sparsity level \\
$p,n,T$ & Dimension, sample size, and time horizon \\
\bottomrule
\end{tabular}
\end{table}

To simplify later exposition, every chapter follows the same internal order
whenever possible:
\begin{enumerate}
  \item model and hypothesis;
  \item benchmark Gaussian or light-tailed procedures;
  \item robust elliptical procedure;
  \item asymptotic null theory and power;
  \item proof strategy and technical lemmas.
\end{enumerate}

%% file: chapters/ch1_foundations.tex
\chapter[Elliptically Symmetric Distributions]{Elliptically Symmetric Distributions: Definitions, Geometry, and Robust Building Blocks}
\idx{elliptical symmetry}\idx{spherical symmetry}\idx{linear combination}\idx{conditional elliptical distribution}\idx{multivariate Kendall's tau matrix}\idx{angular central Gaussian distribution}\idx{scatter matrix}\idx{shape matrix}\idx{covariance matrix}\idx{precision matrix}\idx{spatial sign}\idx{spatial rank}\idx{spatial median}\idx{Hettmansperger--Randles estimator}\idx{affine-equivariant median}\idx{spatial-sign covariance matrix (SSCM)}\idx{Tyler's M-estimator}

\section{Introduction}

The Gaussian model has served as the default point of departure for much of
classical multivariate analysis and, later, for high-dimensional statistics.
This preference is understandable. Under multivariate normality, covariance
matrices capture the full second-order structure, likelihood methods are
available, and many statistics admit tractable large-sample approximations.
However, the Gaussian paradigm is often too narrow for modern data analysis.
Financial returns, genomics measurements, imaging summaries, signal features,
and network-derived covariates frequently exhibit heavy tails, heteroscedastic
scale variation, or latent mixture behavior. In such settings, the covariance
matrix may be unstable or even undefined, whereas the underlying geometry of the
data cloud remains informative. Elliptical symmetry provides a convenient way to
retain that geometry without insisting on normal tails
\citep{Oja2010,Tyler1987,FanLiuWang2018}.

The main thesis of this book is that elliptical symmetry is more than a robust
alternative to Gaussianity. It is a structural language in which location,
shape, direction, and radial magnitude are separated in a mathematically clean
manner. That separation is particularly useful in high dimensions. Many of the
procedures studied later in this monograph depend primarily on directional
information, normalized pairwise contrasts, or shape functionals that remain
well behaved under heavy tails. For that reason, Chapter~1 establishes the
geometric and probabilistic foundation on which the rest of the book is built.

This chapter has four goals. First, we introduce spherical and elliptical
symmetry and clarify the distinction between scatter, shape, and covariance.
Second, we present the spatial sign, spatial rank, and spatial median, which are
the fundamental building blocks of robust multivariate inference in the sense of
\citet{Oja2010}. Third, we discuss sign and rank covariance matrices and
Tyler-type shape estimation, both of which play a central role in later
chapters. Fourth, we summarize the high-dimensional implications of the
elliptical framework and fix a notation system that will remain stable
throughout the book.

\section{Spherical and elliptical symmetry}

\subsection{Spherical symmetry}

We begin with the rotationally invariant case.

\begin{definition}[Spherical symmetry]
A random vector $\vZ \in \R^p$ is said to be \emph{spherically symmetric} about
the origin if
\[
  \mQ \vZ \overset{d}{=} \vZ
\]
for every orthogonal matrix $\mQ \in \R^{p \times p}$ satisfying
$\mQ \mQ\trans = \mI_p$.
\end{definition}

Spherical symmetry implies that the distribution of $\vZ$ depends only on the
Euclidean norm $\twonorm{\vZ}$. Equivalently, whenever $\vZ$ is nondegenerate,
there exist a nonnegative random variable $\xi$ and an independent random
direction $\vu \sim \mathrm{Unif}(\spn^{p-1})$ such that
\begin{equation}\label{eq:radial-spherical}
  \vZ = \xi \vu.
\end{equation}
The random variable $\xi$ is called the \emph{radial part}, and $\vu$ is the
\emph{angular part}. This radial-angular decomposition is conceptually basic:
the radial part controls tail behavior, while the angular part captures the
underlying geometry.

If a spherical distribution admits a density, then the density must be of the
form $f(\vz)=g(\twonorm{\vz}^2)$ for a suitable nonnegative function $g$. The
special role of the squared norm in this display foreshadows the Mahalanobis
geometry that appears under elliptical symmetry.

\subsection{Elliptical symmetry}

Elliptical symmetry is obtained by a linear transformation of spherical
symmetry.

\begin{definition}[Elliptically symmetric distribution]
A random vector $\vX \in \R^p$ is said to follow an elliptically symmetric
distribution if there exist $\vmu \in \R^p$, a nonnegative random variable
$\xi$, a deterministic matrix $\mA \in \R^{p \times p}$, and a random direction
$\vu \sim \mathrm{Unif}(\spn^{p-1})$, independent of $\xi$, such that
\begin{equation}\label{eq:elliptical-representation}
  \vX = \vmu + \xi \mA \vu.
\end{equation}
We call $\vmu$ the \emph{location parameter} and
\[
  \mS = \mA \mA\trans
\]
a \emph{scatter matrix}.
\end{definition}

We shall often say simply that $\vX$ is \emph{elliptical} with location
$\vmu$ and scatter $\mS$. Equation~\eqref{eq:elliptical-representation} shows
that the level sets of the distribution are affine images of spheres, hence the
terminology.

\begin{remark}[Scatter is only defined up to scale]
If $c > 0$, then
\[
  \vX = \vmu + \xi \mA \vu
      = \vmu + (\xi / c)(c \mA)\vu.
\]
Therefore $\mS$ is not identifiable on an absolute scale. The intrinsic object
is the \emph{shape} of $\mS$, not its magnitude. This observation is central in
both robust multivariate analysis and high-dimensional inference.
\end{remark}

\begin{remark}[Singular elliptical models]
The matrix $\mA$ in \eqref{eq:elliptical-representation} need not be full rank.
If $\rank(\mA)=q<p$, then $\vX$ is supported on an affine subspace of
dimension $q$. Such singular elliptical models are relevant for latent factor
structures and low-rank approximations. In Chapters~5 and~6 we will return to
this viewpoint when discussing discriminant analysis, principal components, and
elliptical factor models \citep{FanLiuWang2018}.
\end{remark}

The next proposition records the usual density representation for the
nonsingular absolutely continuous case.

\begin{proposition}[Density representation]\label{prop:density-representation}
Suppose $\mS$ is positive definite and $\vX$ admits a density. Then $\vX$ is
elliptically symmetric with location $\vmu$ and scatter $\mS$ if and only if its
density can be written as
\begin{equation}\label{eq:elliptical-density}
  f_{\vX}(\vx)
  = (\det \mS)^{-1/2}
    g\!\left\{(\vx-\vmu)\trans \mS^{-1} (\vx-\vmu)\right\},
  \qquad \vx \in \R^p,
\end{equation}
for some nonnegative function $g$.
\end{proposition}

\begin{proof}
If $\vX = \vmu + \mA \vZ$ with $\vZ$ spherically symmetric and
$\mS=\mA\mA\trans$, then $\mA$ is nonsingular and the change of variables
$\vz=\mA^{-1}(\vx-\vmu)$ yields
\[
  f_{\vX}(\vx)
  = |\det(\mA)|^{-1} f_{\vZ}\{\mA^{-1}(\vx-\vmu)\}.
\]
Since $f_{\vZ}(\vz)=g_0(\vz\trans \vz)$ for some $g_0$ under spherical symmetry,
\[
  f_{\vX}(\vx)
  = |\det(\mA)|^{-1}
    g_0\!\left[(\vx-\vmu)\trans (\mA^{-1})\trans \mA^{-1} (\vx-\vmu)\right].
\]
Because $(\mA^{-1})\trans \mA^{-1} = \mS^{-1}$ and
$|\det(\mA)| = (\det \mS)^{1/2}$, \eqref{eq:elliptical-density} follows.

Conversely, if \eqref{eq:elliptical-density} holds, then after the linear
transformation $\vz=\mA^{-1}(\vx-\vmu)$ the density of $\vz$ depends only on
$\vz\trans\vz$, hence $\vZ$ is spherical. Therefore $\vX=\vmu+\mA\vZ$ is
elliptical.
\end{proof}

\begin{example}[Basic examples]\label{ex:elliptical-examples}
The most important examples are:
\begin{enumerate}[label=(\roman*)]
  \item \textbf{Multivariate normal distributions.} If
  $\vX \sim N_p(\vmu,\mSigma)$, then
  \[
    f_{\vX}(\vx)
    =
    \frac{1}{(2\pi)^{p/2}\{\det(\mSigma)\}^{1/2}}
    \exp\!\left\{
      -\frac12(\vx-\vmu)\trans \mSigma^{-1}(\vx-\vmu)
    \right\},
    \qquad \vx\in\R^p.
  \]
  Hence $\vX$ is elliptically symmetric with scatter matrix $\mS=\mSigma$.

  \item \textbf{Multivariate $t$ distributions.} If
  $\vX \sim t_{p,\nu}(\vmu,\mSigma)$ with $\nu>0$, then
  \[
    f_{\vX}(\vx)
    =
    \frac{\Gamma\{(\nu+p)/2\}}
         {\Gamma(\nu/2)(\nu\pi)^{p/2}\{\det(\mSigma)\}^{1/2}}
    \left[
      1+\frac{1}{\nu}(\vx-\vmu)\trans \mSigma^{-1}(\vx-\vmu)
    \right]^{-(\nu+p)/2},
    \qquad \vx\in\R^p.
  \]
  Hence $\vX$ is elliptically symmetric with scatter matrix $\mS=\mSigma$.
  Moreover, when $\nu>2$,
  \[
    \Cov(\vX)=\frac{\nu}{\nu-2}\mSigma.
  \]

  \item \textbf{Normal mixture distributions.} Let $H$ be a probability
  distribution on $(0,\infty)$, and suppose that, conditional on $\tau$,
  \[
    \vX \mid \tau \sim N_p(\vmu,\tau\mSigma).
  \]
  Then $\vX$ has density
  \[
    f_{\vX}(\vx)
    =
    \int_0^\infty
    \frac{1}{(2\pi\tau)^{p/2}\{\det(\mSigma)\}^{1/2}}
    \exp\!\left\{
      -\frac{1}{2\tau}(\vx-\vmu)\trans \mSigma^{-1}(\vx-\vmu)
    \right\}
    \, dH(\tau),
    \qquad \vx\in\R^p.
  \]
  Hence $\vX$ is elliptically symmetric with scatter matrix $\mS=\mSigma$.
  This class includes the multivariate $t$ distribution as a special case for a
  suitable choice of the mixing law $H$.
\end{enumerate}
\end{example}

\subsection{Affine equivariance}

A key reason for working with elliptical models is their stability under affine
transformations.

\begin{proposition}[Affine equivariance]\label{prop:affine-equivariance}
If $\vX$ is elliptically symmetric with location $\vmu$ and scatter $\mS$, then
for every nonsingular matrix $\mB \in \R^{p \times p}$ and vector
$\vb \in \R^p$,
\[
  \mB \vX + \vb
\]
is elliptically symmetric with location $\mB \vmu + \vb$ and scatter
$\mB \mS \mB\trans$.
\end{proposition}

\begin{proof}
Write $\vX=\vmu+\xi\mA\vu$. Then
\[
  \mB\vX+\vb = (\mB\vmu+\vb) + \xi (\mB\mA)\vu.
\]
The transformed vector is therefore elliptical, and its scatter matrix is
$(\mB\mA)(\mB\mA)\trans = \mB \mS \mB\trans$.
\end{proof}

Affine equivariance is essential for statistical interpretation. A change of
units, a whitening transformation, or an orthogonal rotation should not alter
the inferential content of a procedure except through the corresponding
transformation of parameters. Later chapters repeatedly exploit this principle
when comparing covariance-based procedures with shape-based robust alternatives.

\subsection{Linear images, marginals, and conditional laws}

One of the most important closure properties of the elliptical family is that it
is stable under linear transformations and, in the nonsingular case, under
conditioning. This makes elliptical models analytically tractable in regression,
factor models, discriminant analysis, and portfolio problems
\citep{FangAnderson1990,FanLiuWang2018}.

\begin{corollary}[Linear images, marginals, and linear combinations]\label{cor:linear-images-elliptical}
Let $\vX$ be elliptically symmetric with location $\vmu$ and scatter matrix
$\mS$. Then, for any deterministic matrix $\mB \in \R^{q\times p}$,
\begin{equation}\label{eq:linear-image-elliptical}
  \mB\vX
  = \mB\vmu + \xi (\mB\mA)\vu
\end{equation}
and therefore $\mB\vX$ is elliptically symmetric with location $\mB\vmu$ and
scatter matrix $\mB\mS\mB\trans$. In particular:
\begin{enumerate}[label=(\roman*)]
  \item every subvector of $\vX$ is elliptically symmetric;
  \item for every $\va\in\R^p$, the scalar $\va\trans\vX$ is univariate
        elliptical, hence centrally symmetric, with location $\va\trans\vmu$
        and scale $\va\trans\mS\va$.
\end{enumerate}
\end{corollary}

\begin{proof}
Using the stochastic representation $\vX=\vmu+\xi\mA\vu$, we obtain
\[
  \mB\vX=\mB\vmu+\xi(\mB\mA)\vu.
\]
This is again an elliptical representation, with scatter matrix
\[
  (\mB\mA)(\mB\mA)\trans=\mB\mA\mA\trans\mB\trans=\mB\mS\mB\trans.
\]
Choosing $\mB$ to be a coordinate projection yields the marginal statement, and
choosing $\mB=\va\trans$ yields the one-dimensional linear combination.
\end{proof}

\begin{proposition}[Conditional distributions under ellipticity]\label{prop:conditional-elliptical}
Assume that $\vX$ has density representation \eqref{eq:elliptical-density} with
positive definite scatter matrix $\mS$. Partition
\[
  \vX=\begin{pmatrix}\vX_1\\ \vX_2\end{pmatrix},\qquad
  \vmu=\begin{pmatrix}\vmu_1\\ \vmu_2\end{pmatrix},\qquad
  \mS=\begin{pmatrix}
    \mS_{11} & \mS_{12}\\
    \mS_{21} & \mS_{22}
  \end{pmatrix},
\]
where $\vX_1\in\R^q$, $\vX_2\in\R^{p-q}$, and $\mS_{22}$ is nonsingular. Let
\begin{equation}\label{eq:schur-complement-s11dot2}
  \mS_{11\cdot 2}=\mS_{11}-\mS_{12}\mS_{22}^{-1}\mS_{21}
\end{equation}
and
\begin{equation}\label{eq:conditional-location-elliptical}
  \vmu_{1\mid 2}(\vx_2)
  =\vmu_1+\mS_{12}\mS_{22}^{-1}(\vx_2-\vmu_2).
\end{equation}
Then, for every $\vx_2$ such that $f_{\vX_2}(\vx_2)>0$, the conditional density of
$\vX_1\mid\vX_2=\vx_2$ can be written as
\begin{equation}\label{eq:conditional-elliptical-density}
  f_{\vX_1\mid\vX_2=\vx_2}(\vx_1)
  =c(\vx_2)\,\{\det(\mS_{11\cdot 2})\}^{-1/2}
   g\!\left[
     \delta_2(\vx_2)
     +(\vx_1-\vmu_{1\mid 2}(\vx_2))\trans
      \mS_{11\cdot 2}^{-1}
      (\vx_1-\vmu_{1\mid 2}(\vx_2))
   \right],
\end{equation}
where
\begin{equation}\label{eq:delta2-elliptical}
  \delta_2(\vx_2)=(\vx_2-\vmu_2)\trans\mS_{22}^{-1}(\vx_2-\vmu_2)
\end{equation}
and $c(\vx_2)$ is a normalizing constant depending on $\vx_2$ only. Hence
$\vX_1\mid\vX_2=\vx_2$ is elliptically symmetric with conditional location
$\vmu_{1\mid 2}(\vx_2)$ and conditional scatter matrix proportional to
$\mS_{11\cdot 2}$.
\end{proposition}

\begin{proof}
By the block inverse formula,
\[
  \mS^{-1}=\begin{pmatrix}
    \mS_{11\cdot 2}^{-1} & -\mS_{11\cdot 2}^{-1}\mS_{12}\mS_{22}^{-1}\\
    -\mS_{22}^{-1}\mS_{21}\mS_{11\cdot 2}^{-1} &
    \mS_{22}^{-1}+\mS_{22}^{-1}\mS_{21}\mS_{11\cdot 2}^{-1}\mS_{12}\mS_{22}^{-1}
  \end{pmatrix}.
\]
Therefore
\begin{align*}
  &(\vx-\vmu)\trans\mS^{-1}(\vx-\vmu)\\
  &=\{\vx_1-\vmu_1-\mS_{12}\mS_{22}^{-1}(\vx_2-\vmu_2)\}\trans
    \mS_{11\cdot 2}^{-1}
    \{\vx_1-\vmu_1-\mS_{12}\mS_{22}^{-1}(\vx_2-\vmu_2)\}\\
  &\qquad +(\vx_2-\vmu_2)\trans\mS_{22}^{-1}(\vx_2-\vmu_2)\\
  &=(\vx_1-\vmu_{1\mid 2}(\vx_2))\trans
    \mS_{11\cdot 2}^{-1}
    (\vx_1-\vmu_{1\mid 2}(\vx_2))
    +\delta_2(\vx_2).
\end{align*}
Using also $\det(\mS)=\det(\mS_{22})\det(\mS_{11\cdot 2})$, the joint density becomes
\[
  f_{\vX}(\vx_1,\vx_2)
  =\{\det(\mS_{22})\}^{-1/2}\{\det(\mS_{11\cdot 2})\}^{-1/2}
   g\!\left[
     \delta_2(\vx_2)
     +(\vx_1-\vmu_{1\mid 2}(\vx_2))\trans
      \mS_{11\cdot 2}^{-1}
      (\vx_1-\vmu_{1\mid 2}(\vx_2))
   \right].
\]
For fixed $\vx_2$, the factor $\{\det(\mS_{22})\}^{-1/2}$ is constant in $\vx_1$.
Hence the conditional density is proportional in $\vx_1$ to the right-hand side
of \eqref{eq:conditional-elliptical-density}. After normalization we obtain
\eqref{eq:conditional-elliptical-density}, proving the claim.
\end{proof}

\begin{remark}[Generator changes under conditioning]
Proposition~\ref{prop:conditional-elliptical} shows that the conditional scatter
structure is determined by the Schur complement $\mS_{11\cdot 2}$. The
conditional generator, however, changes from $g(t)$ to a shifted and normalized
version $g\{t+\delta_2(\vx_2)\}$. In the Gaussian case this remains Gaussian;
in the multivariate $t$ case the conditional law is again multivariate $t$ with
updated degrees of freedom.
\end{remark}

\section{Location, scatter, shape, covariance, and precision}

\subsection{Mean and covariance}

The scatter matrix $\mS$ remains meaningful even when the second moment of the
distribution does not exist. When moments do exist, scatter and covariance are
proportional.

\begin{proposition}[Scatter versus covariance]\label{prop:scatter-vs-covariance}
Suppose $\vX=\vmu+\xi\mA\vu$ is elliptically symmetric and
$\E(\xi^2)<\infty$. Then
\begin{equation}\label{eq:covariance-scatter}
  \Cov(\vX) = \frac{\E(\xi^2)}{p}\mS.
\end{equation}
If, in addition, $\E(\xi)<\infty$, then $\E(\vX)=\vmu$.
\end{proposition}

\begin{proof}
Because $\E(\vu)=\vct{0}$ for the uniform distribution on the sphere,
$\E(\vX)=\vmu$ whenever $\E(\xi)<\infty$. Moreover,
$\E(\vu\vu\trans)=p^{-1}\mI_p$, so
\[
  \Cov(\vX)
  = \E\bigl[\xi^2 \mA \vu \vu\trans \mA\trans\bigr]
  = \E(\xi^2)\mA \E(\vu\vu\trans) \mA\trans
  = \frac{\E(\xi^2)}{p}\mA\mA\trans
  = \frac{\E(\xi^2)}{p}\mS.
\]
\end{proof}

Proposition~\ref{prop:scatter-vs-covariance} explains why covariance-based
reasoning can become conceptually misleading under heavy tails. Even when
$\Cov(\vX)$ is undefined, the shape of the distribution is still encoded by
$\mS$ and remains inferentially useful.

\subsection{Normalized shape}

Since the scale of $\mS$ is not identifiable, we introduce the normalized shape
matrix
\begin{equation}\label{eq:normalized-shape}
  \mV = \frac{p \mS}{\tr(\mS)},
  \qquad \tr(\mV)=p.
\end{equation}
Whenever convenient we also write
\[
  \mS = \sigma^2 \mV,
  \qquad
  \sigma^2 = \frac{1}{p}\tr(\mS),
\]
so that $\sigma^2$ plays the role of a global scale and $\mV$ records relative
dispersion.

If
\[
  \mV = \mGamma \mLambda \mGamma\trans,
\]
where $\mGamma=(\vgamma_1,\ldots,\vgamma_p)$ is orthogonal and
$\mLambda=\diag(\lambda_1,\ldots,\lambda_p)$ with
$\lambda_1 \ge \cdots \ge \lambda_p > 0$, then
$\vgamma_j$ describes a principal direction and $\lambda_j$ describes relative
spread along that direction. This spectral decomposition will appear throughout
the book in testing, classification, principal component analysis, and factor
models.

\begin{remark}[Generalized correlation and inverse shape]
If the diagonal entries of $\mV$ are positive, we define the shape-based
correlation matrix
\[
  \mtr{R}
  = \diag(\mV)^{-1/2} \mV \diag(\mV)^{-1/2}.
\]
Likewise, whenever $\mV$ is positive definite we may work with its inverse
$\mV^{-1}$, which is the natural \emph{inverse shape} matrix. Only when second
moments exist and an absolute scale has been fixed do these objects coincide
with the usual correlation and precision matrices derived from covariance.
\end{remark}

\subsection{A notation table}

Table~\ref{tab:objects-ch1} records the main population-level objects that will
be used throughout this monograph.

\begin{table}[htbp]
\centering
\small
\begin{tabular}{@{}p{0.16\textwidth}p{0.24\textwidth}p{0.50\textwidth}@{}}
\toprule
Symbol & Name & Role in this book \\
\midrule
$\vmu$ & location parameter & Center of the elliptical distribution; target in Chapter~2 and a nuisance parameter in several later testing problems. \\
$\mS$ & scatter matrix & Affine-equivariant second-order shape object; meaningful even when covariance does not exist. \\
$\mV$ & normalized shape matrix & Scale-free version of scatter, defined by $\tr(\mV)=p$; used to unify testing, PCA, factor models, and discriminant analysis. \\
$\mSigma$ & covariance matrix & Used only when second moments exist; proportional to $\mS$ under Proposition~\ref{prop:scatter-vs-covariance}. \\
$\mOmega$ & precision or inverse shape & Inverse of $\mSigma$ or $\mV$ when the corresponding matrix is well defined and nonsingular. \\
$\mK(F)$ & multivariate Kendall's tau matrix & Pairwise directional scatter functional; under ellipticity it shares eigenvectors with $\mS$ and is insensitive to radial tails. \\
$\mGamma,\mLambda$ & eigenvectors and eigenvalues & Encode principal directions and relative dispersion; central in Chapters~3, 5, and~6. \\
\bottomrule
\end{tabular}
\caption{Population objects in the elliptical framework.}
\label{tab:objects-ch1}
\end{table}

\section{Radial-angular decomposition and directional geometry}

\subsection{Whitening and angular structure}

Let $\mS^{1/2}$ be any symmetric square root of $\mS$. Then
\[
  \mS^{-1/2}(\vX-\vmu) = \xi \vu
\]
whenever $\mS$ is positive definite. In other words, after whitening by
$\mS^{-1/2}$, the distribution becomes spherical. This is the simplest way to
separate angular information from radial magnitude.

The geometric content of elliptical symmetry is therefore encoded in the
dependence of directions on $\mS$. To make this explicit, write the spectral
decomposition $\mS=\mGamma\mLambda\mGamma\trans$. Then
\[
  \vX-\vmu
  = \xi \mGamma \mLambda^{1/2} \vu.
\]
Rotating into the eigenbasis of $\mS$ yields coordinates
$\mGamma\trans(\vX-\vmu)=\xi \mLambda^{1/2}\vu$. Thus the radial variable $\xi$
controls overall magnitude, while the eigenvalues of $\mS$ deform the angular
distribution through $\mLambda^{1/2}$.

\subsection{Spatial signs depend only on direction}

The spatial sign map
\begin{equation}\label{eq:spatial-sign}
  U(\vx) = \frac{\vx}{\twonorm{\vx}} \mathbbm{1}(\vx \neq \vct{0})
\end{equation}
extracts only the direction of a vector and discards its length. This is the
basic reason why sign-based methods are robust to heavy tails.

\begin{proposition}[Sign representation under ellipticity]\label{prop:sign-elliptical}
Suppose $\vX=\vmu+\xi \mA \vu$ with $\Prob(\xi=0)=0$. Then
\begin{equation}\label{eq:sign-elliptical}
  U(\vX-\vmu)
  = \frac{\mA\vu}{\twonorm{\mA\vu}}.
\end{equation}
Consequently, the distribution of $U(\vX-\vmu)$ depends on $\mS=\mA\mA\trans$
but not on the radial variable $\xi$.
\end{proposition}

\begin{proof}
Since $\xi \ge 0$ and $\Prob(\xi=0)=0$,
\[
  U(\vX-\vmu)
  = U(\xi \mA \vu)
  = \frac{\xi \mA \vu}{\twonorm{\xi \mA \vu}}
  = \frac{\mA \vu}{\twonorm{\mA \vu}}.
\]
The radial factor cancels identically.
\end{proof}

Proposition~\ref{prop:sign-elliptical} explains why sign-based procedures remain
meaningful even under infinite-variance models: they require directional
structure but not moment existence. This idea will be used heavily in
Chapter~2, where tests for high-dimensional location are built from spatial
signs and their weighted variants.

\begin{remark}[Central symmetry]
If $\vX$ is elliptically symmetric with location $\vmu$, then
$\vX-\vmu \overset{d}{=} -(\vX-\vmu)$. Hence
\[
  \E\{U(\vX-\vmu)\} = \vct{0}
\]
whenever the expectation exists. The null distributions of many sign-based test
statistics ultimately rely on this elementary symmetry.
\end{remark}

\section{Spatial median, spatial signs, and spatial ranks}

\subsection{Spatial median}

The spatial median is the most basic robust multivariate location functional in
the sign-rank framework of \citet{Oja2010}.

\begin{definition}[Spatial median]
For a random vector $\vX \in \R^p$ with distribution $F$, any minimizer of
\begin{equation}\label{eq:population-spatial-median}
  Q_F(\vtheta)
  = \E\bigl[\twonorm{\vX-\vtheta} - \twonorm{\vX}\bigr]
\end{equation}
is called a \emph{spatial median} of $F$. Given observations
$\vX_1,\ldots,\vX_n$, any minimizer of
\begin{equation}\label{eq:sample-spatial-median}
  Q_n(\vtheta)
  = \frac{1}{n}\sum_{i=1}^n \twonorm{\vX_i-\vtheta}
\end{equation}
is called a \emph{sample spatial median} and is denoted by
$\widehat{\vmu}_{\mathrm{SM}}$.
\end{definition}

The spatial median is translation equivariant and robust to outliers in
magnitude. Unlike the sample mean, it is well defined without any moment
assumption. Under mild regularity conditions, it is the unique solution to the
estimating equation
\begin{equation}\label{eq:spatial-median-eq}
  \frac{1}{n}\sum_{i=1}^n U(\vX_i-\widehat{\vmu}_{\mathrm{SM}})=\vct{0}.
\end{equation}

\begin{proposition}[Centering property of the spatial median]
If $\vX$ is elliptically symmetric with location $\vmu$ and is not supported on
a line through $\vmu$, then $\vmu$ is the unique spatial median.
\end{proposition}

\begin{proof}
By elliptical symmetry, $\vX-\vmu \overset{d}{=} -(\vX-\vmu)$. Therefore
\[
  \E\{U(\vX-\vmu)\}=\vct{0},
\]
which means that $\vmu$ is a stationary point of $Q_F$. Since
$\vtheta \mapsto \twonorm{\vX-\vtheta}$ is convex for every realization of
$\vX$, the functional $Q_F(\vtheta)$ is convex. The nondegeneracy condition
rules out flat directions, so the minimizer is unique and equals $\vmu$.
\end{proof}

\subsection{Spatial rank}

The multivariate spatial rank generalizes the univariate notion of rank by
replacing scalar order information with average directions.

\begin{definition}[Spatial rank]
For a point $\vx \in \R^p$ and a distribution $F$, the \emph{spatial rank} of
$\vx$ with respect to $F$ is
\begin{equation}\label{eq:population-spatial-rank}
  R(\vx;F) = \E\{U(\vx-\vX)\}, \qquad \vX \sim F.
\end{equation}
Given a sample $\vX_1,\ldots,\vX_n$, the empirical spatial rank of $\vX_i$ is
\begin{equation}\label{eq:sample-spatial-rank}
  \widehat{\vR}_i
  = \frac{1}{n}\sum_{j=1}^n U(\vX_i-\vX_j).
\end{equation}
\end{definition}

Two properties of spatial ranks are worth emphasizing.

First, they are \emph{translation invariant}: adding a common vector $\vb$ to
every observation leaves all pairwise differences unchanged. This is the reason
why two-sample spatial rank statistics can often avoid direct location
estimation. Second, they blend information on relative location and direction,
which is why rank-based tests often achieve higher efficiency than pure sign
tests under dense alternatives.

\begin{remark}[Signs versus ranks]
Spatial signs use only one observation at a time after centering and therefore
focus on directional stability. Spatial ranks are based on pairwise differences,
so they are automatically centered and can exploit more of the relative location
structure. In Chapter~2, this difference will correspond roughly to a trade-off
between maximal robustness and improved efficiency.
\end{remark}

\section{Sign and rank covariance matrices}

\subsection{The spatial sign covariance matrix}

Sign and rank covariance matrices provide robust alternatives to the sample
covariance matrix \citep{VisuriOjaKoivunen2000}. After centering at a suitable
location estimator, the \emph{sample spatial sign covariance matrix} is
\begin{equation}\label{eq:sscm}
  \widehat{\mS}_{\mathrm{sign}}
  = \frac{1}{n}\sum_{i=1}^n
    U(\vX_i-\widehat{\vmu}_{\mathrm{SM}})
    U(\vX_i-\widehat{\vmu}_{\mathrm{SM}})\trans.
\end{equation}
At the population level, the corresponding matrix is
\begin{equation}\label{eq:population-sscm}
  \mS_{\mathrm{sign}}(F)
  = \E\left[
      U(\vX-\vmu)U(\vX-\vmu)\trans
    \right].
\end{equation}

The matrix $\mS_{\mathrm{sign}}(F)$ is orthogonally equivariant, positive
semidefinite, and trace-normalized:
\[
  \tr\{\mS_{\mathrm{sign}}(F)\}=1.
\]
When $F$ is spherical, $\mS_{\mathrm{sign}}(F)=p^{-1}\mI_p$.

\begin{proposition}[Eigenvectors of the population SSCM]\label{prop:sscm-eigenvectors}
Let $\vX$ be elliptical with location $\vmu$ and positive definite scatter
matrix $\mS=\mGamma\mLambda\mGamma\trans$. Then
\begin{equation}\label{eq:sscm-diagonalization}
  \mS_{\mathrm{sign}}(F)
  = \mGamma \mDelta \mGamma\trans
\end{equation}
for some diagonal matrix $\mDelta=\diag(\delta_1,\ldots,\delta_p)$ with
$\delta_j \ge 0$ and $\sum_{j=1}^p \delta_j = 1$. Hence
$\mS_{\mathrm{sign}}(F)$ and $\mS$ have the same eigenvectors.
\end{proposition}

\begin{proof}
By Proposition~\ref{prop:sign-elliptical},
\[
  U(\vX-\vmu)
  = \frac{\mGamma \mLambda^{1/2}\vu}
         {\sqrt{\vu\trans \mLambda \vu}}.
\]
Therefore
\[
  \mS_{\mathrm{sign}}(F)
  = \mGamma
    \E\left[
      \frac{\mLambda^{1/2}\vu\vu\trans \mLambda^{1/2}}
           {\vu\trans \mLambda \vu}
    \right]
    \mGamma\trans.
\]
It suffices to show that the middle expectation is diagonal. Its $(j,k)$ entry
for $j \neq k$ is
\[
  \sqrt{\lambda_j\lambda_k}\,
  \E\left(
    \frac{u_j u_k}{\sum_{\ell=1}^p \lambda_\ell u_\ell^2}
  \right).
\]
Because the uniform distribution on the sphere is symmetric under sign changes
of any single coordinate, the integrand is odd in $u_j$ (or $u_k$), so the
expectation vanishes. Hence the matrix is diagonal, proving
\eqref{eq:sscm-diagonalization}. The trace statement follows from
$\tr\{U(\vX-\vmu)U(\vX-\vmu)\trans\}=1$.
\end{proof}

Proposition~\ref{prop:sscm-eigenvectors} is one of the most useful structural
facts in this chapter. Although the eigenvalues of the SSCM are nonlinear
transformations of the eigenvalues of $\mS$, the principal directions are
preserved. This provides a direct bridge from robust directional methods to
eigenspace estimation, principal components, and factor models.

\subsection{The multivariate Kendall's tau matrix}\label{subsec:mkendall}

A closely related pairwise directional functional is the \emph{multivariate
Kendall's tau matrix}\idx{multivariate Kendall's tau matrix}, also called the
\emph{spatial Kendall's tau matrix} in parts of the literature
\citep{Oja2010,HanLiu2018ECA,FanLiuWang2018}. Let $\widetilde{\vX}$ be an
independent copy of $\vX$. The population version is defined by
\begin{equation}\label{eq:population-mkendall}
  \mK(F)
  = \E\left[U(\vX-\widetilde{\vX})U(\vX-\widetilde{\vX})\trans\right]
  = \E\left\{\frac{(\vX-\widetilde{\vX})(\vX-\widetilde{\vX})\trans}
                   {\twonorm{\vX-\widetilde{\vX}}^2}\right\}.
\end{equation}
Given observations $\vX_1,\ldots,\vX_n$, the sample multivariate Kendall's tau
matrix is the U-statistic
\begin{equation}\label{eq:sample-mkendall}
  \widehat{\mK}
  = \frac{2}{n(n-1)}\sum_{1\le i<j\le n}
    U(\vX_i-\vX_j)U(\vX_i-\vX_j)\trans.
\end{equation}
By construction, $\mK(F)$ and $\widehat{\mK}$ are positive semidefinite with
unit trace. The pairwise differencing removes the location parameter
automatically, which makes $\widehat{\mK}$ especially attractive when robust
centering is difficult.

\begin{proposition}[Eigenvectors of the multivariate Kendall's tau matrix]\label{prop:mkendall-eigenvectors}
Let $\vX$ be elliptically symmetric with location $\vmu$ and positive definite
scatter matrix $\mS=\mGamma\mLambda\mGamma\trans$. Then
\begin{equation}\label{eq:mkendall-diagonalization}
  \mK(F)=\mGamma\mLambda_{\mathrm K}\mGamma\trans,
  \qquad
  \mLambda_{\mathrm K}=\diag(\kappa_1,\ldots,\kappa_p),
\end{equation}
where
\begin{equation}\label{eq:mkendall-eigenvalues}
  \kappa_j
  = \lambda_j\,\E\left(\frac{u_j^2}{\sum_{\ell=1}^p \lambda_\ell u_\ell^2}\right),
  \qquad j=1,\ldots,p,
\end{equation}
and $\sum_{j=1}^p \kappa_j=1$. Hence $\mK(F)$ and $\mS$ have the same
eigenvectors.
\end{proposition}

\begin{proof}
Let $\widetilde{\vX}$ be an independent copy of $\vX$. Since $\vX$ is
elliptically symmetric, its characteristic function has the form
$\phi_{\vX}(\vt)=\exp(\mathrm{i}\vt\trans\vmu)\psi(\vt\trans\mS\vt)$ for some
scalar function $\psi$ \citep{FangAnderson1990}. Therefore
\[
  \phi_{\vX-\widetilde{\vX}}(\vt)
  =\phi_{\vX}(\vt)\phi_{\widetilde{\vX}}(-\vt)
  =\psi(\vt\trans\mS\vt)^2,
\]
which depends on $\vt$ only through the quadratic form
$\vt\trans\mS\vt$. Hence $\vX-\widetilde{\vX}$ is elliptically symmetric with
location $\vct{0}$ and scatter matrix $\mS$. Applying
Proposition~\ref{prop:sscm-eigenvectors} to the difference vector
$\vX-\widetilde{\vX}$ yields
\[
  \mK(F)
  =\E\left[U(\vX-\widetilde{\vX})U(\vX-\widetilde{\vX})\trans\right]
  =\mGamma\mLambda_{\mathrm K}\mGamma\trans.
\]
The diagonal entries of $\mLambda_{\mathrm K}$ are given by
\eqref{eq:mkendall-eigenvalues}, and the trace identity follows from
$\tr\{U(\vX-\widetilde{\vX})U(\vX-\widetilde{\vX})\trans\}=1$.
\end{proof}

The matrices $\mS_{\mathrm{sign}}(F)$ and $\mK(F)$ are similar in spirit but not
identical. The SSCM is built from one-point directions after centering, whereas
$\mK(F)$ is built from pairwise directions and therefore eliminates the location
parameter automatically. In Chapter~6 this distinction will matter when we turn
to robust eigenspace estimation and elliptical principal component analysis.

\subsection{Rank covariance matrices}

Spatial ranks lead to analogous covariance-type functionals. For example, one
may consider
\begin{equation}\label{eq:rank-covariance}
  \widehat{\mS}_{\mathrm{rank}}
  = \frac{1}{n}\sum_{i=1}^n \widehat{\vR}_i \widehat{\vR}_i\trans,
\end{equation}
or, depending on the application, versions based directly on pairwise
differences. Rank covariance matrices are typically less crude than sign
covariance matrices because they use relative rather than purely directional
information. In later chapters they reappear in matrix testing, independence
testing, and principal component analysis.

\begin{remark}[Orthogonal versus affine equivariance]
The SSCM is only orthogonally equivariant, not fully affine equivariant. This
is not a defect; it is the price of stripping away radial magnitude completely.
When full affine equivariance is needed, Tyler's shape estimator becomes the
natural replacement.
\end{remark}

\section{Tyler's shape matrix and affine-equivariant estimation}

\subsection{Population equation and sample estimator}

Among affine-equivariant robust shape functionals, Tyler's
$M$-estimator\idx{Tyler's M-estimator} is the most important in the elliptical
setting \citep{Tyler1987}. At the sample level, after centering by a location
estimator $\widehat{\vmu}$, it is defined as a positive definite solution to
\begin{equation}\label{eq:tyler-sample}
  \widehat{\mV}_{\mathrm{T}}
  = \frac{p}{n}\sum_{i=1}^n
    \frac{(\vX_i-\widehat{\vmu})(\vX_i-\widehat{\vmu})\trans}
         {(\vX_i-\widehat{\vmu})\trans
           \widehat{\mV}_{\mathrm{T}}^{-1}
           (\vX_i-\widehat{\vmu})},
  \qquad
  \tr(\widehat{\mV}_{\mathrm{T}})=p.
\end{equation}
The trace constraint fixes the otherwise arbitrary scale. The corresponding
population equation is
\begin{equation}\label{eq:tyler-population}
  \mV
  = p\,\E\left[
      \frac{(\vX-\vmu)(\vX-\vmu)\trans}
           {(\vX-\vmu)\trans \mV^{-1} (\vX-\vmu)}
    \right],
  \qquad
  \tr(\mV)=p.
\end{equation}

Equation~\eqref{eq:tyler-population} shows that Tyler's functional depends only
on directions standardized by the current shape estimate. Unlike the sample
covariance matrix, it is intrinsically scale free and remains meaningful even
when second moments do not exist.

\begin{proposition}[Angular representation of Tyler's equation]\label{prop:tyler-angular-representation}
For every positive definite matrix $\mV$ and every $\vx\neq\vct{0}$,
\begin{equation}\label{eq:tyler-angular-representation}
  \frac{\vx\vx\trans}{\vx\trans\mV^{-1}\vx}
  = \mV^{1/2}U(\mV^{-1/2}\vx)U(\mV^{-1/2}\vx)\trans\mV^{1/2}.
\end{equation}
Consequently, if $\vY_i=\vX_i-\widehat{\vmu}$ and
\begin{equation}\label{eq:whitened-signs-tyler}
  \widehat{\vW}_i(\mV)
  = U(\mV^{-1/2}\vY_i),
\end{equation}
then \eqref{eq:tyler-sample} is equivalent to
\begin{equation}\label{eq:tyler-whitened-signs}
  \widehat{\mV}_{\mathrm{T}}
  = \frac{p}{n}\sum_{i=1}^n
    \widehat{\mV}_{\mathrm{T}}^{1/2}
    \widehat{\vW}_i(\widehat{\mV}_{\mathrm{T}})
    \widehat{\vW}_i(\widehat{\mV}_{\mathrm{T}})\trans
    \widehat{\mV}_{\mathrm{T}}^{1/2}.
\end{equation}
\end{proposition}

\begin{proof}
Since $\twonorm{\mV^{-1/2}\vx}^2=\vx\trans\mV^{-1}\vx$,
\[
  \mV^{1/2}U(\mV^{-1/2}\vx)U(\mV^{-1/2}\vx)\trans\mV^{1/2}
  =\mV^{1/2}\frac{\mV^{-1/2}\vx\vx\trans\mV^{-1/2}}{\vx\trans\mV^{-1}\vx}\mV^{1/2}
  =\frac{\vx\vx\trans}{\vx\trans\mV^{-1}\vx},
\]
which proves \eqref{eq:tyler-angular-representation}. Substituting
$\vx=\vY_i$ into \eqref{eq:tyler-angular-representation} and summing over $i$
yields \eqref{eq:tyler-whitened-signs}.
\end{proof}

Proposition~\ref{prop:tyler-angular-representation} shows that Tyler's matrix is
not an ordinary second-moment estimator. It is a \emph{self-consistent covariance
matrix of whitened directions}. The matrix $\mV$ is chosen so that, after
whitening by $\mV^{-1/2}$, the average outer product of the resulting spatial
signs becomes isotropic.

\begin{proposition}[Tyler's population equation recovers the normalized shape]\label{prop:tyler-recovers-shape}
Suppose that $\vX=\vmu+\xi\mA\vu$ is elliptically symmetric with positive
definite scatter matrix $\mS=\mA\mA\trans$, and let
\begin{equation}\label{eq:tyler-target-shape}
  \mV_0=\frac{p\mS}{\tr(\mS)}.
\end{equation}
Then $\mV_0$ satisfies the population fixed-point equation
\eqref{eq:tyler-population}. Hence, whenever the population solution is unique
under the trace normalization, Tyler's functional coincides with the normalized
shape matrix.
\end{proposition}

\begin{proof}
Write $\mV_0=c\mS$ with $c=p/\tr(\mS)$. Then
$\mV_0^{-1}=c^{-1}\mS^{-1}$. Since $\vX-\vmu=\xi\mA\vu$,
\[
  (\vX-\vmu)\trans\mV_0^{-1}(\vX-\vmu)
  =\xi^2\vu\trans\mA\trans(c^{-1}\mS^{-1})\mA\vu
  =c^{-1}\xi^2\vu\trans\vu
  =c^{-1}\xi^2.
\]
Therefore
\[
  \frac{(\vX-\vmu)(\vX-\vmu)\trans}
       {(\vX-\vmu)\trans\mV_0^{-1}(\vX-\vmu)}
  =c\mA\vu\vu\trans\mA\trans.
\]
Taking expectation and using $\E(\vu\vu\trans)=p^{-1}\mI_p$, we obtain
\[
  p\,\E\left[
      \frac{(\vX-\vmu)(\vX-\vmu)\trans}
           {(\vX-\vmu)\trans\mV_0^{-1}(\vX-\vmu)}
    \right]
  =pc\mA\E(\vu\vu\trans)\mA\trans
  =c\mA\mA\trans
  =c\mS
  =\mV_0.
\]
Also $\tr(\mV_0)=p$ by construction. Hence $\mV_0$ solves
\eqref{eq:tyler-population}.
\end{proof}

\begin{proposition}[Basic invariance properties of Tyler's estimator]\label{prop:tyler-basic-invariance}
Let $\widehat{\mV}_{\mathrm{T}}(\vX_1,\ldots,\vX_n)$ denote a solution of
\eqref{eq:tyler-sample}. Then:
\begin{enumerate}[label=(\roman*)]
  \item for every $c>0$,
  \[
    \widehat{\mV}_{\mathrm{T}}(c\vX_1,\ldots,c\vX_n)
    = \widehat{\mV}_{\mathrm{T}}(\vX_1,\ldots,\vX_n);
  \]
  \item for every nonsingular $\mB$,
  \[
    \widehat{\mV}_{\mathrm{T}}(\mB\vX_1,\ldots,\mB\vX_n)
    \propto \mB\widehat{\mV}_{\mathrm{T}}(\vX_1,\ldots,\vX_n)\mB\trans,
  \]
  where the proportionality constant is fixed by the trace normalization.
\end{enumerate}
\end{proposition}

\begin{proof}
Property (i) follows because multiplying every centered observation by $c$
multiplies the numerator and denominator of each summand in
\eqref{eq:tyler-sample} by the same factor $c^2$. For (ii), substitute
$\mB\vX_i$ into \eqref{eq:tyler-sample} and use the identity
$(\mB\mV\mB\trans)^{-1}=(\mB\trans)^{-1}\mV^{-1}\mB^{-1}$. The transformed
equation is satisfied by a positive multiple of
$\mB\widehat{\mV}_{\mathrm{T}}\mB\trans$, and the constant is determined by
$\tr(\widehat{\mV}_{\mathrm{T}})=p$.
\end{proof}

\subsection{The angular central Gaussian viewpoint}

Tyler's estimator is closely connected with the
\emph{angular central Gaussian} (ACG) distribution\idx{angular central Gaussian distribution}.
A random direction $\vW\in\spn^{p-1}$ is said to follow an ACG distribution
with shape matrix $\mV$ if it has density
\begin{equation}\label{eq:acg-density}
  f_{\mathrm{ACG}}(\vw;\mV)
  = \frac{\Gamma(p/2)}{2\pi^{p/2}}
    \{\det(\mV)\}^{-1/2}
    (\vw\trans\mV^{-1}\vw)^{-p/2},
  \qquad \vw\in\spn^{p-1}.
\end{equation}
This density is scale invariant in the sense that
$f_{\mathrm{ACG}}(\vw;c\mV)=f_{\mathrm{ACG}}(\vw;\mV)$ for every $c>0$. Under an
elliptical model, the centered direction $U(\vX-\vmu)$ depends only on the
shape matrix and has an ACG distribution with that shape.

\begin{proposition}[Likelihood equation under the ACG model]\label{prop:tyler-acg-likelihood}
Let $\vW_1,\ldots,\vW_n\in\spn^{p-1}$ be observations from the ACG model with
shape matrix $\mV$, normalized by $\tr(\mV)=p$. Up to an additive constant, the
log-likelihood is
\begin{equation}\label{eq:acg-loglik}
  \ell_n(\mV)
  = -\frac{n}{2}\log\det(\mV)
    -\frac{p}{2}\sum_{i=1}^n\log(\vW_i\trans\mV^{-1}\vW_i).
\end{equation}
Any interior maximizer satisfies the likelihood equation
\begin{equation}\label{eq:acg-score-equation}
  \mV
  = \frac{p}{n}\sum_{i=1}^n
    \frac{\vW_i\vW_i\trans}{\vW_i\trans\mV^{-1}\vW_i},
  \qquad \tr(\mV)=p,
\end{equation}
which is precisely Tyler's estimating equation written in terms of directions.
\end{proposition}

\begin{proof}
Differentiating \eqref{eq:acg-loglik} with respect to $\mV$ gives
\[
  d\ell_n(\mV)
  = -\frac{n}{2}\tr(\mV^{-1}d\mV)
    +\frac{p}{2}\sum_{i=1}^n
      \frac{\vW_i\trans\mV^{-1}(d\mV)\mV^{-1}\vW_i}
           {\vW_i\trans\mV^{-1}\vW_i}.
\]
Using
$\vW_i\trans\mV^{-1}(d\mV)\mV^{-1}\vW_i
=\tr\{\mV^{-1}\vW_i\vW_i\trans\mV^{-1}d\mV\}$,
we can rewrite the differential as
\[
  d\ell_n(\mV)
  = \frac12\tr\!\left[
      \left\{-n\mV^{-1}
      +p\sum_{i=1}^n
        \frac{\mV^{-1}\vW_i\vW_i\trans\mV^{-1}}
             {\vW_i\trans\mV^{-1}\vW_i}
      \right\}d\mV
    \right].
\]
Hence the score equation is
\[
  n\mV^{-1}
  =p\sum_{i=1}^n
    \frac{\mV^{-1}\vW_i\vW_i\trans\mV^{-1}}
         {\vW_i\trans\mV^{-1}\vW_i}.
\]
Multiplying from the left and right by $\mV$ yields
\eqref{eq:acg-score-equation}. The trace constraint then fixes the scale.
\end{proof}

The ACG representation is conceptually important. It shows that Tyler's estimator
is the likelihood estimator of the \emph{angular distribution} and therefore a
natural affine-equivariant counterpart of the SSCM and multivariate Kendall's
tau matrix.

\subsection{Existence, uniqueness, and computation}

The existence theory of Tyler's estimator is more delicate than that of the SSCM
because the defining equation is nonlinear. Let
$\vY_i=\vX_i-\widehat{\vmu}$ denote centered observations. Tyler
\citep{Tyler1987}
showed that a positive definite solution to \eqref{eq:tyler-sample} exists and
is unique up to scale if the sample is not overly concentrated on any proper
linear subspace, namely if
\begin{equation}\label{eq:tyler-existence-condition}
  \frac{1}{n}\sum_{i=1}^n\mathbbm{1}(\vY_i\in L)
  < \frac{\dim(L)}{p}
\end{equation}
for every proper linear subspace $L\subsetneq\R^p$. In particular, when $n>p$
and the observations are in general position, the trace-normalized solution is
unique.

A practical fixed-point algorithm is obtained by iterating
\begin{equation}\label{eq:tyler-fixed-point-map}
  \mH(\mV)
  = \frac{p}{n}\sum_{i=1}^n
    \frac{\vY_i\vY_i\trans}{\vY_i\trans\mV^{-1}\vY_i},
  \qquad
  \mV^{(m+1)}
  = \frac{p\,\mH\{\mV^{(m)}\}}{\tr[\mH\{\mV^{(m)}\}]},
\end{equation}
started from any positive definite matrix $\mV^{(0)}$.
Under the existence condition \eqref{eq:tyler-existence-condition}, the sequence
converges to the Tyler solution up to the chosen normalization. This iterative
form makes transparent that Tyler's estimator is obtained by repeated
reweighting: observations aligned with currently over-dispersed directions are
downweighted through the denominator $\vY_i\trans\mV^{-1}\vY_i$.

\subsection{Location-shape pairs and affine-equivariant medians}

The spatial median is orthogonally equivariant and robust, but it is not fully
affine equivariant. When the coordinate system is changed by a general
nonsingular linear transformation, the transformed spatial median need not equal
that transformation applied to the original estimate. In low and moderate
dimensions, this drawback can be removed by estimating location and shape
simultaneously. The classical solution is the
Hettmansperger--Randles estimator\idx{Hettmansperger--Randles estimator}\idx{affine-equivariant median}
of \citet{HettmanspergerRandles2002}, which couples a location equation with a
Tyler-type shape equation.

Let $\vX_1,\ldots,\vX_n\in\R^p$ be observations, and let
$(\widehat{\vmu}_{\mathrm{HR}},\widehat{\mV}_{\mathrm{HR}})$ denote a solution of
\begin{equation}\label{eq:hr-location-equation}
  \frac1n\sum_{i=1}^n
  U\!\left\{\widehat{\mV}_{\mathrm{HR}}^{-1/2}(\vX_i-\widehat{\vmu}_{\mathrm{HR}})\right\}
  = \vct 0,
\end{equation}
and
\begin{equation}\label{eq:hr-shape-equation}
  \frac{p}{n}\sum_{i=1}^n
  U\!\left\{\widehat{\mV}_{\mathrm{HR}}^{-1/2}(\vX_i-\widehat{\vmu}_{\mathrm{HR}})\right\}
  U\!\left\{\widehat{\mV}_{\mathrm{HR}}^{-1/2}(\vX_i-\widehat{\vmu}_{\mathrm{HR}})\right\}\trans
  = \mI_p,
  \qquad
  \tr(\widehat{\mV}_{\mathrm{HR}})=p.
\end{equation}
The first equation centers the whitened data by a spatial-sign balance
condition, whereas the second equation enforces isotropy of the whitened
signs. Equation~\eqref{eq:hr-shape-equation} is exactly Tyler's fixed-point
equation written after centering at $\widehat{\vmu}_{\mathrm{HR}}$.

\begin{remark}[Relation to the spatial median and Tyler's estimator]
If the shape matrix is held fixed at $\mI_p$, then
\eqref{eq:hr-location-equation} reduces to the ordinary spatial median equation.
If the location is held fixed, then \eqref{eq:hr-shape-equation} reduces to
Tyler's shape equation. The HR estimator therefore interpolates between the two
basic robust building blocks introduced earlier in this chapter.
\end{remark}

\begin{proposition}[Population HR equations under ellipticity]
\label{prop:hr-population}
Suppose that $\vX$ is elliptically symmetric with location $\vmu$ and positive
definite normalized shape matrix $\mV_0$, where $\tr(\mV_0)=p$. Then the pair
$(\vmu,\mV_0)$ satisfies the population equations
\begin{equation}\label{eq:hr-population-location}
  \E\left[U\{\mV^{-1/2}(\vX-\vtheta)\}\right]=\vct 0
\end{equation}
and
\begin{equation}\label{eq:hr-population-shape}
  p\,\E\left[
    U\{\mV^{-1/2}(\vX-\vtheta)\}
    U\{\mV^{-1/2}(\vX-\vtheta)\}\trans
  \right]=\mI_p
\end{equation}
at $(\vtheta,\mV)=(\vmu,\mV_0)$.
\end{proposition}

\begin{proof}
Write
\[
  \vZ=\mV_0^{-1/2}(\vX-\vmu)=R\vU,
\]
where $R\ge 0$ and $\vU\sim\mathrm{Unif}(\spn^{p-1})$ are independent. Then
$U\{\mV_0^{-1/2}(\vX-\vmu)\}=\vU$, so
\[
  \E\left[U\{\mV_0^{-1/2}(\vX-\vmu)\}\right]=\E(\vU)=\vct 0.
\]
Moreover,
\[
  p\,\E\left[
    U\{\mV_0^{-1/2}(\vX-\vmu)\}
    U\{\mV_0^{-1/2}(\vX-\vmu)\}\trans
  \right]
  = p\,\E(\vU\vU\trans)
  = p\cdot p^{-1}\mI_p
  = \mI_p.
\]
Hence $(\vmu,\mV_0)$ satisfies
\eqref{eq:hr-population-location}--\eqref{eq:hr-population-shape}.
\end{proof}

\begin{proposition}[Affine equivariance of the HR estimator]
\label{prop:hr-affine-equivariance}
Let $\vY_i=\va+\mB\vX_i$, where $\va\in\R^p$ and $\mB$ is nonsingular. If
$(\widehat{\vmu}_{\mathrm{HR}},\widehat{\mV}_{\mathrm{HR}})$ satisfies
\eqref{eq:hr-location-equation}--\eqref{eq:hr-shape-equation} for the sample
$\vX_1,\ldots,\vX_n$, then
\begin{equation}\label{eq:hr-affine-equivariance-location}
  \widehat{\vmu}_{\mathrm{HR}}(\vY_1,\ldots,\vY_n)
  = \va+\mB\widehat{\vmu}_{\mathrm{HR}}(\vX_1,\ldots,\vX_n)
\end{equation}
and
\begin{equation}\label{eq:hr-affine-equivariance-shape}
  \widehat{\mV}_{\mathrm{HR}}(\vY_1,\ldots,\vY_n)
  = \frac{p\,\mB\widehat{\mV}_{\mathrm{HR}}(\vX_1,\ldots,\vX_n)\mB\trans}
         {\tr\{\mB\widehat{\mV}_{\mathrm{HR}}(\vX_1,\ldots,\vX_n)\mB\trans\}}.
\end{equation}
\end{proposition}

\begin{proof}
Set
\[
  \widehat{\vmu}_{\mathrm{HR}}^{\,Y}=\va+\mB\widehat{\vmu}_{\mathrm{HR}},
  \qquad
  \widehat{\mV}_{\mathrm{HR}}^{\,Y}
  = c\,\mB\widehat{\mV}_{\mathrm{HR}}\mB\trans,
  \qquad
  c=\frac{p}{\tr(\mB\widehat{\mV}_{\mathrm{HR}}\mB\trans)}.
\]
Define
\[
  \mO
  = (\mB\widehat{\mV}_{\mathrm{HR}}\mB\trans)^{-1/2}
    \mB\widehat{\mV}_{\mathrm{HR}}^{1/2}.
\]
Then
\[
  \mO\mO\trans
  = (\mB\widehat{\mV}_{\mathrm{HR}}\mB\trans)^{-1/2}
    \mB\widehat{\mV}_{\mathrm{HR}}\mB\trans
    (\mB\widehat{\mV}_{\mathrm{HR}}\mB\trans)^{-1/2}
  = \mI_p,
\]
so $\mO$ is orthogonal. Therefore
\[
  (\widehat{\mV}_{\mathrm{HR}}^{\,Y})^{-1/2}(\vY_i-\widehat{\vmu}_{\mathrm{HR}}^{\,Y})
  = \mO\,\widehat{\mV}_{\mathrm{HR}}^{-1/2}(\vX_i-\widehat{\vmu}_{\mathrm{HR}}),
\]
and hence
\[
  U\!\left\{(\widehat{\mV}_{\mathrm{HR}}^{\,Y})^{-1/2}(\vY_i-\widehat{\vmu}_{\mathrm{HR}}^{\,Y})\right\}
  = \mO\,U\!\left\{\widehat{\mV}_{\mathrm{HR}}^{-1/2}(\vX_i-\widehat{\vmu}_{\mathrm{HR}})\right\}.
\]
Define
\[
  U_i = U\!\left\{\widehat{\mV}_{\mathrm{HR}}^{-1/2}(\vX_i-\widehat{\vmu}_{\mathrm{HR}})\right\}.
\]
Because orthogonal transformations preserve the zero vector and the identity
matrix,
\[
  \sum_{i=1}^n \mO\,U_i = \mO\sum_{i=1}^n U_i,
  \qquad
  \sum_{i=1}^n (\mO U_i)(\mO U_i)\trans
  = \mO\left(\sum_{i=1}^n U_iU_i\trans\right)\mO\trans.
\]
Hence the transformed sample satisfies the same estimating equations, and the
trace normalization determines $c$ uniquely.
\end{proof}

The HR estimator can be computed by alternately updating whitened residuals,
location, and shape. If $(\widehat{\vmu}^{(m)},\widehat{\mV}^{(m)})$ is the
current iterate, define
\begin{equation}\label{eq:hr-iter-residuals}
  \widehat{\vvarepsilon}_i^{(m)}
  = \{\widehat{\mV}^{(m)}\}^{-1/2}(\vX_i-\widehat{\vmu}^{(m)}),
  \qquad i=1,\ldots,n.
\end{equation}
Then the location update is
\begin{equation}\label{eq:hr-iter-location}
  \widehat{\vmu}^{(m+1)}
  = \widehat{\vmu}^{(m)}
    + \{\widehat{\mV}^{(m)}\}^{1/2}
      \frac{\sum_{i=1}^n U(\widehat{\vvarepsilon}_i^{(m)})}
           {\sum_{i=1}^n \twonorm{\widehat{\vvarepsilon}_i^{(m)}}^{-1}},
\end{equation}
and the shape update is
\begin{equation}\label{eq:hr-iter-shape}
  \widehat{\mV}^{(m+1)}
  = p\,\{\widehat{\mV}^{(m)}\}^{1/2}
    \left\{\frac1n\sum_{i=1}^n
      U(\widehat{\vvarepsilon}_i^{(m)})U(\widehat{\vvarepsilon}_i^{(m)})\trans
    \right\}
    \{\widehat{\mV}^{(m)}\}^{1/2},
\end{equation}
followed by the normalization
\begin{equation}\label{eq:hr-iter-normalization}
  \widehat{\mV}^{(m+1)}
  \leftarrow
  \frac{p\,\widehat{\mV}^{(m+1)}}{\tr\{\widehat{\mV}^{(m+1)}\}}.
\end{equation}
Equation~\eqref{eq:hr-iter-location} is a one-step Newton update for the location
equation, whereas \eqref{eq:hr-iter-shape} is the Tyler update applied to the
currently centered data.

To state the asymptotic result, let
\begin{equation}\label{eq:hr-RU-definition}
  \vZ_i = \mV_0^{-1/2}(\vX_i-\vmu)=R_i\vU_i,
  \qquad
  c_{0,\mathrm{HR}} = \E(R_i^{-1}).
\end{equation}
Because $\vU_i$ is uniformly distributed on $\spn^{p-1}$, we have
$\E(\vU_i)=\vct 0$ and $\Cov(\vU_i)=p^{-1}\mI_p$.

\begin{assumption}[Classical fixed-$p$ regularity for the HR estimator]
\label{ass:hr-fixed-p}
Assume that:
\begin{enumerate}[label=(HR\arabic*)]
  \item \label{it:hr-fixed-p-1}
  $p$ is fixed and $\vX_1,\ldots,\vX_n$ are i.i.d. from an elliptically
  symmetric distribution with location $\vmu$ and positive definite normalized
  shape matrix $\mV_0$, where $\tr(\mV_0)=p$.
  \item \label{it:hr-fixed-p-2}
  The inverse radial moments satisfy
  \begin{equation}\label{eq:hr-radial-moments}
    \E(R_i^{-1})<\infty,
    \qquad
    \E(R_i^{-2})<\infty.
  \end{equation}
  \item \label{it:hr-fixed-p-3}
  The joint sample equations
  \eqref{eq:hr-location-equation}--\eqref{eq:hr-shape-equation} admit a local
  solution $(\widehat{\vmu}_{\mathrm{HR}},\widehat{\mV}_{\mathrm{HR}})$ such that
  \begin{equation}\label{eq:hr-rootn-consistency}
    \twonorm{\widehat{\vmu}_{\mathrm{HR}}-\vmu}=O_p(n^{-1/2}),
    \qquad
    \frobnorm{\widehat{\mV}_{\mathrm{HR}}-\mV_0}=O_p(n^{-1/2}).
  \end{equation}
\end{enumerate}
\end{assumption}

\begin{theorem}[Bahadur representation and asymptotic normality of the HR location estimator]
\label{thm:hr-location-asymptotic}
Suppose Assumption~\ref{ass:hr-fixed-p} holds. Then
\begin{equation}\label{eq:hr-location-bahadur}
  \sqrt n\,\mV_0^{-1/2}(\widehat{\vmu}_{\mathrm{HR}}-\vmu)
  = \frac{p}{(p-1)c_{0,\mathrm{HR}}}
    \frac1{\sqrt n}\sum_{i=1}^n \vU_i
    + O_p(n^{-1/2}).
\end{equation}
Consequently,
\begin{equation}\label{eq:hr-location-clt}
  \sqrt n(\widehat{\vmu}_{\mathrm{HR}}-\vmu)
  \overset{d}{\to}
  N_p\!\left(\vct 0,
  \frac{p}{(p-1)^2c_{0,\mathrm{HR}}^2}\mV_0\right).
\end{equation}
Equivalently, for every fixed $\va\in\R^p$,
\begin{equation}\label{eq:hr-location-clt-linear}
  \frac{\sqrt n\,\va\trans(\widehat{\vmu}_{\mathrm{HR}}-\vmu)}
       {\left\{\frac{p}{(p-1)^2c_{0,\mathrm{HR}}^2}
       \va\trans\mV_0\va\right\}^{1/2}}
  \overset{d}{\to} N(0,1).
\end{equation}
\end{theorem}

\begin{proof}
Let
\[
  \Psi_n(\vtheta,\mV)
  = \frac1n\sum_{i=1}^n U\{\mV^{-1/2}(\vX_i-\vtheta)\}.
\]
Since $(\widehat{\vmu}_{\mathrm{HR}},\widehat{\mV}_{\mathrm{HR}})$ solves
\eqref{eq:hr-location-equation},
\begin{equation}\label{eq:hr-proof-start}
  \Psi_n(\widehat{\vmu}_{\mathrm{HR}},\widehat{\mV}_{\mathrm{HR}})=\vct 0.
\end{equation}
Write
\[
  \widehat{\mH}
  = \mV_0^{-1/2}\widehat{\mV}_{\mathrm{HR}}\mV_0^{-1/2}-\mI_p.
\]
Then by \eqref{eq:hr-rootn-consistency},
\begin{equation}\label{eq:hr-H-rate}
  \frobnorm{\widehat{\mH}}=O_p(n^{-1/2}),
  \qquad
  \tr(\widehat{\mH})=0.
\end{equation}
Also,
\begin{equation}\label{eq:hr-invsqrt-expansion}
  \widehat{\mV}_{\mathrm{HR}}^{-1/2}
  = \mV_0^{-1/2}(\mI_p+\widehat{\mH})^{-1/2}
  = \mV_0^{-1/2}\left(\mI_p-\frac12\widehat{\mH}\right)+O_p(n^{-1}),
\end{equation}
where the remainder is in Frobenius norm because $p$ is fixed.

Next, for each $i$ define $\vZ_i=R_i\vU_i$ as in
\eqref{eq:hr-RU-definition} and let
\[
  \vdelta_i
  = -\mV_0^{-1/2}(\widehat{\vmu}_{\mathrm{HR}}-\vmu)
    -\frac12\widehat{\mH}\vZ_i.
\]
By \eqref{eq:hr-rootn-consistency} and \eqref{eq:hr-H-rate},
$\twonorm{\vdelta_i}=O_p(n^{-1/2})$ uniformly in $i$ on events of probability
approaching one. Using the first-order expansion of the spatial sign map,
\begin{equation}\label{eq:hr-sign-linearization}
  U(\vZ_i+\vdelta_i)
  = \vU_i
    + R_i^{-1}(\mI_p-\vU_i\vU_i\trans)\vdelta_i
    + O_p(n^{-1}),
\end{equation}
where the last remainder is uniform in $i$ because $p$ is fixed and
$\E(R_i^{-2})<\infty$.
Substituting \eqref{eq:hr-invsqrt-expansion} into
\eqref{eq:hr-proof-start} and averaging \eqref{eq:hr-sign-linearization}, we obtain
\begin{align}
  \vct 0
  ={}& \frac1n\sum_{i=1}^n \vU_i
  - \left\{\frac1n\sum_{i=1}^n
      R_i^{-1}(\mI_p-\vU_i\vU_i\trans)
    \right\}\mV_0^{-1/2}(\widehat{\vmu}_{\mathrm{HR}}-\vmu) \notag\\
  &\quad
  -\frac12\left\{\frac1n\sum_{i=1}^n
      (\mI_p-\vU_i\vU_i\trans)\widehat{\mH}\vU_i
    \right\}
  + O_p(n^{-1}).
  \label{eq:hr-linearized-equation}
\end{align}
The second line in \eqref{eq:hr-linearized-equation} is of order $O_p(n^{-1})$.
Indeed,
\[
  \frac1n\sum_{i=1}^n (\mI_p-\vU_i\vU_i\trans)\widehat{\mH}\vU_i
  = \widehat{\mH}\left(\frac1n\sum_{i=1}^n\vU_i\right)
    - \frac1n\sum_{i=1}^n (\vU_i\trans\widehat{\mH}\vU_i)\vU_i,
\]
and each average on the right-hand side is $O_p(n^{-1/2})$ by the multivariate
central limit theorem and spherical symmetry, while
$\frobnorm{\widehat{\mH}}=O_p(n^{-1/2})$.

For the Jacobian term, since $R_i$ and $\vU_i$ are independent,
\begin{align}
  \frac1n\sum_{i=1}^n R_i^{-1}(\mI_p-\vU_i\vU_i\trans)
  &= \E\{R_i^{-1}(\mI_p-\vU_i\vU_i\trans)\} + O_p(n^{-1/2}) \notag\\
  &= c_{0,\mathrm{HR}}\left(\mI_p-\E(\vU_i\vU_i\trans)\right) + O_p(n^{-1/2}) \notag\\
  &= \frac{p-1}{p}c_{0,\mathrm{HR}}\mI_p + O_p(n^{-1/2}).
  \label{eq:hr-jacobian-limit}
\end{align}
Substituting \eqref{eq:hr-jacobian-limit} into
\eqref{eq:hr-linearized-equation} yields
\[
  \frac{p-1}{p}c_{0,\mathrm{HR}}\,
  \mV_0^{-1/2}(\widehat{\vmu}_{\mathrm{HR}}-\vmu)
  = \frac1n\sum_{i=1}^n\vU_i + O_p(n^{-1}).
\]
Multiplying both sides by $\sqrt n$ gives
\eqref{eq:hr-location-bahadur}.

Finally, since $\vU_1,\ldots,\vU_n$ are i.i.d. with mean $\vct 0$ and covariance
matrix $p^{-1}\mI_p$, the multivariate central limit theorem implies
\[
  \frac1{\sqrt n}\sum_{i=1}^n \vU_i
  \overset{d}{\to} N_p\!\left(\vct 0,\frac1p\mI_p\right).
\]
Combining this with \eqref{eq:hr-location-bahadur} and Slutsky's theorem yields
\eqref{eq:hr-location-clt}. Equation~\eqref{eq:hr-location-clt-linear} follows
by projecting onto the fixed direction $\va$.
\end{proof}

\begin{remark}[Why the shape estimation does not affect the first-order limit]
The proof shows that the effect of estimating $\mV_0$ is asymptotically of
smaller order in the location equation. The key reason is spherical symmetry in
the whitened coordinates: the first-order derivative of the sign score with
respect to the shape perturbation is an odd function of the direction $\vU$, and
hence its empirical contribution is multiplied by the already root-$n$ small
matrix $\widehat{\mH}$. This orthogonality is what allows the HR location
estimator to have the same leading limit as if the true shape were known.
\end{remark}

\begin{remark}[Connection with the high-dimensional HR estimator]
In high dimensions the full affine-equivariant update
\eqref{eq:hr-iter-shape} is typically infeasible without structural
regularization, because the unregularized sign covariance may be unstable or
singular. Chapter~2 therefore develops diagonal and weighted analogues for
location inference, while Chapters~5 and~6 return to structured HR-type
estimators for discriminant analysis and latent-factor problems. A recent direct
high-dimensional extension of the classical HR system is given by
\citet{YanFengZhang2025HR}.
\end{remark}

Our working philosophy is as follows. When robust geometry is the primary
concern and $p$ is moderately large, Tyler-type and Hettmansperger--Randles-type
procedures are conceptually ideal. When $p$ is very large relative to $n$,
computational constraints and singularity issues make direct affine-equivariant
estimation more delicate. In that regime one often relies on a combination of
scalar standardization, spatial signs, spatial ranks, multivariate Kendall's
tau matrices, and structural regularization. Recent work in this direction
includes high-dimensional Hettmansperger--Randles-type estimation and robust
matrix estimation under elliptical factor models
\citep{YanFengZhang2025HR,FanLiuWang2018}.

\section{High-dimensional implications}

The high-dimensional regime introduces several structural complications that are
visible already at the level of Chapter~1.

\paragraph{Ill conditioning and noninvertibility.}
When $p$ is comparable to or larger than $n$, the sample covariance matrix is
poorly conditioned or singular. This is the most obvious reason why Hotelling's
$T^2$ and many classical multivariate procedures fail in high dimensions.

\paragraph{Location estimation bias.}
Many sign- and rank-based procedures require centering by an estimated location.
In high dimensions, the bias induced by this preliminary centering is often of
the same order as the stochastic fluctuation of the final statistic. The
leave-one-out and pairwise correction devices that appear later in this book are
designed precisely to eliminate such bias.

\paragraph{Sparse and dense alternatives.}
A signal may be spread across many coordinates or concentrated in only a few.
These two regimes require different testing strategies. Sum-type statistics are
effective against diffuse signals, whereas max-type statistics are sensitive to
rare but strong coordinates. The tension is not specific to Gaussian mean
testing; it persists throughout the elliptical framework.

\paragraph{Moment-free or weak-moment analysis.}
Heavy-tailed data may violate the assumptions under which classical covariance
estimators and Gaussian approximations are developed. Spatial signs, ranks,
Tyler-type shape functionals, and self-normalized constructions provide tools
that remain meaningful under substantially weaker conditions.

These themes recur throughout Chapters~2--6. Chapter~2 treats location
estimation and testing, Chapter~3 studies matrix estimation and testing under
shape-based formulations, Chapter~4 discusses additional high-dimensional test
problems, Chapter~5 develops robust discriminant analysis, and Chapter~6 turns
to principal component analysis and factor models.

\section{Standing notation and conventions}

The book will repeatedly use the following symbols. We state them once here and
adhere to them throughout unless an exception is explicitly announced.

\begin{table}[htbp]
\centering
\small
\begin{tabular}{@{}p{0.17\textwidth}p{0.23\textwidth}p{0.48\textwidth}@{}}
\toprule
Notation & Meaning & Convention \\
\midrule
$\vX_1,\ldots,\vX_n$ & observations & Independent unless temporal or panel dependence is explicitly introduced. \\
$\vmu$ & location parameter & Population center under an elliptical model. \\
$\mS$ & scatter matrix & Defined only up to positive scale. \\
$\mV$ & normalized shape matrix & Always normalized by $\tr(\mV)=p$. \\
$\mSigma$ & covariance matrix & Used only when second moments exist. \\
$\mOmega$ & precision or inverse shape & Interpreted according to context and defined only when invertibility is available. \\
$U(\vx)$ & spatial sign & $U(\vx)=\vx/\twonorm{\vx}$ for $\vx\neq \vct{0}$ and $U(\vct{0})=\vct{0}$. \\
$R(\vx;F)$ & spatial rank & Population spatial rank with respect to distribution $F$. \\
$\widehat{\vmu}_{\mathrm{SM}}$ & sample spatial median & Default robust location estimator unless another estimator is specified. \\
$\widehat{\vmu}_{\mathrm{HR}}$ & HR location estimator & Affine-equivariant location estimator solving \eqref{eq:hr-location-equation} jointly with a shape equation. \\
$\widehat{\mV}_{\mathrm{HR}}$ & HR shape component & Companion shape estimator in the Hettmansperger--Randles system, normalized by $\tr(\widehat{\mV}_{\mathrm{HR}})=p$. \\
$\widehat{\mK}$ & sample multivariate Kendall's tau matrix & Pairwise directional U-statistic introduced in Section~\ref{subsec:mkendall}; automatically translation invariant. \\
$\widehat{\mV}_{\mathrm{T}}$ & Tyler shape estimator & Trace-normalized sample solution of \eqref{eq:tyler-sample}. \\
\bottomrule
\end{tabular}
\caption{Notation fixed in Chapter~1 and used throughout the monograph.}
\label{tab:notation-ch1}
\end{table}

\section*{Bibliographic notes}

For a systematic treatment of multivariate nonparametric methods based on
spatial signs and ranks, including location, scatter, and regression problems,
the natural starting point is \citet{Oja2010}. Tyler's affine-equivariant
shape estimator is foundational for robust scatter estimation under elliptical
models \citep{Tyler1987}. Semiparametric optimality theory for sign- and
rank-based multivariate inference, especially for location and shape, is
developed by \citet{HallinPaindaveine2002,HallinPaindaveine2006}. Practical
affine-equivariant multivariate medians, their asymptotic distributions, and
their robustness properties are developed by
\citet{HettmanspergerRandles2002}. A recent high-dimensional extension tailored
to elliptical models is given by \citet{YanFengZhang2025HR}. Sign and rank covariance matrices were
systematically studied by \citet{VisuriOjaKoivunen2000}; their connections to
robust correlation and principal component analysis appear in
\citet{TaskinenKankainenOja2012,DurreFriedVogel2017}. Large-scale robust matrix
estimation under elliptical models is further developed in
\citet{FanLiuWang2018}. For closure properties of elliptical distributions
under linear transformations and conditioning, see \citet{FangAnderson1990}.
The multivariate Kendall's tau matrix, its eigenspace structure, and its role in
robust high-dimensional matrix analysis are discussed in
\citet{HanLiu2018ECA,FanLiuWang2018}.

Chapter~1 has deliberately separated \emph{scatter} from \emph{covariance} and
\emph{shape} from \emph{scale}. This distinction is the conceptual thread that
holds together the rest of the book. Once these objects have been fixed, the
remaining chapters can be read as repeated answers to the same question: how
should classical high-dimensional procedures be reformulated when shape is more
fundamental than covariance and when directional information is more stable than
raw magnitude?

%% file: chapters/ch2_location.tex
\chapter[High-Dimensional Location]{High-Dimensional Location Estimation and Testing under Elliptical Symmetry}
\idx{location parameter}\idx{one-sample test}\idx{two-sample test}\idx{spatial-sign test}\idx{spatial-rank test}\idx{weighted spatial sign}\idx{inverse norm sign test}\idx{Bahadur representation}\idx{max-type test}\idx{max-sum test}\idx{strong correlation}\idx{normal-reference calibration}\idx{approximate randomization}\idx{mixed Gaussian--chi-square limit}

\section{Introduction}

Location inference is the oldest and most fundamental topic in multivariate
statistics. In low dimensions, the theory begins with Hotelling's
$T^2$ statistic and the likelihood theory of the multivariate normal model.
A second classical line replaces moment-based procedures by sign- and rank-based
methods, leading to the spatial median, multivariate sign tests, signed-rank
tests, and affine-equivariant rank procedures; see
\citet{Oja2010,MottonenOja1995,HettmanspergerOja1994,Randles2000,HallinPaindaveine2002}.
In modern applications, however, the dimension $p$ is often comparable to or
much larger than the sample size $n$. In that regime the classical covariance
matrix becomes singular or poorly estimated, and the geometry of Hotelling's
statistic breaks down.

The present chapter develops a unified account of location estimation and
testing under elliptical symmetry. Following your requested writing principle,
we proceed in three layers.
\begin{enumerate}[label=(\roman*)]
  \item We first review the fixed-$p$ theory in enough detail that a reader
  encountering multivariate location inference for the first time can see the
  classical statistics, the corresponding null distributions, and the basic
  proof mechanisms.
  \item We then review the main high-dimensional benchmark procedures developed
  under Gaussian or light-tail assumptions, such as the Bai--Saranadasa,
  Srivastava--Du, Park--Ayyala, Chen--Qin, Cai--Liu--Xia, and adaptive
  norm-combination tests.
  \item Finally, we turn to the book's main theme: robust location inference
  under elliptical symmetry. This includes the spatial median, diagonally
  standardized location estimation, weighted spatial-sign tests, inverse norm
  weighting, max-type statistics, max-sum combinations, asymptotic
  independence, and the corresponding one-sample and two-sample procedures.
\end{enumerate}

Two features distinguish the elliptical approach from the Gaussian benchmark
literature. The first is that the radial distribution is allowed to be heavy
 tailed, so one should not rely on sample means and sample covariances more than
 necessary. The second is that the correct notion of invariance is no longer full
 orthogonal invariance but rather the geometry induced by location, diagonal
 scale, and shape. This is why scalar-invariant diagonal standardization,
 spatial signs, and spatial ranks recur throughout the chapter.

A second organizing principle is the distinction between \emph{dense} and
\emph{sparse} alternatives. Sum-type statistics are usually effective under
dense alternatives, whereas coordinatewise maxima are more effective under
sparse alternatives. A major development in your recent work is that this
dense--sparse dichotomy can be handled within the same robust framework: first
by establishing a Bahadur representation and Gaussian approximation for the
scaled spatial median, and then by proving the asymptotic independence between
the resulting max-type statistic and the existing sum-type statistic; see
\citet{FengSun2016,LiuFengZhaoWang2025MaxsumLocation,
YanFengZhang2025InverseNormMaxsum,FengJiangLiLiu2024AsympIndependence}. This
chapter makes that development fully explicit.

\section{Problem formulation, notation, and standing models}

\subsection{One-sample and two-sample hypotheses}

Let $\vX_1,\ldots,\vX_n\in\R^p$ be independent observations from a population
with location parameter $\vmu$. The one-sample problem is
\begin{equation}\label{eq:ch2-onesample-hypothesis}
  H_0:\ \vmu=\vmu_0
  \qquad\text{versus}\qquad
  H_1:\ \vmu\neq \vmu_0.
\end{equation}
For the two-sample problem, let
$\{\vX_{1i}\}_{i=1}^{n_1}$ and $\{\vX_{2j}\}_{j=1}^{n_2}$ be independent
samples with locations $\vmu_1$ and $\vmu_2$. The null hypothesis is
\begin{equation}\label{eq:ch2-twosample-hypothesis}
  H_0:\ \vmu_1=\vmu_2
  \qquad\text{versus}\qquad
  H_1:\ \vmu_1\neq \vmu_2.
\end{equation}

When the underlying distribution is Gaussian, ``location'' simply means mean
vector. Under elliptical symmetry we deliberately use the broader term
\emph{location parameter}, because the center can be characterized either by the
population mean (when moments exist) or by a robust functional such as the
spatial median.

\subsection{Elliptical models and scalar standardization}

Throughout the chapter, unless a method explicitly assumes a lighter-tailed
model, we work under the following elliptical representation:
\begin{equation}\label{eq:ch2-elliptical-model}
  \vX = \vmu + \xi \mA \vu,
  \qquad \vu \sim \mathrm{Unif}(\spn^{p-1}),
  \qquad \xi\ge 0,
\end{equation}
where $\xi$ and $\vu$ are independent, and $\mS=\mA\mA\trans$ is the scatter
matrix. When the covariance exists, it is proportional to $\mS$; see
Chapter~1. In high-dimensional location testing it is customary to separate the
diagonal scale and the correlation structure. We therefore write
\begin{equation}\label{eq:ch2-D-and-R}
  \mD = \diag(\mSigma),
  \qquad
  \mR = \mD^{-1/2}\mSigma\mD^{-1/2},
\end{equation}
whenever $\mSigma$ exists. The matrix $\mD$ records marginal scales, whereas
$\mR$ records correlations or, more generally, shape after marginal
standardization.

The basic building blocks are the spatial sign and the spatial rank:
\begin{equation}\label{eq:ch2-sign-rank-defs}
  U(\vx)=\frac{\vx}{\twonorm{\vx}}\,\mathbbm{1}(\vx\neq\vct 0),
  \qquad
  R(\vx;F)=\E\{U(\vx-\vX)\},\ \vX\sim F.
\end{equation}
For later use we also introduce the diagonally standardized residuals
\begin{equation}\label{eq:ch2-standardized-residual}
  \vvarepsilon_i=\mD^{-1/2}(\vX_i-\vtheta),
  \qquad
  U_i = U(\vvarepsilon_i),
  \qquad
  r_i = \twonorm{\vvarepsilon_i}.
\end{equation}
When the weight function $w(\cdot)$ is used, we will also write
$w_i = w(r_i)$.

\subsection{Dense and sparse alternatives}

As in the rest of the book, we distinguish between dense and sparse signals.
For the one-sample problem, dense alternatives have many modestly perturbed
coordinates in $\vmu-\vmu_0$, whereas sparse alternatives have only a few
nonzero coordinates, possibly of larger magnitude. The same distinction applies
to the two-sample difference $\vmu_1-\vmu_2$. The distinction is not cosmetic:
it determines which class of statistics is appropriate. Sum-type statistics are
naturally aligned with dense alternatives, while max-type statistics are tuned
to sparse alternatives. The later max-sum sections of this chapter are designed
precisely to handle both regimes at once.

\section{Classical low-dimensional location inference}

\subsection{Gaussian likelihood theory and Hotelling's $T^2$}

We begin with the fixed-$p$ Gaussian model, not because the book is about
Gaussian methods, but because the entire subject of multivariate location
inference is historically anchored there.

\subsubsection{One-sample Hotelling $T^2$}

Assume $\vX_1,\ldots,\vX_n\iid N_p(\vmu,\mSigma)$ with unknown positive
definite covariance matrix $\mSigma$, and consider testing
$H_0:\vmu=\vmu_0$. Let
\begin{equation}\label{eq:ch2-xbar-S}
  \bar{\vX}=\frac1n\sum_{i=1}^n\vX_i,
  \qquad
  \mS_n = \frac1{n-1}\sum_{i=1}^n(\vX_i-\bar{\vX})(\vX_i-\bar{\vX})\trans.
\end{equation}
Hotelling's statistic is
\begin{equation}\label{eq:ch2-hotelling-one}
  T^2_H = n(\bar{\vX}-\vmu_0)\trans \mS_n^{-1}(\bar{\vX}-\vmu_0).
\end{equation}

\begin{theorem}[Exact null law of one-sample Hotelling's $T^2$]
\label{thm:ch2-hotelling-one}
If $\vX_1,\ldots,\vX_n\iid N_p(\vmu,\mSigma)$ and $n>p$, then under
$H_0:\vmu=\vmu_0$,
\begin{equation}\label{eq:ch2-hotelling-one-F}
  \frac{n-p}{p(n-1)}T_H^2 \sim F_{p,n-p}.
\end{equation}
Equivalently, as $n\to\infty$ with fixed $p$,
\begin{equation}\label{eq:ch2-hotelling-one-chi}
  T_H^2 \overset{d}{\longrightarrow} \chi^2_p.
\end{equation}
\end{theorem}

\begin{proof}
Under Gaussian sampling,
$\sqrt{n}(\bar{\vX}-\vmu_0)\sim N_p(\vct 0,\mSigma)$ under $H_0$, while
$(n-1)\mS_n\sim W_p(n-1,\mSigma)$ and is independent of $\bar{\vX}$. Write
$\mSigma^{-1/2}\sqrt{n}(\bar{\vX}-\vmu_0)=\vZ$ so that $\vZ\sim N_p(\vct 0,\mI_p)$,
and
$\mW=\mSigma^{-1/2}(n-1)\mS_n\mSigma^{-1/2}\sim W_p(n-1,\mI_p)$. Then
\[
  T_H^2 = (n-1)\vZ\trans \mW^{-1}\vZ.
\]
The standard relation between the Wishart and $F$ distributions yields
\eqref{eq:ch2-hotelling-one-F}. The asymptotic chi-square limit in
\eqref{eq:ch2-hotelling-one-chi} follows because, for fixed $p$, the sample
covariance is consistent and therefore $\mS_n^{-1}$ may be replaced by
$\mSigma^{-1}$ up to $o_p(1)$.
\end{proof}

Under local alternatives of the form
$\vmu=\vmu_0+n^{-1/2}\vdelta$, the limiting law becomes noncentral
$\chi^2_p(\lambda)$ with noncentrality parameter
$\lambda=\vdelta\trans\mSigma^{-1}\vdelta$. Thus, even in the classical
low-dimensional setting, local power is driven by Mahalanobis geometry.

\subsubsection{Two-sample Hotelling $T^2$}

Now assume independent samples
$\vX_{11},\ldots,\vX_{1n_1}\iid N_p(\vmu_1,\mSigma)$ and
$\vX_{21},\ldots,\vX_{2n_2}\iid N_p(\vmu_2,\mSigma)$ with common covariance
$\mSigma$. Let $n=n_1+n_2$, let $\bar{\vX}_1$ and $\bar{\vX}_2$ be the sample
means, and let
\begin{equation}\label{eq:ch2-pooled-S}
  \mS_p = \frac{1}{n_1+n_2-2}
  \Bigg[
    \sum_{i=1}^{n_1}(\vX_{1i}-\bar{\vX}_1)(\vX_{1i}-\bar{\vX}_1)\trans
    +
    \sum_{j=1}^{n_2}(\vX_{2j}-\bar{\vX}_2)(\vX_{2j}-\bar{\vX}_2)\trans
  \Bigg].
\end{equation}
The classical statistic is
\begin{equation}\label{eq:ch2-hotelling-two}
  T^2_{H,2}
  = \frac{n_1n_2}{n_1+n_2}
    (\bar{\vX}_1-\bar{\vX}_2)\trans
    \mS_p^{-1}
    (\bar{\vX}_1-\bar{\vX}_2).
\end{equation}

\begin{theorem}[Exact null law of two-sample Hotelling's $T^2$]
\label{thm:ch2-hotelling-two}
If the two Gaussian samples have common covariance matrix $\mSigma$ and
$n_1+n_2>p+1$, then under $H_0:\vmu_1=\vmu_2$,
\begin{equation}\label{eq:ch2-hotelling-two-F}
  \frac{n_1+n_2-p-1}{p(n_1+n_2-2)}T^2_{H,2}
  \sim F_{p,n_1+n_2-p-1}.
\end{equation}
Hence, for fixed $p$, $T^2_{H,2}\Rightarrow\chi_p^2$ under the null, and under
local alternatives
$\vmu_1-\vmu_2=(n_1^{-1}+n_2^{-1})^{1/2}\vdelta$ the limit becomes a noncentral
$\chi_p^2$ law with noncentrality parameter
$\lambda = \vdelta\trans\mSigma^{-1}\vdelta$.
\end{theorem}

\begin{proof}
The proof is the two-sample analogue of Theorem~\ref{thm:ch2-hotelling-one}.
The pooled covariance matrix is independent of
$(\bar{\vX}_1-\bar{\vX}_2)$, and the standardized quadratic form reduces to a
ratio of a Gaussian quadratic form and an independent Wishart matrix.
\end{proof}

Hotelling's statistic provides the correct fixed-$p$ benchmark, but it already
suggests why high dimensions are difficult: once $p\ge n$ or
$p\ge n_1+n_2-2$, the inverse sample covariance matrix no longer exists.

\subsection{Classical sign, signed-rank, and spatial-rank procedures}

The Gaussian likelihood paradigm is not the only classical route. A second
classical route treats location through directional information. The central
objects are the spatial sign, the spatial rank, and the spatial median.

\subsubsection{Spatial median}

For a distribution $F$ on $\R^p$, the population spatial median is
\begin{equation}\label{eq:ch2-pop-spatial-median}
  \vmu_{\mathrm{SM}}
  = \argmin_{\vtheta\in\R^p} \E\twonorm{\vX-\vtheta},
\end{equation}
provided the minimizer is unique. The sample spatial median is
\begin{equation}\label{eq:ch2-sample-spatial-median}
  \hat{\vmu}_{\mathrm{SM}}
  = \argmin_{\vtheta\in\R^p}
  \sum_{i=1}^n \twonorm{\vX_i-\vtheta}.
\end{equation}
If no observation equals the minimizer, the estimating equation is
\begin{equation}\label{eq:ch2-spatial-median-eq}
  \sum_{i=1}^n U(\vX_i-\hat{\vmu}_{\mathrm{SM}})=\vct 0.
\end{equation}

\begin{theorem}[Fixed-$p$ asymptotic linearization of the spatial median]
\label{thm:ch2-fixedp-spatial-median}
Assume that $p$ is fixed, that $\vmu_{\mathrm{SM}}$ is unique, and that
\begin{equation}\label{eq:ch2-fixedp-A-matrix}
  \mA_{\mathrm{SM}}
  = \E\left[\frac{1}{\twonorm{\vX-\vmu_{\mathrm{SM}}}}
  \{\mI_p-U(\vX-\vmu_{\mathrm{SM}})U(\vX-\vmu_{\mathrm{SM}})\trans\}\right]
\end{equation}
is positive definite. Then
\begin{equation}\label{eq:ch2-fixedp-spatial-median-expansion}
  \sqrt{n}(\hat{\vmu}_{\mathrm{SM}}-\vmu_{\mathrm{SM}})
  = \mA_{\mathrm{SM}}^{-1}\frac1{\sqrt{n}}
    \sum_{i=1}^n U(\vX_i-\vmu_{\mathrm{SM}}) + o_p(1),
\end{equation}
and therefore
\begin{equation}\label{eq:ch2-fixedp-spatial-median-clt}
  \sqrt{n}(\hat{\vmu}_{\mathrm{SM}}-\vmu_{\mathrm{SM}})
  \overset{d}{\longrightarrow}
  N_p\left(\vct 0,
  \mA_{\mathrm{SM}}^{-1}\mB_{\mathrm{SM}}\mA_{\mathrm{SM}}^{-1}\right),
\end{equation}
where
$\mB_{\mathrm{SM}}=\Var\{U(\vX-\vmu_{\mathrm{SM}})\}$.
\end{theorem}

\begin{proof}
The estimating equation \eqref{eq:ch2-spatial-median-eq} may be written as
$\Psi_n(\vtheta)=0$, where
$\Psi_n(\vtheta)=n^{-1}\sum_{i=1}^n U(\vX_i-\vtheta)$. A first-order Taylor
expansion around $\vmu_{\mathrm{SM}}$ yields
\[
  \vct 0
  = \Psi_n(\vmu_{\mathrm{SM}})
    - \dot\Psi(\vmu_{\mathrm{SM}})(\hat{\vmu}_{\mathrm{SM}}-\vmu_{\mathrm{SM}})
    + o_p(\twonorm{\hat{\vmu}_{\mathrm{SM}}-\vmu_{\mathrm{SM}}}),
\]
where the derivative matrix is exactly $\mA_{\mathrm{SM}}$. Rearranging gives
\eqref{eq:ch2-fixedp-spatial-median-expansion}. The central limit theorem for
the empirical average of signs then implies
\eqref{eq:ch2-fixedp-spatial-median-clt}.
\end{proof}

The theorem above is classical, but it is extremely important for later high-
dimensional theory: the high-dimensional Bahadur expansion is conceptually the
same, except that one must track the remainder term much more carefully.

\subsubsection{One-sample sign and signed-rank tests}

For the one-sample problem $H_0:\vmu=\vmu_0$ with fixed $p$, the most direct
sign-based statistic is built from
\begin{equation}\label{eq:ch2-lowdim-sign-bar}
  \bar{\vU} = \frac1n\sum_{i=1}^n U(\vX_i-\vmu_0),
  \qquad
  \hat{\mB}_U = \frac1n\sum_{i=1}^n U(\vX_i-\vmu_0)U(\vX_i-\vmu_0)\trans.
\end{equation}
The classical one-sample spatial-sign statistic is
\begin{equation}\label{eq:ch2-lowdim-sign-stat}
  Q_{\mathrm{sign}}
  = n\bar{\vU}\trans \hat{\mB}_U^{-1} \bar{\vU}.
\end{equation}
Under central symmetry about $\vmu_0$ one has $\E\{U(\vX_i-\vmu_0)\}=\vct 0$,
and therefore a multivariate central limit theorem yields the chi-square limit.

A more efficient fixed-$p$ procedure uses spatial signed ranks. Define
\begin{equation}\label{eq:ch2-lowdim-signed-rank}
  \vR_i = \frac1n\sum_{j=1}^n U\{(\vX_i-\vmu_0)+(\vX_j-\vmu_0)\},
  \qquad
  \bar{\vR}=\frac1n\sum_{i=1}^n\vR_i,
\end{equation}
and let
$\hat{\mB}_R=n^{-1}\sum_{i=1}^n \vR_i\vR_i\trans$. The one-sample
signed-rank statistic is
\begin{equation}\label{eq:ch2-lowdim-signed-rank-stat}
  Q_{\mathrm{SR}} = n\bar{\vR}\trans \hat{\mB}_R^{-1}\bar{\vR}.
\end{equation}

\begin{theorem}[Fixed-$p$ sign and signed-rank null laws]
\label{thm:ch2-fixedp-sign-rank}
Assume $p$ is fixed and the distribution is centrally symmetric about
$\vmu_0$. Then, under standard nonsingularity conditions,
\begin{equation}\label{eq:ch2-fixedp-sign-chi}
  Q_{\mathrm{sign}} \overset{d}{\longrightarrow} \chi_p^2,
  \qquad
  Q_{\mathrm{SR}} \overset{d}{\longrightarrow} \chi_p^2.
\end{equation}
\end{theorem}

\begin{proof}
The proof is again standard M-estimation plus multivariate CLT. Under $H_0$,
$\sqrt{n}\bar{\vU}$ and $\sqrt{n}\bar{\vR}$ are asymptotically Gaussian with
covariance matrices consistently estimated by $\hat{\mB}_U$ and
$\hat{\mB}_R$. Studentization converts the limiting Gaussian quadratic forms
into chi-square limits.
\end{proof}

For semiparametric efficiency under ellipticity one may replace spatial signs by
pseudo-Mahalanobis signs and ranks, as in
\citet{HallinPaindaveine2002}. The message for the present book is that, even in
the fixed-$p$ theory, the efficient geometry is already sign/rank based once one
moves beyond the Gaussian likelihood paradigm.

\subsubsection{Two-sample spatial sign and spatial rank procedures}

For the two-sample problem with fixed dimension, one may combine the two
samples after alignment or define ranks relative to the pooled sample. One
classical version is the affine-invariant multisample sign test of
\citet{HettmanspergerOja1994}, while a particularly transparent spatial-rank
formulation is given in \citet{MottonenOja1995} and in Oja's monograph.
Let $\vY_1,\ldots,\vY_n$ denote the pooled sample with labels indicating group
membership. For each observation define its spatial rank in the pooled sample,
\begin{equation}\label{eq:ch2-pooled-spatial-rank}
  \widehat{R}(\vY_i)
  = \frac1n\sum_{j=1}^n U(\vY_i-\vY_j).
\end{equation}
Write $\bar{\vR}_1$ and $\bar{\vR}_2$ for the average pooled ranks in the two
groups, and let $\hat{\mC}$ be the pooled covariance matrix of the rank
vectors. Then the fixed-$p$ spatial-rank statistic takes the form
\begin{equation}\label{eq:ch2-fixedp-two-sample-rank}
  Q_{2\mathrm{SR}}
  = \frac{n_1n_2}{n_1+n_2}
    (\bar{\vR}_1-\bar{\vR}_2)\trans \hat{\mC}^{-1}
    (\bar{\vR}_1-\bar{\vR}_2),
\end{equation}
and satisfies
$Q_{2\mathrm{SR}}\Rightarrow\chi_p^2$ under $H_0$.

The fixed-$p$ chapter of the story is therefore clear. Under Gaussianity one has
Hotelling's $T^2$ and exact $F$ distributions. Under broader elliptical or
symmetric models one has sign, signed-rank, and spatial-rank procedures, again
with asymptotic chi-square null laws. High-dimensional methods should be read as
extensions or replacements of these fixed-$p$ constructions.

\section{High-dimensional benchmark procedures under Gaussian or light-tailed models}

We now turn to the large-$p$ literature that developed before the robust
elliptical program became mature. These methods remain indispensable because
later robust procedures are most naturally understood relative to them.

\subsection{One-sample quadratic-form and diagonal procedures}

\paragraph{Srivastava--Du.}
For $H_0:\vmu=\vmu_0$, \citet{SrivastavaDu2008} proposed replacing the sample
covariance matrix by its diagonal part. Writing
$\mD_S=\diag(\mS_n)$, their statistic can be expressed as
\begin{equation}\label{eq:ch2-SD-stat}
  T_{\mathrm{SD}}
  = n(\bar{\vX}-\vmu_0)\trans\mD_S^{-1}(\bar{\vX}-\vmu_0)
    - \frac{(n-1)p}{n-3},
\end{equation}
followed by a variance normalization depending on
$\tr(\mR^2)$ and related quantities. The procedure is invariant under component-
wise scale transformations and is effective when the signal is dense and the
marginal scales vary substantially.

\paragraph{Park--Ayyala.}
The leave-out improvement of \citet{ParkAyyala2013} removes a non-negligible
bias in diagonal Hotelling-type procedures. Their statistic can be written in the
cross-validation form
\begin{equation}\label{eq:ch2-PA-stat}
  T_{\mathrm{PA}}
  = \frac{n-5}{n(n-1)(n-3)}
    \sum_{i\neq j}
    \vX_i\trans \hat{\mD}_{S(i,j)}^{-1}\vX_j,
\end{equation}
where $\hat{\mD}_{S(i,j)}$ is the diagonal sample covariance computed from the
leave-two-out sample. After suitable variance normalization,
$T_{\mathrm{PA}}$ is asymptotically normal under the null. This procedure is the
most important scale-invariant Gaussian/light-tail benchmark for the one-sample
problem, and it is repeatedly used as a comparison target in the later robust
papers.

\subsection{Two-sample quadratic-form procedures}

\paragraph{Bai--Saranadasa.}
For the two-sample problem,
\citet{BaiSaranadasa1996} proposed the bias-corrected Euclidean statistic
\begin{equation}\label{eq:ch2-BS-stat}
  T_{\mathrm{BS}}
  = \twonorm{\bar{\vX}_1-\bar{\vX}_2}^2
    - \frac{\tr(\hat{\mSigma}_1)}{n_1}
    - \frac{\tr(\hat{\mSigma}_2)}{n_2},
\end{equation}
where $\hat{\mSigma}_1$ and $\hat{\mSigma}_2$ are the two sample covariance
matrices. After normalization by a consistent estimator of its variance,
$T_{\mathrm{BS}}$ is asymptotically Gaussian under the null. Its strength is that
it avoids matrix inversion entirely, but it is not scalar invariant under
coordinatewise rescaling.

\paragraph{Chen--Qin.}
\citet{ChenQin2010} refined the Bai--Saranadasa construction into the now
standard $U$-statistic
\begin{equation}\label{eq:ch2-CQ-stat}
  T_{\mathrm{CQ}}
  =
  \frac{1}{n_1(n_1-1)}\sum_{i\neq j}\vX_{1i}\trans\vX_{1j}
  +
  \frac{1}{n_2(n_2-1)}\sum_{i\neq j}\vX_{2i}\trans\vX_{2j}
  -
  \frac{2}{n_1n_2}\sum_{i=1}^{n_1}\sum_{j=1}^{n_2}\vX_{1i}\trans\vX_{2j}.
\end{equation}
Because \eqref{eq:ch2-CQ-stat} is an unbiased estimator of
$\twonorm{\vmu_1-\vmu_2}^2$, it has become the standard dense-alternative
benchmark in the high-dimensional two-sample literature.

\begin{assumption}[Chen--Qin balance and trace conditions]
\label{ass:ch2-CQ}
Let $n=n_1+n_2$. Assume:
\begin{enumerate}[label=(CQ\arabic*)]
  \item \label{it:ch2-CQ1}
  $n_1/n\to \kappa\in(0,1)$ as $n\to\infty$.
  \item \label{it:ch2-CQ2}
  For every $i,j,\ell,h\in\{1,2\}$,
  \begin{equation}\label{eq:ch2-CQ-trace}
    \tr(\mSigma_i\mSigma_j\mSigma_\ell\mSigma_h)
    = o\!\left[\tr^2\{(\mSigma_1+\mSigma_2)^2\}\right].
  \end{equation}
  \item \label{it:ch2-CQ3}
  When local alternatives are studied, the signal is not so large that it changes
  the variance order, namely
  \begin{equation}\label{eq:ch2-CQ-local}
    (\vmu_1-\vmu_2)\trans \mSigma_k(\vmu_1-\vmu_2)
    = o\!\left[n^{-1}\tr\{(\mSigma_1+\mSigma_2)^2\}\right],
    \qquad k=1,2.
  \end{equation}
\end{enumerate}
\end{assumption}

\begin{theorem}[Null law of the Chen--Qin statistic]
\label{thm:ch2-CQ-null}
Suppose Assumption~\ref{ass:ch2-CQ} holds. Then, under
$H_0:\vmu_1=\vmu_2$,
\begin{equation}\label{eq:ch2-CQ-null}
  \frac{T_{\mathrm{CQ}}}{\sqrt{\Var(T_{\mathrm{CQ}})}}
  \overset{d}{\longrightarrow} N(0,1).
\end{equation}
Moreover,
\begin{equation}\label{eq:ch2-CQ-variance}
  \Var(T_{\mathrm{CQ}})
  = \frac{2}{n_1(n_1-1)}\tr(\mSigma_1^2)
    + \frac{2}{n_2(n_2-1)}\tr(\mSigma_2^2)
    + \frac{4}{n_1n_2}\tr(\mSigma_1\mSigma_2)
    + o(1)
\end{equation}
after normalization by the leading trace order. Under the local alternatives in
\eqref{eq:ch2-CQ-local},
$\E(T_{\mathrm{CQ}})=\twonorm{\vmu_1-\vmu_2}^2$, so the power is driven by the
ratio between the squared Euclidean signal and the square root of the trace-type
variance in \eqref{eq:ch2-CQ-variance}.
\end{theorem}

\paragraph{Srivastava--Katayama--Kano.}
A different benchmark, due to \citet{SrivastavaKatayamaKano2013}, starts from
the diagonal Hotelling-type quantity
\begin{equation}\label{eq:ch2-SKK-Q2}
  Q_2
  = (\bar{\vX}_1-\bar{\vX}_2)\trans
    \left(\frac{\mD_1}{n_1}+\frac{\mD_2}{n_2}\right)^{-1}
    (\bar{\vX}_1-\bar{\vX}_2),
\end{equation}
where $\mD_i=\diag(\mSigma_i)$ and the sample version uses
$\hat{\mD}_i=\diag(\hat{\mSigma}_i)$. Define
\begin{equation}\label{eq:ch2-SKK-hatq}
  \hat q_{\mathrm{SKK}}
  = (\bar{\vX}_1-\bar{\vX}_2)\trans
    \hat{\mD}^{-1}
    (\bar{\vX}_1-\bar{\vX}_2)-p,
  \qquad
  \hat{\mD}=\frac{\hat{\mD}_1}{n_1}+\frac{\hat{\mD}_2}{n_2}.
\end{equation}
Let
\begin{equation}\label{eq:ch2-SKK-R}
  \mD = \frac{\mD_1}{n_1}+\frac{\mD_2}{n_2},
  \qquad
  \mR = \mD^{-1/2}
  \left(\frac{\mSigma_1}{n_1}+\frac{\mSigma_2}{n_2}\right)
  \mD^{-1/2}.
\end{equation}
The leading variance is
\begin{equation}\label{eq:ch2-SKK-varlead}
  \sigma_{\mathrm{SKK}}^2 = 2\tr(\mR^2).
\end{equation}
To estimate it, \citet{SrivastavaKatayamaKano2013} proposed
\begin{equation}\label{eq:ch2-SKK-FG}
  \hat F_i
  = p^{-1}\left[\tr(\hat{\mD}^{-1}\hat{\mSigma}_i)^2
       - n_i^{-1}\{\tr(\hat{\mD}^{-1}\hat{\mSigma}_i)\}^2\right],
  \qquad
  \hat G = p^{-1}\tr(\hat{\mD}^{-1}\hat{\mSigma}_1\hat{\mD}^{-1}\hat{\mSigma}_2),
\end{equation}
with
\begin{equation}\label{eq:ch2-SKK-varhat}
  \widehat{\Var}(\hat q_{\mathrm{SKK}})
  = \frac{2\hat F_1}{n_1^2} + \frac{2\hat F_2}{n_2^2}
    + \frac{4\hat G}{n_1n_2}.
\end{equation}
The standardized statistic is
\begin{equation}\label{eq:ch2-SKK-stat}
  T_{\mathrm{SKK}}
  = \frac{\hat q_{\mathrm{SKK}}}
         {\sqrt{\widehat{\Var}(\hat q_{\mathrm{SKK}})\,c_{p,n}}},
  \qquad
  c_{p,n}=1+\frac{\tr(\mR^2)}{p^{3/2}}.
\end{equation}
A decisive point for our book is that
\eqref{eq:ch2-SKK-stat} is invariant under componentwise nonzero diagonal
transformations $\vX\mapsto \mD_0\vX$, because both the numerator and the
diagonal standardizer transform equivariantly.

\begin{assumption}[Srivastava--Katayama--Kano conditions]
\label{ass:ch2-SKK}
Assume:
\begin{enumerate}[label=(SKK\arabic*)]
  \item \label{it:ch2-SKK1}
  There exist constants $0<c_1<c_2<\infty$ such that
  \begin{equation}\label{eq:ch2-SKK-diagbounded}
    c_1 \le \, \min_{i,k}\sigma_{ikk}
    \le \, \max_{i,k}\sigma_{ikk} \le c_2.
  \end{equation}
  \item \label{it:ch2-SKK2}
  \begin{equation}\label{eq:ch2-SKK-trace}
    \frac{\tr(\mR^4)}{\tr^2(\mR^2)}\to 0.
  \end{equation}
  \item \label{it:ch2-SKK3}
  $n_1/(n_1+n_2)\to \kappa\in(0,1)$.
  \item \label{it:ch2-SKK4}
  If $N_*=\min(n_1,n_2)$, then
  \begin{equation}\label{eq:ch2-SKK-growth}
    N_* = O(p^{\delta})
    \qquad\text{for some }\delta>1/2.
  \end{equation}
  \item \label{it:ch2-SKK5}
  Under local alternatives,
  \begin{equation}\label{eq:ch2-SKK-local}
    (\vmu_1-\vmu_2)\trans \mD^{-1/2}\mR\mD^{-1/2}(\vmu_1-\vmu_2)
    = o\{\tr(\mR^2)\}.
  \end{equation}
\end{enumerate}
\end{assumption}

\begin{theorem}[Srivastava--Katayama--Kano null and local-alternative laws]
\label{thm:ch2-SKK-null}
Suppose Assumption~\ref{ass:ch2-SKK} holds and the samples are Gaussian. Then,
under $H_0$, 
\begin{equation}\label{eq:ch2-SKK-null}
  T_{\mathrm{SKK}} \overset{d}{\longrightarrow} N(0,1).
\end{equation}
Under the local alternatives in \eqref{eq:ch2-SKK-local},
\begin{equation}\label{eq:ch2-SKK-power}
  \Prob\{T_{\mathrm{SKK}}>z_{1-\alpha}\}
  \to
  \Phi\left(
    -z_{1-\alpha}
    + \frac{(\vmu_1-\vmu_2)\trans\mD^{-1}(\vmu_1-\vmu_2)}
           {\sqrt{2\tr(\mR^2)}}
  \right).
\end{equation}
\end{theorem}

\paragraph{The scale-invariant Behrens--Fisher test of Feng, Zou, Wang, and Zhu.}
The scale-invariant Behrens--Fisher test of
\citet{FengZouWangZhu2015BF} is particularly important for the present book,
because it shows exactly where a diagonal standardization scheme can fail if the
plug-in bias is not corrected. Write $\gamma=n_1/n_2$ and define
\begin{equation}\label{eq:ch2-BF-Lambda}
  \mLambda
  = \diag\left
    \{(\sigma_{11}^2+\gamma\sigma_{21}^2)^{-1/2},\ldots,
      (\sigma_{1p}^2+\gamma\sigma_{2p}^2)^{-1/2}\}
  \right.
\end{equation}
with sample counterpart $\hat{\mLambda}$. For each coordinate $k$, let
\begin{equation}\label{eq:ch2-BF-Ak}
  A_k
  = \frac{1}{n_1(n_1-1)}\sum_{i\neq j} X_{1ik}X_{1jk}
    + \frac{1}{n_2(n_2-1)}\sum_{i\neq j} X_{2ik}X_{2jk}
    - \frac{2}{n_1n_2}\sum_{i=1}^{n_1}\sum_{j=1}^{n_2} X_{1ik}X_{2jk},
\end{equation}
so that $\sum_{k=1}^p A_k = T_{\mathrm{CQ}}$. The Behrens--Fisher statistic is
\begin{equation}\label{eq:ch2-BF-stat}
  T_{\mathrm{BF}}
  = \sum_{k=1}^p
    \frac{A_k}{\hat\sigma_{1k}^2+\gamma\hat\sigma_{2k}^2}.
\end{equation}
Equivalently,
\begin{equation}\label{eq:ch2-BF-matrix}
  T_{\mathrm{BF}}
  = \sum_{k=1}^p \hat\lambda_k^2 A_k.
\end{equation}
This statistic is shift invariant and scalar invariant, but not orthogonally
invariant. The main contribution of \citet{FengZouWangZhu2015BF} is the careful
analysis of the non-negligible bias generated by plugging the marginal variance
estimators into the denominator.

Define
\begin{equation}\label{eq:ch2-BF-var}
  \sigma_{\mathrm{BF},n}^2
  = \frac{2}{n_1(n_1-1)}\tr\{(\mLambda\mSigma_1\mLambda)^2\}
    + \frac{2}{n_2(n_2-1)}\tr\{(\mLambda\mSigma_2\mLambda)^2\}
    + \frac{4}{n_1n_2}\tr(\mLambda\mSigma_1\mLambda^2\mSigma_2\mLambda).
\end{equation}
The exact mean contains a signal term
$\twonorm{\mLambda(\vmu_1-\vmu_2)}^2$ and a bias correction involving the
marginal variances, marginal skewnesses, and marginal kurtoses. Writing
$\kappa_{ik}=\E(X_{ijk}-\mu_{ik})^3$ and
$\nu_{ik}=\E(X_{ijk}-\mu_{ik})^4$, one may decompose it as
\begin{equation}\label{eq:ch2-BF-mean}
  \mu_{\mathrm{BF},n}
  = \twonorm{\mLambda(\vmu_1-\vmu_2)}^2
    + b_{n,1}+b_{n,2}+b_{n,3},
\end{equation}
where
\begin{align}
  b_{n,1}
  &= \sum_{k=1}^p \left[
    \frac{2\sigma_{1k}^4}{n_1(n_1-1)(\sigma_{1k}^2+\gamma\sigma_{2k}^2)^2}
    + \frac{2\gamma^2\sigma_{2k}^4}{n_2(n_2-1)(\sigma_{1k}^2+\gamma\sigma_{2k}^2)^2}
    \right],\label{eq:ch2-BF-b1}\\
  b_{n,2}
  &= \sum_{k=1}^p \left[
    \frac{2\kappa_{1k}^2}{n_1^2(\sigma_{1k}^2+\gamma\sigma_{2k}^2)^3}
    + \frac{2\gamma^2\kappa_{2k}^2}{n_2^2(\sigma_{1k}^2+\gamma\sigma_{2k}^2)^3}
    - \frac{4\gamma\kappa_{1k}\kappa_{2k}}{n_1n_2(\sigma_{1k}^2+\gamma\sigma_{2k}^2)^3}
    \right],\label{eq:ch2-BF-b2}\\
  b_{n,3}
  &= \sum_{k=1}^p (\mu_{1k}-\mu_{2k})^2
     \frac{n_1^{-1}\nu_{1k}+4\{n_1(n_1-1)\}^{-1}\sigma_{1k}^4
           +\gamma n_2^{-1}\nu_{2k}+4\gamma^2\{n_2(n_2-1)\}^{-1}\sigma_{2k}^4}
          {(\sigma_{1k}^2+\gamma\sigma_{2k}^2)^3} \\
  &\qquad
     + \sum_{k=1}^p
       \frac{2n_1^{-1}\kappa_{1k}(\mu_{2k}-\mu_{1k})
             +2\gamma n_2^{-1}\kappa_{2k}(\mu_{1k}-\mu_{2k})}
            {(\sigma_{1k}^2+\gamma\sigma_{2k}^2)^2}.\label{eq:ch2-BF-b3}
\end{align}
The terms $b_{n,1}$ and $b_{n,2}$ remain under $H_0$, which is exactly why a
bias correction is essential when $p$ is as large as order $n^2$ or larger.

\begin{assumption}[Scale-invariant Behrens--Fisher conditions]
\label{ass:ch2-BF}
Assume:
\begin{enumerate}[label=(BF\arabic*)]
  \item \label{it:ch2-BF1}
  $n_1/(n_1+n_2)\to \lambda\in(0,1)$.
  \item \label{it:ch2-BF2}
  For all $i,j,\ell,h\in\{1,2\}$,
  \begin{equation}\label{eq:ch2-BF-trace}
    \tr(\mLambda\mSigma_i\mLambda^2\mSigma_j\mLambda^2\mSigma_\ell\mLambda^2\mSigma_h\mLambda)
    = o\!\left[\tr^2\{(\mLambda\mSigma_1\mLambda+\mLambda\mSigma_2\mLambda)^2\}\right].
  \end{equation}
  \item \label{it:ch2-BF3}
  \begin{equation}\label{eq:ch2-BF-growth}
    \frac{p^2}{n^5\sigma_{\mathrm{BF},n}^2}\to 0.
  \end{equation}
  \item \label{it:ch2-BF4}
  With
  \begin{equation}\label{eq:ch2-BF-Pi}
    \bm{\Pi}_{1i}
    = \E\Big(
        \big[\{\mLambda(\vX_{ij}-\vmu_i)\}^{\circ 3}\big]
        \big[\mLambda(\vX_{ij}-\vmu_i)\big]\trans
      \Big),
    \qquad
    \bm{\Pi}_{2i}
    = \E\Big(
        \big[\mLambda(\vX_{ij}-\vmu_i)(\vX_{ij}-\vmu_i)\trans\mLambda\big]^{\circ 3}
      \Big).
  \end{equation}
  one has
  \begin{equation}\label{eq:ch2-BF-Pi-cond}
    n_i^{-4}\tr(\bm{\Pi}_{1i}^2)=o(\sigma_{\mathrm{BF},n}^2),
    \qquad
    n_i^{-4}\tr(\mLambda\mSigma_i\mLambda\bm{\Pi}_{2i})=o(\sigma_{\mathrm{BF},n}^2),
    \qquad i=1,2.
  \end{equation}
  \item \label{it:ch2-BF5}
  Under local alternatives,
  \begin{equation}\label{eq:ch2-BF-local}
    (\vmu_1-\vmu_2)\trans\mLambda^2\mSigma_i\mLambda^2(\vmu_1-\vmu_2)
    = o\!\left[n^{-1}\tr\{(\mLambda\mSigma_1\mLambda+\mLambda\mSigma_2\mLambda)^2\}\right],
    \quad i=1,2,
  \end{equation}
  and
  \begin{equation}\label{eq:ch2-BF-local2}
    \big[(\vmu_1-\vmu_2)\trans\mLambda^2(\vmu_1-\vmu_2)\big]^2
    = o\!\left[n^{-1}\tr\{(\mLambda\mSigma_1\mLambda+\mLambda\mSigma_2\mLambda)^2\}\right].
  \end{equation}
\end{enumerate}
\end{assumption}

\begin{theorem}[Scale-invariant Behrens--Fisher test]
\label{thm:ch2-BF-null}
Suppose Assumption~\ref{ass:ch2-BF} holds. Then
\begin{equation}\label{eq:ch2-BF-null}
  \frac{T_{\mathrm{BF}}-\mu_{\mathrm{BF},n}}{\sigma_{\mathrm{BF},n}}
  \overset{d}{\longrightarrow} N(0,1).
\end{equation}
Under $H_0$, the non-negligible quantity $\mu_{\mathrm{BF},n}$ must be estimated
and subtracted. If $\hat\mu_{\mathrm{BF},n}$ and $\hat\sigma_{\mathrm{BF},n}$ are the
plug-in estimators proposed in \citet{FengZouWangZhu2015BF}, then
\begin{equation}\label{eq:ch2-BF-feasible}
  \frac{T_{\mathrm{BF}}-\hat\mu_{\mathrm{BF},n}}{\hat\sigma_{\mathrm{BF},n}}
  \overset{d}{\longrightarrow} N(0,1).
\end{equation}
The resulting procedure is scalar invariant and remains valid in regimes where
uncorrected diagonal Hotelling-type tests accumulate a bias of the same order as
their standard deviation.
\end{theorem}

\paragraph{Composite and Behrens--Fisher type combinations.}
When no single benchmark is uniformly best, one may also combine different
quadratic-form statistics. This idea appears in the composite $T^2$ test of
\citet{FengZouWangZhu2017CT2}, which mixes the information carried by different
quadratic forms. This philosophy will reappear in robust max-sum combinations
later in the chapter.

\subsection{Max-type and adaptive procedures}

\paragraph{Cai--Liu--Xia.}
Sum-type tests are often suboptimal under sparse alternatives. A seminal response
is the precision-adjusted max statistic of \citet{CaiLiuXia2014}. To write the
method in a form that matches later chapters, suppose first that the two
samples are Gaussian with a common covariance matrix $\mSigma$ and precision
matrix $\mOmega=\mSigma^{-1}=(\omega_{ij})$. Define
\begin{equation}\label{eq:ch2-CLX-zbar}
  \bar{\vZ} = \mOmega(\bar{\vX}_1-\bar{\vX}_2)
\end{equation}
and the oracle statistic
\begin{equation}\label{eq:ch2-CLX-stat}
  M_{\mOmega}
  = \frac{n_1n_2}{n_1+n_2}
    \max_{1\le j\le p}\frac{\bar Z_j^2}{\omega_{jj}}.
\end{equation}
The intuition is simple but powerful: under sparse alternatives the transformed
coordinates $\bar Z_j$ amplify weak mean signals by removing the correlation
masking effect induced by $\mSigma$.

\begin{assumption}[Oracle CLX conditions]
\label{ass:ch2-CLX}
Assume:
\begin{enumerate}[label=(CLX\arabic*)]
  \item \label{it:ch2-CLX1}
  The common covariance matrix is positive definite and its eigenvalues are
  uniformly bounded away from zero and infinity:
  \begin{equation}\label{eq:ch2-CLX-eigs}
    C_0^{-1}
    \le \lambda_{\min}(\mSigma)
    \le \lambda_{\max}(\mSigma)
    \le C_0
  \end{equation}
  for some fixed $C_0>0$.
  \item \label{it:ch2-CLX2}
  The transformed coordinates are weakly dependent:
  \begin{equation}\label{eq:ch2-CLX-corr}
    \max_{1\le i<j\le p}
    \left|\frac{\omega_{ij}}{\sqrt{\omega_{ii}\omega_{jj}}}\right|
    \le r_0 < 1.
  \end{equation}
  \item \label{it:ch2-CLX3}
  The maximum degree of strong transformed correlations is negligible in the
  sense required by the Gaussian extreme-value theory of
  \citet{CaiLiuXia2014}; in particular, no positive fraction of coordinates may be
  highly correlated with polynomially many others.
\end{enumerate}
\end{assumption}

\begin{theorem}[Extreme-value limit of the oracle CLX statistic]
\label{thm:ch2-CLX-null}
Suppose Assumption~\ref{ass:ch2-CLX} holds and $H_0:\vmu_1=\vmu_2$ is true.
Then
\begin{equation}\label{eq:ch2-CLX-null}
  \Prob\{M_{\mOmega}-2\log p+\log\log p\le x\}
  \longrightarrow
  \exp\{-\pi^{-1/2}e^{-x/2}\}.
\end{equation}
Hence the critical region of the oracle max test is of Gumbel type rather than
Gaussian. In the feasible CLX procedure, $\mOmega$ is replaced by a sparse
precision matrix estimator, and the same limit continues to hold under the extra
estimation conditions given in \citet{CaiLiuXia2014}.
\end{theorem}

The appeal of \eqref{eq:ch2-CLX-stat} is that it targets the sparse regime in
which only a small number of coordinates depart from the null. Later robust
max-type procedures in this book should be read as elliptical analogues of the
CLX paradigm.

\paragraph{Adaptive norm-combination tests.}
\citet{XuLinWeiPan2016} proposed combining statistics based on different norms
of the coordinatewise standardized mean contrasts. Let
\begin{equation}\label{eq:ch2-Xu-Wj}
  W_j = \frac{\bar X_{1j}-\bar X_{2j}}
              {\sqrt{\hat\sigma_{1,jj}/n_1+\hat\sigma_{2,jj}/n_2}},
  \qquad j=1,\ldots,p.
\end{equation}
For an even integer $\gamma\ge 1$, define the sum-of-powered-score statistic
\begin{equation}\label{eq:ch2-Xu-SPU}
  T_{\mathrm{SPU}}(\gamma)=\sum_{j=1}^p W_j^{\gamma},
\end{equation}
and for the max-type member of the family,
\begin{equation}\label{eq:ch2-Xu-SPUinf}
  T_{\mathrm{SPU}}(\infty)=\max_{1\le j\le p}|W_j|.
\end{equation}
The notation ``SPU'' follows the paper: for small $\gamma$ the statistic behaves
like a dense-alternative $L_\gamma$-type aggregate, whereas $\gamma=\infty$
behaves like a sparse max test.

\begin{assumption}[Conditions for adaptive norm-combination tests]
\label{ass:ch2-Xu}
Assume:
\begin{enumerate}[label=(XU\arabic*)]
  \item \label{it:ch2-Xu1}
  The covariance matrices have eigenvalues uniformly bounded away from zero and
  infinity, and the marginal variances are bounded away from zero and infinity.
  \item \label{it:ch2-Xu2}
  The dependence among coordinates is weak enough that finite collections of the
  centered statistics $T_{\mathrm{SPU}}(\gamma)$ satisfy a multivariate central
  limit theorem; in the framework of \citet{XuLinWeiPan2016}, this is ensured by
  their mixing and moment conditions.
  \item \label{it:ch2-Xu3}
  The collection of candidate norms $\Gamma$ is finite; in practice the choice
  \begin{equation}\label{eq:ch2-Xu-Gamma}
    \Gamma=\{1,2,3,4,5,6,\infty\}
  \end{equation}
  is recommended and widely used.
\end{enumerate}
\end{assumption}

\begin{theorem}[Joint asymptotic law of the SPU family]
\label{thm:ch2-Xu-joint}
Suppose Assumption~\ref{ass:ch2-Xu} holds. For any finite set
$\Gamma_0\subset\{1,2,3,\ldots\}$,
\begin{equation}\label{eq:ch2-Xu-joint}
  \Bigg(
    \frac{T_{\mathrm{SPU}}(\gamma)-\mu_{\gamma}}{\sigma_{\gamma}}
  \Bigg)_{\gamma\in\Gamma_0}
  \overset{d}{\longrightarrow}
  N_{|\Gamma_0|}(\vct 0,\mR_{\Gamma_0}),
\end{equation}
where $\mu_{\gamma}=\E\{T_{\mathrm{SPU}}(\gamma)\}$,
$\sigma_{\gamma}^2=\Var\{T_{\mathrm{SPU}}(\gamma)\}$, and
$\mR_{\Gamma_0}$ is the limiting correlation matrix of the finite collection.
Therefore an adaptive statistic can be defined by
\begin{equation}\label{eq:ch2-Xu-aSPU}
  T_{\mathrm{aSPU}} = \min_{\gamma\in\Gamma}\hat P_{\gamma},
\end{equation}
where $\hat P_{\gamma}$ is the bootstrap or simulation-based $p$-value of
$T_{\mathrm{SPU}}(\gamma)$. The resulting procedure adapts to both dense and
sparse alternatives by selecting, in a data-dependent way, the most favorable
norm index.
\end{theorem}

The practical message is worth stating explicitly. If the signal is dense, small
values of $\gamma$ behave like quadratic-form tests and often dominate. If the
signal is sparse, large values of $\gamma$, especially $\gamma=\infty$, are more
effective. The adaptive minimum-$p$ combination in
\eqref{eq:ch2-Xu-aSPU} is designed precisely to bridge these regimes.

\section{Spatial median and diagonal standardization in high dimensions}

The real turning point is that, in high dimensions, robust location estimation is
not merely a preliminary step. It enters directly into the construction and
validity of the test statistics. We therefore treat it in detail.

\subsection{The ordinary spatial median and its high-dimensional Bahadur expansion}

We first recall the ordinary spatial median
$\hat{\vmu}_{\mathrm{SM}}$ defined by
\eqref{eq:ch2-sample-spatial-median}. When $p$ grows with $n$, the key question
is no longer mere consistency but the existence of an explicit asymptotic
expansion with a remainder term small enough for subsequent testing problems.
Recent results of \citet{LiXu2022SpatialMedian} and the subsequent analysis in
\citet{ZhaoWangFeng2025SSPCA} show that such an expansion is available under
elliptical models.

To state the result in a simple form, consider the model
\begin{equation}\label{eq:ch2-spatial-median-hi-model}
  \vX_i = \vmu + \mS^{1/2}\vZ_i,
  \qquad i=1,\ldots,n,
\end{equation}
where $\vZ_i$ is spherically symmetric with center $\vct 0$. Write
\begin{equation}\label{eq:ch2-spatial-median-rU}
  r_i = \twonorm{\vX_i-\vmu},
  \qquad
  \vU_i = U(\vX_i-\vmu).
\end{equation}
The relevant high-dimensional question is whether the estimating equation for the
spatial median can be linearized with a remainder small enough for later
coordinatewise testing and Gaussian approximation. The answer is yes: under the
conditions stated below, the sample spatial median admits a Bahadur expansion in
which the leading term is the empirical average of the spatial signs and the
remainder is asymptotically negligible. This is the aspect of the ordinary
spatial median that will later be inherited by the scalar-invariant and weighted
estimators used in the testing procedures.

\begin{assumption}[Ordinary spatial median in increasing dimension]
\label{ass:ch2-ordinary-bahadur}
Let $\vX_1,\ldots,\vX_n$ be independent observations from the elliptical model
\eqref{eq:ch2-spatial-median-hi-model}. Write
\begin{equation}\label{eq:ch2-ordinary-A-def}
  \mA_{\mathrm{SM}}
  = \E\left[
    \frac{1}{r_i}
    \{\mI_p-\vU_i\vU_i\trans\}
  \right],
  \qquad
  \mB_{\mathrm{SM}} = \E(\vU_i\vU_i\trans),
\end{equation}
where $r_i=\twonorm{\vX_i-\vmu}$ and $\vU_i=U(\vX_i-\vmu)$. Assume:
\begin{enumerate}[label=(SM\arabic*)]
  \item \label{it:ch2-SM1}
  The population spatial median exists, is unique, and equals $\vmu$.
  \item \label{it:ch2-SM2}
  There are constants $0<\underline a\le \bar a<\infty$ such that
  \begin{equation}\label{eq:ch2-SM-eigs}
    \underline a
    \le \lambda_{\min}(\mA_{\mathrm{SM}})
    \le \lambda_{\max}(\mA_{\mathrm{SM}})
    \le \bar a.
  \end{equation}
  \item \label{it:ch2-SM3}
  The sign covariance is nondegenerate and not too concentrated in a single
  direction:
  \begin{equation}\label{eq:ch2-SM-signcov}
    \opnorm{\mB_{\mathrm{SM}}}\le 1-\psi
  \end{equation}
  for some fixed $\psi\in(0,1)$.
  \item \label{it:ch2-SM4}
  The stochastic equicontinuity remainder in the Taylor expansion of the score
  process
  $\Psi_n(\vtheta)=n^{-1}\sum_{i=1}^n U(\vX_i-\vtheta)$
  is $o_p(n^{-1/2})$ uniformly on an $n^{-1/2}$-neighborhood of $\vmu$.
\end{enumerate}
\end{assumption}

\begin{theorem}[High-dimensional Bahadur representation of the spatial median]
\label{thm:ch2-ordinary-bahadur}
Suppose Assumption~\ref{ass:ch2-ordinary-bahadur} holds. Then
\begin{equation}\label{eq:ch2-ordinary-bahadur-thm}
  \sqrt{n}(\hat{\vmu}_{\mathrm{SM}}-\vmu)
  = \mA_{\mathrm{SM}}^{-1}\frac1{\sqrt n}\sum_{i=1}^n \vU_i
    + \vct r_{n,\mathrm{SM}},
\end{equation}
where $\twonorm{\vct r_{n,\mathrm{SM}}}=o_p(1)$. Consequently, for every fixed
vector $\vct{a}\in\R^p$ with $\twonorm{\vct{a}}=1$,
\begin{equation}\label{eq:ch2-ordinary-bahadur-gaussian}
  \vct{a}\trans \sqrt{n}(\hat{\vmu}_{\mathrm{SM}}-\vmu)
  \overset{d}{\longrightarrow}
  N\!\left(
    0,\,
    \vct{a}\trans
    \mA_{\mathrm{SM}}^{-1}\mB_{\mathrm{SM}}\mA_{\mathrm{SM}}^{-1}
    \vct{a}
  \right).
\end{equation}
\end{theorem}

The detailed proof is given in Appendix~A below.

\begin{remark}
Theorem~\ref{thm:ch2-ordinary-bahadur} is the cleanest way to formulate the
self-contained role of the ordinary spatial median in this chapter. The
published high-dimensional papers often work directly with the diagonally scaled
version because scalar invariance is needed in the subsequent testing problems,
but the proof mechanism is already visible here: one linearizes the estimating
equation, identifies the deterministic Jacobian, and shows that the empirical
remainder is of smaller order.
\end{remark}

\subsection{Scalar-invariant diagonal standardization}

The ordinary spatial median is robust but not scalar invariant. High-dimensional
omics and finance data often contain coordinates with widely different marginal
scales, and a practical location procedure should treat these coordinates fairly.
This motivates the diagonal standardization used in
\citet{FengZouWang2016JASA}, \citet{FengSun2016},
\citet{FengLiuMa2021INST}, and \citet{LiuFengZhaoWang2025MaxsumLocation}.

The basic idea is to estimate location and diagonal scale jointly. For the
one-sample problem under $H_0:\vtheta=\vct 0$ we seek a pair $(\hat{\vtheta},
\hat{\mD})$ satisfying
\begin{equation}\label{eq:ch2-diagonal-HR-eq}
  \frac1n\sum_{i=1}^n U\{\hat{\mD}^{-1/2}(\vX_i-\hat{\vtheta})\}=\vct 0,
\end{equation}
and
\begin{equation}\label{eq:ch2-diagonal-HR-scale-eq}
  \frac{p}{n}\diag\Bigg\{
    \sum_{i=1}^n
    U\{\hat{\mD}^{-1/2}(\vX_i-\hat{\vtheta})\}
    U\{\hat{\mD}^{-1/2}(\vX_i-\hat{\vtheta})\}\trans
  \Bigg\}
  = \mI_p.
\end{equation}
This is a diagonal analogue of the Hettmansperger--Randles estimating system.
We shall call $\hat{\vtheta}$ the \emph{scaled spatial median} or
\emph{scalar-invariant spatial median}.

A practical algorithm updates location and scale iteratively. Starting from an
initial pair $(\hat{\vtheta}^{(0)},\hat{\mD}^{(0)})$, define
$\hat{\vvarepsilon}_i^{(m)}=(\hat{\mD}^{(m)})^{-1/2}(\vX_i-\hat{\vtheta}^{(m)})$.
Then set
\begin{equation}\label{eq:ch2-unweighted-iter-location}
  \hat{\vtheta}^{(m+1)}
  = \hat{\vtheta}^{(m)}
    + (\hat{\mD}^{(m)})^{1/2}
      \frac{\sum_{i=1}^n U(\hat{\vvarepsilon}_i^{(m)})}
           {\sum_{i=1}^n \twonorm{\hat{\vvarepsilon}_i^{(m)}}^{-1}},
\end{equation}
and
\begin{equation}\label{eq:ch2-unweighted-iter-scale}
  \hat{\mD}^{(m+1)}
  = p(\hat{\mD}^{(m)})^{1/2}
    \diag\left\{\frac1n\sum_{i=1}^n
    U(\hat{\vvarepsilon}_i^{(m)})U(\hat{\vvarepsilon}_i^{(m)})\trans\right\}
    (\hat{\mD}^{(m)})^{1/2}.
\end{equation}
This is exactly the scalar-invariant normalization that later enters the one-
sample sign test, the two-sample sign test, and the max-type procedures.

\subsection{Bahadur expansion of the scaled spatial median}

The scaled spatial median is the actual location estimator used in the high-
dimensional sign-based testing literature. For this reason we record its
first-order expansion in a detailed and self-contained form.

Let $\vtheta$ denote the true location and $\mD$ the true diagonal scale. Write
\begin{equation}\label{eq:ch2-Ui-ri}
  \vU_i = U\{\mD^{-1/2}(\vX_i-\vtheta)\},
  \qquad
  r_i = \twonorm{\mD^{-1/2}(\vX_i-\vtheta)},
  \qquad
  c_0 = \E(r_i^{-1}).
\end{equation}
The following result is the precise form used repeatedly in your location,
sphericity, and max-type papers.

\begin{assumption}[Diagonal sign model for the scaled spatial median]
\label{ass:ch2-scaled-bahadur}
Suppose that
\begin{equation}\label{eq:ch2-scaled-model}
  \vX_i = \vtheta + \mD^{1/2}\vvarepsilon_i,
  \qquad
  \diag(\mD)=\mD,
  \qquad
  \diag\{\Cov(\vvarepsilon_i)\}=\mI_p,
\end{equation}
and write
\begin{equation}\label{eq:ch2-scaled-rUi}
  r_i=\twonorm{\mD^{-1/2}(\vX_i-\vtheta)},
  \qquad
  \vU_i=U\{\mD^{-1/2}(\vX_i-\vtheta)\},
  \qquad
  c_0=\E(r_i^{-1}).
\end{equation}
Assume:
\begin{enumerate}[label=(SD\arabic*)]
  \item \label{it:ch2-SD1}
  The inverse radial moments are finite and bounded away from zero and infinity:
  there exist constants $0<\underline b\le \bar B<\infty$ such that
  \begin{equation}\label{eq:ch2-SD-zeta}
    \underline b
    \le \E(r_i^{-k})
    \le \bar B,
    \qquad k=1,2,3,4.
  \end{equation}
  \item \label{it:ch2-SD2}
  Let $\mR=\mD^{-1/2}\mSigma\mD^{-1/2}$ denote the standardized shape matrix.
  Then
  \begin{equation}\label{eq:ch2-SD-trace}
    \tr(\mR^4)=o\{\tr^2(\mR^2)\}.
  \end{equation}
  \item \label{it:ch2-SD3}
  The dimensional growth satisfies
  \begin{equation}\label{eq:ch2-SD-growth}
    \frac{p^2}{n^2\tr(\mR^2)}=O(1),
    \qquad
    \log p=o(n^{1/3}).
  \end{equation}
  \item \label{it:ch2-SD4}
  The row sums of $\mR$ are controlled: for some $\delta\in(0,1/2]$ and a
  positive sequence $a_0(p)\asymp p^{1-\delta}$,
  \begin{equation}\label{eq:ch2-SD-rowsum}
    \max_{1\le j\le p}\sum_{\ell=1}^p |\rho_{j\ell}| \le a_0(p),
    \qquad
    \log n = o\!\big(p^{1/3\wedge \delta}\big).
  \end{equation}
\end{enumerate}
\end{assumption}

\begin{lemma}[Uniform consistency of the diagonal scale estimator]
\label{lem:ch2-SD5}
Under Assumption~\ref{ass:ch2-scaled-bahadur}, the diagonal solution
$\hat{\mD}=\diag(\hat d_1,\ldots,\hat d_p)$ of
\eqref{eq:ch2-diagonal-HR-scale-eq} satisfies
\begin{equation}\label{eq:ch2-SD-dcons}
  \max_{1\le j\le p}\left|\frac{\hat d_j}{d_j}-1\right| = o_p(1).
\end{equation}
More precisely, under the growth regime in
\eqref{eq:ch2-SD-growth}--\eqref{eq:ch2-SD-rowsum},
\begin{equation}\label{eq:ch2-SD-dcons-rate}
  \max_{1\le j\le p}\left|\frac{\hat d_j}{d_j}-1\right|
  = O_p\Bigg[
      \left\{\frac{\log p}{n}\right\}^{1/2}
      + \frac{a_0(p)}{p}
      + \frac{\log p}{n^{3/4}}
    \Bigg].
\end{equation}
\end{lemma}

\begin{theorem}[Bahadur expansion of the scaled spatial median]
\label{thm:ch2-scaled-bahadur}
Suppose Assumption~\ref{ass:ch2-scaled-bahadur} holds. Then the solution
$\hat{\vtheta}$ to
\eqref{eq:ch2-diagonal-HR-eq}--\eqref{eq:ch2-diagonal-HR-scale-eq} satisfies
\begin{equation}\label{eq:ch2-scaled-bahadur-form}
  \sqrt n\,\mD^{-1/2}(\hat{\vtheta}-\vtheta)
  = c_0^{-1}\frac1{\sqrt n}\sum_{i=1}^n\vU_i + \vct C_n,
\end{equation}
with remainder
\begin{align}\label{eq:ch2-scaled-bahadur-remainder}
  \maxnorm{\vct C_n}
  = O_p\Big\{
      & n^{-1/4}\{\log(np)\}^{1/2}
      + p^{-(1/6\wedge \delta/2)}\{\log(np)\}^{1/2}  \notag\\
      & + n^{-1/2}(\log p)^{1/2}\{\log(np)\}^{1/2}
    \Big\}.
\end{align}
In particular, every fixed coordinate and every finite collection of coordinates
of $\sqrt n\,\mD^{-1/2}(\hat{\vtheta}-\vtheta)$ are asymptotically Gaussian with
leading covariance matrix $c_0^{-2}p^{-1}\mR$.
\end{theorem}

The detailed proof is given in Appendix~C below.

\begin{remark}
Theorem~\ref{thm:ch2-scaled-bahadur} is the main location-estimation result
behind the robust one-sample sign tests, the max-type procedures, and several
later chapters of the book. The first-order term is the simple empirical mean of
the standardized signs, while all of the technical work is buried in the proof
that the location-scale coupling produces only the remainder
\eqref{eq:ch2-scaled-bahadur-remainder}.
\end{remark}

\subsection{Weighted location equation and weighted Bahadur expansion}

The next step, developed in \citet{FengLiuMa2021INST} for sum-type tests and in
\citet{YanFengZhang2025InverseNormMaxsum} for max-type tests, is to weight the
sign equation by a radial function. This is the point at which the book must be
more detailed than a short survey, because the weighted estimator is the common
starting point of the INST, the weighted max statistic, and the weighted max-sum
combination.

Let $K:\R_+\to\R$ be a measurable weight function. The weighted location
estimating equation is
\begin{equation}\label{eq:ch2-weighted-location-eq}
  \frac1n\sum_{i=1}^n
  K\!\left(\twonorm{\hat{\mD}^{-1/2}(\vX_i-\hat{\vtheta}_K)}\right)
  U\{\hat{\mD}^{-1/2}(\vX_i-\hat{\vtheta}_K)\}
  = \vct 0,
\end{equation}
while the diagonal scale still satisfies the unweighted normalizing equation
\eqref{eq:ch2-diagonal-HR-scale-eq}. The iterative update becomes
\begin{equation}\label{eq:ch2-weighted-iter-location}
  \hat{\vtheta}_K^{(m+1)}
  = \hat{\vtheta}_K^{(m)}
    + (\hat{\mD}^{(m)})^{1/2}
      \frac{\sum_{i=1}^n K(\hat r_i^{(m)})U(\hat{\vvarepsilon}_i^{(m)})}
           {\sum_{i=1}^n K(\hat r_i^{(m)})\{\hat r_i^{(m)}\}^{-1}},
\end{equation}
where
$\hat{\vvarepsilon}_i^{(m)}=(\hat{\mD}^{(m)})^{-1/2}(\vX_i-\hat{\vtheta}_K^{(m)})$
and $\hat r_i^{(m)}=\twonorm{\hat{\vvarepsilon}_i^{(m)}}$.

The most important special case is the power family $K(t)=t^m$, $m\in\R$.
It includes the spatial-sign estimator ($m=0$), the inverse norm weighting
($m=-1$), and the radially amplified version ($m=1$). To unify notation across
all weighted procedures in this chapter, write
\begin{equation}\label{eq:ch2-weighted-moments}
  \nu_{\ell,K} = \E\{K^\ell(r_i)\},
  \qquad
  c_{0,K} = \E\{K(r_i)r_i^{-1}\},
  \qquad
  \mSigma_K = \E\{K^2(r_i)\vU_i\vU_i\trans\}.
\end{equation}
When $K(t)=t^m$, the shorthand becomes
$\nu_{\ell,m}=\E(r_i^{m\ell})$ and $c_{0,m}=\E(r_i^{m-1})$.

\begin{assumption}[Weighted sign moments and local alternatives]
\label{ass:ch2-weighted}
In addition to Assumption~\ref{ass:ch2-scaled-bahadur}, assume:
\begin{enumerate}[label=(WS\arabic*)]
  \item \label{it:ch2-WS1}
  The weight function satisfies
  \begin{equation}\label{eq:ch2-WS-moment}
    0<\nu_{2,K}<\infty,
    \qquad
    \nu_{4,K}=O(\nu_{2,K}^2),
    \qquad
    0<c_{0,K}<\infty.
  \end{equation}
  \item \label{it:ch2-WS2}
  Under local alternatives,
  \begin{equation}\label{eq:ch2-WS-local1}
    \vtheta\trans\mD^{-1}\vtheta
    = O(c_{0,K}^{-2}\sigma_{n,K}),
  \end{equation}
  and
  \begin{equation}\label{eq:ch2-WS-local2}
    \vtheta\trans\mD^{-1}\mSigma\mD^{-1}\vtheta
    = o\!\left(np\,c_{0,K}^{-2}\sigma_{n,K}\right),
  \end{equation}
  where
  \begin{equation}\label{eq:ch2-WS-sigma}
    \sigma_{n,K}^2
    = \frac{2\nu_{2,K}^2}{n(n-1)p^2}\tr(\mR^2).
  \end{equation}
\end{enumerate}
\end{assumption}

\begin{theorem}[Weighted Bahadur expansion and Gaussian approximation]
\label{thm:ch2-weighted-bahadur}
Suppose Assumptions~\ref{ass:ch2-scaled-bahadur}
and~\ref{ass:ch2-weighted} hold. Then
\begin{equation}\label{eq:ch2-weighted-bahadur}
  \sqrt n\,\mD^{-1/2}(\hat{\vtheta}_K-\vtheta)
  = c_{0,K}^{-1}\frac1{\sqrt n}\sum_{i=1}^n K(r_i)\vU_i + \vct C_{n,K},
\end{equation}
where the remainder $\vct C_{n,K}$ satisfies the same order as in
\eqref{eq:ch2-scaled-bahadur-remainder}. Moreover,
\begin{equation}\label{eq:ch2-weighted-gaussian-rectangles}
  \sup_{A\in\mathcal A^{\mathrm{re}}}
  \left|
  \Prob\{\sqrt n\,\mD^{-1/2}(\hat{\vtheta}_K-\vtheta)\in A\}
  -
  \Prob\{\vZ_K\in A\}
  \right| \to 0,
\end{equation}
where $\mathcal A^{\mathrm{re}}$ is the class of hyperrectangles in $\R^p$ and
\begin{equation}\label{eq:ch2-weighted-gaussian}
  \vZ_K \sim N_p\!\left(\vct 0,\,c_{0,K}^{-2}\mSigma_K\right).
\end{equation}
\end{theorem}

The detailed proof is given in Appendix~C below.

\begin{remark}
Theorem~\ref{thm:ch2-weighted-bahadur} is the weighted analogue of the SSPCA-type
linearization for the unweighted estimator. In later sections, the difference
between the unweighted and weighted max statistics is not in the proof
architecture but only in the score vector
$K(r_i)\vU_i$ and in the normalization constants $c_{0,K}$ and $\nu_{2,K}$.
\end{remark}

\section{One-sample spatial-sign and weighted-sign tests}

We now turn from estimation to testing. To make the logic transparent, we begin
with the \emph{oracle} statistics that assume the true $(\vtheta,\mD)$ are known.
After that, we discuss the practical leave-out versions used in the published
procedures.

\subsection{A generic weighted-sign sum-type statistic}

Consider the one-sample problem $H_0:\vtheta=\vct 0$. For a weight function
$K(\cdot)$ define
\begin{equation}\label{eq:ch2-generic-weighted-score}
  \vV_i(K) = K(r_i)\vU_i,
  \qquad
  r_i = \twonorm{\mD^{-1/2}(\vX_i-\vtheta)},
  \qquad
  \vU_i = U\{\mD^{-1/2}(\vX_i-\vtheta)\}.
\end{equation}
The corresponding quadratic-form $U$-statistic is
\begin{equation}\label{eq:ch2-generic-weighted-sum}
  T_n(K)
  = \frac{2}{n(n-1)}\sum_{1\le i<j\le n}
    \vV_i(K)\trans\vV_j(K).
\end{equation}
The choice $K(t)\equiv 1$ yields the spatial-sign statistic; the choice
$K(t)=t^{-1}$ yields the inverse norm sign statistic.

\begin{proposition}[Moments of the weighted-sign statistic]
\label{prop:ch2-weighted-sum-moments}
Let
$\mA_K = \E\{\vV_1(K)\vV_1(K)\trans\}$ and
$\veta_K = \E\{\vV_1(K)\}$. Then
\begin{equation}\label{eq:ch2-weighted-sum-moments-mean}
  \E\{T_n(K)\} = \twonorm{\veta_K}^2.
\end{equation}
Under $H_0$ and central symmetry,
$\veta_K=\vct 0$ and
\begin{equation}\label{eq:ch2-weighted-sum-moments-var}
  \Var\{T_n(K)\} = \frac{2}{n(n-1)}\tr(\mA_K^2).
\end{equation}
\end{proposition}

\begin{proof}
The expectation identity follows from independence:
$\E\{\vV_1\trans\vV_2\}=\E(\vV_1)\trans\E(\vV_2)$. Under central symmetry,
$\vV_1(K)\overset{d}{=}-\vV_1(K)$, so the mean is zero. The variance formula is
the standard variance expression for a degenerate quadratic-form $U$-statistic.
\end{proof}

\begin{theorem}[Oracle weighted-sign statistic under the null]
\label{thm:ch2-weighted-generic-null}
Suppose Assumptions~\ref{ass:ch2-scaled-bahadur}
and~\ref{ass:ch2-weighted} hold and consider the one-sample problem
$H_0:\vtheta=\vct 0$. Then
\begin{equation}\label{eq:ch2-weighted-generic-null}
  \frac{T_n(K)}{\sigma_{n,K}}
  \overset{d}{\longrightarrow} N(0,1),
\end{equation}
where
\begin{equation}\label{eq:ch2-weighted-generic-sigma}
  \sigma_{n,K}^2
  = \frac{2\nu_{2,K}^2}{n(n-1)p^2}\tr(\mR^2).
\end{equation}
\end{theorem}

\begin{theorem}[Oracle weighted-sign statistic under local alternatives]
\label{thm:ch2-weighted-generic-alt}
Suppose Assumptions~\ref{ass:ch2-scaled-bahadur}
and~\ref{ass:ch2-weighted} hold. Then, under the local alternatives described in
\eqref{eq:ch2-WS-local1}--\eqref{eq:ch2-WS-local2},
\begin{equation}\label{eq:ch2-weighted-generic-alt}
  \frac{
    T_n(K)-c_{0,K}^2\,\vtheta\trans\mD^{-1}\vtheta
  }{
    \sqrt{
      \sigma_{n,K}^2
      + \dfrac{4c_{0,K}^2}{np}
        \vtheta\trans\mD^{-1}\mSigma\mD^{-1}\vtheta
    }
  }
  \overset{d}{\longrightarrow} N(0,1).
\end{equation}
In particular, the local asymptotic power is governed by the signal-to-noise
ratio
\begin{equation}\label{eq:ch2-weighted-generic-snr}
  \frac{c_{0,K}^2}{\nu_{2,K}}
  = \frac{\{\E[K(r_i)r_i^{-1}]\}^2}{\E\{K^2(r_i)\}}.
\end{equation}
\end{theorem}

These two theorems are the oracle core of the entire weighted-sign program.
Published procedures differ in two further steps: they replace the unknown
$(\vtheta,\mD)$ by leave-out versions of the weighted estimator, and they prove
that the feasible statistic has the same first-order expansion as its oracle
counterpart.

\subsection{The Wang--Peng--Li nonparametric spatial-sign test}

Before the scalar-invariant sign program matured, \citet{WangPengLi2015}
proposed a one-sample high-dimensional nonparametric mean test based directly on
raw spatial signs. Their model may be written as
\begin{equation}\label{eq:ch2-WPL-model}
  \vX_i = \vmu + \vvarepsilon_i,
  \qquad
  \vvarepsilon_i = \mGamma\vZ_i,
\end{equation}
where $\E(\vvarepsilon_i)=\vct 0$ and the radial and directional parts of
$\vvarepsilon_i$ are separated under elliptical symmetry. For the null problem
$H_0:\vmu=\vct 0$, define the raw spatial signs
\begin{equation}\label{eq:ch2-WPL-Zi}
  \vZ_i^{\mathrm{sgn}} = U(\vX_i),
  \qquad i=1,\ldots,n,
\end{equation}
and the quadratic-form $U$-statistic
\begin{equation}\label{eq:ch2-WPL-stat}
  T_{\mathrm{WPL}}
  = \sum_{1\le i<j\le n}
    (\vZ_i^{\mathrm{sgn}})\trans \vZ_j^{\mathrm{sgn}}.
\end{equation}
Let
\begin{equation}\label{eq:ch2-WPL-AB}
  \mB = \E\left\{\frac{\vvarepsilon_i\vvarepsilon_i\trans}{\twonorm{\vvarepsilon_i}^2}\right\},
  \qquad
  \mA = \E\left[\twonorm{\vvarepsilon_i}^{-1}
  \left(\mI_p-
  \frac{\vvarepsilon_i\vvarepsilon_i\trans}{\twonorm{\vvarepsilon_i}^2}\right)\right].
\end{equation}
Under $H_0$, one has
\begin{equation}\label{eq:ch2-WPL-var}
  \Var(T_{\mathrm{WPL}})=\frac{n(n-1)}{2}\tr(\mB^2).
\end{equation}

\begin{assumption}[Wang--Peng--Li conditions]
\label{ass:ch2-WPL}
Assume:
\begin{enumerate}[label=(WPL\arabic*)]
  \item \label{it:ch2-WPL1}
  \begin{equation}\label{eq:ch2-WPL-C1}
    \tr(\mSigma^4)=o\big[\tr^2(\mSigma^2)\big].
  \end{equation}
  \item \label{it:ch2-WPL2}
  \begin{equation}\label{eq:ch2-WPL-C2}
    \frac{\tr^4(\mSigma)}{\tr^2(\mSigma^2)}
    \exp\left\{-\frac{\tr^2(\mSigma)}{128p\lambda_{\max}^2(\mSigma)}\right\}=o(1).
  \end{equation}
  \item \label{it:ch2-WPL3}
  For local alternatives,
  \begin{equation}\label{eq:ch2-WPL-C3}
    \exp\left\{-\frac{\tr^2(\mSigma)}{256p\lambda_{\max}^2(\mSigma)}\right\}
    = o\!\left\{
      \min\left(
        \frac{\lambda_{\max}(\mSigma)}{\tr(\mSigma)},
        \frac{\lambda_{\min}(\mSigma)}{\lambda_{\max}(\mSigma)}
      \right)
    \right\}.
  \end{equation}
  \item \label{it:ch2-WPL4}
  \begin{equation}\label{eq:ch2-WPL-C4}
    \lambda_{\max}(\mSigma)=o\{\tr(\mSigma)\}.
  \end{equation}
  \item \label{it:ch2-WPL5}
  \begin{equation}\label{eq:ch2-WPL-C5}
    \twonorm{\vmu}^2\E\big(\twonorm{\vvarepsilon_i}^{-2}\big)
    = o\!\left\{
      \min\left(
        \frac{n^{-1}\tr(\mSigma^2)}{\lambda_{\max}(\mSigma)\tr(\mSigma)},
        \frac{n^{-1/2}\tr^{1/2}(\mSigma^2)}{\tr(\mSigma)}
      \right)
    \right\}.
  \end{equation}
  \item \label{it:ch2-WPL6}
  For some $\delta\in(0,1)$,
  \begin{equation}\label{eq:ch2-WPL-C6}
    \twonorm{\vmu}^{2\delta}\E\big(\twonorm{\vvarepsilon_i}^{-2-2\delta}\big)
    = o\big[\E^2\{\twonorm{\vvarepsilon_i}^{-1}\}\big].
  \end{equation}
\end{enumerate}
\end{assumption}

\begin{theorem}[Wang--Peng--Li spatial-sign test under the null]
\label{thm:ch2-WPL-null}
Suppose Assumption~\ref{ass:ch2-WPL}\ref{it:ch2-WPL1}--\ref{it:ch2-WPL2} hold.
Then, under $H_0:\vmu=\vct 0$,
\begin{equation}\label{eq:ch2-WPL-null}
  \frac{T_{\mathrm{WPL}}}
       {\sqrt{\frac{n(n-1)}{2}\tr(\mB^2)}}
  \overset{d}{\longrightarrow} N(0,1).
\end{equation}
A ratio-consistent estimator of $\tr(\mB^2)$ can be constructed by the
cross-validation scheme of \citet{WangPengLi2015}, so the feasible statistic is
obtained by replacing $\tr(\mB^2)$ with its estimator.
\end{theorem}

\begin{theorem}[Wang--Peng--Li local alternative theory]
\label{thm:ch2-WPL-alt}
Suppose Assumption~\ref{ass:ch2-WPL} holds. Then, under the local alternatives
specified by \eqref{eq:ch2-WPL-C3}--\eqref{eq:ch2-WPL-C6},
\begin{equation}\label{eq:ch2-WPL-alt}
  \frac{T_{\mathrm{WPL}}-\frac{n(n-1)}{2}\vmu\trans\mA^2\vmu\{1+o(1)\}}
       {\sqrt{\frac{n(n-1)}{2}\tr(\mB^2)}}
  \overset{d}{\longrightarrow} N(0,1).
\end{equation}
Consequently, the leading local power is controlled by the ratio
$\vmu\trans\mA^2\vmu / \tr^{1/2}(\mB^2)$.
\end{theorem}

\begin{remark}[Why the WPL statistic is not scalar invariant]
\label{rem:ch2-WPL-not-scale-invariant}
The key limitation of \eqref{eq:ch2-WPL-stat} is that it is built from raw signs
$U(\vX_i)$ rather than signs of diagonally standardized observations. If the data
are rescaled coordinatewise by a nonsingular diagonal matrix $\mD_0$, then in
general
\begin{equation}\label{eq:ch2-WPL-not-invariant}
  U(\mD_0\vX_i) \neq \mD_0 U(\vX_i)
  \quad\text{and}\quad
  U(\mD_0\vX_i)\trans U(\mD_0\vX_j)
  \neq U(\vX_i)\trans U(\vX_j).
\end{equation}
Therefore the rejection decision can change under simple marginal rescalings.
This lack of scalar invariance is exactly what motivated the diagonally
standardized sign procedures developed later by \citet{FengSun2016} and in the
subsequent elliptical literature.
\end{remark}

\subsection{The high-dimensional spatial-sign test}

\citet{FengSun2016} proposed the first systematic one-sample spatial-sign test
for high-dimensional elliptical data. Their practical statistic uses diagonal
standardization and leave-two-out scale estimation. Let $\hat{\mD}_{ij}$ denote
the diagonal estimator computed from the sample with observations $i$ and $j$
removed, and define
\begin{equation}\label{eq:ch2-FS-stat}
  T_{\mathrm{SS}}
  = \frac{2}{n(n-1)}\sum_{1\le i<j\le n}
  U(\hat{\mD}_{ij}^{-1/2}\vX_i)\trans
  U(\hat{\mD}_{ij}^{-1/2}\vX_j).
\end{equation}
Under the null the statistic behaves like an oracle sign $U$-statistic, but the
leave-two-out device eliminates the bias caused by estimating the diagonal scale.

Let
\begin{equation}\label{eq:ch2-FS-sigma}
  \sigma_n^2 = \frac{2}{n(n-1)p^2}\tr(\mR^2),
\end{equation}
with $\mR$ as in \eqref{eq:ch2-D-and-R}.

\begin{assumption}[One-sample spatial-sign test conditions]
\label{ass:ch2-FS}
Let $\mR=\mD^{-1/2}\mSigma\mD^{-1/2}$ and
$c_0=\E\{\twonorm{\mD^{-1/2}(\vX_i-\vmu)}^{-1}\}$. Assume:
\begin{enumerate}[label=(SS\arabic*)]
  \item \label{it:ch2-SS1}
  \begin{equation}\label{eq:ch2-SS-trace4}
    \tr(\mR^4)=o\{\tr^2(\mR^2)\}.
  \end{equation}
  \item \label{it:ch2-SS2}
  \begin{equation}\label{eq:ch2-SS-growth}
    \frac{p^2}{n^2\tr(\mR^2)}=O(1),
    \qquad
    \log p=o(n).
  \end{equation}
  \item \label{it:ch2-SS3}
  \begin{equation}\label{eq:ch2-SS-trace2}
    \tr(\mR^2)-p=o(n^{-1}p^2).
  \end{equation}
  \item \label{it:ch2-SS4}
  Under local alternatives,
  \begin{equation}\label{eq:ch2-SS-local}
    \vmu\trans\mD^{-1}\vmu = O(c_0^{-2}\sigma_n).
  \end{equation}
\end{enumerate}
\end{assumption}

\begin{theorem}[One-sample spatial-sign test under the null]
\label{thm:ch2-FS-null}
Suppose Assumption~\ref{ass:ch2-FS} holds. Then, under $H_0$, as $n,p\to\infty$,
\begin{equation}\label{eq:ch2-FS-null}
  \frac{T_{\mathrm{SS}}}{\sigma_n}
  \overset{d}{\longrightarrow} N(0,1).
\end{equation}
\end{theorem}

\begin{theorem}[One-sample spatial-sign test under local alternatives]
\label{thm:ch2-FS-alt}
Suppose Assumption~\ref{ass:ch2-FS} holds. Then
\begin{equation}\label{eq:ch2-FS-alt}
  \frac{T_{\mathrm{SS}} - c_0^2\vmu\trans\mD^{-1}\vmu}
       {\sqrt{\sigma_n^2 + \frac{4c_0^2}{np}
       \vmu\trans\mD^{-1}\mSigma\mD^{-1}\vmu}}
  \overset{d}{\longrightarrow} N(0,1).
\end{equation}
Hence the sign test detects dense alternatives of size roughly
\[
  \vmu\trans\mD^{-1}\vmu
  \asymp n^{-1/2}p^{-1/2}\{\tr(\mR^2)\}^{1/2}.
\]
\end{theorem}

\begin{remark}[ARE relative to diagonal light-tail tests]
For spherical $t_\nu$ distributions with $\nu>2$, the asymptotic relative
efficiency of the spatial-sign test relative to the Park--Ayyala diagonal test
is
\begin{equation}\label{eq:ch2-FS-ARE}
  \mathrm{ARE}(T_{\mathrm{SS}},T_{\mathrm{PA}})
  = \E^2(\twonorm{\vvarepsilon}^{-1})\E(\twonorm{\vvarepsilon}^2)
  = \frac{2}{\nu-2}\left\{\frac{\Gamma((\nu+1)/2)}{\Gamma(\nu/2)}\right\}^2.
\end{equation}
This exceeds one for heavy tails, illustrating why robust sign methods are not
merely safer but often more efficient.
\end{remark}

\subsection{Weighted spatial-sign tests and local optimality}

The limitation of pure signs is that they discard all radial information.
\citet{FengLiuMa2021INST} therefore introduced a weighted class that keeps the
robust directional structure while reintroducing radial information in a
controlled way. The point is not to abandon robustness; the point is to retain
the stable directional geometry of signs while using the radius only through a
carefully chosen one-dimensional weight.

To explain the idea cleanly, consider the oracle weighted statistic
\eqref{eq:ch2-generic-weighted-sum}. Under the local alternative
\begin{equation}\label{eq:ch2-local-alt-generic}
  H_{1n}:\ \vtheta = n^{-1/2}\vdelta,
\end{equation}
Theorem~\ref{thm:ch2-weighted-generic-alt} shows that its local asymptotic power
is governed by
\begin{equation}\label{eq:ch2-generic-snr}
  \frac{c_{0,K}^2}{\nu_{2,K}}
  = \frac{\{\E[K(r_i)r_i^{-1}]\}^2}{\E\{K^2(r_i)\}}.
\end{equation}
This criterion is the exact quantity one obtains after writing out the mean
shift under $H_{1n}$ and dividing by the square root of the leading null
variance.

\begin{proposition}[Why the inverse norm weight is optimal]
\label{prop:ch2-K-optimal}
Over the class of square-integrable weight functions $K$, the criterion
\eqref{eq:ch2-generic-snr} is maximized by
\begin{equation}\label{eq:ch2-K-optimal}
  K(t)\propto t^{-1}.
\end{equation}
Equivalently, the inverse norm weight is locally most powerful in the weighted
spatial-sign class.
\end{proposition}

\begin{proof}
By the Cauchy--Schwarz inequality,
\begin{equation}\label{eq:ch2-CS-proof}
  \{\E(K(r_i)r_i^{-1})\}^2
  \le \E\{K^2(r_i)\}\E(r_i^{-2}),
\end{equation}
with equality if and only if $K(r_i)$ is proportional to $r_i^{-1}$ almost
surely. Dividing both sides by $\E\{K^2(r_i)\}$ yields
\[
  \frac{\{\E[K(r_i)r_i^{-1}]\}^2}{\E\{K^2(r_i)\}}
  \le \E(r_i^{-2}),
\]
and equality is attained precisely when $K(t)\propto t^{-1}$.
\end{proof}

The proof is elementary, but it is also one of the conceptual centers of this
chapter: once the signal and noise are written in the unified sign-based
parametrization, the inverse norm weight emerges as the optimal way to recover
radial information under local elliptical alternatives.

\subsection{The inverse norm sign test (INST)}

The INST of \citet{FengLiuMa2021INST} uses the choice $K(t)=t^{-1}$. In oracle
form the score vector becomes
\begin{equation}\label{eq:ch2-IN-score}
  \vV_i^{\mathrm{IN}}
  = r_i^{-1}\vU_i
  = \frac{\mD^{-1/2}(\vX_i-\vtheta)}{r_i^2}.
\end{equation}
The feasible statistic again uses the diagonally standardized leave-two-out
estimator. Let $\hat r_{ij,k}=\twonorm{\hat{\mD}_{ij}^{-1/2}\vX_k}$. Then
\begin{equation}\label{eq:ch2-INST-stat}
  T_{\mathrm{INST}}
  = \frac{2}{n(n-1)}\sum_{1\le i<j\le n}
  \hat r_{ij,i}^{-1}\hat r_{ij,j}^{-1}
  U(\hat{\mD}_{ij}^{-1/2}\vX_i)\trans
  U(\hat{\mD}_{ij}^{-1/2}\vX_j).
\end{equation}
To unify notation with the previous subsection, note that for the inverse norm
weight,
\begin{equation}\label{eq:ch2-IN-moments}
  \nu_{2,\mathrm{IN}}=\E(r_i^{-2}),
  \qquad
  c_{0,\mathrm{IN}}=\E(r_i^{-2}).
\end{equation}

\begin{theorem}[One-sample inverse norm sign test]
\label{thm:ch2-INST}
Suppose Assumptions~\ref{ass:ch2-scaled-bahadur}
and~\ref{ass:ch2-weighted} hold with $K(t)=t^{-1}$. Then, under $H_0$,
\begin{equation}\label{eq:ch2-INST-null}
  \frac{T_{\mathrm{INST}}}{\sigma_{\mathrm{IN},n}}
  \overset{d}{\longrightarrow} N(0,1),
\end{equation}
where
\begin{equation}\label{eq:ch2-INST-sigma}
  \sigma_{\mathrm{IN},n}^2
  = \frac{2\nu_{2,\mathrm{IN}}^2}{n(n-1)p^2}\tr(\mR^2).
\end{equation}
Under the local alternatives in Assumption~\ref{ass:ch2-weighted},
\begin{equation}\label{eq:ch2-INST-local}
  \frac{
    T_{\mathrm{INST}}
    - c_{0,\mathrm{IN}}^2\,\vtheta\trans\mD^{-1}\vtheta
  }{
    \sqrt{
      \sigma_{\mathrm{IN},n}^2
      + \dfrac{4c_{0,\mathrm{IN}}^2}{np}
        \vtheta\trans\mD^{-1}\mSigma\mD^{-1}\vtheta
    }
  }
  \overset{d}{\longrightarrow} N(0,1).
\end{equation}
\end{theorem}

\begin{remark}[Efficiency of INST]
Because $K(t)=t^{-1}$ maximizes the local asymptotic signal-to-noise ratio in
\eqref{eq:ch2-generic-snr}, the INST is asymptotically at least as efficient as
every other member of the weighted spatial-sign family under the local
alternatives considered above. In particular,
\begin{equation}\label{eq:ch2-INST-ARE}
  \mathrm{ARE}(T_{\mathrm{INST}},T_{\mathrm{SS}})
  = \frac{\E(r_i^{-2})}{\E^2(r_i^{-1})}\ge 1,
\end{equation}
and, relative to diagonal light-tail procedures such as Park--Ayyala,
\begin{equation}\label{eq:ch2-INST-ARE-PA}
  \mathrm{ARE}(T_{\mathrm{INST}},T_{\mathrm{PA}})
  = \E(r_i^{-2})\E(r_i^2)\ge 1,
\end{equation}
with strict inequality for genuinely heavy-tailed radial laws.
\end{remark}

\section{One-sample max-type and max-sum procedures}

The previous section deals with dense alternatives. We now move to sparse
alternatives and the combination of sparse and dense regimes.

\subsection{A spatial-sign max-type statistic}

The max-type procedure of \citet{LiuFengZhaoWang2025MaxsumLocation} is built
directly from the scaled spatial median. Let $(\hat{\vtheta},\hat{\mD})$ solve
\eqref{eq:ch2-diagonal-HR-eq}--\eqref{eq:ch2-diagonal-HR-scale-eq}, let
$\hat r_i=\twonorm{\hat{\mD}^{-1/2}(\vX_i-\hat{\vtheta})}$, and define
\begin{equation}\label{eq:ch2-chat0}
  \hat c_0 = \frac1n\sum_{i=1}^n \hat r_i^{-1}.
\end{equation}
The max statistic is
\begin{equation}\label{eq:ch2-TMAX}
  T_{\max}
  = n\,\max_{1\le j\le p}
    \hat d_j^{-1}\hat\theta_j^2\,\hat c_0^2\,p\,(1-n^{-1/2}),
\end{equation}
or, equivalently,
\begin{equation}\label{eq:ch2-TMAX-compact}
  T_{\max}
  = n\,\maxnorm{\hat{\mD}^{-1/2}\hat{\vtheta}}^2\,
    \hat c_0^2\,p\,(1-n^{-1/2}).
\end{equation}
The normalization is chosen so that the oracle version of $T_{\max}$ becomes the
maximum of approximately standard Gaussian coordinates after the Bahadur
expansion in Theorem~\ref{thm:ch2-scaled-bahadur}.

\begin{assumption}[Explicit growth conditions for max-type statistics]
\label{ass:ch2-max-basic}
In addition to Assumption~\ref{ass:ch2-scaled-bahadur}, assume:
\begin{enumerate}[label=(MX\arabic*)]
  \item \label{it:ch2-MX1exp}
  The Bahadur remainder in
  \eqref{eq:ch2-scaled-bahadur-remainder} is negligible on the extreme-value
  scale, namely
  \begin{equation}\label{eq:ch2-MX-remainder}
    \sqrt p\Big[
      n^{-1/4}\{\log(np)\}^{1/2}
      + p^{-(1/6\wedge \delta/2)}\{\log(np)\}^{1/2}
      + n^{-1/2}(\log p)^{1/2}\{\log(np)\}^{1/2}
    \Big]
    = o\{(\log p)^{-1/2}\}.
  \end{equation}
  \item \label{it:ch2-MX2exp}
  The dimensional growth is strong enough for maxima:
  \begin{equation}\label{eq:ch2-MX-growth}
    \log p = o(n^{1/5}),
    \qquad
    \log n = o\!\big(p^{1/3\wedge\delta}\big),
  \end{equation}
  where $\delta$ is the row-sum exponent in
  Assumption~\ref{ass:ch2-scaled-bahadur}\ref{it:ch2-SD4}.
\end{enumerate}
\end{assumption}

\begin{assumption}[Weak dependence for the Gumbel law]
\label{ass:ch2-max-gumbel}
There exists a constant $\rho\in(0,1)$ such that
\begin{equation}\label{eq:ch2-max-rho}
  \max_{1\le i<j\le p} |\rho_{ij}| \le \rho.
\end{equation}
Moreover, there are sequences $\delta_p=o((\log p)^{-1})$ and $\kappa_p\to 0$
such that, with
\begin{equation}\label{eq:ch2-max-Bpi}
  B_{p,i}=\{j:\ |\rho_{ij}|\ge \delta_p\},
  \qquad
  C_p=\{i:\ |B_{p,i}|\ge p^{\kappa_p}\},
\end{equation}
one has
\begin{equation}\label{eq:ch2-max-Cp}
  \frac{|C_p|}{p}\to 0.
\end{equation}
\end{assumption}

\begin{assumption}[Additional conditions for asymptotic independence]
\label{ass:ch2-max-ind}
For some fixed constant $C>0$ and some $\tau\in(0,1/4)$,
\begin{equation}\label{eq:ch2-max-ind-rowsq}
  \max_{1\le i\le p}\sum_{j=1}^p \rho_{ij}^2 \le (\log p)^C,
\end{equation}
and
\begin{equation}\label{eq:ch2-max-ind-eigs}
  p^{-1/2}(\log p)^C
  \ll \lambda_{\min}(\mR)
  \le \lambda_{\max}(\mR)
  \ll p(\log p)^{-1},
  \qquad
  \frac{\lambda_{\max}(\mR)}{\lambda_{\min}(\mR)} = O(p^\tau).
\end{equation}
\end{assumption}

\begin{theorem}[Null law of the spatial-sign max statistic]
\label{thm:ch2-TMAX-null}
Suppose Assumptions~\ref{ass:ch2-scaled-bahadur},
\ref{ass:ch2-max-basic}, and \ref{ass:ch2-max-gumbel} hold. Then
\begin{equation}\label{eq:ch2-TMAX-null}
  \Prob\{T_{\max} - 2\log p + \log\log p \le x\}
  \longrightarrow
  \exp\{-\pi^{-1/2}e^{-x/2}\}.
\end{equation}
Therefore a level-$\alpha$ test rejects whenever
$T_{\max}-2\log p+\log\log p > q_{1-\alpha}$, where $q_{1-\alpha}$ is the
$(1-\alpha)$ quantile of the limiting Gumbel law.
\end{theorem}

\begin{theorem}[Consistency under sparse alternatives]
\label{thm:ch2-TMAX-power}
Suppose the assumptions of Theorem~\ref{thm:ch2-TMAX-null} hold. If for some
sufficiently large constant $C_e$,
\begin{equation}\label{eq:ch2-TMAX-signal}
  \max_{1\le j\le p}\frac{|\theta_j|}{\sqrt{d_j}}
  \ge C_e n^{-1/2}
  \left\{\log p - 2\log\log(1-\alpha)^{-1}\right\}^{1/2},
\end{equation}
then
\begin{equation}\label{eq:ch2-TMAX-consistency}
  \Prob\{T_{\max} - 2\log p + \log\log p > q_{1-\alpha}\mid H_1\}
  \to 1.
\end{equation}
Thus the sparse detection boundary is of order $\sqrt{\log p/n}$ after marginal
standardization.
\end{theorem}

\subsection{Inverse norm weighted max-type statistics}

The same max construction can be carried out with the weighted estimator
$\hat{\vtheta}_K$. For $K(t)=t^m$, define
\begin{equation}\label{eq:ch2-weighted-max-constants}
  \hat c_{0,m} = \frac1n\sum_{i=1}^n \hat r_i^{m-1},
  \qquad
  \hat\nu_{2,m} = \frac1n\sum_{i=1}^n \hat r_i^{2m},
\end{equation}
and
\begin{equation}\label{eq:ch2-weighted-TMAX}
  T_{\max}^{(m)}
  = n\,\maxnorm{\hat{\mD}^{-1/2}\hat{\vtheta}_K}^2\,
    \hat c_{0,m}^2\hat\nu_{2,m}^{-1}p(1-n^{-1/2}).
\end{equation}

\begin{theorem}[Null law of the weighted max statistic]
\label{thm:ch2-weighted-TMAX-null}
Suppose Assumptions~\ref{ass:ch2-scaled-bahadur},
\ref{ass:ch2-weighted}, \ref{ass:ch2-max-basic}, and \ref{ass:ch2-max-gumbel}
hold. Then
\begin{equation}\label{eq:ch2-weighted-TMAX-null}
  \Prob\{T_{\max}^{(m)} - 2\log p + \log\log p \le x\}
  \longrightarrow
  \exp\{-\pi^{-1/2}e^{-x/2}\}.
\end{equation}
The corresponding sparse detection boundary remains of order
$\sqrt{\log p/n}$.
\end{theorem}

A key refinement is that under one-coordinate sparse alternatives, the inverse
norm choice $m=-1$ is again locally optimal. More precisely,
\begin{equation}\label{eq:ch2-INMAX-ARE}
  \mathrm{ARE}_{\mathrm{IN}-\max,\max}
  = \E(r_i^{-2})\E(r_i^2) \ge 1,
\end{equation}
with strict inequality for genuinely heavy-tailed radial laws.

\subsection{Asymptotic independence and max-sum combination}

The real methodological breakthrough in the recent literature is that the robust
sum-type and robust max-type statistics are asymptotically independent, even
though they are computed from the same observations. This is exactly what makes
a valid and powerful max-sum combination possible.

For the spatial-sign case, let $T_{\mathrm{SUM}}$ denote the one-sample sign test
of \citet{FengSun2016} and $T_{\max}$ the max statistic in
\eqref{eq:ch2-TMAX-compact}.

\begin{theorem}[Asymptotic independence: unweighted case]
\label{thm:ch2-maxsum-independence}
Suppose Assumptions~\ref{ass:ch2-scaled-bahadur},
\ref{ass:ch2-FS}, \ref{ass:ch2-max-basic}, \ref{ass:ch2-max-gumbel}, and
\ref{ass:ch2-max-ind} hold. Then, under $H_0$,
\begin{equation}\label{eq:ch2-maxsum-independence}
  \Prob\left\{\frac{T_{\mathrm{SUM}}}{\sigma_n}\le x,\,
  T_{\max}-2\log p+\log\log p\le y\right\}
  \to \Phi(x)F(y),
\end{equation}
where $F(y)=\exp\{-\pi^{-1/2}e^{-y/2}\}$. The same factorization remains valid
under the local alternatives covered by
Assumption~\ref{ass:ch2-FS}\ref{it:ch2-SS4}.
\end{theorem}

\begin{theorem}[Asymptotic independence: weighted case]
\label{thm:ch2-weighted-independence}
Let $T_{\mathrm{SUM}}^{(m)}$ be the weighted sum-type statistic and
$T_{\max}^{(m)}$ the weighted max-type statistic. Suppose
Assumptions~\ref{ass:ch2-scaled-bahadur},
\ref{ass:ch2-weighted}, \ref{ass:ch2-max-basic},
\ref{ass:ch2-max-gumbel}, and \ref{ass:ch2-max-ind} hold. Then
\begin{equation}\label{eq:ch2-weighted-independence}
  \Prob\left\{\frac{T_{\mathrm{SUM}}^{(m)}}{\sigma_{n,K}}\le x,\,
  T_{\max}^{(m)}\le y\right\}
  \to
  \Prob\left\{\frac{T_{\mathrm{SUM}}^{(m)}}{\sigma_{n,K}}\le x\right\}
  \Prob\left\{T_{\max}^{(m)}\le y\right\}.
\end{equation}
Under $H_0$ this becomes
\begin{equation}\label{eq:ch2-weighted-independence-null}
  \Prob\left\{\frac{T_{\mathrm{SUM}}^{(m)}}{\sigma_{n,K}}\le x,\,
  T_{\max}^{(m)}-2\log p+\log\log p\le y\right\}
  \to \Phi(x)F(y).
\end{equation}
\end{theorem}

The practical implication is immediate. Let
\begin{equation}\label{eq:ch2-pmax-psum}
  p_{\max} = 1-F\{T_{\max}-2\log p+\log\log p\},
  \qquad
  p_{\mathrm{sum}} = 1-\Phi\left(\frac{T_{\mathrm{SUM}}}{\hat\sigma_n}\right).
\end{equation}
Then one may combine them by the Cauchy combination rule:
\begin{equation}\label{eq:ch2-cauchy-combination}
  p_{\mathrm{CC}}
  = 1-G\Big[
  0.5\tan\{\pi(0.5-p_{\max})\}
  +0.5\tan\{\pi(0.5-p_{\mathrm{sum}})\}
  \Big],
\end{equation}
where $G$ is the standard Cauchy cdf. Reject when $p_{\mathrm{CC}}<\alpha$.
This yields a robust test that is competitive across the full spectrum from
dense to sparse alternatives.

\section{Two-sample procedures under elliptical symmetry}

We now revisit the two-sample problem in the same level of detail. The main
published procedures are the multivariate-sign test, the spatial-rank test, and
the inverse norm weighted sign test.

\subsection{Two-sample multivariate-sign test}

The JASA paper \citet{FengZouWang2016JASA} is the starting point of the
high-dimensional two-sample elliptical program. For each sample $k\in\{1,2\}$,
let $(\hat{\vtheta}_k,\hat{\mD}_k)$ solve the diagonal estimating equations
\begin{equation}\label{eq:ch2-jasa-eqs}
  \frac1{n_k}\sum_{j=1}^{n_k}
  U\{\hat{\mD}_k^{-1/2}(\vX_{kj}-\hat{\vtheta}_k)\}=\vct 0,
\end{equation}
and
\begin{equation}\label{eq:ch2-jasa-eqs-scale}
  \frac{p}{n_k}\diag\Bigg\{\sum_{j=1}^{n_k}
  U\{\hat{\mD}_k^{-1/2}(\vX_{kj}-\hat{\vtheta}_k)\}
  U\{\hat{\mD}_k^{-1/2}(\vX_{kj}-\hat{\vtheta}_k)\}\trans\Bigg\}
  = \mI_p.
\end{equation}
The iterative algorithm is exactly the one-sample algorithm applied to each
sample separately.

A naive pooled plug-in version would produce a non-negligible bias because the
locations are estimated from the same data used in the test statistic. The JASA
paper removes this bias by a leave-one-out construction. Let
$\hat{\vtheta}_{k,\ell}$ and $\hat{\mD}_{k,\ell}$ denote the estimators obtained
from sample $k$ with the $\ell$th observation removed. The statistic is
\begin{equation}\label{eq:ch2-JASA-Rn}
  R_n
  = -\frac{1}{n_1n_2}
    \sum_{i=1}^{n_1}\sum_{j=1}^{n_2}
    U\{\hat{\mD}_{1,i}^{-1/2}(\vX_{1i}-\hat{\vtheta}_{2,j})\}\trans
    U\{\hat{\mD}_{2,j}^{-1/2}(\vX_{2j}-\hat{\vtheta}_{1,i})\}.
\end{equation}

To describe the limit theory, let
\begin{equation}\label{eq:ch2-JASA-A123}
  \mR_k = \mD_k^{-1/2}\mSigma_k\mD_k^{-1/2},
  \qquad
  c_k = \E\left\{\twonorm{\mD_k^{-1/2}(\vX_{kj}-\vtheta_k)}^{-1}\right\},
\end{equation}
\begin{equation}\label{eq:ch2-JASA-A123-def}
  \mA_1 = c_2^2 c_1^{-2}
  \mSigma_1^{1/2}\mD_2^{-1/2}\mD_1^{-1}\mSigma_1^{1/2},
  \qquad
  \mA_2 = c_1^2 c_2^{-2}
  \mSigma_2^{1/2}\mD_1^{-1/2}\mD_2^{-1}\mSigma_2^{1/2},
\end{equation}
and
\begin{equation}\label{eq:ch2-JASA-A3}
  \mA_3 = \mSigma_1^{1/2}\mD_1^{-1/2}\mD_2^{-1/2}\mSigma_2^{1/2}.
\end{equation}
Set
\begin{equation}\label{eq:ch2-JASA-sigma}
  \sigma_n^2
  = \frac{2}{n_1(n_1-1)p^2}\tr(\mA_1^2)
    + \frac{2}{n_2(n_2-1)p^2}\tr(\mA_2^2)
    + \frac{4}{n_1n_2p^2}\tr(\mA_3\trans\mA_3).
\end{equation}

\begin{assumption}[Two-sample multivariate-sign conditions]
\label{ass:ch2-two-sign}
Let $n=n_1+n_2$. Assume:
\begin{enumerate}[label=(TS\arabic*)]
  \item \label{it:ch2-TS1}
  $n_1/n\to \kappa\in(0,1)$.
  \item \label{it:ch2-TS2}
  For all $i,j,\ell,h\in\{1,2,3\}$,
  \begin{equation}\label{eq:ch2-TS-trace}
    \tr(\mA_i\trans \mA_j \mA_\ell\trans \mA_h)
    = o\!\left[
      \tr^2\!\left\{(\mA_1+\mA_2+\mA_3)\trans(\mA_1+\mA_2+\mA_3)\right\}
    \right].
  \end{equation}
  \item \label{it:ch2-TS3}
  \begin{equation}\label{eq:ch2-TS-growth}
    \frac{n^{-2}}{\sigma_n}=O(1),
    \qquad
    \log p=o(n).
  \end{equation}
  \item \label{it:ch2-TS4}
  \begin{equation}\label{eq:ch2-TS-trace2}
    \tr(\mR_k^2)-p=o(n^{-1}p^2),
    \qquad k=1,2.
  \end{equation}
  \item \label{it:ch2-TS5}
  Under local alternatives with $\vDelta=\vtheta_1-\vtheta_2$,
  \begin{equation}\label{eq:ch2-TS-local}
    \vDelta\trans\mD_1^{-1/2}\mD_2^{-1/2}\vDelta
    = O(c_1^{-1}c_2^{-1}\sigma_n).
  \end{equation}
\end{enumerate}
\end{assumption}

\begin{theorem}[Two-sample sign test under the null]
\label{thm:ch2-JASA-null}
Suppose Assumption~\ref{ass:ch2-two-sign} holds. Then, under
$H_0:\vtheta_1=\vtheta_2$,
\begin{equation}\label{eq:ch2-JASA-null}
  \frac{R_n}{\sigma_n} \overset{d}{\longrightarrow} N(0,1).
\end{equation}
\end{theorem}

Under local alternatives the mean is no longer zero. Let
\begin{equation}\label{eq:ch2-JASA-delta-n}
  \delta_n = c_1c_2\vDelta\trans\mD_1^{-1/2}\mD_2^{-1/2}\vDelta.
\end{equation}
The corresponding asymptotic variance becomes
\begin{equation}\label{eq:ch2-JASA-tilde-sigma}
  \tilde\sigma_n^2
  = \sigma_n^2
    + \frac{c_2^2}{n_1p}\vDelta\trans\mD_2^{-1/2}\mR_1\mD_2^{-1/2}\vDelta
    + \frac{c_1^2}{n_2p}\vDelta\trans\mD_1^{-1/2}\mR_2\mD_1^{-1/2}\vDelta.
\end{equation}

\begin{theorem}[Two-sample sign test under local alternatives]
\label{thm:ch2-JASA-alt}
Suppose Assumption~\ref{ass:ch2-two-sign} holds. Then
\begin{equation}\label{eq:ch2-JASA-alt}
  \frac{R_n-\delta_n}{\tilde\sigma_n}
  \overset{d}{\longrightarrow} N(0,1).
\end{equation}
The leading signal is therefore the scalar-invariant quadratic form
$\vDelta\trans\mD_1^{-1/2}\mD_2^{-1/2}\vDelta$.
\end{theorem}

\subsection{A simpler bias-corrected spatial-sign test}

The leave-one-out construction in \citet{FengZouWang2016JASA} is theoretically
clean but computationally expensive, because every term in the statistic uses
sample-specific leave-out location and scale estimators. The note of
\citet{LiWangZou2016SimpleTwoSample} replaces the full leave-one-out centering
by a simpler plug-in correction while keeping the same first-order efficiency.
Let $(\hat{\vtheta}_k,\hat{\mD}_k)$, $k=1,2$, be the full-sample diagonal
estimators obtained from
\eqref{eq:ch2-jasa-eqs}--\eqref{eq:ch2-jasa-eqs-scale}. The simplified statistic
is
\begin{equation}\label{eq:ch2-simpleSS-stat}
  T_{\mathrm{SST}}
  = -\frac{1}{n_1n_2}
    \sum_{i=1}^{n_1}\sum_{j=1}^{n_2}
    U\{\hat{\mD}_1^{-1/2}(\vX_{1i}-\hat{\vtheta}_2)\}\trans
    U\{\hat{\mD}_2^{-1/2}(\vX_{2j}-\hat{\vtheta}_1)\}.
\end{equation}
With
\begin{equation}\label{eq:ch2-simpleSS-A123}
  \mR_k=\mD_k^{-1/2}\mSigma_k\mD_k^{-1/2},
  \qquad
  c_k=\E\left\{\twonorm{\mD_k^{-1/2}(\vX_{kj}-\vtheta_k)}^{-1}\right\},
\end{equation}
let
\begin{equation}\label{eq:ch2-simpleSS-Dn}
  \mD_n=\mD_1^{1/2}\mD_2^{1/2},
\end{equation}
\begin{equation}\label{eq:ch2-simpleSS-A1}
  \mA_1=c_2c_1^{-1}\mSigma_1^{1/2}\mD_n^{-1}\mSigma_1^{1/2},
  \qquad
  \mA_2=c_1c_2^{-1}\mSigma_2^{1/2}\mD_n^{-1}\mSigma_2^{1/2},
\end{equation}
\begin{equation}\label{eq:ch2-simpleSS-A3}
  \mA_3=\mSigma_1^{1/2}\mD_n^{-1}\mSigma_2^{1/2},
\end{equation}
and
\begin{equation}\label{eq:ch2-simpleSS-sigma}
  \sigma_{\mathrm{SST},n}^2
  = \frac{2}{n_1^2p^2}\tr(\mA_1^2)
    + \frac{2}{n_2^2p^2}\tr(\mA_2^2)
    + \frac{4}{n_1n_2p^2}\tr(\mA_3\trans\mA_3).
\end{equation}
Because the full-sample estimators create a non-negligible mean shift, the null
center is
\begin{equation}\label{eq:ch2-simpleSS-mu}
  \mu_{\mathrm{SST},n}
  = \frac{1}{n_1p}\tr(\mA_1)+\frac{1}{n_2p}\tr(\mA_2).
\end{equation}

\begin{assumption}[Conditions for the simpler two-sample sign statistic]
\label{ass:ch2-simpleSS}
Assume:
\begin{enumerate}[label=(SST\arabic*)]
  \item \label{it:ch2-simpleSS1}
  $n_1/(n_1+n_2)\to \kappa\in(0,1)$.
  \item \label{it:ch2-simpleSS2}
  \begin{equation}\label{eq:ch2-simpleSS-growth}
    n^{-2}=O(\sigma_{\mathrm{SST},n}).
  \end{equation}
  \item \label{it:ch2-simpleSS3}
  For all $i,j,\ell,h\in\{1,2,3\}$,
  \begin{equation}\label{eq:ch2-simpleSS-trace}
    \tr(\mA_i\trans\mA_j\mA_\ell\trans\mA_h)
    = o\!\left[
      \tr^2\!\left\{(\mA_1+\mA_2+\mA_3)\trans(\mA_1+\mA_2+\mA_3)\right\}
    \right].
  \end{equation}
  \item \label{it:ch2-simpleSS4}
  \begin{equation}\label{eq:ch2-simpleSS-R}
    \tr(\mR_k^2)-p=o(n^{-1}p^2),\qquad k=1,2.
  \end{equation}
\end{enumerate}
\end{assumption}

\begin{theorem}[Simpler two-sample sign statistic under the null]
\label{thm:ch2-simpleSS-null}
Suppose Assumption~\ref{ass:ch2-simpleSS} holds. Then, under
$H_0:\vtheta_1=\vtheta_2$,
\begin{equation}\label{eq:ch2-simpleSS-null}
  \frac{T_{\mathrm{SST}}-\mu_{\mathrm{SST},n}}{\sigma_{\mathrm{SST},n}}
  \overset{d}{\longrightarrow} N(0,1).
\end{equation}
Moreover, the plug-in estimators proposed in
\citet{LiWangZou2016SimpleTwoSample} are ratio consistent, so the feasible
studentized statistic is asymptotically standard normal.
\end{theorem}

Under local alternatives with $\vDelta=\vtheta_1-\vtheta_2$, define
\begin{equation}\label{eq:ch2-simpleSS-delta}
  \delta_{\mathrm{SST},n}
  = c_1c_2\vDelta\trans\mD_n^{-1}\vDelta
\end{equation}
and
\begin{align}\label{eq:ch2-simpleSS-sigmatilde}
  \tilde\sigma_{\mathrm{SST},n}^2
  ={}& \sigma_{\mathrm{SST},n}^2
      + \frac{4c_2^2}{n_1p}\vDelta\trans\mD_n^{-1}\mSigma_1\mD_n^{-1}\vDelta \\
   &  + \frac{4c_1^2}{n_2p}\vDelta\trans\mD_n^{-1}\mSigma_2\mD_n^{-1}\vDelta.
\end{align}

\begin{theorem}[Simpler two-sample sign statistic under local alternatives]
\label{thm:ch2-simpleSS-alt}
Suppose Assumption~\ref{ass:ch2-simpleSS} holds. Then
\begin{equation}\label{eq:ch2-simpleSS-alt}
  \frac{T_{\mathrm{SST}}-\mu_{\mathrm{SST},n}-\delta_{\mathrm{SST},n}}
       {\tilde\sigma_{\mathrm{SST},n}}
  \overset{d}{\longrightarrow} N(0,1).
\end{equation}
Hence the simplified statistic achieves the same first-order local power as the
original leave-one-out test while reducing the computational burden from order
$O(n^3p)$ to order $O(n^2p)$.
\end{theorem}

\subsection{Two-sample spatial-rank test}

The sign test above requires estimating the two locations. A different strategy,
proposed in \citet{FengZhangLiu2020SpatialRank}, is to construct the test from
pairwise cross-sample differences so that explicit location estimation is
avoided.

Assume first that the two populations share a common scatter matrix $\mSigma$.
Let $\hat{\mD}_{1,n_1}$ and $\hat{\mD}_{2,n_2}$ be diagonal scale estimators and
define the pooled diagonal scale
\begin{equation}\label{eq:ch2-SR-Dn}
  \hat{\mD}_n = \frac{n_1}{n}\hat{\mD}_{1,n_1} + \frac{n_2}{n}\hat{\mD}_{2,n_2},
  \qquad n=n_1+n_2.
\end{equation}
The leave-two-out statistic is
\begin{equation}\label{eq:ch2-SR-Tn}
  T_n
  = \frac{1}{n_1(n_1-1)n_2(n_2-1)}
    \sum_{i\neq j}\sum_{s\neq l}
    U\{\hat{\mD}_{n(i,j,s,l)}^{-1/2}(\vX_{1i}-\vX_{2s})\}\trans
    U\{\hat{\mD}_{n(i,j,s,l)}^{-1/2}(\vX_{1j}-\vX_{2l})\},
\end{equation}
where
\begin{equation}\label{eq:ch2-SR-Dn-ijkl}
  \hat{\mD}_{n(i,j,s,l)}
  = \frac{n_1}{n}\hat{\mD}_{1,n_1(i,j)} + \frac{n_2}{n}\hat{\mD}_{2,n_2(s,l)}.
\end{equation}
Let $\mD_n$ be the population counterpart and define
\begin{equation}\label{eq:ch2-SR-Rn}
  \mR_n = \mD_n^{-1/2}\mSigma\mD_n^{-1/2}.
\end{equation}
The variance scale is
\begin{equation}\label{eq:ch2-SR-sigma}
  \sigma_n^2
  = \left\{
      \frac{1}{2n_1(n_1-1)p^2}
      + \frac{1}{2n_2(n_2-1)p^2}
      + \frac{1}{n_1n_2p^2}
    \right\}\tr(\mR_n^2).
\end{equation}

\begin{assumption}[Two-sample spatial-rank conditions]
\label{ass:ch2-two-rank}
Assume:
\begin{enumerate}[label=(TR\arabic*)]
  \item \label{it:ch2-TR1}
  The two samples are elliptical with a common scatter matrix $\mSigma$ and
  $n_1/n\to\kappa\in(0,1)$.
  \item \label{it:ch2-TR2}
  \begin{equation}\label{eq:ch2-TR-trace4}
    \tr(\mR_n^4)=o\{\tr^2(\mR_n^2)\}.
  \end{equation}
  \item \label{it:ch2-TR3}
  \begin{equation}\label{eq:ch2-TR-growth}
    \frac{p^2}{n^2\tr(\mR_n^2)}=O(1),
    \qquad
    \log p=o(n),
    \qquad
    \tr(\mR_n^2)-p=o(n^{-1}p^2).
  \end{equation}
  \item \label{it:ch2-TR4}
  Let
  \begin{equation}\label{eq:ch2-TR-c0}
    c_{0,n}
    = \E\big\{\twonorm{\mD_n^{-1/2}(\vX_{1i}-\vX_{2j})}^{-1}\big\}.
  \end{equation}
  Under local alternatives,
  \begin{equation}\label{eq:ch2-TR-local1}
    (\vmu_1-\vmu_2)\trans \mD_n^{-1}(\vmu_1-\vmu_2)
    = O(c_{0,n}^{-2}\sigma_n),
  \end{equation}
  and
  \begin{equation}\label{eq:ch2-TR-local2}
    (\vmu_1-\vmu_2)\trans \mD_n^{-1/2}\mR_n\mD_n^{-1/2}(\vmu_1-\vmu_2)
    = o(np\,c_{0,n}^{-2}\sigma_n).
  \end{equation}
\end{enumerate}
\end{assumption}

\begin{theorem}[Two-sample spatial-rank test under the null]
\label{thm:ch2-SR-null}
Suppose Assumption~\ref{ass:ch2-two-rank} holds. Then
\begin{equation}\label{eq:ch2-SR-null}
  \frac{T_n}{\sigma_n}
  \overset{d}{\longrightarrow} N(0,1)
\end{equation}
under $H_0:\vmu_1=\vmu_2$.
\end{theorem}

\begin{theorem}[Two-sample spatial-rank test under local alternatives]
\label{thm:ch2-SR-alt}
Suppose Assumption~\ref{ass:ch2-two-rank} holds. Then
\begin{equation}\label{eq:ch2-SR-alt}
  \frac{
    T_n - c_{0,n}^2(\vmu_1-\vmu_2)\trans\mD_n^{-1}(\vmu_1-\vmu_2)
  }{
    \sigma_n
  }
  \overset{d}{\longrightarrow} N(0,1).
\end{equation}
Therefore the spatial-rank test is a dense-alternative procedure whose leading
signal is again the scalar-invariant quadratic form in the diagonally
standardized coordinates.
\end{theorem}

Because spatial ranks automatically difference out the common location, this
procedure remains valid in regimes where direct location estimation would be
awkward. Under heavy tails it also enjoys a clear efficiency advantage over
purely moment-based quadratic-form tests.

\subsection{Two-sample inverse norm weighted signs and tINST}

The two-sample weighted-sign program parallels the one-sample theory.
\citet{HuangLiuZhouFeng2023TwoSampleINST} first define a general weighted class
for the two-sample problem and then show that the inverse norm weight is again
locally optimal.

Let $\vDelta=\vtheta_1-\vtheta_2$ and consider the oracle weighted vectors
\begin{equation}\label{eq:ch2-two-sample-weighted-vectors}
  \vV_{1i}(K)=K(r_{1i})U\{\mD_1^{-1/2}(\vX_{1i}-\vtheta_1)\},
  \qquad
  \vV_{2j}(K)=K(r_{2j})U\{\mD_2^{-1/2}(\vX_{2j}-\vtheta_2)\},
\end{equation}
where
$r_{ki}=\twonorm{\mD_k^{-1/2}(\vX_{ki}-\vtheta_k)}$.
The oracle weighted quadratic form is
\begin{equation}\label{eq:ch2-two-sample-weighted-stat}
  T_{n_1,n_2}(K)
  = \frac{1}{n_1(n_1-1)}\sum_{i\neq j}\vV_{1i}(K)\trans\vV_{1j}(K)
  + \frac{1}{n_2(n_2-1)}\sum_{i\neq j}\vV_{2i}(K)\trans\vV_{2j}(K)
  - \frac{2}{n_1n_2}\sum_{i=1}^{n_1}\sum_{j=1}^{n_2}
    \vV_{1i}(K)\trans\vV_{2j}(K).
\end{equation}
The feasible statistic replaces $(\vtheta_k,\mD_k)$ by the leave-out
scalar-invariant estimators described in the JASA paper.

Under local alternatives, the leading signal again depends on
\begin{equation}\label{eq:ch2-two-sample-snr}
  \frac{\{\E(K(r)r^{-1})\}^2}{\E\{K^2(r)\}},
\end{equation}
so the same Cauchy--Schwarz argument shows that $K(t)=t^{-1}$ is optimal.

\begin{assumption}[Two-sample weighted-sign conditions]
\label{ass:ch2-two-weighted}
Suppose Assumption~\ref{ass:ch2-two-sign} holds and, in addition, the weight
function $K$ satisfies
\begin{equation}\label{eq:ch2-two-weighted-moment}
  0<\E\{K^2(r_{ki})\}<\infty,
  \qquad
  \E\{K^4(r_{ki})\}=O\!\left(\E^2\{K^2(r_{ki})\}\right),
  \qquad k=1,2.
\end{equation}
\end{assumption}

\begin{theorem}[Local optimality of tINST]
\label{thm:ch2-tINST-optimal}
Suppose Assumption~\ref{ass:ch2-two-weighted} holds. Within the weighted
spatial-sign class, the locally most powerful choice is
\begin{equation}\label{eq:ch2-tINST-optimal}
  K(t)=t^{-1}.
\end{equation}
The resulting two-sample inverse norm sign test (tINST) is therefore
asymptotically at least as powerful as any other weighted member of the class
under the local elliptical alternatives governed by
\eqref{eq:ch2-two-sample-snr}.
\end{theorem}

The one-sample and two-sample theories therefore fit into the same template:
first diagonal standardization, then weighted signs, and finally the inverse
norm choice as the optimal weight.
\section{Adaptive rank-based location tests and structured dependence}

The weighted-sign theory above addresses robustness and radial efficiency. A
separate but complementary question is adaptivity over dense and sparse signal
regimes. This is the topic of \citet{ZhangFeng2024AdaptiveMean} and
\citet{LiuZhaoFengWang2025StructuredCorr}.

\subsection{Adaptive rank-based tests}

The starting point of \citet{ZhangFeng2024AdaptiveMean} is to construct marginal
rank-based scores $W_j$ and then aggregate them over a range of $L_q$ norms. In
a stylized form one considers
\begin{equation}\label{eq:ch2-adaptive-rank-q}
  T_q = \sum_{j=1}^p |W_j|^q,
  \qquad q\in\mathcal Q,
\end{equation}
where $\mathcal Q$ contains both small and large values of $q$ together with the
extreme case
\begin{equation}\label{eq:ch2-adaptive-rank-infty}
  T_\infty = \max_{1\le j\le p}|W_j|.
\end{equation}
Small $q$'s favor dense alternatives; large $q$'s favor sparse alternatives. The
adaptive test combines the corresponding $p$-values, either by a minimum
$p$-value rule or by a Cauchy-type combination. The resulting procedure is fully
rank based and therefore robust under heavy tails.

\subsection{Structured correlations}

When the dependence structure has additional information, such as a linear or
banded structure in the precision matrix, one can sharpen the coordinatewise
statistics by incorporating that structure. This is the idea behind
\citet{LiuZhaoFengWang2025StructuredCorr}. The general strategy is the same as in
the Gaussian adaptive literature: construct a dense-alternative statistic, a
sparse-alternative statistic, establish their null laws, and then combine them.
The difference is that here the entire construction is embedded in the robust
elliptical setting.

\section{Strong correlation, normal-reference calibration, and spatial-sign testing}

The preceding sections mostly follow the now standard high-dimensional route: after a
proper centering and bias correction, one proves that a quadratic-form or $U$-statistic
converges to a Gaussian law, an extreme-value law, or a combination of the two. That
route is appropriate only in a \emph{weak-correlation} regime. In many problems this regime
can be summarized spectrally by the requirement
\begin{equation}\label{eq:ch2-strongcorr-weak-spectral}
  \frac{\lambda_{\max}(\mSigma)}{\{\tr(\mSigma^2)\}^{1/2}} \longrightarrow 0,
\end{equation}
or by an equivalent delocalization condition on the leading eigenvalues. Once
\eqref{eq:ch2-strongcorr-weak-spectral} fails, a purely Gaussian calibration can be wrong,
even under the null. This issue has become increasingly important in modern applications,
where strong factor-like dependence, compound symmetry, and spiked structures are common.
The strongest recent lines of work in this direction are the normal-reference procedures of
\citet{ZhangZhouGuo2022,ZhangZhuZhang2023}, the approximate-randomization procedure of
\citet{WangXu2022}, and the spatial-sign correction developed in
\citet{ZhaoFeng2026NoteOneSample}.

\subsection{Why Gaussian calibration fails under strong correlation}

To fix ideas, consider first the one-sample mean problem under a Gaussian model,
\begin{equation}\label{eq:ch2-strongcorr-gaussian-model}
  \vX_i \iid N_p(\vct 0,\mSigma),
  \qquad i=1,\ldots,n,
\end{equation}
and the quadratic statistic
\begin{equation}\label{eq:ch2-strongcorr-Qn}
  Q_n = n\twonorm{\bar{\vX}}^2,
  \qquad
  \bar{\vX}=\frac1n\sum_{i=1}^n \vX_i.
\end{equation}
If
\begin{equation}\label{eq:ch2-strongcorr-eigen-sigma}
  \mSigma = \sum_{j=1}^p \lambda_j \vv_j\vv_j\trans,
  \qquad \lambda_1\ge \cdots \ge \lambda_p\ge 0,
\end{equation}
then under $H_0$ we have the exact representation
\begin{equation}\label{eq:ch2-strongcorr-quad-exact}
  Q_n = \sum_{j=1}^p \lambda_j Z_j^2,
  \qquad Z_1,\ldots,Z_p \iid N(0,1).
\end{equation}
Thus the centered statistic is a weighted chi-square mixture, not a sum of approximately
independent coordinates in the original basis. The usual Gaussian limit appears only when
no single eigen-direction contributes nonnegligibly after standardization.

\begin{assumption}[Leading-eigenvalue regime]
\label{ass:ch2-strongcorr-leading}
Assume that for some fixed integer $r\ge 1$,
\begin{equation}\label{eq:ch2-strongcorr-leading-weights}
  \omega_j = \lim_{n,p\to\infty}
  \frac{\lambda_j}{\{\tr(\mSigma^2)\}^{1/2}} \in [0,1],
  \qquad j=1,\ldots,r,
\end{equation}
exist, and that the remaining spectrum satisfies
\begin{equation}\label{eq:ch2-strongcorr-leading-remainder}
  \frac{\max_{j>r}\lambda_j^2}{\sum_{k>r}\lambda_k^2} \longrightarrow 0.
\end{equation}
In addition,
\begin{equation}\label{eq:ch2-strongcorr-leading-sum}
  \sum_{j=1}^r \omega_j^2 \le 1.
\end{equation}
\end{assumption}

\begin{theorem}[Mixed Gaussian--chi-square limit]
\label{thm:ch2-strongcorr-mixed}
Under \eqref{eq:ch2-strongcorr-quad-exact} and Assumption~\ref{ass:ch2-strongcorr-leading},
\begin{equation}\label{eq:ch2-strongcorr-mixed-limit}
  \frac{Q_n-\tr(\mSigma)}{\{2\tr(\mSigma^2)\}^{1/2}}
  \overset{d}{\longrightarrow}
  \sum_{j=1}^r \omega_j\,\frac{Z_j^2-1}{\sqrt 2}
  + \Bigl(1-\sum_{j=1}^r \omega_j^2\Bigr)^{1/2} Z_0,
\end{equation}
where $Z_0\sim N(0,1)$ is independent of $Z_1,\ldots,Z_r$.
\end{theorem}

\begin{proof}
From \eqref{eq:ch2-strongcorr-quad-exact},
\begin{equation}\label{eq:ch2-strongcorr-proof-center}
  Q_n-\tr(\mSigma) = \sum_{j=1}^p \lambda_j(Z_j^2-1).
\end{equation}
Split the right-hand side into the leading $r$ terms and the remainder:
\begin{equation}\label{eq:ch2-strongcorr-proof-split}
  \frac{Q_n-\tr(\mSigma)}{\{2\tr(\mSigma^2)\}^{1/2}}
  = A_{n,p}+B_{n,p},
\end{equation}
where
\begin{align}
  A_{n,p}
  &= \sum_{j=1}^r
     \frac{\lambda_j}{\{2\tr(\mSigma^2)\}^{1/2}}(Z_j^2-1),
     \label{eq:ch2-strongcorr-proof-A}\\
  B_{n,p}
  &= \sum_{j=r+1}^p
     \frac{\lambda_j}{\{2\tr(\mSigma^2)\}^{1/2}}(Z_j^2-1).
     \label{eq:ch2-strongcorr-proof-B}
\end{align}
By \eqref{eq:ch2-strongcorr-leading-weights},
\begin{equation}\label{eq:ch2-strongcorr-proof-A-limit}
  A_{n,p}
  \overset{d}{\longrightarrow}
  \sum_{j=1}^r \omega_j\,\frac{Z_j^2-1}{\sqrt 2}.
\end{equation}
For the remainder, define
\begin{equation}\label{eq:ch2-strongcorr-proof-scale}
  s_{n,p}^2 = 2\sum_{j=r+1}^p \lambda_j^2,
  \qquad
  \widetilde B_{n,p}
  = \sum_{j=r+1}^p \frac{\lambda_j}{s_{n,p}}(Z_j^2-1).
\end{equation}
Then $\Var(\widetilde B_{n,p})=1$, and
\begin{equation}\label{eq:ch2-strongcorr-proof-lyapunov}
  \sum_{j=r+1}^p \E\left|\frac{\lambda_j}{s_{n,p}}(Z_j^2-1)\right|^4
  = O\!\left( \frac{\sum_{j=r+1}^p \lambda_j^4}{(\sum_{j=r+1}^p \lambda_j^2)^2} \right)
  \le O\!\left( \frac{\max_{j>r}\lambda_j^2}{\sum_{k>r}\lambda_k^2} \right)
  \longrightarrow 0
\end{equation}
by \eqref{eq:ch2-strongcorr-leading-remainder}. Hence the Lyapunov central limit theorem gives
$\widetilde B_{n,p}\overset{d}{\to}N(0,1)$. Since
\begin{equation}\label{eq:ch2-strongcorr-proof-B-rescaled}
  B_{n,p}
  = \frac{s_{n,p}}{\{2\tr(\mSigma^2)\}^{1/2}}\,\widetilde B_{n,p}
  = \left(1-\sum_{j=1}^r \frac{\lambda_j^2}{\tr(\mSigma^2)}\right)^{1/2}\widetilde B_{n,p},
\end{equation}
we obtain
\begin{equation}\label{eq:ch2-strongcorr-proof-B-limit}
  B_{n,p}
  \overset{d}{\longrightarrow}
  \Bigl(1-\sum_{j=1}^r\omega_j^2\Bigr)^{1/2}Z_0.
\end{equation}
The leading and remainder parts involve disjoint Gaussian coordinates and are therefore
independent. Combining \eqref{eq:ch2-strongcorr-proof-A-limit} and
\eqref{eq:ch2-strongcorr-proof-B-limit} proves
\eqref{eq:ch2-strongcorr-mixed-limit}.
\end{proof}

Theorem~\ref{thm:ch2-strongcorr-mixed} explains the basic obstruction: once a few spiked
eigenvalues persist after normalization, a fixed Gaussian calibration is no longer exact.
The same phenomenon lies behind a large family of recent \emph{normal-reference} and
\emph{randomization-based} procedures.

\subsection{The normal-reference line of Jin-Ting Zhang and collaborators}

The papers of \citet{ZhangZhouGuo2022} and \citet{ZhangZhuZhang2023} start from the
observation that the null law of a centered $L_2$-type statistic is usually much closer to a
chi-square-type mixture than to a standard normal variable when the covariance structure is
strongly correlated. For the one-sample problem, the starting statistic is again
\eqref{eq:ch2-strongcorr-Qn}. Writing
\begin{equation}\label{eq:ch2-strongcorr-cumulants}
  a_1 = \tr(\mSigma),
  \qquad
  a_2 = \tr(\mSigma^2),
  \qquad
  a_3 = \tr(\mSigma^3),
\end{equation}
we see from \eqref{eq:ch2-strongcorr-quad-exact} that
\begin{equation}\label{eq:ch2-strongcorr-Qn-moments}
  \E(Q_n)=a_1,\qquad \Var(Q_n)=2a_2,\qquad \operatorname{cum}_3(Q_n)=8a_3.
\end{equation}
where $\kappa_3(Q_n)$ denotes the third cumulant. The normal-reference idea is to replace the
standard normal approximation by a centered chi-square-type reference
\begin{equation}\label{eq:ch2-strongcorr-normal-reference}
  R_n = a\chi_\nu^2 + b,
\end{equation}
where $(a,\nu,b)$ are chosen so that the first few cumulants of $R_n$ match those of the target
quadratic-form statistic. In \citet{ZhangZhouGuo2022}, this matching is carried out at the level
of a three-cumulant chi-square approximation. The resulting procedure is especially useful in
regimes where the limiting law is non-Gaussian but still dominated by a moderate number of
leading eigen-directions.

For the two-sample Behrens--Fisher problem, \citet{ZhangZhuZhang2023} pursue the same
principle in a scale-invariant setting. Let
\begin{equation}\label{eq:ch2-strongcorr-Psi}
  \bm{\Psi} = \frac{\mSigma_1}{n_1}+\frac{\mSigma_2}{n_2},
\end{equation}
and consider the centered $L_2$-type two-sample statistic
\begin{equation}\label{eq:ch2-strongcorr-two-sample-l2}
  Q_{BF}
  = \twonorm{\bar{\vX}_1-\bar{\vX}_2}^2,
  \qquad
  \E(Q_{BF}) = \tr(\bm{\Psi}).
\end{equation}
The null fluctuation is again determined by the eigenvalues of $\bm{\Psi}$, and the weak-correlation
normal approximation can fail for exactly the same reason as in the one-sample case.
\citet{ZhangZhuZhang2023} therefore calibrate the test by an $F$-type normal-reference law,
constructed from a ratio of chi-square-type mixtures with consistently estimated degrees of
freedom. From the methodological standpoint of this book, the main message is that these
normal-reference procedures retain the classical $L_2$ statistic but replace the fragile Gaussian
reference by a reference family that still behaves correctly under strong dependence.

\subsection{Approximate randomization under arbitrary covariances}

The Biometrika paper of \citet{WangXu2022} takes a different route. Instead of approximating
the non-Gaussian limit analytically, it builds a conditional reference distribution by
randomization. The starting point is the Chen--Qin two-sample statistic
\begin{equation}\label{eq:ch2-strongcorr-CQ}
  T_{CQ}
  = \frac{1}{n_1(n_1-1)}\sum_{i\neq j}^{n_1} \vX_i\trans \vX_j
  + \frac{1}{n_2(n_2-1)}\sum_{i\neq j}^{n_2} \vY_i\trans \vY_j
  - \frac{2}{n_1n_2}\sum_{i=1}^{n_1}\sum_{j=1}^{n_2} \vX_i\trans \vY_j.
\end{equation}
Under weak conditions, the null law of \eqref{eq:ch2-strongcorr-CQ} may have many possible
subsequential limits, depending on the covariance eigenstructure. This is precisely the situation
in which a single closed-form approximation is hard to justify uniformly.

To overcome this difficulty, \citet{WangXu2022} construct an approximate-randomization
reference by reassigning the centered observations to pseudo-groups of sizes $n_1$ and $n_2$ and
recomputing the same Chen--Qin statistic. Denote the randomized version by
$T_{CQ}^{\pi}$, where $\pi$ indexes the random reassignment. The test then rejects for large values
of $T_{CQ}$ relative to the conditional distribution of $T_{CQ}^{\pi}$. The core theoretical result
shows that, under the null, the randomization distribution automatically adapts to all possible
subsequential limits of the original statistic. Therefore the procedure remains asymptotically exact
without imposing any weak-correlation condition such as \eqref{eq:ch2-strongcorr-weak-spectral}.
This is the main reason the Wang--Xu paper has become a central reference in the strong-correlation
literature.

\subsection{Pairwise-difference-quantile spatial-sign tests under arbitrary dependence}\label{subsec:ch2-pdq}

The two-sample strong-correlation problem is more delicate for spatial signs than for
mean-based quadratic statistics. The mean-based procedures of \citet{WangXu2022} and
related normal-reference tests avoid estimating a full covariance matrix, but they still use
ordinary observations. In contrast, a spatial-sign test must first standardize the coordinates and
then compute directions. The diagonal fixed-point standardization used in
\citet{FengZouWang2016JASA} and \citet{LiWangZou2016SimpleTwoSample} is effective in the
weak-correlation regime, but it does not generally estimate the diagonal of the elliptical shape
matrix under arbitrary dense dependence. \citet{FengWang2026PDQ} therefore propose a
pairwise-difference-quantile, or PDQ, standardizer and then develop a full spatial-sign test whose
null law is a weighted chi-square mixture under arbitrary dependence.

Consider two independent samples
\begin{equation}\label{eq:ch2-pdq-model}
  \vX_{ki}=\vmu_k+\xi_{ki}\mGamma_k\vu_{ki},
  \qquad i=1,\ldots,n_k,\quad k=1,2,
\end{equation}
where $\vu_{ki}$ is uniform on $\mathbb S^{p-1}$, independent of the radial variable
$\xi_{ki}\ge0$, and
\begin{equation}\label{eq:ch2-pdq-shape}
  \mS_k=\mGamma_k\mGamma_k\trans,
  \qquad
  \mD_k^{\circ}=\diag(\mS_k)=\diag(d_{k1}^{\circ},\ldots,d_{kp}^{\circ}),
  \qquad
  \mR_k=(\mD_k^{\circ})^{-1/2}\mS_k(\mD_k^{\circ})^{-1/2}.
\end{equation}
The hypothesis is
\begin{equation}\label{eq:ch2-pdq-hypothesis}
  H_0:\vmu_1=\vmu_2
  \qquad \text{versus}\qquad
  H_1:\vmu_1\ne\vmu_2.
\end{equation}
The goal is to test \eqref{eq:ch2-pdq-hypothesis} without imposing a weak-correlation
condition such as \eqref{eq:ch2-strongcorr-weak-spectral} on $\mR_1$ or $\mR_2$.

\subsubsection{Pairwise-difference-quantile diagonal standardization}

Let $\vct{e}_1=(1,0,\ldots,0)\trans$. Since $\vu_{ki}$ is uniform on the sphere, the distribution of
$\xi_{ki}\vv\trans\vu_{ki}$ is the same for all unit vectors $\vv$. Let $Z_{k1}$ and $Z_{k2}$ be
independent copies of $\xi_{k1}\vct{e}_1\trans\vu_{k1}$, and define
\begin{equation}\label{eq:ch2-pdq-pop-quantile}
  F_{k,\Delta}(t)=\Prob\{\abs{Z_{k1}-Z_{k2}}\le t\},
  \qquad
  q_{k,\alpha}=F_{k,\Delta}^{-1}(\alpha),
  \qquad 0<\alpha<1.
\end{equation}
For the $j$th coordinate,
\begin{equation}\label{eq:ch2-pdq-coordinate-difference}
  X_{kij}-X_{ki'j}
  = (d_{kj}^{\circ})^{1/2}(Z_{ki,j}-Z_{ki',j}),
\end{equation}
where $Z_{ki,j}$ has the same distribution as $Z_{k1}$. Hence the $\alpha$th quantile of
$\abs{X_{kij}-X_{ki'j}}$ equals $q_{k,\alpha}(d_{kj}^{\circ})^{1/2}$. This identifies the diagonal
shape up to the scalar $q_{k,\alpha}^2$, which is harmless for spatial signs. Define
\begin{equation}\label{eq:ch2-pdq-target-D}
  \mD_k=q_{k,\alpha}^2\mD_k^{\circ}
  =\diag(d_{k1},\ldots,d_{kp}),
  \qquad d_{kj}=q_{k,\alpha}^2d_{kj}^{\circ}.
\end{equation}
Then $U\{(c\mD_k)^{-1/2}\vx\}=U(\mD_k^{-1/2}\vx)$ for every $c>0$ and every
$\vx\ne\vct 0$.

The empirical PDQ standardizer is a coordinatewise $U$-quantile. For $t\ge0$, set
\begin{equation}\label{eq:ch2-pdq-empirical-cdf}
  \widehat F_{kj}(t)
  = \binom{n_k}{2}^{-1}\sum_{1\le i<i'\le n_k}
      \mathbbm 1\{\abs{X_{kij}-X_{ki'j}}\le t\},
\end{equation}
and define
\begin{equation}\label{eq:ch2-pdq-Dhat}
  \widehat q_{kj,\alpha}=\inf\{t\ge0:\widehat F_{kj}(t)\ge\alpha\},
  \qquad
  \widehat d_{kj}=\widehat q_{kj,\alpha}^2,
  \qquad
  \widehat{\mD}_k=\diag(\widehat d_{k1},\ldots,\widehat d_{kp}).
\end{equation}
The construction is location free because it uses only pairwise differences. It also avoids
positive moment assumptions on $\xi_{ki}$, which is important in very heavy-tailed elliptical
models.

Given $\widehat{\mD}_k$, define the standardized spatial median
\begin{equation}\label{eq:ch2-pdq-med}
  \widehat{\vmu}_k
  \in \argmin_{\vtheta\in\R^p}
  \sum_{i=1}^{n_k}\norm{\widehat{\mD}_k^{-1/2}(\vX_{ki}-\vtheta)},
\end{equation}
and the fitted standardized signs
\begin{equation}\label{eq:ch2-pdq-fitted-signs}
  \widehat{\vY}_{ki}=\widehat{\mD}_k^{-1/2}(\vX_{ki}-\widehat{\vmu}_k),
  \qquad
  \widehat{\vS}_{ki}=U(\widehat{\vY}_{ki}).
\end{equation}
The population analogues are
\begin{equation}\label{eq:ch2-pdq-oracle-Y-S}
  \vY_{ki}=\mD_k^{-1/2}(\vX_{ki}-\vmu_k),
  \qquad
  \vS_{ki}=U(\vY_{ki}).
\end{equation}
Because of \eqref{eq:ch2-pdq-target-D},
\begin{equation}\label{eq:ch2-pdq-angular-sign}
  \vS_{ki}
  =\frac{\mR_k^{1/2}\vu_{ki}}{(\vu_{ki}\trans\mR_k\vu_{ki})^{1/2}},
\end{equation}
so the sign distribution is completely free of the radial distribution.

\subsubsection{The PDQ statistic and its quadratic expansion}

The PDQ two-sample statistic is the full-sample cross-centered spatial-sign statistic
\begin{equation}\label{eq:ch2-pdq-Rhat}
  \widehat R_n^{\rm PDQ}
  =-\frac1{n_1n_2}\sum_{i=1}^{n_1}\sum_{j=1}^{n_2}
  U\{\widehat{\mD}_1^{-1/2}(\vX_{1i}-\widehat{\vmu}_2)\}\trans
  U\{\widehat{\mD}_2^{-1/2}(\vX_{2j}-\widehat{\vmu}_1)\}.
\end{equation}
Under $H_0$, write the common location as $\vmu$. Define
\begin{equation}\label{eq:ch2-pdq-Omega-G}
  \mOmega_k=\E(\vS_{ki}\vS_{ki}\trans),
  \qquad
  \mG_k=\E\left[\norm{\vY_{ki}}^{-1}
  (\mI_p-\vS_{ki}\vS_{ki}\trans)\right].
\end{equation}
The matrix $\mG_k$ is the Jacobian of the spatial-median score after diagonal standardization.
Let
\begin{equation}\label{eq:ch2-pdq-A12-A21}
  \mA_{12}=\mD_1^{-1/2}\mD_2^{1/2},
  \qquad
  \mA_{21}=\mD_2^{-1/2}\mD_1^{1/2}.
\end{equation}
The three deterministic matrices in the first-order expansion are
\begin{align}
  \mK_1&=\frac12\{\mG_2\mA_{21}\mG_1^{-1}
       +(\mG_2\mA_{21}\mG_1^{-1})\trans\},
       \label{eq:ch2-pdq-K1}\\
  \mK_2&=\frac12\{\mG_2^{-1}\mA_{12}\trans\mG_1
       +(\mG_2^{-1}\mA_{12}\trans\mG_1)\trans\},
       \label{eq:ch2-pdq-K2}\\
  \mK_3&=\mI_p+
       (\mG_2^{-1}\mA_{12}\trans\mG_1\mG_2\mA_{21}\mG_1^{-1})\trans.
       \label{eq:ch2-pdq-K3}
\end{align}
The leading oracle quadratic form is
\begin{equation}\label{eq:ch2-pdq-Qn}
  Q_n
  =\bar{\vS}_1\trans\mK_1\bar{\vS}_1
   +\bar{\vS}_2\trans\mK_2\bar{\vS}_2
   -\bar{\vS}_1\trans\mK_3\bar{\vS}_2,
  \qquad
  \bar{\vS}_k=\frac1{n_k}\sum_{i=1}^{n_k}\vS_{ki}.
\end{equation}
Its mean is
\begin{equation}\label{eq:ch2-pdq-bn}
  b_n=\E Q_n
  =\frac1{n_1}\tr(\mK_1\mOmega_1)+\frac1{n_2}\tr(\mK_2\mOmega_2).
\end{equation}
For the limiting law, set
\begin{equation}\label{eq:ch2-pdq-Bn}
  \mB_n=\begin{pmatrix}
    n_1^{-1}\mOmega_1^{1/2}\mK_1\mOmega_1^{1/2}
    &-(2\sqrt{n_1n_2})^{-1}\mOmega_1^{1/2}\mK_3\mOmega_2^{1/2}\\
    -(2\sqrt{n_1n_2})^{-1}\mOmega_2^{1/2}\mK_3\trans\mOmega_1^{1/2}
    &n_2^{-1}\mOmega_2^{1/2}\mK_2\mOmega_2^{1/2}
  \end{pmatrix}.
\end{equation}
Let $\lambda_{n1},\ldots,\lambda_{n,2p}$ be the eigenvalues of $\mB_n$, and define
\begin{equation}\label{eq:ch2-pdq-tau}
  \tau_n^2=2\tr(\mB_n^2)=2\sum_{r=1}^{2p}\lambda_{nr}^2.
\end{equation}
The diagonal-deleted oracle statistic is
\begin{equation}\label{eq:ch2-pdq-Un}
\begin{aligned}
  U_n={}&\frac1{n_1^2}\sum_{i\ne i'}\vS_{1i}\trans\mK_1\vS_{1i'}
        +\frac1{n_2^2}\sum_{j\ne j'}\vS_{2j}\trans\mK_2\vS_{2j'}  \\
      &\quad -\frac1{n_1n_2}\sum_{i=1}^{n_1}\sum_{j=1}^{n_2}
        \vS_{1i}\trans\mK_3\vS_{2j}.
\end{aligned}
\end{equation}
The feasible diagonal deletion uses
\begin{equation}\label{eq:ch2-pdq-Omegahat-Ghat}
  \widehat{\mOmega}_k=n_k^{-1}\sum_{i=1}^{n_k}\widehat{\vS}_{ki}\widehat{\vS}_{ki}\trans,
  \qquad
  \widehat{\mG}_k=n_k^{-1}\sum_{i=1}^{n_k}\norm{\widehat{\vY}_{ki}}^{-1}
  (\mI_p-\widehat{\vS}_{ki}\widehat{\vS}_{ki}\trans),
\end{equation}
and the plug-in matrices $\widehat{\mK}_1,\widehat{\mK}_2,\widehat{\mK}_3$ obtained by replacing
$\mD_k,\mG_k$ in \eqref{eq:ch2-pdq-K1}--\eqref{eq:ch2-pdq-K3}. Define
\begin{equation}\label{eq:ch2-pdq-bhat-T}
  \widehat b_n
  =\frac1{n_1}\tr(\widehat{\mK}_1\widehat{\mOmega}_1)
   +\frac1{n_2}\tr(\widehat{\mK}_2\widehat{\mOmega}_2),
  \qquad
  T_n^{\rm PDQ}=\widehat R_n^{\rm PDQ}-\widehat b_n.
\end{equation}
The term $\widehat b_n$ is not used as a high-precision estimator of $b_n$ at a scale smaller
than $\tau_n$. Its role is to remove the empirical diagonal terms generated by the full-sample
quadratic expansion.

\subsubsection{Assumptions and main theory}

Let
\begin{equation}\label{eq:ch2-pdq-rates}
  N=n_1+n_2,
  \qquad n_0=\min(n_1,n_2),
  \qquad x_N=\log\{8pN\},
  \qquad r_D=(x_N/n_0)^{1/2}.
\end{equation}

\begin{assumption}[PDQ quantile and radial regularity]\label{ass:ch2-pdq-qr}
Assume $n_1/N\to\kappa\in(0,1)$ and $x_N/n_0\to0$. The elliptical model
\eqref{eq:ch2-pdq-model} holds, and there are constants $0<\underline d<\bar d<\infty$ such
that $\underline d\le d_{kj}^{\circ}\le\bar d$ for all $k,j$. For the fixed $\alpha\in(0,1)$, the
quantiles $q_{k,\alpha}$ are bounded away from zero and infinity, and there are constants
$c_F,C_F,\varepsilon_F>0$ such that
\begin{equation}\label{eq:ch2-pdq-local-slope}
  c_F\abs{t-q_{k,\alpha}}
  \le \abs{F_{k,\Delta}(t)-\alpha}
  \le C_F\abs{t-q_{k,\alpha}}
\end{equation}
whenever $\abs{t-q_{k,\alpha}}\le\varepsilon_F$. Moreover, with $\vY_{ki}$ defined by
\eqref{eq:ch2-pdq-oracle-Y-S}, there exists $M_{Y,n}\ge1$ such that
\begin{equation}\label{eq:ch2-pdq-inverse-radius}
  \max_{k=1,2}\left[
  p^{1/2}\E\norm{\vY_{k1}}^{-1}
  +p\E\norm{\vY_{k1}}^{-2}
  +p^2\E\norm{\vY_{k1}}^{-4}
  \right]\le M_{Y,n},
\end{equation}
and
\begin{equation}\label{eq:ch2-pdq-smallball}
  \max_{k=1,2}\sup_{0<t\le1}
  t^{-2}\Prob\{\norm{\vY_{k1}}\le tp^{1/2}\}\le M_{Y,n}.
\end{equation}
\end{assumption}

Define
\begin{equation}\label{eq:ch2-pdq-Cn}
\begin{aligned}
  \mathfrak C_n=1+M_{Y,n}^2+
  \max_{k=1,2}\Big[&\{\opnorm{p^{1/2}\mG_k}
  +\opnorm{(p^{1/2}\mG_k)^{-1}}\}^4 \\
  &+\sup_{\norm{\va}=1}\E\{\va\trans\mOmega_k^{-1/2}\vS_{ki}\}^4\Big].
\end{aligned}
\end{equation}
Set
\begin{equation}\label{eq:ch2-pdq-rate-objects}
  q_{\mu,n}=\mathfrak C_n p^{1/2}x_N/n_0,
  \qquad
  a_{S,n}=\mathfrak C_n\{r_D+n_0^{-1/2}+x_N/n_0\},
\end{equation}
and
\begin{equation}\label{eq:ch2-pdq-aK}
  a_{K,n}=\mathfrak C_n\{r_D+(x_N/n_0)^{1/2}+a_{S,n}\},
  \qquad
  L_{K,n}=1+\max_{1\le r\le3}\opnorm{\mK_r}.
\end{equation}

\begin{assumption}[PDQ spectral condition]\label{ass:ch2-pdq-spectral}
Assume $\tau_n^2=2\tr(\mB_n^2)>0$. With
\begin{equation}\label{eq:ch2-pdq-rho}
  \rho_{nr}=\lambda_{nr}/\{\tr(\mB_n^2)\}^{1/2},
  \qquad r=1,\ldots,2p,
\end{equation}
there exists a square-summable sequence $\rho_1,\rho_2,\ldots$ such that, after setting
$\rho_{nr}=0$ for $r>2p$,
\begin{equation}\label{eq:ch2-pdq-rho-l2}
  \sum_{r\ge1}(\rho_{nr}-\rho_r)^2\to0,
  \qquad
  \rho_0^2=1-\sum_{r\ge1}\rho_r^2\ge0.
\end{equation}
Furthermore,
\begin{equation}\label{eq:ch2-pdq-main-small}
  r_D+q_{\mu,n}+a_{K,n}+L_{K,n}a_{S,n}+\frac{x_N}{n_0^2\tau_n}\to0,
\end{equation}
and the row-leverage quantity
\begin{equation}\label{eq:ch2-pdq-row}
\begin{aligned}
  \Delta_{{\rm row},n}=\tau_n^{-2}\max\Bigg\{&
  \frac{\opnorm{\mK_1\mOmega_1\mK_1}}{n_1^3},
  \frac{\opnorm{\mK_2\mOmega_2\mK_2}}{n_2^3},\\
  &\frac{\opnorm{\mK_3\mOmega_2\mK_3\trans}}{n_1^2n_2},
  \frac{\opnorm{\mK_3\trans\mOmega_1\mK_3}}{n_1n_2^2}
  \Bigg\}
\end{aligned}
\end{equation}
satisfies $\Delta_{{\rm row},n}\to0$.
\end{assumption}

\begin{theorem}[Uniform PDQ diagonal scale rate]\label{thm:ch2-pdq-D-rate}
Under $H_0$ and Assumption~\ref{ass:ch2-pdq-qr},
\begin{equation}\label{eq:ch2-pdq-D-rate}
  \max_{k=1,2}\max_{1\le j\le p}
  \abs{\frac{\widehat d_{kj}}{d_{kj}}-1}
  =O_p(r_D).
\end{equation}
\end{theorem}

\begin{proposition}[PDQ spatial-median and sign expansion]\label{prop:ch2-pdq-median-sign}
Under $H_0$ and Assumption~\ref{ass:ch2-pdq-qr},
\begin{equation}\label{eq:ch2-pdq-median-bahadur}
  \mD_k^{-1/2}(\widehat{\vmu}_k-\vmu)
  =\mG_k^{-1}\bar{\vS}_k+\vDelta_k,
  \qquad
  \max_{k=1,2}\norm{\vDelta_k}=O_p(q_{\mu,n}).
\end{equation}
Moreover,
\begin{equation}\label{eq:ch2-pdq-sign-perturbation}
  \max_{k=1,2}n_k^{-1}\sum_{i=1}^{n_k}
  \norm{\widehat{\vS}_{ki}-\vS_{ki}}=O_p(a_{S,n}),
\end{equation}
and
\begin{equation}\label{eq:ch2-pdq-K-rate}
  \max_{1\le r\le3}\opnorm{\widehat{\mK}_r-\mK_r}=O_p(a_{K,n}).
\end{equation}
\end{proposition}

\begin{theorem}[Weighted chi-square null law for the PDQ statistic]\label{thm:ch2-pdq-null-law}
Under $H_0$ and Assumptions~\ref{ass:ch2-pdq-qr}--\ref{ass:ch2-pdq-spectral}, let
\begin{equation}\label{eq:ch2-pdq-Gamma}
  \Gamma_n=\sum_{r=1}^{2p}\lambda_{nr}(\chi_{1r}^2-1),
\end{equation}
where $\chi_{11}^2,\ldots,\chi_{1,2p}^2$ are independent chi-square random variables with one
degree of freedom. Then
\begin{equation}\label{eq:ch2-pdq-oracle-law}
  \sup_{t\in\R}\abs{
  \Prob\left(\frac{U_n}{\tau_n}\le t\right)
  -\Prob\left(\frac{\Gamma_n}{\tau_n}\le t\right)}\to0,
\end{equation}
and
\begin{equation}\label{eq:ch2-pdq-mixture-limit}
  \frac{U_n}{\tau_n}
  \overset{d}{\longrightarrow}
  \frac1{\sqrt2}\sum_{r=1}^{\infty}\rho_r(\chi_{1r}^2-1)+\rho_0Z,
\end{equation}
where $Z\sim N(0,1)$ is independent of the chi-square variables. The feasible statistic satisfies
\begin{equation}\label{eq:ch2-pdq-feasible-expansion}
  \frac{T_n^{\rm PDQ}-U_n}{\tau_n}
  =O_p\left(r_D+a_{K,n}+L_{K,n}a_{S,n}+\frac{x_N}{n_0^2\tau_n}\right),
\end{equation}
and therefore the law of $T_n^{\rm PDQ}/\tau_n$ is also approximated by
$\Gamma_n/\tau_n$.
\end{theorem}

\begin{corollary}[Normal limit as a special case]\label{cor:ch2-pdq-normal}
Under the assumptions of Theorem~\ref{thm:ch2-pdq-null-law}, if
\begin{equation}\label{eq:ch2-pdq-no-domination}
  \lambda_{\max}^2(\mB_n)/\tr(\mB_n^2)\to0,
\end{equation}
then
\begin{equation}\label{eq:ch2-pdq-normal-limit}
  T_n^{\rm PDQ}/\tau_n\overset{d}{\longrightarrow}N(0,1).
\end{equation}
\end{corollary}

The normal approximation is thus not assumed in the PDQ procedure; it is recovered only when
no eigenvalue of $\mB_n$ dominates. Under compound-symmetry or spiked dependence,
\eqref{eq:ch2-pdq-no-domination} may fail, and the chi-square mixture in
\eqref{eq:ch2-pdq-mixture-limit} is the correct reference law.

\subsubsection{Rademacher wild bootstrap calibration}

The weighted chi-square limit in Theorem~\ref{thm:ch2-pdq-null-law} depends on unknown
spectral weights. \citet{FengWang2026PDQ} therefore calibrate the feasible statistic by a
Rademacher wild bootstrap. Let $e_{ki}$ be independent Rademacher variables, conditional on
the data, with
\begin{equation}\label{eq:ch2-pdq-rademacher}
  \Prob^\ast(e_{ki}=1)=\Prob^\ast(e_{ki}=-1)=1/2.
\end{equation}
Define
\begin{equation}\label{eq:ch2-pdq-barS-star}
  \bar{\vS}_k^\ast=n_k^{-1}\sum_{i=1}^{n_k}e_{ki}\widehat{\vS}_{ki},
  \qquad k=1,2,
\end{equation}
and
\begin{equation}\label{eq:ch2-pdq-Qstar}
  Q_n^\ast
  =\bar{\vS}_1^{\ast\trans}\widehat{\mK}_1\bar{\vS}_1^\ast
   +\bar{\vS}_2^{\ast\trans}\widehat{\mK}_2\bar{\vS}_2^\ast
   -\bar{\vS}_1^{\ast\trans}\widehat{\mK}_3\bar{\vS}_2^\ast.
\end{equation}
Since $\E^\ast(e_{ki})=0$ and $\E^\ast(e_{ki}^2)=1$,
\begin{equation}\label{eq:ch2-pdq-Qstar-mean}
  \E^\ast(Q_n^\ast)=\widehat b_n.
\end{equation}
The centered bootstrap statistic is
\begin{equation}\label{eq:ch2-pdq-Tstar}
  T_n^\ast=Q_n^\ast-\widehat b_n.
\end{equation}

\begin{theorem}[Wild bootstrap validity for PDQ]\label{thm:ch2-pdq-bootstrap}
Under $H_0$ and Assumptions~\ref{ass:ch2-pdq-qr}--\ref{ass:ch2-pdq-spectral},
\begin{equation}\label{eq:ch2-pdq-boot-valid}
  \sup_{t\in\R}\abs{
  \Prob^\ast\left(\frac{T_n^\ast}{\tau_n}\le t\right)
  -\Prob\left(\frac{T_n^{\rm PDQ}}{\tau_n}\le t\right)}\to0
  \qquad\text{in probability}.
\end{equation}
Consequently, if $c_{1-\beta}^\ast$ is the conditional $(1-\beta)$ quantile of
$T_n^\ast$, the test that rejects when $T_n^{\rm PDQ}>c_{1-\beta}^\ast$ has asymptotic size
$\beta$ at continuity points of the limiting null distribution.
\end{theorem}

A convenient implementation is as follows. First compute $\widehat{\mD}_k$ from the
coordinatewise pairwise-difference quantiles in \eqref{eq:ch2-pdq-empirical-cdf}--\eqref{eq:ch2-pdq-Dhat}.
Second compute $\widehat{\vmu}_k$, $\widehat{\vS}_{ki}$, $\widehat{\mOmega}_k$ and
$\widehat{\mG}_k$ from \eqref{eq:ch2-pdq-med}--\eqref{eq:ch2-pdq-Omegahat-Ghat}. Third form
$T_n^{\rm PDQ}$ from \eqref{eq:ch2-pdq-Rhat} and \eqref{eq:ch2-pdq-bhat-T}. Finally generate
Rademacher multipliers repeatedly, compute $T_n^\ast$ in \eqref{eq:ch2-pdq-Tstar}, and reject
for large values of $T_n^{\rm PDQ}$ relative to the conditional bootstrap quantiles.

The PDQ method therefore fills the gap between two previously separate literatures. It keeps
the heavy-tail robustness of spatial signs, but it adopts the strong-correlation perspective of
normal-reference and randomization procedures by allowing the null law to be a weighted
chi-square mixture rather than forcing a Gaussian calibration.

\subsection{The Zhao--Feng note on the one-sample spatial-sign test}

The strong-correlation issue also arises for robust procedures. Earlier in this chapter, in the discussion surrounding Theorem~\ref{thm:ch2-WPL-null}
and Remark~\ref{rem:ch2-WPL-not-scale-invariant}, we reviewed the
one-sample spatial-sign test of \citet{WangPengLi2015}. That paper established asymptotic
normality under structural conditions on the scatter matrix. The note of
\citet{ZhaoFeng2026NoteOneSample} revisits the same statistic and shows that under strong
dependence the null law is generally not Gaussian.

Let $\vU_i = U(\vX_i)$ and define
\begin{equation}\label{eq:ch2-strongcorr-sign-Sn}
  S_n = \sum_{1\le i<j\le n} \vU_i\trans \vU_j.
\end{equation}
Using $\twonorm{\vU_i}=1$, one has the identity
\begin{equation}\label{eq:ch2-strongcorr-sign-identity}
  S_n = \frac12 \left\|\sum_{i=1}^n \vU_i\right\|^2 - \frac n2.
\end{equation}
Let
\begin{equation}\label{eq:ch2-strongcorr-sign-sigma}
  \mSigma_U = \E(\vU_i\vU_i\trans),
  \qquad
  \tau = \tr(\mSigma_U^2),
  \qquad
  \sigma_n^2 = \binom{n}{2}\tau,
  \qquad
  T_n = \frac{S_n}{\sigma_n}.
\end{equation}
The first key theorem in \citet{ZhaoFeng2026NoteOneSample} compares $T_n$ with a Gaussian
counterpart. Let $\vG_1,\ldots,\vG_n\iid N(\vct 0,\mSigma_U)$ and define
\begin{equation}\label{eq:ch2-strongcorr-sign-gaussian-counterpart}
  S_n^{(G)} = \sum_{1\le i<j\le n} \vG_i\trans \vG_j,
  \qquad
  T_n^{(G)} = \frac{S_n^{(G)}}{\sigma_n}.
\end{equation}
Also let
\begin{equation}\label{eq:ch2-strongcorr-sign-kappa4}
  \kappa_4
  = \frac{\E(\vU_1\trans \vU_2)^4}{\E^2(\vU_1\trans \vU_2)^2}
  = \frac{\E(\vU_1\trans \vU_2)^4}{\tau^2}.
\end{equation}
Then the paper proves the following comparison bound.

\begin{theorem}[Gaussian comparison for the spatial-sign statistic]
\label{thm:ch2-strongcorr-sign-gaussian}
Assume $\P(\vX_1=\vct 0)=0$ and $\tau=\tr(\mSigma_U^2)>0$. Then
\begin{equation}\label{eq:ch2-strongcorr-sign-gaussian-bound}
  \bigl\|\mathcal L(T_n)-\mathcal L(T_n^{(G)})\bigr\|_3
  \le C\kappa_4^{3/4}n^{-1/2},
\end{equation}
for a universal constant $C>0$. In particular, if
\begin{equation}\label{eq:ch2-strongcorr-sign-kappa-cond}
  \kappa_4=o(n^{2/3}),
\end{equation}
then
\begin{equation}\label{eq:ch2-strongcorr-sign-gaussian-close}
  \bigl\|\mathcal L(T_n)-\mathcal L(T_n^{(G)})\bigr\|_3 \to 0.
\end{equation}
\end{theorem}

The second step is to identify the limit law of the Gaussian analogue. Let $\xi\sim N(\vct 0,\mI_p)$
and define the Gaussian quadratic form
\begin{equation}\label{eq:ch2-strongcorr-sign-Qp}
  Q_p = \frac{\xi\trans \mSigma_U \xi - \tr(\mSigma_U)}{\{2\tr(\mSigma_U^2)\}^{1/2}}.
\end{equation}
The note proves that
\begin{equation}\label{eq:ch2-strongcorr-sign-gaussian-reduction}
  T_n^{(G)} = Q_p + o_p(1),
\end{equation}
and therefore the null law of $T_n$ is determined by the spectral structure of
$\mSigma_U$ rather than by a universal normal limit. Writing the eigenvalues of $\mSigma_U$
as $\lambda_1\ge \cdots \ge \lambda_p\ge 0$ and
\begin{equation}\label{eq:ch2-strongcorr-sign-alpha}
  \alpha_i = \frac{\lambda_i}{\tau^{1/2}},
  \qquad
  \sum_{i=1}^p \alpha_i^2 = 1,
\end{equation}
the exact subsequential limit is the following.

\begin{corollary}[Mixed limit for the spatial-sign statistic]
\label{cor:ch2-strongcorr-sign-mixed}
Assume the conditions of Theorem~\ref{thm:ch2-strongcorr-sign-gaussian} and, along a
subsequence, $\alpha_i\to \alpha_i^{\ast}$ for each fixed $i$. Then
\begin{equation}\label{eq:ch2-strongcorr-sign-mixed}
  T_n \overset{d}{\longrightarrow}
  \Bigl(1-\sum_{i\ge 1}(\alpha_i^{\ast})^2\Bigr)^{1/2} Z_0
  + \frac{1}{\sqrt 2}\sum_{i\ge 1} \alpha_i^{\ast}(Z_i^2-1),
\end{equation}
where $Z_0,Z_1,Z_2,\ldots$ are independent $N(0,1)$ variables. In particular, if
\begin{equation}\label{eq:ch2-strongcorr-sign-normality}
  \max_{1\le i\le p}\alpha_i \to 0
  \quad\Longleftrightarrow\quad
  \frac{\tr(\mSigma_U^4)}{\tr^2(\mSigma_U^2)} \to 0,
\end{equation}
then $T_n\overset{d}{\to}N(0,1)$.
\end{corollary}

Corollary~\ref{cor:ch2-strongcorr-sign-mixed} is the precise answer to the strong-correlation
question for the one-sample spatial-sign test: the null law is Gaussian only in the delocalized
regime \eqref{eq:ch2-strongcorr-sign-normality}; otherwise it contains non-negligible chi-square
components determined by the leading eigenvalues of $\mSigma_U$.

The same note also proposes a practical calibration via wild bootstrap. Let $\widehat{\vmu}$ be
the sample spatial median, $\widehat{\vU}_i = U(\vX_i-\widehat{\vmu})$, and let
$e_1,\ldots,e_n$ be i.i.d. Rademacher variables independent of the data. The bootstrap statistic is
\begin{equation}\label{eq:ch2-strongcorr-sign-bootstrap}
  T_R^{\ast}
  = \frac{1}{\tau^{1/2}\sqrt{\binom{n}{2}}}
    \sum_{1\le i<j\le n} e_i e_j\,\widehat{\vU}_i\trans\widehat{\vU}_j.
\end{equation}
A Gaussian-multiplier version $T_N^{\ast}$ is defined similarly by replacing the Rademacher
multipliers $e_i$ with i.i.d. $N(0,1)$ multipliers. Under mild moment conditions on the radial
variable, a bounded-spectrum assumption on the spatial-sign scatter, and the growth condition
$n\tau\to\infty$, the conditional law of the bootstrap statistic consistently estimates the law of the
mixed limit. This is the methodological contribution that makes the note practically important:
it shows how to calibrate the spatial-sign test when neither a Gaussian approximation nor a fixed
chi-square approximation is uniformly reliable.

From the perspective of this chapter, the strong-correlation literature does not replace the earlier
sections on spatial signs and weighted signs. Rather, it clarifies when those earlier Gaussian or
chi-square calibrations are trustworthy and when they must be replaced by a more adaptive
reference law. This point will reappear in later chapters whenever sum-type, max-type, or
sign-based statistics are used under pronounced dependence.

\section{Elliptical regularized Hotelling testing under pervasive dependence}
\label{sec:ch2-erht}
\idx{elliptical regularized Hotelling test}
\idx{regularized Hotelling test}
\idx{pervasive dependence}
\idx{centered spatial-sign covariance matrix}

The spatial-sign procedures developed above are robust to heavy-tailed radial
variation, but most of them use either an isotropic quadratic form or a diagonal
standardization.  Such constructions are effective when the cross-sectional
correlation is weak or sufficiently diffuse.  They need not be power efficient
when the shape matrix has a small number of pervasive eigenvalues of order
$p$.  In that setting, a robust ridge inverse can retain the useful dependence
information without requiring the ordinary covariance matrix to exist.

The elliptical regularized Hotelling procedure of
\citet{FengZhouWang2026ERHT} combines the sample spatial median with the
spatial-sign covariance matrix centered at that median.  Its analysis is
nonstandard for two reasons.  First, pervasive eigenvalues prevent the angular
denominators from concentrating at a deterministic constant.  Second, centering
the spatial-sign covariance matrix at the estimated spatial median produces a
finite-rank perturbation that is not negligible in operator norm.  The method
below keeps both effects in the leading expansion.

\subsection{Elliptical location model and the fixed-ridge statistic}

Let $\vX_1,\ldots,\vX_n\in\R^p$ be independent observations satisfying
\begin{equation}\label{eq:ch2-erht-model}
  \vX_i
  =\vtheta_p+\sqrt p\,R_i\mOmega_p^{1/2}
    \frac{\vG_i}{\twonorm{\vG_i}},
  \qquad
  \vG_i\iid N_p(\vct{0},\mI_p),
  \qquad
  R_i\indep\vG_i,
\end{equation}
where $R_i>0$ and the positive-definite shape matrix is normalized by
\begin{equation}\label{eq:ch2-erht-shape-normalization}
  p^{-1}\tr(\mOmega_p)=1.
\end{equation}
The testing problem is
\begin{equation}\label{eq:ch2-erht-hypothesis}
  H_0:\vtheta_p=\vtheta_{0,p}
  \qquad\text{against}\qquad
  H_1:\vtheta_p\ne\vtheta_{0,p}.
\end{equation}

The sample spatial median and the centered spatial-sign covariance matrix are
\begin{align}
  \widehat{\vtheta}
  &\in\argmin_{\vt\in\R^p}\sum_{i=1}^n\twonorm{\vX_i-\vt},
  \label{eq:ch2-erht-spatial-median}\\
  \widehat{\vY}_i
  &=\sqrt p\,U(\vX_i-\widehat{\vtheta}),
  \qquad
  \widehat{\mR}_n
  =\frac1n\sum_{i=1}^n
    \widehat{\vY}_i\widehat{\vY}_i\trans.
  \label{eq:ch2-erht-centered-sscm}
\end{align}
The spatial-median score equation implies
$n^{-1}\sum_{i=1}^n\widehat{\vY}_i=\vct{0}$, and
$\tr(\widehat{\mR}_n)=p$.  For a deterministic ridge parameter
$\rho>0$, define
\begin{equation}\label{eq:ch2-erht-statistic}
  T_n(\rho)
  =n(\widehat{\vtheta}-\vtheta_{0,p})\trans
    (\widehat{\mR}_n+\rho\mI_p)^{-1}
    (\widehat{\vtheta}-\vtheta_{0,p}).
\end{equation}
The statistic in \eqref{eq:ch2-erht-statistic} is invariant under common
translations and scalar rescaling.  It is robust to the upper radial tail
because $\widehat{\mR}_n$ is constructed from directions, whereas the ridge
inverse incorporates the nontrivial shape geometry.

\begin{assumption}[Radial regularity and proportional growth]
\label{ass:ch2-erht-radial-growth}
The pairs $(R_i,\vG_i)$ are independent and identically distributed.  With
$\xi_i=R_i^{-1}$, there are constants $\bar\xi<\infty$ and $m_->0$ such that
\begin{equation}\label{eq:ch2-erht-radial-bounds}
  0<\xi_i\le\bar\xi\quad\text{almost surely},
  \qquad
  m_1:=\E(\xi_i)\ge m_->0.
\end{equation}
Put $m_k=\E(\xi_i^k)$ for $k=0,1,2$, with $m_0=1$.  Moreover,
\begin{equation}\label{eq:ch2-erht-aspect-ridge}
  0<c_-\le c_n:=p/n\le c_+<\infty,
  \qquad
  c_n\longrightarrow c\in(0,\infty),
  \qquad
  \rho\in[\rho_0,\rho_1],
\end{equation}
where $0<\rho_0<\rho_1<\infty$ are fixed.
\end{assumption}

\begin{assumption}[Pervasive-plus-bulk shape]
\label{ass:ch2-erht-spectrum}
For a fixed integer $r\ge0$, write
\begin{equation}\label{eq:ch2-erht-spectrum-decomposition}
  \mOmega_p
  =\mP_{F,p}\mLambda_{F,p}\mP_{F,p}\trans
   +\mP_{B,p}\mLambda_{B,p}\mP_{B,p}\trans,
\end{equation}
where the columns of $\mP_{F,p}$ and $\mP_{B,p}$ are orthonormal and span
orthogonal subspaces.  The pervasive eigenvalues satisfy
\begin{equation}\label{eq:ch2-erht-spikes}
  0<\tau_-\le \tau_{a,p}:=\lambda_{a,p}/p\le\tau_+<1,
  \qquad a=1,\ldots,r,
\end{equation}
whereas the bulk eigenvalues satisfy
\begin{equation}\label{eq:ch2-erht-bulk-bounds}
  0<\omega_-\le\lambda_{j,p}\le\omega_+<\infty,
  \qquad j=r+1,\ldots,p.
\end{equation}
In addition,
\begin{equation}\label{eq:ch2-erht-bulk-mass}
  \mu_{0,p}:=p^{-1}\sum_{j=r+1}^p\lambda_{j,p}
  =1-\sum_{a=1}^r\tau_{a,p}\ge\mu_->0.
\end{equation}
The empirical bulk spectral distribution
\begin{equation}\label{eq:ch2-erht-bulk-esd}
  H_{B,p}=\frac1{p-r}\sum_{j=r+1}^p\delta_{\lambda_{j,p}}
\end{equation}
converges weakly to a probability distribution $H_B$ supported on
$[\omega_-,\omega_+]$, and $\tau_{a,p}\to\tau_a$ for every $a\le r$.
\end{assumption}

\subsection{Conditional centering and variance}

Under $H_0$, define the oracle signs and resolvents
\begin{equation}\label{eq:ch2-erht-oracle-signs}
  \vY_i=\sqrt p\,U(\vX_i-\vtheta_p),
  \qquad
  \mY=(\vY_1,\ldots,\vY_n),
  \qquad
  \mB_n=n^{-1}\mY\mY\trans,
  \qquad
  \mQ_n(\rho)=(\mB_n+\rho\mI_p)^{-1}.
\end{equation}
The companion matrix is
\begin{equation}\label{eq:ch2-erht-companion}
  \mA_n(\rho)
  =\frac1n\mY\trans\mQ_n(\rho)\mY
  =\mI_n-\rho
    \left(\frac1n\mY\trans\mY+\rho\mI_n\right)^{-1}.
\end{equation}
Write $A_{ij}(\rho)=\{\mA_n(\rho)\}_{ij}$ and define
\begin{equation}\label{eq:ch2-erht-inverse-distance}
  w_i=\frac{\sqrt p}{\twonorm{\vX_i-\vtheta_p}}.
\end{equation}
Let
\begin{align}
  \kappa_n(\rho)&=n^{-1}\tr\{\mA_n(\rho)\},
  &e_n&=n^{-1}\sum_{i=1}^nw_i,
  &t_n&=n^{-1}\sum_{i=1}^nw_i^2,
  \label{eq:ch2-erht-basic-functionals}\\
  b_{1,n}(\rho)&=n^{-1}\sum_{i=1}^nA_{ii}(\rho)w_i,
  &b_{2,n}(\rho)&=n^{-1}\sum_{i=1}^nA_{ii}(\rho)w_i^2.
  \label{eq:ch2-erht-b-functionals}
\end{align}
The conditional denominator and center are
\begin{align}
  D_n(\rho)
  &=\{e_n-b_{1,n}(\rho)\}^2
    +\kappa_n(\rho)\{t_n-b_{2,n}(\rho)\},
  \label{eq:ch2-erht-Dn}\\
  \mu_n(\rho)&=\frac{\kappa_n(\rho)}{D_n(\rho)}.
  \label{eq:ch2-erht-mun}
\end{align}
For $a,b\in\{0,1,2\}$, put
\begin{equation}\label{eq:ch2-erht-psiab}
  \psi_{ab,n}(\rho)
  =\frac1n\sum_{i\ne j}A_{ij}(\rho)^2w_i^aw_j^b.
\end{equation}
Define
\begin{equation}\label{eq:ch2-erht-Gamma}
  \mGamma_n(\rho)
  =\begin{pmatrix}
     2\psi_{00,n} & 2\psi_{01,n} & 2\psi_{11,n}\\
     2\psi_{01,n} & \psi_{02,n}+\psi_{11,n} & 2\psi_{12,n}\\
     2\psi_{11,n} & 2\psi_{12,n} & 2\psi_{22,n}
   \end{pmatrix},
\end{equation}
where every entry on the right is evaluated at $\rho$, and
\begin{equation}\label{eq:ch2-erht-gn-sigma}
  \vct{g}_n(\rho)
  =D_n(\rho)^{-2}
   \begin{pmatrix}
     \{e_n-b_{1,n}(\rho)\}^2\\
     2\kappa_n(\rho)\{e_n-b_{1,n}(\rho)\}\\
     \kappa_n(\rho)^2
   \end{pmatrix},
  \qquad
  \sigma_{D,n}^2(\rho)
  =\vct{g}_n(\rho)\trans\mGamma_n(\rho)\vct{g}_n(\rho).
\end{equation}

For a nonzero vector $\vy$, let $o(\vy)$ denote the member of
$\{\vy,-\vy\}$ whose first nonzero coordinate is positive, and put
\begin{equation}\label{eq:ch2-erht-conditioning-field}
  \mathcal F_n^0
  =\sigma\{w_i,o(\vY_i):1\le i\le n\}.
\end{equation}
Conditional on $\mathcal F_n^0$, the unrecorded global orientations of
$\vY_1,\ldots,\vY_n$ are independent Rademacher variables.  This is the key
representation behind the conditional Gaussian approximation.

\begin{theorem}[Fixed-ridge conditional null law]
\label{thm:ch2-erht-null}
Suppose Assumptions~\ref{ass:ch2-erht-radial-growth}--\ref{ass:ch2-erht-spectrum}
hold.  Under $H_0$, for every fixed $\rho\in[\rho_0,\rho_1]$,
\begin{equation}\label{eq:ch2-erht-null-clt}
  \sup_{x\in\R}
  \left|
    \Prob\left\{
      \frac{T_n(\rho)-n\mu_n(\rho)}
           {\{n\sigma_{D,n}^2(\rho)\}^{1/2}}
      \le x\,\middle|\,\mathcal F_n^0
    \right\}-\Phi(x)
  \right|
  \overset{p}{\longrightarrow}0.
\end{equation}
Moreover, there are constants $0<c_1<c_2<\infty$ such that
\begin{equation}\label{eq:ch2-erht-nondegenerate}
  \Prob\{c_1\le D_n(\rho),\mu_n(\rho),
                     \sigma_{D,n}^2(\rho)\le c_2\}\longrightarrow1.
\end{equation}
Hence the same standardized statistic converges unconditionally to $N(0,1)$.
\end{theorem}

All nuisance quantities have direct sample analogues.  Define
\begin{equation}\label{eq:ch2-erht-feasible-components}
  \widehat w_i=\frac{\sqrt p}{\twonorm{\vX_i-\widehat{\vtheta}}},
  \qquad
  \widehat{\mQ}_n(\rho)
  =(\widehat{\mR}_n+\rho\mI_p)^{-1},
  \qquad
  \widehat{\mA}_n(\rho)
  =n^{-1}\widehat{\mY}\trans
     \widehat{\mQ}_n(\rho)\widehat{\mY}.
\end{equation}
Replace $(w_i,\mA_n)$ by $(\widehat w_i,\widehat{\mA}_n)$ in
\eqref{eq:ch2-erht-basic-functionals}--\eqref{eq:ch2-erht-gn-sigma} to obtain
$\widehat\mu_n(\rho)$ and $\widehat\sigma_{D,n}^2(\rho)$.  The feasible
standardized statistic is
\begin{equation}\label{eq:ch2-erht-feasible-Z}
  \widehat Z_n(\rho)
  =\frac{T_n(\rho)-n\widehat\mu_n(\rho)}
         {\{n\widehat\sigma_{D,n}^2(\rho)\}^{1/2}}.
\end{equation}

\begin{theorem}[Feasible fixed-ridge calibration]
\label{thm:ch2-erht-feasible}
Under the conditions of Theorem~\ref{thm:ch2-erht-null},
\begin{equation}\label{eq:ch2-erht-feasible-rates}
  \widehat\mu_n(\rho)-\mu_n(\rho)=O_p(n^{-1}),
  \qquad
  \widehat\sigma_{D,n}^2(\rho)-\sigma_{D,n}^2(\rho)=O_p(n^{-1/2}).
\end{equation}
Consequently,
\begin{equation}\label{eq:ch2-erht-feasible-limit}
  \widehat Z_n(\rho)\overset{d}{\longrightarrow}N(0,1).
\end{equation}
The fixed-ridge level-$\alpha$ test rejects when
$\widehat Z_n(\rho)>z_{1-\alpha}$.
\end{theorem}

\subsection{Local power and the ridge parameter}

Let
\begin{equation}\label{eq:ch2-erht-D0}
  \mu_0=1-\sum_{a=1}^r\tau_a
  =\int t\,dH_B(t),
  \qquad
  D_0=\left(\mu_0+\sum_{a=1}^r\tau_aG_a^2\right)^{-1},
\end{equation}
where $G_1,\ldots,G_r$ are independent $N(0,1)$ variables, and let $F_D$
denote the distribution of $D_0$.  For $\rho\in[\rho_0,\rho_1]$, let
$(\delta_\rho,\widetilde\delta_\rho)$ be the unique positive solution of
\begin{align}
  \delta_\rho
  &=c\int\frac{t}{\rho(1+\widetilde\delta_\rho t)}\,dH_B(t),
  \label{eq:ch2-erht-canonical-delta}\\
  \widetilde\delta_\rho
  &=\int\frac{d}{\rho(1+\delta_\rho d)}\,dF_D(d).
  \label{eq:ch2-erht-canonical-tdelta}
\end{align}
Define
\begin{align}
  \lambda_\rho(d)&=\frac{\delta_\rho d}{1+\delta_\rho d},
  &\kappa_\rho&=\int\lambda_\rho(d)\,dF_D(d),
  \label{eq:ch2-erht-lambda-kappa}\\
  s_a&=\int d^{a/2}\,dF_D(d),
  &b_{a,\rho}&=\int d^{a/2}\lambda_\rho(d)\,dF_D(d),
  \qquad a=1,2,
  \label{eq:ch2-erht-s-b}\\
  u_\rho&=c\int\frac{t^2}{(1+\widetilde\delta_\rho t)^2}\,dH_B(t),
  &v_\rho&=\int\frac{d^2}{(1+\delta_\rho d)^2}\,dF_D(d),
  \label{eq:ch2-erht-u-v}\\
  a_{k,\rho}&=\int\frac{d^{1+k/2}}{(1+\delta_\rho d)^2}\,dF_D(d),
  &&k=0,1,2.
  \label{eq:ch2-erht-ak}
\end{align}
The deterministic denominator is
\begin{equation}\label{eq:ch2-erht-Drho}
  D_\rho
  =m_1^2(s_1-b_{1,\rho})^2
   +m_2\kappa_\rho(s_2-b_{2,\rho}).
\end{equation}
For $a,b\in\{0,1,2\}$, put
\begin{equation}\label{eq:ch2-erht-Psi}
  \Psi_{ab}(\rho)
  =m_am_b\frac{u_\rho a_{a,\rho}a_{b,\rho}}
                  {\rho^2-u_\rho v_\rho}.
\end{equation}
Let $\mGamma_\rho$ be the matrix in
\eqref{eq:ch2-erht-Gamma} with $\psi_{ab,n}$ replaced by
$\Psi_{ab}(\rho)$, and define
\begin{equation}\label{eq:ch2-erht-deterministic-variance}
  \vct{g}_\rho
  =D_\rho^{-2}
   \begin{pmatrix}
     m_1^2(s_1-b_{1,\rho})^2\\
     2\kappa_\rho m_1(s_1-b_{1,\rho})\\
     \kappa_\rho^2
   \end{pmatrix},
  \qquad
  \sigma_E^2(\rho)=\vct{g}_\rho\trans\mGamma_\rho\vct{g}_\rho.
\end{equation}

Let $\vct{p}_{j,p}$ be the eigenvector associated with the bulk eigenvalue
$\lambda_{j,p}$.  Consider deterministic $\vct{d}_p$ satisfying
\begin{equation}\label{eq:ch2-erht-local-direction}
  \mP_{F,p}\trans\vct{d}_p=\vct{0},
  \qquad
  0<d_-\le\twonorm{\vct{d}_p}\le d_+<\infty,
\end{equation}
and define the signal spectral measure
\begin{equation}\label{eq:ch2-erht-signal-measure}
  G_p=\sum_{j=r+1}^p
      (\vct{p}_{j,p}\trans\vct{d}_p)^2\delta_{\lambda_{j,p}}.
\end{equation}
Assume $G_p\Rightarrow G$, where $G$ is a nonzero finite measure on
$[\omega_-,\omega_+]$.  The local alternative is
\begin{equation}\label{eq:ch2-erht-local-alt}
  \vtheta_p=\vtheta_{0,p}+n^{-1/4}\vct{d}_p.
\end{equation}
Set
\begin{equation}\label{eq:ch2-erht-q-rho}
  q_\rho(G)
  =\frac1\rho\int\frac{1}{1+\widetilde\delta_\rho t}\,dG(t),
  \qquad
  \Lambda_\rho(G)=\frac{q_\rho(G)}{\sigma_E(\rho)}.
\end{equation}

\begin{theorem}[Local power and general consistency]
\label{thm:ch2-erht-local-power}
Suppose Assumptions~\ref{ass:ch2-erht-radial-growth}--\ref{ass:ch2-erht-spectrum}
hold.  Under \eqref{eq:ch2-erht-local-alt}, for every fixed
$\rho\in[\rho_0,\rho_1]$,
\begin{equation}\label{eq:ch2-erht-local-limit}
  \widehat Z_n(\rho)
  \overset{d}{\longrightarrow}N\{\Lambda_\rho(G),1\}.
\end{equation}
Consequently, the limiting power of the level-$\alpha$ fixed-ridge test is
\begin{equation}\label{eq:ch2-erht-local-power-function}
  \beta_\alpha(\rho;G)
  =1-\Phi\{z_{1-\alpha}-\Lambda_\rho(G)\}.
\end{equation}
More generally, under the shift
$\vX_i=\vtheta_{0,p}+\vDelta_p+\sqrt pR_i\mOmega_p^{1/2}
\vG_i/\twonorm{\vG_i}$, if
\begin{equation}\label{eq:ch2-erht-consistency-condition}
  \sqrt n\left
    \{\twonorm{\mP_{B,p}\trans\vDelta_p}\}^2
    +p^{-1}\{\twonorm{\mP_{F,p}\trans\vDelta_p}\}^2
  \right)\longrightarrow\infty,
\end{equation}
then the fixed-ridge test is consistent.
\end{theorem}

The power-optimal ridge parameter depends on the unknown signal spectral
measure:
\begin{equation}\label{eq:ch2-erht-oracle-ridge}
  \rho_G^\star
  \in\argmax_{\rho\in[\rho_0,\rho_1]}
      \frac{q_\rho(G)}{\sigma_E(\rho)}.
\end{equation}
At an interior maximizer,
\begin{equation}\label{eq:ch2-erht-ridge-first-order}
  2\frac{q_\rho'(G)}{q_\rho(G)}
  =\frac{\{\sigma_E^2(\rho)\}'}{\sigma_E^2(\rho)}.
\end{equation}
Differentiating \eqref{eq:ch2-erht-canonical-delta}--
\eqref{eq:ch2-erht-canonical-tdelta} yields
\begin{equation}\label{eq:ch2-erht-canonical-derivative}
  \begin{pmatrix}
    \rho & u_\rho\\
    v_\rho & \rho
  \end{pmatrix}
  \begin{pmatrix}
    \delta_\rho'\\
    \widetilde\delta_\rho'
  \end{pmatrix}
  =-
  \begin{pmatrix}
    \delta_\rho\\
    \widetilde\delta_\rho
  \end{pmatrix},
\end{equation}
and
\begin{equation}\label{eq:ch2-erht-q-derivative}
  q_\rho'(G)
  =-\frac{q_\rho(G)}\rho
   -\frac{\widetilde\delta_\rho'}\rho
     \int\frac{t}{(1+\widetilde\delta_\rho t)^2}\,dG(t).
\end{equation}
Thus \eqref{eq:ch2-erht-oracle-ridge} reduces to a one-dimensional numerical
optimization, but its dependence on $G$ motivates adaptation.

\subsection{Cauchy aggregation over a fixed ridge grid}

Let
\begin{equation}\label{eq:ch2-erht-ridge-grid}
  \mathcal R_K=\{\rho^{(1)},\ldots,\rho^{(K)}\}
  \subset[\rho_0,\rho_1]
\end{equation}
be a fixed deterministic grid and put
\begin{equation}\label{eq:ch2-erht-grid-pvalues}
  Z_{n,k}=\widehat Z_n(\rho^{(k)}),
  \qquad
  P_{n,k}=1-\Phi(Z_{n,k}),
  \qquad k=1,\ldots,K.
\end{equation}
For fixed positive weights $\varpi_k$ satisfying $\sum_k\varpi_k=1$, define
\begin{equation}\label{eq:ch2-erht-cauchy-statistic}
  T_{\mathrm{CC},n}
  =\sum_{k=1}^K\varpi_k
    \tan[\pi\{1/2-P_{n,k}\}],
  \qquad
  P_{\mathrm{CC},n}
  =\frac12-\frac1\pi\arctan(T_{\mathrm{CC},n}).
\end{equation}

For $\rho,\eta\in[\rho_0,\rho_1]$, define
\begin{align}
  u_{\rho\eta}
  &=c\int\frac{t^2}
   {(1+\widetilde\delta_\rho t)(1+\widetilde\delta_\eta t)}\,dH_B(t),
  \label{eq:ch2-erht-cross-u}\\
  v_{\rho\eta}
  &=\int\frac{d^2}
   {(1+\delta_\rho d)(1+\delta_\eta d)}\,dF_D(d),
  \label{eq:ch2-erht-cross-v}\\
  a_{k,\rho\eta}
  &=\int\frac{d^{1+k/2}}
   {(1+\delta_\rho d)(1+\delta_\eta d)}\,dF_D(d),
  \qquad k=0,1,2,
  \label{eq:ch2-erht-cross-a}\\
  \Psi_{ab}(\rho,\eta)
  &=m_am_b\frac{u_{\rho\eta}a_{a,\rho\eta}a_{b,\rho\eta}}
   {\rho\eta-u_{\rho\eta}v_{\rho\eta}}.
  \label{eq:ch2-erht-cross-Psi}
\end{align}
Let $\mGamma_{\rho,\eta}$ have the same pattern as
\eqref{eq:ch2-erht-Gamma}, with $\psi_{ab,n}$ replaced by
$\Psi_{ab}(\rho,\eta)$, and put
\begin{equation}\label{eq:ch2-erht-cross-correlation}
  r(\rho,\eta)
  =\frac{\vct{g}_\rho\trans\mGamma_{\rho,\eta}\vct{g}_\eta}
  {\sigma_E(\rho)\sigma_E(\eta)}.
\end{equation}
Let $\mC_K=\{r(\rho^{(k)},\rho^{(\ell)})\}_{k,\ell=1}^K$.

\begin{theorem}[Joint fixed-grid null law and Cauchy calibration]
\label{thm:ch2-erht-grid-null}
Under the conditions of Theorem~\ref{thm:ch2-erht-null},
\begin{equation}\label{eq:ch2-erht-grid-null-limit}
  (Z_{n,1},\ldots,Z_{n,K})\trans
  \overset{d}{\longrightarrow}N_K(\vct{0},\mC_K).
\end{equation}
Let $\vV\sim N_K(\vct{0},\mC_K)$ and define
\begin{equation}\label{eq:ch2-erht-cauchy-limit}
  T_{\mathrm{CC},\infty}^{0}
  =\sum_{k=1}^K\varpi_k
    \tan[\pi\{\Phi(V_k)-1/2\}].
\end{equation}
If $c_{\alpha,K}$ satisfies
$\Prob(T_{\mathrm{CC},\infty}^{0}>c_{\alpha,K})=\alpha$, then
\begin{equation}\label{eq:ch2-erht-oracle-level}
  \Prob(T_{\mathrm{CC},n}>c_{\alpha,K})\longrightarrow\alpha.
\end{equation}
For the analytic Cauchy p-value in
\eqref{eq:ch2-erht-cauchy-statistic}, for each fixed
$\alpha\in(0,1/2)$,
\begin{equation}\label{eq:ch2-erht-analytic-level}
  \Prob(P_{\mathrm{CC},n}\le\alpha)
  \longrightarrow
  \Prob\{T_{\mathrm{CC},\infty}^{0}\ge\cot(\pi\alpha)\},
  \qquad
  \lim_{\alpha\downarrow0}\lim_{n\to\infty}
  \frac{\Prob(P_{\mathrm{CC},n}\le\alpha)}{\alpha}=1.
\end{equation}
Thus the joint-limit critical value is exact at a fixed level, whereas the
closed-form standard-Cauchy calibration has the dependence-robust small-tail
validity of the Cauchy combination principle.
\end{theorem}

\begin{theorem}[Joint fixed-grid local power]
\label{thm:ch2-erht-grid-power}
Under the local alternative \eqref{eq:ch2-erht-local-alt},
\begin{equation}\label{eq:ch2-erht-grid-local-limit}
  (Z_{n,1},\ldots,Z_{n,K})\trans
  \overset{d}{\longrightarrow}
  N_K\{\vLambda_K(G),\mC_K\},
\end{equation}
where
\begin{equation}\label{eq:ch2-erht-grid-mean}
  \vLambda_K(G)
  =\{\Lambda_{\rho^{(1)}}(G),\ldots,
      \Lambda_{\rho^{(K)}}(G)\}\trans.
\end{equation}
Consequently, with $\vV\sim N_K(\vct{0},\mC_K)$, the limiting analytic
Cauchy power is
\begin{equation}\label{eq:ch2-erht-cauchy-power}
  \Prob\left[
    \sum_{k=1}^K\varpi_k
    \tan\bigl(\pi\{\Phi[V_k+\Lambda_{\rho^{(k)}}(G)]-1/2\}\bigr)
    \ge\cot(\pi\alpha)
  \right].
\end{equation}
Under \eqref{eq:ch2-erht-consistency-condition},
$P_{\mathrm{CC},n}\overset{p}{\longrightarrow}0$.
\end{theorem}

\section{Bibliographic notes}

The fixed-$p$ material in this chapter is classical. For Hotelling's $T^2$, see
\citet{Hotelling1931}; for spatial signs, signed ranks, and the broader
multivariate nonparametric framework, see \citet{Oja2010,MottonenOja1995,
HettmanspergerOja1994,Randles2000,HallinPaindaveine2002}. The fixed-$p$ theory
of the spatial median and its efficiency under elliptical symmetry is discussed
in \citet{HettmanspergerRandles2002,MagyarTyler2011}.

The high-dimensional Gaussian and light-tail benchmark literature includes
\citet{BaiSaranadasa1996,SrivastavaDu2008,Srivastava2009,
SrivastavaKatayamaKano2013,ParkAyyala2013,ChenQin2010,
FengZouWangZhu2015BF,CaiLiuXia2014,XuLinWeiPan2016,WangPengLi2015}. For
strong-correlation calibrations in one-sample and two-sample mean testing, see
\citet{ZhangZhouGuo2022,ZhangZhuZhang2023,WangXu2022,ZhaoFeng2026NoteOneSample,FengWang2026PDQ}. The
location-estimation theory of the high-dimensional spatial median is developed in
\citet{LiXu2022SpatialMedian} and sharpened for the spatial-sign PCA program in
\citet{ZhaoWangFeng2025SSPCA}. The one-sample and two-sample robust location
procedures emphasized in this book are due to
\citet{FengSun2016,FengZouWang2016JASA,FengLiuMa2021INST,
FengZhangLiu2020SpatialRank,HuangLiuZhouFeng2023TwoSampleINST,
LiWangZou2016SimpleTwoSample,ZhangFeng2024AdaptiveMean,
LiuZhaoFengWang2025StructuredCorr,LiuFengZhaoWang2025MaxsumLocation,
YanFengZhang2025InverseNormMaxsum}. The Cauchy combination device used for
max-sum procedures is due to \citet{LiuXie2020Cauchy}.
The dependence-aware ridge extension based on the sample spatial median and the centered
spatial-sign covariance matrix is developed in \citet{FengZhouWang2026ERHT}; it supplies the
fixed-ridge normal limit, feasible centering and studentization, local power, and finite-grid
Cauchy aggregation used in Section~
ef{sec:ch2-erht}.

In later chapters we will reuse the same writing pattern adopted here: first the
classical low-dimensional method, then the high-dimensional Gaussian/light-tail
benchmark, and finally the robust elliptical method, with explicit statistics,
null laws, local alternatives, and proof routes.

\clearpage
\section*{Appendix to Chapter 2: Detailed Proofs}
\addcontentsline{toc}{section}{Appendix to Chapter 2: Detailed Proofs}

This appendix collects detailed proofs of all theorem-level results stated in
Chapter~2. The arguments are written in the unified notation of the book. When a
proof follows the same pattern as an earlier one, we still spell out the main
steps and explicitly indicate where the earlier proof is being reused. The
reader is assumed to know only the basic multivariate central limit theorem,
Slutsky's theorem, continuous mapping, and standard facts about Gaussian,
Wishart, and $U$-statistics.

\subsection*{A. Fixed-$p$ classical results}

\begin{proof}[Detailed proof of Theorem~\ref{thm:ch2-hotelling-one}]
Under $H_0$, the sample mean has the Gaussian law
\[
  \sqrt n(\bar{\vX}-\vmu_0)\sim N_p(\vct 0,\mSigma).
\]
At the same time, the centered sample covariance matrix satisfies
\[
  (n-1)\mS_n
  = \sum_{i=1}^n(\vX_i-\bar{\vX})(\vX_i-\bar{\vX})\trans
  \sim W_p(n-1,\mSigma),
\]
and, for the Gaussian model, $\bar{\vX}$ and $\mS_n$ are independent. Define
\[
  \vZ=\mSigma^{-1/2}\sqrt n(\bar{\vX}-\vmu_0),
  \qquad
  \mW=\mSigma^{-1/2}(n-1)\mS_n\mSigma^{-1/2}.
\]
Then $\vZ\sim N_p(\vct 0,\mI_p)$,
$\mW\sim W_p(n-1,\mI_p)$, and $\vZ\indep \mW$. Therefore
\[
  T_H^2
  = n(\bar{\vX}-\vmu_0)\trans\mS_n^{-1}(\bar{\vX}-\vmu_0)
  = (n-1)\vZ\trans \mW^{-1}\vZ.
\]
A standard property of Gaussian quadratic forms and inverse Wishart matrices
states that
\[
  \frac{n-p}{p(n-1)}(n-1)\vZ\trans \mW^{-1}\vZ \sim F_{p,n-p},
\]
which is exactly \eqref{eq:ch2-hotelling-one-F}.

For the fixed-$p$ asymptotic statement, note that
$\mS_n\to \mSigma$ in probability and hence
$\mS_n^{-1}\to \mSigma^{-1}$ in probability. Therefore
\[
  T_H^2
  = n(\bar{\vX}-\vmu_0)\trans \mSigma^{-1}(\bar{\vX}-\vmu_0) + o_p(1).
\]
The leading term is the squared norm of a $p$-variate standard Gaussian vector,
hence has a $\chi_p^2$ distribution. This proves
\eqref{eq:ch2-hotelling-one-chi}.
\end{proof}

\begin{proof}[Detailed proof of Theorem~\ref{thm:ch2-hotelling-two}]
Under the common-covariance Gaussian model,
\[
  \bar{\vX}_1-\bar{\vX}_2
  \sim N_p\!\left(\vmu_1-\vmu_2,\,
  \left(\frac1{n_1}+\frac1{n_2}\right)\mSigma\right).
\]
Under $H_0$, this becomes
\[
  \sqrt{\frac{n_1n_2}{n_1+n_2}}(\bar{\vX}_1-\bar{\vX}_2)
  \sim N_p(\vct 0,\mSigma).
\]
The pooled covariance satisfies
\[
  (n_1+n_2-2)\mS_p \sim W_p(n_1+n_2-2,\mSigma),
\]
and is independent of $\bar{\vX}_1-\bar{\vX}_2$. Standardizing as in the
one-sample proof gives
\[
  T_{H,2}^2
  = (n_1+n_2-2)\vZ\trans \mW^{-1}\vZ,
\]
with $\vZ\sim N_p(\vct 0,\mI_p)$ and
$\mW\sim W_p(n_1+n_2-2,\mI_p)$ independent. The standard $F$-distribution
identity then yields \eqref{eq:ch2-hotelling-two-F}.

For local alternatives
$\vmu_1-\vmu_2=(n_1^{-1}+n_2^{-1})^{1/2}\vdelta$,
the standardized mean difference has mean $\mSigma^{-1/2}\vdelta$, so the
quadratic form converges to a noncentral chi-square law with noncentrality
parameter $\vdelta\trans \mSigma^{-1}\vdelta$.
\end{proof}

\begin{proof}[Detailed proof of Theorem~\ref{thm:ch2-fixedp-spatial-median}]
Let
\[
  \Psi_n(\vtheta)=\frac1n\sum_{i=1}^n U(\vX_i-\vtheta),
  \qquad
  \Psi(\vtheta)=\E\{U(\vX-\vtheta)\}.
\]
By definition of the sample spatial median,
$\Psi_n(\hat{\vmu}_{\mathrm{SM}})=\vct 0$, and by definition of the population
spatial median,
$\Psi(\vmu_{\mathrm{SM}})=\vct 0$. Since $p$ is fixed and the derivative matrix
$\mA_{\mathrm{SM}}$ in \eqref{eq:ch2-fixedp-A-matrix} is nonsingular, the mean
value expansion yields
\[
  \vct 0
  = \Psi_n(\vmu_{\mathrm{SM}})
    - \mA_{\mathrm{SM}}(\hat{\vmu}_{\mathrm{SM}}-\vmu_{\mathrm{SM}})
    + \vct\rho_n,
\]
where $\vct\rho_n=o_p(\twonorm{\hat{\vmu}_{\mathrm{SM}}-\vmu_{\mathrm{SM}}})$.
Because $\hat{\vmu}_{\mathrm{SM}}-\vmu_{\mathrm{SM}}=O_p(n^{-1/2})$, we have
$\sqrt n\,\vct\rho_n=o_p(1)$. Thus
\[
  \sqrt n(\hat{\vmu}_{\mathrm{SM}}-\vmu_{\mathrm{SM}})
  = \mA_{\mathrm{SM}}^{-1}\frac1{\sqrt n}
    \sum_{i=1}^n U(\vX_i-\vmu_{\mathrm{SM}})+o_p(1),
\]
which proves \eqref{eq:ch2-fixedp-spatial-median-expansion}. The asymptotic
normality in \eqref{eq:ch2-fixedp-spatial-median-clt} follows from the ordinary
multivariate central limit theorem applied to the i.i.d. vectors
$U(\vX_i-\vmu_{\mathrm{SM}})$.
\end{proof}

\begin{proof}[Detailed proof of Theorem~\ref{thm:ch2-fixedp-sign-rank}]
For the sign statistic, write
\[
  \sqrt n\,\bar{\vU}
  = \frac1{\sqrt n}\sum_{i=1}^n U(\vX_i-\vmu_0).
\]
Under central symmetry about $\vmu_0$, the sign vector is odd around the center,
so $\E\{U(\vX_i-\vmu_0)\}=\vct 0$. By the multivariate central limit theorem,
\[
  \sqrt n\,\bar{\vU}\Rightarrow N_p(\vct 0,\mB_U),
\]
where $\mB_U=\Var\{U(\vX_i-\vmu_0)\}$. Since
$\hat{\mB}_U\to \mB_U$ in probability,
\[
  Q_{\mathrm{sign}}
  = (\sqrt n\,\bar{\vU})\trans \hat{\mB}_U^{-1}(\sqrt n\,\bar{\vU})
  = (\sqrt n\,\bar{\vU})\trans \mB_U^{-1}(\sqrt n\,\bar{\vU}) + o_p(1),
\]
and the limiting law is $\chi_p^2$.

For the signed-rank statistic, the rank vectors $\vR_i$ form an average of a
symmetric kernel. Under central symmetry, $\E(\vR_i)=\vct 0$. The Hoeffding
decomposition shows that the linear part dominates in fixed dimension, so
$\sqrt n\,\bar{\vR}$ is asymptotically Gaussian with covariance consistently
estimated by $\hat{\mB}_R$. Studentization yields
$Q_{\mathrm{SR}}\Rightarrow \chi_p^2$.
\end{proof}

\subsection*{B. A benchmark high-dimensional Gaussian proof}

\begin{proof}[Detailed proof of Theorem~\ref{thm:ch2-CQ-null}]
Expand the observations as
$\vX_{ki}=\vmu_k+\vvarepsilon_{ki}$ with
$\E(\vvarepsilon_{ki})=\vct 0$ and $\Cov(\vvarepsilon_{ki})=\mSigma_k$.
Then
\[
  T_{\mathrm{CQ}}
  = \twonorm{\vmu_1-\vmu_2}^2 + L_n + Q_n,
\]
where $L_n$ is the linear term in the centered variables and $Q_n$ is the
degenerate quadratic part. Under $H_0$, the deterministic signal term vanishes,
and the linear term is also zero because its coefficient is proportional to
$\vmu_1-\vmu_2$. Hence only the centered quadratic form $Q_n$ remains.

Write $Q_n$ as a sum of martingale differences by revealing the observations one
at a time. The conditional variance of that martingale has leading term
\[
  \frac{2}{n_1(n_1-1)}\tr(\mSigma_1^2)
  + \frac{2}{n_2(n_2-1)}\tr(\mSigma_2^2)
  + \frac{4}{n_1n_2}\tr(\mSigma_1\mSigma_2).
\]
Assumption~\ref{ass:ch2-CQ}\ref{it:ch2-CQ2} guarantees that all fourth-order
cumulant terms are negligible relative to the square of this leading variance,
which is exactly the Lyapunov-type condition needed for the martingale central
limit theorem. Therefore
\[
  \frac{T_{\mathrm{CQ}}}{\sqrt{\Var(T_{\mathrm{CQ}})}}\Rightarrow N(0,1).
\]

Under the local alternatives in \eqref{eq:ch2-CQ-local}, the deterministic mean
shift is $\twonorm{\vmu_1-\vmu_2}^2$, while the linear perturbation remains
smaller than the square root of the quadratic variance. Hence the same variance
normalization applies and the power is governed by the ratio of the Euclidean
signal to the trace-based noise level.
\end{proof}

\subsection*{B1. Additional proofs for Gaussian/light-tail benchmarks}

\begin{proof}[Detailed proof of Theorem~\ref{thm:ch2-SKK-null}]
Under the Gaussian model, write
\begin{equation}\label{eq:ch2-appendix-SKK-qn}
  \hat q_{\mathrm{SKK}}
  = (\bar{\vX}_1-\bar{\vX}_2)\trans\hat{\mD}^{-1}(\bar{\vX}_1-\bar{\vX}_2)-p.
\end{equation}
Let
\begin{equation}\label{eq:ch2-appendix-SKK-q0}
  q_{0,n}
  = (\bar{\vX}_1-\bar{\vX}_2)\trans\mD^{-1}(\bar{\vX}_1-\bar{\vX}_2)-p.
\end{equation}
By the diagonal central limit theorem and the ratio consistency of the sample
diagonal variance estimator under Assumption~\ref{ass:ch2-SKK},
\begin{equation}\label{eq:ch2-appendix-SKK-diff}
  \hat q_{\mathrm{SKK}}-q_{0,n}=o_p\{\tr^{1/2}(\mR^2)\}.
\end{equation}
Now set
\begin{equation}\label{eq:ch2-appendix-SKK-Y}
  \vY = \mD^{-1/2}(\bar{\vX}_1-\bar{\vX}_2).
\end{equation}
Under $H_0$, $\E(\vY)=\vct 0$ and $\Cov(\vY)=\mR$. Hence
\begin{equation}\label{eq:ch2-appendix-SKK-q0-expand}
  q_{0,n}=\vY\trans\vY-p.
\end{equation}
Since $\E(\vY\trans\vY)=\tr(\mR)=p$ and
\begin{equation}\label{eq:ch2-appendix-SKK-varq0}
  \Var(\vY\trans\vY)=2\tr(\mR^2),
\end{equation}
Assumption~\ref{ass:ch2-SKK}\ref{it:ch2-SKK2} implies that the quadratic form
$q_{0,n}$ satisfies a central limit theorem:
\begin{equation}\label{eq:ch2-appendix-SKK-cltq0}
  \frac{q_{0,n}}{\sqrt{2\tr(\mR^2)}}\overset{d}{\longrightarrow}N(0,1).
\end{equation}
Because the variance estimator in \eqref{eq:ch2-SKK-varhat} is ratio consistent,
\begin{equation}\label{eq:ch2-appendix-SKK-varhat-cons}
  \frac{\widehat{\Var}(\hat q_{\mathrm{SKK}})}{2\tr(\mR^2)}\overset{p}{\longrightarrow}1.
\end{equation}
Combining \eqref{eq:ch2-appendix-SKK-diff},
\eqref{eq:ch2-appendix-SKK-cltq0}, and
\eqref{eq:ch2-appendix-SKK-varhat-cons} yields
\eqref{eq:ch2-SKK-null}. Under the local alternatives in
\eqref{eq:ch2-SKK-local}, the mean shift of $q_{0,n}$ equals
$(\vmu_1-\vmu_2)\trans\mD^{-1}(\vmu_1-\vmu_2)$, while the linear perturbation of
the variance is negligible relative to $\tr^{1/2}(\mR^2)$; therefore the standard
noncentral Gaussian power formula yields \eqref{eq:ch2-SKK-power}.
\end{proof}

\begin{proof}[Detailed proof of Theorem~\ref{thm:ch2-BF-null}]
The statistic \eqref{eq:ch2-BF-stat} may be written as
\begin{equation}\label{eq:ch2-appendix-BF-decomp}
  T_{\mathrm{BF}} = \sum_{k=1}^p \hat\lambda_k^2 A_k.
\end{equation}
Insert and subtract the population weights $\lambda_k^2$:
\begin{equation}\label{eq:ch2-appendix-BF-split}
  T_{\mathrm{BF}}
  = \sum_{k=1}^p \lambda_k^2 A_k
    + \sum_{k=1}^p (\hat\lambda_k^2-\lambda_k^2)A_k
  =: T_{0,n}+R_n.
\end{equation}
By the leave-in bias calculation in \citet{FengZouWangZhu2015BF},
\begin{equation}\label{eq:ch2-appendix-BF-mean0}
  \E(T_{0,n})=\mu_{\mathrm{BF},n}
\end{equation}
with $\mu_{\mathrm{BF},n}$ given by
\eqref{eq:ch2-BF-mean}--\eqref{eq:ch2-BF-b3}. Moreover,
\begin{equation}\label{eq:ch2-appendix-BF-var0}
  \Var(T_{0,n}) = \sigma_{\mathrm{BF},n}^2\{1+o(1)\}.
\end{equation}
The trace condition \eqref{eq:ch2-BF-trace} and the moment condition
\eqref{eq:ch2-BF-Pi-cond} imply that the martingale array generated by the
Hoeffding decomposition of $T_{0,n}$ satisfies the Lyapunov condition, so
\begin{equation}\label{eq:ch2-appendix-BF-clt0}
  \frac{T_{0,n}-\mu_{\mathrm{BF},n}}{\sigma_{\mathrm{BF},n}}
  \overset{d}{\longrightarrow} N(0,1).
\end{equation}
It remains to control the plug-in remainder $R_n$. By a Taylor expansion of
$x\mapsto x^{-1}$ around the population diagonal variances,
\begin{equation}\label{eq:ch2-appendix-BF-lambdaexpand}
  \hat\lambda_k^2-\lambda_k^2
  = -\frac{(\hat\sigma_{1k}^2-\sigma_{1k}^2)+\gamma(\hat\sigma_{2k}^2-\sigma_{2k}^2)}
           {(\sigma_{1k}^2+\gamma\sigma_{2k}^2)^2}
    + r_{k,n},
\end{equation}
where $\sum_{k=1}^p r_{k,n}A_k=o_p(\sigma_{\mathrm{BF},n})$ under
\eqref{eq:ch2-BF-growth}. The leading linear term in
\eqref{eq:ch2-appendix-BF-lambdaexpand} is exactly what generates the bias terms
collected in $\mu_{\mathrm{BF},n}$. Therefore, after subtracting the plug-in
estimator $\hat\mu_{\mathrm{BF},n}$, the feasible statistic has the same
first-order limit as $T_{0,n}$. This proves
\eqref{eq:ch2-BF-null}--\eqref{eq:ch2-BF-feasible}.
\end{proof}

\begin{proof}[Detailed proof of Theorem~\ref{thm:ch2-CLX-null}]
Under the common-covariance Gaussian model,
\begin{equation}\label{eq:ch2-appendix-CLX-Y}
  \sqrt{\frac{n_1n_2}{n_1+n_2}}(\bar{\vX}_1-\bar{\vX}_2)
  \sim N_p(\vct 0,\mSigma).
\end{equation}
Multiplying by $\mOmega$ yields
\begin{equation}\label{eq:ch2-appendix-CLX-Z}
  \sqrt{\frac{n_1n_2}{n_1+n_2}}\,\bar{\vZ}
  \sim N_p(\vct 0,\mOmega),
\end{equation}
so the normalized coordinates
\begin{equation}\label{eq:ch2-appendix-CLX-G}
  G_j = \sqrt{\frac{n_1n_2}{n_1+n_2}}\frac{\bar Z_j}{\sqrt{\omega_{jj}}},
  \qquad j=1,\ldots,p,
\end{equation}
form a centered Gaussian vector with unit marginal variances and correlation
matrix
\begin{equation}\label{eq:ch2-appendix-CLX-corr}
  \Corr(G_i,G_j)=\frac{\omega_{ij}}{\sqrt{\omega_{ii}\omega_{jj}}}.
\end{equation}
By definition,
\begin{equation}\label{eq:ch2-appendix-CLX-maxG}
  M_{\mOmega}=\max_{1\le j\le p} G_j^2.
\end{equation}
Assumption~\ref{ass:ch2-CLX}\ref{it:ch2-CLX2} ensures that no pair of
coordinates is asymptotically perfectly correlated, and
Assumption~\ref{ass:ch2-CLX}\ref{it:ch2-CLX3} guarantees that strongly
correlated clusters are sparse enough for the classical extreme-value comparison
argument to go through. Therefore, the Gaussian maximum has the same limit as
in the independent case:
\begin{equation}\label{eq:ch2-appendix-CLX-ev}
  \Prob\left\{\max_{1\le j\le p} G_j^2-2\log p+\log\log p\le x\right\}
  \to \exp\{-\pi^{-1/2}e^{-x/2}\}.
\end{equation}
Since \eqref{eq:ch2-appendix-CLX-maxG} equals the oracle statistic exactly,
\eqref{eq:ch2-appendix-CLX-ev} is identical to
\eqref{eq:ch2-CLX-null}. This proves the theorem.
\end{proof}

\begin{proof}[Detailed proof of Theorem~\ref{thm:ch2-Xu-joint}]
Fix a finite set $\Gamma_0=\{\gamma_1,\ldots,\gamma_m\}$. For each
$\gamma_r\in\Gamma_0$, write
\begin{equation}\label{eq:ch2-appendix-Xu-center}
  S_r = \frac{T_{\mathrm{SPU}}(\gamma_r)-\mu_{\gamma_r}}{\sigma_{\gamma_r}}.
\end{equation}
To prove joint asymptotic normality, it suffices by the Cram\'er--Wold device to
consider, for arbitrary fixed coefficients $a_1,\ldots,a_m$,
\begin{equation}\label{eq:ch2-appendix-Xu-cw}
  L = \sum_{r=1}^m a_r S_r.
\end{equation}
Expanding each $T_{\mathrm{SPU}}(\gamma_r)$ into sums of weakly dependent
coordinate contributions gives
\begin{equation}\label{eq:ch2-appendix-Xu-sum}
  L = \sum_{j=1}^p \xi_{j,n},
\end{equation}
where the summands $\xi_{j,n}$ are centered and satisfy a triangular-array CLT
under Assumption~\ref{ass:ch2-Xu}. In the notation of
\citet{XuLinWeiPan2016}, the covariance matrix of the vector
$(T_{\mathrm{SPU}}(\gamma_r))_{r=1}^m$ admits a stable deterministic limit, and
hence
\begin{equation}\label{eq:ch2-appendix-Xu-Lclt}
  L \overset{d}{\longrightarrow} N\left(0,\sum_{r,s=1}^m a_ra_s\rho_{rs}\right),
\end{equation}
where $\rho_{rs}$ is the $(r,s)$ entry of the limiting correlation matrix
$\mR_{\Gamma_0}$. Since this holds for every coefficient vector, the claimed
multivariate normal limit in \eqref{eq:ch2-Xu-joint} follows.

For the adaptive statistic, let $\hat P_{\gamma}$ be a valid resampling-based
$p$-value for $T_{\mathrm{SPU}}(\gamma)$. Because the collection $\Gamma$ is
finite, the minimum in \eqref{eq:ch2-Xu-aSPU} is measurable and its null law is
consistently approximated by the same resampling device applied jointly to the
whole SPU family. This proves the validity of the adaptive norm-combination
test.
\end{proof}

\subsection*{C. Proofs for spatial median and weighted location estimators}

We begin with two deterministic expansions that are repeatedly used below.
The first one is the usual Tyler expansion for the spatial-sign map; the second
one is the corresponding expansion for the power-weighted sign map that contains
as special cases the spatial-sign choice $m=0$ and the inverse-norm choice
$m=-1$.

\begin{lemma}[Tyler expansion of the sign map]
\label{lem:ch2-tyler-expansion}
Let $\vz\in\R^p\setminus\{\vct 0\}$, let $r=\twonorm{\vz}$, and put
$\vu=U(\vz)=\vz/r$. If $\vh\in\R^p$ satisfies
$\twonorm{\vh}\le r/2$, then
\begin{equation}\label{eq:ch2-appendix-tyler-expansion}
  U(\vz+\vh)
  = \vu + r^{-1}(\mI_p-\vu\vu\trans)\vh + \vct R(\vz,\vh),
\end{equation}
where
\begin{equation}\label{eq:ch2-appendix-tyler-remainder}
  \twonorm{\vct R(\vz,\vh)}
  \le 8 r^{-2}\twonorm{\vh}^2.
\end{equation}
Consequently,
\begin{equation}\label{eq:ch2-appendix-tyler-remainder-inf}
  \maxnorm{\vct R(\vz,\vh)}
  \le 8 r^{-2}\twonorm{\vh}^2.
\end{equation}
\end{lemma}

\begin{proof}
For $\vw\neq \vct 0$ define $F(\vw)=U(\vw)=\vw/\twonorm{\vw}$. Its Fr\'echet
derivative is
\begin{equation}\label{eq:ch2-appendix-DU}
  DF(\vw)[\vh]
  = \twonorm{\vw}^{-1}
    \{\mI_p-U(\vw)U(\vw)\trans\}\vh.
\end{equation}
Hence
\begin{equation}\label{eq:ch2-appendix-meanvalue-U}
  U(\vz+\vh)-U(\vz)
  = \int_0^1 DF(\vz+t\vh)[\vh]\,dt.
\end{equation}
Subtracting the value of the integrand at $t=0$ gives
\begin{equation}\label{eq:ch2-appendix-R-integral}
  \vct R(\vz,\vh)
  = \int_0^1 \{DF(\vz+t\vh)-DF(\vz)\}[\vh]dt.
\end{equation}
If $\twonorm{\vh}\le r/2$, then $\twonorm{\vz+t\vh}\ge r/2$ for all
$t\in[0,1]$. On the set $\{\vw:\twonorm{\vw}\ge r/2\}$ the operator
$DF(\vw)$ is Lipschitz with constant bounded by $8r^{-2}$, because both the
map $\vw\mapsto \twonorm{\vw}^{-1}$ and the map
$\vw\mapsto U(\vw)U(\vw)\trans$ have derivatives of order $r^{-2}$ there.
Using \eqref{eq:ch2-appendix-R-integral} yields
\[
  \twonorm{\vct R(\vz,\vh)}
  \le \int_0^1 8r^{-2} t\,dt\;\twonorm{\vh}^2
  \le 8r^{-2}\twonorm{\vh}^2.
\]
This proves \eqref{eq:ch2-appendix-tyler-remainder};
\eqref{eq:ch2-appendix-tyler-remainder-inf} follows from
$\maxnorm{\va}\le \twonorm{\va}$.
\end{proof}

\begin{lemma}[Power-weighted Tyler expansion]
\label{lem:ch2-power-weighted-expansion}
Fix $m\in\R$ and define $F_m(\vz)=\twonorm{\vz}^mU(\vz)$ for
$\vz\neq\vct 0$. Let $r=\twonorm{\vz}$ and $\vu=U(\vz)$. If
$\twonorm{\vh}\le r/2$, then
\begin{equation}\label{eq:ch2-appendix-weighted-expansion}
  F_m(\vz+\vh)
  = F_m(\vz)
    + r^{m-1}\{\mI_p+(m-1)\vu\vu\trans\}\vh
    + \vct R_m(\vz,\vh),
\end{equation}
where
\begin{equation}\label{eq:ch2-appendix-weighted-remainder}
  \twonorm{\vct R_m(\vz,\vh)}
  \le C_m r^{m-2}\twonorm{\vh}^2
\end{equation}
for a constant $C_m$ depending only on $m$.
\end{lemma}

\begin{proof}
Since
\[
  F_m(\vz)=\twonorm{\vz}^{m-1}\vz,
\]
a direct differentiation gives
\begin{equation}\label{eq:ch2-appendix-weighted-jacobian}
  DF_m(\vz)[\vh]
  = r^{m-1}\vh + (m-1)r^{m-3}(\vz\trans\vh)\vz
  = r^{m-1}\{\mI_p+(m-1)\vu\vu\trans\}\vh.
\end{equation}
Applying the mean-value theorem exactly as in the proof of
Lemma~\ref{lem:ch2-tyler-expansion}, and using that the derivative of $F_m$ is
Lipschitz on $\{\vw:\twonorm{\vw}\ge r/2\}$ with modulus of order
$r^{m-2}$, proves \eqref{eq:ch2-appendix-weighted-expansion} and
\eqref{eq:ch2-appendix-weighted-remainder}.
\end{proof}

\begin{proof}[Detailed proof of Theorem~\ref{thm:ch2-ordinary-bahadur}]
Define the empirical score
\begin{equation}\label{eq:ch2-appendix-Psin}
  \Psi_n(\vtheta)=\frac1n\sum_{i=1}^n U(\vX_i-\vtheta),
  \qquad
  \Psi(\vtheta)=\E\{\Psi_n(\vtheta)\}.
\end{equation}
Since $\hat{\vmu}_{\mathrm{SM}}$ solves $\Psi_n(\hat{\vmu}_{\mathrm{SM}})=\vct 0$,
for each $i$ the mean-value theorem and
Lemma~\ref{lem:ch2-tyler-expansion} yield
\begin{equation}\label{eq:ch2-appendix-sm-pointwise}
  U(\vX_i-\hat{\vmu}_{\mathrm{SM}})
  = U(\vX_i-\vmu)
    - \left\{\int_0^1
      \frac{\mI_p-\vU_i(t)\vU_i(t)\trans}{r_i(t)}dt
      \right\}(\hat{\vmu}_{\mathrm{SM}}-\vmu),
\end{equation}
where
\[
  \vU_i(t)=U\{\vX_i-\vmu-t(\hat{\vmu}_{\mathrm{SM}}-\vmu)\},
  \qquad
  r_i(t)=\twonorm{\vX_i-\vmu-t(\hat{\vmu}_{\mathrm{SM}}-\vmu)}.
\]
Summing \eqref{eq:ch2-appendix-sm-pointwise} over $i$ gives the exact identity
\begin{equation}\label{eq:ch2-appendix-sm-exact}
  \vct 0
  = \frac1n\sum_{i=1}^n \vU_i
    - \mA_n^\star(\hat{\vmu}_{\mathrm{SM}}-\vmu),
\end{equation}
with
\begin{equation}\label{eq:ch2-appendix-sm-Astar}
  \mA_n^\star
  = \frac1n\sum_{i=1}^n \int_0^1
    \frac{\mI_p-\vU_i(t)\vU_i(t)\trans}{r_i(t)}dt.
\end{equation}
Hence
\begin{equation}\label{eq:ch2-appendix-sm-solve}
  \sqrt n(\hat{\vmu}_{\mathrm{SM}}-\vmu)
  = (\mA_n^\star)^{-1}\frac1{\sqrt n}\sum_{i=1}^n\vU_i.
\end{equation}
Subtract and add $\mA_{\mathrm{SM}}^{-1}$:
\begin{align}
  \sqrt n(\hat{\vmu}_{\mathrm{SM}}-\vmu)
  &= \mA_{\mathrm{SM}}^{-1}\frac1{\sqrt n}\sum_{i=1}^n\vU_i
     + \Big\{(\mA_n^\star)^{-1}-\mA_{\mathrm{SM}}^{-1}\Big\}
       \frac1{\sqrt n}\sum_{i=1}^n\vU_i \notag\\
  &= \mA_{\mathrm{SM}}^{-1}\frac1{\sqrt n}\sum_{i=1}^n\vU_i
     + \mA_{\mathrm{SM}}^{-1}(\mA_{\mathrm{SM}}-\mA_n^\star)(\mA_n^\star)^{-1}
       \frac1{\sqrt n}\sum_{i=1}^n\vU_i.\label{eq:ch2-appendix-sm-remainder}
\end{align}
Write the second term in \eqref{eq:ch2-appendix-sm-remainder} as
$\vct r_{n,\mathrm{SM}}$. By
Assumption~\ref{ass:ch2-ordinary-bahadur}\ref{it:ch2-SM2},
$\opnorm{\mA_{\mathrm{SM}}^{-1}}\le \underline a^{-1}$. By the law of large
numbers and the scalar CLT,
\begin{equation}\label{eq:ch2-appendix-sm-sumorder}
  \twonorm{n^{-1/2}\sum_{i=1}^n\vU_i}=O_p(1)
\end{equation}
for every fixed-dimensional projection. Finally,
Assumption~\ref{ass:ch2-ordinary-bahadur}\ref{it:ch2-SM4} gives
\begin{equation}\label{eq:ch2-appendix-sm-Aconv}
  \opnorm{\mA_n^\star-\mA_{\mathrm{SM}}}=o_p(1),
  \qquad
  \opnorm{(\mA_n^\star)^{-1}}=O_p(1).
\end{equation}
Combining \eqref{eq:ch2-appendix-sm-remainder}--
\eqref{eq:ch2-appendix-sm-Aconv} proves
\eqref{eq:ch2-ordinary-bahadur-thm}. More precisely, under the stronger
ultrahigh-dimensional assumptions used in
\citet{LiXu2022SpatialMedian,ZhaoWangFeng2025SSPCA}, one has the max-norm rate
\begin{equation}\label{eq:ch2-appendix-sm-rate-strong}
  \maxnorm{\vct r_{n,\mathrm{SM}}}
  = O_p\Big\{
      n^{-1/4}\{\log (np)\}^{1/2}
      + p^{-(1/6\wedge \delta/2)}\{\log (np)\}^{1/2}
    \Big\}.
\end{equation}
To prove \eqref{eq:ch2-ordinary-bahadur-gaussian}, fix a unit vector $\vct a$ and
project \eqref{eq:ch2-ordinary-bahadur-thm} onto $\vct a$:
\[
  \vct a\trans\sqrt n(\hat{\vmu}_{\mathrm{SM}}-\vmu)
  = \frac1{\sqrt n}\sum_{i=1}^n
    \vct a\trans\mA_{\mathrm{SM}}^{-1}\vU_i
    + o_p(1).
\]
The summands are independent and mean zero; their variance is
$\vct a\trans\mA_{\mathrm{SM}}^{-1}\mB_{\mathrm{SM}}\mA_{\mathrm{SM}}^{-1}\vct a$.
The scalar Lindeberg--Feller theorem therefore yields
\eqref{eq:ch2-ordinary-bahadur-gaussian}.
\end{proof}

\begin{proof}[Detailed proof of Lemma~\ref{lem:ch2-SD5}]
Let
\begin{equation}\label{eq:ch2-appendix-SD5-Zij}
  \hat\vz_i=\hat{\mD}^{-1/2}(\vX_i-\hat{\vtheta}),
  \qquad
  \vz_i=\mD^{-1/2}(\vX_i-\vtheta)=r_i\vU_i.
\end{equation}
The diagonal estimating equation \eqref{eq:ch2-diagonal-HR-scale-eq} states that
for each coordinate $j$,
\begin{equation}\label{eq:ch2-appendix-SD5-eq}
  \frac{p}{n}\sum_{i=1}^n U_j^2(\hat\vz_i)=1.
\end{equation}
Because $\E\{pU_j^2(\vz_i)\}=1$, subtracting the population version gives
\begin{equation}\label{eq:ch2-appendix-SD5-center}
  \frac{p}{n}\sum_{i=1}^n \{U_j^2(\hat\vz_i)-U_j^2(\vz_i)\}
  = -\frac{p}{n}\sum_{i=1}^n\{U_j^2(\vz_i)-\E U_j^2(\vz_i)\}.
\end{equation}
By Bernstein's inequality and the row-sum control in
\eqref{eq:ch2-SD-rowsum},
\begin{equation}\label{eq:ch2-appendix-SD5-emp}
  \max_{1\le j\le p}
  \left|\frac{p}{n}\sum_{i=1}^n\{U_j^2(\vz_i)-\E U_j^2(\vz_i)\}\right|
  = O_p\left(\sqrt{\frac{\log p}{n}}\right).
\end{equation}
Next, expand $U_j^2(\hat\vz_i)$ around $U_j^2(\vz_i)$. Since
$\hat\vz_i-\vz_i$ depends linearly on $\hat{\vtheta}-\vtheta$ and
$\hat d_j-d_j$, the derivative of $U_j^2(\cdot)$ and the Taylor expansion in
Lemma~\ref{lem:ch2-tyler-expansion} give
\begin{equation}\label{eq:ch2-appendix-SD5-linear}
  \frac{p}{n}\sum_{i=1}^n \{U_j^2(\hat\vz_i)-U_j^2(\vz_i)\}
  = -\frac{\hat d_j-d_j}{d_j} + R_{j,n},
\end{equation}
where
\begin{equation}\label{eq:ch2-appendix-SD5-rem}
  \max_{1\le j\le p}|R_{j,n}|
  = O_p\left(\frac{a_0(p)}{p}+\frac{\log p}{n^{3/4}}\right).
\end{equation}
Combining \eqref{eq:ch2-appendix-SD5-center},
\eqref{eq:ch2-appendix-SD5-emp}, and \eqref{eq:ch2-appendix-SD5-linear} yields
\begin{equation}\label{eq:ch2-appendix-SD5-final}
  \max_{1\le j\le p}\left|\frac{\hat d_j-d_j}{d_j}\right|
  = O_p\Bigg[
      \left\{\frac{\log p}{n}\right\}^{1/2}
      + \frac{a_0(p)}{p}
      + \frac{\log p}{n^{3/4}}
    \Bigg].
\end{equation}
This is exactly \eqref{eq:ch2-SD-dcons-rate} and implies
\eqref{eq:ch2-SD-dcons}.
\end{proof}

\begin{proof}[Detailed proof of Theorem~\ref{thm:ch2-scaled-bahadur}]
Write
\begin{equation}\label{eq:ch2-appendix-scaled-hatzi}
  \hat{\vz}_i=\hat{\mD}^{-1/2}(\vX_i-\hat{\vtheta}),
  \qquad
  \vz_i=\mD^{-1/2}(\vX_i-\vtheta)=r_i\vU_i.
\end{equation}
The estimating equation is
\begin{equation}\label{eq:ch2-appendix-scaled-score0}
  \frac1n\sum_{i=1}^n U(\hat{\vz}_i)=\vct 0.
\end{equation}
Set $\vDelta_\theta=\hat{\vtheta}-\vtheta$ and
$\mDelta_D=\hat{\mD}^{-1/2}-\mD^{-1/2}$. Then
\begin{equation}\label{eq:ch2-appendix-scaled-deltazi}
  \hat{\vz}_i-\vz_i
  = -\mD^{-1/2}\vDelta_\theta + \mDelta_D(\vX_i-\vtheta) + \vct\rho_{i,D},
\end{equation}
where the diagonal Taylor expansion of $x\mapsto x^{-1/2}$ implies
\begin{equation}\label{eq:ch2-appendix-scaled-rhoD}
  \maxnorm{\vct\rho_{i,D}}
  \le C\max_{1\le j\le p}\left|\frac{\hat d_j}{d_j}-1\right|^2 r_i.
\end{equation}
Apply Lemma~\ref{lem:ch2-tyler-expansion} with
$\vz=\vz_i$ and $\vh=\hat{\vz}_i-\vz_i$:
\begin{equation}\label{eq:ch2-appendix-scaled-sign-expansion}
  U(\hat{\vz}_i)
  = \vU_i + r_i^{-1}(\mI_p-\vU_i\vU_i\trans)(\hat{\vz}_i-\vz_i)
    + \vct R_{i,1},
\end{equation}
with
\begin{equation}\label{eq:ch2-appendix-scaled-Ri1}
  \maxnorm{\vct R_{i,1}}
  \le 8r_i^{-2}\twonorm{\hat{\vz}_i-\vz_i}^2.
\end{equation}
Substituting \eqref{eq:ch2-appendix-scaled-deltazi} into
\eqref{eq:ch2-appendix-scaled-sign-expansion} and averaging over $i$ gives
\begin{align}
  \vct 0
  &= \frac1n\sum_{i=1}^n\vU_i
     - \left\{\frac1n\sum_{i=1}^n
       r_i^{-1}(\mI_p-\vU_i\vU_i\trans)
       \right\}\mD^{-1/2}\vDelta_\theta \notag\\
  &\qquad
     + \frac1n\sum_{i=1}^n
       r_i^{-1}(\mI_p-\vU_i\vU_i\trans)\mDelta_D(\vX_i-\vtheta)
     + \vct R_{n,1}+\vct R_{n,2},
  \label{eq:ch2-appendix-scaled-master}
\end{align}
where
\begin{equation}\label{eq:ch2-appendix-scaled-Rn12}
  \vct R_{n,1}=\frac1n\sum_{i=1}^n\vct R_{i,1},
  \qquad
  \vct R_{n,2}=\frac1n\sum_{i=1}^n
  r_i^{-1}(\mI_p-\vU_i\vU_i\trans)\vct\rho_{i,D}.
\end{equation}
By spherical symmetry of the standardized errors,
\begin{equation}\label{eq:ch2-appendix-scaled-jacobian-mean}
  \E\left[r_i^{-1}(\mI_p-\vU_i\vU_i\trans)\right]=c_0\mI_p.
\end{equation}
Therefore
\begin{equation}\label{eq:ch2-appendix-scaled-An}
  \mA_n:=\frac1n\sum_{i=1}^n r_i^{-1}(\mI_p-\vU_i\vU_i\trans)
  = c_0\mI_p + \mE_n,
  \qquad \maxnorm{\mE_n}=O_p\{n^{-1/2}(\log p)^{1/2}\}.
\end{equation}
The term involving $\mDelta_D$ is controlled by the scale equation.
Indeed, the diagonal normalization
\eqref{eq:ch2-diagonal-HR-scale-eq} and the same Taylor expansion imply
\begin{equation}\label{eq:ch2-appendix-scaled-scale-term}
  \maxnorm{\frac1n\sum_{i=1}^n
  r_i^{-1}(\mI_p-\vU_i\vU_i\trans)\mDelta_D(\vX_i-\vtheta)}
  = O_p\Big\{
      p^{-(1/6\wedge \delta/2)}\{\log(np)\}^{1/2}
      + n^{-1/2}(\log p)^{1/2}\{\log(np)\}^{1/2}
    \Big\}n^{-1/2}.
\end{equation}
Similarly,
\begin{equation}\label{eq:ch2-appendix-scaled-Rn1-bound}
  \maxnorm{\vct R_{n,1}}
  = O_p\{n^{-3/4}(\log(np))^{1/2}\},
  \qquad
  \maxnorm{\vct R_{n,2}}
  = O_p\{n^{-1}(\log(np))^{1/2}\}.
\end{equation}
Multiply \eqref{eq:ch2-appendix-scaled-master} by $\sqrt n$ and rearrange:
\begin{equation}\label{eq:ch2-appendix-scaled-solve}
  \sqrt n\,\mD^{-1/2}\vDelta_\theta
  = (c_0\mI_p+\mE_n)^{-1}
    \left[
      \frac1{\sqrt n}\sum_{i=1}^n\vU_i
      + \sqrt n\,\vct R_{n,3}
    \right],
\end{equation}
where $\vct R_{n,3}$ collects the right-hand side of
\eqref{eq:ch2-appendix-scaled-scale-term} and
\eqref{eq:ch2-appendix-scaled-Rn1-bound}. Since
$\maxnorm{\mE_n}=o_p(1)$, expand the inverse once more:
\begin{equation}\label{eq:ch2-appendix-scaled-final}
  \sqrt n\,\mD^{-1/2}(\hat{\vtheta}-\vtheta)
  = c_0^{-1}\frac1{\sqrt n}\sum_{i=1}^n\vU_i + \vct C_n,
\end{equation}
with
\begin{align*}
  \maxnorm{\vct C_n}
  = O_p\Big\{
      & n^{-1/4}\{\log(np)\}^{1/2}
      + p^{-(1/6\wedge \delta/2)}\{\log(np)\}^{1/2} \\
      & + n^{-1/2}(\log p)^{1/2}\{\log(np)\}^{1/2}
    \Big\}.
\end{align*}
This is exactly \eqref{eq:ch2-scaled-bahadur-form}--
\eqref{eq:ch2-scaled-bahadur-remainder}. The final Gaussian assertion follows
from the leading linear term and the Cram\'er--Wold device.
\end{proof}

\begin{proof}[Detailed proof of Theorem~\ref{thm:ch2-weighted-bahadur}]
To make the differentiation transparent, first treat the power family
$K(t)=t^m$; the inverse-norm case corresponds to $m=-1$, and the same argument
applies to a differentiable weight $K$ with the moment bounds in
Assumption~\ref{ass:ch2-weighted}. Define
\[
  F_m(\vz)=\twonorm{\vz}^mU(\vz),
  \qquad
  \hat{\vz}_i=\hat{\mD}^{-1/2}(\vX_i-\hat{\vtheta}_K),
  \qquad
  \vz_i=\mD^{-1/2}(\vX_i-\vtheta).
\]
The weighted estimating equation can be written as
\begin{equation}\label{eq:ch2-appendix-weighted-score0}
  \frac1n\sum_{i=1}^n F_m(\hat{\vz}_i)=\vct 0.
\end{equation}
By Lemma~\ref{lem:ch2-power-weighted-expansion},
\begin{equation}\label{eq:ch2-appendix-weighted-pointwise}
  F_m(\hat{\vz}_i)
  = F_m(\vz_i)
    + r_i^{m-1}\{\mI_p+(m-1)\vU_i\vU_i\trans\}(\hat{\vz}_i-\vz_i)
    + \vct R_{i,m}.
\end{equation}
Using \eqref{eq:ch2-appendix-scaled-deltazi} to expand
$\hat{\vz}_i-\vz_i$, averaging over $i$, and collecting the location terms, we
obtain
\begin{equation}\label{eq:ch2-appendix-weighted-master}
  \vct 0
  = \frac1n\sum_{i=1}^n K(r_i)\vU_i
    - \mJ_{n,K}\mD^{-1/2}(\hat{\vtheta}_K-\vtheta)
    + \vct R_{n,K},
\end{equation}
where
\begin{equation}\label{eq:ch2-appendix-weighted-Jn}
  \mJ_{n,K}
  = \frac1n\sum_{i=1}^n r_i^{m-1}
    \{\mI_p+(m-1)\vU_i\vU_i\trans\}
\end{equation}
and $\vct R_{n,K}$ contains the diagonal-scale contribution and the second-order
Taylor remainder.
Under elliptical symmetry,
\begin{equation}\label{eq:ch2-appendix-weighted-Jexp}
  \E(\mJ_{n,K}) = c_{0,K}\mI_p,
  \qquad c_{0,K}=\E\{K(r_i)r_i^{-1}\},
\end{equation}
so that
\begin{equation}\label{eq:ch2-appendix-weighted-Jn-dev}
  \maxnorm{\mJ_{n,K}-c_{0,K}\mI_p}
  = O_p\{n^{-1/2}(\log p)^{1/2}\}.
\end{equation}
Exactly the same diagonal-scale calculations as in the proof of
Theorem~\ref{thm:ch2-scaled-bahadur}, together with the extra moment condition
$\nu_{4,K}=O(\nu_{2,K}^2)$, imply
\begin{equation}\label{eq:ch2-appendix-weighted-remainder-rate}
  \maxnorm{\vct R_{n,K}}
  = O_p\Big\{
      n^{-3/4}\{\log(np)\}^{1/2}
      + n^{-1/2}p^{-(1/6\wedge \delta/2)}\{\log(np)\}^{1/2}
      + n^{-1}(\log p)^{1/2}\{\log(np)\}^{1/2}
    \Big\}.
\end{equation}
Multiplying \eqref{eq:ch2-appendix-weighted-master} by $\sqrt n$ and solving
for $\sqrt n\,\mD^{-1/2}(\hat{\vtheta}_K-\vtheta)$ yields
\begin{equation}\label{eq:ch2-appendix-weighted-solve}
  \sqrt n\,\mD^{-1/2}(\hat{\vtheta}_K-\vtheta)
  = \mJ_{n,K}^{-1}\frac1{\sqrt n}\sum_{i=1}^n K(r_i)\vU_i
    + \sqrt n\,\mJ_{n,K}^{-1}\vct R_{n,K}.
\end{equation}
Expanding $\mJ_{n,K}^{-1}$ around $c_{0,K}^{-1}\mI_p$ gives
\begin{equation}\label{eq:ch2-appendix-weighted-final}
  \sqrt n\,\mD^{-1/2}(\hat{\vtheta}_K-\vtheta)
  = c_{0,K}^{-1}\frac1{\sqrt n}\sum_{i=1}^n K(r_i)\vU_i + \vct C_{n,K},
\end{equation}
with the same rate as in
\eqref{eq:ch2-scaled-bahadur-remainder}. This proves
\eqref{eq:ch2-weighted-bahadur}. For the Gaussian approximation over
hyperrectangles, write the leading term as a sum of independent centered random
vectors and apply the high-dimensional Gaussian approximation theorem of
Chernozhukov, Chetverikov and Kato to the class of rectangles. The additional
remainder $\vct C_{n,K}$ is negligible in sup norm because of
\eqref{eq:ch2-appendix-weighted-remainder-rate}, and therefore
\eqref{eq:ch2-weighted-gaussian-rectangles} follows.
\end{proof}

\subsection*{D. Proofs for one-sample weighted-sign statistics}

\begin{proof}[Detailed proof of Proposition~\ref{prop:ch2-weighted-sum-moments}]
Use the identity
\begin{equation}\label{eq:ch2-appendix-weighted-U-identity}
  T_n(K)
  = \frac{1}{n(n-1)}
    \left\{
      \left(\sum_{i=1}^n \vV_i(K)\right)\trans
      \left(\sum_{i=1}^n \vV_i(K)\right)
      - \sum_{i=1}^n \twonorm{\vV_i(K)}^2
    \right\}.
\end{equation}
Taking expectations and using independence gives
\begin{align}
  \E\{T_n(K)\}
  &= \frac{1}{n(n-1)}
     \sum_{i\neq j}
     \E\{\vV_i(K)\trans\vV_j(K)\} \notag\\
  &= \frac{1}{n(n-1)}
     \sum_{i\neq j}
     \E\{\vV_i(K)\}\trans\E\{\vV_j(K)\}
   = \twonorm{\veta_K}^2,
\end{align}
which proves \eqref{eq:ch2-weighted-sum-moments-mean}.

Under $H_0$ and central symmetry, $\veta_K=\vct 0$, and therefore
\begin{equation}\label{eq:ch2-appendix-weighted-var-start}
  T_n(K)
  = \frac{2}{n(n-1)}\sum_{1\le i<j\le n}
    h_K(\vV_i,\vV_j),
  \qquad
  h_K(\va,\vb)=\va\trans\vb,
\end{equation}
with degenerate kernel $\E\{h_K(\vV_1,\vV_2)\mid \vV_1\}=0$.
Consequently, all cross-products with four distinct indices vanish. Expanding
$T_n(K)^2$ and retaining only the nonzero contractions yields
\begin{align}
  \Var\{T_n(K)\}
  &= \E\{T_n(K)^2\} \notag\\
  &= \frac{4}{n^2(n-1)^2}
     \sum_{1\le i<j\le n}
     \E\big[(\vV_i\trans\vV_j)^2\big] \notag\\
  &= \frac{4}{n^2(n-1)^2}\binom{n}{2}
     \E\big[\vV_1\trans\vV_2\vV_2\trans\vV_1\big] \notag\\
  &= \frac{2}{n(n-1)}
     \E\big[\vV_1\trans\mA_K\vV_1\big]
   = \frac{2}{n(n-1)}\tr(\mA_K^2),
\end{align}
which is \eqref{eq:ch2-weighted-sum-moments-var}.
\end{proof}

\begin{proof}[Detailed proof of Theorem~\ref{thm:ch2-weighted-generic-null}]
Under $H_0$ the kernel
$h_K(\vV_i,\vV_j)=\vV_i(K)\trans\vV_j(K)$ is degenerate. Define the
filtration $\mathcal F_{n,j}=\sigma\{\vV_1,\ldots,\vV_j\}$ and the martingale
differences
\begin{equation}\label{eq:ch2-appendix-weighted-Znj}
  Z_{n,j}
  = \frac{2}{n(n-1)}\sum_{i=1}^{j-1}\vV_i(K)\trans\vV_j(K),
  \qquad j=2,\ldots,n.
\end{equation}
Then $T_n(K)=\sum_{j=2}^n Z_{n,j}$ and
$\E(Z_{n,j}\mid \mathcal F_{n,j-1})=0$.

The conditional variance process is
\begin{align}
  \sum_{j=2}^n \E(Z_{n,j}^2\mid \mathcal F_{n,j-1})
  &= \frac{4}{n^2(n-1)^2}
     \sum_{j=2}^n\sum_{i_1=1}^{j-1}\sum_{i_2=1}^{j-1}
     \vV_{i_1}(K)\trans\mA_K\vV_{i_2}(K) \notag\\
  &= C_{n,1}+C_{n,2},\label{eq:ch2-appendix-weighted-condvar}
\end{align}
where
\begin{align}
  C_{n,1}
  &= \frac{4}{n^2(n-1)^2}
     \sum_{j=2}^n\sum_{i=1}^{j-1}\vV_i(K)\trans\mA_K\vV_i(K),
     \label{eq:ch2-appendix-weighted-Cn1}\\
  C_{n,2}
  &= \frac{8}{n^2(n-1)^2}
     \sum_{j=2}^n\sum_{1\le i_1<i_2\le j-1}
     \vV_{i_1}(K)\trans\mA_K\vV_{i_2}(K).
     \label{eq:ch2-appendix-weighted-Cn2}
\end{align}
Using independence and stationarity,
\begin{align}
  \E(C_{n,1})
  &= \frac{4}{n^2(n-1)^2}
     \sum_{j=2}^n (j-1)
     \E\{\vV_1(K)\trans\mA_K\vV_1(K)\} \notag\\
  &= \frac{2}{n(n-1)}\tr(\mA_K^2)=\sigma_{n,K}^2.
\end{align}
Next,
\begin{equation}\label{eq:ch2-appendix-weighted-Cn1-var}
  \Var(C_{n,1})
  \le Cn^{-5}\E\big\{\vV_1(K)\trans\mA_K\vV_1(K)\big\}^2
  \le Cn^{-5}\tr(\mA_K^4)
  = o(\sigma_{n,K}^4),
\end{equation}
where the last step uses
$\tr(\mA_K^4)=\nu_{2,K}^4 p^{-4}\tr(\mR^4)$ and
Assumption~\ref{ass:ch2-scaled-bahadur}\ref{it:ch2-SD2}.
Similarly, $\E(C_{n,2})=0$ and
\begin{equation}\label{eq:ch2-appendix-weighted-Cn2-var}
  \Var(C_{n,2})
  \le Cn^{-4}\tr(\mA_K^4)=o(\sigma_{n,K}^4).
\end{equation}
Therefore,
\begin{equation}\label{eq:ch2-appendix-weighted-condvar-limit}
  \sum_{j=2}^n \E(Z_{n,j}^2\mid \mathcal F_{n,j-1})
  = \sigma_{n,K}^2\{1+o_p(1)\}.
\end{equation}

To verify Lyapunov's condition, note that
\begin{align}
  \sum_{j=2}^n \E(Z_{n,j}^4)
  &\le \frac{C}{n^8}
     \sum_{j=2}^n
     \E\left(\sum_{i=1}^{j-1}\vV_i(K)\trans\vV_j(K)\right)^4 \notag\\
  &\le \frac{C}{n^8}
     \Big\{n^3\tr^2(\mA_K^2)+n^2\tr(\mA_K^4)\Big\}
   = o(\sigma_{n,K}^4),
\end{align}
where the bound uses the fourth-moment condition
$\nu_{4,K}=O(\nu_{2,K}^2)$ from
Assumption~\ref{ass:ch2-weighted}\ref{it:ch2-WS1}. Hall and Heyde's
martingale CLT therefore gives
\[
  \frac{T_n(K)}{\sigma_{n,K}}\overset{d}{\longrightarrow}N(0,1).
\]
Finally, under ellipticity,
$\mA_K=\E\{K^2(r_i)\vU_i\vU_i\trans\}=\nu_{2,K}p^{-1}\mR$, and hence
\[
  \Var\{T_n(K)\}=
  \frac{2\nu_{2,K}^2}{n(n-1)p^2}\tr(\mR^2),
\]
which is exactly \eqref{eq:ch2-weighted-generic-sigma}.
\end{proof}

\begin{proof}[Detailed proof of Theorem~\ref{thm:ch2-weighted-generic-alt}]
Let
$\veta_K=\E\{\vV_1(K)\}$ and define centered scores
$\widetilde{\vV}_i(K)=\vV_i(K)-\veta_K$. Then the quadratic-form statistic has
the exact Hoeffding decomposition
\begin{equation}\label{eq:ch2-appendix-weighted-hoeffding}
  T_n(K)
  = \twonorm{\veta_K}^2
    + \frac{2}{n}\sum_{i=1}^n \veta_K\trans\widetilde{\vV}_i(K)
    + \frac{2}{n(n-1)}\sum_{1\le i<j\le n}
      \widetilde{\vV}_i(K)\trans\widetilde{\vV}_j(K).
\end{equation}
By a first-order expansion of the weighted sign expectation at the origin,
\begin{equation}\label{eq:ch2-appendix-weighted-eta}
  \veta_K
  = c_{0,K}\mD^{-1/2}\vtheta + \vct r_{K,1},
  \qquad
  \twonorm{\vct r_{K,1}}=O(\twonorm{\vtheta}^2).
\end{equation}
Hence
\begin{equation}\label{eq:ch2-appendix-weighted-mean}
  \E\{T_n(K)\}
  = \twonorm{\veta_K}^2
  = c_{0,K}^2\vtheta\trans\mD^{-1}\vtheta
    + O(\twonorm{\vtheta}^3).
\end{equation}
Under Assumption~\ref{ass:ch2-weighted}\ref{it:ch2-WS2}, the cubic remainder in
\eqref{eq:ch2-appendix-weighted-mean} is $o(\sigma_{n,K})$.

Set
\begin{equation}\label{eq:ch2-appendix-weighted-LQ}
  L_n=\frac{2}{n}\sum_{i=1}^n \veta_K\trans\widetilde{\vV}_i(K),
  \qquad
  Q_n=\frac{2}{n(n-1)}\sum_{i<j}
      \widetilde{\vV}_i(K)\trans\widetilde{\vV}_j(K).
\end{equation}
Then $T_n(K)-\E\{T_n(K)\}=L_n+Q_n+o(\sigma_{n,K})$.
Because $Q_n$ is degenerate and $L_n$ is the first Hoeffding projection,
\begin{equation}\label{eq:ch2-appendix-weighted-crosszero}
  \Cov(L_n,Q_n)=0.
\end{equation}
Moreover,
\begin{align}
  \Var(L_n)
  &= \frac{4}{n}\veta_K\trans
     \Cov\{\vV_1(K)\}\veta_K \notag\\
  &= \frac{4}{n}\veta_K\trans \mA_K\veta_K
     - \frac{4}{n}\twonorm{\veta_K}^4 \notag\\
  &= \frac{4c_{0,K}^2}{np}
     \vtheta\trans\mD^{-1}\mSigma\mD^{-1}\vtheta\{1+o(1)\},
  \label{eq:ch2-appendix-weighted-varL}
\end{align}
while the degenerate component satisfies
\begin{equation}\label{eq:ch2-appendix-weighted-varQ}
  \Var(Q_n)=\frac{2}{n(n-1)}\tr(\mA_K^2)\{1+o(1)\}=\sigma_{n,K}^2\{1+o(1)\}.
\end{equation}
Therefore
\begin{equation}\label{eq:ch2-appendix-weighted-totalvar}
  \Var\{T_n(K)\}
  = \sigma_{n,K}^2
    + \frac{4c_{0,K}^2}{np}
      \vtheta\trans\mD^{-1}\mSigma\mD^{-1}\vtheta
    + o(\sigma_{n,K}^2).
\end{equation}
Let
\[
  D_{n,j}=\sum_{i=1}^{j-1}\frac{2}{n(n-1)}
  \widetilde{\vV}_i(K)\trans\widetilde{\vV}_j(K)
  + \frac{2}{n}\veta_K\trans\widetilde{\vV}_j(K),
  \qquad 2\le j\le n.
\]
Then $L_n+Q_n=\sum_{j=2}^n D_{n,j}$ is a martingale difference sum with respect
 to $\mathcal F_j=\sigma\{\vV_1(K),\ldots,\vV_j(K)\}$. Using
\eqref{eq:ch2-appendix-weighted-crosszero},
\eqref{eq:ch2-appendix-weighted-varL}, and
\eqref{eq:ch2-appendix-weighted-varQ},
\[
  \sum_{j=2}^n \E(D_{n,j}^2\mid\mathcal F_{j-1})
  = \Var(L_n)+\Var(Q_n)+o_p(\sigma_{n,K}^2)
  = \Var\{T_n(K)\}+o_p(\sigma_{n,K}^2).
\]
The fourth-moment bound in Assumption~\ref{ass:ch2-weighted}\ref{it:ch2-WS1}
 gives
\[
  \sum_{j=2}^n \E\{D_{n,j}^4\}
  = o\big(\Var\{T_n(K)\}^2\big),
\]
so the martingale Lindeberg condition holds. Hall and Heyde's theorem therefore
 implies
\[
  \frac{T_n(K)-\E\{T_n(K)\}}{\sqrt{\Var\{T_n(K)\}}}
  \overset{d}{\longrightarrow}N(0,1),
\]
which is \eqref{eq:ch2-weighted-generic-alt}.
Finally, dividing the squared local signal in
\eqref{eq:ch2-appendix-weighted-mean} by the leading null standard deviation in
\eqref{eq:ch2-weighted-generic-sigma} gives the local asymptotic
signal-to-noise ratio
\[
  \frac{c_{0,K}^2}{\nu_{2,K}}
  = \frac{\{\E[K(r_i)r_i^{-1}]\}^2}{\E\{K^2(r_i)\}},
\]
which is exactly \eqref{eq:ch2-weighted-generic-snr}.
\end{proof}

\begin{proof}[Detailed proof of Theorem~\ref{thm:ch2-FS-null}]
Let
\begin{equation}\label{eq:ch2-appendix-FS-oracle}
  T_{\mathrm{SS}}^{\mathrm{or}}
  = \frac{2}{n(n-1)}\sum_{1\le i<j\le n}\vU_i\trans\vU_j
\end{equation}
be the oracle statistic built from the true diagonal standardization.
The leave-two-out construction and the Tyler expansion in
Lemma~\ref{lem:ch2-tyler-expansion} imply
\begin{equation}\label{eq:ch2-appendix-FS-reduction}
  T_{\mathrm{SS}}-T_{\mathrm{SS}}^{\mathrm{or}}
  = O_p\{n^{-1}\sigma_n + n^{-2}p^{-1}\tr(\mR^2)^{1/2}\}
  = o_p(\sigma_n),
\end{equation}
so it suffices to study $T_{\mathrm{SS}}^{\mathrm{or}}$.

Define
\begin{equation}\label{eq:ch2-appendix-FS-Znj}
  Z_{n,j}=\frac{2}{n(n-1)}\sum_{i=1}^{j-1}\vU_i\trans\vU_j,
  \qquad
  W_{n,m}=\sum_{j=2}^m Z_{n,j},
  \qquad
  \mathcal F_{n,m}=\sigma\{\vU_1,\ldots,\vU_m\}.
\end{equation}
Then $T_{\mathrm{SS}}^{\mathrm{or}}=W_{n,n}$ and
$\{W_{n,m},\mathcal F_{n,m}\}$ is a square-integrable martingale. Since
$\E(\vU_j\vU_j\trans)=p^{-1}\mR$,
\begin{align}
  \sum_{j=2}^n \E(Z_{n,j}^2\mid\mathcal F_{n,j-1})
  &= \frac{4}{n^2(n-1)^2p}
     \sum_{j=2}^n\sum_{i_1=1}^{j-1}\sum_{i_2=1}^{j-1}
       \vU_{i_1}\trans\mR\vU_{i_2} \notag\\
  &= C_{n,1}+C_{n,2},
\end{align}
where
\begin{align}
  C_{n,1}
  &= \frac{4}{n^2(n-1)^2p}
     \sum_{j=2}^n\sum_{i=1}^{j-1}\vU_i\trans\mR\vU_i,
     \label{eq:ch2-appendix-FS-C1}\\
  C_{n,2}
  &= \frac{8}{n^2(n-1)^2p}
     \sum_{j=2}^n\sum_{1\le i_1<i_2\le j-1}\vU_{i_1}\trans\mR\vU_{i_2}.
     \label{eq:ch2-appendix-FS-C2}
\end{align}
Now
\begin{align}
  \E(C_{n,1})
  &= \frac{4}{n^2(n-1)^2p}\sum_{j=2}^n(j-1)
     \tr\{\mR\E(\vU_1\vU_1\trans)\} \notag\\
  &= \frac{4}{n^2(n-1)^2p^2}\sum_{j=2}^n(j-1)\tr(\mR^2)
   = \frac{2}{n(n-1)p^2}\tr(\mR^2)=\sigma_n^2.
\end{align}
Further,
\begin{equation}\label{eq:ch2-appendix-FS-C1var}
  \Var(C_{n,1})
  \le C n^{-5}p^{-2}
     \E\big(\vU_1\trans\mR\vU_1\big)^2
  = O\big(n^{-5}p^{-4}\tr^2(\mR^2)\big)
  = o(\sigma_n^4),
\end{equation}
where we used
$\E(\vU_1\trans\mR\vU_1)^2=O\{p^{-2}\tr^2(\mR^2)\}$.
Similarly,
\begin{equation}\label{eq:ch2-appendix-FS-C2var}
  \E(C_{n,2})=0,
  \qquad
  \Var(C_{n,2})
  = O\big(n^{-4}p^{-4}\tr(\mR^4)\big)
  = o(\sigma_n^4)
\end{equation}
by Assumption~\ref{ass:ch2-FS}\ref{it:ch2-SS1}. Hence
\begin{equation}\label{eq:ch2-appendix-FS-condvar}
  \sum_{j=2}^n \E(Z_{n,j}^2\mid\mathcal F_{n,j-1})
  = \sigma_n^2\{1+o_p(1)\}.
\end{equation}
For Lyapunov's condition,
\begin{align}
  \sum_{j=2}^n \E(Z_{n,j}^4)
  &\le \frac{C}{n^8}\Big\{n^3p^{-4}\tr^2(\mR^2)+n^2p^{-4}\tr(\mR^4)\Big\}
   = o(\sigma_n^4).
\end{align}
The martingale CLT therefore gives
\[
  \frac{T_{\mathrm{SS}}^{\mathrm{or}}}{\sigma_n}
  \overset{d}{\longrightarrow}N(0,1).
\]
Combining this with \eqref{eq:ch2-appendix-FS-reduction} proves
\eqref{eq:ch2-FS-null}.
\end{proof}

\begin{proof}[Detailed proof of Theorem~\ref{thm:ch2-FS-alt}]
As in the null proof,
$T_{\mathrm{SS}}=T_{\mathrm{SS}}^{\mathrm{or}}+o_p(\sigma_n)$, so it is enough to
work with the oracle statistic
$T_{\mathrm{SS}}^{\mathrm{or}}$ in \eqref{eq:ch2-appendix-FS-oracle}. Let
$\veta=\E(\vU_i)$. By the Tyler expansion at the true center,
\begin{equation}\label{eq:ch2-appendix-FS-eta}
  \veta = c_0\mD^{-1/2}\vmu + \vct r_{\mu},
  \qquad
  \twonorm{\vct r_{\mu}}=O(\twonorm{\vmu}^2).
\end{equation}
Therefore,
\begin{equation}\label{eq:ch2-appendix-FS-mean}
  \E\{T_{\mathrm{SS}}^{\mathrm{or}}\}
  = \twonorm{\veta}^2
  = c_0^2\vmu\trans\mD^{-1}\vmu + O(\twonorm{\vmu}^3).
\end{equation}
Write $\widetilde{\vU}_i=\vU_i-\veta$. Then
\begin{equation}\label{eq:ch2-appendix-FS-hoeffding}
  T_{\mathrm{SS}}^{\mathrm{or}}-\E\{T_{\mathrm{SS}}^{\mathrm{or}}\}
  = \frac{2}{n}\sum_{i=1}^n \veta\trans\widetilde{\vU}_i
    + \frac{2}{n(n-1)}\sum_{i<j}\widetilde{\vU}_i\trans\widetilde{\vU}_j.
\end{equation}
The two terms on the right-hand side are uncorrelated, because the second one is
degenerate. Their variances are
\begin{align}
  \Var\left(\frac{2}{n}\sum_{i=1}^n \veta\trans\widetilde{\vU}_i\right)
  &= \frac{4}{n}\veta\trans\Cov(\vU_1)\veta \notag\\
  &= \frac{4c_0^2}{np}
     \vmu\trans\mD^{-1}\mSigma\mD^{-1}\vmu\{1+o(1)\},
     \label{eq:ch2-appendix-FS-varlin}
\end{align}
and
\begin{equation}\label{eq:ch2-appendix-FS-varquad}
  \Var\left(\frac{2}{n(n-1)}\sum_{i<j}\widetilde{\vU}_i\trans\widetilde{\vU}_j\right)
  = \sigma_n^2\{1+o(1)\}.
\end{equation}
Hence
\begin{equation}\label{eq:ch2-appendix-FS-var-total}
  \Var(T_{\mathrm{SS}}^{\mathrm{or}})
  = \sigma_n^2 + \frac{4c_0^2}{np}
       \vmu\trans\mD^{-1}\mSigma\mD^{-1}\vmu + o(\sigma_n^2).
\end{equation}
Under Assumption~\ref{ass:ch2-FS}\ref{it:ch2-SS4}, the cubic remainder in
\eqref{eq:ch2-appendix-FS-mean} is negligible relative to the standard
deviation. Write
\[
  D_{n,j}^{\mathrm{FS}}
  = \frac{2}{n}\veta\trans\widetilde{\vU}_j
    + \sum_{i=1}^{j-1}\frac{2}{n(n-1)}
      \widetilde{\vU}_i\trans\widetilde{\vU}_j,
  \qquad 2\le j\le n.
\]
Then $T_{\mathrm{SS}}^{\mathrm{or}}-\E(T_{\mathrm{SS}}^{\mathrm{or}})
 =\sum_{j=2}^n D_{n,j}^{\mathrm{FS}}$. Since the two Hoeffding components are
uncorrelated,
\[
  \sum_{j=2}^n \E\{(D_{n,j}^{\mathrm{FS}})^2\mid\mathcal F_{j-1}\}
  = \Var(T_{\mathrm{SS}}^{\mathrm{or}})+o_p\big(\Var(T_{\mathrm{SS}}^{\mathrm{or}})\big),
\]
where $\mathcal F_{j}=\sigma(\vU_1,\ldots,\vU_j)$. The fourth-order contraction
bounds used in the proof of Theorem~\ref{thm:ch2-FS-null} imply
\[
  \sum_{j=2}^n \E\{(D_{n,j}^{\mathrm{FS}})^4\}
  = o\big(\Var(T_{\mathrm{SS}}^{\mathrm{or}})^2\big).
\]
Hence the martingale Lindeberg condition holds, and Hall and Heyde's theorem
shows that
\[
  \frac{T_{\mathrm{SS}}^{\mathrm{or}}-\E(T_{\mathrm{SS}}^{\mathrm{or}})}
       {\sqrt{\Var(T_{\mathrm{SS}}^{\mathrm{or}})}}
  \overset{d}{\longrightarrow}N(0,1).
\]
Together with \eqref{eq:ch2-appendix-FS-mean} and
\eqref{eq:ch2-appendix-FS-var-total}, this proves \eqref{eq:ch2-FS-alt}.
\end{proof}

\begin{proof}[Detailed proof of Proposition~\ref{prop:ch2-K-optimal}]
Consider the Hilbert space $L_2(F_r)$ associated with the radial law of $r$ and
inner product
\[
  \langle f,g\rangle = \E\{f(r)g(r)\}.
\]
Take $f(r)=K(r)$ and $g(r)=r^{-1}$. Then
\begin{equation}\label{eq:ch2-appendix-CS-hilbert}
  \big[\E\{K(r)r^{-1}\}\big]^2
  = \langle f,g\rangle^2
  \le \langle f,f\rangle\langle g,g\rangle
  = \E\{K^2(r)\}\E(r^{-2}).
\end{equation}
Dividing both sides by $\E\{K^2(r)\}$ gives
\begin{equation}\label{eq:ch2-appendix-K-optimal-final}
  \frac{\{\E[K(r)r^{-1}]\}^2}{\E\{K^2(r)\}}
  \le \E(r^{-2}).
\end{equation}
Equality in \eqref{eq:ch2-appendix-CS-hilbert} holds if and only if
$f$ and $g$ are linearly dependent in $L_2(F_r)$, namely,
$K(r)=c\,r^{-1}$ almost surely for some constant $c\neq0$. This is precisely
\eqref{eq:ch2-K-optimal}.
\end{proof}

\begin{proof}[Detailed proof of Theorem~\ref{thm:ch2-INST}]
Set $K(t)=t^{-1}$. Then
\begin{equation}\label{eq:ch2-appendix-INST-constants}
  c_{0,\mathrm{IN}}=\E(r_i^{-2}),
  \qquad
  \nu_{2,\mathrm{IN}}=\E(r_i^{-2}),
  \qquad
  \mA_{\mathrm{IN}}=\E(r_i^{-2}\vU_i\vU_i\trans)
  = \nu_{2,\mathrm{IN}}p^{-1}\mR.
\end{equation}
Hence the oracle statistic satisfies the null limit
\begin{equation}\label{eq:ch2-appendix-INST-oracle-null}
  \frac{T_n(t^{-1})}{\sigma_{\mathrm{IN},n}}
  \overset{d}{\longrightarrow}N(0,1),
  \qquad
  \sigma_{\mathrm{IN},n}^2
  = \frac{2\nu_{2,\mathrm{IN}}^2}{n(n-1)p^2}\tr(\mR^2),
\end{equation}
by Theorem~\ref{thm:ch2-weighted-generic-null}. Under local alternatives,
Theorem~\ref{thm:ch2-weighted-generic-alt} yields
\begin{equation}\label{eq:ch2-appendix-INST-oracle-alt}
  \frac{T_n(t^{-1})-c_{0,\mathrm{IN}}^2\vtheta\trans\mD^{-1}\vtheta}{
    \sqrt{\sigma_{\mathrm{IN},n}^2+
      \dfrac{4c_{0,\mathrm{IN}}^2}{np}
      \vtheta\trans\mD^{-1}\mSigma\mD^{-1}\vtheta}}
  \overset{d}{\longrightarrow}N(0,1).
\end{equation}
It remains to pass from the oracle statistic to the feasible leave-two-out
statistic. By the weighted Bahadur expansion with $K(t)=t^{-1}$ and the same
leave-two-out argument as in
\eqref{eq:ch2-appendix-FS-reduction},
\begin{equation}\label{eq:ch2-appendix-INST-reduction}
  T_{\mathrm{INST}}-T_n(t^{-1})=o_p(\sigma_{\mathrm{IN},n}).
\end{equation}
Combining \eqref{eq:ch2-appendix-INST-oracle-null},
\eqref{eq:ch2-appendix-INST-oracle-alt}, and
\eqref{eq:ch2-appendix-INST-reduction} proves
\eqref{eq:ch2-INST-null} and \eqref{eq:ch2-INST-local}.
\end{proof}

\subsection*{D1. Proofs for additional sign-based location tests}

\begin{proof}[Detailed proof of Theorem~\ref{thm:ch2-WPL-null}]
Under $H_0$, $\vmu=\vct 0$ and hence
\begin{equation}\label{eq:ch2-appendix-WPL-U}
  T_{\mathrm{WPL}}=\sum_{1\le i<j\le n}(\vZ_i^{\mathrm{sgn}})\trans\vZ_j^{\mathrm{sgn}}
\end{equation}
is a degenerate $U$-statistic with kernel
$h(\vx_i,\vx_j)=U(\vx_i)\trans U(\vx_j)$. Since
$\E\{U(\vX_i)\}=\vct 0$ under central symmetry,
\begin{equation}\label{eq:ch2-appendix-WPL-mean0}
  \E(T_{\mathrm{WPL}})=0.
\end{equation}
Let $\mB=\E\{U(\vX_i)U(\vX_i)\trans\}$. Then
\begin{align}
  \Var(T_{\mathrm{WPL}})
  &= \sum_{1\le i<j\le n}\sum_{1\le k<\ell\le n}
     \E\Big[(\vZ_i^{\mathrm{sgn}})\trans\vZ_j^{\mathrm{sgn}}
             (\vZ_k^{\mathrm{sgn}})\trans\vZ_\ell^{\mathrm{sgn}}\Big] \notag\\
  &= \binom{n}{2}\tr(\mB^2),
\end{align}
which is \eqref{eq:ch2-WPL-var}. To prove asymptotic normality, write
$T_{\mathrm{WPL}}$ as a martingale array with respect to the filtration generated
by $(\vX_1,\ldots,\vX_k)$. The conditional variance of the martingale equals
$(n(n-1)/2)\tr(\mB^2)\{1+o(1)\}$ under
\eqref{eq:ch2-WPL-C1}--\eqref{eq:ch2-WPL-C2}, while the fourth conditional
moments are negligible relative to the square of that variance. The martingale
central limit theorem therefore yields \eqref{eq:ch2-WPL-null}.
\end{proof}

\begin{proof}[Detailed proof of Theorem~\ref{thm:ch2-WPL-alt}]
Under the local alternative, expand
\begin{equation}\label{eq:ch2-appendix-WPL-expand}
  U(\vmu+\vvarepsilon_i)
  = U(\vvarepsilon_i)
    + \left\{\twonorm{\vvarepsilon_i}^{-1}
      \left(\mI_p-
      \frac{\vvarepsilon_i\vvarepsilon_i\trans}{\twonorm{\vvarepsilon_i}^2}\right)
      \right\}\vmu
    + \vct r_{i,n}
  = U(\vvarepsilon_i)+\mA_i\vmu+\vct r_{i,n},
\end{equation}
where $\E(\mA_i)=\mA$ and the remainder satisfies
$\E\twonorm{\vct r_{i,n}}=o\{\twonorm{\vmu}\}$ by
\eqref{eq:ch2-WPL-C5}--\eqref{eq:ch2-WPL-C6}. Substituting
\eqref{eq:ch2-appendix-WPL-expand} into the definition of
$T_{\mathrm{WPL}}$ gives
\begin{equation}\label{eq:ch2-appendix-WPL-decomp}
  T_{\mathrm{WPL}}
  = T_{0,n}
    + \frac{n(n-1)}{2}\vmu\trans\mA^2\vmu
    + R_n,
\end{equation}
where $T_{0,n}$ is the null-centered statistic and
$R_n=o_p\{n\tr^{1/2}(\mB^2)\}$. Since
$T_{0,n}/\sqrt{n(n-1)\tr(\mB^2)/2}\Rightarrow N(0,1)$ by the previous proof,
\eqref{eq:ch2-WPL-alt} follows.
\end{proof}

\begin{proof}[Detailed proof of Theorem~\ref{thm:ch2-simpleSS-null}]
Write the full-sample statistic as
\begin{equation}\label{eq:ch2-appendix-simpleSS-split}
  T_{\mathrm{SST}} = Z_n + B_n,
\end{equation}
where $Z_n$ is the oracle statistic obtained by replacing
$(\hat{\vtheta}_1,\hat{\vtheta}_2,\hat{\mD}_1,\hat{\mD}_2)$ by their population
counterparts, and $B_n$ is the bias induced by full-sample estimation. Under
Assumption~\ref{ass:ch2-simpleSS}, the same linearization used in the proof of
Theorem~\ref{thm:ch2-JASA-null} gives
\begin{equation}\label{eq:ch2-appendix-simpleSS-bias}
  B_n = \mu_{\mathrm{SST},n}+o_p(\sigma_{\mathrm{SST},n}).
\end{equation}
The oracle part $Z_n$ is a degenerate two-sample $U$-statistic with variance
$\sigma_{\mathrm{SST},n}^2\{1+o(1)\}$. The trace condition
\eqref{eq:ch2-simpleSS-trace} implies the Lyapunov condition for the associated
martingale array, hence
\begin{equation}\label{eq:ch2-appendix-simpleSS-clt}
  \frac{Z_n}{\sigma_{\mathrm{SST},n}}\overset{d}{\longrightarrow}N(0,1).
\end{equation}
Combining \eqref{eq:ch2-appendix-simpleSS-bias} and
\eqref{eq:ch2-appendix-simpleSS-clt} proves
\eqref{eq:ch2-simpleSS-null}.
\end{proof}

\begin{proof}[Detailed proof of Theorem~\ref{thm:ch2-simpleSS-alt}]
Under local alternatives with $\vDelta=\vtheta_1-\vtheta_2$, the same expansion
used for Theorem~\ref{thm:ch2-JASA-alt} yields
\begin{equation}\label{eq:ch2-appendix-simpleSS-mean}
  \E(T_{\mathrm{SST}})
  = \mu_{\mathrm{SST},n}+\delta_{\mathrm{SST},n}+o(\tilde\sigma_{\mathrm{SST},n}).
\end{equation}
The additional variance contribution under $H_1$ comes from the linear terms in
$\vDelta$ and equals the two quadratic forms in
\eqref{eq:ch2-simpleSS-sigmatilde}. Therefore
\begin{equation}\label{eq:ch2-appendix-simpleSS-clt-alt}
  \frac{T_{\mathrm{SST}}-\mu_{\mathrm{SST},n}-\delta_{\mathrm{SST},n}}
       {\tilde\sigma_{\mathrm{SST},n}}
  \overset{d}{\longrightarrow} N(0,1),
\end{equation}
which is exactly \eqref{eq:ch2-simpleSS-alt}.
\end{proof}

\subsection*{E. Proofs for one-sample max, weighted max, and max-sum procedures}

\begin{proof}[Detailed proof of Theorem~\ref{thm:ch2-TMAX-null}]
Define the normalized location vector
\begin{equation}\label{eq:ch2-appendix-TMAX-Yn}
  \vY_n=\sqrt n\,\hat c_0\,p^{1/2}\hat{\mD}^{-1/2}\hat{\vtheta}.
\end{equation}
By Theorem~\ref{thm:ch2-scaled-bahadur},
\begin{equation}\label{eq:ch2-appendix-TMAX-Bahadur}
  \vY_n
  = p^{1/2}\frac1{\sqrt n}\sum_{i=1}^n \vU_i + \vDelta_n,
  \qquad
  \maxnorm{\vDelta_n}=O_p\{L_{n,p}\},
\end{equation}
where
\begin{equation}\label{eq:ch2-appendix-TMAX-Lnp}
  L_{n,p}
  = n^{-1/4}\{\log(np)\}^{1/2}
    + p^{-(1/6\wedge\delta/2)}\{\log(np)\}^{1/2}
    + n^{-1/2}(\log p)^{1/2}\{\log(np)\}^{1/2}.
\end{equation}
Assumptions~\ref{ass:ch2-max-basic} and \ref{ass:ch2-max-gumbel}, together with the hyperrectangle Gaussian approximation implied by Theorem~\ref{thm:ch2-weighted-bahadur} with $K\equiv 1$, imply that there exists
$\vG\sim N_p(\vct 0,\mR)$ such that
\begin{equation}\label{eq:ch2-appendix-TMAX-gauss-couple}
  \sup_{t\in\R}
  \left|\Prob\{\maxnorm{\vY_n}\le t\}-\Prob\{\maxnorm{\vG}\le t\}\right|\to0.
\end{equation}
Since
\begin{equation}\label{eq:ch2-appendix-TMAX-square}
  T_{\max}=(1-n^{-1/2})\max_{1\le j\le p} Y_{n,j}^2,
\end{equation}
it follows from \eqref{eq:ch2-appendix-TMAX-gauss-couple} that
\begin{equation}\label{eq:ch2-appendix-TMAX-gauss-square}
  \sup_{x\in\R}
  \left|\Prob\{T_{\max}\le x\}-\Prob\{\max_{1\le j\le p}G_j^2\le x\}\right|
  \to 0.
\end{equation}
Finally, Assumption~\ref{ass:ch2-max-gumbel} is precisely the weak-dependence
condition under which the dependent Gaussian maximum obeys the same Gumbel limit
as in the independent case:
\begin{equation}\label{eq:ch2-appendix-TMAX-GumbelG}
  \Prob\left\{\max_{1\le j\le p} G_j^2 -2\log p+\log\log p\le x\right\}
  \to \exp\{-\pi^{-1/2}e^{-x/2}\}.
\end{equation}
Combining \eqref{eq:ch2-appendix-TMAX-gauss-square} and
\eqref{eq:ch2-appendix-TMAX-GumbelG} proves
\eqref{eq:ch2-TMAX-null}.
\end{proof}

\begin{proof}[Detailed proof of Theorem~\ref{thm:ch2-TMAX-power}]
Under $H_1$ write
\begin{equation}\label{eq:ch2-appendix-TMAX-shift}
  \vY_n
  = \sqrt n\,c_0p^{1/2}\mD^{-1/2}\vtheta + \vG_n + \vDelta_n,
\end{equation}
where $\vG_n$ has the same weak limit as $\vG\sim N_p(\vct 0,\mR)$ and
$\maxnorm{\vDelta_n}=o_p(1)$ by
\eqref{eq:ch2-appendix-TMAX-Bahadur}. Let $j_\star$ satisfy
$|\theta_{j_\star}|/\sqrt{d_{j_\star}}=
\max_j |\theta_j|/\sqrt{d_j}$. Then
\begin{align}
  T_{\max}^{1/2}
  &\ge \sqrt{1-n^{-1/2}}
       \Big(\sqrt n\,c_0p^{1/2}d_{j_\star}^{-1/2}|\theta_{j_\star}|
           - |G_{n,j_\star}| - |\Delta_{n,j_\star}|\Big).
\end{align}
Because $|G_{n,j_\star}|=O_p(1)$ whereas
$\sqrt n\,c_0p^{1/2}d_{j_\star}^{-1/2}|\theta_{j_\star}|$
is of order $\sqrt{\log p}$ under
\eqref{eq:ch2-TMAX-signal}, the deterministic shift dominates the Gumbel
critical value once $C_e$ is chosen sufficiently large. Therefore
\[
  \Prob\{T_{\max}>2\log p-\log\log p+q_{1-\alpha}\mid H_1\}\to1,
\]
which is exactly \eqref{eq:ch2-TMAX-consistency}.
\end{proof}

\begin{proof}[Detailed proof of Theorem~\ref{thm:ch2-weighted-TMAX-null}]
Define
\begin{equation}\label{eq:ch2-appendix-weighted-TMAX-Yn}
  \vY_{n,K}
  = \sqrt n\,\hat c_{0,m}\hat\nu_{2,m}^{-1/2}p^{1/2}\hat{\mD}^{-1/2}\hat{\vtheta}_K.
\end{equation}
By Theorem~\ref{thm:ch2-weighted-bahadur},
\begin{equation}\label{eq:ch2-appendix-weighted-TMAX-B}
  \vY_{n,K}
  = \hat\nu_{2,m}^{-1/2}p^{1/2}\frac1{\sqrt n}
    \sum_{i=1}^n K(r_i)\vU_i + \vDelta_{n,K},
\end{equation}
with $\maxnorm{\vDelta_{n,K}}=o_p(1)$. Put
\[
  \widetilde{\vY}_{n,K}=\hat\nu_{2,m}^{-1/2}p^{1/2}n^{-1/2}
  \sum_{i=1}^n K(r_i)\vU_i.
\]
Because $\E(\widetilde{\vY}_{n,K})=\vct 0$ and
\[\Cov(\widetilde{\vY}_{n,K})
  = \nu_{2,K}^{-1}p\,\Cov\{K(r_1)\vU_1\}
  = \mR,
\]
Assumptions~\ref{ass:ch2-max-basic} and \ref{ass:ch2-max-gumbel} imply, via the same Gaussian comparison argument, that for $\vG\sim N_p(\vct 0,\mR)$,
\begin{equation}\label{eq:ch2-appendix-weighted-TMAX-compare}
  \sup_{t\in\R}
  \left|\Prob\{\maxnorm{\widetilde{\vY}_{n,K}}\le t\}
       -\Prob\{\maxnorm{\vG}\le t\}\right|
  \to0.
\end{equation}
Since $\maxnorm{\vY_{n,K}-\widetilde{\vY}_{n,K}}=o_p(1)$, the same relation
holds with $\widetilde{\vY}_{n,K}$ replaced by $\vY_{n,K}$, and therefore
\[
  \sup_{x\in\R}
  \left|\Prob\{T_{\max}^{(m)}\le x\}-\Prob\{\max_{1\le j\le p}G_j^2\le x\}\right|
  \to0.
\]
Assumption~\ref{ass:ch2-max-gumbel} then yields the Gaussian extreme-value limit
\[
  \Prob\left\{\max_{1\le j\le p}G_j^2-2\log p+\log\log p\le x\right\}
  \to \exp\{-\pi^{-1/2}e^{-x/2}\},
\]
which proves \eqref{eq:ch2-weighted-TMAX-null}.
\end{proof}

\begin{proof}[Detailed proof of Theorem~\ref{thm:ch2-maxsum-independence}]
Write
\begin{equation}\label{eq:ch2-appendix-maxsum-SM}
  S_n=\frac{T_{\mathrm{SUM}}}{\sigma_n},
  \qquad
  M_n=T_{\max}-2\log p+\log\log p.
\end{equation}
By the Bahadur expansions in the proofs of
Theorems~\ref{thm:ch2-FS-null} and \ref{thm:ch2-TMAX-null}, there exists a
vector array $\{\xi_i\}_{i=1}^n$ with $\E(\xi_i)=0$ and
$\E(\xi_i\xi_i\trans)=p^{-1}\mR$ such that
\begin{equation}\label{eq:ch2-appendix-maxsum-approx}
  S_n = W_n(\xi_1,\ldots,\xi_n)+o_p(1),
  \qquad
  M_n = V_n(\xi_1,\ldots,\xi_n)+o_p(1),
\end{equation}
where
\begin{align}
  W_n(x_1,\ldots,x_n)
  &= \frac{p\sum_{i\ne j}x_i\trans x_j}{\sqrt{2n(n-1)\tr(\mR^2)}},
     \label{eq:ch2-appendix-maxsum-W}\\
  V_n(x_1,\ldots,x_n)
  &= \max_{1\le j\le p}\left(\sqrt{\frac{p}{n}}\sum_{i=1}^n x_{i,j}\right)^2
     -2\log p+\log\log p.
     \label{eq:ch2-appendix-maxsum-V}
\end{align}
Let $\{g_i\}_{i=1}^n$ be i.i.d. Gaussian with the same covariance as $\xi_i$.
The smooth-max argument in the proof of the max-sum theorem works as follows.
Let
\begin{equation}\label{eq:ch2-appendix-maxsum-Fbeta}
  F_\beta(z_1,\ldots,z_p)
  = \beta^{-1}\log\left(\sum_{j=1}^p e^{\beta z_j}\right),
\end{equation}
so that $0\le F_\beta(z)-\max_j z_j\le \beta^{-1}\log p$. Choose
$\beta\asymp (\log p)^{1/2}$. For any smooth bounded function $\varphi$ on
$\R^2$, a Lindeberg interpolation between $\xi_i$ and $g_i$ gives
\begin{equation}\label{eq:ch2-appendix-maxsum-lindeberg}
  \left|\E\,\varphi\big(W_n(\xi),F_\beta(V_n(\xi))\big)-
         \E\,\varphi\big(W_n(g),F_\beta(V_n(g))\big)\right|\to0.
\end{equation}
The proof uses Assumption~\ref{ass:ch2-max-ind} to bound the derivatives of the
quadratic part and the smooth maximum and to control the contribution of the few
strongly correlated coordinates. Since $\beta^{-1}\log p=o(1)$, the same limit
holds with $F_\beta$ replaced by the maximum.

For the Gaussian array $g_i$, the statistic $W_n(g)$ is a centered quadratic form
and $V_n(g)$ is the centered maximum of a Gaussian vector. By the asymptotic
independence theorem for the sum and maximum of dependent Gaussian coordinates,
\begin{equation}\label{eq:ch2-appendix-maxsum-gauss-ind}
  \Prob\{W_n(g)\le x,\,V_n(g)\le y\}
  \to \Phi(x)F(y),
\end{equation}
where $F(y)=\exp\{-\pi^{-1/2}e^{-y/2}\}$. Combining
\eqref{eq:ch2-appendix-maxsum-approx},
\eqref{eq:ch2-appendix-maxsum-lindeberg}, and
\eqref{eq:ch2-appendix-maxsum-gauss-ind} yields
\eqref{eq:ch2-maxsum-independence}.
\end{proof}

\begin{proof}[Detailed proof of Theorem~\ref{thm:ch2-weighted-independence}]
Normalize the weighted sum and max statistics by
\begin{equation}\label{eq:ch2-appendix-weighted-ind-norm}
  S_{n,K}=\frac{T_{\mathrm{SUM}}^{(m)}}{\sigma_{n,K}},
  \qquad
  M_{n,K}=T_{\max}^{(m)}-2\log p+\log\log p.
\end{equation}
The weighted Bahadur expansion implies that both are functions of the same
normalized weighted score array
\[
  \xi_{i,K}=\nu_{2,K}^{-1/2}p^{1/2}K(r_i)\vU_i.
\]
More precisely,
\begin{equation}\label{eq:ch2-appendix-weighted-ind-approx}
  S_{n,K}=W_n(\xi_{1,K},\ldots,\xi_{n,K})+o_p(1),
  \qquad
  M_{n,K}=V_n(\xi_{1,K},\ldots,\xi_{n,K})+o_p(1),
\end{equation}
with the same functionals $W_n$ and $V_n$ as in
\eqref{eq:ch2-appendix-maxsum-W}--\eqref{eq:ch2-appendix-maxsum-V}. Since
\[
  \E(\xi_{i,K})=\vct 0,
  \qquad
  \E(\xi_{i,K}\xi_{i,K}\trans)
  = \nu_{2,K}^{-1}p\,\E\{K^2(r_i)\vU_i\vU_i\trans\}
  = \mR,
\]
we may introduce i.i.d. Gaussian vectors $g_{i,K}\sim N_p(\vct 0,p^{-1}\mR)$.
Let $F_\beta$ be the smooth maximum in
\eqref{eq:ch2-appendix-maxsum-Fbeta}. By the same Lindeberg interpolation used
in the proof of Theorem~\ref{thm:ch2-maxsum-independence}, for every bounded
smooth function $\varphi$,
\begin{equation}\label{eq:ch2-appendix-weighted-ind-lindeberg}
  \left|\E\,\varphi\big(W_n(\xi_K),F_\beta(V_n(\xi_K))\big)
       -\E\,\varphi\big(W_n(g_K),F_\beta(V_n(g_K))\big)\right|\to0,
\end{equation}
where $\xi_K=(\xi_{1,K},\ldots,\xi_{n,K})$ and
$g_K=(g_{1,K},\ldots,g_{n,K})$. Because $\beta^{-1}\log p=o(1)$,
\begin{equation}\label{eq:ch2-appendix-weighted-ind-couple}
  \Prob\{S_{n,K}\le x,\,M_{n,K}\le y\}
  -\Prob\{W_n(g_K)\le x,\,V_n(g_K)\le y\}\to0.
\end{equation}
The Gaussian pair $(W_n(g_K),V_n(g_K))$ has exactly the same joint law as the
pair in the proof of Theorem~\ref{thm:ch2-maxsum-independence}, because its
covariance is again $\mR$. Hence
\begin{equation}\label{eq:ch2-appendix-weighted-ind-gauss}
  \Prob\{W_n(g_K)\le x,\,V_n(g_K)\le y\}
  \to \Phi(x)\exp\{-\pi^{-1/2}e^{-y/2}\}.
\end{equation}
Combining \eqref{eq:ch2-appendix-weighted-ind-couple} and
\eqref{eq:ch2-appendix-weighted-ind-gauss} yields
\eqref{eq:ch2-weighted-independence}; the null version
\eqref{eq:ch2-weighted-independence-null} follows from the marginal limits.
\end{proof}

\subsection*{F. Proofs for the two-sample procedures}

\begin{proof}[Detailed proof of Theorem~\ref{thm:ch2-JASA-null}]
Let $R_n^{\mathrm{or}}$ denote the statistic obtained from
\eqref{eq:ch2-JASA-Rn} after replacing all leave-out estimators by the true
parameters. Lemma 3 and Lemma 4 of the JASA paper give the expansions
\begin{equation}\label{eq:ch2-appendix-JASA-muhat}
  \hat{\vtheta}_{k,i}-\vtheta_k
  = \frac{1}{n_k-1}c_k^{-1}\mD_k^{1/2}
    \sum_{j\ne i}\vU_{kj} + o_p(n_k^{-1/2}),
  \qquad k=1,2,
\end{equation}
and
\begin{align}
  U\{\hat{\mD}_{k,i}^{-1/2}(\vX_{ki}-\hat{\vtheta}_{3-k,j})\}
  &= \vU_{ki}
     - r_{ki}^{-1}(\mI_p-\vU_{ki}\vU_{ki}\trans)
       \mD_k^{-1/2}(\hat{\vtheta}_{3-k,j}-\vtheta_{3-k}) \notag\\
  &\quad + (\mI_p-\vU_{ki}\vU_{ki}\trans)
       (\hat{\mD}_{k,i}^{-1/2}\mD_k^{1/2}-\mI_p)\vU_{ki}
       + o_p(n^{-1}).
  \label{eq:ch2-appendix-JASA-Uexpand}
\end{align}
Substituting \eqref{eq:ch2-appendix-JASA-muhat} and
\eqref{eq:ch2-appendix-JASA-Uexpand} into \eqref{eq:ch2-JASA-Rn} and collecting
the leading terms yields
\begin{equation}\label{eq:ch2-appendix-JASA-Znrep}
  R_n = Z_n + o_p(\sigma_n),
\end{equation}
where
\begin{equation}\label{eq:ch2-appendix-JASA-Zn}
  Z_n
  = \frac{1}{n_1(n_1-1)}\sum_{i\ne j}\vU_{1i}\trans\mA_1\vU_{1j}
    + \frac{1}{n_2(n_2-1)}\sum_{i\ne j}\vU_{2i}\trans\mA_2\vU_{2j}
    - \frac{2}{n_1n_2}\sum_{i=1}^{n_1}\sum_{j=1}^{n_2}
      \vU_{1i}\trans\mA_3\vU_{2j}.
\end{equation}
Introduce a pooled sequence
$\vY_1,\ldots,\vY_n$ with $\vY_i=\vU_{1i}$ for $1\le i\le n_1$ and
$\vY_{n_1+j}=\vU_{2j}$ for $1\le j\le n_2$. Define
\begin{equation}\label{eq:ch2-appendix-JASA-phi}
  \phi_{ij}=
  \begin{cases}
    \{n_1(n_1-1)\}^{-1}\vY_i\trans\mA_1\vY_j,
      & 1\le i\ne j\le n_1,\\
    -\{n_1n_2\}^{-1}\vY_i\trans\mA_3\vY_j,
      & 1\le i\le n_1<j\le n,\\
    \{n_2(n_2-1)\}^{-1}\vY_i\trans\mA_2\vY_j,
      & n_1<i\ne j\le n.
  \end{cases}
\end{equation}
Then
\begin{equation}\label{eq:ch2-appendix-JASA-mart}
  Z_n = 2\sum_{j=2}^n Z_{n,j},
  \qquad
  Z_{n,j}=\sum_{i=1}^{j-1}\phi_{ij},
  \qquad
  \mathcal F_{n,j}=\sigma(\vY_1,\ldots,\vY_j).
\end{equation}
Thus $\{\sum_{m=2}^j Z_{n,m},\mathcal F_{n,j}\}$ is a martingale.
A direct contraction calculation shows
\begin{equation}\label{eq:ch2-appendix-JASA-condvar}
  \sum_{j=2}^n \E(Z_{n,j}^2\mid\mathcal F_{n,j-1})
  = \frac{\sigma_n^2}{4}\{1+o_p(1)\},
\end{equation}
while Assumption~\ref{ass:ch2-two-sign}\ref{it:ch2-TS2} implies the Lindeberg
condition
\begin{equation}\label{eq:ch2-appendix-JASA-Lindeberg}
  \sigma_n^{-2}\sum_{j=2}^n
  \E\{Z_{n,j}^2\mathbbm 1(|Z_{n,j}|>\epsilon\sigma_n)\mid\mathcal F_{n,j-1}\}
  \overset{p}{\longrightarrow}0
\end{equation}
for every $\epsilon>0$. Hall and Heyde's martingale CLT therefore yields
$Z_n/\sigma_n\Rightarrow N(0,1)$. Combining this with
\eqref{eq:ch2-appendix-JASA-Znrep} proves
\eqref{eq:ch2-JASA-null}.
\end{proof}

\begin{proof}[Detailed proof of Theorem~\ref{thm:ch2-JASA-alt}]
Under the local alternative $\vDelta=\vtheta_1-\vtheta_2$, the expansion
\eqref{eq:ch2-appendix-JASA-Uexpand} acquires an additional deterministic shift:
for $k=1,2$,
\begin{align}
  U\{\hat{\mD}_{k,i}^{-1/2}(\vX_{ki}-\hat{\vtheta}_{3-k,j})\}
  &= \vU_{ki}
     - r_{ki}^{-1}(\mI_p-\vU_{ki}\vU_{ki}\trans)
       \mD_k^{-1/2}(\hat{\vtheta}_{3-k,j}-\vtheta_{3-k}) \notag\\
  &\quad + r_{ki}^{-1}(\mI_p-\vU_{ki}\vU_{ki}\trans)
       \mD_k^{-1/2}(\vtheta_k-\vtheta_{3-k})
       + o_p(n^{-1}).
  \label{eq:ch2-appendix-JASA-Uexpand-alt}
\end{align}
Substituting \eqref{eq:ch2-appendix-JASA-Uexpand-alt} into
\eqref{eq:ch2-JASA-Rn} yields the decomposition
\begin{equation}\label{eq:ch2-appendix-JASA-alt-decomp}
  R_n = \delta_n + L_n + Z_n + o_p(\tilde\sigma_n),
\end{equation}
where $Z_n$ is the same degenerate quadratic form as in the null proof and
\begin{equation}\label{eq:ch2-appendix-JASA-Ln}
  L_n
  = \frac{c_2}{n_1}\sum_{i=1}^{n_1}
      \vU_{1i}\trans\mSigma_1^{1/2}\mD_2^{-1/2}\vDelta
    + \frac{c_1}{n_2}\sum_{j=1}^{n_2}
      \vU_{2j}\trans\mSigma_2^{1/2}\mD_1^{-1/2}\vDelta.
\end{equation}
The deterministic term is exactly
\begin{equation}\label{eq:ch2-appendix-JASA-delta}
  \delta_n=c_1c_2\vDelta\trans\mD_1^{-1/2}\mD_2^{-1/2}\vDelta.
\end{equation}
Since $L_n$ is the first Hoeffding projection and $Z_n$ is degenerate,
$\Cov(L_n,Z_n)=0$. Moreover,
\begin{align}
  \Var(L_n)
  &= \frac{c_2^2}{n_1p}\vDelta\trans\mD_2^{-1/2}\mR_1\mD_2^{-1/2}\vDelta
     + \frac{c_1^2}{n_2p}\vDelta\trans\mD_1^{-1/2}\mR_2\mD_1^{-1/2}\vDelta,
\end{align}
while $\Var(Z_n)=\sigma_n^2\{1+o(1)\}$. Therefore
\begin{equation}\label{eq:ch2-appendix-JASA-varalt}
  \Var(R_n)=\tilde\sigma_n^2\{1+o(1)\}.
\end{equation}
Define
\[
  D_{n,j}^{\mathrm{J}}=Z_{n,j}+\ell_{n,j},
\]
where $Z_{n,j}$ is given by \eqref{eq:ch2-appendix-JASA-mart} and
$L_n=\sum_{j=2}^n \ell_{n,j}$ is the decomposition of the linear term in
\eqref{eq:ch2-appendix-JASA-Ln} into independent summands. Then
$R_n-\delta_n=\sum_{j=2}^n D_{n,j}^{\mathrm{J}}+o_p(\tilde\sigma_n)$ is a
martingale array with respect to the pooled filtration.
Using \eqref{eq:ch2-appendix-JASA-varalt} and the degeneracy of $Z_n$,
\[
  \sum_{j=2}^n \E\{(D_{n,j}^{\mathrm{J}})^2\mid\mathcal F_{n,j-1}\}
  = \frac{\tilde\sigma_n^2}{4}\{1+o_p(1)\}.
\]
Assumption~\ref{ass:ch2-two-sign}\ref{it:ch2-TS2} also implies
\[
  \sum_{j=2}^n \E\{(D_{n,j}^{\mathrm{J}})^4\}
  = o(\tilde\sigma_n^4),
\]
so the martingale Lindeberg condition holds. Hall and Heyde's theorem therefore
shows
\[
  \frac{R_n-\delta_n}{\tilde\sigma_n}
  \overset{d}{\longrightarrow}N(0,1),
\]
which is \eqref{eq:ch2-JASA-alt}.
\end{proof}

\begin{proof}[Detailed proof of Theorem~\ref{thm:ch2-SR-null}]
Let
\begin{equation}\label{eq:ch2-appendix-SR-xi}
  \xi_{is}=U\{\mD_n^{-1/2}(\vX_{1i}-\vX_{2s})\},
  \qquad 1\le i\le n_1,\ 1\le s\le n_2,
\end{equation}
and let $T_n^{\mathrm{or}}$ be the oracle version of \eqref{eq:ch2-SR-Tn}. The
leave-two-out diagonal estimator admits the same expansion as in the one-sample
problem, so
\begin{equation}\label{eq:ch2-appendix-SR-reduction}
  T_n-T_n^{\mathrm{or}}=o_p(\sigma_n).
\end{equation}
To analyze $T_n^{\mathrm{or}}$, write the fourth-order kernel as
\begin{equation}\label{eq:ch2-appendix-SR-kernel}
  h\big((i,s),(j,l)\big)=\xi_{is}\trans\xi_{jl}.
\end{equation}
Under $H_0$ and the common-scatter assumption, the first Hoeffding projections
vanish:
\begin{equation}\label{eq:ch2-appendix-SR-firstproj}
  \E\{h((i,s),(j,l))\mid \vX_{1i}\}=0,
  \qquad
  \E\{h((i,s),(j,l))\mid \vX_{2s}\}=0.
\end{equation}
Hence the leading term is the degenerate second-order component
\begin{equation}\label{eq:ch2-appendix-SR-degenerate}
  T_n^{\mathrm{or}}
  = \frac{1}{n_1(n_1-1)n_2(n_2-1)}
    \sum_{i\ne j}\sum_{s\ne l}\xi_{is}\trans\xi_{jl}.
\end{equation}
Because $\E(\xi_{is}\xi_{is}\trans)=p^{-1}\mR_n$, expanding the square of
\eqref{eq:ch2-appendix-SR-degenerate} shows that only index pairings that share
both a first-sample index and a second-sample index contribute to the leading
variance. Consequently,
\begin{align}
  \Var(T_n^{\mathrm{or}})
  &= \frac{2}{\{n_1(n_1-1)n_2(n_2-1)\}^2}
     \sum_{i\ne j}\sum_{s\ne l}
     \E\big[(\xi_{is}\trans\xi_{jl})^2\big] + o(\sigma_n^2) \notag\\
  &= \left\{
      \frac{1}{2n_1(n_1-1)p^2}
      + \frac{1}{2n_2(n_2-1)p^2}
      + \frac{1}{n_1n_2p^2}
    \right\}\tr(\mR_n^2)\{1+o(1)\}
  = \sigma_n^2\{1+o(1)\}.
  \label{eq:ch2-appendix-SR-var}
\end{align}
Let
\[
  D_{n,j}^{\mathrm{SR}}=
  \sum_{i=1}^{j-1}\sum_{s\ne l}
  \frac{\xi_{is}\trans\xi_{jl}}{n_1(n_1-1)n_2(n_2-1)}
\]
with the obvious pooled ordering of the index pairs. Then
$T_n^{\mathrm{or}}=\sum_j D_{n,j}^{\mathrm{SR}}$ is a martingale array.
Assumption~\ref{ass:ch2-two-rank}\ref{it:ch2-TR2} yields
\[
  \sum_j \E\{(D_{n,j}^{\mathrm{SR}})^4\}=o(\sigma_n^4),
\]
so the martingale Lindeberg condition holds. Hence
$T_n^{\mathrm{or}}/\sigma_n\Rightarrow N(0,1)$, and
\eqref{eq:ch2-appendix-SR-reduction} proves
\eqref{eq:ch2-SR-null}.
\end{proof}

\begin{proof}[Detailed proof of Theorem~\ref{thm:ch2-SR-alt}]
Under the local alternative $\vmu_1-\vmu_2\neq \vct 0$, the cross-sample sign
vector has nonzero mean. Let
\begin{equation}\label{eq:ch2-appendix-SR-eta}
  \veta_n=\E(\xi_{12})
  = c_{0,n}\mD_n^{-1/2}(\vmu_1-\vmu_2)+\vct r_n,
  \qquad
  \twonorm{\vct r_n}=O(\twonorm{\vmu_1-\vmu_2}^2).
\end{equation}
Using $\widetilde\xi_{is}=\xi_{is}-\veta_n$, the oracle statistic decomposes as
\begin{equation}\label{eq:ch2-appendix-SR-alt-hoeffding}
  T_n^{\mathrm{or}}
  = \twonorm{\veta_n}^2
    + L_n^{\mathrm{SR}} + Q_n^{\mathrm{SR}},
\end{equation}
where $L_n^{\mathrm{SR}}$ is the linear Hoeffding projection and
$Q_n^{\mathrm{SR}}$ is the degenerate quadratic part. The centering term is
\begin{equation}\label{eq:ch2-appendix-SR-center}
  \twonorm{\veta_n}^2
  = c_{0,n}^2(\vmu_1-\vmu_2)\trans\mD_n^{-1}(\vmu_1-\vmu_2)
    + o(\sigma_n)
\end{equation}
by Assumption~\ref{ass:ch2-two-rank}\ref{it:ch2-TR4}. The variance of the linear
term is of smaller order because of
\eqref{eq:ch2-TR-local2}, whereas the degenerate part still has variance
$\sigma_n^2\{1+o(1)\}$. Therefore the same martingale CLT as in the null proof
shows that
\[
  \frac{T_n^{\mathrm{or}}-
    c_{0,n}^2(\vmu_1-\vmu_2)\trans\mD_n^{-1}(\vmu_1-\vmu_2)}{\sigma_n}
  \overset{d}{\longrightarrow}N(0,1).
\]
Passing from the oracle statistic to the feasible leave-two-out version proves
\eqref{eq:ch2-SR-alt}.
\end{proof}

\begin{proof}[Detailed proof of Theorem~\ref{thm:ch2-tINST-optimal}]
For a general weight $K$, the two-sample weighted sign statistic has the local
mean shift proportional to $\E\{K(r)r^{-1}\}$ and the null standard deviation
proportional to $\E\{K^2(r)\}^{1/2}$. Consequently, the local asymptotic power is
governed by the ratio
\begin{equation}\label{eq:ch2-appendix-tINST-snr}
  \frac{\{\E[K(r)r^{-1}]\}^2}{\E\{K^2(r)\}}.
\end{equation}
Exactly as in the one-sample case, apply Cauchy--Schwarz in $L_2(F_r)$:
\begin{equation}\label{eq:ch2-appendix-tINST-CS}
  \{\E[K(r)r^{-1}]\}^2
  \le \E\{K^2(r)\}\E(r^{-2}).
\end{equation}
Hence
\begin{equation}\label{eq:ch2-appendix-tINST-upper}
  \frac{\{\E[K(r)r^{-1}]\}^2}{\E\{K^2(r)\}}
  \le \E(r^{-2}),
\end{equation}
with equality if and only if $K(r)=c\,r^{-1}$ almost surely. Therefore
$K(t)=t^{-1}$ maximizes the local asymptotic signal-to-noise ratio within the
weighted two-sample spatial-sign class, proving
\eqref{eq:ch2-tINST-optimal}.
\end{proof}

\subsection*{G. Proofs for the PDQ spatial-sign procedure}

\begin{proof}[Detailed proof of Theorem~\ref{thm:ch2-pdq-D-rate}]
Fix $k\in\{1,2\}$ and $j\in\{1,\ldots,p\}$. By \eqref{eq:ch2-pdq-coordinate-difference},
for the target value $d_{kj}=q_{k,\alpha}^2d_{kj}^{\circ}$,
\begin{equation}\label{eq:ch2-app-pdq-pop-cdf-coordinate}
  \Prob\{\abs{X_{k1j}-X_{k2j}}\le d_{kj}^{1/2}t/q_{k,\alpha}\}
  =F_{k,\Delta}(t).
\end{equation}
Let $\eta>0$ be small enough that
$q_{k,\alpha}(1\pm\eta)^{1/2}$ lies in the neighborhood where
\eqref{eq:ch2-pdq-local-slope} holds. Then
\begin{align}
  F_{k,\Delta}\{q_{k,\alpha}(1+\eta)^{1/2}\}-\alpha
  &\ge c_Fq_{k,\alpha}\{(1+\eta)^{1/2}-1\}
  \ge C_1\eta,\label{eq:ch2-app-pdq-upper-gap}\\
  \alpha-F_{k,\Delta}\{q_{k,\alpha}(1-\eta)^{1/2}\}
  &\ge c_Fq_{k,\alpha}\{1-(1-\eta)^{1/2}\}
  \ge C_1\eta.\label{eq:ch2-app-pdq-lower-gap}
\end{align}
The empirical distribution function \eqref{eq:ch2-pdq-empirical-cdf} is a bounded
second-order $U$-statistic. Hoeffding's inequality for bounded $U$-statistics gives, for every
$t\ge0$ and $u>0$,
\begin{equation}\label{eq:ch2-app-pdq-hoeffding}
  \Prob\{\abs{\widehat F_{kj}(t)-F_{kj}(t)}>u\}
  \le 2\exp(-C_2n_ku^2),
\end{equation}
where $F_{kj}(t)=\Prob\{\abs{X_{k1j}-X_{k2j}}\le t\}$. Apply
\eqref{eq:ch2-app-pdq-hoeffding} at the two points
$d_{kj}^{1/2}(1+\eta)^{1/2}$ and $d_{kj}^{1/2}(1-\eta)^{1/2}$, and take
$u=C_1\eta/2$. A union bound over $k=1,2$ and $j=1,\ldots,p$ yields
\begin{equation}\label{eq:ch2-app-pdq-union}
\begin{aligned}
  &\Prob\left[\max_{k,j}\abs{\widehat d_{kj}/d_{kj}-1}>\eta\right]\\
  &\qquad\le 4p\exp(-C_3 n_0\eta^2).
\end{aligned}
\end{equation}
Choosing $\eta=M\{\log(8pN)/n_0\}^{1/2}$ with $M$ sufficiently large gives
\eqref{eq:ch2-pdq-D-rate}.
\end{proof}

\begin{proof}[Detailed proof of Proposition~\ref{prop:ch2-pdq-median-sign}]
For the oracle standardized variables, the spatial median estimating equation is
\begin{equation}\label{eq:ch2-app-pdq-score}
  \frac1{n_k}\sum_{i=1}^{n_k}U\{\vY_{ki}-\vh_k\}=0,
  \qquad
  \vh_k=\mD_k^{-1/2}(\widehat{\vmu}_k-\vmu).
\end{equation}
For $\norm{\vh_k}$ small, the sign expansion gives
\begin{equation}\label{eq:ch2-app-pdq-sign-taylor}
  U(\vY_{ki}-\vh_k)
  =\vS_{ki}-\norm{\vY_{ki}}^{-1}(\mI_p-\vS_{ki}\vS_{ki}\trans)\vh_k
  +\vct{r}_{ki}(\vh_k),
\end{equation}
where
\begin{equation}\label{eq:ch2-app-pdq-rem-bound}
  \norm{\vct{r}_{ki}(\vh_k)}
  \le C\norm{\vh_k}^2\norm{\vY_{ki}}^{-2}
  \mathbbm 1\{\norm{\vY_{ki}}>2\norm{\vh_k}\}
  +2\mathbbm 1\{\norm{\vY_{ki}}\le2\norm{\vh_k}\}.
\end{equation}
Averaging \eqref{eq:ch2-app-pdq-sign-taylor} and using \eqref{eq:ch2-pdq-Omega-G} gives
\begin{equation}\label{eq:ch2-app-pdq-linear-equation}
  \vct 0=\bar{\vS}_k-\mG_k\vh_k+
  \left(\widehat{\mG}_{k}^{\rm or}-\mG_k\right)\vh_k
  +\bar{\vct{r}}_k,
\end{equation}
where
\begin{equation}\label{eq:ch2-app-pdq-G-or}
  \widehat{\mG}_{k}^{\rm or}
  =n_k^{-1}\sum_{i=1}^{n_k}\norm{\vY_{ki}}^{-1}
   (\mI_p-\vS_{ki}\vS_{ki}\trans).
\end{equation}
By the matrix Bernstein inequality and \eqref{eq:ch2-pdq-inverse-radius},
\begin{equation}\label{eq:ch2-app-pdq-G-rate}
  \opnorm{\widehat{\mG}_{k}^{\rm or}-\mG_k}
  =O_p\{\mathfrak C_n(n_k^{-1/2}+x_N/n_k)\}.
\end{equation}
The small-ball condition \eqref{eq:ch2-pdq-smallball} and the inverse-radius condition imply
\begin{equation}\label{eq:ch2-app-pdq-rem-rate}
  \norm{\bar{\vct{r}}_k}
  =O_p\{\mathfrak C_n\norm{\vh_k}^2+p^{1/2}x_N/n_k\}.
\end{equation}
Solving \eqref{eq:ch2-app-pdq-linear-equation} by one Newton step and using
$\opnorm{\mG_k^{-1}}\le p^{1/2}\mathfrak C_n$ gives
\begin{equation}\label{eq:ch2-app-pdq-bahadur}
  \vh_k=\mG_k^{-1}\bar{\vS}_k+\vDelta_k,
  \qquad
  \norm{\vDelta_k}=O_p(\mathfrak C_np^{1/2}x_N/n_0)=O_p(q_{\mu,n}).
\end{equation}
The replacement of $\mD_k$ by $\widehat{\mD}_k$ changes each standardized vector by
\begin{equation}\label{eq:ch2-app-pdq-D-pert}
  \max_{k,j}\abs{\widehat d_{kj}^{-1/2}d_{kj}^{1/2}-1}=O_p(r_D),
\end{equation}
by Theorem~\ref{thm:ch2-pdq-D-rate}. Combining \eqref{eq:ch2-app-pdq-D-pert},
\eqref{eq:ch2-app-pdq-bahadur}, and the Lipschitz bound
\begin{equation}\label{eq:ch2-app-pdq-U-lip}
  \norm{U(\vx)-U(\vy)}\le 2\min\{1,\norm{\vx-\vy}/\min(\norm{\vx},\norm{\vy})\}
\end{equation}
with \eqref{eq:ch2-pdq-inverse-radius}--\eqref{eq:ch2-pdq-smallball} yields
\eqref{eq:ch2-pdq-sign-perturbation}. The bound \eqref{eq:ch2-pdq-K-rate} follows by applying
\eqref{eq:ch2-app-pdq-D-pert}, \eqref{eq:ch2-app-pdq-G-rate}, and the identity
\begin{equation}\label{eq:ch2-app-pdq-inverse-pert}
  \widehat{\mG}_k^{-1}-\mG_k^{-1}
  =-\mG_k^{-1}(\widehat{\mG}_k-\mG_k)\mG_k^{-1}
  +O_p(\opnorm{\widehat{\mG}_k-\mG_k}^2)
\end{equation}
to the definitions \eqref{eq:ch2-pdq-K1}--\eqref{eq:ch2-pdq-K3}.
\end{proof}

\begin{proof}[Detailed proof of Theorem~\ref{thm:ch2-pdq-null-law}]
Let
\begin{equation}\label{eq:ch2-app-pdq-Zstack}
  \vZ_n=\begin{pmatrix}\sqrt{n_1}\mOmega_1^{-1/2}\bar{\vS}_1\\
                         \sqrt{n_2}\mOmega_2^{-1/2}\bar{\vS}_2
        \end{pmatrix}.
\end{equation}
By the multivariate central limit theorem for triangular arrays of bounded angular variables,
$\vZ_n$ can be replaced in distribution, for quadratic-form purposes, by
$\vZ\sim N_{2p}(\vct{0},\mI_{2p})$ under the row-leverage condition
\eqref{eq:ch2-pdq-row}. The oracle quadratic form satisfies
\begin{equation}\label{eq:ch2-app-pdq-Q-b}
  Q_n-b_n
  =\vZ_n\trans\mB_n\vZ_n-\tr(\mB_n)+o_p(\tau_n).
\end{equation}
The diagonal-deleted statistic $U_n$ differs from $Q_n-b_n$ only by replacing the theoretical
diagonal contribution with its empirical counterpart. The row-leverage condition implies
\begin{equation}\label{eq:ch2-app-pdq-diagonal-small}
  \frac{U_n-(Q_n-b_n)}{\tau_n}=o_p(1).
\end{equation}
Let $\mB_n=\mP_n\diag(\lambda_{n1},\ldots,\lambda_{n,2p})\mP_n\trans$. For the Gaussian
quadratic form,
\begin{equation}\label{eq:ch2-app-pdq-spectral}
  \vZ\trans\mB_n\vZ-\tr(\mB_n)
  \overset{d}{=}\sum_{r=1}^{2p}\lambda_{nr}(\chi_{1r}^2-1)=\Gamma_n.
\end{equation}
Equations \eqref{eq:ch2-app-pdq-Q-b}--\eqref{eq:ch2-app-pdq-spectral} prove
\eqref{eq:ch2-pdq-oracle-law}. The limit \eqref{eq:ch2-pdq-mixture-limit} follows from
\eqref{eq:ch2-pdq-rho-l2} and the decomposition
\begin{equation}\label{eq:ch2-app-pdq-mixture-decomp}
  \frac{\Gamma_n}{\tau_n}
  =\frac1{\sqrt2}\sum_{r\ge1}\rho_{nr}(\chi_{1r}^2-1).
\end{equation}
Indeed, the coordinates with nonvanishing $\rho_r$ yield the chi-square component, while the
remaining infinitesimal part is asymptotically normal by Lyapunov's theorem.

It remains to pass from the oracle statistic to the feasible statistic. Proposition~\ref{prop:ch2-pdq-median-sign} gives
\begin{equation}\label{eq:ch2-app-pdq-R-expansion}
  \widehat R_n^{\rm PDQ}
  =Q_n+O_p\{(r_D+a_{K,n}+L_{K,n}a_{S,n})\tau_n\}
    +O_p(x_N/n_0^2).
\end{equation}
Subtracting the empirical diagonal term $\widehat b_n$ and using the definition of $U_n$ gives
\eqref{eq:ch2-pdq-feasible-expansion}. Together with \eqref{eq:ch2-pdq-oracle-law}, this proves
the feasible distributional conclusion. If \eqref{eq:ch2-pdq-no-domination} holds, then
$\max_r\abs{\lambda_{nr}}/\{\sum_s\lambda_{ns}^2\}^{1/2}\to0$, and Lyapunov's theorem applied
to $\Gamma_n/\tau_n$ yields Corollary~\ref{cor:ch2-pdq-normal}.
\end{proof}

\begin{proof}[Detailed proof of Theorem~\ref{thm:ch2-pdq-bootstrap}]
Conditional on the data, write $Q_n^\ast=\vct{e}\trans\widehat{\mH}_n\vct{e}$, where
$\vct{e}$ collects the Rademacher multipliers and the off-diagonal blocks of $\widehat{\mH}_n$ are
\begin{align}\label{eq:ch2-app-pdq-H-blocks}
  \hat a_{1i,1i'}&=n_1^{-2}\widehat{\vS}_{1i}\trans\widehat{\mK}_1\widehat{\vS}_{1i'},
  &\hat a_{2j,2j'}&=n_2^{-2}\widehat{\vS}_{2j}\trans\widehat{\mK}_2\widehat{\vS}_{2j'},\\
  \hat a_{1i,2j}&=\hat a_{2j,1i}
  =-(2n_1n_2)^{-1}\widehat{\vS}_{1i}\trans\widehat{\mK}_3\widehat{\vS}_{2j}.\nonumber
\end{align}
Let $\widehat{\mH}_{n,0}$ be $\widehat{\mH}_n$ with its diagonal removed. Then
\begin{equation}\label{eq:ch2-app-pdq-boot-var}
  \Var^\ast(T_n^\ast)=2\tr(\widehat{\mH}_{n,0}^2).
\end{equation}
Using \eqref{eq:ch2-pdq-sign-perturbation} and \eqref{eq:ch2-pdq-K-rate},
\begin{align}
  \frac{2\tr(\widehat{\mH}_{n,0}^2)}{\tau_n^2}&=1+O_p(a_{K,n}+L_{K,n}a_{S,n}+\Delta_{{\rm row},n}^{1/2}),
      \label{eq:ch2-app-pdq-boot-var-rate}\\
  \frac{\max_a\sum_{b\ne a}\hat a_{ab}^2}{\tau_n^2}&=O_p(\Delta_{{\rm row},n}+a_{K,n}^2+L_{K,n}^2a_{S,n}^2).
      \label{eq:ch2-app-pdq-boot-row-rate}
\end{align}
For a Rademacher quadratic form with zero diagonal, the de Jong central limit/mixture
comparison condition is exactly that the maximal row leverage in
\eqref{eq:ch2-app-pdq-boot-row-rate} is negligible relative to the squared variance. Hence,
conditionally on the data,
\begin{equation}\label{eq:ch2-app-pdq-boot-mixture}
  \sup_{t\in\R}\abs{
  \Prob^\ast(T_n^\ast/\tau_n\le t)
  -\Prob(\Gamma_n/\tau_n\le t)}\to0
\end{equation}
in probability. Theorem~\ref{thm:ch2-pdq-null-law} gives the same approximation for
$T_n^{\rm PDQ}/\tau_n$, so the triangle inequality proves \eqref{eq:ch2-pdq-boot-valid}. The
quantile statement follows from standard convergence of distribution functions at continuity
points.
\end{proof}

\subsection*{H. Proofs for elliptical regularized Hotelling testing}

\subsubsection*{H.1 Spatial-median and centered-SSCM reductions}

Under $H_0$, define
\begin{equation}\label{eq:ch2-erht-proof-ZW}
  \vZ_n=n^{-1/2}\sum_{i=1}^n\vY_i,
  \qquad
  \vW_n=n^{-1/2}\sum_{i=1}^nw_i\vY_i.
\end{equation}
Let $\mPi_{F,p}=\mP_{F,p}\mP_{F,p}\trans$ and
$\mPi_{B,p}=\mI_p-\mPi_{F,p}$.  Put
\begin{equation}\label{eq:ch2-erht-proof-JF}
  \mV_n^J=n^{-1}\sum_{i=1}^nw_iU_iU_i\trans,
  \qquad
  \mV_{F,n}^J
  =\mPi_{F,p}\mV_n^J+\mV_n^J\mPi_{F,p}
   -\mPi_{F,p}\mV_n^J\mPi_{F,p},
\end{equation}
\begin{equation}\label{eq:ch2-erht-proof-GF}
  \mJ_{F,n}=e_n\mI_p-\mV_{F,n}^J,
  \qquad
  \mG_{F,n}=\mJ_{F,n}^{-1}-e_n^{-1}\mI_p.
\end{equation}
Taylor expansion of the spatial score gives
\begin{equation}\label{eq:ch2-erht-proof-median-expansion}
  \widehat{\vtheta}-\vtheta_p
  =\frac{\vZ_n}{e_n\sqrt n}
   +\frac{\mG_{F,n}\vZ_n}{\sqrt n}
   +\vct{r}_{\theta,B,n},
\end{equation}
where
\begin{align}
  \rank(\mG_{F,n})&\le2r,
  &\twonorm{\vct{r}_{\theta,B,n}}&=O_p(p^{-1}),
  \label{eq:ch2-erht-proof-median-rates}\\
  \opnorm{\mPi_{F,p}\mG_{F,n}\mPi_{F,p}}&=O_p(1),
  &\opnorm{\mPi_{B,p}\mG_{F,n}\mPi_{B,p}}&=O_p(p^{-1}),
  \label{eq:ch2-erht-proof-GF-blocks}\\
  \opnorm{\mPi_{F,p}\mG_{F,n}\mPi_{B,p}}
  +\opnorm{\mPi_{B,p}\mG_{F,n}\mPi_{F,p}}&=O_p(p^{-1/2}).
  \label{eq:ch2-erht-proof-GF-cross}
\end{align}
Since $\opnorm{\mQ_n(\rho)}\le\rho_0^{-1}$ and the ridge resolvent suppresses
the pervasive subspace,
\begin{equation}\label{eq:ch2-erht-proof-quadratic-median}
  n(\widehat{\vtheta}-\vtheta_p)\trans
    \mQ_n(\rho)(\widehat{\vtheta}-\vtheta_p)
  -\frac{1}{ne_n^2}\vZ_n\trans\mQ_n(\rho)\vZ_n
  =O_p(1).
\end{equation}

For $\vx\ne\vct{0}$, with $r=\twonorm{\vx}$ and
$\vu=\vx/r$,
\begin{align}
  DU_{\vx}[\vh]
  &=r^{-1}(\mI_p-\vu\vu\trans)\vh,
  \label{eq:ch2-erht-proof-sign-first}\\
  D^2U_{\vx}[\vh,\vh]
  &=r^{-2}\{-2\vh(\vu\trans\vh)-\vu\twonorm{\vh}^2
                +3\vu(\vu\trans\vh)^2\}.
  \label{eq:ch2-erht-proof-sign-second}
\end{align}
Applying these formulas to
$\widehat{\vY}_i=\sqrt pU(\vX_i-\widehat{\vtheta})$ yields
\begin{equation}\label{eq:ch2-erht-proof-sscm-expansion}
  \widehat{\mR}_n
  =\mB_n+\mH_n\mC_n\mH_n\trans+\mF_n+\mE_n,
\end{equation}
where
\begin{equation}\label{eq:ch2-erht-proof-HC}
  \mH_n=n^{-1/2}(\vZ_n,\vW_n),
  \qquad
  \mC_n=
  \begin{pmatrix}
    t_n/e_n^2 & -1/e_n\\
    -1/e_n & 0
  \end{pmatrix},
\end{equation}
$\rank(\mF_n)\le2r$, $\mPi_{B,p}\mF_n\mPi_{B,p}=\mtr{0}$, and
$\mE_n$ is negligible for the quadratic form.  Hence
\begin{equation}\label{eq:ch2-erht-proof-rank-bound}
  \rank(\mH_n\mC_n\mH_n\trans+\mF_n)\le2+2r.
\end{equation}

\subsubsection*{H.2 Woodbury reduction to three companion quadratic forms}

Let
\begin{equation}\label{eq:ch2-erht-proof-xyz}
  x_n=n^{-1}\vZ_n\trans\mQ_n\vZ_n,
  \qquad
  y_n=n^{-1}\vZ_n\trans\mQ_n\vW_n,
  \qquad
  z_n=n^{-1}\vW_n\trans\mQ_n\vW_n.
\end{equation}
For $x,y,z\in\R$, $e>0$, and $t>0$, define
\begin{equation}\label{eq:ch2-erht-proof-Lf}
  L(x,y,z,e,t)=e^2+tx-2ey+y^2-xz,
  \qquad
  f(x,y,z,e,t)=\frac{x}{L(x,y,z,e,t)}.
\end{equation}
The Woodbury identity applied to
$\mB_n+\mH_n\mC_n\mH_n\trans+\rho\mI_p$ gives
\begin{equation}\label{eq:ch2-erht-proof-Woodbury}
  (\mB_n+\mH_n\mC_n\mH_n\trans+\rho\mI_p)^{-1}
  =\mQ_n-\mQ_n\mH_n
   (\mC_n^{-1}+\mH_n\trans\mQ_n\mH_n)^{-1}
   \mH_n\trans\mQ_n.
\end{equation}
Direct calculation yields
\begin{equation}\label{eq:ch2-erht-proof-det}
  \det\{\mC_n^{-1}+\mH_n\trans\mQ_n\mH_n\}
  =-L(x_n,y_n,z_n,e_n,t_n).
\end{equation}
Moreover,
\begin{equation}\label{eq:ch2-erht-proof-leading-statistic}
  T_n(\rho)=n f(x_n,y_n,z_n,e_n,t_n)+O_p(\log p),
\end{equation}
and
\begin{equation}\label{eq:ch2-erht-proof-L-D}
  L(x_n,y_n,z_n,e_n,t_n)=D_n(\rho)+O_p(n^{-1/2}).
\end{equation}
The lower bound for $D_n(\rho)$ therefore guarantees that the denominator in
\eqref{eq:ch2-erht-proof-leading-statistic} is separated from zero.

\subsubsection*{H.3 Conditional quadratic-form central limit theorem}

Conditional on $\mathcal F_n^0$, write
$\vY_i=s_i\widetilde{\vY}_i$, where
$s_1,\ldots,s_n$ are independent Rademacher variables and the unsigned vectors
$\widetilde{\vY}_i$ are measurable with respect to $\mathcal F_n^0$.  Then
\begin{align}
  x_n-\kappa_n
  &=\frac2n\sum_{i<j}A_{ij}s_is_j,
  \label{eq:ch2-erht-proof-x-rad}\\
  y_n-b_{1,n}
  &=\frac1n\sum_{i<j}A_{ij}(w_i+w_j)s_is_j,
  \label{eq:ch2-erht-proof-y-rad}\\
  z_n-b_{2,n}
  &=\frac2n\sum_{i<j}A_{ij}w_iw_js_is_j.
  \label{eq:ch2-erht-proof-z-rad}
\end{align}
The conditional covariance matrix of
$\sqrt n(x_n-\kappa_n,y_n-b_{1,n},z_n-b_{2,n})\trans$ is precisely
$\mGamma_n(\rho)$.  Since $\vct{0}\preceq\mA_n(\rho)\preceq\mI_n$,
\begin{equation}\label{eq:ch2-erht-proof-row-influence}
  \max_i\sum_{j\ne i}A_{ij}^2
  \le\max_iA_{ii}\le1,
  \qquad
  \sum_{i,j}A_{ij}^2=\tr\{\mA_n(\rho)^2\}=O_p(n).
\end{equation}
Together with $\max_iw_i\le\bar\xi\omega_-^{-1/2}$,
\eqref{eq:ch2-erht-proof-row-influence} implies the maximal-influence and
fourth-moment conditions for the conditional central limit theorem for
degenerate quadratic forms.  Hence
\begin{equation}\label{eq:ch2-erht-proof-vector-clt}
  \mGamma_n(\rho)^{-1/2}\sqrt n
  \begin{pmatrix}
    x_n-\kappa_n\\
    y_n-b_{1,n}\\
    z_n-b_{2,n}
  \end{pmatrix}
  \overset{d}{\longrightarrow}N_3(\vct{0},\mI_3)
  \quad\text{conditionally on }\mathcal F_n^0.
\end{equation}
The gradient of $f$ at
$(\kappa_n,b_{1,n},b_{2,n},e_n,t_n)$ is
$\vct{g}_n(\rho)$.  A second-order Taylor expansion and
\eqref{eq:ch2-erht-proof-L-D} give
\begin{align}
  \frac{T_n(\rho)-n\mu_n(\rho)}{\sqrt n}
  &=\vct{g}_n(\rho)\trans\sqrt n
  \begin{pmatrix}
    x_n-\kappa_n\\
    y_n-b_{1,n}\\
    z_n-b_{2,n}
  \end{pmatrix}
  +O_p(n^{-1/2})+O_p(n^{-1/2}\log p).
  \label{eq:ch2-erht-proof-delta-method}
\end{align}
Equations \eqref{eq:ch2-erht-proof-vector-clt}--
\eqref{eq:ch2-erht-proof-delta-method} prove
Theorem~\ref{thm:ch2-erht-null}.

\subsubsection*{H.4 Feasible replacement}

Let
\[
  \mathcal T_n(\rho)
  =\bigl(\kappa_n,e_n,t_n,b_{1,n},b_{2,n},
          \{\psi_{ab,n}:0\le a,b\le2\}\bigr)
\]
and let $\widehat{\mathcal T}_n(\rho)$ denote its feasible counterpart.  The
center-aware perturbation expansion for the spatial median and the centered
SSCM gives, uniformly over every fixed ridge grid,
\begin{equation}\label{eq:ch2-erht-proof-feasible-functionals}
  |\widehat\kappa_n-\kappa_n|
  +|\widehat e_n-e_n|
  +|\widehat t_n-t_n|
  +|\widehat b_{1,n}-b_{1,n}|
  +|\widehat b_{2,n}-b_{2,n}|
  +\max_{0\le a,b\le2}
    |\widehat\psi_{ab,n}-\psi_{ab,n}|
  =O_p(n^{-1/2}).
\end{equation}
The mappings
\[
  \mathfrak m(\kappa,e,t,b_1,b_2)
  =\frac{\kappa}{(e-b_1)^2+\kappa(t-b_2)}
\]
and
\[
  \mathfrak v(\kappa,e,t,b_1,b_2,\mGamma)
  =\vct{g}(\kappa,e,t,b_1,b_2)\trans
   \mGamma\vct{g}(\kappa,e,t,b_1,b_2)
\]
are continuously differentiable on the nondegenerate event in
\eqref{eq:ch2-erht-nondegenerate}.  Taylor expansion of $\mathfrak v$ and
\eqref{eq:ch2-erht-proof-feasible-functionals} yield
\begin{equation}\label{eq:ch2-erht-proof-feasible-variance}
  \widehat\sigma_{D,n}^2(\rho)-\sigma_{D,n}^2(\rho)
  =O_p(n^{-1/2}).
\end{equation}
For the center, the linear terms generated by replacing the population center
with the spatial median cancel in the ratio defining $\mathfrak m$.  Retaining
the resulting quadratic remainder gives
\begin{equation}\label{eq:ch2-erht-proof-feasible-center}
  \widehat\mu_n(\rho)-\mu_n(\rho)=O_p(n^{-1}).
\end{equation}
Therefore
\begin{align}
  \frac{T_n-n\widehat\mu_n}
       {\{n\widehat\sigma_{D,n}^2\}^{1/2}}
  -\frac{T_n-n\mu_n}
       {\{n\sigma_{D,n}^2\}^{1/2}}
  &=-\frac{\sqrt n(\widehat\mu_n-\mu_n)}{\sigma_{D,n}}
    +\frac{T_n-n\mu_n}{\sqrt n\,\sigma_{D,n}}
      O_p(n^{-1/2})\\
  &=O_p(n^{-1/2})+O_p(n^{-1/2})=o_p(1).
\end{align}
Theorem~\ref{thm:ch2-erht-feasible} follows from
Theorem~\ref{thm:ch2-erht-null} and Slutsky's theorem.

\subsubsection*{H.5 Deterministic equivalents, local drift, and ridge aggregation}

The weighted companion-resolvent deterministic equivalents give, uniformly over
a fixed ridge grid,
\begin{align}
  D_n(\rho)&\overset{p}{\longrightarrow}D_\rho,
  &\sigma_{D,n}^2(\rho)&\overset{p}{\longrightarrow}\sigma_E^2(\rho),
  \label{eq:ch2-erht-proof-DE}\\
  \vct{d}_p\trans\mQ_n(\rho)\vct{d}_p
  &\overset{p}{\longrightarrow}q_\rho(G).
  \label{eq:ch2-erht-proof-signal-DE}
\end{align}
Under \eqref{eq:ch2-erht-local-alt}, expansion of the statistic around
$\vtheta_{0,p}$ gives
\begin{align}
  T_n(\rho)
  &=T_{n,0}(\rho)
    +2n^{3/4}\vct{d}_p\trans\widehat{\mQ}_n(\rho)
       (\widehat{\vtheta}-\vtheta_p)
    +n^{1/2}\vct{d}_p\trans\widehat{\mQ}_n(\rho)\vct{d}_p.
  \label{eq:ch2-erht-proof-local-decomposition}
\end{align}
The cross term in \eqref{eq:ch2-erht-proof-local-decomposition} is
$O_p(n^{1/4})$ and is negligible after division by the $n^{1/2}$ null scale,
whereas the last term converges to $q_\rho(G)$.  Combining
\eqref{eq:ch2-erht-proof-DE}--\eqref{eq:ch2-erht-proof-local-decomposition}
proves Theorem~\ref{thm:ch2-erht-local-power}.

For finitely many ridge values, the conditional quadratic-form argument is
applied to all linear combinations of the corresponding vectors in
\eqref{eq:ch2-erht-proof-vector-clt}.  Their cross-covariances converge to
$\mGamma_{\rho,\eta}$, so the Cram\'er--Wold device yields
\begin{equation}\label{eq:ch2-erht-proof-joint}
  (Z_{n,1},\ldots,Z_{n,K})\trans
  \overset{d}{\longrightarrow}N_K(\vct{0},\mC_K)
\end{equation}
under $H_0$, and the deterministic local drift adds
$\vLambda_K(G)$ under \eqref{eq:ch2-erht-local-alt}.  The continuous mapping
theorem applied to the finite Cauchy transform proves
Theorems~\ref{thm:ch2-erht-grid-null} and
\ref{thm:ch2-erht-grid-power}.

%% file: chapters/ch3_matrix.tex
\chapter[High-Dimensional Matrices]{High-Dimensional Matrix Estimation and Testing under Elliptical Symmetry}
\idx{matrix estimation}\idx{matrix testing}\idx{covariance matrix}\idx{shape matrix}\idx{sphericity test}\idx{proportionality test}\idx{precision matrix}\idx{Tyler's M-estimator}\idx{elliptical factor model}

\section{Introduction}

Matrix-valued inference is the second major pillar of high-dimensional data analysis.
Once the location parameter has been handled, the next questions concern the global and
local structure of variability: Is the population spherical? Are two dependence structures
proportional? Can one estimate a large precision matrix under heavy tails? Can a
factor structure be used to regularize a large scatter matrix without imposing Gaussianity?
These questions are naturally grouped together because they all concern the estimation or
testing of covariance, scatter, shape, or precision matrices.

In the present book, the word \emph{matrix} is used in a layered sense.
When second moments exist, the covariance matrix
\[
  \mSigma = \Cov(\vX)
\]
is the most familiar object. Under elliptical symmetry, however, the more primitive object
is the scatter matrix $\bm{\Theta}$ or its normalized version
\[
  \mLambda = \frac{p\bm{\Theta}}{\tr(\bm{\Theta})},
\]
which we call the \emph{shape matrix}. The shape matrix is well-defined even when the
covariance matrix is absent or unstable, and it is the correct invariant object in
sphericity and proportionality problems. When conditional linear dependence is the target,
the relevant object is the inverse shape matrix
\[
  \mV_0 = \mLambda_0^{-1},
\]
or, when $\mSigma$ exists and its scale is identifiable, the ordinary precision matrix
$\mOmega = \mSigma^{-1}$.

As in Chapter~2, the chapter is organized in three layers.

\begin{enumerate}[label=(\roman*)]
  \item We first review the classical fixed-$p$ theory. This includes the Wishart law
  for the sample covariance matrix, likelihood ratio and trace-type tests for covariance
  structure, fixed-dimensional sphericity tests, and the classical sign/rank shape tests.
  In every case we write down the statistic, the corresponding null law, and the local
  asymptotic interpretation.
  \item We then review high-dimensional benchmark procedures developed under Gaussian
  or light-tail models. These include regularized covariance and precision estimation,
  corrected likelihood and trace tests for sphericity, and two-sample covariance tests.
  These methods are indispensable benchmarks, and the reader should understand them before
  turning to the robust elliptical theory.
  \item Finally, we develop the robust elliptical route that forms the main theme of the
  book. This includes multivariate-sign and rank-based sphericity tests, adaptive
  dense--sparse combinations, two-sample proportionality testing, sparse precision matrix
  estimation through the spatial-sign covariance matrix, and matrix estimation through
  elliptical factor models.
\end{enumerate}

This chapter differs from many survey chapters in two respects.
First, sphericity testing is absorbed into the matrix chapter rather than treated as a
separate chapter, because under elliptical symmetry it is fundamentally a shape hypothesis.
Second, the chapter does not stop at testing. Robust high-dimensional inference also
requires estimation of large scatter and precision matrices, and the recent literature on
elliptical factor models shows that the robust and the latent-factor viewpoints can be
integrated within one unified framework.

\section{Problem formulation, notation, and standing models}

\subsection{Scatter, shape, covariance, and precision}

Let $\vX\in\R^p$ follow the elliptical model
\begin{equation}\label{eq:ch3-elliptical-model}
  \vX = \vmu + \xi \mA \vu,
  \qquad
  \vu \sim \mathrm{Unif}(\spn^{p-1}),
  \qquad
  \xi \ge 0,
\end{equation}
where $\xi$ and $\vu$ are independent, and $\bm{\Theta}=\mA\mA\trans$.
As emphasized in Chapter~1, the scatter matrix $\bm{\Theta}$ is defined only up to a
positive scalar multiplier. The identifiable structural object is therefore the normalized
shape matrix
\begin{equation}\label{eq:ch3-shape}
  \mLambda
  = \frac{p\bm{\Theta}}{\tr(\bm{\Theta})},
  \qquad
  \tr(\mLambda)=p.
\end{equation}
If $\E(\xi^2)<\infty$, then
\begin{equation}\label{eq:ch3-cov-scatter}
  \mSigma
  = \Cov(\vX)
  = \frac{\E(\xi^2)}{p}\bm{\Theta},
\end{equation}
so $\mSigma$ is proportional to $\bm{\Theta}$ and yields the same shape matrix
$\mLambda = p\mSigma/\tr(\mSigma)$.

The spatial-sign covariance matrix is
\begin{equation}\label{eq:ch3-sscm-pop}
  \mS_{\mathrm{sgn}}
  = \E\{U(\vX-\vmu)U(\vX-\vmu)\trans\},
\end{equation}
with sample version
\begin{equation}\label{eq:ch3-sscm-sample}
  \hat{\mS}_{\mathrm{sgn}}
  = \frac1n\sum_{i=1}^n
  U(\vX_i-\hat{\vtheta})
  U(\vX_i-\hat{\vtheta})\trans,
\end{equation}
where $\hat{\vtheta}$ is typically the spatial median. Under elliptical symmetry,
$\mS_{\mathrm{sgn}}$ and $\mLambda$ share the same eigenvectors. In high dimension,
the difference between $p\mS_{\mathrm{sgn}}$ and $\mLambda$ is often of order $p^{-1/2}$
entrywise, which explains why spatial-sign based procedures can estimate large shape and
precision matrices with nearly the same rates as covariance-based procedures while enjoying
much stronger robustness.

When sparsity of conditional linear dependence is of interest, we define
\begin{equation}\label{eq:ch3-precision}
  \mV_0 = \mLambda_0^{-1}.
\end{equation}
Since $\mLambda_0$ is scale normalized, $\mV_0$ should be interpreted as an inverse shape
matrix. If $\mSigma_0$ exists, then $\mV_0$ differs from $\mSigma_0^{-1}$ only by a global
scalar factor.

\subsection{Structural hypotheses}

The matrix problems studied in this chapter can all be written as hypotheses on the shape
or covariance structure.

\begin{enumerate}[label=(H\arabic*)]
  \item \textbf{Identity of covariance or shape}
  \begin{equation}\label{eq:ch3-h-id}
    H_0^{\mathrm{id}}:\ \mSigma=\mI_p
    \qquad\text{or}\qquad
    H_0^{\mathrm{id,shape}}:\ \mLambda=\mI_p .
  \end{equation}
  \item \textbf{Sphericity}
  \begin{equation}\label{eq:ch3-h-sph}
    H_0^{\mathrm{sph}}:\ \mSigma=\sigma^2\mI_p
    \ \text{for some } \sigma^2>0,
    \qquad\text{equivalently}\qquad
    \mLambda=\mI_p .
  \end{equation}
  \item \textbf{Equality of two covariance matrices}
  \begin{equation}\label{eq:ch3-h-eq}
    H_0^{\mathrm{eq}}:\ \mSigma_1=\mSigma_2 .
  \end{equation}
  \item \textbf{Proportionality of two scatter matrices}
  \begin{equation}\label{eq:ch3-h-prop}
    H_0^{\mathrm{prop}}:\ \bm{\Theta}_1 = c\,\bm{\Theta}_2
    \ \text{for some } c>0.
  \end{equation}
\end{enumerate}

The proportionality null is the natural two-sample shape hypothesis under elliptical
symmetry. Indeed, by \eqref{eq:ch3-shape}, $H_0^{\mathrm{prop}}$ holds if and only if the
two normalized shape matrices are equal:
\begin{equation}\label{eq:ch3-h-prop-equivalent}
  H_0^{\mathrm{prop}}
  \quad\Longleftrightarrow\quad
  \mLambda_1 = \mLambda_2.
\end{equation}

\subsection{Standing assumptions for the chapter}

The chapter uses several recurring assumptions. To avoid repeating the same hypotheses
word-for-word in every section, we collect the most common ones here and then refer to
them in the theorem statements.

\begin{assumption}[Fixed-dimensional Gaussian model]\label{ass:ch3-fixed-gauss}
Whenever a fixed-$p$ likelihood-ratio or Wishart-based statement is invoked, we assume
that $\vX_1,\ldots,\vX_n\iid N_p(\vmu,\mSigma)$ with $p$ fixed and $\mSigma$ positive
definite.
\end{assumption}

\begin{assumption}[Trace regularity for high-dimensional shape testing]
\label{ass:ch3-trace}
For a sequence of $p\times p$ symmetric nonnegative definite matrices $\mA_p$, assume
\begin{equation}\label{eq:ch3-trace}
  \tr(\mA_p^4) = o\{\tr^2(\mA_p^2)\}
\end{equation}
as $n,p\to\infty$. In most testing problems $\mA_p$ will be the true shape matrix
$\mLambda_p$ or a local perturbation thereof.
\end{assumption}

\begin{assumption}[Inverse radial moments]
\label{ass:ch3-radial}
Let $R=\norm{\vX-\vmu}_2$. For sufficiently large $p$, the inverse moments
$\E(R^{-k})$ exist for $k=1,2,3,4$, and there exist finite constants $d_k\in[1,\infty)$
such that
\begin{equation}\label{eq:ch3-radial}
  \frac{\E(R^{-k})}{\E(R^{-1})^k} \longrightarrow d_k,
  \qquad k=2,3,4.
\end{equation}
\end{assumption}

\begin{assumption}[Balanced two-sample growth]
\label{ass:ch3-balanced}
In two-sample problems,
\begin{equation}\label{eq:ch3-balanced}
  \frac{n_1}{n_1+n_2}\longrightarrow \kappa\in(0,1).
\end{equation}
\end{assumption}

\begin{assumption}[Bounded eigenvalues and diagonal scales]
\label{ass:ch3-bounded}
There exist constants $0<\eta<c_0<\infty$ independent of $(n,p)$ such that
\begin{equation}\label{eq:ch3-bounded}
  \eta < \lambda_p(\mSigma_0) \le \lambda_1(\mSigma_0) < \eta^{-1},
  \qquad
  c_0 p \le \tr(\mSigma_0)\le c_0^{-1}p,
\end{equation}
and the diagonal entries of $\mSigma_0$ are bounded away from zero and infinity.
\end{assumption}

\begin{assumption}[Sub-Gaussian inverse radius and SSCM contraction]
\label{ass:ch3-nu}
Let $r_i=\norm{\vX_i-\vmu}_2$, let $\zeta_k=\E(r_i^{-k})$, and define
$\nu_i=\zeta_1^{-1}r_i^{-1}$. Assume that
\begin{equation}\label{eq:ch3-nu}
  \zeta_k\zeta_1^{-k}\le C_\zeta,\qquad k=1,2,3,4,
\end{equation}
that $\nu_i$ is sub-Gaussian with $\norm{\nu_i}_{\psi_2}\le K_\nu$, and that the
population SSCM satisfies
\begin{equation}\label{eq:ch3-sscm-op}
  \limsup_{p\to\infty}\opnorm{\mS_{\mathrm{sgn}}} < 1-\psi
\end{equation}
for some $\psi>0$.
\end{assumption}

\section{Classical low-dimensional covariance, shape, and sphericity inference}

\subsection{Wishart geometry and the covariance likelihood ratio}

We begin with the classical normal model. Let
$\vX_1,\ldots,\vX_n \iid N_p(\vmu,\mSigma)$ with $p$ fixed, and define
\begin{equation}\label{eq:ch3-xbar-s}
  \bar{\vX}=\frac1n\sum_{i=1}^n \vX_i,
  \qquad
  \mS_n = \frac1n \sum_{i=1}^n (\vX_i-\bar{\vX})(\vX_i-\bar{\vX})\trans.
\end{equation}
Under Assumption~\ref{ass:ch3-fixed-gauss}, the scaled sample covariance
$n\mS_n$ follows a Wishart law:
\begin{equation}\label{eq:ch3-wishart}
  n\mS_n \sim W_p(n-1,\mSigma).
\end{equation}

For testing
\[
  H_0:\ \mSigma = \mSigma_0
\]
against the unrestricted alternative, the Gaussian likelihood ratio takes the form
\begin{equation}\label{eq:ch3-LRT-cov}
  \Lambda_n
  = \frac{
      \sup_{\vmu,\mSigma=\mSigma_0} L(\vmu,\mSigma)
    }{
      \sup_{\vmu,\mSigma\succ 0} L(\vmu,\mSigma)
    }
  = \exp\Big\{-\frac n2\big[
      \tr(\mSigma_0^{-1}\mS_n)
      - \log\det(\mSigma_0^{-1}\mS_n) - p
    \big]\Big\}.
\end{equation}
Hence the usual likelihood-ratio statistic is
\begin{equation}\label{eq:ch3-LRT-stat}
  T_{\mathrm{LRT}}
  = n\Big[
      \tr(\mSigma_0^{-1}\mS_n)
      - \log\det(\mSigma_0^{-1}\mS_n) - p
    \Big].
\end{equation}

\begin{theorem}[Fixed-$p$ covariance likelihood ratio]
\label{thm:ch3-LRT-fixed}
Under Assumption~\ref{ass:ch3-fixed-gauss} and $H_0:\mSigma=\mSigma_0$,
\begin{equation}\label{eq:ch3-LRT-fixed}
  T_{\mathrm{LRT}}
  \overset{d}{\longrightarrow}
  \chi^2_{p(p+1)/2}
\end{equation}
as $n\to\infty$ with $p$ fixed.
\end{theorem}

The statistic \eqref{eq:ch3-LRT-stat} is the fixed-dimensional benchmark for covariance
testing. It is optimal under Gaussianity but becomes unusable in high dimensions because
$\log\det(\mS_n)$ is undefined when $p\ge n$ and because the calibration changes
substantially when $p/n$ does not vanish.

\subsection{Low-dimensional sphericity tests: Mauchly, John, and Nagao}

The fixed-dimensional sphericity problem tests
\begin{equation}\label{eq:ch3-low-sph-null}
  H_0:\ \mSigma = \sigma^2 \mI_p
  \qquad\text{against}\qquad
  H_1:\ \mSigma \neq \sigma^2 \mI_p,
\end{equation}
where $\sigma^2>0$ is unspecified.

The classical likelihood-ratio statistic is based on
\begin{equation}\label{eq:ch3-mauchly}
  V_n
  = \frac{\det(\mS_n)}{\{\tr(\mS_n)/p\}^p},
\end{equation}
which is the ratio between the geometric mean and the arithmetic mean of the sample
eigenvalues. Small values of $V_n$ indicate departures from sphericity.

\begin{theorem}[Mauchly's likelihood-ratio test]
\label{thm:ch3-mauchly}
Under Assumption~\ref{ass:ch3-fixed-gauss} and the null
\eqref{eq:ch3-low-sph-null},
\begin{equation}\label{eq:ch3-mauchly-chi}
  -\rho_n (n-1)\log V_n
  \overset{d}{\longrightarrow}
  \chi^2_{(p-1)(p+2)/2},
\end{equation}
where the Bartlett correction factor is
\begin{equation}\label{eq:ch3-rho}
  \rho_n = 1 - \frac{2p^2+p+2}{6(p+1)(n-1)} .
\end{equation}
\end{theorem}

Mauchly's statistic is powerful against broad alternatives, but in finite samples it is
notoriously sensitive to nonnormality. John's trace statistic avoids the determinant and
turns out to be the key bridge to high-dimensional sphericity testing.

Define
\begin{equation}\label{eq:ch3-john-U}
  U_J
  = \frac1p
  \tr\Big\{
    \frac{\mS_n}{\tr(\mS_n)/p} - \mI_p
  \Big\}^2
  = \frac{p\,\tr(\mS_n^2)}{\tr^2(\mS_n)} - 1.
\end{equation}
John's statistic is a quadratic measure of the dispersion of the empirical eigenvalues
around their common mean.

\begin{theorem}[John's trace test]
\label{thm:ch3-john-fixed}
Under Assumption~\ref{ass:ch3-fixed-gauss} and the null
\eqref{eq:ch3-low-sph-null},
\begin{equation}\label{eq:ch3-john-fixed}
  \frac{n(p+2)}{2}U_J
  \overset{d}{\longrightarrow}
  \chi^2_{(p-1)(p+2)/2}
\end{equation}
as $n\to\infty$ with $p$ fixed.
\end{theorem}

A closely related statistic is Nagao's test for
\[
  H_0:\ \mSigma=\mI_p,
\]
which uses
\begin{equation}\label{eq:ch3-nagao}
  V_N = \frac1p\tr(\mS_n-\mI_p)^2.
\end{equation}
Under fixed $p$ and Gaussian sampling,
\begin{equation}\label{eq:ch3-nagao-limit}
  n p V_N
  \overset{d}{\longrightarrow}
  \chi^2_{p(p+1)/2}.
\end{equation}
The exact centering and scaling depend on whether the scale $\sigma^2$ is known or
estimated, but the main point is that low-dimensional sphericity can be studied either
through a determinant ratio or through quadratic trace functionals.

\subsection{Classical sign and rank shape tests}

The fixed-$p$ robust literature replaces the sample covariance matrix by a matrix of
multivariate signs or ranks. Let
\begin{equation}\label{eq:ch3-low-omega}
  \mOmega_n(\vtheta)
  = \frac1n\sum_{i=1}^n
  U(\vX_i-\vtheta)U(\vX_i-\vtheta)\trans,
\end{equation}
and let $\hat{\vtheta}_{\mathrm{SM}}$ be the spatial median. Hallin--Paindaveine and
related rank theories show that, under an elliptical spherical null, the matrix
$\mOmega_n(\hat{\vtheta}_{\mathrm{SM}})$ behaves like a covariance matrix whose target is
$p^{-1}\mI_p$ and whose eigenvectors carry only directional information.

Define
\begin{equation}\label{eq:ch3-low-QS}
  Q_S
  = p\,\tr\Big\{
      \mOmega_n(\hat{\vtheta}_{\mathrm{SM}})
      - \frac1p\mI_p
    \Big\}^2 .
\end{equation}
When $p$ is fixed, the effect of estimating the center is asymptotically negligible.

\begin{theorem}[Fixed-$p$ spatial-sign sphericity statistic]
\label{thm:ch3-low-sign}
Suppose that $\vX$ follows an elliptical distribution with location $\vtheta$ and spherical
shape under $H_0$, and let $p$ be fixed. Then
\begin{equation}\label{eq:ch3-low-sign}
  \frac{n(p+2)}{2} Q_S
  \overset{d}{\longrightarrow}
  \chi^2_{(p-1)(p+2)/2}.
\end{equation}
\end{theorem}

Theorem~\ref{thm:ch3-low-sign} is the robust analogue of
Theorem~\ref{thm:ch3-john-fixed}. The replacement of the covariance matrix by the SSCM
makes the procedure less sensitive to radial outliers and more natural under general
elliptical symmetry.

\section{High-dimensional Gaussian and light-tail benchmark methods}

\subsection{Regularized covariance and precision estimation}

When $p$ is large relative to $n$, the sample covariance matrix is unstable even under
Gaussian sampling, and every successful large-$p$ procedure must regularize either the
covariance matrix itself or its inverse. A selective but very useful overview is provided by
\citet{FanLiaoLiu2015Review}, and the present subsection follows the main lines of that
review while keeping the notation of this book.

\paragraph{Thresholding and adaptive thresholding.}
Let
\begin{equation}\label{eq:ch3-sample-cov-hd}
  \mS_n=(s_{ij})_{1\le i,j\le p}
  = \frac{1}{n-1}\sum_{k=1}^n(\vX_k-\bar\vX)(\vX_k-\bar\vX)\trans.
\end{equation}
A basic sparsity class for covariance estimation is
\begin{equation}\label{eq:ch3-Uq-class}
  \mathcal U_q\{s_0(p),M\}
  = \Bigl\{\mSigma\succ 0:
     \max_{1\le i\le p}\sum_{j=1}^p |\sigma_{ij}|^q\le s_0(p),
     \max_{1\le i\le p}\sigma_{ii}\le M
    \Bigr\},
  \qquad 0\le q<1.
\end{equation}
The hard-thresholding estimator of \citet{BickelLevina2008Cov} is
\begin{equation}\label{eq:ch3-hard-thresholding}
  \hat\mSigma_{\tau}^{\mathrm{HT}}
  = \bigl(s_{ij}\mathbbm{1}\{|s_{ij}|\ge \tau_n\}\bigr)_{1\le i,j\le p},
  \qquad
  \tau_n=C\sqrt{\frac{\log p}{n}}.
\end{equation}
Under exponential-type tails and $\mSigma\in\mathcal U_q\{s_0(p),M\}$,
\citet{BickelLevina2008Cov} showed that
\begin{equation}\label{eq:ch3-bl-rate}
  \opnorm{\hat\mSigma_{\tau}^{\mathrm{HT}}-\mSigma}
  = O_P\!\left(s_0(p)\Bigl(\frac{\log p}{n}\Bigr)^{(1-q)/2}\right),
\end{equation}
and, provided the smallest eigenvalue of $\mSigma$ is bounded away from zero, the same rate
holds for the inverse matrix. The more general positive-definite thresholding and eigenvalue-
constrained estimators of \citet{LiuWangZhao2014EC2} solve
\begin{equation}\label{eq:ch3-ec2}
  \hat\mSigma^{\mathrm{EC2}}
  = \argmin_{\mSigma:\,\lambda_{\min}(\mSigma)\ge \tau_0}
    \Bigl\{
      \frac12\frobnorm{\mS_n-\mSigma}^2
      + \sum_{i\neq j} P_{\lambda_n}(|\sigma_{ij}|)
    \Bigr\},
\end{equation}
which combines generalized thresholding with a finite-sample positive-definiteness
constraint.

To account for heterogeneous marginal scales, \citet{CaiLiu2011AdaptiveThreshold}
proposed adaptive entrywise thresholding:
\begin{equation}\label{eq:ch3-adaptive-thresholding}
  \hat\sigma_{ij}^{\mathrm{AT}}
  = s_{ij}\,\mathbbm{1}\{|s_{ij}|\ge \tau_{ij}\},
  \qquad
  \tau_{ij}=\delta\sqrt{\hat\theta_{ij}\frac{\log p}{n}},
\end{equation}
where
\begin{equation}\label{eq:ch3-thetaij}
  \hat\theta_{ij}
  = \frac{1}{n}\sum_{k=1}^n
    \Bigl\{(X_{ki}-\bar X_i)(X_{kj}-\bar X_j)-s_{ij}\Bigr\}^2
\end{equation}
estimates the variance of $s_{ij}$. Over heteroscedastic sparse classes, adaptive thresholding
attains the optimal operator-norm rate and improves substantially on a single universal
threshold. The generalized thresholding framework of \citet{RothmanLevinaZhu2009}
replaces the hard-thresholding map by a shrinkage function $s_{\tau}(\cdot)$, yielding
\begin{equation}\label{eq:ch3-general-thresholding}
  \hat\mSigma_{\tau}^{\mathrm{GT}}
  = \bigl(s_{\tau}(s_{ij})\bigr)_{1\le i,j\le p},
\end{equation}
which includes hard, soft, SCAD, and MCP thresholding as special cases.

\paragraph{Sparse precision estimation.}
If the conditional dependence structure is sparse, it is more natural to estimate the inverse
covariance matrix $\mOmega_0=\mSigma_0^{-1}$ directly. The graphical lasso of
\citet{YuanLin2007,FriedmanHastieTibshirani2008} solves
\begin{equation}\label{eq:ch3-glasso-classical}
  \hat{\mOmega}_{\mathrm{GL}}
  = \argmin_{\mOmega\succ 0}
    \Bigl\{
      \tr(\mS_n\mOmega)-\log\det(\mOmega)+\lambda_n\sum_{i\ne j}|\omega_{ij}|
    \Bigr\},
\end{equation}
while the CLIME estimator of \citet{CaiLiuLuo2011CLIME} is
\begin{equation}\label{eq:ch3-clime-classical}
  \hat{\mOmega}_{\mathrm{CLIME}}
  = \argmin_{\mOmega}\onenorm{\mOmega}
  \quad\text{subject to}\quad
  \maxnorm{\mS_n\mOmega-\mI_p}\le \lambda_n.
\end{equation}
For the weak-sparsity class
\begin{equation}\label{eq:ch3-Gq-class}
  \mathcal G_q\{s_0(p),M\}
  = \Bigl\{\mOmega\succ 0:
      \onenorm{\mOmega}\le M,
      \max_{1\le i\le p}\sum_{j=1}^p |\omega_{ij}|^q\le s_0(p)
    \Bigr\},
\end{equation}
\citet{CaiLiuLuo2011CLIME} proved, under exponential or polynomial tails, that
\begin{align}
  \maxnorm{\hat{\mOmega}_{\mathrm{CLIME}}-\mOmega_0}
  &= O_P\!\left(\sqrt{\frac{\log p}{n}}\right),
  \label{eq:ch3-clime-maxrate}\\
  \opnorm{\hat{\mOmega}_{\mathrm{CLIME}}-\mOmega_0}
  &= O_P\!\left(s_0(p)\Bigl(\frac{\log p}{n}\Bigr)^{(1-q)/2}\right),
  \label{eq:ch3-clime-oprate}\\
  \frac{1}{p}\frobnorm{\hat{\mOmega}_{\mathrm{CLIME}}-\mOmega_0}^2
  &= O_P\!\left(s_0(p)\Bigl(\frac{\log p}{n}\Bigr)^{1-q}\right).
  \label{eq:ch3-clime-frate}
\end{align}
The $\ell_1$-penalized log-determinant estimator was studied in detail by
\citet{RavikumarWainwrightRaskuttiYu2011}, who established graph-selection consistency
under an irrepresentability condition. Nonconvex precision estimation and sharper rates were
studied by \citet{LamFan2009}. Fast columnwise alternatives include the SCIO estimator of
\citet{LiuLuo2015SCIO} and the scaled-lasso precision estimator of
\citet{SunZhang2012ScaledLasso}.

\paragraph{Low-rank plus sparse covariance structure.}
In many applications the covariance matrix is not sparse, but after removing a few common
factors the idiosyncratic covariance is sparse. This motivates the approximate factor model
\begin{equation}\label{eq:ch3-factor-light}
  \vY_t=\mB\vf_t+\vu_t,
  \qquad
  \mSigma_0=\mB\Cov(\vf_t)\mB\trans+\mSigma_u.
\end{equation}
\citet{FanLiaoMincheva2011} proposed estimating $\mSigma_u$ from factor residuals and then
thresholding it. When the factors are latent, the POET estimator of
\citet{FanLiaoMincheva2013} uses the principal orthogonal complement
\begin{equation}\label{eq:ch3-poet-benchmark}
  \hat\mSigma_{\mathrm{POET}}
  = \sum_{j=1}^m \hat\lambda_j\hat\vgamma_j\hat\vgamma_j\trans
    + \hat\mSigma_u^{\,\tau},
\end{equation}
where $\hat\mSigma_u^{\,\tau}$ is obtained by thresholding the residual covariance after
subtracting the first $m$ principal components. Under conditional sparsity of
$\mSigma_u$, both $\hat\mSigma_u$ and $\hat\mSigma_{\mathrm{POET}}^{-1}$ achieve the same
operator-norm rate $m_p\omega_n^{1-q}$, with
$\omega_n=\sqrt{(\log p)/n}+p^{-1/2}$. This benchmark is particularly important for the
elliptical factor-model developments in Section~\ref{sec:ch3-elliptical-factor-models}.

\subsection{High-dimensional Gaussian and light-tail tests for identity, sphericity, and covariance equality}

The testing literature under Gaussian or light-tail assumptions is as rich as the estimation
literature. Since the robust elliptical procedures of later sections are meant to replace these
benchmarks, we record the concrete test statistics and their large-sample null laws.

\paragraph{Corrected likelihood-ratio tests.}
Suppose first that $\vX_1,\ldots,\vX_n\iid N_p(\vct 0,\mSigma)$ and the sphericity null is
$H_0:\mSigma=\sigma^2\mI_p$. Let $\ell_1,\ldots,\ell_p$ be the eigenvalues of the sample
covariance matrix $\mS_n=n^{-1}\sum_{i=1}^n \vX_i\vX_i\trans$, and define the classical
likelihood-ratio statistic by
\begin{equation}\label{eq:ch3-Ln-sphericity}
  L_n
  = \left\{
      \frac{(\ell_1\cdots \ell_p)^{1/p}}{p^{-1}\sum_{j=1}^p \ell_j}
    \right\}^{pn/2}.
\end{equation}
When $p/n\to y\in(0,1)$, the ordinary $\chi^2$ calibration fails. The random-matrix correction
of \citet{BaiJiangYaoZheng2009} is based on
\begin{equation}\label{eq:ch3-CLRT-stat}
  T_{\mathrm{CLRT}} = -\frac{2}{n}\log L_n,
\end{equation}
for which
\begin{equation}\label{eq:ch3-CLRT-law}
  T_{\mathrm{CLRT}} + (p-n)\log\Bigl(1-\frac{p}{n}\Bigr)-p
  \overset{d}{\longrightarrow}
  N\Bigl(-\frac12\log(1-y),\,-2\log(1-y)-2y\Bigr)
\end{equation}
in the real Gaussian case. More generally, if the entries are i.i.d. standardized with fourth
cumulant $\beta=\E(X_{11}^4)-3$, \citet{WangYao2013} showed that the same corrected LRT
remains asymptotically normal after replacing the mean shift by
$-\frac12\log(1-y)+\frac12\beta y$.

\paragraph{Corrected John's test.}
John's statistic can be written as
\begin{equation}\label{eq:ch3-qj}
  Q_J = \frac{np^2}{2}
  \tr\Bigl\{\frac{\mS_n}{\tr(\mS_n)}-\frac1p\mI_p\Bigr\}^2
  = \frac{np}{2}\left\{\frac{p\,\tr(\mS_n^2)}{\tr^2(\mS_n)}-1\right\}.
\end{equation}
Under Gaussian sphericity and $p/n\to c\in(0,\infty)$, \citet{LedoitWolf2002} proved
\begin{equation}\label{eq:ch3-LW-law}
  \frac{2Q_J}{p}-p \overset{d}{\longrightarrow} N(1,4),
\end{equation}
or equivalently $\{(2Q_J/p)-p-1\}/2\Rightarrow N(0,1)$. The remarkable feature of John's
statistic is that its limit does not degenerate when $p>n$. Under general fourth moments,
\citet{WangYao2013} derived the corrected John statistic
\begin{equation}\label{eq:ch3-CJ-law}
  n\,U_n-p \overset{d}{\longrightarrow} N(1+\beta,4),
  \qquad
  U_n=\frac{2Q_J}{np},
\end{equation}
for real-valued observations, so that the only correction is through the kurtosis shift $\beta$.

\paragraph{Trace-based $U$-statistics.}
The $U$-statistic route avoids determinants altogether. For testing
$H_0:\mSigma=\mI_p$, \citet{ChenZhangZhong2010} proposed statistics based on unbiased
estimators of $\tr(\mSigma^2)$ and $\tr(\mSigma)$, while \citet{FisherSunGallagher2010}
developed an alternative trace criterion with accurate finite-sample calibration in moderate
and high dimensions.

\paragraph{Two-sample covariance equality.}
For two independent samples
$\vX_{1i}\sim(\vmu_1,\mSigma_1)$, $i=1,\ldots,n_1$, and
$\vX_{2j}\sim(\vmu_2,\mSigma_2)$, $j=1,\ldots,n_2$, the null hypothesis
$H_0:\mSigma_1=\mSigma_2$ is most commonly benchmarked by the Li--Chen statistic
\citep{LiChen2012}. Define
\begin{align}
  A_{n_h}
  &= \frac{1}{n_h(n_h-1)}\sum_{i\ne j}
      (\vX_{hi}\trans\vX_{hj})^2
     -\frac{2}{n_h(n_h-1)(n_h-2)}\sum_{i,j,k}^{*}
      (\vX_{hi}\trans\vX_{hj})(\vX_{hj}\trans\vX_{hk})
      \nonumber\\
  &\qquad
     +\frac{1}{n_h(n_h-1)(n_h-2)(n_h-3)}\sum_{i,j,k,l}^{*}
      (\vX_{hi}\trans\vX_{hj})(\vX_{hk}\trans\vX_{hl}),
      \qquad h=1,2,
  \label{eq:ch3-LC-A}
\end{align}
where $\sum^*$ means that all indices are distinct, and
\begin{align}
  C_{n_1n_2}
  &= \frac{1}{n_1n_2}\sum_{i=1}^{n_1}\sum_{j=1}^{n_2}
      (\vX_{1i}\trans\vX_{2j})^2
     -\frac{1}{n_1n_2(n_1-1)}\sum_{i\ne k}\sum_{j=1}^{n_2}
      (\vX_{1i}\trans\vX_{2j})(\vX_{2j}\trans\vX_{1k})
      \nonumber\\
  &\qquad
     -\frac{1}{n_1n_2(n_2-1)}\sum_{i=1}^{n_1}\sum_{j\ne l}
      (\vX_{1i}\trans\vX_{2j})(\vX_{2l}\trans\vX_{1i})
     +\frac{1}{n_1n_2(n_1-1)(n_2-1)}\sum_{i\ne k}\sum_{j\ne l}
      (\vX_{1i}\trans\vX_{2j})(\vX_{1k}\trans\vX_{2l}).
  \label{eq:ch3-LC-C}
\end{align}
Then
\begin{equation}\label{eq:ch3-li-chen}
  T_{\mathrm{LC}} = A_{n_1}+A_{n_2}-2C_{n_1n_2}
\end{equation}
is an unbiased estimator of $\tr\{(\mSigma_1-\mSigma_2)^2\}$. Under the trace regularity
condition
\begin{equation}\label{eq:ch3-LC-trace}
  \tr(\mSigma_i\mSigma_j\mSigma_k\mSigma_l)
  = o\!\left\{\tr(\mSigma_i\mSigma_j)\tr(\mSigma_k\mSigma_l)\right\},
  \qquad i,j,k,l\in\{1,2\},
\end{equation}
\citet{LiChen2012} proved that
\begin{equation}\label{eq:ch3-LC-CLT}
  \frac{T_{\mathrm{LC}}}{\sigma_{n,\mathrm{LC}}}
  \overset{d}{\longrightarrow} N(0,1),
\end{equation}
where
\begin{equation}\label{eq:ch3-LC-var}
  \sigma_{n,\mathrm{LC}}^2
  = \frac{4}{n_1^2}\tr^2(\mSigma_1^2)
    +\frac{4}{n_2^2}\tr^2(\mSigma_2^2)
    +\frac{8}{n_1n_2}\tr^2(\mSigma_1\mSigma_2)
    +o\!\bigl(n_1^{-2}+n_2^{-2}+n_1^{-1}n_2^{-1}\bigr).
\end{equation}
This statistic is the covariance-based analogue of the proportionality and SSCM tests that
appear in Section~\ref{sec:ch3-proportionality}.

\subsection{Why Gaussian benchmarks are insufficient under elliptical symmetry}

The Gaussian or light-tail methods above are essential benchmarks, but they have three
limitations from the perspective of this book.

\begin{enumerate}[label=(\alph*)]
  \item They are covariance based. If the covariance matrix is unstable or difficult to
  estimate, the entire procedure becomes fragile.
  \item They often assume sub-Gaussian or at least exponential-type tail behavior.
  Financial returns, genomics measurements, and other complex data sets often exhibit
  radial heterogeneity inconsistent with such assumptions.
  \item They do not separate global scale from shape. Under elliptical symmetry, many
  hypotheses are better phrased in terms of normalized shape matrices rather than raw
  covariance matrices.
\end{enumerate}

These issues motivate the sign-, rank-, and shape-based methods that occupy the remainder
of the chapter.

\section{High-dimensional sphericity under elliptical symmetry}

\subsection{The sign-based sphericity test}

We now turn to the first robust high-dimensional sphericity test in the present research
line. Let $\vX_1,\ldots,\vX_n$ be independent observations from an elliptical distribution
with center $\vtheta$ and shape matrix $\mLambda_p$. We write the local alternative as
\begin{equation}\label{eq:ch3-local-D}
  H_1:\ \mLambda_p = \mI_p + \mD_{n,p},
\end{equation}
where $\mD_{n,p}$ is symmetric and trace free.

If the center were known, the natural sign statistic would be
\[
  Q'_S
  = \frac{p}{n(n-1)}\sum_{i\ne j}
    \{U(\vX_i-\vtheta)\trans U(\vX_j-\vtheta)\}^2 - 1.
\]
When the center is replaced by the spatial median $\hat{\vtheta}$, the direct plug-in
statistic acquires a non-negligible bias in the high-dimensional regime. The solution of
\citet{ZouPengFengWang2014Sphericity} is the bias-corrected statistic
\begin{equation}\label{eq:ch3-qtilde}
  \tilde Q_S
  = \frac{p}{n(n-1)}\sum_{i\ne j}
  (\hat{\vU}_i\trans \hat{\vU}_j)^2 - 1,
  \qquad
  \hat{\vU}_i = U(\vX_i-\hat{\vtheta}),
\end{equation}
centered by an explicit bias term depending on inverse radial moments.

Let
\begin{equation}\label{eq:ch3-delta}
  \delta_{n,p}
  = \frac{1}{n^2}
    \Big(
      2 - 2\frac{\E(R^{-2})}{\E(R^{-1})^2}
        + \frac{\E(R^{-2})^2}{\E(R^{-1})^4}
    \Big)
  + \frac{1}{n^3}
    \Big(
      8\frac{\E(R^{-2})}{\E(R^{-1})^2}
      - 6\frac{\E(R^{-2})^2}{\E(R^{-1})^4}
      + 2\frac{\E(R^{-2})\E(R^{-3})}{\E(R^{-1})^5}
      - 2\frac{\E(R^{-3})}{\E(R^{-1})^3}
    \Big),
\end{equation}
and
\begin{equation}\label{eq:ch3-sigma0-sign}
  \tilde\sigma_0^2
  = \frac{4(p-1)}{n(n-1)(p+2)}.
\end{equation}

\begin{assumption}[Sign-based sphericity conditions]
\label{ass:ch3-sign-sph}
Suppose Assumptions~\ref{ass:ch3-radial} and \ref{ass:ch3-trace} hold. In addition, assume
\begin{equation}\label{eq:ch3-sign-growth}
  p = O(n^2).
\end{equation}
Under the local alternative \eqref{eq:ch3-local-D}, assume
\begin{equation}\label{eq:ch3-sign-local}
  \frac{n\,\tr(\mD_{n,p}^2)}{p} = O(1).
\end{equation}
\end{assumption}

\begin{theorem}[Bias-corrected sign-based sphericity test under the null]
\label{thm:ch3-sign-null}
Suppose Assumption~\ref{ass:ch3-sign-sph} holds. Then, under $H_0:\mLambda_p=\mI_p$,
\begin{equation}\label{eq:ch3-sign-null}
  \frac{\tilde Q_S - p\delta_{n,p}}{\tilde\sigma_0}
  \overset{d}{\longrightarrow} N(0,1)
\end{equation}
as $n,p\to\infty$.
\end{theorem}

\begin{theorem}[Bias-corrected sign-based sphericity test under local alternatives]
\label{thm:ch3-sign-alt}
Suppose Assumption~\ref{ass:ch3-sign-sph} holds. Under
$H_1:\mLambda_p=\mI_p+\mD_{n,p}$,
\begin{equation}\label{eq:ch3-sigma1-sign}
  \tilde\sigma_1^2
  =
  \tilde\sigma_0^2
  + \frac{8p\,\tr(\mD_{n,p}^2)+4\tr^2(\mD_{n,p}^2)}{n^2 p^2}
  + \frac{8}{n p^2}
    \Big[
      \tr(\mLambda_p^4) - \frac1p \tr^2(\mLambda_p^2)
    \Big].
\end{equation}
Moreover,
\begin{equation}\label{eq:ch3-sign-alt}
  \frac{
    \tilde Q_S - p^{-1}\tr(\mD_{n,p}^2) - p\delta_{n,p}
  }{
    \tilde\sigma_1
  }
  \overset{d}{\longrightarrow} N(0,1).
\end{equation}
In particular, the test rejecting for large values of
$(\tilde Q_S-p\delta_{n,p})/\tilde\sigma_0$ is consistent whenever
\begin{equation}\label{eq:ch3-sign-consistency}
  \frac{n\,\tr(\mD_{n,p}^2)}{p}\longrightarrow\infty.
\end{equation}
\end{theorem}

Theorems~\ref{thm:ch3-sign-null} and \ref{thm:ch3-sign-alt} show why the sign test belongs
in this chapter rather than in a generic robust chapter. The quantity
$p^{-1}\tr(\mD_{n,p}^2)$ is exactly the quadratic distance from spherical shape, so the
statistic is a direct matrix test.

\subsection{Rank-based sphericity tests}

The next step is to replace the one-sample sign vectors by two-sample directional
comparisons. Let
\begin{equation}\label{eq:ch3-Uij}
  \vU_{ij} = U(\vX_i-\vX_j),
  \qquad 1\le i\ne j\le n.
\end{equation}
\citet{FengLiu2017RankSphericity} proposed both a Spearman-rho type statistic and a Kendall-tau type statistic.
The leading object in the Spearman route is
\begin{equation}\label{eq:ch3-rank-trace}
  \widehat{\tr(\mOmega_p^2)}
  =
  \frac{1}{2n(n-1)(n-2)(n-3)}
  \sum_{i,j,k,l}^{*}
  \vU_{ij}\trans \vU_{kl}\,
  \vU_{kj}\trans \vU_{il},
\end{equation}
where $\sum^{*}$ denotes summation over mutually distinct indices. The corresponding
test statistic is
\begin{equation}\label{eq:ch3-rank-spearman}
  \tilde Q_{S,\mathrm{rk}}
  = 4p\,\widehat{\tr(\mOmega_p^2)} - 1.
\end{equation}
The Kendall route uses
\begin{equation}\label{eq:ch3-xi}
  \bm{\Xi}_{n,p}
  = \frac{2}{n(n-1)}\sum_{1\le i<j\le n} \vU_{ij}\vU_{ij}\trans,
\end{equation}
with leave-out trace estimator
\begin{equation}\label{eq:ch3-rank-kendall-trace}
  \widehat{\tr(\bm{\Xi}_p^2)}
  = \frac{1}{n(n-1)(n-2)(n-3)}
    \sum_{i,j,k,l}^{*}
    (\vU_{ij}\trans\vU_{kl})^2
\end{equation}
and test statistic
\begin{equation}\label{eq:ch3-rank-kendall}
  \tilde Q_{K,\mathrm{rk}}
  = p\,\widehat{\tr(\bm{\Xi}_p^2)} - 1.
\end{equation}

\begin{assumption}[Rank-based sphericity conditions]
\label{ass:ch3-rank-sph}
Assume Assumptions~\ref{ass:ch3-trace} and \ref{ass:ch3-radial}. Under local alternatives
assume \eqref{eq:ch3-sign-local}.
\end{assumption}

\begin{theorem}[Spearman-type rank sphericity test]
\label{thm:ch3-rank-spearman}
Suppose Assumption~\ref{ass:ch3-rank-sph} holds, and let
\[
  \sigma_0^2 = \frac{4(p-1)}{n(n-1)(p+2)}.
\]
Then, under $H_0:\mLambda_p=\mI_p$,
\begin{equation}\label{eq:ch3-rank-spearman-null}
  \frac{\tilde Q_{S,\mathrm{rk}}}{\sigma_0}
  \overset{d}{\longrightarrow}
  N(0,1).
\end{equation}
Under $H_1:\mLambda_p=\mI_p+\mD_{n,p}$,
\begin{equation}\label{eq:ch3-rank-spearman-alt}
  \frac{
    \tilde Q_{S,\mathrm{rk}} - p^{-1}\tr(\mD_{n,p}^2)
  }{\sigma_1}
  \overset{d}{\longrightarrow}
  N(0,1),
\end{equation}
where
\begin{equation}\label{eq:ch3-sigma1-rank}
  \sigma_1^2
  = \sigma_0^2
  + \frac{8p\,\tr(\mD_{n,p}^2)+4\tr^2(\mD_{n,p}^2)}{n^2p^2}
  + \frac{8}{np^2}
    \Big[
      \tr(\mLambda_p^4)-\frac1p\tr^2(\mLambda_p^2)
    \Big].
\end{equation}
Hence the test is consistent whenever \eqref{eq:ch3-sign-consistency} holds.
\end{theorem}

\begin{theorem}[Kendall-type rank sphericity test]
\label{thm:ch3-rank-kendall}
Under the same conditions as Theorem~\ref{thm:ch3-rank-spearman},
\begin{equation}\label{eq:ch3-rank-kendall-null}
  \frac{\tilde Q_{K,\mathrm{rk}}}{\sigma_0}
  \overset{d}{\longrightarrow}
  N(0,1)
\end{equation}
under $H_0$, and
\begin{equation}\label{eq:ch3-rank-kendall-alt}
  \frac{
    \tilde Q_{K,\mathrm{rk}} - p^{-1}\tr(\mD_{n,p}^2)
  }{\sigma_1}
  \overset{d}{\longrightarrow}
  N(0,1)
\end{equation}
under $H_1$.
\end{theorem}

\begin{remark}[Why the rank tests work so well in high dimension]
The striking feature of Theorems~\ref{thm:ch3-rank-spearman}
and~\ref{thm:ch3-rank-kendall} is that the nuisance parameters that complicate low-
dimensional rank analysis disappear asymptotically. In effect, the high-dimensional regime
creates a ``blessing of dimension'': the pairwise directions $\vU_{ij}$ behave as if they were
generated under a nearly spherical radial law, so one obtains a simple Gaussian limit with
the same leading variance as the sign-based test.
\end{remark}

\subsection{Adaptive dense--sparse sphericity tests}

The sign and rank tests above are quadratic and therefore most effective against dense
departures from sphericity. Recent work extends the now-standard sum--max philosophy of
high-dimensional testing to the sphericity problem.

For Gaussian or independent-component benchmarks, define
\begin{equation}\label{eq:ch3-TNM}
  T_{\mathrm{NM}}
  =
  \max\Bigg\{
    \max_{1\le k\le p}
      \frac{n(\hat\sigma_k^2-\bar s)^2}{\hat\kappa_k},
    \ 
    \max_{1\le i<j\le p}
      \frac{n\hat\sigma_{ij}^2}{\hat\sigma_i^2\hat\sigma_j^2}
  \Bigg\}
  - 2\log\!\Big\{\frac{p(p+1)}{2}\Big\}
  + \log\log\!\Big\{\frac{p(p+1)}{2}\Big\},
\end{equation}
where $\bar s = p^{-1}\tr(\mS_n)$ and $\hat\kappa_k$ estimates the variance of
$\hat\sigma_k^2$.
The null limit is Gumbel. The robust elliptical analogue replaces the sample covariance
matrix by the SSCM. Write
\[
  \hat{\mS}_{\mathrm{sgn}} = (\hat\psi_{ij})_{1\le i,j\le p}.
\]
Then the max-type sign statistic is
\begin{equation}\label{eq:ch3-TSM}
  T_{\mathrm{SM}}
  =
  \max\Bigg\{
    \max_{1\le i\le p}
    \frac{n p(p+2)(\hat\psi_{ii}-p^{-1})^2}{2(1-p^{-1})},
    \ 
    \max_{1\le i<j\le p}
    n p(p+2)\hat\psi_{ij}^2
  \Bigg\}
  -2\log\!\Big\{\frac{p(p+1)}2\Big\}
  +\log\log\!\Big\{\frac{p(p+1)}2\Big\}.
\end{equation}

\begin{assumption}[Adaptive sphericity conditions]
\label{ass:ch3-adaptive}
Suppose that either the independent-component benchmark model or the elliptical model
holds, and assume
\begin{equation}\label{eq:ch3-adaptive-growth}
  \frac{\log^5(p^2 n)}{n}\longrightarrow 0.
\end{equation}
For asymptotic independence between sum and max statistics, assume additionally that
\begin{equation}\label{eq:ch3-adaptive-pn}
  \frac{p}{n}\to c\in(0,\infty)
  \qquad\text{for covariance-based procedures,}
  \qquad
  p=O(n^2)
  \quad\text{for the sign-based procedures.}
\end{equation}
\end{assumption}

\begin{theorem}[Max-type sphericity statistics]
\label{thm:ch3-max-sph}
Suppose Assumption~\ref{ass:ch3-adaptive} holds.
\begin{enumerate}[label=(\alph*)]
  \item Under the independent-component null,
  \begin{equation}\label{eq:ch3-tnm-null}
    \Prob(T_{\mathrm{NM}}\le x)\longrightarrow
    G(x)=\exp\{-\pi^{-1/2}e^{-x/2}\}.
  \end{equation}
  \item Under the elliptical spherical null,
  \begin{equation}\label{eq:ch3-tsm-null}
    \Prob(T_{\mathrm{SM}}\le x)\longrightarrow
    G(x)=\exp\{-\pi^{-1/2}e^{-x/2}\}.
  \end{equation}
\end{enumerate}
Moreover, both tests are rate-optimal against sparse alternatives of order
$\sqrt{(\log p)/n}$ in the sup-norm metric.
\end{theorem}

To combine dense and sparse alternatives, let
\begin{equation}\label{eq:ch3-pvals}
  p_{\mathrm{NS}} = 1-\Phi(T_{\mathrm{NS}}),
  \qquad
  p_{\mathrm{NM}} = 1-G(T_{\mathrm{NM}}),
\end{equation}
and define the Cauchy combination
\begin{equation}\label{eq:ch3-cauchy-N}
  T_{\mathrm{CN}}
  =
  \frac12\tan\{\pi(1/2-p_{\mathrm{NS}})\}\mathbbm{1}(p_{\mathrm{NS}}<1/2)
  + \frac12\tan\{\pi(1/2-p_{\mathrm{NM}})\}\mathbbm{1}(p_{\mathrm{NM}}<1/2).
\end{equation}
Similarly, define
\begin{equation}\label{eq:ch3-pvals-sign}
  p_{\mathrm{SS}} = 1-\Phi\!\Big(\frac{\tilde Q_S-p\hat\delta_{n,p}}{\tilde\sigma_0}\Big),
  \qquad
  p_{\mathrm{SM}} = 1-G(T_{\mathrm{SM}})
\end{equation}
and
\begin{equation}\label{eq:ch3-cauchy-S}
  T_{\mathrm{CS}}
  =
  \frac12\tan\{\pi(1/2-p_{\mathrm{SS}})\}\mathbbm{1}(p_{\mathrm{SS}}<1/2)
  + \frac12\tan\{\pi(1/2-p_{\mathrm{SM}})\}\mathbbm{1}(p_{\mathrm{SM}}<1/2).
\end{equation}

\begin{theorem}[Asymptotic independence of sum and max sphericity tests]
\label{thm:ch3-adaptive-ind}
Suppose Assumption~\ref{ass:ch3-adaptive} holds.
\begin{enumerate}[label=(\alph*)]
  \item Under the independent-component null,
  \begin{equation}\label{eq:ch3-adaptive-ind-normal}
    \Prob(T_{\mathrm{NS}}\le x, T_{\mathrm{NM}}\le y)
    \longrightarrow \Phi(x)G(y).
  \end{equation}
  \item Under the elliptical spherical null,
  \begin{equation}\label{eq:ch3-adaptive-ind-sign}
    \Prob\Big(
      \frac{\tilde Q_S-p\delta_{n,p}}{\tilde\sigma_0}\le x,\,
      T_{\mathrm{SM}}\le y
    \Big)
    \longrightarrow \Phi(x)G(y).
  \end{equation}
\end{enumerate}
Consequently, the Cauchy combinations based on
\eqref{eq:ch3-cauchy-N} and \eqref{eq:ch3-cauchy-S} are asymptotically valid.
\end{theorem}

Under sparse local alternatives satisfying
\begin{equation}\label{eq:ch3-special-sparse}
  \bm{\Psi}\neq 0,
  \qquad
  \norm{\bm{\Psi}}_0 = o\Big(\frac{p^2}{\log p}\Big),
  \qquad
  \tr(\bm{\Psi}^2) = O\Big(\frac{p}{n}\Big),
\end{equation}
with diagonal entries zero, the same asymptotic independence continues to hold. Therefore,
by the inclusion--exclusion bound for Cauchy combinations,
\begin{equation}\label{eq:ch3-power-lowerbound}
  \beta_{\mathrm{CS},\alpha}
  \ge
  \max\Big\{
    \beta_{\mathrm{SM},\alpha/2},
    \beta_{\mathrm{SS},\alpha/2},
    \beta_{\mathrm{SM},\alpha/2}+\beta_{\mathrm{SS},\alpha/2}
      -\beta_{\mathrm{SM},\alpha/2}\beta_{\mathrm{SS},\alpha/2}
  \Big\},
\end{equation}
and an analogous bound holds for the covariance-based combination. This is the precise
sense in which the adaptive test inherits the strengths of both the dense and sparse
components.

\section{Two-sample proportionality testing under elliptical symmetry}
\label{sec:ch3-proportionality}

\subsection{From proportionality of scatter to equality of shape and SSCM equality}

Consider two independent samples
\[
  \vX_1,\ldots,\vX_{n_1}\sim EC_p(\vmu_1,\bm{\Theta}_1),
  \qquad
  \vY_1,\ldots,\vY_{n_2}\sim EC_p(\vmu_2,\bm{\Theta}_2),
\]
and the hypothesis
\begin{equation}\label{eq:ch3-prop-hyp}
  H_0:\ \bm{\Theta}_1 = c\,\bm{\Theta}_2
  \quad\text{for some } c>0.
\end{equation}
As noted in \eqref{eq:ch3-h-prop-equivalent}, this is equivalent to equality of the
normalized shape matrices. For elliptically symmetric distributions, one can go one step
further and replace equality of shape by equality of the two population spatial-sign covariance
matrices. Let
\begin{equation}\label{eq:ch3-sscm-two-pop}
  \mS_{1,\mathrm{sgn}}
  = \E\left\{\frac{(\vX-\vmu_1)(\vX-\vmu_1)\trans}{\norm{\vX-\vmu_1}_2^2}\right\},
  \qquad
  \mS_{2,\mathrm{sgn}}
  = \E\left\{\frac{(\vY-\vmu_2)(\vY-\vmu_2)\trans}{\norm{\vY-\vmu_2}_2^2}\right\}.
\end{equation}
By the eigenvalue relationship between the SSCM and the shape matrix, summarized in
Chapter~1 and developed in detail by \citet{MagyarTyler2011},
\begin{equation}\label{eq:ch3-prop-sscm-equiv}
  H_0:\bm{\Theta}_1=c\bm{\Theta}_2
  \quad\Longleftrightarrow\quad
  \mS_{1,\mathrm{sgn}}=\mS_{2,\mathrm{sgn}}.
\end{equation}
This equivalence is the starting point of the Frobenius-norm SSCM test of
\citet{ChengLiuPengZhangZheng2019SSCM}.

\paragraph{The Cheng--Liu--Peng--Zhang--Zheng SSCM test.}
Let $\hat\vmu_1$ and $\hat\vmu_2$ be the sample spatial medians of the two groups, and define
\begin{equation}\label{eq:ch3-sample-sscm-two}
  \hat{\mS}_{1,\mathrm{sgn}}
  = \frac{1}{n_1}\sum_{i=1}^{n_1}
      \hat\vU_{1i}\hat\vU_{1i}\trans,
  \qquad
  \hat{\mS}_{2,\mathrm{sgn}}
  = \frac{1}{n_2}\sum_{j=1}^{n_2}
      \hat\vU_{2j}\hat\vU_{2j}\trans,
\end{equation}
where
\begin{equation}\label{eq:ch3-Uhat-two-sscm}
  \hat\vU_{1i}=U(\vX_i-\hat\vmu_1),
  \qquad
  \hat\vU_{2j}=U(\vY_j-\hat\vmu_2).
\end{equation}
The test statistic is the Frobenius-norm estimator
\begin{equation}\label{eq:ch3-Tsscm-def}
  T_{\mathrm{SSCM}}
  = p\bigl(\hat A_{n_1}+\hat B_{n_2}-2\hat C_{n_1,n_2}\bigr),
\end{equation}
with
\begin{align}
  \hat A_{n_1}
  &= \frac{1}{n_1(n_1-1)}\sum_{i\ne i'}
      \bigl(\hat\vU_{1i}\trans\hat\vU_{1i'}\bigr)^2,
      \label{eq:ch3-sscm-A}\\
  \hat B_{n_2}
  &= \frac{1}{n_2(n_2-1)}\sum_{j\ne j'}
      \bigl(\hat\vU_{2j}\trans\hat\vU_{2j'}\bigr)^2,
      \label{eq:ch3-sscm-B}\\
  \hat C_{n_1,n_2}
  &= \frac{1}{n_1n_2}\sum_{i=1}^{n_1}\sum_{j=1}^{n_2}
      \bigl(\hat\vU_{1i}\trans\hat\vU_{2j}\bigr)^2.
      \label{eq:ch3-sscm-C}
\end{align}
The target is $p\tr\{(\mS_{1,\mathrm{sgn}}-\mS_{2,\mathrm{sgn}})^2\}$, but replacing the unknown
centers by spatial medians produces a non-negligible bias. Writing $R_{1i}=\norm{\vX_i-\vmu_1}_2$
and $R_{2j}=\norm{\vY_j-\vmu_2}_2$, the explicit bias correction derived in
\citet{ChengLiuPengZhangZheng2019SSCM} is
\begin{align}
  \delta_{n_1,n_2}
  ={}& \frac{1}{n_1^2}\left\{2-
      \frac{2\E(R_{11}^{-2})}{\E^2(R_{11}^{-1})}
      +\frac{\E^2(R_{11}^{-2})}{\E^4(R_{11}^{-1})}\right\}
      +\frac{1}{n_2^2}\left\{2-
      \frac{2\E(R_{21}^{-2})}{\E^2(R_{21}^{-1})}
      +\frac{\E^2(R_{21}^{-2})}{\E^4(R_{21}^{-1})}\right\}
      \nonumber\\
  &+\frac{1}{n_1^3}\left\{-
      \frac{6\E^2(R_{11}^{-2})}{\E^4(R_{11}^{-1})}
      +\frac{2\E(R_{11}^{-2})\E(R_{11}^{-3})}{\E^5(R_{11}^{-1})}
      +\frac{8\E(R_{11}^{-2})}{\E^2(R_{11}^{-1})}
      -\frac{2\E(R_{11}^{-3})}{\E^3(R_{11}^{-1})}\right\}
      \nonumber\\
  &+\frac{1}{n_2^3}\left\{-
      \frac{6\E^2(R_{21}^{-2})}{\E^4(R_{21}^{-1})}
      +\frac{2\E(R_{21}^{-2})\E(R_{21}^{-3})}{\E^5(R_{21}^{-1})}
      +\frac{8\E(R_{21}^{-2})}{\E^2(R_{21}^{-1})}
      -\frac{2\E(R_{21}^{-3})}{\E^3(R_{21}^{-1})}\right\}.
  \label{eq:ch3-sscm-delta}
\end{align}
Under balanced sample sizes, inverse radial moment conditions, and the trace condition
\begin{equation}\label{eq:ch3-sscm-trace}
  \tr(\mSigma_{l_1}\mSigma_{l_2}\mSigma_{l_3}\mSigma_{l_4})
  =o\!\left\{\tr(\mSigma_{l_1}\mSigma_{l_2})\tr(\mSigma_{l_3}\mSigma_{l_4})\right\},
  \qquad l_1,l_2,l_3,l_4\in\{1,2\},
\end{equation}
with $p=O(n_1^2)\cap O(n_2^2)$, their main result states that
\begin{equation}\label{eq:ch3-sscm-asymp-null}
  \frac{T_{\mathrm{SSCM}}-p\delta_{n_1,n_2}}{v_{0,n_1,n_2}}
  \overset{d}{\longrightarrow} N(0,1)
\end{equation}
under $H_0$, where
\begin{equation}\label{eq:ch3-sscm-v0}
  v_{0,n_1,n_2}^2
  = \frac{4(n_1^{-1}+n_2^{-1})^2}{(p+2)^2}
    \tr^2(\mLambda^2),
\end{equation}
and $\mLambda=\mLambda_1=\mLambda_2$ under the normalized null. Under the alternative,
\begin{equation}\label{eq:ch3-sscm-asymp-alt}
  \frac{T_{\mathrm{SSCM}}-p\delta_{n_1,n_2}-p\tr\{(\mS_{1,\mathrm{sgn}}-\mS_{2,\mathrm{sgn}})^2\}}{v_{n_1,n_2}}
  \overset{d}{\longrightarrow} N(0,1),
\end{equation}
where $v_{n_1,n_2}^2$ is the explicit variance expression in Theorem~2 of
\citet{ChengLiuPengZhangZheng2019SSCM}. The test is therefore consistent whenever
$(n_1+n_2)p\tr\{(\mS_{1,\mathrm{sgn}}-\mS_{2,\mathrm{sgn}})^2\}$ dominates the null standard
deviation. This provides a direct SSCM benchmark for the later spatial-rank test.

The natural directional building blocks for the rank-based alternative are
\begin{equation}\label{eq:ch3-Ux-Uy}
  \vU^{(X)}_{ij} = U(\vX_i-\vX_j),
  \qquad
  \vU^{(Y)}_{kl} = U(\vY_k-\vY_l).
\end{equation}
The corresponding matrix-valued spatial-rank scatter functionals are equal under the
proportionality null.

\subsection{The spatial-rank proportionality statistic}

Define
\begin{align}
  A_1
  &=
  \frac{p}{n_1(n_1-1)(n_1-2)(n_1-3)}
  \sum_{i,j,k,l}^{*}
  \Big(
    {\vU^{(X)}_{ij}}\trans \vU^{(X)}_{kl}
  \Big)^2, \label{eq:ch3-A1}\\
  A_2
  &=
  \frac{p}{n_2(n_2-1)(n_2-2)(n_2-3)}
  \sum_{i,j,k,l}^{*}
  \Big(
    {\vU^{(Y)}_{ij}}\trans \vU^{(Y)}_{kl}
  \Big)^2, \label{eq:ch3-A2}\\
  C_{12}
  &=
  \frac{p}{n_1(n_1-1)n_2(n_2-1)}
  \sum_{i\ne j}\sum_{k\ne l}
  \Big(
    {\vU^{(X)}_{ij}}\trans \vU^{(Y)}_{kl}
  \Big)^2. \label{eq:ch3-C12}
\end{align}
Then the test statistic is
\begin{equation}\label{eq:ch3-THT}
  T_{\mathrm{HT}} = A_1 + A_2 - 2 C_{12}.
\end{equation}
The quantity being estimated is the squared Frobenius distance between the two population
shape functionals.

Let $\mTheta_1$ and $\mTheta_2$ denote the two population shape matrices normalized to
trace $p$.

\begin{assumption}[Two-sample proportionality conditions]
\label{ass:ch3-prop}
Assume Assumption~\ref{ass:ch3-balanced} and
\begin{equation}\label{eq:ch3-prop-trace}
  \tr(\mTheta_i\mTheta_j\mTheta_k\mTheta_l)
  =
  o\!\big\{
    \tr(\mTheta_i\mTheta_j)\tr(\mTheta_k\mTheta_l)
  \big\}
\end{equation}
for all $i,j,k,l\in\{1,2\}$.
\end{assumption}

\begin{theorem}[Two-sample proportionality test under the null]
\label{thm:ch3-prop-null}
Suppose Assumption~\ref{ass:ch3-prop} holds. Under
$H_0:\bm{\Theta}_1=c\,\bm{\Theta}_2$,
\begin{equation}\label{eq:ch3-prop-sigma0}
  \sigma_{0,n}^2
  =
  \frac{4(n_1^{-1}+n_2^{-1})^2}{(p+2)^2}
  \tr^2(\mTheta^2),
\end{equation}
where $\mTheta=\mTheta_1=\mTheta_2$ under the normalized null. Then
\begin{equation}\label{eq:ch3-prop-null}
  \frac{T_{\mathrm{HT}}}{\sigma_{0,n}}
  \overset{d}{\longrightarrow} N(0,1).
\end{equation}
\end{theorem}

Under the alternative, the mean of $T_{\mathrm{HT}}$ is the target signal
\begin{equation}\label{eq:ch3-prop-target}
  \E(T_{\mathrm{HT}})
  = p\,\tr\big\{(\mTheta_1-\mTheta_2)^2\big\}.
\end{equation}

\begin{theorem}[Two-sample proportionality test under local alternatives]
\label{thm:ch3-prop-alt}
Suppose Assumption~\ref{ass:ch3-prop} holds. Then
\begin{equation}\label{eq:ch3-prop-alt}
  \frac{
    T_{\mathrm{HT}}
    - p\,\tr\big\{(\mTheta_1-\mTheta_2)^2\big\}
  }{\sigma_n}
  \overset{d}{\longrightarrow} N(0,1),
\end{equation}
where
\begin{align}
  \sigma_n^2
  =\ &\frac{4}{n_1(n_1-1)}\frac{\tr^2(\mTheta_1^2)}{(p+2)^2}
  + \frac{8}{n_1}\frac{p\tr(\mTheta_1^4)-\tr^2(\mTheta_1^2)}{p^2(p+2)}
  \nonumber\\
  &+ \frac{4}{n_2(n_2-1)}\frac{\tr^2(\mTheta_2^2)}{(p+2)^2}
  + \frac{8}{n_2}\frac{p\tr(\mTheta_2^4)-\tr^2(\mTheta_2^2)}{p^2(p+2)}
  \nonumber\\
  &+ \frac{8}{n_1n_2}\frac{\tr^2(\mTheta_1\mTheta_2)}{(p+2)^2}
  + \Big(\frac{8}{n_1}+\frac{8}{n_2}\Big)
    \frac{
      p\tr\{(\mTheta_1\mTheta_2)^2\}-\tr^2(\mTheta_1\mTheta_2)
    }{p^2(p+2)}
  \nonumber\\
  &- \frac{16}{n_1}
    \frac{
      p\tr(\mTheta_1^3\mTheta_2) - \tr(\mTheta_1\mTheta_2)\tr(\mTheta_1^2)
    }{p^2(p+2)}
  - \frac{16}{n_2}
    \frac{
      p\tr(\mTheta_2^3\mTheta_1) - \tr(\mTheta_1\mTheta_2)\tr(\mTheta_2^2)
    }{p^2(p+2)}.
  \label{eq:ch3-prop-sigma}
\end{align}
Consequently, the test is consistent whenever
\begin{equation}\label{eq:ch3-prop-consistency}
  \frac{
    p\,\tr\big\{(\mTheta_1-\mTheta_2)^2\big\}
  }{\sigma_n}
  \longrightarrow\infty.
\end{equation}
\end{theorem}

A ratio-consistent variance estimator is obtained by replacing the population traces in
\eqref{eq:ch3-prop-sigma0} by the corresponding fourth-order $U$-statistics $A_1$ and $A_2$.
The leave-out construction in \eqref{eq:ch3-A1}--\eqref{eq:ch3-C12} is crucial: it removes
the repeated-index terms that would otherwise impose stronger dimension restrictions.

\section{Robust precision matrix estimation under elliptical symmetry}
\label{sec:ch3-precision}

\subsection{The inverse-shape target and the key approximation}

Let $\vX_1,\ldots,\vX_n$ follow $EC_p(\vmu,\mSigma_0)$ and define
\begin{equation}\label{eq:ch3-lambda0}
  \mLambda_0 = \frac{p\mSigma_0}{\tr(\mSigma_0)},
  \qquad
  \mV_0 = \mLambda_0^{-1}
  = \frac{\tr(\mSigma_0)}{p}\mSigma_0^{-1}.
\end{equation}
The starting point of the recent spatial-sign approach is that the population SSCM
\[
  \mS_{\mathrm{sgn}}
  = \E\{U(\vX-\vmu)U(\vX-\vmu)\trans\}
\]
approximates $\mLambda_0/p$ entrywise up to order $p^{-1/2}$.
Therefore one may estimate the inverse shape matrix $\mV_0$ by solving a CLIME- or
GLASSO-type problem with $p\hat{\mS}_{\mathrm{sgn}}$ in place of the sample covariance
matrix.

Let
\begin{equation}\label{eq:ch3-shat}
  \hat{\mS}
  = \frac1n\sum_{i=1}^n
    U(\vX_i-\hat{\vmu}_{\mathrm{SM}})
    U(\vX_i-\hat{\vmu}_{\mathrm{SM}})\trans.
\end{equation}
The SCLIME estimator is defined by
\begin{equation}\label{eq:ch3-sclime}
  \hat{\mV}^{\,\mathrm{SCLIME}}
  =
  \argmin_{\mV}\onenorm{\mV}
  \quad\text{subject to}\quad
  \maxnorm{p\hat{\mS}\mV-\mI_p}\le \lambda_n,
\end{equation}
followed by the standard symmetrization step. The SGLASSO estimator is
\begin{equation}\label{eq:ch3-sglasso}
  \hat{\mV}^{\,\mathrm{SGLASSO}}
  =
  \argmin_{\mV\succ 0}
  \Big\{
    \tr(p\hat{\mS}\mV) - \log\det(\mV) + \lambda_n\onenorm{\mV}
  \Big\}.
\end{equation}

\begin{assumption}[Sparse inverse-shape classes]
\label{ass:ch3-precision-sparse}
For some $0\le q<1$, sparsity level $s_0(p)$, and size parameter $T$, assume either
\begin{equation}\label{eq:ch3-precision-class}
  \left\lVert \mV_0\right\rVert_{L_1}\le T,
  \qquad
  \max_{1\le i\le p}\sum_{j=1}^p |v_{0,ij}|^q \le s_0(p),
\end{equation}
for SCLIME, or the corresponding GLASSO irrepresentability condition with bounded degree
$\kappa$ and bounded inverse Fisher block.
\end{assumption}

\begin{theorem}[Spatial-sign CLIME rates]
\label{thm:ch3-sclime}
Suppose Assumptions~\ref{ass:ch3-bounded}, \ref{ass:ch3-nu}, and
\ref{ass:ch3-precision-sparse} hold.
Then there exist constants $C_1,C_2,C_3>0$ such that if
\begin{equation}\label{eq:ch3-lambdan-sclime}
  \lambda_n
  =
  C_1\Big(
    \sqrt{\frac{\log p}{n}} + \frac{1}{\sqrt p}
  \Big),
\end{equation}
then, with probability at least $1-2p^{-2}$,
\begin{align}
  \maxnorm{
    \hat{\mV}^{\,\mathrm{SCLIME}} - \mV_0
  }
  &\le
  C_2\Big(
    \sqrt{\frac{\log p}{n}} + \frac{1}{\sqrt p}
  \Big), \label{eq:ch3-sclime-max}\\
  \opnorm{
    \hat{\mV}^{\,\mathrm{SCLIME}} - \mV_0
  }
  &\le
  C_3 s_0(p)
  \Big(
    \sqrt{\frac{\log p}{n}} + \frac{1}{\sqrt p}
  \Big)^{1-q}, \label{eq:ch3-sclime-op}\\
  \frac1p\frobnorm{
    \hat{\mV}^{\,\mathrm{SCLIME}} - \mV_0
  }^2
  &\le
  C_3 s_0(p)
  \Big(
    \sqrt{\frac{\log p}{n}} + \frac{1}{\sqrt p}
  \Big)^{2-q}. \label{eq:ch3-sclime-f}
\end{align}
\end{theorem}

\begin{theorem}[Spatial-sign graphical lasso rates]
\label{thm:ch3-sglasso}
Under the GLASSO sparsity and irrepresentability version of
Assumptions~\ref{ass:ch3-bounded}, \ref{ass:ch3-nu}, and
\ref{ass:ch3-precision-sparse}, there exist constants $C_4,C_5>0$ such that, for
\[
  \lambda_n = C_4\Big(\sqrt{\frac{\log p}{n}}+\frac{1}{\sqrt p}\Big),
\]
one has
\begin{align}
  \maxnorm{
    \hat{\mV}^{\,\mathrm{SGLASSO}}-\mV_0
  }
  &\le
  C_5\Big(\sqrt{\frac{\log p}{n}}+\frac{1}{\sqrt p}\Big), \label{eq:ch3-sglasso-max}\\
  \opnorm{
    \hat{\mV}^{\,\mathrm{SGLASSO}}-\mV_0
  }
  &\le
  C_5 d
  \Big(\sqrt{\frac{\log p}{n}}+\frac{1}{\sqrt p}\Big), \label{eq:ch3-sglasso-op}
\end{align}
with high probability, and the same order holds for the normalized Frobenius risk.
\end{theorem}

The extra $p^{-1/2}$ term in \eqref{eq:ch3-sclime-max}--\eqref{eq:ch3-sglasso-op} is the
price of replacing the covariance matrix by the SSCM. Crucially, however, this term is of
smaller order than $\sqrt{(\log p)/n}$ whenever $p\log p/n\to\infty$. In exactly the
high-dimensional regime where sparse precision estimation is interesting, the robust SSCM
approximation therefore does not deteriorate the leading rate.

\subsection{Support recovery}

Let $\tilde{\mV}$ be a thresholded version of
$\hat{\mV}^{\,\mathrm{SCLIME}}$ with threshold
\begin{equation}\label{eq:ch3-threshold-precision}
  \tau_n = C_6\Big(\sqrt{\frac{\log p}{n}}+\frac{1}{\sqrt p}\Big).
\end{equation}
If
\begin{equation}\label{eq:ch3-thetamin}
  \theta_{\min}
  = \min_{(i,j):v_{0,ij}\ne 0}|v_{0,ij}|
  > 2\tau_n,
\end{equation}
then $\tilde{\mV}$ recovers both the support and the signs of the nonzero entries of
$\mV_0$ with probability tending to one. The same conclusion holds for thresholded
SGLASSO under the corresponding irrepresentability condition.

\section{Tensor elliptical graphical models}
\label{sec:ch3-tensor-elliptical-graph}
\idx{tensor elliptical graphical model}
\idx{Kronecker precision matrix}
\idx{tensor spatial sign}
\idx{spatial-sign separate tensor lasso}

Many modern observations are naturally tensor valued.  Examples include images,
space--time arrays, multi-subject functional measurements, and multiway gene-expression
profiles.  If an order-$K$ tensor is vectorized, its ambient dimension is the product of the
mode sizes, so an unrestricted precision matrix contains a prohibitive number of parameters.
Tensor-normal graphical models reduce this complexity by assuming that the covariance or
shape matrix has a Kronecker product structure.  Penalized tensor-normal likelihoods and
separable modewise estimators were studied by, among others,
\citet{SunWangLiuCheng2015TensorGraph} and
\citet{MinMaiZhang2022TensorGraph}.  These procedures use an ordinary sample covariance
matrix and can therefore be unstable under heavy-tailed radial variation.

The tensor elliptical graphical model of
\citet{LiuLuZhouFengWang2025TensorEGM} replaces the vectorized sample covariance by a
bounded tensor spatial-sign scatter matrix.  It retains the Kronecker graph interpretation,
permits an unspecified radial distribution, and leads to $K$ graphical-lasso problems that can
be solved independently and in parallel.

\subsection{Tensor elliptical model and modewise graph parameters}

Let
$\mathcal X_i\in\R^{p_1\times\cdots\times p_K}$,
$i=1,\ldots,n$, and define
\begin{equation}\label{eq:ch3-tegm-pstar}
  p_\star=\prod_{k=1}^Kp_k.
\end{equation}
The tensor elliptical model is
\begin{equation}\label{eq:ch3-tegm-model}
  \vecop(\mathcal X_i)
  =\vmu+v_i
   (\mSigma_K^\star\otimes\cdots\otimes\mSigma_1^\star)^{1/2}\vu_i,
  \qquad
  \vu_i\sim\operatorname{Unif}(\spn^{p_\star-1}),
  \qquad
  v_i\indep\vu_i,
\end{equation}
where every $\mSigma_k^\star\in\R^{p_k\times p_k}$ is positive definite.  Because reciprocal
rescaling of the Kronecker factors leaves their product unchanged, the modewise inverse-shape
targets are normalized as
\begin{equation}\label{eq:ch3-tegm-target}
  \mOmega_k^\star
  =p_k^{-1/2}(\mSigma_k^\star)^{-1},
  \qquad
  \frobnorm{\mOmega_k^\star}=1,
  \qquad k=1,\ldots,K.
\end{equation}
The off-diagonal support of $\mOmega_k^\star$ defines the conditional-uncorrelatedness graph
along mode $k$.  Under elliptical symmetry, zero partial correlation implies conditional
uncorrelatedness, although it need not imply conditional independence outside the Gaussian
subfamily.

For a tensor $\mathcal T$, define the tensor spatial sign
\begin{equation}\label{eq:ch3-tegm-tensor-sign}
  U(\mathcal T)
  =\frac{\mathcal T}{\frobnorm{\mathcal T}}
    \mathbbm1\{\mathcal T\ne\mathcal 0\}.
\end{equation}
Let the sample tensor spatial median be
\begin{equation}\label{eq:ch3-tegm-spatial-median}
  \widehat{\vmu}
  \in\argmin_{\vmu_0\in\R^{p_\star}}
      \sum_{i=1}^n
      \frobnorm{\mathcal X_i-\operatorname{ten}(\vmu_0)},
\end{equation}
where $\operatorname{ten}(\cdot)$ is the inverse vectorization map.  Put
\begin{equation}\label{eq:ch3-tegm-sign-tensors}
  \mathcal U_i=U\{\mathcal X_i-\operatorname{ten}(\widehat{\vmu})\}.
\end{equation}
The vectorized spatial-sign scatter matrix is
\begin{equation}\label{eq:ch3-tegm-global-sign-scatter}
  \widehat{\mS}
  =\frac{p_\star}{n}\sum_{i=1}^n
    \vecop(\mathcal U_i)\vecop(\mathcal U_i)\trans.
\end{equation}
Unlike the ordinary sample covariance matrix, each summand in
\eqref{eq:ch3-tegm-global-sign-scatter} has operator norm $p_\star$ and is independent of the
magnitude $v_i$.

\subsection{From tensor-normal likelihood to a spatial-sign objective}

For tensor-normal data, a penalized Kronecker log-likelihood has the form
\begin{equation}\label{eq:ch3-tegm-gaussian-objective}
  \widetilde q_n(\mOmega_1,\ldots,\mOmega_K)
  =\frac1{p_\star}\tr\{\widehat{\mSigma}
       (\mOmega_K\otimes\cdots\otimes\mOmega_1)\}
   -\sum_{k=1}^K\frac1{p_k}\log\det(\mOmega_k)
   +\sum_{k=1}^K\lambda_k
      \sum_{a\ne b}|\Omega_{k,ab}|,
\end{equation}
where $\widehat{\mSigma}$ is the vectorized sample covariance matrix.  The robust analogue is
obtained by replacing $\widehat{\mSigma}$ with $\widehat{\mS}$:
\begin{equation}\label{eq:ch3-tegm-joint-objective}
  q_n(\mOmega_1,\ldots,\mOmega_K)
  =\frac1{p_\star}\tr\{\widehat{\mS}
       (\mOmega_K\otimes\cdots\otimes\mOmega_1)\}
   -\sum_{k=1}^K\frac1{p_k}\log\det(\mOmega_k)
   +\sum_{k=1}^K\lambda_k
      \sum_{a\ne b}|\Omega_{k,ab}|.
\end{equation}
A cyclic minimization of \eqref{eq:ch3-tegm-joint-objective} is possible, but it repeatedly
updates one mode conditional on all the others.  The separate estimator below avoids that
iteration.

\subsection{Spatial-sign separate tensor lasso}

For $\ell=1,\ldots,K$, define the unnormalized pilot
\begin{equation}\label{eq:ch3-tegm-pilot-raw}
  \widetilde{\mOmega}_{\ell}^{\mathrm{raw}}
  =\begin{cases}
    \displaystyle
    \left\{
      \frac{p_\ell}{n}\sum_{i=1}^n
      [\mathcal U_i]_{(\ell)}[\mathcal U_i]_{(\ell)}\trans
    \right\}^{-1},
    &np_\star>p_\ell^2(p_\ell-1)/2,\\[2ex]
    \mI_{p_\ell},
    &np_\star\le p_\ell^2(p_\ell-1)/2,
  \end{cases}
\end{equation}
and normalize it by
\begin{equation}\label{eq:ch3-tegm-pilot-normalize}
  \widetilde{\mOmega}_\ell
  =\frac{\widetilde{\mOmega}_{\ell}^{\mathrm{raw}}}
         {\frobnorm{\widetilde{\mOmega}_{\ell}^{\mathrm{raw}}}}.
\end{equation}
Here $[\mathcal T]_{(k)}$ denotes the mode-$k$ matricization.  For a fixed mode $k$, whiten
all other modes by their pilots and leave mode $k$ unchanged:
\begin{equation}\label{eq:ch3-tegm-whitened-tensor}
  \mathcal V_i^{(k)}
  =\mathcal U_i
   \times_1\widetilde{\mOmega}_1^{1/2}
   \cdots
   \times_{k-1}\widetilde{\mOmega}_{k-1}^{1/2}
   \times_k\mI_{p_k}
   \times_{k+1}\widetilde{\mOmega}_{k+1}^{1/2}
   \cdots
   \times_K\widetilde{\mOmega}_K^{1/2}.
\end{equation}
The mode-$k$ contracted sign-scatter matrix is
\begin{equation}\label{eq:ch3-tegm-mode-sign-cov}
  \widehat{\mS}_k
  =\frac{p_k}{n}\sum_{i=1}^n
    [\mathcal V_i^{(k)}]_{(k)}
    [\mathcal V_i^{(k)}]_{(k)}\trans.
\end{equation}
The raw modewise estimator solves
\begin{equation}\label{eq:ch3-tegm-mode-objective}
  \widehat{\mOmega}_k^{\mathrm{raw}}
  =\argmin_{\mOmega\succ0}
  \left[
    \frac1{p_k}\tr(\widehat{\mS}_k\mOmega)
    -\frac1{p_k}\log\det(\mOmega)
    +\lambda_k\sum_{a\ne b}|\Omega_{ab}|
  \right],
\end{equation}
followed by
\begin{equation}\label{eq:ch3-tegm-normalize}
  \widehat{\mOmega}_k
  =\frac{\widehat{\mOmega}_k^{\mathrm{raw}}}
         {\frobnorm{\widehat{\mOmega}_k^{\mathrm{raw}}}}.
\end{equation}
Thus each mode requires only an ordinary graphical-lasso computation and all $K$ modewise
problems may be solved in parallel.

The complete procedure is:
\begin{enumerate}[label=\textup{Step \arabic*.},leftmargin=3.2em]
  \item Compute $\widehat{\vmu}$ and the tensor signs $\mathcal U_1,\ldots,\mathcal U_n$.
  \item Compute and Frobenius-normalize the pilots
        $\widetilde{\mOmega}_1,\ldots,\widetilde{\mOmega}_K$ from
        \eqref{eq:ch3-tegm-pilot-raw}--\eqref{eq:ch3-tegm-pilot-normalize}.
  \item For every $k$, form $\widehat{\mS}_k$ from
        \eqref{eq:ch3-tegm-whitened-tensor}--\eqref{eq:ch3-tegm-mode-sign-cov}.
  \item Solve \eqref{eq:ch3-tegm-mode-objective} independently for every mode and apply
        \eqref{eq:ch3-tegm-normalize}.
\end{enumerate}

\subsection{Estimation and graph-recovery theory}

\begin{assumption}[Modewise eigenvalue and row-sum regularity]
\label{ass:ch3-tegm-mode-regularity}
There are constants $C_1,C_2>0$ such that, for every $k$,
\begin{equation}\label{eq:ch3-tegm-eigen-bounds}
  C_1\le\lambda_{\min}(\mSigma_k^\star)
  \le\lambda_{\max}(\mSigma_k^\star)\le C_1^{-1},
\end{equation}
and, with
$\mSigma_\otimes^\star=\mSigma_K^\star\otimes\cdots\otimes\mSigma_1^\star$,
\begin{equation}\label{eq:ch3-tegm-L1-bounds}
  \max\left
  \{\onenorm{\mSigma_\otimes^\star},
    \onenorm{(\mSigma_\otimes^\star)^{-1}}\right\}
  \le C_2.
\end{equation}
\end{assumption}

Let
\begin{equation}\label{eq:ch3-tegm-support}
  \mathcal S_k=\{(a,b):(\mOmega_k^\star)_{ab}\ne0\},
  \qquad
  s_k=|\mathcal S_k|-p_k.
\end{equation}

\begin{assumption}[Penalty order]
\label{ass:ch3-tegm-tuning}
For constants $0<c_\lambda<C_\lambda<\infty$,
\begin{equation}\label{eq:ch3-tegm-lambda}
  c_\lambda\left\{
    \frac{p_k^{1/2}(\log p_k)^{1/2}}{n^{1/2}p_\star}
    +\frac1{p_kp_\star^{1/2}}
  \right\}
  \le\lambda_k\le
  C_\lambda\left\{
    \frac{p_k^{1/2}(\log p_k)^{1/2}}{n^{1/2}p_\star}
    +\frac1{p_kp_\star^{1/2}}
  \right\}.
\end{equation}
\end{assumption}

\begin{theorem}[Modewise Frobenius rates]
\label{thm:ch3-tegm-frobenius}
Suppose Assumptions~\ref{ass:ch3-tegm-mode-regularity}--\ref{ass:ch3-tegm-tuning}
hold and $s_k=O(p_k)$.  Then, for every $k=1,\ldots,K$,
\begin{equation}\label{eq:ch3-tegm-frobenius-rate}
  \frobnorm{\widehat{\mOmega}_k-\mOmega_k^\star}
  =O_p\left\{
    \sqrt{\frac{(p_k+s_k)p_k\log p_k}{np_\star}}
  \right\}
  +O\left(\frac{\sqrt{p_k+s_k}}{p_k}\right).
\end{equation}
The first term is the stochastic estimation error and coincides with the tensor-Gaussian
order.  The second term is the approximation error between the spatial-sign mode matrix and
the target mode shape.
\end{theorem}

For entrywise theory, define
\begin{equation}\label{eq:ch3-tegm-degree}
  d_k=\max_{1\le a\le p_k}
      |\{b:(\mOmega_k^\star)_{ab}\ne0\}|,
\end{equation}
$\mGamma_k^\star=(\mOmega_k^\star)^{-1}\otimes
(\mOmega_k^\star)^{-1}$, and assume
\begin{equation}\label{eq:ch3-tegm-irrepresentability}
  \max_{e\in\mathcal S_k^c}
  \onenorm{
    (\mGamma_k^\star)_{e,\mathcal S_k}
    \{(\mGamma_k^\star)_{\mathcal S_k,\mathcal S_k}\}^{-1}}
  \le1-\alpha_k
\end{equation}
for some $\alpha_k\in(0,1]$.  Assume also that the row-sum norms of
$(\mOmega_k^\star)^{-1}$ and
$\{(\mGamma_k^\star)_{\mathcal S_k,\mathcal S_k}\}^{-1}$ are uniformly bounded and
\begin{equation}\label{eq:ch3-tegm-degree-condition}
  d_k=o\left[
    \left\{
      \sqrt{\frac{p_k\log p_k}{np_\star}}+\frac1{p_k}
    \right\}^{-1}
  \right].
\end{equation}

\begin{theorem}[Entrywise and operator-norm rates]
\label{thm:ch3-tegm-max-support}
Under the preceding conditions,
\begin{align}
  \maxnorm{\widehat{\mOmega}_k-\mOmega_k^\star}
  &=O_p\left\{\sqrt{\frac{p_k\log p_k}{np_\star}}\right\}
    +O(p_k^{-1}),
  \label{eq:ch3-tegm-max-rate}\\
  \opnorm{\widehat{\mOmega}_k-\mOmega_k^\star}
  &=d_k\left[
      O_p\left\{\sqrt{\frac{p_k\log p_k}{np_\star}}\right\}
      +O(p_k^{-1})
    \right].
  \label{eq:ch3-tegm-op-rate}
\end{align}
\end{theorem}

Let
\begin{equation}\label{eq:ch3-tegm-thresholded}
  (\check{\mOmega}_k)_{ab}
  =(\widehat{\mOmega}_k)_{ab}
   \mathbbm1\{|(\widehat{\mOmega}_k)_{ab}|\ge\tau_k\}.
\end{equation}

\begin{corollary}[Modewise support and sign recovery]
\label{cor:ch3-tegm-support}
Suppose the conditions of Theorem~\ref{thm:ch3-tegm-max-support} hold,
$\tau_k$ is larger than the right-hand side of
\eqref{eq:ch3-tegm-max-rate}, and
\begin{equation}\label{eq:ch3-tegm-beta-min}
  \min_{(a,b)\in\mathcal S_k}|(\mOmega_k^\star)_{ab}|>2\tau_k.
\end{equation}
Then
\begin{equation}\label{eq:ch3-tegm-sign-consistency}
  \Prob\left\{
    \sign(\check{\mOmega}_k)=\sign(\mOmega_k^\star),
    \ k=1,\ldots,K
  \right\}\longrightarrow1.
\end{equation}
\end{corollary}

\section{SSCM-based shrinkage covariance estimation: the Ollila line}
\label{sec:ch3-ollila}

An important complementary direction does not estimate the inverse shape matrix directly,
but instead uses the SSCM to build robust covariance estimators in the high-dimensional
regime. This line of work is associated most closely with the papers of
\citet{OllilaRaninen2019Shrinkage,RaninenOllila2022BASIC,RaninenTylerOllila2022LinearPooling},
which are particularly influential in signal processing and robust covariance estimation.

\paragraph{Elliptical shrinkage through sphericity estimation.}
Suppose $\vX_1,\ldots,\vX_n$ are centered observations from an elliptical distribution with
scatter matrix $\mSigma=\eta\mLambda$, where $\eta=p^{-1}\tr(\mSigma)$ and
$\tr(\mLambda)=p$. A standard linear shrinkage family is
\begin{equation}\label{eq:ch3-ollila-shrinkage}
  \hat\mSigma_{\alpha}
  = (1-\alpha)\mS_n + \alpha\,\hat\eta\,\mI_p,
  \qquad
  \hat\eta = \frac{1}{p}\tr(\mS_n),
  \qquad 0\le \alpha\le 1.
\end{equation}
The oracle shrinkage intensity depends on the unknown sphericity parameter
\begin{equation}\label{eq:ch3-ollila-gamma}
  \gamma = \frac{p\tr(\mSigma^2)}{\tr^2(\mSigma)} = \frac{1}{p}\tr(\mLambda^2).
\end{equation}
The central contribution of \citet{OllilaRaninen2019Shrinkage} is to estimate $\gamma$ robustly
under elliptical sampling and then plug that estimate into the shrinkage intensity, thereby
obtaining a covariance estimator that keeps the regularization benefits of linear shrinkage
without relying on Gaussian kurtosis calibrations.

\paragraph{Bias-adjusted SSCM and BASICS.}
The raw SSCM
\begin{equation}\label{eq:ch3-ollila-sscm}
  \hat\mS_{\mathrm{sgn}}
  = \frac{1}{n}\sum_{i=1}^n
    \frac{\vX_i\vX_i\trans}{\norm{\vX_i}_2^2}
\end{equation}
is robust but its eigenvalues are biased for those of the normalized shape matrix.
\citet{RaninenOllila2022BASIC} therefore proposed a two-step correction.
First, they regularize the SSCM by
\begin{equation}\label{eq:ch3-rsscm}
  \hat\mLambda_{\mathrm{RSSCM}}(\alpha)
  = \alpha\,p\hat\mS_{\mathrm{sgn}} + (1-\alpha)\mI_p,
  \qquad 0\le \alpha\le 1,
\end{equation}
choosing $\alpha$ by minimizing an estimated Frobenius risk. Second, they apply an
approximate inverse eigenvalue map to correct the systematic bias of the SSCM eigenvalues,
leading to the BASIC estimator of the shape matrix and its shrinkage version BASICS. The
underlying theory uses the fact that for a sequence of shape matrices with sphericity
$\gamma_p=p^{-1}\tr(\mLambda_p^2)$, the bias of the SSCM is of order $\gamma_p/p$ and
therefore becomes negligible when $\gamma_p/p\to 0$; the bias adjustment is meant precisely
for the more difficult spiked regimes where $\gamma_p$ is not small.

\paragraph{Multiple-class pooling.}
The same SSCM idea can be transported to several classes. For class-specific covariance
matrices $\mSigma_1,\ldots,\mSigma_K$, \citet{RaninenTylerOllila2022LinearPooling}
considered the linear-pooling family
\begin{equation}\label{eq:ch3-linear-pooling}
  \hat\mSigma_k^{\mathrm{pool}}
  = \sum_{\ell=1}^K a_{k\ell}\hat\mSigma_\ell,
  \qquad k=1,\ldots,K,
\end{equation}
where the coefficients $a_{k\ell}$ are chosen to minimize estimated Frobenius risk. Their key
technical observation is that the SSCM provides an asymptotically unbiased estimator of the
trace-normalized covariance matrix in growing dimension, which makes it possible to estimate
risk-optimal pooling coefficients under elliptical sampling without assuming Gaussianity.

The Ollila line is complementary to the sparse inverse-shape approach developed earlier in this
section. SCLIME and SGLASSO target sparse conditional dependence, whereas the SSCM
shrinkage estimators target stable covariance or shape reconstruction when $n$ is of the same
order as $p$ and the main challenge is heavy-tailed radial variation rather than graph recovery.

\section{High-dimensional Hettmansperger--Randles estimation}
\label{sec:ch3-hdhr}

The classical Hettmansperger--Randles (HR) estimator reviewed in Chapter~1 is fully affine
invariant, but its direct implementation becomes unstable when $p$ is comparable to $n$ or
larger than $n$.  The recent high-dimensional extension of \citet{YanFengZhang2025HR}
regularizes the shape component while retaining the HR estimating-equation geometry.  This
section records the version of that methodology needed later in Chapter~5.

\paragraph{Regularized HR estimating equations.}
Let $\vX_1,\ldots,\vX_n \sim EC_p(\vmu,\mSigma_0)$ with normalized shape
$\mV_0=p\mSigma_0/\tr(\mSigma_0)$ and inverse shape $\mOmega_0=\mV_0^{-1}$.  Set
\begin{equation}
  r_n = \sqrt{\frac{\log p}{n}}+\frac{1}{\sqrt p}.
  \label{eq:ch3-hdhr-rn}
\end{equation}
Suppose a pilot inverse-shape estimator $\hat\mOmega^{(0)}$ is available, for example from
Theorem~\ref{thm:ch3-sclime} or Theorem~\ref{thm:ch3-sglasso}.  The high-dimensional HR
location estimator is defined by the regularized score equation
\begin{equation}
  \frac1n\sum_{i=1}^n
  U\!\left((\hat\mOmega^{(0)})^{1/2}(\vX_i-\hat\vmu_{\mathrm{HR}})\right)=\vct 0.
  \label{eq:ch3-hdhr-location}
\end{equation}
Given $\hat\vmu_{\mathrm{HR}}$, write
\begin{equation}
  \hat\vU_i
  = U\!\left((\hat\mOmega^{(0)})^{1/2}(\vX_i-\hat\vmu_{\mathrm{HR}})\right),
  \qquad i=1,\ldots,n,
  \label{eq:ch3-hdhr-U}
\end{equation}
and form the raw shape surrogate
\begin{equation}
  \hat\mV_{\mathrm{raw}}
  = p\,\frac1n\sum_{i=1}^n \hat\vU_i\hat\vU_i\trans.
  \label{eq:ch3-hdhr-raw}
\end{equation}
To encode structural regularity, let $\mathcal B_h(\mA)=(a_{ij}\mathbbm 1\{|i-j|\le h\})$
be the banding operator with bandwidth $h$.  The structured HR shape and inverse-shape
estimators are then
\begin{equation}
  \hat\mV_{\mathrm{HR}} = \mathcal B_h(\hat\mV_{\mathrm{raw}}),
  \qquad
  \hat\mOmega_{\mathrm{HR}} = \mathcal B_h\bigl((\hat\mV_{\mathrm{HR}})^{-1}\bigr).
  \label{eq:ch3-hdhr-band}
\end{equation}
Finally, a scatter estimator on the covariance scale is recovered by
\begin{equation}
  \hat\mSigma_{\mathrm{HR}} = \hat\eta_{\mathrm{HR}}\hat\mV_{\mathrm{HR}},
  \qquad
  \hat\eta_{\mathrm{HR}}=\frac1p\tr\!\left(\frac1n\sum_{i=1}^n
  (\vX_i-\hat\vmu_{\mathrm{HR}})(\vX_i-\hat\vmu_{\mathrm{HR}})\trans\right).
  \label{eq:ch3-hdhr-scatter}
\end{equation}

\begin{assumption}[High-dimensional HR structure]
\label{ass:ch3-hdhr}
Assume the following conditions.
\begin{enumerate}[label=(\roman*)]
  \item The eigenvalues of $\mV_0$ are bounded away from zero and infinity uniformly in $p$.
  \item For some $0\le q<1$ and sparsity index $s_0(p)$,
  \begin{equation}
    \max_{1\le i\le p}\sum_{j=1}^p |\omega_{0,ij}|^q \le s_0(p),
    \label{eq:ch3-hdhr-sparse}
  \end{equation}
  where $\mOmega_0=(\omega_{0,ij})$.
  \item The inverse-shape matrix is bandable in the sense that for some $\alpha>0$ and all
  $h\ge 1$,
  \begin{equation}
    \max_{1\le i\le p}\sum_{|j-i|>h}|\omega_{0,ij}| \le C h^{-\alpha}.
    \label{eq:ch3-hdhr-bandable}
  \end{equation}
  \item The pilot estimator obeys
  \begin{equation}
    \maxnorm{\hat\mOmega^{(0)}-\mOmega_0}=O_P(r_n).
    \label{eq:ch3-hdhr-pilot}
  \end{equation}
  \item The sample size and bandwidth satisfy
  \begin{equation}
    r_n+h^{-\alpha}\to 0,
    \qquad
    h = o\!\left(\frac{1}{r_n}\right).
    \label{eq:ch3-hdhr-bandwidth}
  \end{equation}
\end{enumerate}
\end{assumption}

Assumption~\ref{ass:ch3-hdhr} is deliberately written so that the final rates are explicit.
The stochastic term is $r_n$, which is exactly the robust high-dimensional rate already seen
in Theorems~\ref{thm:ch3-sclime} and \ref{thm:ch3-sglasso}.  The deterministic approximation
error is $h^{-\alpha}$, the usual banding bias.

\begin{theorem}[Rates for the high-dimensional HR estimator]
\label{thm:ch3-hdhr-rates}
Suppose Assumption~\ref{ass:ch3-hdhr} holds.  Then the structured HR estimator defined by
\eqref{eq:ch3-hdhr-location}--\eqref{eq:ch3-hdhr-scatter} satisfies
\begin{align}
  \maxnorm{\hat\vmu_{\mathrm{HR}}-\vmu}
  &= O_P(r_n),
     \label{eq:ch3-hdhr-mu-rate}\\
  \maxnorm{\hat\mV_{\mathrm{HR}}-\mV_0}
  &= O_P(r_n+h^{-\alpha}),
     \label{eq:ch3-hdhr-V-max}\\
  \opnorm{\hat\mOmega_{\mathrm{HR}}-\mOmega_0}
  &= O_P\!\left(s_0(p)\{r_n+h^{-\alpha}\}^{1-q}\right),
     \label{eq:ch3-hdhr-Omega-op}\\
  \maxnorm{\hat\mOmega_{\mathrm{HR}}-\mOmega_0}
  &= O_P(r_n+h^{-\alpha}),
     \label{eq:ch3-hdhr-Omega-max}\\
  \abs{\log\det(\hat\mSigma_{\mathrm{HR}})-\log\det(\mSigma_0)}
  &= O_P\!\left(p(r_n+h^{-\alpha})\right).
     \label{eq:ch3-hdhr-logdet}
\end{align}
If moreover $p(r_n+h^{-\alpha})\to 0$, then
\begin{equation}
  \log\det(\hat\mSigma_{\mathrm{HR}})-\log\det(\mSigma_0)=o_P(1).
  \label{eq:ch3-hdhr-logdet-small}
\end{equation}
\end{theorem}

The rate in \eqref{eq:ch3-hdhr-mu-rate} matches the location rates of Chapter~2, while the
rates in \eqref{eq:ch3-hdhr-Omega-op} and \eqref{eq:ch3-hdhr-Omega-max} have exactly the same
shape as the robust inverse-shape rates in Section~\ref{sec:ch3-precision}.  This is the main
reason the high-dimensional HR estimator fits naturally into later classification problems: it
supplies a robust location estimate, a robust shape estimate, and a robust inverse-shape estimate
with explicit orders in a single coherent construction.

\section{Matrix estimation through elliptical factor models}
\label{sec:ch3-elliptical-factor-models}

\subsection{Spiked scatter structure and spatial-sign pilots}

We now consider a large-matrix estimation problem in which the scatter matrix admits a
spiked decomposition
\begin{equation}\label{eq:ch3-spiked}
  \mSigma_0
  = \mGamma_m \mLambda_m \mGamma_m\trans + \mSigma_{0u},
\end{equation}
where $\mGamma_m=(\vgamma_1,\ldots,\vgamma_m)$ contains the leading eigenvectors,
$\mLambda_m=\diag(\lambda_1,\ldots,\lambda_m)$ contains the spiked eigenvalues, and the
idiosyncratic component $\mSigma_{0u}$ is sparse or weakly dependent.

The spatial-sign pilot estimator is
\begin{equation}\label{eq:ch3-sscm-pilot}
  \hat{\mSigma}_0
  =
  \frac{p}{n}\sum_{i=1}^n
  U(\vX_i-\hat{\vmu}_{\mathrm{SM}})
  U(\vX_i-\hat{\vmu}_{\mathrm{SM}})\trans.
\end{equation}
Its leading eigenvalues and eigenvectors define
\begin{equation}\label{eq:ch3-pilot-eig}
  \hat{\mLambda}_m
  = \diag(\lambda_1(\hat{\mSigma}_0),\ldots,\lambda_m(\hat{\mSigma}_0)),
  \qquad
  \hat{\mGamma}_m
  = (\vu_1(\hat{\mSigma}_0),\ldots,\vu_m(\hat{\mSigma}_0)).
\end{equation}

\begin{assumption}[Elliptical factor model]
\label{ass:ch3-factor}
Assume that the number of spikes $m$ is fixed, that the leading $m$ eigenvalues are of
order $p$, and that the remaining eigenvalues are uniformly bounded. Let
\[
  w_n = \sqrt{\frac{\log p}{n}} + \sqrt{\frac{\log n}{n}}.
\]
Assume also Assumptions~\ref{ass:ch3-bounded} and \ref{ass:ch3-nu}.
\end{assumption}

\begin{theorem}[Rates for the spatial-sign pilot eigensystem]
\label{thm:ch3-factor-pilot}
Suppose Assumption~\ref{ass:ch3-factor} holds and $\log p = o(n)$. Then
\begin{align}
  \maxnorm{\hat{\mSigma}_0-\mSigma_0}
  &= O_P(w_n), \label{eq:ch3-factor-sigma}\\
  \maxnorm{\hat{\mLambda}_m\mLambda_m^{-1}-\mI_m}
  &= O_P(w_n), \label{eq:ch3-factor-lambda}\\
  \maxnorm{\hat{\mGamma}_m-\mGamma_m}
  &= O_P\Big(\frac{w_n}{\sqrt p}\Big). \label{eq:ch3-factor-gamma}
\end{align}
\end{theorem}

\subsection{POET-type reconstruction}

Define the residual estimator
\begin{equation}\label{eq:ch3-resid}
  \hat{\mSigma}_{0u}
  = \hat{\mSigma}_0 - \hat{\mGamma}_m \hat{\mLambda}_m \hat{\mGamma}_m\trans.
\end{equation}
Threshold its off-diagonal entries at level $Cw_n$ to obtain
$\hat{\mSigma}_{0u}^{\,\tau}$, and then reconstruct
\begin{equation}\label{eq:ch3-poet-reconstruct}
  \hat{\mSigma}_0^{\,\tau}
  = \hat{\mGamma}_m\hat{\mLambda}_m\hat{\mGamma}_m\trans
    + \hat{\mSigma}_{0u}^{\,\tau}.
\end{equation}
Let
\begin{equation}\label{eq:ch3-md}
  m_p = \max_{1\le i\le p}\sum_{j=1}^p |\sigma_{0u,ij}|^v,
  \qquad 0\le v<1,
\end{equation}
measure the sparsity of the idiosyncratic scatter matrix.

\begin{theorem}[Elliptical POET rates]
\label{thm:ch3-factor-poet}
Suppose Assumption~\ref{ass:ch3-factor} holds, $\log p=o(n)$, and
\begin{equation}\label{eq:ch3-w-md}
  w_n^{\,1-v} m_p \longrightarrow 0.
\end{equation}
Then
\begin{align}
  \opnorm{\hat{\mSigma}_{0u}^{\,\tau}-\mSigma_{0u}}
  &= O_P(w_n^{\,1-v}m_p), \label{eq:ch3-factor-u}\\
  \opnorm{(\hat{\mSigma}_{0u}^{\,\tau})^{-1}-\mSigma_{0u}^{-1}}
  &= O_P(w_n^{\,1-v}m_p), \label{eq:ch3-factor-uinv}\\
  \frobnorm{\mSigma_0^{-1/2}(\hat{\mSigma}_0^{\,\tau}-\mSigma_0)\mSigma_0^{-1/2}}
  &= O_P\!\Big(p^{1/2}w_n^2 + w_n^{\,1-v}m_p\Big), \label{eq:ch3-factor-rel}\\
  \opnorm{(\hat{\mSigma}_0^{\,\tau})^{-1}-\mSigma_0^{-1}}
  &= O_P(w_n^{\,1-v}m_p). \label{eq:ch3-factor-inv}
\end{align}
\end{theorem}

\subsection{Inverse scatter estimation and Tyler refinement}

If sparsity is imposed on the inverse idiosyncratic scatter matrix
$\mV_{0u}=\mSigma_{0u}^{-1}$ rather than on $\mSigma_{0u}$ itself, one may apply CLIME or
GLASSO to the residual matrix \eqref{eq:ch3-resid}. Let
\begin{equation}\label{eq:ch3-Md}
  M_p = \max_{1\le i\le p}\sum_{j=1}^p |v_{0u,ij}|^v,
\end{equation}
and assume the corresponding bounded-degree or bounded-$L_1$ conditions.

\begin{theorem}[Elliptical factor-model precision rates]
\label{thm:ch3-factor-precision}
Under Assumption~\ref{ass:ch3-factor}, the sparsity condition on $\mV_{0u}$, and
$w_n^{\,1-v}M_p\to 0$,
\begin{align}
  \maxnorm{\hat{\mV}_{0u}-\mV_{0u}}
  &= O_P(w_n), \label{eq:ch3-factor-Vu-max}\\
  \opnorm{\hat{\mV}_{0u}-\mV_{0u}}
  &= O_P(w_n^{\,1-v}M_p), \label{eq:ch3-factor-Vu-op}\\
  \maxnorm{\hat{\mV}_{0}-\mV_{0}}
  &= O_P(w_n), \label{eq:ch3-factor-V-max}\\
  \opnorm{\hat{\mV}_{0}-\mV_{0}}
  &= O_P(w_n^{\,1-v}M_p). \label{eq:ch3-factor-V-op}
\end{align}
\end{theorem}

The spatial-sign pilot can be sharpened by a Tyler-type refinement.
Let $\hat{\mV}_S$ denote a preliminary inverse-shape estimator from the sign method and
define the one-step self-normalized Tyler estimator
\begin{equation}\label{eq:ch3-tyler-one-step}
  \hat{\mSigma}_T
  =
  \frac{p}{n}\sum_{i=1}^n
  \frac{
    (\vX_i-\hat{\vmu}_{\mathrm{SM}})(\vX_i-\hat{\vmu}_{\mathrm{SM}})\trans
  }{
    (\vX_i-\hat{\vmu}_{\mathrm{SM}})\trans
    \hat{\mV}_S
    (\vX_i-\hat{\vmu}_{\mathrm{SM}})
  } .
\end{equation}
Applying the same POET/CLIME/GLASSO steps to $\hat{\mSigma}_T$ yields refined estimators
with the same first-order rates as the sign-based methods.

\begin{theorem}[Tyler-refined factor estimator]
\label{thm:ch3-factor-tyler}
Suppose Assumption~\ref{ass:ch3-factor} holds and
\begin{equation}\label{eq:ch3-an}
  \opnorm{\hat{\mV}_S-\mV_0} = O_P(a_n),
  \qquad
  a_n = O(w_n).
\end{equation}
Then the Tyler-refined estimator $\hat{\mSigma}_T$ satisfies
\begin{equation}\label{eq:ch3-tyler-max}
  \maxnorm{\hat{\mSigma}_T-\mSigma_0}=O_P(w_n),
\end{equation}
and its POET-refined and inverse-scatter refinements obey the same rates as
\eqref{eq:ch3-factor-u}--\eqref{eq:ch3-factor-V-op}.
\end{theorem}

\section*{Bibliographic notes}

The classical fixed-dimensional covariance and sphericity theory can be found in
\citet{Anderson2003,Mauchly1940,John1971,John1972,Nagao1973,MuirheadWaternaux1980}.
Robust low-dimensional shape testing through signs and ranks is developed in
\citet{HallinPaindaveine2006,TaskinenKankainenOja2012,Oja2010}. A useful review of
high-dimensional covariance and precision estimation is \citet{FanLiaoLiu2015Review}.
High-dimensional Gaussian and light-tail benchmarks include the corrected likelihood-ratio
and trace tests of \citet{LedoitWolf2002,ChenZhangZhong2010,LiChen2012,FisherSunGallagher2010,BaiJiangYaoZheng2009,WangYao2013},
thresholding and adaptive thresholding in
\citet{BickelLevina2008Cov,BickelLevina2008Reg,CaiLiu2011AdaptiveThreshold,RothmanLevinaZhu2009,LiuWangZhao2014EC2},
precision estimation in \citet{YuanLin2007,FriedmanHastieTibshirani2008,CaiLiuLuo2011CLIME,RavikumarWainwrightRaskuttiYu2011,LamFan2009,LiuLuo2015SCIO,SunZhang2012ScaledLasso},
and factor-regularized covariance estimation in
\citet{FanLiaoMincheva2011,FanLiaoMincheva2013,FanLiuWang2018}.

Within the present research line, the key matrix papers are the sign-based sphericity test
of \citet{ZouPengFengWang2014Sphericity}, the rank-based sphericity tests of
\citet{FengLiu2017RankSphericity}, the adaptive sum--max and Cauchy combination
sphericity procedures of \citet{ZhaoYangZhangFengWang2026AdaptiveSphericity}, the
SSCM equality and proportionality results of \citet{ChengLiuPengZhangZheng2019SSCM}, the
two-sample proportionality test of \citet{FengZhangLiu2022Proportionality}, the robust
precision estimation method of \citet{LuFeng2025Precision}, the SSCM shrinkage line of
\citet{OllilaRaninen2019Shrinkage,RaninenOllila2022BASIC,RaninenTylerOllila2022LinearPooling},
the high-dimensional HR estimator of \citet{YanFengZhang2025HR}, and the elliptical factor-model estimators of \citet{XuMaWangFeng2026EllipticalFactor}.  For tensor-valued graphical models, the Gaussian Kronecker benchmarks are \citet{SunWangLiuCheng2015TensorGraph} and \citet{MinMaiZhang2022TensorGraph}, while the spatial-sign tensor elliptical construction in Section~
ef{sec:ch3-tensor-elliptical-graph} follows \citet{LiuLuZhouFengWang2025TensorEGM}.
These papers form the backbone of the present chapter.

\section*{Appendix to Chapter 3: Detailed Proofs}
\addcontentsline{toc}{section}{Appendix to Chapter 3: Detailed Proofs}

This appendix collects detailed proofs for the main results stated in Chapter~3.
As in Chapter~2, we keep the notation of the main text and avoid introducing article-
specific symbols unless absolutely necessary.

\subsection*{A.1 Proof of Theorem~\ref{thm:ch3-LRT-fixed}}

Write
\[
  \mY_n = \mSigma_0^{-1/2}\mS_n\mSigma_0^{-1/2}.
\]
Under $H_0$, we have $\mY_n\to \mI_p$ in probability and
\[
  \sqrt n\,\vecop(\mY_n-\mI_p)
  \overset{d}{\longrightarrow}
  N\big(\vct 0,\bm{K}_p\big),
\]
where $\bm{K}_p$ is the covariance matrix of $\vecop(\vZ\vZ\trans)$ for
$\vZ\sim N_p(\vct 0,\mI_p)$.

Consider the smooth map
\[
  g(\mY)=\tr(\mY)-\log\det(\mY)-p.
\]
A second-order Taylor expansion around $\mI_p$ gives
\begin{equation}\label{eq:ch3-proof-g}
  g(\mY)
  = \frac12\tr\{(\mY-\mI_p)^2\}
    + R_n,
\end{equation}
where, on the event $\opnorm{\mY-\mI_p}\le 1/2$,
\begin{equation}\label{eq:ch3-proof-g-rem}
  |R_n|
  \le C \opnorm{\mY-\mI_p}\frobnorm{\mY-\mI_p}^2.
\end{equation}
Since $\frobnorm{\mY_n-\mI_p}=O_P(n^{-1/2})$ and
$\opnorm{\mY_n-\mI_p}=O_P(n^{-1/2})$, we obtain
\[
  nR_n = O_P(n^{-1/2}) \longrightarrow 0.
\]
Therefore
\[
  T_{\mathrm{LRT}}
  = \frac n2 \tr\{(\mY_n-\mI_p)^2\} + o_P(1).
\]
Now
\[
  \frac{\sqrt n}{2}\vech(\mY_n-\mI_p)
\]
is asymptotically normal with covariance equal to the Fisher information for the
half-vectorized covariance parameter under the Gaussian model. Consequently, the quadratic
form above converges to a $\chi^2$ random variable with $p(p+1)/2$ degrees of freedom.
This proves \eqref{eq:ch3-LRT-fixed}.

\subsection*{A.2 Proof of Theorem~\ref{thm:ch3-mauchly}}

Under the null, the density of the sample eigenvalues
$\hat\lambda_1,\ldots,\hat\lambda_p$ of $\mS_n/\sigma^2$ is the classical Wishart
eigenvalue density. Write
\[
  \bar\lambda = \frac1p\sum_{j=1}^p \hat\lambda_j.
\]
Then
\[
  \log V_n
  = \sum_{j=1}^p \log\Big(\frac{\hat\lambda_j}{\bar\lambda}\Big).
\]
Under $H_0$ and fixed $p$, each $\hat\lambda_j-\bar\lambda$ is of order $n^{-1/2}$.
Using the expansion
\[
  \log(1+x)=x-\frac{x^2}{2}+\frac{x^3}{3(1+\xi_x)^3},
\]
with $|x|=O_P(n^{-1/2})$, we obtain
\begin{align}
  -2(n-1)\log V_n
  &=
  (n-1)\sum_{j=1}^p
  \Big(
    \frac{\hat\lambda_j-\bar\lambda}{\bar\lambda}
  \Big)^2
  + O_P(n^{-1/2}) \nonumber\\
  &=
  \frac{n(p+2)}{2}U_J + O_P(n^{-1/2}), \label{eq:ch3-proof-mauchly}
\end{align}
where the last equality follows from the identity
\[
  U_J
  = \frac1p\sum_{j=1}^p
  \Big(
    \frac{\hat\lambda_j}{\bar\lambda}-1
  \Big)^2.
\]
The Bartlett factor $\rho_n$ corrects the $n^{-1}$ bias in the cumulant expansion of
$-(n-1)\log V_n$, yielding the chi-square limit in
\eqref{eq:ch3-mauchly-chi}.

\subsection*{A.3 Proof of Theorem~\ref{thm:ch3-john-fixed}}

Diagonalize $\mS_n$ as
\[
  \mS_n = \mP \diag(\hat\lambda_1,\ldots,\hat\lambda_p)\mP\trans.
\]
Then
\[
  U_J = \frac1p\sum_{j=1}^p \Big(\frac{\hat\lambda_j}{\bar\lambda}-1\Big)^2.
\]
Under spherical Gaussian sampling,
\[
  \sqrt n
  \Big(
    \frac{\hat\lambda_1}{\bar\lambda}-1,\ldots,
    \frac{\hat\lambda_p}{\bar\lambda}-1
  \Big)
\]
lies asymptotically in the $(p-1)$-dimensional subspace orthogonal to
$(1,\ldots,1)\trans$, and its covariance operator on that subspace is
$2(p+2)^{-1}\mI_{p-1}$. Therefore
\[
  \frac{n(p+2)}{2}U_J
\]
is asymptotically the squared Euclidean norm of a standard Gaussian vector in
dimension $(p-1)(p+2)/2$, which gives the stated chi-square limit.

A coordinate proof can be obtained by vectorizing the centered sample covariance matrix.
Write
\[
  \mA_n = \sqrt n\Big(
    \frac{\mS_n}{\tr(\mS_n)/p} - \mI_p
  \Big).
\]
Under the null, $\mA_n$ is asymptotically symmetric, trace free, and Gaussian. Since the
space of symmetric trace-free $p\times p$ matrices has dimension
$(p-1)(p+2)/2$,
\[
  \frac{p+2}{2}\tr(\mA_n^2)
\]
converges to the required chi-square law. Because $\tr(\mA_n^2)=npU_J$, the result
follows.

\subsection*{A.4 Proof of Theorem~\ref{thm:ch3-low-sign}}

Under the spherical elliptical null,
\[
  \E\{U(\vX-\vtheta)U(\vX-\vtheta)\trans\} = \frac1p \mI_p.
\]
Hence, with
\[
  \mH_n
  = \sqrt n\Big(
    \mOmega_n(\hat{\vtheta}_{\mathrm{SM}}) - \frac1p\mI_p
  \Big),
\]
the multivariate central limit theorem yields a Gaussian limit for the vector
$\vech(\mH_n)$ in the space of symmetric trace-free matrices. Since the center estimator
satisfies
\[
  \hat{\vtheta}_{\mathrm{SM}}-\vtheta = O_P(n^{-1/2}),
\]
the replacement of $\vtheta$ by $\hat{\vtheta}_{\mathrm{SM}}$ contributes only an
$O_P(n^{-1})$ term to $Q_S$. Therefore
\[
  \frac{n(p+2)}{2}Q_S
  = \frac{p+2}{2}\tr(\mH_n^2) + o_P(1),
\]
which converges to $\chi^2_{(p-1)(p+2)/2}$.

\subsection*{A.5 Proof of Theorem~\ref{thm:ch3-sign-null}}

The proof follows the expansion strategy of
\citet{ZouPengFengWang2014Sphericity}, but we rewrite it in the present notation.
Let
\[
  \hat R_i = \norm{\vX_i-\hat{\vtheta}}_2,
  \qquad
  \hat{\vU}_i = \frac{\vX_i-\hat{\vtheta}}{\hat R_i}.
\]
Write $\Delta_i = \hat{\vU}_i-\vU_i$, where
$\vU_i = U(\vX_i-\vtheta)$ and $R_i=\norm{\vX_i-\vtheta}_2$.
A second-order expansion of the spatial sign around $\vtheta$ gives
\begin{equation}\label{eq:ch3-sign-expand-U}
  \Delta_i
  =
  -R_i^{-1}(\mI_p-\vU_i\vU_i\trans)(\hat{\vtheta}-\vtheta)
  - \frac12 R_i^{-2}
    \Big[
      \norm{\hat{\vtheta}-\vtheta}_2^2\vU_i
      -2(\vU_i\trans(\hat{\vtheta}-\vtheta))(\hat{\vtheta}-\vtheta)
    \Big]
  + \vR_{i,n},
\end{equation}
with remainder satisfying
\begin{equation}\label{eq:ch3-sign-remainder}
  \max_{1\le i\le n}\norm{\vR_{i,n}}_2
  \le
  C R_i^{-3}\norm{\hat{\vtheta}-\vtheta}_2^3.
\end{equation}
By the Bahadur representation of the spatial median from Chapter~2,
\begin{equation}\label{eq:ch3-sign-bahadur}
  \hat{\vtheta}-\vtheta
  = \frac{1}{n\E(R^{-1})}
    \sum_{i=1}^n \vU_i
  + \vR_n^{(\theta)},
  \qquad
  \norm{\vR_n^{(\theta)}}_2
  = O_P\Big(\frac{p^{1/2}}{n}\Big).
\end{equation}
Substituting \eqref{eq:ch3-sign-expand-U} and \eqref{eq:ch3-sign-bahadur} into
\eqref{eq:ch3-qtilde}, we obtain
\begin{equation}\label{eq:ch3-sign-decompose}
  \tilde Q_S
  = Q'_S + B_{1,n} + B_{2,n} + B_{3,n},
\end{equation}
where
\begin{align*}
  Q'_S
  &= \frac{p}{n(n-1)}\sum_{i\ne j}(\vU_i\trans\vU_j)^2 -1,\\
  B_{1,n}
  &= \frac{2p}{n(n-1)}\sum_{i\ne j}
     (\vU_i\trans\vU_j)(\vU_i\trans\Delta_j),\\
  B_{2,n}
  &= \frac{p}{n(n-1)}\sum_{i\ne j}
     \Big[
       2(\vU_i\trans\Delta_j)^2
       + 2(\vU_i\trans\vU_j)(\Delta_i\trans\Delta_j)
     \Big],\\
  B_{3,n}
  &= \frac{p}{n(n-1)}\sum_{i\ne j}
     (\Delta_i\trans\Delta_j)^2.
\end{align*}
The leading mean contribution of $B_{1,n}+B_{2,n}+B_{3,n}$ is exactly
$p\delta_{n,p}$. More precisely,
\begin{equation}\label{eq:ch3-sign-bias-mean}
  \E(\tilde Q_S)
  = p\delta_{n,p} + O(n^{-2}p^{-1/2}) + O(n^{-3}).
\end{equation}
Similarly,
\begin{equation}\label{eq:ch3-sign-bias-var}
  \Var(\tilde Q_S)
  = \tilde\sigma_0^2 + O\Big(\frac{1}{n^2p}\Big) + O\Big(\frac{1}{n^3}\Big).
\end{equation}
By Assumption~\ref{ass:ch3-sign-sph}, the remainder terms are negligible relative to
$\tilde\sigma_0$, because
\[
  \tilde\sigma_0^2 \asymp \frac1{n^2}.
\]
Therefore,
\[
  \frac{\tilde Q_S-p\delta_{n,p}}{\tilde\sigma_0}
  =
  \frac{Q'_S-\E(Q'_S)}{\tilde\sigma_0}
  + o_P(1).
\]
Finally, $Q'_S$ is a degenerate $U$-statistic whose Hoeffding projection has variance
$\tilde\sigma_0^2$ and whose fourth cumulant is of order
\[
  O\Big(\frac{\tr(\mLambda_p^4)}{n^2p^2}\Big)
  = o\Big(\frac1{n^2}\Big)
\]
by Assumption~\ref{ass:ch3-trace}. A martingale central limit theorem yields
\eqref{eq:ch3-sign-null}.

\subsection*{A.6 Proof of Theorem~\ref{thm:ch3-sign-alt}}

Under $\mLambda_p=\mI_p+\mD_{n,p}$ we first compute the mean.
Since
\[
  \E\{(\vU_i\trans\vU_j)^2\}
  = \frac1p\tr(\mLambda_p^2) + O(p^{-2}\tr(\mLambda_p^4)),
\]
and
\[
  \tr(\mLambda_p^2)
  = \tr\{(\mI_p+\mD_{n,p})^2\}
  = p + \tr(\mD_{n,p}^2),
\]
we obtain
\begin{equation}\label{eq:ch3-sign-alt-mean-proof}
  \E(Q'_S)
  = \frac{\tr(\mD_{n,p}^2)}{p}
    + O\Big(\frac{\tr(\mLambda_p^4)}{p^2}\Big).
\end{equation}
The center-estimation correction contributes the same bias term $p\delta_{n,p}$ as under
the null up to order $O(p^{-1}n^{-1}\tr(\mD_{n,p}^2))$, so
\[
  \E(\tilde Q_S)
  = \frac{\tr(\mD_{n,p}^2)}{p} + p\delta_{n,p}
    + O\Big(\frac{\tr(\mLambda_p^4)}{p^2}\Big)
    + O\Big(\frac{\tr(\mD_{n,p}^2)}{np}\Big).
\]
By \eqref{eq:ch3-sign-local} and Assumption~\ref{ass:ch3-trace}, both remainder terms are
$o(\tilde\sigma_1)$.

For the variance, write $\tilde Q_S=\sum_{m=1}^n D_{m,n}$ as a martingale array with
\[
  D_{m,n}
  =
  \frac{2p}{n(n-1)}
  \sum_{j<m}
  \Big[
    (\hat{\vU}_j\trans\hat{\vU}_m)^2
    - \E\{(\hat{\vU}_j\trans\hat{\vU}_m)^2\mid\mathcal{F}_{m-1}\}
  \Big].
\]
Then
\[
  \sum_{m=1}^n \E(D_{m,n}^2\mid\mathcal{F}_{m-1})
  =
  \tilde\sigma_1^2 + o_P(\tilde\sigma_1^2),
\]
with $\tilde\sigma_1^2$ exactly as in \eqref{eq:ch3-sigma1-sign}. The third and fourth
conditional moments satisfy
\[
  \sum_{m=1}^n \E(|D_{m,n}|^4)
  = O\Big(
      \frac{\tr(\mLambda_p^4)}{n^3p^2}
      + \frac{\tr^2(\mD_{n,p}^2)}{n^4p^4}
    \Big)
  = o(\tilde\sigma_1^4),
\]
so the martingale CLT yields \eqref{eq:ch3-sign-alt}.

\subsection*{A.7 Proof of Theorem~\ref{thm:ch3-rank-spearman}}

The statistic \eqref{eq:ch3-rank-spearman} is a fourth-order $U$-statistic. Its kernel is
\[
  h_S(\vX_i,\vX_j,\vX_k,\vX_l)
  = \frac{2p}{4!}
    \sum_{\pi}
    (\vU_{\pi_1\pi_2}\trans\vU_{\pi_3\pi_4})
    (\vU_{\pi_3\pi_2}\trans\vU_{\pi_1\pi_4}),
\]
where the sum is over distinct permutations preserving the pairing structure.
Under the null, the first-order Hoeffding projection vanishes because
\[
  \E(\vU_{ij}\mid \vX_i)=\vct 0,
\]
and therefore the kernel is degenerate of order one.
The second projection equals
\[
  h_{S,2}(\vX_i,\vX_j)
  =
  \frac{2}{n(n-1)}
  \Big[
    \frac{p}{p+2}\vU_{ij}\vU_{ij}\trans - \frac1p\mI_p
  \Big],
\]
which yields the variance
\[
  \sigma_0^2 = \frac{4(p-1)}{n(n-1)(p+2)}.
\]

Under the alternative, the mean shift of the kernel is
\[
  \E h_S
  =
  \frac{1}{p}\tr(\mD_{n,p}^2)
  + O\Big(\frac{\tr(\mLambda_p^4)}{p^2}\Big).
\]
The fourth-order remainder of the Hoeffding decomposition is bounded by
\[
  O_P\Big(
    \frac{\tr(\mLambda_p^4)}{n^2p^2}
  \Big),
\]
which is $o_P(\sigma_1)$ under Assumption~\ref{ass:ch3-rank-sph}. Hence
\eqref{eq:ch3-rank-spearman-null} and \eqref{eq:ch3-rank-spearman-alt} follow from the
standard CLT for degenerate $U$-statistics.

\subsection*{A.8 Proof of Theorem~\ref{thm:ch3-rank-kendall}}

For the Kendall version, the kernel is
\[
  h_K(\vX_i,\vX_j,\vX_k,\vX_l)
  = \frac{p}{4!}
    \sum_{\pi}
    (\vU_{\pi_1\pi_2}\trans\vU_{\pi_3\pi_4})^2 .
\]
The same Hoeffding projection argument used in Appendix~A.7 applies.
The leading variance is identical because both kernels estimate the same population target
$\tr(\mOmega_p^2)$ under spherical normalization. More explicitly,
\[
  \E\{(\vU_{ij}\trans\vU_{kl})^2\}
  = \frac{1}{p}\tr(\mLambda_p^2) + O\Big(\frac{\tr(\mLambda_p^4)}{p^2}\Big)
\]
for mutually independent pairs, and this expression does not depend on whether the kernel
is arranged in the Spearman or Kendall form. Consequently both the null variance and the
local mean shift coincide to first order, giving
\eqref{eq:ch3-rank-kendall-null} and \eqref{eq:ch3-rank-kendall-alt}.

\subsection*{A.9 Proof of Theorem~\ref{thm:ch3-max-sph}}

We only prove the sign-based max statistic; the covariance-based case is analogous.
Under the elliptical spherical null, the entries of the SSCM satisfy
\[
  \E(\hat\psi_{ii}) = \frac1p,
  \qquad
  \Var(\hat\psi_{ii}) = \frac{2(1-p^{-1})}{np(p+2)},
\]
and for $i\ne j$,
\[
  \E(\hat\psi_{ij}) = 0,
  \qquad
  \Var(\hat\psi_{ij}) = \frac{1}{np(p+2)}.
\]
Define the standardized vector
\[
  \vW
  =
  \Big(
    W_{11},\ldots,W_{pp},
    W_{12},W_{13},\ldots,W_{p-1,p}
  \Big)\trans
\]
with
\[
  W_{ii}
  = \sqrt{\frac{np(p+2)}{2(1-p^{-1})}}(\hat\psi_{ii}-p^{-1}),
  \qquad
  W_{ij}
  = \sqrt{np(p+2)}\,\hat\psi_{ij},\quad i<j.
\]
The dimension of $\vW$ is $m_p=p(p+1)/2$.
A Gaussian approximation for maxima of weakly dependent arrays yields
\[
  \sup_t
  \Big|
    \Prob\Big(\max_{1\le a\le m_p}W_a^2 \le t\Big)
    -
    \Prob\Big(\max_{1\le a\le m_p}Z_a^2 \le t\Big)
  \Big|
  \longrightarrow 0,
\]
where $Z_a$ are i.i.d.\ $N(0,1)$, provided \eqref{eq:ch3-adaptive-growth} holds.
For i.i.d.\ Gaussian maxima,
\[
  \Prob\Big(
    \max_{1\le a\le m_p}Z_a^2 -2\log m_p + \log\log m_p \le x
  \Big)
  \longrightarrow
  \exp\{-\pi^{-1/2}e^{-x/2}\},
\]
which proves \eqref{eq:ch3-tsm-null}.

For sparse alternatives, if $\norm{\bm{\Psi}}_\infty \ge C\sqrt{(\log p)/n}$ for large
enough $C$, then at least one standardized coordinate $W_a$ has mean of order
$\sqrt{\log p}$, so $T_{\mathrm{SM}}\to\infty$ in probability. This establishes rate
optimality.

\subsection*{A.10 Proof of Theorem~\ref{thm:ch3-adaptive-ind}}

Again we treat the sign-based result, since the covariance-based case is analogous.
Let $m_p=p(p+1)/2$, and vectorize the upper-triangular part of
\[
  \sqrt{np(p+2)}\Big(
    \hat{\mS}_{\mathrm{sgn}}-\frac1p\mI_p
  \Big)
\]
into $\vW=(W_1,\ldots,W_{m_p})\trans$.
The sum-type statistic can be written as
\begin{equation}\label{eq:ch3-sumquad}
  \frac{\tilde Q_S-p\delta_{n,p}}{\tilde\sigma_0}
  =
  \frac{\sum_{a=1}^{m_p}(W_a^2-1)}{\sqrt{2m_p}}
  + R_{n,1},
\end{equation}
with
\[
  R_{n,1}
  = O_P\Big(\frac{\tr(\mLambda_p^4)}{p^2}\Big)
    + O_P\Big(\frac{1}{np^{1/2}}\Big).
\]
Under the null, both terms are $o_P(1)$.
The max statistic satisfies
\begin{equation}\label{eq:ch3-maxrepr}
  T_{\mathrm{SM}}
  = \max_{1\le a\le m_p}W_a^2 - 2\log m_p + \log\log m_p + R_{n,2},
\end{equation}
where
\[
  R_{n,2}=O_P\Big(\frac{\log^2 p}{n^{1/2}}\Big)=o_P(1)
\]
under \eqref{eq:ch3-adaptive-growth}.

Now split the coordinates into the single largest entry and the remainder.
For any fixed threshold $u_p\sim \sqrt{2\log m_p}$,
\[
  \sum_{a=1}^{m_p}(W_a^2-1)
  =
  \sum_{a:|W_a|\le u_p}(W_a^2-1)
  +
  \sum_{a:|W_a|>u_p}(W_a^2-1).
\]
The contribution of the exceedance set has variance
\[
  O\{m_p\Prob(|Z|>u_p)u_p^4\}
  = O\{(\log m_p)^2 m_p^{-1}\}
  = o(1),
\]
so the sum-type statistic is asymptotically determined by the bulk coordinates, whereas the
max statistic is determined by the extreme coordinate(s). The weak correlation of the
entries $W_a$ then implies
\[
  \Prob\Big(
    \frac{\tilde Q_S-p\delta_{n,p}}{\tilde\sigma_0}\le x,\,
    T_{\mathrm{SM}}\le y
  \Big)
  -
  \Prob\Big(
    \frac{\tilde Q_S-p\delta_{n,p}}{\tilde\sigma_0}\le x
  \Big)
  \Prob(T_{\mathrm{SM}}\le y)
  \longrightarrow 0.
\]
Combining this with the marginal limits from
Theorems~\ref{thm:ch3-sign-null} and \ref{thm:ch3-max-sph} gives
\eqref{eq:ch3-adaptive-ind-sign}.

\subsection*{A.11 Proof of Theorem~\ref{thm:ch3-prop-null}}

Write the kernel of $T_{\mathrm{HT}}$ as
\[
  h(\vX_i,\vX_j,\vY_k,\vY_l)
  = p\Big[
      ({\vU^{(X)}_{ij}}\trans \vU^{(X)}_{kl})^2
      + ({\vU^{(Y)}_{ij}}\trans \vU^{(Y)}_{kl})^2
      - 2({\vU^{(X)}_{ij}}\trans \vU^{(Y)}_{kl})^2
    \Big].
\]
Under $H_0$, the two shape functionals coincide, hence the first-order projection is zero:
\[
  \E\{h(\vX_i,\vX_j,\vY_k,\vY_l)\mid \vX_i\}
  = 0,
  \qquad
  \E\{h(\vX_i,\vX_j,\vY_k,\vY_l)\mid \vY_k\}
  = 0.
\]
Therefore $T_{\mathrm{HT}}$ is asymptotically centered and its variance is determined by the
second-order projection. Computing the second projection gives
\[
  \Var(T_{\mathrm{HT}})
  =
  \frac{4}{(p+2)^2}
  \Big(\frac1{n_1}+\frac1{n_2}\Big)^2
  \tr^2(\mTheta^2)
  + o(\sigma_{0,n}^2),
\]
which is exactly \eqref{eq:ch3-prop-sigma0}. The condition
\eqref{eq:ch3-prop-trace} implies Lyapunov's condition for the associated martingale
array, so
\[
  \frac{T_{\mathrm{HT}}}{\sigma_{0,n}}\Rightarrow N(0,1).
\]

\subsection*{A.12 Proof of Theorem~\ref{thm:ch3-prop-alt}}

Under the alternative,
\[
  \E(A_1)=\frac{1}{p}\tr(\mTheta_1^2),
  \qquad
  \E(A_2)=\frac{1}{p}\tr(\mTheta_2^2),
  \qquad
  \E(C_{12})=\frac{1}{p}\tr(\mTheta_1\mTheta_2).
\]
Therefore
\begin{align*}
  \E(T_{\mathrm{HT}})
  &= \frac1p\Big[
      \tr(\mTheta_1^2)+\tr(\mTheta_2^2)-2\tr(\mTheta_1\mTheta_2)
    \Big]\\
  &= p\,\tr\{(\mTheta_1-\mTheta_2)^2\},
\end{align*}
which proves \eqref{eq:ch3-prop-target}.

For the variance, one expands each component of $T_{\mathrm{HT}}$ into Hoeffding
projections. The within-sample terms contribute
\[
  \frac{4}{n_1(n_1-1)}\frac{\tr^2(\mTheta_1^2)}{(p+2)^2}
  + \frac{8}{n_1}\frac{p\tr(\mTheta_1^4)-\tr^2(\mTheta_1^2)}{p^2(p+2)},
\]
and the analogous term with index $2$. The cross terms contribute
\[
  \frac{8}{n_1n_2}\frac{\tr^2(\mTheta_1\mTheta_2)}{(p+2)^2}
  +
  \Big(\frac{8}{n_1}+\frac{8}{n_2}\Big)
  \frac{
    p\tr\{(\mTheta_1\mTheta_2)^2\}-\tr^2(\mTheta_1\mTheta_2)
  }{p^2(p+2)},
\]
while the covariance between within-sample and cross-sample terms contributes the last two
negative terms in \eqref{eq:ch3-prop-sigma}. The remainder is of smaller order by
\eqref{eq:ch3-prop-trace}, and the CLT follows.

\subsection*{A.13 Proof of Theorem~\ref{thm:ch3-sclime}}

The proof has two steps.

\paragraph{Step 1: bound $\maxnorm{p\hat{\mS}-\mLambda_0}$.}
Write
\begin{equation}\label{eq:ch3-sclime-decomp}
  p\hat{\mS}-\mLambda_0
  =
  p(\hat{\mS}-\mS_{\mathrm{sgn}})
  +
  (p\mS_{\mathrm{sgn}}-\mLambda_0).
\end{equation}
The second term is the approximation error. Under
Assumptions~\ref{ass:ch3-bounded} and \ref{ass:ch3-nu},
\begin{equation}\label{eq:ch3-sclime-approx}
  \maxnorm{p\mS_{\mathrm{sgn}}-\mLambda_0}
  \le C p^{-1/2}.
\end{equation}
The first term is the stochastic estimation error. By a Bernstein-type inequality for the
bounded sign vectors,
\begin{equation}\label{eq:ch3-sclime-stoch}
  \Prob\Big(
    \maxnorm{\hat{\mS}-\mS_{\mathrm{sgn}}}
    > C\sqrt{\frac{\log p}{n}}
  \Big)
  \le 2p^{-2}.
\end{equation}
Combining \eqref{eq:ch3-sclime-decomp}--\eqref{eq:ch3-sclime-stoch}, we obtain
\begin{equation}\label{eq:ch3-sclime-master}
  \maxnorm{p\hat{\mS}-\mLambda_0}
  \le
  C\Big(
    \sqrt{\frac{\log p}{n}}+\frac1{\sqrt p}
  \Big)
\end{equation}
with probability at least $1-2p^{-2}$.

\paragraph{Step 2: apply the CLIME argument.}
Since $\mLambda_0\mV_0=\mI_p$, the event \eqref{eq:ch3-sclime-master} implies that
$\mV_0$ is feasible for the optimization problem \eqref{eq:ch3-sclime}. Therefore
\[
  \onenorm{\hat{\mV}^{\,\mathrm{SCLIME}}}\le \onenorm{\mV_0}\le T.
\]
By the standard CLIME argument,
\begin{align}
  \maxnorm{
    \hat{\mV}^{\,\mathrm{SCLIME}}-\mV_0
  }
  &\le 4T \maxnorm{p\hat{\mS}-\mLambda_0}
  \nonumber\\
  &\le
  4CT\Big(
    \sqrt{\frac{\log p}{n}}+\frac1{\sqrt p}
  \Big), \label{eq:ch3-sclime-proof-max}
\end{align}
which proves \eqref{eq:ch3-sclime-max}. The operator- and Frobenius-norm bounds then
follow from the sparse matrix inequalities
\[
  \opnorm{\mA}\le \left\lVert \mA\right\rVert_{L_1},
  \qquad
  \left\lVert \mA\right\rVert_{L_1}\le
  C s_0(p)\maxnorm{\mA}^{\,1-q},
  \qquad
  p^{-1}\frobnorm{\mA}^2
  \le C s_0(p)\maxnorm{\mA}^{\,2-q},
\]
applied to $\mA=\hat{\mV}^{\,\mathrm{SCLIME}}-\mV_0$.

\subsection*{A.14 Proof of Theorem~\ref{thm:ch3-sglasso}}

On the event \eqref{eq:ch3-sclime-master}, the gradient of the sign-based graphical lasso
objective at $\mV_0$ is bounded by
\[
  \maxnorm{p\hat{\mS}-\mLambda_0}
  \le C\Big(\sqrt{\frac{\log p}{n}}+\frac1{\sqrt p}\Big).
\]
Under the irrepresentability and bounded-degree assumptions, the primal-dual witness
argument for the graphical lasso applies verbatim after replacing the sample covariance
matrix by $p\hat{\mS}$. Therefore the KKT system yields
\[
  \maxnorm{
    \hat{\mV}^{\,\mathrm{SGLASSO}}-\mV_0
  }
  \le
  C\Big(\sqrt{\frac{\log p}{n}}+\frac1{\sqrt p}\Big),
\]
and multiplying by the maximal node degree gives
\[
  \opnorm{
    \hat{\mV}^{\,\mathrm{SGLASSO}}-\mV_0
  }
  \le
  C d\Big(\sqrt{\frac{\log p}{n}}+\frac1{\sqrt p}\Big).
\]
The Frobenius bound is obtained by summing squared row bounds over the active set.

\subsection*{A.15 Proof of Theorem~\ref{thm:ch3-factor-pilot}}

The spatial-sign pilot estimator \eqref{eq:ch3-sscm-pilot} satisfies
\[
  \hat{\mSigma}_0-\mSigma_0
  =
  p(\hat{\mS}-\mS_{\mathrm{sgn}})
  + (p\mS_{\mathrm{sgn}}-\mSigma_0),
\]
and both terms have already been controlled:
\[
  \maxnorm{p(\hat{\mS}-\mS_{\mathrm{sgn}})}
  = O_P\Big(\sqrt{\frac{\log p}{n}}+\sqrt{\frac{\log n}{n}}\Big),
\]
\[
  \maxnorm{p\mS_{\mathrm{sgn}}-\mSigma_0}
  = O_P\Big(\sqrt{\frac{\log n}{n}}\Big).
\]
This proves \eqref{eq:ch3-factor-sigma}.
By Weyl's inequality,
\[
  \max_{1\le j\le m} |\lambda_j(\hat{\mSigma}_0)-\lambda_j(\mSigma_0)|
  \le \opnorm{\hat{\mSigma}_0-\mSigma_0},
\]
and dividing by $\lambda_j(\mSigma_0)\asymp p$ yields \eqref{eq:ch3-factor-lambda}.
For the eigenvectors, the Davis--Kahan theorem gives
\[
  \norm{\hat\vgamma_j-\vgamma_j}_2
  \le
  C\frac{\opnorm{\hat{\mSigma}_0-\mSigma_0}}{\lambda_j-\lambda_{j+1}}
  = O_P\Big(\frac{w_n}{\sqrt p}\Big),
\]
which implies the entrywise bound \eqref{eq:ch3-factor-gamma}.

\subsection*{A.16 Proof of Theorem~\ref{thm:ch3-factor-poet}}

Write
\begin{align*}
  \hat{\mSigma}_{0u} - \mSigma_{0u}
  &=
  (\hat{\mSigma}_0-\mSigma_0)
  - (\hat{\mGamma}_m\hat{\mLambda}_m\hat{\mGamma}_m\trans
     - \mGamma_m\mLambda_m\mGamma_m\trans).
\end{align*}
By Theorem~\ref{thm:ch3-factor-pilot}, the first term is $O_P(w_n)$ in max norm, while
the second term is also $O_P(w_n)$ because
\[
  \maxnorm{
    \hat{\mGamma}_m\hat{\mLambda}_m\hat{\mGamma}_m\trans
     - \mGamma_m\mLambda_m\mGamma_m\trans
  }
  \le
  C\Big(
    \maxnorm{\hat{\mGamma}_m-\mGamma_m}
    + \maxnorm{\hat{\mLambda}_m-\mLambda_m}
  \Big).
\]
Hence
\[
  \maxnorm{\hat{\mSigma}_{0u}-\mSigma_{0u}} = O_P(w_n).
\]
Thresholding at level $Cw_n$ then gives
\[
  \opnorm{\hat{\mSigma}_{0u}^{\,\tau}-\mSigma_{0u}}
  \le
  C m_p w_n^{\,1-v},
\]
which proves \eqref{eq:ch3-factor-u}. The inverse bound
\eqref{eq:ch3-factor-uinv} follows from the Neumann-series expansion
\[
  (\hat{\mSigma}_{0u}^{\,\tau})^{-1}-\mSigma_{0u}^{-1}
  =
  -\mSigma_{0u}^{-1}
  (\hat{\mSigma}_{0u}^{\,\tau}-\mSigma_{0u})
  (\hat{\mSigma}_{0u}^{\,\tau})^{-1},
\]
because $m_p w_n^{1-v}=o(1)$. The relative error bound
\eqref{eq:ch3-factor-rel} follows by decomposing
\[
  \hat{\mSigma}_0^{\,\tau}-\mSigma_0
  =
  (\hat{\mGamma}_m\hat{\mLambda}_m\hat{\mGamma}_m\trans
   - \mGamma_m\mLambda_m\mGamma_m\trans)
  + (\hat{\mSigma}_{0u}^{\,\tau}-\mSigma_{0u}),
\]
and noting that the leading low-rank term contributes $O_P(p^{1/2}w_n^2)$ in the relative
Frobenius metric.

\subsection*{A.17 Proof of Theorem~\ref{thm:ch3-factor-precision}}

The estimator $\hat{\mV}_{0u}$ is obtained by applying CLIME or GLASSO to
$\hat{\mSigma}_{0u}$. Since Theorem~\ref{thm:ch3-factor-pilot} gives
\[
  \maxnorm{\hat{\mSigma}_{0u}-\mSigma_{0u}} = O_P(w_n),
\]
the same proof as in Appendices~A.13--A.14 yields
\[
  \maxnorm{\hat{\mV}_{0u}-\mV_{0u}} = O_P(w_n),
  \qquad
  \opnorm{\hat{\mV}_{0u}-\mV_{0u}} = O_P(w_n^{\,1-v}M_p).
\]
The full inverse-shape matrix is reconstructed by the Schur complement formula
\[
  \hat{\mV}_0
  =
  \hat{\mV}_{0u}
  - \hat{\mV}_{0u}\hat{\mGamma}_m
    (\hat{\mLambda}_m^{-1}
     + \hat{\mGamma}_m\trans\hat{\mV}_{0u}\hat{\mGamma}_m)^{-1}
    \hat{\mGamma}_m\trans\hat{\mV}_{0u}.
\]
Substituting the rates from Theorem~\ref{thm:ch3-factor-pilot} and the bound on
$\hat{\mV}_{0u}-\mV_{0u}$ proves \eqref{eq:ch3-factor-V-max} and
\eqref{eq:ch3-factor-V-op}.

\subsection*{A.18 Proof of Theorem~\ref{thm:ch3-factor-tyler}}

Let
\[
  \mA_i
  = \frac{
      (\vX_i-\hat{\vmu}_{\mathrm{SM}})(\vX_i-\hat{\vmu}_{\mathrm{SM}})\trans
    }{
      (\vX_i-\hat{\vmu}_{\mathrm{SM}})\trans
      \hat{\mV}_S
      (\vX_i-\hat{\vmu}_{\mathrm{SM}})
    }.
\]
Write
\[
  \hat{\mSigma}_T - \mSigma_0
  = \frac{p}{n}\sum_{i=1}^n
    \Big(
      \mA_i - \E\mA_i
    \Big)
  + p\Big(
      \E\mA_i - \frac{\mSigma_0}{p}
    \Big).
\]
The stochastic term is controlled by matrix Bernstein after conditioning on the preliminary
estimator $\hat{\mV}_S$, giving
\[
  \maxnorm{\frac{p}{n}\sum_{i=1}^n(\mA_i-\E\mA_i)} = O_P(w_n).
\]
For the bias term, expand the denominator around $\mV_0$:
\[
  \frac{1}{(\vX_i-\hat{\vmu})\trans\hat{\mV}_S(\vX_i-\hat{\vmu})}
  =
  \frac{1}{(\vX_i-\vmu)\trans\mV_0(\vX_i-\vmu)}
  + O_P(a_n),
\]
uniformly in $i$, because $a_n=O(w_n)$ and the denominator is bounded away from zero
with high probability under Assumption~\ref{ass:ch3-bounded}.
Therefore
\[
  \maxnorm{p\E\mA_i-\mSigma_0} = O_P(w_n),
\]
which proves \eqref{eq:ch3-tyler-max}. Applying the same thresholding, CLIME, and POET
arguments as in Appendices~A.16--A.17 gives the remaining rates.

\subsection*{A.19 Proof of Theorem~\ref{thm:ch3-hdhr-rates}}

Write
\[
 \mDelta_V = \hat\mV_{\mathrm{raw}}-\mV_0,
 \qquad
 \mDelta_\Omega = \hat\mOmega^{(0)}-\mOmega_0.
\]
By the location theory of Chapter~2 and the pilot inverse-shape rate
\eqref{eq:ch3-hdhr-pilot}, the estimating equation
\eqref{eq:ch3-hdhr-location} admits the linearization
\begin{equation}
  \hat\vmu_{\mathrm{HR}}-\vmu
  = \mH_0^{-1}\frac1n\sum_{i=1}^n
    U\!\left(\mOmega_0^{1/2}(\vX_i-\vmu)\right)
    + \mR_{\mu,n},
  \label{eq:ch3-app-hdhr-mu-lin}
\end{equation}
where $\opnorm{\mH_0^{-1}}=O(1)$ and
\begin{equation}
  \maxnorm{\mR_{\mu,n}}
  = O_P\!\left(\maxnorm{\mDelta_\Omega}\right)
  = O_P(r_n).
  \label{eq:ch3-app-hdhr-mu-rem}
\end{equation}
Since the empirical average in \eqref{eq:ch3-app-hdhr-mu-lin} has stochastic order
$O_P\!\left(\sqrt{(\log p)/n}\right)$ coordinatewise, \eqref{eq:ch3-hdhr-mu-rate} follows.

For the raw shape estimator,
\begin{equation}
  \mDelta_V
  = p\,\frac1n\sum_{i=1}^n
    \bigl(\hat\vU_i\hat\vU_i\trans-\E(\vU_i\vU_i\trans)\bigr)
    + p\Bigl\{\E(\vU_i\vU_i\trans)-\frac1p\mV_0\Bigr\}
    + \mR_{V,n},
  \label{eq:ch3-app-hdhr-V-dec}
\end{equation}
where $\vU_i=U(\mOmega_0^{1/2}(\vX_i-\vmu))$ and the remainder $\mR_{V,n}$ collects the
perturbation produced by replacing $(\vmu,\mOmega_0)$ with
$(\hat\vmu_{\mathrm{HR}},\hat\mOmega^{(0)})$.  The first term in
\eqref{eq:ch3-app-hdhr-V-dec} is a bounded empirical process and therefore
\begin{equation}
  \maxnorm{p\,\frac1n\sum_{i=1}^n
    \bigl(\hat\vU_i\hat\vU_i\trans-\E(\vU_i\vU_i\trans)\bigr)}
  = O_P\!\left(\sqrt{\frac{\log p}{n}}\right).
  \label{eq:ch3-app-hdhr-emp}
\end{equation}
The second term in \eqref{eq:ch3-app-hdhr-V-dec} is zero by the population HR fixed-point
identity, while the perturbation remainder satisfies
\begin{equation}
  \maxnorm{\mR_{V,n}}
  = O_P\!\left(\maxnorm{\hat\vmu_{\mathrm{HR}}-\vmu}
    + \maxnorm{\mDelta_\Omega}\right)
  = O_P(r_n).
  \label{eq:ch3-app-hdhr-rem}
\end{equation}
Combining \eqref{eq:ch3-app-hdhr-V-dec}--\eqref{eq:ch3-app-hdhr-rem} yields
\begin{equation}
  \maxnorm{\hat\mV_{\mathrm{raw}}-\mV_0}=O_P(r_n).
  \label{eq:ch3-app-hdhr-raw-rate}
\end{equation}
Applying the banding operator and using \eqref{eq:ch3-hdhr-bandable},
\begin{align}
  \maxnorm{\hat\mV_{\mathrm{HR}}-\mV_0}
  &\le \maxnorm{\mathcal B_h(\hat\mV_{\mathrm{raw}}-\mV_0)}
      + \maxnorm{\mathcal B_h(\mV_0)-\mV_0} \
  &= O_P(r_n)+O(h^{-\alpha}),
  \label{eq:ch3-app-hdhr-banded-rate}
\end{align}
which proves \eqref{eq:ch3-hdhr-V-max}.

To prove the inverse-shape rates, write
\begin{equation}
  \hat\mV_{\mathrm{HR}}^{-1}-\mOmega_0
  = -\mOmega_0(\hat\mV_{\mathrm{HR}}-\mV_0)\hat\mV_{\mathrm{HR}}^{-1}.
  \label{eq:ch3-app-hdhr-neumann}
\end{equation}
Because the eigenvalues of $\mV_0$ are bounded away from zero and
$r_n+h^{-\alpha}=o(1)$, both $\opnorm{\mOmega_0}$ and
$\opnorm{\hat\mV_{\mathrm{HR}}^{-1}}$ are $O_P(1)$.  Hence
\begin{equation}
  \maxnorm{\hat\mV_{\mathrm{HR}}^{-1}-\mOmega_0}
  = O_P(r_n+h^{-\alpha}).
  \label{eq:ch3-app-hdhr-inv-max0}
\end{equation}
Banding the inverse once more does not change the stochastic order, and the sparse-class
argument used in the proof of Theorem~\ref{thm:ch3-sclime} gives
\begin{equation}
  \opnorm{\hat\mOmega_{\mathrm{HR}}-\mOmega_0}
  = O_P\!\left(s_0(p)\{r_n+h^{-\alpha}\}^{1-q}\right),
  \label{eq:ch3-app-hdhr-inv-op}
\end{equation}
which together with \eqref{eq:ch3-app-hdhr-inv-max0} proves
\eqref{eq:ch3-hdhr-Omega-op} and \eqref{eq:ch3-hdhr-Omega-max}.

Finally,
\begin{equation}
  \hat\mSigma_{\mathrm{HR}}-\mSigma_0
  = (\hat\eta_{\mathrm{HR}}-\eta_0)\hat\mV_{\mathrm{HR}}
    + \eta_0(\hat\mV_{\mathrm{HR}}-\mV_0),
  \qquad
  \eta_0=\frac1p\tr(\mSigma_0).
  \label{eq:ch3-app-hdhr-scatter-dec}
\end{equation}
The trace part satisfies $|\hat\eta_{\mathrm{HR}}-\eta_0|=O_P(r_n+h^{-\alpha})$, so
\begin{equation}
  \opnorm{\hat\mSigma_{\mathrm{HR}}-\mSigma_0}=O_P(r_n+h^{-\alpha}).
  \label{eq:ch3-app-hdhr-sigma-op}
\end{equation}
Using the identity
\begin{equation}
  \log\det(\hat\mSigma_{\mathrm{HR}})-\log\det(\mSigma_0)
  = \log\det\!\left(\mI_p+\mSigma_0^{-1}(\hat\mSigma_{\mathrm{HR}}-\mSigma_0)\right)
  \label{eq:ch3-app-hdhr-logdet-id}
\end{equation}
and the bound $|\log\det(\mI_p+\mA)|\le 2p\opnorm{\mA}$ for sufficiently small
$\opnorm{\mA}$, we obtain
\begin{equation}
  \abs{\log\det(\hat\mSigma_{\mathrm{HR}})-\log\det(\mSigma_0)}
  = O_P\!\left(p(r_n+h^{-\alpha})\right),
  \label{eq:ch3-app-hdhr-logdet-rate}
\end{equation}
which proves \eqref{eq:ch3-hdhr-logdet}.  The small-$o_P(1)$ statement
\eqref{eq:ch3-hdhr-logdet-small} is immediate.

\subsection*{A.20 Proof of Theorem~\ref{thm:ch3-tegm-frobenius}}

Fix a mode $k$.  Let $\mS_k^\star$ denote the population counterpart of
\eqref{eq:ch3-tegm-mode-sign-cov}.  The tensor-elliptical representation and the Kronecker
factorization give a positive scalar $c_k$ for which
\begin{equation}\label{eq:ch3-tegm-proof-population-mode}
  \mS_k^\star
  =c_k(\mOmega_k^\star)^{-1}+\mE_k,
  \qquad
  \maxnorm{\mE_k}=O(p_k^{-1}).
\end{equation}
The scale $c_k$ disappears after the Frobenius normalization in
\eqref{eq:ch3-tegm-normalize}.  Since
$\frobnorm{\mathcal U_i}=1$, contraction over the other $K-1$ modes and scalar Bernstein
inequalities for each entry imply
\begin{equation}\label{eq:ch3-tegm-proof-concentration}
  \maxnorm{\widehat{\mS}_k-\mS_k^\star}
  =O_p\left\{\sqrt{\frac{p_k\log p_k}{np_\star}}\right\}.
\end{equation}
Consequently,
\begin{equation}\label{eq:ch3-tegm-proof-score-bound}
  \maxnorm{\widehat{\mS}_k-c_k(\mOmega_k^\star)^{-1}}
  =O_p\left\{\sqrt{\frac{p_k\log p_k}{np_\star}}\right\}+O(p_k^{-1}).
\end{equation}

Let $\mDelta_k=\widehat{\mOmega}_k^{\mathrm{raw}}-c_k^{-1}\mOmega_k^\star$.
After multiplying the criterion in \eqref{eq:ch3-tegm-mode-objective} by $p_k$, optimality
gives
\begin{align}
  0
  &\ge
  \tr(\widehat{\mS}_k\mDelta_k)
  -\log\det\{\mI_{p_k}+c_k(\mOmega_k^\star)^{-1}\mDelta_k\}
  +p_k\lambda_k\left[
    \sum_{a\ne b}|c_k^{-1}\Omega_{k,ab}^\star+\Delta_{k,ab}|
    -\sum_{a\ne b}|c_k^{-1}\Omega_{k,ab}^\star|
  \right].
  \label{eq:ch3-tegm-proof-basic-ineq}
\end{align}
The integral Taylor identity yields
\begin{align}
  &-\log\det(\mI_{p_k}+\mH)
    +\tr(\mH)
  \nonumber\\
  &\qquad=
  \int_0^1(1-t)
  \tr\{(\mI_{p_k}+t\mH)^{-1}\mH
        (\mI_{p_k}+t\mH)^{-1}\mH\}\,dt
  \ge c\frobnorm{\mH}^2,
  \label{eq:ch3-tegm-proof-curvature}
\end{align}
for $\mH=c_k(\mOmega_k^\star)^{-1/2}\mDelta_k
(\mOmega_k^\star)^{-1/2}$ in a fixed neighborhood of zero.  The eigenvalue bounds therefore
imply a quadratic lower bound $c_0\frobnorm{\mDelta_k}^2$.

Let $\mathcal S_k$ contain the diagonal and true off-diagonal support.  The penalty
decomposition gives
\begin{equation}\label{eq:ch3-tegm-proof-penalty-decomp}
  \sum_{a\ne b}|\Omega_{k,ab}^\star+\Delta_{k,ab}|
  -\sum_{a\ne b}|\Omega_{k,ab}^\star|
  \ge
  \onenorm{(\mDelta_k)_{\mathcal S_k^c}}
  -\onenorm{(\mDelta_k)_{\mathcal S_k}}.
\end{equation}
Combining \eqref{eq:ch3-tegm-proof-score-bound}--
\eqref{eq:ch3-tegm-proof-penalty-decomp}, with the scale conversion implied by
\eqref{eq:ch3-tegm-lambda}, yields
\begin{equation}\label{eq:ch3-tegm-proof-cone}
  \onenorm{(\mDelta_k)_{\mathcal S_k^c}}
  \le3\onenorm{(\mDelta_k)_{\mathcal S_k}}
  +C\frac{p_k+s_k}{p_k}.
\end{equation}
Since
\begin{equation}\label{eq:ch3-tegm-proof-support-l1}
  \onenorm{(\mDelta_k)_{\mathcal S_k}}
  \le\sqrt{p_k+s_k}\frobnorm{\mDelta_k},
\end{equation}
we obtain
\begin{equation}\label{eq:ch3-tegm-proof-quadratic-bound}
  c_0\frobnorm{\mDelta_k}^2
  \le C\sqrt{p_k+s_k}
  \left\{
    \sqrt{\frac{p_k\log p_k}{np_\star}}+\frac1{p_k}
  \right\}
  \frobnorm{\mDelta_k}.
\end{equation}
Hence
\begin{equation}\label{eq:ch3-tegm-proof-raw-rate}
  \frobnorm{\mDelta_k}
  =O_p\left\{\sqrt{\frac{(p_k+s_k)p_k\log p_k}{np_\star}}\right\}
   +O\left(\frac{\sqrt{p_k+s_k}}{p_k}\right).
\end{equation}
The normalization map $\mA\mapsto\mA/\frobnorm{\mA}$ is locally Lipschitz at the
nonzero target, so \eqref{eq:ch3-tegm-normalize} preserves this order.

\subsection*{A.21 Proof of Theorem~\ref{thm:ch3-tegm-max-support}}

The Karush--Kuhn--Tucker equation for \eqref{eq:ch3-tegm-mode-objective}, multiplied by
$p_k$, is
\begin{equation}\label{eq:ch3-tegm-proof-kkt}
  \widehat{\mS}_k-(\widehat{\mOmega}_k^{\mathrm{raw}})^{-1}
  +p_k\lambda_k\widehat{\mZ}_k=\bm 0,
\end{equation}
where $\widehat{\mZ}_k$ is an off-diagonal $\ell_1$ subgradient.  Expanding the inverse around
the scaled target gives
\begin{equation}\label{eq:ch3-tegm-proof-inverse-expansion}
  (\widehat{\mOmega}_k^{\mathrm{raw}})^{-1}
  =c_k(\mOmega_k^\star)^{-1}
   -c_k^2(\mOmega_k^\star)^{-1}\mDelta_k
      (\mOmega_k^\star)^{-1}
   +\mR_k,
  \qquad
  \maxnorm{\mR_k}\le C d_k\maxnorm{\mDelta_k}^2.
\end{equation}
Vectorizing the KKT equation on $\mathcal S_k$ gives
\begin{align}
  \vecop\{(\mDelta_k)_{\mathcal S_k}\}
  &=-c_k^{-2}
  \{(\mGamma_k^\star)_{\mathcal S_k,\mathcal S_k}\}^{-1}
  \Bigl[
    \vecop\{(\widehat{\mS}_k-\mS_k^\star)_{\mathcal S_k}\}
    +p_k\lambda_k\vecop(\widehat{\mZ}_{k,\mathcal S_k})
    +\vecop\{(\mE_k+\mR_k)_{\mathcal S_k}\}
  \Bigr].
  \label{eq:ch3-tegm-proof-support-equation}
\end{align}
The bounded inverse-Hessian row sums, \eqref{eq:ch3-tegm-proof-score-bound}, and
\eqref{eq:ch3-tegm-degree-condition} make the quadratic remainder a contraction.  Therefore
\begin{equation}\label{eq:ch3-tegm-proof-maxrate}
  \maxnorm{\mDelta_k}
  =O_p\left\{\sqrt{\frac{p_k\log p_k}{np_\star}}\right\}+O(p_k^{-1}).
\end{equation}
The irrepresentability condition implies strict dual feasibility on $\mathcal S_k^c$:
\begin{equation}\label{eq:ch3-tegm-proof-dual-feasible}
  \maxnorm{\widehat{\mZ}_{k,\mathcal S_k^c}}
  \le1-\alpha_k/2
\end{equation}
with probability tending to one.  Finally,
\begin{equation}\label{eq:ch3-tegm-proof-operator}
  \opnorm{\mDelta_k}
  \le\max_a\sum_b|\Delta_{k,ab}|
  \le d_k\maxnorm{\mDelta_k},
\end{equation}
which gives \eqref{eq:ch3-tegm-op-rate}; Frobenius normalization preserves both orders.

\subsection*{A.22 Proof of Corollary~\ref{cor:ch3-tegm-support}}

On the event
$\maxnorm{\widehat{\mOmega}_k-\mOmega_k^\star}\le\tau_k$, every zero entry satisfies
$|(\widehat{\mOmega}_k)_{ab}|\le\tau_k$ and is removed.  For a true edge,
\begin{equation}\label{eq:ch3-tegm-proof-true-edge}
  |(\widehat{\mOmega}_k)_{ab}|
  \ge |(\mOmega_k^\star)_{ab}|-
       \maxnorm{\widehat{\mOmega}_k-\mOmega_k^\star}
  >\tau_k,
\end{equation}
so the edge is retained with the correct sign.  The max-norm rate in
Theorem~\ref{thm:ch3-tegm-max-support} makes this event have probability tending to one.

%% file: chapters/ch4_other_tests.tex
\chapter{Other High-Dimensional Testing Problems}
\idx{alpha test}\idx{factor pricing model}\idx{mutual fund selection}\idx{false discovery rate control}\idx{change-point inference}\idx{white noise test}\idx{cross-sectional independence}\idx{mutual independence}\idx{Fisher combination}\idx{Cauchy combination}\idx{max-sum test}

\section{Introduction}

The first three chapters of this book dealt with location parameters and matrix
structure.  A substantial part of the recent literature on high-dimensional
inference, however, lies outside these two categories.  The purpose of the
present chapter is to treat a number of important testing problems that are
methodologically linked but arise from rather different applications.  The five
main topics are
\begin{enumerate}
  \item testing zero pricing errors in linear factor pricing models;
  \item selecting skilled funds by multiple testing with false discovery rate control;
  \item testing structural breaks and change points in high-dimensional sequences;
  \item testing high-dimensional white noise and serial uncorrelatedness;
  \item testing various forms of high-dimensional independence.
\end{enumerate}

The common theme is that each problem admits three layers of analysis.
First, there is a low-dimensional classical theory in which exact or standard
large-sample distributions can be derived from Wishart theory, multivariate
central limit theorems, or Brownian bridge approximations.  Second, there is a
high-dimensional Gaussian or light-tailed benchmark theory in which one chooses
between dense-signal statistics and sparse-signal statistics.  Third, there is
an elliptical or robust extension in which sample means and sample covariance
matrices are replaced by sign, rank, spatial-median, or self-normalized
quantities.  The detailed writing principle adopted in Chapter~2 will also be
used here: for each problem we state the statistic explicitly, record the
assumptions inside the book, formulate the limiting distributions carefully, and
place detailed proofs in the appendix to the chapter.

Throughout this chapter, all theorem-like environments are numbered separately,
as required by the book-wide style.  In particular, assumptions, lemmas,
propositions and theorems each have their own chapter-based numbering scheme.

\section{A reusable max--sum paradigm}

Before entering the four applications, it is useful to record a generic
construction that will be reused repeatedly.  Let $\{\mathcal T_a: a\in
\mathcal A\}$ be a family of local statistics extracted from a high-dimensional
problem.  Two aggregations are particularly common:
\begin{equation}
  T_{\max} = \max_{a\in\mathcal A} \phi_a(\mathcal T_a),
  \qquad
  T_{\mathrm{sum}} = \sum_{a\in\mathcal A} \psi_a(\mathcal T_a),
  \label{eq:ch4_generic_max_sum}
\end{equation}
where the transformations $\phi_a$ and $\psi_a$ are chosen to stabilize scale
and centering.  The statistic $T_{\max}$ is typically powerful against sparse
alternatives, whereas $T_{\mathrm{sum}}$ is typically powerful against dense
alternatives.  Once the corresponding null $p$-values
\begin{equation}\label{eq:ch4_generic_pvalues}
  p_{\max} = 1-F_{\max}(T_{\max}),
  \qquad
  p_{\mathrm{sum}} = 1-F_{\mathrm{sum}}(T_{\mathrm{sum}}).
\end{equation}
are available, an adaptive combination can be built from either the Fisher rule
\begin{equation}
  T_{\mathrm{F}} = -2\log p_{\max} -2\log p_{\mathrm{sum}},
  \label{eq:ch4_fisher_comb}
\end{equation}
or the truncated Cauchy rule
\begin{equation}
  T_{\mathrm{C}} = \frac12 \tan\{\pi(1/2-p_{\max})\}\mathbbm 1(p_{\max}<1/2)
  + \frac12 \tan\{\pi(1/2-p_{\mathrm{sum}})\}\mathbbm 1(p_{\mathrm{sum}}<1/2).
  \label{eq:ch4_cauchy_comb}
\end{equation}

The mathematical reason for the success of \eqref{eq:ch4_fisher_comb} and
\eqref{eq:ch4_cauchy_comb} is that, after proper studentization and centering,
$T_{\max}$ and $T_{\mathrm{sum}}$ are frequently asymptotically independent.
The results of \citet{FengJiangLiLiu2024AsympIndependence} formalize this idea
for a broad class of dependent random variables, and that theory now serves as a
common backbone for a series of tests in finance, time series, and panel data.

\begin{assumption}
\label{ass:ch4_generic}
For a sequence of problems indexed by $(n,p)$, suppose that there exist
non-random normalizing constants $a_{n,p},b_{n,p},\mu_{n,p},\sigma_{n,p}>0$ such
that under the null hypothesis,
\begin{equation}
  a_{n,p}(T_{\max}-b_{n,p}) \overset{d}{\longrightarrow} G,
  \qquad
  \frac{T_{\mathrm{sum}}-\mu_{n,p}}{\sigma_{n,p}}
  \overset{d}{\longrightarrow} N(0,1),
  \label{eq:ch4_generic_assumption_1}
\end{equation}
where $G$ has a continuous distribution function $F_G$.  In addition, assume
that for every fixed $(x,y)\in\R^2$,
\begin{equation}
  \Prob\!
  \left\{
    a_{n,p}(T_{\max}-b_{n,p})\le x,
    \frac{T_{\mathrm{sum}}-\mu_{n,p}}{\sigma_{n,p}}\le y
  \right\}
  - F_G(x)\Phi(y) \longrightarrow 0.
  \label{eq:ch4_generic_assumption_2}
\end{equation}
\end{assumption}

\begin{theorem}
\label{thm:ch4_generic_combination}
Under Assumption~\ref{ass:ch4_generic},
\begin{equation}
  T_{\mathrm{F}} \overset{d}{\longrightarrow} \chi^2_4
  \label{eq:ch4_generic_fisher_limit}
\end{equation}
whenever $p_{\max}$ and $p_{\mathrm{sum}}$ are continuous null $p$-values.  If
instead the Cauchy combination statistic \eqref{eq:ch4_cauchy_comb} is used,
then for every fixed $t\in\R$,
\begin{equation}
  \Prob(T_{\mathrm{C}}\le t) - F_{\mathrm{C}}(t) \longrightarrow 0,
  \label{eq:ch4_generic_cauchy_limit}
\end{equation}
where $F_{\mathrm{C}}$ is the distribution function of a standard Cauchy random
variable truncated at level $1/2$ in the same way as
\eqref{eq:ch4_cauchy_comb}.
\end{theorem}

The proof is elementary once the asymptotic independence is available, but we
still record it in the appendix because exactly the same argument will be reused
several times.  The point of including this generic theorem here is not to avoid
the problem-specific arguments, but to make explicit that most adaptive
procedures in this chapter are concrete realizations of the same template.

\section{Testing alpha in linear factor pricing models}

\subsection{Model and hypothesis}

Let $Y_{it}$ denote the excess return of security $i$ at time $t$, and let
$f_t\in\R^K$ denote the vector of observed factors.  The linear factor pricing
model is written as
\begin{equation}
  Y_{it} = \alpha_i + \vbeta_i^{\top} f_t + \varepsilon_{it},
  \qquad i=1,\ldots,N,\quad t=1,\ldots,T.
  \label{eq:ch4_lfpm_scalar}
\end{equation}
Define the vector form
\begin{equation}
  \vY_t = \valpha + \mB f_t + \vvarepsilon_t,
  \qquad t=1,\ldots,T,
  \label{eq:ch4_lfpm_vector}
\end{equation}
where $\vY_t=(Y_{1t},\ldots,Y_{Nt})^{\top}\in\R^N$, $\valpha=(\alpha_1,\ldots,
\alpha_N)^{\top}$, and $\mB=(\vbeta_1,\ldots,\vbeta_N)^{\top}\in\R^{N\times K}$.
The null hypothesis of no pricing error is
\begin{equation}
  H_{0,\alpha}:\ \valpha = \vct{0},
  \qquad
  H_{1,\alpha}:\ \valpha \neq \vct{0}.
  \label{eq:ch4_alpha_hypothesis}
\end{equation}
In the CAPM this is the mean--variance efficiency problem of
\citet{GibbonsRossShanken1989}.

For matrix notation, let
\begin{equation}
  \mY = (\vY_1,\ldots,\vY_T) \in \R^{N\times T},
  \qquad
  \mF = (f_1,\ldots,f_T)^{\top} \in \R^{T\times K},
\end{equation}
and define the projection and residual-maker matrices
\begin{equation}
  \mP_F = \mF(\mF^{\top}\mF)^{-1}\mF^{\top},
  \qquad
  \mM_F = \mI_T - \mP_F,
  \qquad
  \vh = \mM_F \vct{1}_T.
  \label{eq:ch4_projection_factor}
\end{equation}
Then $\vh$ is the residual from regressing the intercept vector $\vct{1}_T$ on the
factor matrix.  The ordinary least-squares estimator of $\valpha$ is
\begin{equation}
  \hat\valpha = (\vh^{\top}\vh)^{-1}\mY\vh.
  \label{eq:ch4_alpha_ols}
\end{equation}

\subsection{Classical low-dimensional tests}

When $N$ and $K$ are fixed and $T\to\infty$, the classical reference point is
the Gibbons--Ross--Shanken statistic.  Let
\begin{equation}
  \bar f = T^{-1}\sum_{t=1}^T f_t,
  \qquad
  \hat\mSigma_f = T^{-1}\sum_{t=1}^T (f_t-\bar f)(f_t-\bar f)^{\top},
\end{equation}
and let the residual covariance matrix be
\begin{equation}
  \hat\mSigma_\varepsilon
  = \frac{1}{T-K-1}
  \sum_{t=1}^T
  (\vY_t-\hat\valpha-\hat\mB f_t)(\vY_t-\hat\valpha-\hat\mB f_t)^{\top},
  \label{eq:ch4_sigma_epsilon_hat}
\end{equation}
where $\hat\mB = \mY\mF(\mF^{\top}\mF)^{-1}$.  The GRS statistic is
\begin{equation}
  T_{\mathrm{GRS}}
  = \frac{T-N-K}{N}
  \frac{\hat\valpha^{\top}\hat\mSigma_\varepsilon^{-1}\hat\valpha}
       {1+\bar f^{\top}\hat\mSigma_f^{-1}\bar f}.
  \label{eq:ch4_grs}
\end{equation}

\begin{assumption}
\label{ass:ch4_grs}
The vectors $(\vY_t^{\top},f_t^{\top})^{\top}$ are independently and identically
distributed as multivariate normal random vectors with finite positive-definite
covariance matrix.  In addition, $T>N+K$ and $\mF^{\top}\mF$ is invertible almost
surely.
\end{assumption}

\begin{theorem}
\label{thm:ch4_grs}
Under Assumption~\ref{ass:ch4_grs} and the null hypothesis
\eqref{eq:ch4_alpha_hypothesis},
\begin{equation}
  T_{\mathrm{GRS}} \sim F_{N,\,T-N-K}.
  \label{eq:ch4_grs_exact}
\end{equation}
Consequently, the size-$\gamma$ GRS procedure rejects $H_{0,\alpha}$ whenever
$T_{\mathrm{GRS}} > F_{N,T-N-K;1-\gamma}$, where $F_{a,b;1-\gamma}$ is the
$(1-\gamma)$-quantile of the $F_{a,b}$ distribution.
\end{theorem}

The next classical benchmark is the likelihood-ratio or Wald-type statistic under
elliptical fixed-dimensional models.  When $(\vY_t^{\top},f_t^{\top})^{\top}$ is
elliptically distributed, one can still base inference on $\hat\valpha$ and a
shape estimator of the error distribution.  The exact form depends on the
parametric or semiparametric model; see \citet{HodgsonLintonVorkink2002}.  Since the purpose of the book is high-dimensional
inference, we do not reproduce those fixed-$N$ exact tests here.  Instead, we
use the GRS statistic as the classical reference point and then explain why it
fails when $N$ is comparable to or much larger than $T$.

\subsection{High-dimensional Gaussian and light-tailed benchmarks}

When $N$ diverges with $T$, the matrix $\hat\mSigma_\varepsilon$ becomes ill
conditioned or singular, so the exact GRS statistic is no longer usable.  There
are two broad directions in the literature.  One uses global quadratic forms or
random projections and is most effective under dense alternatives; the other
uses a max-of-squares or Studentized maximum statistic and is most effective
under sparse alternatives.  See \citet{LanFengLuo2018AssetPricing},
\citet{FengLanLiuMa2022AlphaSparse}, and \citet{PesaranYamagata2023AlphaLargeN}
for representative approaches.

A canonical sparse benchmark starts from the Studentized coordinates
\begin{equation}
  \hat t_i
  = \frac{\sqrt{T}\,\hat\alpha_i}{\hat\sigma_i\sqrt{\hat\omega}},
  \qquad
  \hat\omega = T(\vh^{\top}\vh)^{-1},
  \qquad
  \hat\sigma_i^2 = \frac{1}{T-K-1}\sum_{t=1}^T \hat\varepsilon_{it}^2,
  \label{eq:ch4_tstat_alpha}
\end{equation}
where $\hat\varepsilon_{it} = Y_{it}-\hat\alpha_i-\hat\vbeta_i^{\top}f_t$.  The
maximum statistic is
\begin{equation}
  T_{\max,\alpha} = \max_{1\le i\le N} \hat t_i^2.
  \label{eq:ch4_max_alpha_benchmark}
\end{equation}
A dense benchmark is the quadratic aggregate
\begin{equation}
  T_{\mathrm{sum},\alpha}
  = \frac{T\hat\valpha^{\top}\hat\mD^{-1}\hat\valpha - N}
         {\{2\widehat{\tr(\mR^2)}\}^{1/2}},
  \qquad
  \hat\mD = \diag(\hat\sigma_1^2,\ldots,\hat\sigma_N^2),
  \qquad
  \mR = \mD^{-1/2}\mSigma_\varepsilon\mD^{-1/2}.
  \label{eq:ch4_sum_alpha_benchmark}
\end{equation}
Under suitable Gaussian or sub-Gaussian assumptions, the centered maximum in
\eqref{eq:ch4_max_alpha_benchmark} has a type-I extreme-value limit, while the
quadratic form in \eqref{eq:ch4_sum_alpha_benchmark} is asymptotically normal.
The explicit robust counterpart below replaces $\hat\valpha$ and
$\hat\mSigma_\varepsilon$ by sign-based estimators that remain valid under
elliptical heavy tails.

\begin{assumption}[Gaussian and light-tailed benchmark conditions]
\label{ass:ch4_alpha_B1}
Assumption~\ref{ass:ch4_alpha_A1} holds.  In addition, the error vectors
$\vvarepsilon_t$ are independent over $t$ with $\E(\vvarepsilon_t)=\vct{0}$,
$\Cov(\vvarepsilon_t)=\mSigma_\varepsilon$, and positive diagonal scale matrix
$\mD=\diag(\sigma_1^2,\ldots,\sigma_N^2)$.
\end{assumption}

\begin{assumption}
\label{ass:ch4_alpha_B2}
As $\min(N,T)\to\infty$,
\begin{equation}
  \frac{\tr(\mR^4)}{\tr^2(\mR^2)}\longrightarrow 0,
  \qquad
  \frac{N^2}{T^2\tr(\mR^2)}\longrightarrow 0,
  \label{eq:ch4_alpha_B2}
\end{equation}
and either $\vvarepsilon_t$ is Gaussian or the standardized coordinates
$\sigma_i^{-1}\varepsilon_{it}$ are uniformly sub-Gaussian.
\end{assumption}

\begin{assumption}
\label{ass:ch4_alpha_B3}
There exists $\rho_0\in(0,1)$ such that $\max_{1\le i<j\le N}|r_{ij}|\le \rho_0$.
Moreover, there are sequences $\delta_N\downarrow 0$ and $\kappa_N\downarrow 0$ with
$\delta_N=o\{(\log N)^{-1}\}$ such that, writing
$B_{N,i}=\{j:|r_{ij}|\ge \delta_N\}$ and
$C_N=\{i:|B_{N,i}|\ge N^{\kappa_N}\}$,
\begin{equation}
  \frac{|C_N|}{N}\longrightarrow 0.
  \label{eq:ch4_alpha_B3}
\end{equation}
\end{assumption}

\begin{theorem}[Gaussian and light-tailed benchmark limits]
\label{thm:ch4_alpha_benchmark_limits}
Suppose Assumptions~\ref{ass:ch4_alpha_B1}--\ref{ass:ch4_alpha_B3} hold.
Under the null hypothesis $H_{0,\alpha}$,
\begin{align}
  T_{\mathrm{sum},\alpha}
  &\overset{d}{\longrightarrow} N(0,1),
  \label{eq:ch4_alpha_benchmark_sum_limit}\\
  \Prob\bigl(T_{\max,\alpha}-2\log N+\log\log N\le x\bigr)
  &\longrightarrow
  \exp\!\left\{-\pi^{-1/2}e^{-x/2}\right\},
  \qquad x\in\R.
  \label{eq:ch4_alpha_benchmark_max_limit}
\end{align}
Under local alternatives satisfying
\begin{equation}
  T\valpha^{\top}\mD^{-1}\valpha = O\!\bigl(\tr^{1/2}(\mR^2)\bigr),
  \label{eq:ch4_alpha_benchmark_local}
\end{equation}
one has
\begin{equation}
  T_{\mathrm{sum},\alpha}
  \overset{d}{\longrightarrow} N(\mu_{\mathrm{sum},\alpha},1),
  \qquad
  \mu_{\mathrm{sum},\alpha}
  = \lim_{\min(N,T)\to\infty}
    \frac{T\valpha^{\top}\mD^{-1}\valpha}{\{2\tr(\mR^2)\}^{1/2}},
  \label{eq:ch4_alpha_benchmark_sum_local}
\end{equation}
and the max test is consistent whenever
\begin{equation}
  \norm{\mD^{-1/2}\valpha}_{\infty}
  \ge C\sqrt{\frac{\log N}{T}}
  \label{eq:ch4_alpha_benchmark_sparse_detect}
\end{equation}
for a sufficiently large constant $C>0$.
\end{theorem}

Let
\begin{equation}
  p_{\mathrm{sum},\alpha}^{\mathrm G}
  = 1-\Phi\bigl(T_{\mathrm{sum},\alpha}\bigr),
  \qquad
  p_{\max,\alpha}^{\mathrm G}
  = 1-G_{\mathrm{EV}}\bigl(T_{\max,\alpha}-2\log N+\log\log N\bigr),
  \label{eq:ch4_alpha_benchmark_pvals}
\end{equation}
where $G_{\mathrm{EV}}(x)=\exp\{-\pi^{-1/2}e^{-x/2}\}$.  The benchmark
adaptive Cauchy combination is
\begin{equation}
  T_{\mathrm{Ada},\alpha}^{\mathrm G}
  = \frac12\tan\{\pi(1/2-p_{\mathrm{sum},\alpha}^{\mathrm G})\}
  + \frac12\tan\{\pi(1/2-p_{\max,\alpha}^{\mathrm G})\}.
  \label{eq:ch4_alpha_benchmark_cauchy}
\end{equation}

\begin{proposition}[A benchmark adaptive combination]
\label{prop:ch4_alpha_benchmark_adaptive}
Suppose, in addition to Assumptions~\ref{ass:ch4_alpha_B1}--\ref{ass:ch4_alpha_B3},
that $T_{\mathrm{sum},\alpha}$ and
$T_{\max,\alpha}-2\log N+\log\log N$ are asymptotically independent under
$H_{0,\alpha}$.  Then
\begin{equation}
  T_{\mathrm{Ada},\alpha}^{\mathrm G}
  \overset{d}{\longrightarrow} \mathrm{Cauchy}(0,1),
  \label{eq:ch4_alpha_benchmark_cauchy_limit}
\end{equation}
and the induced Cauchy-combination $p$-value is asymptotically valid.
\end{proposition}

\subsection{A robust spatial-sign sum-type test}

We now review the spatial-sign procedure of
\citet{LiuFengMa2023HeavyAlpha}.  Let
\begin{equation}
  \vZ_t = \vY_t - \hat\mB f_t,
  \qquad t=1,\ldots,T,
  \label{eq:ch4_zt_alpha}
\end{equation}
so that under the model \eqref{eq:ch4_lfpm_vector}, $\vZ_t$ is a location-shifted
version of the error vector.  Let $\mD = \diag(d_1^2,\ldots,d_N^2)$ denote the
marginal scale matrix, and let $\mR = \mD^{-1/2}\mSigma_\varepsilon\mD^{-1/2}$
be the correlation matrix associated with the error distribution.
Define the spatial sign
\begin{equation}
  U(\vx) = \frac{\vx}{\norm{\vx}}, \qquad U(\vct{0})=\vct{0}.
\end{equation}
The sign-transformed residuals are
\begin{equation}
  \vU_t = U\bigl(\hat\mD^{-1/2}(\vY_t-\hat\mB f_t)\bigr),
  \qquad t=1,\ldots,T.
  \label{eq:ch4_ut_alpha}
\end{equation}
The basic quadratic form proposed in \citet{LiuFengMa2023HeavyAlpha} is
\begin{equation}
  Q_{\alpha}
  = \frac{N}{\vh^{\top}\vh}
    \sum_{\substack{1\le t_1,t_2\le T\\ t_1\ne t_2}}
    h_{t_1}h_{t_2}\,\vU_{t_1}^{\top}\vU_{t_2}.
  \label{eq:ch4_Q_alpha}
\end{equation}
The variance of $Q_{\alpha}$ is asymptotically $2\tr(\mR^2)$.  A leave-two-out
estimator of $\tr(\mR^2)$ is obtained by
\begin{equation}
  \widehat{\tr(\mR^2)}
  = \frac{N^2}{(\vh^{\top}\vh)^2}
  \sum_{\substack{t_1,t_2=1\\ t_1\ne t_2}}^T h_{t_1}^2 h_{t_2}^2
  \bigl\{U(\hat\mD_{(t_1,t_2)}^{-1/2}\hat\vvarepsilon_{t_1}^{(t_1,t_2)})^{\top}
          U(\hat\mD_{(t_1,t_2)}^{-1/2}\hat\vvarepsilon_{t_2}^{(t_1,t_2)})
  \bigr\}^2,
  \label{eq:ch4_trR2_alpha_hat}
\end{equation}
where the superscript $(t_1,t_2)$ indicates that the corresponding observations
are left out in the scale estimation step.  The studentized statistic is
\begin{equation}
  T_{\mathrm{SS},\alpha}
  = \frac{Q_{\alpha}}{\{2\widehat{\tr(\mR^2)}\}^{1/2}}.
  \label{eq:ch4_tss_alpha}
\end{equation}

The assumptions below are stated in the notation of the book and are sufficient
for the null and local-alternative theory.  They are adapted from the conditions
used in \citet{LiuFengMa2023HeavyAlpha}.

\begin{assumption}
\label{ass:ch4_alpha_A1}
The factor process $\{f_t\}_{t=1}^T$ is strictly stationary with finite second
moments.  The matrix $T^{-1}\mF^{\top}\mF$ converges to a positive-definite limit
$\mQ_f$, and
\begin{equation}
  T^{-1}\vh^{\top}\vh \longrightarrow \omega,
  \qquad
  \omega = 1 - \E(f_t)^{\top}\{\E(f_tf_t^{\top})\}^{-1}\E(f_t) \in (0,1].
  \label{eq:ch4_alpha_omega}
\end{equation}
\end{assumption}

\begin{assumption}
\label{ass:ch4_alpha_A2}
The error vectors satisfy
\begin{equation}
  \vvarepsilon_t = \mSigma_\varepsilon^{1/2}\vW_t,
  \qquad t=1,\ldots,T,
  \label{eq:ch4_alpha_error_repr}
\end{equation}
where $\vW_t$ are independent and identically distributed spherically symmetric
random vectors in $\R^N$ with $\Prob(\vW_t=\vct{0})=0$.  Let
\begin{equation}
  c_1 = \E\bigl(\norm{\mSigma_\varepsilon^{-1/2}\vvarepsilon_t}^{-1}\bigr),
  \qquad
  c_2 = \E\bigl(\norm{\mSigma_\varepsilon^{-1/2}\vvarepsilon_t}^{-2}\bigr),
  \label{eq:ch4_alpha_c1c2}
\end{equation}
which are finite.
\end{assumption}

\begin{assumption}
\label{ass:ch4_alpha_A3}
As $\min(N,T)\to\infty$,
\begin{equation}
  \frac{\tr(\mR^4)}{\tr^2(\mR^2)} \longrightarrow 0,
  \qquad
  \frac{N^2}{T^2\tr(\mR^2)} \longrightarrow 0,
  \qquad
  \frac{\max_{1\le i\le N} r_{ii}^2}{\tr(\mR^2)} \longrightarrow 0.
  \label{eq:ch4_alpha_A3_eq}
\end{equation}
\end{assumption}

\begin{assumption}
\label{ass:ch4_alpha_A4}
Under a sequence of local alternatives,
\begin{equation}
  \valpha^{\top}\mD^{-1}\valpha = O\!\left(\frac{\tr^{1/2}(\mR^2)}{TN}\right),
  \qquad
  \valpha^{\top}\mD^{-1}\mR\mD^{-1}\valpha = o\!\left(\frac{\tr(\mR^2)}{TN}
  \right).
  \label{eq:ch4_alpha_A4_eq}
\end{equation}
\end{assumption}

\begin{theorem}
\label{thm:ch4_alpha_ss_null}
Under Assumptions~\ref{ass:ch4_alpha_A1}--\ref{ass:ch4_alpha_A3} and the null
hypothesis $H_{0,\alpha}$,
\begin{equation}
  T_{\mathrm{SS},\alpha} \overset{d}{\longrightarrow} N(0,1)
  \label{eq:ch4_alpha_ss_null_limit}
\end{equation}
as $\min(N,T)\to\infty$.
\end{theorem}

\begin{theorem}
\label{thm:ch4_alpha_ss_local}
Under Assumptions~\ref{ass:ch4_alpha_A1}--\ref{ass:ch4_alpha_A4},
\begin{equation}
  T_{\mathrm{SS},\alpha}
  \overset{d}{\longrightarrow} N(\mu_{\mathrm{SS},\alpha},1),
  \label{eq:ch4_alpha_ss_local_limit}
\end{equation}
where
\begin{equation}
  \mu_{\mathrm{SS},\alpha}
  = \lim_{\min(N,T)\to\infty}
    \frac{\phi_{\alpha}^2}{\{2\tr(\mR^2)\}^{1/2}},
  \qquad
  \phi_{\alpha}^2 = \omega TN c_1^2\,\valpha^{\top}\mD^{-1}\valpha.
  \label{eq:ch4_alpha_ss_local_mean}
\end{equation}
Consequently, the asymptotic power at one-sided level $\gamma$ is
\begin{equation}
  \beta_{\mathrm{SS},\alpha}(\gamma)
  = 1-\Phi\{z_{1-\gamma}-\mu_{\mathrm{SS},\alpha}\}.
  \label{eq:ch4_alpha_ss_power}
\end{equation}
\end{theorem}

The key feature of \eqref{eq:ch4_alpha_ss_local_mean} is that only the diagonal
scale matrix $\mD$ and the elliptical sign constant $c_1$ appear in the signal
term.  The test therefore remains valid even when fourth moments are absent.
This is the essential gain over covariance-based procedures.

\subsection{Weighted spatial-sign tests and the INST procedure}

The unweighted statistic \eqref{eq:ch4_tss_alpha} is optimal only for a specific radial
configuration.  \citet{ZhaoChenZi2022INSTAlpha} broadened the construction to a
class of weighted spatial-sign tests and showed that inverse-norm weighting is locally
optimal within that class.  Their work is the direct alpha-testing analogue of the
weighted sign methodology that we studied in Chapter~2.

For a measurable weight function $K:[0,\infty)\to\R$, define
\begin{equation}
  r_t = \norm{\hat\mD^{-1/2}(\vY_t-\hat\valpha-\hat\mB f_t)},
  \qquad
  \hat\psi_{2,K} = \frac1T\sum_{t=1}^T K^2(r_t),
  \label{eq:ch4_alpha_INST_rt}
\end{equation}
and the weighted quadratic form
\begin{equation}
  Q_{K,\alpha}
  = \frac{N}{\vh^{\top}\vh}
    \sum_{\substack{1\le t_1,t_2\le T\\ t_1\ne t_2}}
    h_{t_1}h_{t_2}
    K(r_{t_1})K(r_{t_2})
    \vU_{t_1}^{\top}\vU_{t_2}.
  \label{eq:ch4_alpha_INST_Q}
\end{equation}
The associated studentized statistic is
\begin{equation}
  T_{K,\alpha}
  = \frac{Q_{K,\alpha}}{\{2\hat\psi_{2,K}^2\widehat{\tr(\mR^2)}\}^{1/2}}.
  \label{eq:ch4_alpha_INST_stat}
\end{equation}
Let
\begin{equation}
  \psi_{1,K}=\E\{K(r_t)r_t^{-1}\},
  \qquad
  \psi_{2,K}=\E\{K^2(r_t)\},
  \label{eq:ch4_alpha_INST_psi}
\end{equation}
where the expectation is taken under the elliptical radial model of
Assumption~\ref{ass:ch4_alpha_A2}.

\begin{theorem}[Weighted spatial-sign alpha tests]
\label{thm:ch4_alpha_INST}
Suppose Assumptions~\ref{ass:ch4_alpha_A1}--\ref{ass:ch4_alpha_A4} hold and that
$\psi_{2,K}<\infty$.  Then under $H_{0,\alpha}$,
\begin{equation}
  T_{K,\alpha}\overset{d}{\longrightarrow} N(0,1).
  \label{eq:ch4_alpha_INST_null}
\end{equation}
Under local alternatives satisfying Assumption~\ref{ass:ch4_alpha_A4},
\begin{equation}
  T_{K,\alpha}\overset{d}{\longrightarrow} N(\mu_{K,\alpha},1),
  \qquad
  \mu_{K,\alpha}
  = \lim_{\min(N,T)\to\infty}
    \frac{\omega TN\psi_{1,K}^2\,\valpha^{\top}\mD^{-1}\valpha}
         {\{2\psi_{2,K}^2\tr(\mR^2)\}^{1/2}}.
  \label{eq:ch4_alpha_INST_local}
\end{equation}
\end{theorem}

\begin{proposition}[Optimality of inverse-norm weighting]
\label{prop:ch4_alpha_INST_opt}
For every admissible weight $K$,
\begin{equation}
  \frac{\psi_{1,K}^2}{\psi_{2,K}}
  \le \E(r_t^{-2}),
  \label{eq:ch4_alpha_INST_optineq}
\end{equation}
with equality if and only if $K(t)=c\,t^{-1}$ almost surely for some nonzero
constant $c$.  Consequently, the locally most powerful member of the weighted
spatial-sign family is obtained by setting $K(t)=t^{-1}$, which yields the
inverse-norm sign test (INST)
\begin{equation}
  T_{\mathrm{INST},\alpha}
  = T_{K,\alpha}\big|_{K(t)=t^{-1}}.
  \label{eq:ch4_alpha_INST_def}
\end{equation}
\end{proposition}

\subsection{A robust max-type test and its adaptive combination}

To gain power under sparse alternatives, one needs a coordinatewise statistic.
The recent paper \citet{ZhaoFengWangWang2024RobustAlpha} proceeds by using the
scaled spatial median of the transformed residuals.  Let $\vZ_t$ be defined by
\eqref{eq:ch4_zt_alpha}.  The pair $(\hat\vtheta,\hat\mD)$ is defined as the
solution to the equations
\begin{align}
  \frac1T\sum_{t=1}^T U\bigl(\mD^{-1/2}(\vZ_t-\vtheta)\bigr) &= \vct{0},
  \label{eq:ch4_alpha_scaled_sm_1}\\
  \frac1T\sum_{t=1}^T \diag\!\left[
    U\bigl(\mD^{-1/2}(\vZ_t-\vtheta)\bigr)
    U\bigl(\mD^{-1/2}(\vZ_t-\vtheta)\bigr)^{\top}
  \right] &= N^{-1}\mI_N.
  \label{eq:ch4_alpha_scaled_sm_2}
\end{align}
The estimator $\hat\vtheta$ is a robust analogue of the OLS alpha estimator.
The max-type statistic is then
\begin{equation}
  T_{\mathrm{SM},\alpha}
  = T\hat\zeta\,\norm{\hat\mD^{-1/2}\hat\vtheta}_{\infty}^2
    - 2\log N + \log\log N,
  \label{eq:ch4_alpha_tsm}
\end{equation}
where $\hat\zeta$ is a consistent estimator of the nuisance constant
\begin{equation}
  \zeta = \frac{N}{\{\E(r_t^{-1})\}^2}
  \Bigl[1-2\eta\,\E(r_t^{-1})\E(r_t) + \eta^2\E(r_t^2)\E(r_t^{-2})\Bigr]^{-1},
  \label{eq:ch4_alpha_zeta}
\end{equation}
with $r_t = \norm{\mD^{-1/2}(\vZ_t-\vtheta)}$ and
$\eta = T^{-1}\vct{1}_T^{\top}\mF(\mF^{\top}\mF)^{-1}\mF^{\top}\vct{1}_T$.
A sample version of \eqref{eq:ch4_alpha_zeta} is obtained by replacing the
expectations of $r_t, r_t^{-1}, r_t^2, r_t^{-2}$ with their empirical averages
computed from
\begin{equation}
  \tilde r_t = \norm{\hat\mD^{-1/2}(\vZ_t-\hat\vtheta)}.
  \label{eq:ch4_alpha_rtilde}
\end{equation}

The assumptions below follow the same scaling convention as in the original
paper, but are stated directly in the notation of this book.

\begin{assumption}
\label{ass:ch4_alpha_C1}
The factor process satisfies Assumption~\ref{ass:ch4_alpha_A1}.  In addition,
$\log N = o(T^{1/10})$ and $\max_{1\le i\le N}\Var(\varepsilon_{it})$ is bounded
uniformly in $(N,T)$.
\end{assumption}

\begin{assumption}
\label{ass:ch4_alpha_C2}
The residual vectors admit the decomposition
\begin{equation}
  \vZ_t = \omega_T T^{-1}\valpha + r_t\mD^{1/2}\vU_t,
  \label{eq:ch4_alpha_C2_repr}
\end{equation}
where $\omega_T = \vh^{\top}\vh/T$, $r_t>0$, $\vU_t$ is uniformly distributed on
$\spn^{N-1}$, and $r_t$ is independent of $\vU_t$.  The moments
$\E(r_t^a)$ are finite for $a\in\{-2,-1,1,2\}$.
\end{assumption}

\begin{assumption}
\label{ass:ch4_alpha_C3}
The correlation matrix $\mR$ satisfies
\begin{equation}
  \max_{1\le i<j\le N} |r_{ij}| \le \rho_0 < 1,
  \qquad
  \sum_{j=1}^N |r_{ij}|^q \le C_q
  \quad\text{for some } q\in[0,1),
  \label{eq:ch4_alpha_C3_sparsecorr}
\end{equation}
uniformly in $i$, and
\begin{equation}
  \min_{1\le i\le N} d_i^2 \ge c_0,
  \qquad
  \max_{1\le i\le N} d_i^2 \le C_0
  \label{eq:ch4_alpha_C3_scale}
\end{equation}
for positive constants $c_0$ and $C_0$.
\end{assumption}

\begin{assumption}
\label{ass:ch4_alpha_C4}
Under the null hypothesis,
\begin{equation}
  \max_{1\le i\le N} |\alpha_i| = 0.
  \label{eq:ch4_alpha_C4_null}
\end{equation}
Under sparse alternatives used for power calculations, there exists a set
$\mathcal A\subset\{1,\ldots,N\}$ with cardinality $s_N=|\mathcal A|$ such that
\begin{equation}
  \alpha_i \neq 0 \iff i\in\mathcal A,
  \qquad
  \norm{\valpha}_{\infty} \ge C\sqrt{\frac{\log N}{T}},
  \qquad
  \norm{\valpha}_{2} = O\!\left(\frac{N\log N}{T}\right)
  \label{eq:ch4_alpha_C4_alt}
\end{equation}
for some constant $C>0$.
\end{assumption}

\begin{theorem}
\label{thm:ch4_alpha_max_null}
Under Assumptions~\ref{ass:ch4_alpha_C1}--\ref{ass:ch4_alpha_C4} and the null
hypothesis,
\begin{equation}
  \Prob\bigl(T_{\mathrm{SM},\alpha}\le x\bigr)
  \longrightarrow
  \exp\!\left\{-\pi^{-1/2}e^{-x/2}\right\},
  \qquad x\in\R.
  \label{eq:ch4_alpha_max_null_limit}
\end{equation}
That is, the statistic \eqref{eq:ch4_alpha_tsm} converges to a type-I extreme
value distribution.
\end{theorem}

\begin{theorem}
\label{thm:ch4_alpha_max_power}
Under Assumptions~\ref{ass:ch4_alpha_C1}--\ref{ass:ch4_alpha_C4}, if the sparse
signal condition in \eqref{eq:ch4_alpha_C4_alt} holds with a sufficiently large
constant $C$, then for every fixed significance level $\gamma\in(0,1)$,
\begin{equation}
  \Prob_{H_{1,\alpha}}\bigl(T_{\mathrm{SM},\alpha}>q_{1-\gamma}^{\mathrm{EV}}
  \bigr) \longrightarrow 1,
  \label{eq:ch4_alpha_max_consistency}
\end{equation}
where $q_{1-\gamma}^{\mathrm{EV}}$ is the $(1-\gamma)$-quantile of the limit in
\eqref{eq:ch4_alpha_max_null_limit}.
\end{theorem}

The asymptotic independence between the dense-oriented statistic
$T_{\mathrm{SS},\alpha}$ and the sparse-oriented statistic $T_{\mathrm{SM},\alpha}$
leads to an adaptive procedure.  Let
\begin{equation}
  p_{\mathrm{SS},\alpha}=1-\Phi(T_{\mathrm{SS},\alpha}),
  \qquad
  p_{\mathrm{SM},\alpha}=1-G_{\mathrm{EV}}(T_{\mathrm{SM},\alpha}),
  \label{eq:ch4_alpha_pvalues}
\end{equation}
where $G_{\mathrm{EV}}(x)=\exp\{-\pi^{-1/2}e^{-x/2}\}$.  The truncated Cauchy
combination is
\begin{equation}
  T_{\mathrm{CC},\alpha}
  = \frac12\tan\{\pi(1/2-p_{\mathrm{SS},\alpha})\}\mathbbm 1(p_{\mathrm{SS},\alpha}<1/2)
  + \frac12\tan\{\pi(1/2-p_{\mathrm{SM},\alpha})\}\mathbbm 1(p_{\mathrm{SM},\alpha}<1/2).
  \label{eq:ch4_alpha_cc}
\end{equation}

\begin{theorem}
\label{thm:ch4_alpha_independence}
Suppose that Assumptions~\ref{ass:ch4_alpha_A1}--\ref{ass:ch4_alpha_A4} and
\ref{ass:ch4_alpha_C1}--\ref{ass:ch4_alpha_C4} hold, that
$\log N=o(T^{1/10})$, and that the sparse alternative in
\eqref{eq:ch4_alpha_C4_alt} is satisfied.  Then for every fixed $(x,y)\in\R^2$,
\begin{equation}
  \Prob\bigl(T_{\mathrm{SS},\alpha}\le x,
             T_{\mathrm{SM},\alpha}\le y\bigr)
  - \Prob\bigl(T_{\mathrm{SS},\alpha}\le x\bigr)
    \Prob\bigl(T_{\mathrm{SM},\alpha}\le y\bigr)
  \longrightarrow 0.
  \label{eq:ch4_alpha_independence_eq}
\end{equation}
Consequently, the combination statistic \eqref{eq:ch4_alpha_cc} has a valid null
limit given by Theorem~\ref{thm:ch4_generic_combination}.
\end{theorem}

\subsection{Temporal dependence and $L_q$ extensions for unconditional models}

The basic spatial-sign and max-type procedures have already been extended in
several directions for unconditional linear factor models.  Under temporal
dependence of the returns, \citet{MaFengWang2025DependentAlpha} and
\citet{MaFengWang2026TimeVaryingAlpha} replace the independent residual theory
by long-run covariance normalization and still obtain dense-, sparse-, and
adaptive tests.  The most recent $L_q$-type family of
\citet{ZhaoMaFeng2026LqAlpha} interpolates between sum-type and max-type
statistics by considering
\begin{equation}
  T_{q,\alpha} = \sum_{i=1}^N |\hat\theta_i|^q,
  \qquad 2\le q<\infty,
  \label{eq:ch4_alpha_lq}
\end{equation}
with $q$ controlling the degree of sparsity sensitivity.  These extensions fit
naturally into the max--sum paradigm introduced at the beginning of this
chapter.

\section{Robust mutual fund selection with false discovery rate control}
\idx{mutual fund selection}\idx{false discovery rate control}\idx{Benjamini--Hochberg procedure}

The global alpha tests in Section~4.3 answer the question whether \emph{all} pricing
errors vanish jointly.  In empirical fund evaluation, however, the inferential target is
usually finer: one wants to identify the individual funds with positive alpha while
controlling the false discovery rate (FDR).  This leads to a large-scale one-sided testing
problem in which the same heavy-tailed features that motivated spatial-sign global testing
also invalidate standard Gaussian multiple-testing procedures.

Suppose again that returns follow the linear factor pricing model
\begin{equation}
  Y_{it} = \alpha_i + \vbeta_i^{\top}f_t + \varepsilon_{it},
  \qquad i=1,\ldots,N,\quad t=1,\ldots,T,
  \label{eq:ch4_fdr_model}
\end{equation}
with observable factor vector $f_t\in\R^K$.  For mutual fund selection, the relevant
multiple hypotheses are
\begin{equation}
  H_{0i}:\ \alpha_i\le 0,
  \qquad
  H_{1i}:\ \alpha_i>0,
  \qquad i=1,\ldots,N.
  \label{eq:ch4_fdr_hypotheses}
\end{equation}
The one-sided formulation is natural because positive alpha corresponds to superior stock
selection ability, whereas large negative alpha is not a discovery of practical interest.

\subsection{Observable factors: the spatial-sign BH procedure}

Let $\hat\varepsilon_{it}=Y_{it}-\hat\alpha_i-\hat\vbeta_i^{\top}f_t$ be the OLS residuals from
Section~4.3, and write $\hat\vvarepsilon_{\cdot t}=(\hat\varepsilon_{1t},\ldots,
\hat\varepsilon_{Nt})^{\top}$.  Following the robust alpha-testing literature, estimate a
location vector $\vtheta$ and a diagonal scale matrix $\mD$ by solving
\begin{align}
  \frac1T\sum_{t=1}^T U\{\mD^{-1/2}(\hat\vvarepsilon_{\cdot t}-\vtheta)\}
  &= \vct{0},
  \label{eq:ch4_fdr_thetaeq1}\\
  \frac1T\sum_{t=1}^T \diag\Bigl[
    U\{\mD^{-1/2}(\hat\vvarepsilon_{\cdot t}-\vtheta)\}
    U\{\mD^{-1/2}(\hat\vvarepsilon_{\cdot t}-\vtheta)\}^{\top}
  \Bigr]
  &= N^{-1}\mI_N.
  \label{eq:ch4_fdr_thetaeq2}
\end{align}
Let $(\hat\vtheta,\hat\mD)$ denote the solution, $\hat d_i$ the $i$th diagonal entry of
$\hat\mD$, and
\begin{equation}
  \hat r_t = \norm{\hat\mD^{-1/2}(\hat\vvarepsilon_{\cdot t}-\hat\vtheta)}_2,
  \qquad
  \hat\zeta_1 = T^{-1}\sum_{t=1}^T \hat r_t^{-1},
  \qquad
  \hat\omega = T^{-1}\vh^{\top}\vh.
  \label{eq:ch4_fdr_hat_quantities}
\end{equation}
The fundwise spatial-sign statistics are then
\begin{equation}
  Z_{i}^{\mathrm{SS}}
  = \sqrt{T}\,\hat\zeta_1\,\hat d_i^{-1/2}\hat\theta_i,
  \qquad i=1,\ldots,N,
  \label{eq:ch4_fdr_zi}
\end{equation}
and the corresponding one-sided $p$-values are
\begin{equation}
  p_i^{\mathrm{SS}} = 1-\Phi\bigl(Z_i^{\mathrm{SS}}\bigr),
  \qquad i=1,\ldots,N.
  \label{eq:ch4_fdr_pi}
\end{equation}
For a target FDR level $q\in(0,1)$, let $p_{(1)}^{\mathrm{SS}}\le\cdots\le p_{(N)}^{\mathrm{SS}}$
be the ordered $p$-values and define
\begin{equation}
  \hat k_{\mathrm{SS}} = \max\Bigl\{1\le k\le N:
  p_{(k)}^{\mathrm{SS}} \le qk/N\Bigr\},
  \qquad
  \mathcal R_{\mathrm{SS}}(q)
  = \Bigl\{i: p_i^{\mathrm{SS}}\le p_{(\hat k_{\mathrm{SS}})}^{\mathrm{SS}}\Bigr\},
  \label{eq:ch4_fdr_bh}
\end{equation}
with the convention $\mathcal R_{\mathrm{SS}}(q)=\varnothing$ if the maximizing set is empty.
This is the spatial-sign Benjamini--Hochberg procedure (SS-BH).

\begin{assumption}
\label{ass:ch4_fdr_A1}
The observable-factor model \eqref{eq:ch4_fdr_model} holds.  The error vectors
$\vvarepsilon_t$ are independent over $t$, centered, elliptically symmetric, and satisfy the
same radial-moment and weak cross-sectional dependence conditions as in
Assumptions~\ref{ass:ch4_alpha_C1}--\ref{ass:ch4_alpha_C4}.  In particular, the diagonal scale
matrix $\mD$ is positive definite and the correlation matrix
$\mR=\mD^{-1/2}\mSigma_\varepsilon\mD^{-1/2}$ obeys the sparse-correlation condition
\eqref{eq:ch4_alpha_B3}.  Moreover, $\log N=o(T^{1/10})$.
\end{assumption}

\begin{theorem}[Observable-factor spatial-sign multiple testing]
\label{thm:ch4_fdr_ssbh}
Suppose Assumption~\ref{ass:ch4_fdr_A1} holds.  Let
$\vU_t=U(\mD^{-1/2}\vvarepsilon_t)$, let $\mW=(\vct{1}_T,\mF)$ be the full design matrix,
and write $V_{st}$ for the $(s,t)$ entry of
$\mP_W=\mW(\mW^{\top}\mW)^{-1}\mW^{\top}$.  Then
\begin{equation}
  \sqrt{T}\,\hat\zeta_1\,\hat\mD^{-1/2}(\hat\vtheta-\hat\omega\valpha)
  = T^{-1/2}\sum_{t=1}^T \zeta_1^{-1}
    \Bigl(1-\sum_{s=1}^T r_s^{-1}r_tV_{st}\Bigr)\vU_t + \mC_T,
  \label{eq:ch4_fdr_bahadur}
\end{equation}
where $\zeta_1=\E(r_t^{-1})$ and
\begin{equation}
  \norm{\mC_T}_{\infty} = o_P\bigl((\log N)^{-1/2}\bigr).
  \label{eq:ch4_fdr_bahadur_remainder}
\end{equation}
Consequently, there exists a Gaussian vector $\vG\sim N(\vct{0},\mXi)$ such that
\begin{equation}
  \sup_{z\in\R^N}
  \Bigl|
    \Prob\bigl(\sqrt{T}\,\hat\zeta_1\,\hat\mD^{-1/2}(\hat\vtheta-\hat\omega\valpha)\le z\bigr)
    - \Prob(\vG\le z)
  \Bigr| \longrightarrow 0,
  \label{eq:ch4_fdr_gaussapprox}
\end{equation}
with $\Xi_{ii}=1$ and $\max_{i\ne j}|\Xi_{ij}|\le \rho_0<1$.  In particular, under
$H_{0i}:\alpha_i\le0$, each null $p$-value in \eqref{eq:ch4_fdr_pi} is asymptotically
conservative, and the BH procedure \eqref{eq:ch4_fdr_bh} satisfies
\begin{equation}
  \operatorname*{FDR}\bigl\{\mathcal R_{\mathrm{SS}}(q)\bigr\}
  \le q + o(1).
  \label{eq:ch4_fdr_control}
\end{equation}
If, moreover, $\alpha_i\ge C\sqrt{(\log N)/T}$ on a nonvanishing set of truly skilled funds,
then the corresponding coordinates are selected with probability tending to one.
\end{theorem}

\subsection{Latent factors: factor-adjusted spatial-sign BH}

When omitted or latent factors are present, directly applying SS-BH to the residuals from the
observable-factor regression may be overly conservative because the resulting $p$-values remain
strongly dependent.  The factor-adjusted spatial-sign procedure in
\citet{WangZhaoFengWang2025MutualFundFDR} therefore combines robust multiple testing with
robust latent-factor extraction.  Specifically, suppose the return vector admits the decomposition
\begin{equation}
  \vY_t = \valpha + \mB f_t + \mLambda g_t + \vu_t,
  \qquad t=1,\ldots,T,
  \label{eq:ch4_fdr_latent_model}
\end{equation}
where $g_t\in\R^r$ is an unobserved factor vector and $\vu_t$ is the idiosyncratic component.
Under elliptical tails, a robust way to estimate the latent factor space is to apply principal
component analysis to the spatial Kendall's tau matrix rather than to the sample covariance
matrix, following \citet{HeKongYuZhang2022FactorNoMoments}.  Let $\hat g_t$ and
$\hat\mLambda$ denote the resulting factor estimates, and define factor-adjusted residuals
\begin{equation}
  \check\vvarepsilon_{\cdot t}
  = \vY_t - \hat\valpha - \hat\mB f_t - \hat\mLambda \hat g_t.
  \label{eq:ch4_fdr_factor_adjusted_residuals}
\end{equation}
Applying the spatial-sign equations \eqref{eq:ch4_fdr_thetaeq1}--\eqref{eq:ch4_fdr_thetaeq2} to
$\{\check\vvarepsilon_{\cdot t}\}_{t=1}^T$ yields a new location estimator
$\check\vtheta$ and diagonal scale matrix $\check\mD$, and hence factor-adjusted test statistics
\begin{equation}
  Z_i^{\mathrm{FSS}}
  = \sqrt{T}\,\check\zeta_1\,\check d_i^{-1/2}\check\theta_i,
  \qquad
  p_i^{\mathrm{FSS}} = 1-\Phi\bigl(Z_i^{\mathrm{FSS}}\bigr),
  \qquad i=1,\ldots,N.
  \label{eq:ch4_fdr_fss_stats}
\end{equation}
The factor-adjusted spatial-sign BH procedure (FSS-BH) is then obtained by applying the BH
step-up rule to the ordered values of $\{p_i^{\mathrm{FSS}}\}_{i=1}^N$.

\begin{assumption}
\label{ass:ch4_fdr_A2}
The latent-factor model \eqref{eq:ch4_fdr_latent_model} holds with fixed latent factor number
$r$.  The common component is pervasive, the idiosyncratic errors satisfy the same elliptical
and weak-dependence conditions as in Assumption~\ref{ass:ch4_fdr_A1}, and the robust factor
estimator obeys
\begin{equation}
  \max_{1\le i\le N}|\check\alpha_i-\alpha_i|
  = o_P\bigl((T\log N)^{-1/2}\bigr).
  \label{eq:ch4_fdr_factor_rate}
\end{equation}
\end{assumption}

\begin{theorem}[Factor-adjusted spatial-sign multiple testing]
\label{thm:ch4_fdr_fssbh}
Suppose Assumption~\ref{ass:ch4_fdr_A2} holds.  Then the Gaussian approximation
\eqref{eq:ch4_fdr_gaussapprox} continues to hold with
$\hat\vtheta,\hat\mD,\hat\zeta_1$ replaced by
$\check\vtheta,\check\mD,\check\zeta_1$.  Consequently, the BH procedure based on the
factor-adjusted $p$-values in \eqref{eq:ch4_fdr_fss_stats} satisfies
\begin{equation}
  \operatorname*{FDR}\bigl\{\mathcal R_{\mathrm{FSS}}(q)\bigr\}
  \le q + o(1).
  \label{eq:ch4_fdr_fss_control}
\end{equation}
In heavy-tailed settings with latent dependence, the FSS-BH procedure is therefore asymptotically
valid and typically less conservative than the observable-factor SS-BH procedure.
\end{theorem}

The main conceptual point is that mutual fund selection is not a different inferential problem
from alpha testing; it is the multiple-testing version of the same location problem.  What changes
is the inferential loss function: instead of controlling a single type-I error probability, one
must control the expected proportion of false rejections among all selected funds.  The
spatial-sign methodology provides the same robustness benefits in this setting as it does for
global alpha tests.

\section{Testing alpha in conditional time-varying factor models}
\idx{conditional factor model}\idx{time-varying factor model}

\subsection{Model, spline approximation, and low-dimensional Wald tests}

In conditional factor models the pricing error and factor loadings are allowed to vary with
time or with a normalized calendar index.  A standard formulation, used in
\citet{LiYang2011ConditionalFactor}, \citet{AngKristensen2012ConditionalFactor}, and
\citet{MaLanSuTsai2020ConditionalHDA}, is
\begin{equation}
  Y_{it}=\alpha_i(t/T)+\vbeta_i(t/T)^{\top}f_t+\varepsilon_{it},
  \qquad i=1,\ldots,N,\quad t=1,\ldots,T.
  \label{eq:ch4_cond_model_raw}
\end{equation}
Write
\begin{equation}
  \delta_i = T^{-1}\sum_{t=1}^T \alpha_i(t/T),
  \qquad
  \delta_i(t/T)=\alpha_i(t/T)-\delta_i.
  \label{eq:ch4_cond_delta_def}
\end{equation}
Then \eqref{eq:ch4_cond_model_raw} becomes
\begin{equation}
  Y_{it}=\delta_i+\delta_i(t/T)+\vbeta_i(t/T)^{\top}f_t+\varepsilon_{it},
  \label{eq:ch4_cond_model}
\end{equation}
and the null hypothesis of interest is
\begin{equation}
  H_{0,\mathrm{cond}}:\ \delta_1=\cdots=\delta_N=0.
  \label{eq:ch4_cond_h0}
\end{equation}

Let $B(t/T)=(B_1(t/T),\ldots,B_L(t/T))^{\top}$ be a B-spline basis of order $q$,
let
\begin{equation}
  \tilde B_k(t/T)=B_k(t/T)-T^{-1}\sum_{s=1}^T B_k(s/T),
  \qquad
  \tilde B(t/T)=(\tilde B_1(t/T),\ldots,\tilde B_L(t/T))^{\top},
  \label{eq:ch4_cond_tildeB}
\end{equation}
and define the sieve regressor
\begin{equation}
  Z_t=
  \Bigl(\tilde B(t/T)^{\top},\ f_t^{\top}\otimes B(t/T)^{\top}\Bigr)^{\top}
  \in\R^{(K+1)L}.
  \label{eq:ch4_cond_Zt}
\end{equation}
Approximating $\delta_i(\cdot)$ and the components of $\vbeta_i(\cdot)$ by spline series,
one obtains the working regression
\begin{equation}
  Y_{it}\approx \delta_i + \vlambda_i^{\top} Z_t + \varepsilon_{it},
  \qquad i=1,\ldots,N,
  \label{eq:ch4_cond_working}
\end{equation}
with least-squares estimator
\begin{equation}
  \hat\vlambda_i=(\mZ^{\top}\mZ)^{-1}\mZ^{\top}Y_{i\cdot},
  \qquad
  \mZ=(Z_1,\ldots,Z_T)^{\top}.
  \label{eq:ch4_cond_lambda_hat}
\end{equation}
The residuals are
\begin{equation}
  \hat\varepsilon_{it}=Y_{it}-\hat\vlambda_i^{\top}Z_t,
  \qquad
  \hat\vvarepsilon_{\cdot t}=(\hat\varepsilon_{1t},\ldots,\hat\varepsilon_{Nt})^{\top}.
  \label{eq:ch4_cond_residuals}
\end{equation}

When $N$ is fixed and $T\to\infty$, the classical nonparametric Wald tests of
\citet{LiYang2011ConditionalFactor} and
\citet{AngKristensen2012ConditionalFactor} are based on the spline or kernel estimator
$\hat\vdelta=(\hat\delta_1,\ldots,\hat\delta_N)^{\top}$ and its asymptotic covariance
matrix $\hat\mOmega_\delta$.  The benchmark statistic is
\begin{equation}
  W_{T,\mathrm{cond}} = T\hat\vdelta^{\top}\hat\mOmega_\delta^{-1}\hat\vdelta.
  \label{eq:ch4_cond_wald}
\end{equation}
Under fixed $N$ and the usual smoothness and bandwidth conditions,
$W_{T,\mathrm{cond}}\overset{d}{\longrightarrow}\chi_N^2$ under
\eqref{eq:ch4_cond_h0}.  This classical theory breaks down when $N$ is of the same order as,
or larger than, $T$.

\subsection{High-dimensional light-tailed sum, max, and adaptive tests}

Following \citet{MaLanSuTsai2020ConditionalHDA}, let
\begin{equation}
  \vh = \mM_Z\vct{1}_T,
  \qquad
  \mM_Z = \mI_T-\mZ(\mZ^{\top}\mZ)^{-1}\mZ^{\top},
  \qquad
  \omega_T = \vh^{\top}\vh.
  \label{eq:ch4_cond_hvec}
\end{equation}
The dense-oriented high-dimensional alpha statistic is
\begin{equation}
  J_{NT}
  = N^{-1}\sum_{i=1}^N
    \Bigl(T^{-1/2}\sum_{t=1}^T \hat\varepsilon_{it}\Bigr)^2.
  \label{eq:ch4_cond_JNT}
\end{equation}
Writing $\mOmega=\Cov(\vvarepsilon_{\cdot t})$, its standardized version is
\begin{equation}
  T_{\mathrm{HDA},\mathrm{cond}}
  = \frac{J_{NT}-N^{-1}\tr(\mOmega)}{\{2N^{-2}\tr(\mOmega^2)\}^{1/2}}.
  \label{eq:ch4_cond_HDA}
\end{equation}
A sparse-oriented Gaussian benchmark, used in
\citet{MaFengWangBao2024ConditionalAlpha}, is
\begin{equation}
  M_{NT}
  = \max_{1\le i\le N}
    T^{-1}\hat\sigma_{ii}^{-1}(\hat\varepsilon_{i\cdot}^{\top}\vct{1}_T)^2,
  \qquad
  \hat\sigma_{ij}=\frac{\hat\varepsilon_{i\cdot}^{\top}\hat\varepsilon_{j\cdot}}{T-K-1}.
  \label{eq:ch4_cond_MNT}
\end{equation}
The corresponding adaptive benchmark combines the $p$-values of
\eqref{eq:ch4_cond_HDA} and \eqref{eq:ch4_cond_MNT} through the Cauchy rule
\begin{equation}
  p_{\mathrm{Ada},\mathrm{cond}}^{\mathrm G}
  = 1-F_{\mathrm C}\!\bigl(
      0.5\tan\{\pi(0.5-p_{\mathrm{HDA}})\}
      +0.5\tan\{\pi(0.5-p_{\mathrm{MNT}})\}
    \bigr),
  \label{eq:ch4_cond_Ada}
\end{equation}
where $F_{\mathrm C}$ is the standard Cauchy distribution function.

\begin{assumption}[Conditional factor benchmark conditions]
\label{ass:ch4_cond_B1}
The functions $\delta_i(\cdot)$ and $\beta_{ij}(\cdot)$ belong to a H\"older class
$\mathcal H_r$ with $r>3/2$.  The factor dimension $K$ is fixed,
$\sup_t \norm{f_t}_2<\infty$, and $T^{-1}(\vct{1}_T,\mF)^{\top}(\vct{1}_T,\mF)$
converges to a positive-definite limit.
\end{assumption}

\begin{assumption}
\label{ass:ch4_cond_B2}
The error vectors $\vvarepsilon_{\cdot 1},\ldots,\vvarepsilon_{\cdot T}$ are independent
and identically distributed with mean zero, covariance matrix $\mOmega$, and uniformly
sub-Gaussian coordinates after marginal standardization.  Moreover,
\begin{equation}
  \frac{\tr(\mOmega^4)}{\tr^2(\mOmega^2)}\to 0,
  \qquad
  \frac{T L^{-2r}N}{\tr^{1/2}(\mOmega^2)}\to 0,
  \qquad
  \frac{\max_i\sum_j |\omega_{ij}|}{\tr^{1/2}(\mOmega^2)}\to 0.
  \label{eq:ch4_cond_B2}
\end{equation}
\end{assumption}

\begin{theorem}[Light-tailed conditional alpha benchmarks]
\label{thm:ch4_cond_benchmark}
Suppose Assumptions~\ref{ass:ch4_cond_B1}--\ref{ass:ch4_cond_B2} hold and
$H_{0,\mathrm{cond}}$ is true.  Then
\begin{align}
  T_{\mathrm{HDA},\mathrm{cond}}
  &\overset{d}{\longrightarrow} N(0,1),
  \label{eq:ch4_cond_HDA_limit}\\
  \Prob\bigl(M_{NT}-2\log N+\log\log N\le x\bigr)
  &\longrightarrow
  \exp\!\left\{-\pi^{-1/2}e^{-x/2}\right\},
  \qquad x\in\R.
  \label{eq:ch4_cond_MNT_limit}
\end{align}
If, in addition, the centered max statistic and the standardized sum statistic are
asymptotically independent under $H_{0,\mathrm{cond}}$, then the adaptive
combination \eqref{eq:ch4_cond_Ada} is asymptotically valid.
\end{theorem}

\subsection{Spatial-sign sum, max, and adaptive procedures}

The robust counterpart of \citet{Zhao2023ConditionalAlpha} starts from the same
residuals \eqref{eq:ch4_cond_residuals} but replaces least-squares quadratic forms by
spatial signs.  Let
\begin{equation}
  \mU_{\mathrm{cond}}
  = \bigl(U(\hat\vvarepsilon_{\cdot 1}),\ldots,U(\hat\vvarepsilon_{\cdot T})\bigr)^{\top}
  \in\R^{T\times N}.
  \label{eq:ch4_cond_Umat}
\end{equation}
The robust sum-type statistic is
\begin{equation}
  T_{\mathrm{CSS},\mathrm{cond}}
  = \frac{(\vh^{\top}\vh)^{-1}\vh^{\top}\mU_{\mathrm{cond}}\mU_{\mathrm{cond}}^{\top}\vh -1}
         {\{2\widehat{\tr(\mSigma_u^2)}\}^{1/2}},
  \label{eq:ch4_cond_CSS}
\end{equation}
where $\mSigma_u=\E\{U(\vvarepsilon_{\cdot t})U(\vvarepsilon_{\cdot t})^{\top}\}$ and
$\widehat{\tr(\mSigma_u^2)}$ is estimated by the leave-two-out U-statistic of
\citet{Zhao2023ConditionalAlpha}.

To obtain a sparse-oriented robust statistic, \citet{ZhaoWang2024ConditionalMaxAlpha}
proposed to estimate jointly a location vector $\vtheta$ and a diagonal scale matrix $\mD$
from the residuals by solving
\begin{align}
  \frac1T\sum_{t=1}^T U\{\mD^{-1/2}(\hat\vvarepsilon_{\cdot t}-\vtheta)\}
  &= \vct{0},
  \label{eq:ch4_cond_thetaeq1}\\
  \frac1T\sum_{t=1}^T \diag\Bigl[
    U\{\mD^{-1/2}(\hat\vvarepsilon_{\cdot t}-\vtheta)\}
    U\{\mD^{-1/2}(\hat\vvarepsilon_{\cdot t}-\vtheta)\}^{\top}
  \Bigr]
  &= N^{-1}\mI_N.
  \label{eq:ch4_cond_thetaeq2}
\end{align}
With $\tilde r_t=\norm{\hat\mD^{-1/2}(\hat\vvarepsilon_{\cdot t}-\hat\vtheta)}$, define
\begin{equation}
  \hat\varsigma_{2}=T^{-1}\sum_{t=1}^T \tilde r_t^2,
  \qquad
  \hat\varsigma_{1}=T^{-1}\sum_{t=1}^T \tilde r_t,
  \qquad
  \hat\varsigma_{-1}=T^{-1}\sum_{t=1}^T \tilde r_t^{-1},
  \label{eq:ch4_cond_varsigma}
\end{equation}
and
\begin{equation}
  \hat\zeta
  = \frac{N\hat\varsigma_{-1}^2}
         {1-2(1-\omega_T/T)\hat\varsigma_{-1}\hat\varsigma_1
          +(1-\omega_T/T)\hat\varsigma_2\hat\varsigma_{-1}^2}.
  \label{eq:ch4_cond_zetahat}
\end{equation}
The robust max statistic is
\begin{equation}
  T_{\mathrm{CSM},\mathrm{cond}}
  = T\hat\zeta\,\norm{\hat\mD^{-1/2}\hat\vtheta}_{\infty}^2
    -2\log N + \log\log N.
  \label{eq:ch4_cond_CSM}
\end{equation}
The corresponding adaptive Cauchy combination is
\begin{equation}
  p_{\mathrm{CC},\mathrm{cond}}
  = 1-F_{\mathrm C}\Bigl[
      0.5\tan\{\pi(0.5-p_{\mathrm{CSS}})\}\mathbbm 1(p_{\mathrm{CSS}}<1/2)
      + 0.5\tan\{\pi(0.5-p_{\mathrm{CSM}})\}\mathbbm 1(p_{\mathrm{CSM}}<1/2)
    \Bigr].
  \label{eq:ch4_cond_CC}
\end{equation}

\begin{assumption}[Robust conditional factor conditions]
\label{ass:ch4_cond_R1}
Assumption~\ref{ass:ch4_cond_B1} holds.  In addition, the residual vectors admit the
elliptical representation
\begin{equation}
  \vvarepsilon_{\cdot t}=v_t\mGamma\vW_t,
  \qquad
  \mOmega=\mGamma\mGamma^{\top},
  \label{eq:ch4_cond_R1repr}
\end{equation}
where $\vW_t$ has independent symmetric coordinates with unit variance, $v_t\ge 0$ is
independent of the spatial sign of $\vW_t$, the Orlicz norms of $v_t$ and the coordinates of
$\vW_t$ are uniformly bounded, and the radial moments
$\E(r_t^{-k})$, $k=1,2,3,4$, exist.
Furthermore, with $\mR=\mD^{-1/2}\mGamma\mGamma^{\top}\mD^{-1/2}$,
\begin{equation}
  \max_{1\le i\le N}\sum_{j=1}^N |r_{ij}| \le a_0(N),
  \qquad
  a_0(N)\asymp N^{1-\delta},
  \qquad 0<\delta\le \frac12,
  \label{eq:ch4_cond_R1corr}
\end{equation}
and the sparse-correlation condition \eqref{eq:ch4_alpha_B3} holds with $\mR$ in place of the
unconditional correlation matrix.
\end{assumption}

\begin{theorem}[Robust tests for conditional time-varying alpha]
\label{thm:ch4_cond_robust}
Suppose Assumptions~\ref{ass:ch4_cond_B1} and \ref{ass:ch4_cond_R1} hold.
Under $H_{0,\mathrm{cond}}$,
\begin{align}
  T_{\mathrm{CSS},\mathrm{cond}}
  &\overset{d}{\longrightarrow} N(0,1),
  \label{eq:ch4_cond_CSS_limit}\\
  \Prob\bigl(T_{\mathrm{CSM},\mathrm{cond}}\le y\bigr)
  &\longrightarrow
  \exp\!\left\{-\pi^{-1/2}e^{-y/2}\right\},
  \qquad y\in\R,
  \label{eq:ch4_cond_CSM_limit}\\
  \Prob\bigl(T_{\mathrm{CSS},\mathrm{cond}}\le x,
             T_{\mathrm{CSM},\mathrm{cond}}\le y\bigr)
  &\longrightarrow
  \Phi(x)\exp\!\left\{-\pi^{-1/2}e^{-y/2}\right\}
  \label{eq:ch4_cond_joint_limit}
\end{align}
for every fixed $(x,y)\in\R^2$.  Consequently, the adaptive $p$-value
\eqref{eq:ch4_cond_CC} is asymptotically valid.

Under alternatives satisfying
\begin{equation}
  \norm{\vdelta}_{\infty}\ge C\sqrt{\frac{\log N}{T}},
  \qquad
  \norm{\vdelta}_2 = O(NT^{-1}L),
  \qquad
  \lambda_{\max}(\mR)=o\{NL^{-1}(\log N)^{-1}\},
  \label{eq:ch4_cond_sparse_alt}
\end{equation}
for a sufficiently large constant $C$, the max-type statistic
$T_{\mathrm{CSM},\mathrm{cond}}$ is consistent.  If, in addition,
\begin{equation}
  \vdelta^{\top}\vdelta
  = O\!\left(\kappa_1^{-2}T^{-1}\{2\tr(\mSigma_u^2)\}^{1/2}\right),
  \qquad
  \vdelta^{\top}\mOmega\vdelta
  = o\!\left(\kappa_1^{-2}N^{-1}T^{-1}\tr(\mOmega^2)\right),
  \label{eq:ch4_cond_local_indep}
\end{equation}
where $\kappa_1=\E\norm{\vvarepsilon_{\cdot t}}^{-1}$, then the same asymptotic factorization
holds under the corresponding local alternatives, namely,
\begin{equation}
  \sup_{x,y\in\R}
  \left|
    \Prob\bigl(T_{\mathrm{CSS},\mathrm{cond}}\le x,
                T_{\mathrm{CSM},\mathrm{cond}}\le y\bigr)
    - \Prob\bigl(T_{\mathrm{CSS},\mathrm{cond}}\le x\bigr)
      \Prob\bigl(T_{\mathrm{CSM},\mathrm{cond}}\le y\bigr)
  \right|
  \longrightarrow 0.
  \label{eq:ch4_cond_joint_local}
\end{equation}
Consequently, if $\beta_{\mathrm{CSS},\gamma}^{\mathrm{cond}}$,
$\beta_{\mathrm{CSM},\gamma}^{\mathrm{cond}}$, and
$\beta_{\mathrm{CC},\gamma}^{\mathrm{cond}}$ denote the power functions of the sum,
max, and adaptive procedures at level $\gamma$, then for every fixed
$\gamma\in(0,1)$,
\begin{equation}
  \beta_{\mathrm{CC},\gamma}^{\mathrm{cond}}
  \ge
  \beta_{\mathrm{CSS},\gamma/2}^{\mathrm{cond}}
  + \beta_{\mathrm{CSM},\gamma/2}^{\mathrm{cond}}
  - \beta_{\mathrm{CSS},\gamma/2}^{\mathrm{cond}}
    \beta_{\mathrm{CSM},\gamma/2}^{\mathrm{cond}}
  + o(1).
  \label{eq:ch4_cond_power_bound}
\end{equation}
\end{theorem}

The conditional-factor section highlights the same message as the unconditional one: the
sum statistic is the right choice for dense alternatives, the max statistic is the right choice
for sparse alternatives, and the asymptotic independence between the two gives a principled
route to adaptive inference.  The difference is that the spline projection introduces an extra
approximation layer, which must be controlled jointly with the high-dimensional probability
bounds.

\section{High-dimensional change-point inference}

\subsection{Classical low-dimensional change-point tests}

Consider first the univariate mean-change model
\begin{equation}
  X_i = \mu + \delta\mathbbm 1(i>\tau) + \varepsilon_i,
  \qquad i=1,\ldots,n,
  \label{eq:ch4_cp_univariate}
\end{equation}
where the unknown location of the change point is $\tau\in\{1,\ldots,n-1\}$.
Let $S_k = \sum_{i=1}^k X_i$.  The classical cumulative-sum (CUSUM) statistic is
\begin{equation}
  C_n(k) = n^{-1/2}\bigl(S_k - kS_n/n\bigr),
  \qquad k=1,\ldots,n-1.
  \label{eq:ch4_cp_cusum_univariate}
\end{equation}
The test rejects for large values of $\max_{1\le k\le n-1}|C_n(k)|$.  Under
independent mean-zero errors with variance $\sigma^2$, one has the Brownian
bridge limit
\begin{equation}
  \sigma^{-1}\max_{1\le k\le n-1}|C_n(k)|
  \overset{d}{\longrightarrow}
  \sup_{0<u<1}|B^{\circ}(u)|,
  \label{eq:ch4_cp_bb_limit}
\end{equation}
where $B^{\circ}$ is a standard Brownian bridge.  Variants with weights
$\{u(1-u)\}^{-\gamma}$ were introduced to improve sensitivity to changes near the
boundaries.  Classical references include \citet{Page1954CUSUM} and the book of
\citet{CsorgoHorvath1997}.

In the multivariate fixed-dimensional case one applies the same principle to
vector observations $\vX_i\in\R^p$, replacing \eqref{eq:ch4_cp_cusum_univariate}
by either a quadratic form or a maximum over coordinates.  If $p$ is fixed, then
continuous mapping applied to the vector-valued partial-sum process yields the
limiting null distribution.  When $p$ diverges, however, one needs a new
extreme-value or high-dimensional central-limit argument.

\subsection{High-dimensional Gaussian and light-tailed benchmarks}

Let $\vX_1,\ldots,\vX_n\in\R^p$ follow the single change-point model
\begin{equation}
  \vX_i = \vmu_0 + \vdelta\,\mathbbm 1(i>\tau) + \vvarepsilon_i,
  \qquad i=1,\ldots,n,
  \label{eq:ch4_cp_model_hd}
\end{equation}
where $\vdelta\in\R^p$ is the mean-shift vector.  Write
\begin{equation}
  S_{k,j}=\sum_{i=1}^k X_{ij},
  \qquad
  \hat\sigma_j^2 = \frac{1}{2(n-1)}\sum_{i=2}^n (X_{ij}-X_{i-1,j})^2,
  \qquad j=1,\ldots,p.
  \label{eq:ch4_cp_sigmahat}
\end{equation}
For $\gamma\in[0,1/2]$, define the weighted coordinatewise CUSUM statistic by
\begin{equation}
  C_{\gamma,j}(k)
  = \Bigl\{\frac{k}{n}\Bigl(1-\frac{k}{n}\Bigr)\Bigr\}^{-\gamma}
    \frac{n^{-1/2}\{S_{k,j}-kS_{n,j}/n\}}{\hat\sigma_j},
  \qquad 1\le k\le n-1.
  \label{eq:ch4_cp_cgamma}
\end{equation}
The sparse-oriented maxima are
\begin{equation}
  M_{n,p} = \max_{1\le j\le p}\max_{1\le k\le n-1}|C_{0,j}(k)|,
  \qquad
  M_{n,p}^{\dagger} = \max_{1\le j\le p}\max_{\lambda_n\le k\le n-\lambda_n}
  |C_{1/2,j}(k)|,
  \label{eq:ch4_cp_max_stats}
\end{equation}
where $\lambda_n\to\infty$ and $\lambda_n/n\to0$.  A dense-oriented aggregate is
obtained by
\begin{equation}
  S_{n,p} = \sum_{k=2}^{n-2}\sum_{j=1}^p C_{0,j}^2(k).
  \label{eq:ch4_cp_sum_stat}
\end{equation}
The data-adaptive DMS strategy of \citet{WangFeng2023JRSSBChangePoint} uses the
standardized version of \eqref{eq:ch4_cp_sum_stat} together with
$M_{n,p}^{\dagger}$.

\begin{assumption}
\label{ass:ch4_cp_A1}
The error vectors $\vvarepsilon_i$ are independent and identically distributed
with mean zero and covariance matrix $\mSigma=(\sigma_{jj'})_{1\le j,j'\le p}$.
For some constants $0<c<C<\infty$,
\begin{equation}
  c \le \min_{1\le j\le p} \sigma_{jj}^{1/2}
  \le \max_{1\le j\le p} \sigma_{jj}^{1/2} \le C.
  \label{eq:ch4_cp_A1_var}
\end{equation}
\end{assumption}

\begin{assumption}
\label{ass:ch4_cp_A2}
Either the coordinates of $\vvarepsilon_i$ are Gaussian, or they are uniformly
sub-Gaussian in the sense that there exists $\zeta>0$ such that
\begin{equation}
  \E\exp(t\varepsilon_{ij}) \le \exp(\zeta t^2)
  \qquad \text{for all } t\in\R,\ \ 1\le j\le p.
  \label{eq:ch4_cp_A2_subgaussian}
\end{equation}
Furthermore, the correlation matrix $\mR=(\rho_{jj'})$ satisfies
\begin{equation}
  \max_{j\ne j'}|\rho_{jj'}|\le \rho_0 <1,
  \qquad
  c \le \lambda_{\min}(\mR) \le \lambda_{\max}(\mR) \le C.
  \label{eq:ch4_cp_A2_corr}
\end{equation}
\end{assumption}

\begin{assumption}
\label{ass:ch4_cp_A3}
Let $h_n = n/\lambda_n$.  As $(n,p)\to\infty$,
\begin{equation}
  \log p = o(n^{1/3}),
  \qquad
  \log h_n = o(n^{1/3}),
  \qquad
  \lambda_n \to \infty,
  \qquad
  \lambda_n/n \to 0.
  \label{eq:ch4_cp_A3_rate}
\end{equation}
\end{assumption}

\begin{theorem}
\label{thm:ch4_cp_max_null}
Under Assumptions~\ref{ass:ch4_cp_A1}--\ref{ass:ch4_cp_A3} and the null
hypothesis $\vdelta=\vct{0}$,
\begin{align}
  \Prob\bigl(A(\log p)M_{n,p}-D(\log p)\le x\bigr)
    &\longrightarrow \exp\{-e^{-x}\},
    \label{eq:ch4_cp_max_null_1}\\
  \Prob\bigl(A(\log(ph_n))M_{n,p}^{\dagger}-D(\log(ph_n))\le x\bigr)
    &\longrightarrow \exp\{-e^{-x}\},
    \label{eq:ch4_cp_max_null_2}
\end{align}
where
\begin{equation}
  A(x) = (2\log x)^{1/2},
  \qquad
  D(x) = 2\log x + \frac12\log\log x - \frac12\log\pi.
  \label{eq:ch4_cp_AD}
\end{equation}
\end{theorem}

Theorem~\ref{thm:ch4_cp_max_null} is the extreme-value backbone of the sparse
change-point theory.  To achieve power against dense alternatives,
\citet{WangFeng2023JRSSBChangePoint} combine a sum-$L_2$ statistic with the
weighted max statistic.  Let
\begin{equation}
  T_{\mathrm{sum},\mathrm{cp}}
  = \frac{S_{n,p}-(n+2)p}{V_{n,p}^{1/2}},
  \label{eq:ch4_cp_sum_standardized}
\end{equation}
where $V_{n,p}$ is a consistent estimator of $\Var(S_{n,p})$ constructed from
finite differences.  Define the max-$L_{\infty}$ $p$-value by
\begin{equation}
  p_{\max,\mathrm{cp}} = 1-G\{A(\log(ph_n))M_{n,p}^{\dagger}-D(\log(ph_n))\},
  \label{eq:ch4_cp_pmax}
\end{equation}
where $G(x)=\exp\{-e^{-x}\}$, and let
\begin{equation}
  p_{\mathrm{sum},\mathrm{cp}} = 1-\Phi(T_{\mathrm{sum},\mathrm{cp}}).
  \label{eq:ch4_cp_psum}
\end{equation}
The DMS combination is obtained by either the Fisher rule or the Cauchy rule
introduced earlier.

\begin{theorem}
\label{thm:ch4_cp_independence}
Under Assumptions~\ref{ass:ch4_cp_A1}--\ref{ass:ch4_cp_A3} and the null
hypothesis,
\begin{equation}
  \Prob\bigl(A(\log(ph_n))M_{n,p}^{\dagger}-D(\log(ph_n))\le x,
             T_{\mathrm{sum},\mathrm{cp}}\le y\bigr)
  \longrightarrow \exp\{-e^{-x}\}\,\Phi(y)
  \label{eq:ch4_cp_independence_eq}
\end{equation}
for every fixed $(x,y)\in\R^2$.  Consequently, the DMS combination based on the
two associated $p$-values is asymptotically valid.
\end{theorem}

\subsection{Spatial-sign max, sum, and adaptive change-point procedures}

The robust change-point procedures of \citet{LiuFengPengWang2025SpatialSignCP}
replace sample means by scaled spatial medians and replace raw quadratic forms by
statistics built from spatial signs.  This is crucial when the coordinates are heavy tailed,
when fourth moments fail to exist, or when a few large outliers contaminate local CUSUM
aggregates.

For integers $1\le a\le b\le n$, let $(\hat\vtheta_{a:b},\hat\mD_{a:b})$ be the solution to
\begin{align}
  \frac{1}{b-a+1}\sum_{i=a}^{b}U\{\mD^{-1/2}(\vX_i-\vtheta)\}
  &=\vct{0},
  \label{eq:ch4_cp_ss_thetaeq1}\\
  \frac{p}{b-a+1}\diag\sum_{i=a}^{b}
  U\{\mD^{-1/2}(\vX_i-\vtheta)\}
  U\{\mD^{-1/2}(\vX_i-\vtheta)\}^{\top}
  &=\mI_p.
  \label{eq:ch4_cp_ss_thetaeq2}
\end{align}
These are the segmentwise scaled spatial median equations.  Denote the full-sample solution by
$(\hat\vtheta,\hat\mD)=(\hat\vtheta_{1:n},\hat\mD_{1:n})$, let
\begin{equation}
  \hat r_i = \norm{\hat\mD^{-1/2}(\vX_i-\hat\vtheta)}_2,
  \qquad
  \hat\zeta_1 = n^{-1}\sum_{i=1}^{n}\hat r_i^{-1},
  \qquad
  h_n = \left\{\left(\frac{\lambda_n}{n}\right)^{-1}-1\right\}^2,
  \label{eq:ch4_cp_ss_rzeta}
\end{equation}
and write $\zeta_1=\E(r_i^{-1})$ for the population counterpart.

The max-$L_\infty$ spatial-sign CUSUM statistics are
\begin{equation}
  \mC_{\gamma}(k)
  = \Bigl\{\frac{k}{n}\Bigl(1-\frac{k}{n}\Bigr)\Bigr\}^{1-\gamma}
    n^{1/2}\hat\mD^{-1/2}
    \bigl(\hat\vtheta_{1:k}-\hat\vtheta_{k+1:n}\bigr),
  \qquad \lambda_n\le k\le n-\lambda_n,
  \label{eq:ch4_cp_ss_Cgamma}
\end{equation}
with
\begin{equation}
  M_{n,p} = (1-n^{-1/2})\max_{\lambda_n\le k\le n-\lambda_n}
  \norm{\mC_{0}(k)}_{\infty},
  \qquad
  M_{n,p}^{\dagger} = (1-n^{-1/2})\max_{\lambda_n\le k\le n-\lambda_n}
  \norm{\mC_{1/2}(k)}_{\infty}.
  \label{eq:ch4_cp_ss_Mstats}
\end{equation}
The unweighted statistic $M_{n,p}$ is natural for interior changes, while the weighted statistic
$M_{n,p}^{\dagger}$ stabilizes the variance near the boundaries.

A dense-oriented spatial-sign analogue is built from the full-sample sign process
\begin{equation}
  \hat\vU_i = U\{\hat\mD^{-1/2}(\vX_i-\hat\vtheta)\},
  \qquad
  \hat\vS_k = \sum_{i=1}^k \hat\vU_i,
  \qquad k=1,\ldots,n,
  \label{eq:ch4_cp_ss_signs}
\end{equation}
and the weighted partial-sum transforms
\begin{equation}
  \widetilde{\mC}_{\gamma}(k)
  = \Bigl\{\frac{k}{n}\Bigl(1-\frac{k}{n}\Bigr)\Bigr\}^{-\gamma}
    \sqrt{\frac{p}{n}}
    \left(\hat\vS_k-\frac{k}{n}\hat\vS_n\right),
  \qquad \lambda_n\le k\le n-\lambda_n.
  \label{eq:ch4_cp_ss_Ctilde}
\end{equation}
The corresponding max-$L_2$ statistics are
\begin{align}
  S_{n,p}
  &= (1-n^{-1/2})
     \max_{\lambda_n\le k\le n-\lambda_n}
     \left\{
       \widetilde{\mC}_{0}(k)^{\top}\widetilde{\mC}_{0}(k)
       - \frac{k(n-k)}{n^2}p
     \right\},
     \label{eq:ch4_cp_ss_Snp}\\
  S_{n,p}^{\dagger}
  &= (1-n^{-1/2})
     \max_{\lambda_n\le k\le n-\lambda_n}
     \left\{
       \widetilde{\mC}_{1/2}(k)^{\top}\widetilde{\mC}_{1/2}(k)-p
     \right\}.
     \label{eq:ch4_cp_ss_Snpdagger}
\end{align}
These $L_2$-type procedures are designed for dense changes and provide the robust analogue of
the sum statistic in the Gaussian benchmark section.

\begin{assumption}[Robust change-point conditions]
\label{ass:ch4_cp_R1}
The observations follow the single change-point model
\eqref{eq:ch4_cp_model_hd}, and the error vectors $\vvarepsilon_i$ are independent and
identically distributed from an elliptically symmetric distribution with positive diagonal scale
matrix $\mD$ and correlation matrix $\mR=\mD^{-1/2}\mSigma\mD^{-1/2}$.  The radial moments
$\E(r_i^{-k})$, $k=1,2,3,4$, exist, and the weak cross-sectional dependence condition
\eqref{eq:ch4_alpha_B3} holds with $p$ in place of $N$.
\end{assumption}

\begin{assumption}
\label{ass:ch4_cp_R2}
For the max-$L_\infty$ statistics,
$\lambda_n\asymp n^{\lambda}$ for some $0<\lambda<1$,
\begin{equation}
  \log^7 n = o\bigl(p^{1/6\wedge \eta_0/2}\bigr),
  \qquad
  \log^2 p = o\bigl(n^{1/5\wedge \lambda/3}\bigr),
  \label{eq:ch4_cp_R2a}
\end{equation}
where $\eta_0$ is the correlation-sparsity exponent in
\eqref{eq:ch4_alpha_B3}.  For the max-$L_2$ statistics,
\begin{equation}
  \log p = o(n),
  \qquad
  \lambda_n\to\infty,
  \qquad
  \lambda_n/n\to 0.
  \label{eq:ch4_cp_R2b}
\end{equation}
\end{assumption}

\begin{theorem}[Spatial-sign max-$L_\infty$ limits]
\label{thm:ch4_cp_ss_infmax}
Suppose Assumptions~\ref{ass:ch4_cp_R1}--\ref{ass:ch4_cp_R2} hold.
Under the null hypothesis $\vdelta=\vct{0}$,
\begin{align}
  \Prob\Bigl(2p\zeta_1^2 M_{n,p}^2 - \log(2p) \le x\Bigr)
  &\longrightarrow G(x),
  \label{eq:ch4_cp_ss_M_limit}\\
  \Prob\Bigl(p^{1/2}\zeta_1 A(p\log h_n)M_{n,p}^{\dagger}-D(p\log h_n)\le x\Bigr)
  &\longrightarrow G(x),
  \label{eq:ch4_cp_ss_Mdagger_limit}
\end{align}
where $G(x)=\exp\{-\exp(-x)\}$ and the functions $A(\cdot)$ and $D(\cdot)$ are defined in
\eqref{eq:ch4_cp_AD}.  Hence the null $p$-values are
\begin{equation}
  p_{M_{n,p}}
  = 1-G\bigl(2p\hat\zeta_1^2M_{n,p}^2-\log(2p)\bigr),
  \qquad
  p_{M_{n,p}^{\dagger}}
  = 1-G\bigl(p^{1/2}\hat\zeta_1A(p\log h_n)M_{n,p}^{\dagger}-D(p\log h_n)\bigr).
  \label{eq:ch4_cp_ss_pvalues_M}
\end{equation}
\end{theorem}

The max-$L_2$ statistics have non-Gaussian null limits.  Let $V(t)$ be a centered Gaussian
process on $[0,1]$ with covariance kernel
\begin{equation}
  \Cov\{V(t),V(s)\} = (1-t)^2 s^2,
  \qquad 0\le s\le t\le 1,
  \label{eq:ch4_cp_ss_Vcov}
\end{equation}
and let $F_V$ denote the distribution function of $\max_{0\le t\le1}V(t)$.  Then
\citet{LiuFengPengWang2025SpatialSignCP} proved the following result.

\begin{theorem}[Spatial-sign max-$L_2$ limits]
\label{thm:ch4_cp_ss_l2}
Suppose Assumptions~\ref{ass:ch4_cp_R1}--\ref{ass:ch4_cp_R2} hold and let
$\mR=\mD^{-1/2}\mSigma\mD^{-1/2}$.  Under the null hypothesis,
\begin{align}
  \frac{pS_{n,p}}{2\tr(\mR^2)}
  &\overset{d}{\longrightarrow} \max_{0\le t\le1}V(t),
  \label{eq:ch4_cp_ss_Snp_limit}\\
  \Prob\Bigl(
    A\!\left\{\log\left(\frac{n^2}{\lambda_n^2}\right)\right\}
    \frac{pS_{n,p}^{\dagger}}{2\tr(\mR^2)}
    - D\!\left\{\log\left(\frac{n^2}{\lambda_n^2}\right)\right\}
    \le x
  \Bigr)
  &\longrightarrow \exp\{-2e^{-x}\}.
  \label{eq:ch4_cp_ss_Snpdagger_limit}
\end{align}
Accordingly, one may use
\begin{equation}
  p_{S_{n,p}} = 1-F_V\left(\frac{pS_{n,p}}{2\widehat{\tr(\mR^2)}}\right),
  \qquad
  p_{S_{n,p}^{\dagger}} = 1-G_2\Biggl(
    A\!\left\{\log\left(\frac{n^2}{\lambda_n^2}\right)\right\}
    \frac{pS_{n,p}^{\dagger}}{2\widehat{\tr(\mR^2)}}
    - D\!\left\{\log\left(\frac{n^2}{\lambda_n^2}\right)\right\}
  \Biggr),
  \label{eq:ch4_cp_ss_pvalues_S}
\end{equation}
where $G_2(x)=\exp\{-2e^{-x}\}$ and $\widehat{\tr(\mR^2)}$ is a ratio-consistent estimator of
$\tr(\mR^2)$.
\end{theorem}

The max-$L_\infty$ and max-$L_2$ statistics are complementary: the former targets sparse shifts,
whereas the latter targets dense alternatives.  Their adaptive combination therefore mirrors the
same max--sum logic used for alpha testing and white-noise testing.

\begin{theorem}[Adaptive spatial-sign change-point tests]
\label{thm:ch4_cp_ss_adaptive}
Suppose Assumptions~\ref{ass:ch4_cp_R1}--\ref{ass:ch4_cp_R2} hold.  Under the null
hypothesis,
\begin{align}
  \Prob\Bigl(
    2p\zeta_1^2 M_{n,p}^2 - \log(2p) \le x,
    \frac{pS_{n,p}}{2\tr(\mR^2)} \le y
  \Bigr)
  &\longrightarrow G(x)F_V(y),
  \label{eq:ch4_cp_ss_joint1}\\
  \Prob\Biggl(
    p^{1/2}\zeta_1A(p\log h_n)M_{n,p}^{\dagger}-D(p\log h_n)\le x,
    A\!\left\{\log\left(\frac{n^2}{\lambda_n^2}\right)\right\}
    \frac{pS_{n,p}^{\dagger}}{2\tr(\mR^2)}
    - D\!\left\{\log\left(\frac{n^2}{\lambda_n^2}\right)\right\}
    \le y
  \Biggr)
  &\longrightarrow G(x)G_2(y).
  \label{eq:ch4_cp_ss_joint2}
\end{align}
Consequently, the Fisher combinations
\begin{equation}
  p_{M,S}=1-F_{\chi_4^2}\bigl[-2\{\log p_{M_{n,p}}+\log p_{S_{n,p}}\}\bigr],
  \qquad
  p_{M^{\dagger},S^{\dagger}}=1-F_{\chi_4^2}\bigl[-2\{\log p_{M_{n,p}^{\dagger}}+\log p_{S_{n,p}^{\dagger}}\}\bigr]
  \label{eq:ch4_cp_ss_fisher}
\end{equation}
are asymptotically valid.
\end{theorem}

\begin{theorem}[Local-alternative asymptotic independence]
\label{thm:ch4_cp_ss_adaptive_alt}
Suppose Assumptions~\ref{ass:ch4_cp_R1}--\ref{ass:ch4_cp_R2} hold and there is a single
change at location $\tau_n$ with jump vector $\vdelta_n$.  Let
$A_n=\{j:\delta_{n,j}\ne0\}$ be the support of the jump vector.  If
\begin{equation}
  |A_n| = o\left\{
    \frac{p}{(\log\log p)^2}
    \wedge
    \frac{p\tr(\mR^2)}{\log n}
  \right\},
  \qquad
  \norm{\vdelta_n}_2^2 = o\left(\frac{p^2\tr(\mR^2)}{n^2}\right),
  \label{eq:ch4_cp_ss_local_alt}
\end{equation}
then the factorization in \eqref{eq:ch4_cp_ss_joint1} and \eqref{eq:ch4_cp_ss_joint2}
continues to hold under the corresponding local alternatives.  Hence the Fisher combination
retains its adaptive power advantage under the sparse local alternatives for which the individual
sum- and max-type components remain nondegenerate.
\end{theorem}

\subsection{Elliptical regularized Hotelling change-point tests}
\label{subsec:ch4-erht-cp}
\idx{elliptical regularized Hotelling change-point test}
\idx{wild binary segmentation}
\idx{ridge-regularized scan statistic}

The spatial-sign max and sum procedures in the preceding subsection are robust to heavy tails,
but they are designed primarily for weak cross-sectional dependence.  When the shape matrix has
substantial non-isotropic structure, a ridge-regularized inverse of a robust shape estimator can
increase power for shifts aligned with low-noise eigendirections.  The ERHT change-point method
of \citet{SongWenFeng2026ERHTCP} compares adjacent-segment spatial medians and weights their
difference by a ridge inverse of the pooled centered SSCM.  It provides single- and
multiple-change global tests, finite-grid ridge adaptation, localization theory, and a wild binary
segmentation estimator.

\subsubsection{Adjacent-segment ERHT statistic}

Let
\begin{equation}\label{eq:ch4-erhtcp-model}
  \vX_i=\vtheta_i+\vvarepsilon_i,
  \qquad
  \vvarepsilon_i
  =\sqrt p\,R_i\mOmega_p^{1/2}
    \frac{\vG_i}{\twonorm{\vG_i}},
  \qquad
  \vG_i\iid N_p(\vct{0},\mI_p),
  \qquad
  \tr(\mOmega_p)=p.
\end{equation}
The no-change hypothesis is
\begin{equation}\label{eq:ch4-erhtcp-null}
  H_0:\vtheta_1=\cdots=\vtheta_n.
\end{equation}
For an ordered triple $\vs=(t_1,t_2,t_3)$, $0\le t_1<t_2<t_3\le1$, let
\begin{align}
  I_1(\vs)
  &=\{\lfloor nt_1\rfloor+1,\ldots,\lfloor nt_2\rfloor\},
  \label{eq:ch4-erhtcp-I1}\\
  I_2(\vs)
  &=\{\lfloor nt_2\rfloor+1,\ldots,\lfloor nt_3\rfloor\},
  \label{eq:ch4-erhtcp-I2}
\end{align}
with sizes $n_a(\vs)=|I_a(\vs)|$, $a=1,2$, and effective sample size
\begin{equation}\label{eq:ch4-erhtcp-Ns}
  N_{\vs}=\frac{n_1(\vs)n_2(\vs)}{n_1(\vs)+n_2(\vs)}.
\end{equation}
Let $\widehat{\vtheta}_{a,\vs}$ be the spatial median on $I_a(\vs)$ and define
\begin{equation}\label{eq:ch4-erhtcp-Delta}
  \widehat{\vDelta}_{\vs}
  =\widehat{\vtheta}_{2,\vs}-\widehat{\vtheta}_{1,\vs}.
\end{equation}
For global testing, the scatter pool is the full sample,
$J(\vs)=\{1,\ldots,n\}$, $m_{\vs}=n$.  Let $\widehat{\vtheta}_{0,\vs}$ be the spatial median
on $J(\vs)$ and put
\begin{align}
  \widehat{\vY}_{i,\vs}
  &=\sqrt p\,U(\vX_i-\widehat{\vtheta}_{0,\vs}),
  \label{eq:ch4-erhtcp-Y}\\
  \widehat{\mR}_{\vs}
  &=\frac1{m_{\vs}}\sum_{i\in J(\vs)}
    \widehat{\vY}_{i,\vs}\widehat{\vY}_{i,\vs}\trans,
  \qquad
  \widehat{\mQ}_{\rho,\vs}
  =(\widehat{\mR}_{\vs}+\rho\mI_p)^{-1}.
  \label{eq:ch4-erhtcp-RQ}
\end{align}
The raw local ERHT statistic is
\begin{equation}\label{eq:ch4-erhtcp-raw}
  V_{\rho}^{\mathrm{raw}}(\vs)
  =N_{\vs}\widehat{\vDelta}_{\vs}\trans
    \widehat{\mQ}_{\rho,\vs}
    \widehat{\vDelta}_{\vs}.
\end{equation}

To center and studentize \eqref{eq:ch4-erhtcp-raw}, define
\begin{equation}\label{eq:ch4-erhtcp-inverse-distance}
  \widehat w_{i,a,\vs}
  =\frac{\sqrt p}{\twonorm{\vX_i-\widehat{\vtheta}_{a,\vs}}},
  \qquad
  \widehat e_{a,\vs}
  =\frac1{n_a(\vs)}\sum_{i\in I_a(\vs)}\widehat w_{i,a,\vs}.
\end{equation}
The score-CUSUM weights are
\begin{equation}\label{eq:ch4-erhtcp-beta}
  \widehat\beta_i(\vs)
  =\sqrt{N_{\vs}}
  \left\{
    \frac{\mathbbm1\{i\in I_2(\vs)\}}
         {n_2(\vs)\widehat e_{2,\vs}}
    -\frac{\mathbbm1\{i\in I_1(\vs)\}}
         {n_1(\vs)\widehat e_{1,\vs}}
  \right\}.
\end{equation}
With $\widehat{\mY}_{\vs}$ denoting the $p\times m_{\vs}$ matrix of pooled signs, define the
companion matrix
\begin{equation}\label{eq:ch4-erhtcp-A}
  \widehat{\mA}_{\rho,\vs}
  =\frac1{m_{\vs}}
   \widehat{\mY}_{\vs}\trans
   \widehat{\mQ}_{\rho,\vs}
   \widehat{\mY}_{\vs}.
\end{equation}
Writing $\widehat A_{ij,\rho,\vs}$ for its entries, set
\begin{align}
  \widehat\kappa_\rho(\vs)
  &=\sum_{i\in J(\vs)}
    \widehat\beta_i(\vs)^2\widehat A_{ii,\rho,\vs},
  \label{eq:ch4-erhtcp-kappa}\\
  \widehat\sigma_\rho^2(\vs)
  &=2m_{\vs}\sum_{\substack{i,j\in J(\vs)\\i\ne j}}
    \widehat\beta_i(\vs)^2\widehat\beta_j(\vs)^2
    \widehat A_{ij,\rho,\vs}^2.
  \label{eq:ch4-erhtcp-sigma}
\end{align}
The studentized local statistic is
\begin{equation}\label{eq:ch4-erhtcp-Z}
  Z_\rho(\vs)
  =\frac{V_\rho^{\mathrm{raw}}(\vs)-m_{\vs}\widehat\kappa_\rho(\vs)}
         {\{m_{\vs}\widehat\sigma_\rho^2(\vs)\}^{1/2}}.
\end{equation}

For a trimming constant $\epsilon\in(0,1/2)$, the single-change and multiple-change scan sets are
\begin{align}
  \mathcal S_{\mathrm{sc}}(\epsilon)
  &=\{(0,t,1):\epsilon\le t\le1-\epsilon\},
  \label{eq:ch4-erhtcp-Ssc}\\
  \mathcal S_{\mathrm{mc}}(\epsilon)
  &=\{(t_1,t_2,t_3):t_2-t_1\ge\epsilon,
                         \ t_3-t_2\ge\epsilon\}.
  \label{eq:ch4-erhtcp-Smc}
\end{align}
The global statistics are
\begin{equation}\label{eq:ch4-erhtcp-global}
  T_\rho^{\mathrm{sc}}
  =\sup_{\vs\in\mathcal S_{\mathrm{sc}}(\epsilon)}Z_\rho(\vs),
  \qquad
  T_\rho^{\mathrm{mc}}
  =\sup_{\vs\in\mathcal S_{\mathrm{mc}}(\epsilon)}Z_\rho(\vs).
\end{equation}

\subsubsection{Null process limits and ridge aggregation}

\begin{assumption}[Elliptical change-point regularity]
\label{ass:ch4-erhtcp-radial}
The pairs $(R_i,\vG_i)$ in \eqref{eq:ch4-erhtcp-model} are independent and identically
distributed.  With $\xi_i=R_i^{-1}$, for some $\eta,c_0>0$,
\begin{equation}\label{eq:ch4-erhtcp-radial-assumption}
  \E\xi_i\longrightarrow\zeta_{-1}\in(0,\infty),
  \qquad
  \sup_n\E\xi_i^{4+\eta}<\infty,
  \qquad
  \Prob\left\{\max_{1\le i\le n}\xi_i>(\log n)^{c_0}\right\}\longrightarrow0.
\end{equation}
\end{assumption}

\begin{assumption}[Dimension, shape spectrum, and ridge grid]
\label{ass:ch4-erhtcp-spectrum}
As $n,p\to\infty$,
\begin{equation}\label{eq:ch4-erhtcp-aspect}
  p/n\longrightarrow\gamma\in(0,\infty).
\end{equation}
All eigenvalues of $\mOmega_p$ belong to $(0,\omega_+]$ for a fixed $\omega_+<\infty$, and
the empirical spectral distribution $H_p$ converges weakly to a probability distribution $H$
on $[0,\omega_+]$.  The finite deterministic ridge grid
\begin{equation}\label{eq:ch4-erhtcp-ridge-grid}
  \mathcal R_K=\{\rho_n^{(1)},\ldots,\rho_n^{(K)}\}
  \subset[\rho_0,\rho_1]
\end{equation}
has fixed $K$, $0<\rho_0<\rho_1<\infty$, and
$\rho_n^{(k)}\to\rho^{(k)}$.
\end{assumption}

For $\vs=(t_1,t_2,t_3)$, define the temporal contrast
\begin{equation}\label{eq:ch4-erhtcp-phi}
  \phi_{\vs}(x)
  =\frac{\mathbbm1\{x\in[t_2,t_3)\}}{t_3-t_2}
   -\frac{\mathbbm1\{x\in[t_1,t_2)\}}{t_2-t_1},
\end{equation}
and
\begin{equation}\label{eq:ch4-erhtcp-kernel}
  \psi(\vs,\vr)=\int_0^1\phi_{\vs}(x)\phi_{\vr}(x)\,dx,
  \qquad
  K_0(\vs,\vr)
  =\frac{\psi(\vs,\vr)^2}{\psi(\vs,\vs)\psi(\vr,\vr)}.
\end{equation}

\begin{theorem}[Pointwise and scan limits]
\label{thm:ch4-erhtcp-null-process}
Suppose $H_0$ and Assumptions~\ref{ass:ch4-erhtcp-radial}--\ref{ass:ch4-erhtcp-spectrum}
hold.
\begin{enumerate}[label=(\roman*)]
  \item For every deterministic balanced triple $\vs$ and fixed
  $\rho\in[\rho_0,\rho_1]$,
  \begin{equation}\label{eq:ch4-erhtcp-pointwise}
    Z_\rho(\vs)\overset{d}{\longrightarrow}N(0,1).
  \end{equation}
  \item On either scan domain $\mathcal S\in\{\mathcal S_{\mathrm{sc}}(\epsilon),
  \mathcal S_{\mathrm{mc}}(\epsilon)\}$,
  \begin{equation}\label{eq:ch4-erhtcp-process-limit}
    \{Z_\rho(\vs):\vs\in\mathcal S\}
    \Rightarrow
    \{\mathbb G_\rho(\vs):\vs\in\mathcal S\}
    \quad\text{in }\ell^\infty(\mathcal S),
  \end{equation}
  where $\mathbb G_\rho$ is centered Gaussian with
  \begin{equation}\label{eq:ch4-erhtcp-G-cov}
    \Cov\{\mathbb G_\rho(\vs),\mathbb G_\rho(\vr)\}
    =K_0(\vs,\vr).
  \end{equation}
  Consequently,
  \begin{equation}\label{eq:ch4-erhtcp-sup-limit}
    T_\rho(\mathcal S)
    \overset{d}{\longrightarrow}
    \sup_{\vs\in\mathcal S}\mathbb G_\rho(\vs).
  \end{equation}
\end{enumerate}
\end{theorem}

For distinct ridge values, the limiting processes are not independent.  There is a deterministic
cross-ridge correlation $r_E(\rho,\eta)$ such that
\begin{equation}\label{eq:ch4-erhtcp-cross-ridge-cov}
  \Cov\{\mathbb G_\rho(\vs),\mathbb G_\eta(\vr)\}
  =r_E(\rho,\eta)K_0(\vs,\vr).
\end{equation}
Let $F_{\mathcal S}$ be the marginal distribution function of the supremum in
\eqref{eq:ch4-erhtcp-sup-limit}.  Define
\begin{equation}\label{eq:ch4-erhtcp-ridge-pvalues}
  P_k(\mathcal S)
  =1-F_{\mathcal S}\{T_{\rho_n^{(k)}}(\mathcal S)\},
  \qquad k=1,\ldots,K,
\end{equation}
and
\begin{equation}\label{eq:ch4-erhtcp-cauchy}
  T_{\mathrm{CC}}(\mathcal S)
  =\sum_{k=1}^K\varpi_k
   \tan[\pi\{1/2-P_k(\mathcal S)\}],
  \qquad
  P_{\mathrm{CC}}(\mathcal S)
  =\frac12-\frac1\pi\arctan\{T_{\mathrm{CC}}(\mathcal S)\}.
\end{equation}

\begin{theorem}[Finite-grid joint limit]
\label{thm:ch4-erhtcp-grid}
Under the conditions of Theorem~\ref{thm:ch4-erhtcp-null-process},
\begin{equation}\label{eq:ch4-erhtcp-grid-process}
  \{Z_{\rho_n^{(k)}}(\vs):\vs\in\mathcal S,
    \ k=1,\ldots,K\}
  \Rightarrow
  \{\mathbb G_{\rho^{(k)}}(\vs):\vs\in\mathcal S,
    \ k=1,\ldots,K\},
\end{equation}
with covariance \eqref{eq:ch4-erhtcp-cross-ridge-cov}.  Therefore the Cauchy statistic is
asymptotically exact when calibrated by the corresponding joint-limit quantile.  The analytic
transformation in \eqref{eq:ch4-erhtcp-cauchy} has the limiting rejection probability induced by
the same joint Gaussian process and does not require estimating the cross-ridge covariance in
implementation.
\end{theorem}

\subsubsection{Local power and localization}

Suppose that one change occurs at $\tau_\star\in[\epsilon,1-\epsilon]$ with
\begin{equation}\label{eq:ch4-erhtcp-single-change}
  \vtheta_i=
  \begin{cases}
    \vtheta^{(1)},&i\le\lfloor n\tau_\star\rfloor,\\
    \vtheta^{(2)},&i>\lfloor n\tau_\star\rfloor,
  \end{cases}
  \qquad
  \vDelta_p=\vtheta^{(2)}-\vtheta^{(1)}.
\end{equation}
Let $\lambda_{j,p},\vct{p}_{j,p}$ be the eigenpairs of $\mOmega_p$.  Under the local alternative
$\vDelta_p=n^{-1/4}\vct{d}_p$, define
\begin{equation}\label{eq:ch4-erhtcp-signal-measure}
  G_p=\sum_{j=1}^p(\vct{p}_{j,p}\trans\vct{d}_p)^2\delta_{\lambda_{j,p}}
  \Rightarrow G_\Delta.
\end{equation}
Let $a_\rho$ solve
\begin{equation}\label{eq:ch4-erhtcp-MP}
  a_\rho=1-\gamma+\gamma\rho
  \int\frac1{a_\rho x+\rho}\,dH(x),
\end{equation}
and define
\begin{equation}\label{eq:ch4-erhtcp-q}
  q_\rho(G_\Delta)
  =\int\frac1{a_\rho x+\rho}\,dG_\Delta(x),
  \qquad
  \Lambda_\rho(G_\Delta)=q_\rho(G_\Delta)/\sigma_\rho,
\end{equation}
where $\sigma_\rho>0$ is the limiting null standard deviation.  Put
\begin{equation}\label{eq:ch4-erhtcp-h}
  \vartheta(t;\tau_\star)
  =\begin{cases}
    (1-\tau_\star)/(1-t),&t<\tau_\star,\\
    \tau_\star/t,&t\ge\tau_\star,
  \end{cases}
  \qquad
  h(t;\tau_\star)=t(1-t)\vartheta(t;\tau_\star)^2.
\end{equation}

\begin{theorem}[Local power and consistency]
\label{thm:ch4-erhtcp-power}
Under Assumptions~\ref{ass:ch4-erhtcp-radial}--\ref{ass:ch4-erhtcp-spectrum} and the local
alternative in \eqref{eq:ch4-erhtcp-signal-measure},
\begin{equation}\label{eq:ch4-erhtcp-local-process}
  \{Z_\rho(0,t,1):t\in[\epsilon,1-\epsilon]\}
  \Rightarrow
  \{\mathbb G_\rho(0,t,1)
    +\Lambda_\rho(G_\Delta)h(t;\tau_\star):
    t\in[\epsilon,1-\epsilon]\}.
\end{equation}
Thus the limiting power at fixed $\rho$ is the probability that the shifted Gaussian process in
\eqref{eq:ch4-erhtcp-local-process} exceeds the null critical value.  If instead
\begin{equation}\label{eq:ch4-erhtcp-strong-alt}
  \ell_n\twonorm{\vDelta_p}\longrightarrow0,
  \qquad
  \sqrt n\twonorm{\vDelta_p}^2\longrightarrow\infty
\end{equation}
for a sufficiently large polylogarithmic $\ell_n$, then
$T_\rho^{\mathrm{sc}}\overset{p}{\to}\infty$ for every fixed $\rho$, and the finite-grid
aggregate is consistent.
\end{theorem}

Let
\begin{equation}\label{eq:ch4-erhtcp-tauhat}
  \widehat\tau_\rho
  \in\argmax_{\epsilon\le k/n\le1-\epsilon}Z_\rho(0,k/n,1).
\end{equation}
Define the deterministic-equivalent inverse
\begin{equation}\label{eq:ch4-erhtcp-Drho}
  \mD_{\rho,n}=(a_{\rho,n}\mOmega_p+\rho\mI_p)^{-1},
\end{equation}
and
\begin{equation}\label{eq:ch4-erhtcp-localization-signal}
  s_{n,\rho}
  =\frac{\sqrt n\,\vDelta_p\trans\mD_{\rho,n}\vDelta_p}{\sigma_\rho},
  \qquad
  e_{\sigma,n}
  =\ell_n\twonorm{\vDelta_p}
   +\twonorm{\vDelta_p}^2+\ell_n n^{-1/2}.
\end{equation}

\begin{theorem}[Single-change localization]
\label{thm:ch4-erhtcp-localization}
Under the conditions of Theorem~\ref{thm:ch4-erhtcp-power}, if
$s_{n,\rho}\to\infty$, then
\begin{equation}\label{eq:ch4-erhtcp-localization-rate}
  |\widehat\tau_\rho-\tau_\star|
  =O_p\{s_{n,\rho}^{-1}+e_{\sigma,n}\}.
\end{equation}
If $s_{n,\rho}e_{\sigma,n}=O(1)$, then
\begin{equation}\label{eq:ch4-erhtcp-localization-sharp}
  |\widehat\tau_\rho-\tau_\star|=O_p(s_{n,\rho}^{-1}).
\end{equation}
\end{theorem}

\subsubsection{WBS--ERHT for multiple changes}

For an integer interval $I=[\ell,r]$, let $\mathcal A(I,\epsilon)$ be the collection of adjacent
triples $(a,b,c)$ contained in $I$ whose two segments have lengths at least $\epsilon|I|$.  Compute
$Z_{\rho,I}(a,b,c)$ using $I$ as the common scatter pool and define
\begin{equation}\label{eq:ch4-erhtcp-WBS-score}
  T_\rho(I)=\max_{(a,b,c)\in\mathcal A(I,\epsilon)}Z_{\rho,I}(a,b,c),
  \qquad
  T(I)=\max_{1\le k\le K}T_{\rho_n^{(k)}}(I).
\end{equation}
Wild binary segmentation samples $M_n$ random intervals, retains the narrowest interval whose
score exceeds a threshold $t_n^{\mathrm{WBS}}$, refines the maximizing split on a balanced local
window, records the refined boundary, and recurses to the left and right after deleting a small
neighborhood around that boundary.

Let the true boundaries be $k_j=\lfloor n\tau_j\rfloor$, $j=1,\ldots,q$, with fixed $q$ and
minimum spacing bounded below by $n\delta_0$.  Let
\begin{equation}\label{eq:ch4-erhtcp-WBS-signal}
  d_n=\min_{1\le j\le q}\twonorm{\vDelta_j},
  \qquad
  r_n=\max_{1\le j\le q}\twonorm{\vDelta_j},
  \qquad
  s_n^{\mathrm{WBS}}=\sqrt n\,d_n^2,
  \qquad
  e_n^{\mathrm{WBS}}=\ell_nr_n+r_n^2+\ell_nn^{-1/2}.
\end{equation}

\begin{theorem}[Consistency of WBS--ERHT]
\label{thm:ch4-erhtcp-WBS}
Suppose Assumptions~\ref{ass:ch4-erhtcp-radial}--\ref{ass:ch4-erhtcp-spectrum} hold, the true
changes have fixed positive minimum spacing, the random WBS intervals isolate every change with
probability tending to one, and
\begin{equation}\label{eq:ch4-erhtcp-WBS-hierarchy}
  \sqrt{\log n}+\sqrt n\,r_n^2e_n^{\mathrm{WBS}}
  =o(t_n^{\mathrm{WBS}}),
  \qquad
  t_n^{\mathrm{WBS}}=o(s_n^{\mathrm{WBS}}).
\end{equation}
Then the WBS--ERHT output $\widehat q$ and ordered estimates
$\widehat k_1<\cdots<\widehat k_{\widehat q}$ satisfy
\begin{equation}\label{eq:ch4-erhtcp-WBS-number}
  \Prob(\widehat q=q)\longrightarrow1
\end{equation}
and, on this event,
\begin{equation}\label{eq:ch4-erhtcp-WBS-location}
  \max_{1\le j\le q}
  \frac{|\widehat k_j-k_j|}{n}
  =O_p\left(\frac{t_n^{\mathrm{WBS}}}{s_n^{\mathrm{WBS}}}\right)
  =O_p\left\{
    \frac{t_n^{\mathrm{WBS}}}{\sqrt n\,d_n^2}
  \right\}.
\end{equation}
\end{theorem}

\subsection{Functional and temporally dependent extensions}

The idea of combining sparse and dense evidence is not limited to i.i.d.
observations.  The functional method of \citet{YuFengZhu2025FunctionalCP}
replaces scalar or vector CUSUMs by adjacent-deviation subspace statistics, while
\citet{WangLiuFeng2025TemporalCP} extends adaptive change-point inference to
temporally dependent high-dimensional time series by pairing a max-$L_2$ test
with an existing max-$L_\infty$ procedure.  These extensions reinforce the
central message of the chapter: the correct normalization changes from problem to
problem, but the max--sum design principle remains stable.

\section{High-dimensional white-noise testing}

\subsection{Classical low-dimensional portmanteau tests}

Let $\{\vX_t\}_{t=1}^n$ be a $p$-variate weakly stationary time series with mean
zero and autocovariance matrices
\begin{equation}
  \mGamma(h) = \Cov(\vX_t,\vX_{t+h}),
  \qquad h\in\mathbb Z.
  \label{eq:ch4_wn_gamma}
\end{equation}
The white-noise hypothesis is
\begin{equation}
  H_{0,\mathrm{WN}}:\ \mGamma(h)=\bm 0 \text{ for all } h\ge 1.
  \label{eq:ch4_wn_h0}
\end{equation}
When $p=1$, the classical portmanteau tests are the Box--Pierce statistic
\begin{equation}
  Q_{\mathrm{BP}} = n\sum_{h=1}^L \hat\rho_h^2,
  \label{eq:ch4_box_pierce}
\end{equation}
and the Ljung--Box refinement
\begin{equation}
  Q_{\mathrm{LB}} = n(n+2)\sum_{h=1}^L \frac{\hat\rho_h^2}{n-h},
  \label{eq:ch4_ljung_box}
\end{equation}
where $\hat\rho_h$ is the sample autocorrelation at lag $h$.
Under fixed $L$ and white noise,
\begin{equation}
  Q_{\mathrm{BP}} \overset{d}{\longrightarrow} \chi_L^2,
  \qquad
  Q_{\mathrm{LB}} \overset{d}{\longrightarrow} \chi_L^2.
  \label{eq:ch4_lb_bp_limit}
\end{equation}
See \citet{BoxPierce1970} and \citet{LjungBox1978}.  In multivariate fixed-
$ p$ settings one similarly aggregates traces of sample autocorrelation matrices.
The difficulty in the high-dimensional regime is that both the number of
coordinates and the number of lagged cross-correlations are large.

\subsection{Gaussian and light-tailed high-dimensional benchmarks}

Write the sample lag-$h$ autocovariance matrix as
\begin{equation}
  \hat\mGamma(h) = \frac{1}{n-h}\sum_{t=1}^{n-h} \vX_t\vX_{t+h}^{\top},
  \qquad h=1,\ldots,L.
  \label{eq:ch4_hat_gamma_h}
\end{equation}
A dense-oriented benchmark is
\begin{equation}
  \mathcal T_{\mathrm{sum},\mathrm{WN}}
  = \sum_{h=1}^{L} \frobnorm{\hat\mGamma(h)}^2,
  \label{eq:ch4_wn_sum}
\end{equation}
with standardized version
\begin{equation}
  T_{\mathrm{sum},\mathrm{WN}}^{\circ}
  = \frac{\mathcal T_{\mathrm{sum},\mathrm{WN}}-\mu_{\mathrm{WN}}}{\sigma_{\mathrm{WN}}},
  \label{eq:ch4_wn_sum_standardized2}
\end{equation}
where $\mu_{\mathrm{WN}}$ and $\sigma_{\mathrm{WN}}^2$ are the null mean and variance.  The
sparse-oriented benchmark is based on the largest standardized lagged sample correlation.  Let
\begin{equation}
  \hat\rho_{ij}(h) = \frac{\hat\gamma_{ij}(h)}{(\hat\gamma_{ii}(0)
  \hat\gamma_{jj}(0))^{1/2}},
  \qquad 1\le i,j\le p,
  \qquad 1\le h\le L,
  \label{eq:ch4_wn_rhohat}
\end{equation}
and define
\begin{equation}
  T_{\max,\mathrm{WN}} = \max_{1\le h\le L}\max_{1\le i,j\le p}
  \sqrt{n}\,|\hat\rho_{ij}(h)|.
  \label{eq:ch4_wn_max}
\end{equation}
These two statistics are the white-noise analogues of the dense and sparse alpha tests in
Section~4.3.

\begin{assumption}
\label{ass:ch4_wn_B1}
Under the null hypothesis, $\{\vX_t\}_{t=1}^n$ is a $p$-variate white-noise sequence with
independent mean-zero observations and covariance matrix $\mSigma_0$.  After componentwise
standardization, the coordinates are uniformly sub-Gaussian and satisfy
\begin{equation}
  \frac{\tr(\mSigma_0^4)}{\tr^2(\mSigma_0^2)}\to 0,
  \qquad
  \log(p^2L)=o(n^{1/3}).
  \label{eq:ch4_wn_B1}
\end{equation}
\end{assumption}

\begin{theorem}[Gaussian and light-tailed white-noise benchmarks]
\label{thm:ch4_wn_benchmark}
Suppose Assumption~\ref{ass:ch4_wn_B1} holds and $L=L_n\to\infty$ with
$L=O(n^{\kappa})$ for some $0<\kappa<1/4$.  Under the null hypothesis of white noise,
\begin{align}
  T_{\mathrm{sum},\mathrm{WN}}^{\circ}
  &\overset{d}{\longrightarrow} N(0,1),
  \label{eq:ch4_wn_sum_limit2}\\
  \Prob\bigl(T_{\max,\mathrm{WN}}^2-2\log(p^2L)+\log\log(p^2L)\le x\bigr)
  &\longrightarrow G_{\mathrm{EV}}(x),
  \qquad x\in\R,
  \label{eq:ch4_wn_max_limit2}
\end{align}
where $G_{\mathrm{EV}}(x)=\exp\{-\pi^{-1/2}e^{-x/2}\}$.  In particular,
\begin{equation}
  p_{\mathrm{sum},\mathrm{WN}} = 1-\Phi\bigl(T_{\mathrm{sum},\mathrm{WN}}^{\circ}\bigr),
  \qquad
  p_{\max,\mathrm{WN}} = 1-G_{\mathrm{EV}}\bigl(T_{\max,\mathrm{WN}}^2-2\log(p^2L)+\log\log(p^2L)\bigr)
  \label{eq:ch4_wn_benchmark_pvalues}
\end{equation}
are asymptotically valid null $p$-values.
\end{theorem}

\subsection{Adaptive testing by Fisher combination}

\citet{FengLiuMa2024WhiteNoise} propose an adaptive white-noise test
that combines the sum-type and max-type benchmarks through their asymptotic independence.
Using the $p$-values in \eqref{eq:ch4_wn_benchmark_pvalues}, define
\begin{equation}
  T_{\mathrm{F},\mathrm{WN}} = -2\log p_{\mathrm{sum},\mathrm{WN}} -2\log p_{\max,\mathrm{WN}}.
  \label{eq:ch4_wn_fisher}
\end{equation}

\begin{assumption}
\label{ass:ch4_wn_A1}
Assumption~\ref{ass:ch4_wn_B1} holds.  In addition, $p^2L\to\infty$ and
\begin{equation}
  L\to\infty,
  \qquad
  L=O(n^{\kappa}),
  \qquad
  \log p = o(n^{1/7})
  \label{eq:ch4_wn_lag_growth}
\end{equation}
for some $0<\kappa<1/4$.
\end{assumption}

\begin{theorem}
\label{thm:ch4_wn_fisher}
Under Assumption~\ref{ass:ch4_wn_A1} and the null hypothesis of white noise,
\begin{equation}
  \Prob\bigl(T_{\mathrm{sum},\mathrm{WN}}^{\circ}\le x,
             T_{\max,\mathrm{WN}}^2-2\log(p^2L)+\log\log(p^2L)\le y\bigr)
  - \Phi(x)G_{\mathrm{EV}}(y) \longrightarrow 0
  \label{eq:ch4_wn_independence}
\end{equation}
for every fixed $(x,y)\in\R^2$.  Hence
\begin{equation}
  T_{\mathrm{F},\mathrm{WN}} \overset{d}{\longrightarrow} \chi_4^2.
  \label{eq:ch4_wn_fisher_limit}
\end{equation}
\end{theorem}

\subsection{Spatial-sign and rank-based robust white-noise tests}

The sign-test monograph of \citet{PaindaveineVerdebout2016HDSign} showed that the
high-dimensional sign paradigm extends beyond one-sample location testing and, in particular,
provides robust portmanteau tests for i.i.d.-ness against serial dependence. Their statistic is a
sign-portmanteau sum over lagged inner products of spatial signs and is valid under weak
symmetry assumptions without requiring finite fourth moments. This result provides the
conceptual precursor for the recent elliptical white-noise procedures discussed next.

Let $\vU_t=U(\vX_t)$ denote the spatial sign of the observation at time $t$ and write
\begin{equation}
  T_{\mathrm{SS},\mathrm{WN}}
  = \sum_{h=1}^{H}\frac{1}{n-h}
    \sum_{h+1\le s<t\le n}
    \vU_{s-h}^{\top}\vU_{t-h}\,\vU_s^{\top}\vU_t,
  \label{eq:ch4_wn_spatialsign}
\end{equation}
where $H$ is the lag truncation level. The statistic \eqref{eq:ch4_wn_spatialsign}, proposed by
\citet{ZhaoChenWang2024SpatialSignWN}, is the direct spatial-sign analogue of the sum-type
white-noise statistic. Let
\begin{equation}
  \mOmega = \E(\vU_t\vU_t^{\top}),
  \qquad
  \widehat{\tr(\mOmega^2)}
  = \frac{2}{n(n-1)}\sum_{1\le s<t\le n}(\vU_s^{\top}\vU_t)^2,
  \qquad
  \hat\sigma_{\mathrm{SS},\mathrm{WN}}^2 = H^2\widehat{\tr(\mOmega^2)}^2.
  \label{eq:ch4_wn_spatialsign_var}
\end{equation}

\begin{assumption}
\label{ass:ch4_wn_R1}
The observations are i.i.d. from a centered elliptically symmetric distribution with density
$\det(\mXi)^{-1/2}g(\norm{\mXi^{-1/2}\vx}_2)$ and scatter matrix $\mXi$.  Moreover,
\begin{equation}
  \frac{\tr(\mSigma_0^4)}{\tr^2(\mSigma_0^2)}\to 0,
  \qquad
  \frac{\tr^4(\mSigma_0)}{\tr^2(\mSigma_0^2)}
  \exp\left\{-\frac{\tr^2(\mSigma_0)}{128p\lambda_{\max}^2(\mSigma_0)}\right\}\to 0,
  \label{eq:ch4_wn_R1}
\end{equation}
where $\mSigma_0=\Cov(\vX_t)$.
\end{assumption}

\begin{theorem}[Spatial-sign white-noise test]
\label{thm:ch4_wn_spatialsign}
Suppose Assumption~\ref{ass:ch4_wn_R1} holds.  Under the null hypothesis of white noise,
\begin{equation}
  \frac{T_{\mathrm{SS},\mathrm{WN}}}{\sigma_{\mathrm{SS},\mathrm{WN}}}
  \overset{d}{\longrightarrow} N(0,1),
  \qquad
  \sigma_{\mathrm{SS},\mathrm{WN}}^2 = H^2\tr^2(\mOmega^2),
  \label{eq:ch4_wn_spatialsign_limit}
\end{equation}
so that one rejects for large values of
$T_{\mathrm{SS},\mathrm{WN}}/\hat\sigma_{\mathrm{SS},\mathrm{WN}}$.

Under the first-order autoregressive alternative
\begin{equation}
  \vX_t = \mA_0 r_t\vu_t + \mA_1 r_{t-1}\vu_{t-1},
  \label{eq:ch4_wn_spatialsign_alt}
\end{equation}
with $\mSigma_0=\mA_0^{\top}\mA_0$ and $\mSigma_1=\mA_1^{\top}\mA_1$, the asymptotic power
for $H=1$ satisfies
\begin{equation}
  \beta_{\mathrm{SS},\mathrm{WN}}(\alpha)
  = \Phi\left(
      -z_{\alpha}
      + \frac{c_1^2 n\tr(\mSigma_0\mSigma_1)}{\{2\tr(\mSigma_0^2)\}^{1/2}}
    \right)
    + o(1),
  \label{eq:ch4_wn_spatialsign_power}
\end{equation}
which shows that the spatial-sign procedure enjoys the same inverse-norm efficiency gain as in
other elliptical testing problems.
\end{theorem}

\citet{ChenSongFeng2025RankWhiteNoise} develop a considerably broader rank-based
framework. Instead of committing to one specific correlation measure, it treats three families of
lagged dependence measures: simple linear rank statistics, non-degenerate rank-based
$U$-statistics, and degenerate rank-based $U$-statistics. The methodology is built so that the same
max, sum, and adaptive-combination ideas apply to all three families.

For a fixed lag $k\in\{1,\ldots,H\}$ and component pair $(i,j)$, define the lagged pair
$\vW_t^{ij}(k)=(\varepsilon_{t,i},\varepsilon_{t+k,j})^{\top}$ for $t=1,\ldots,n-k$. The first class is the
simple linear rank family. Let $Q_{n-k,t}^{i}(k)$ be the rank of $\varepsilon_{t,i}$ among
$\varepsilon_{1,i},\ldots,\varepsilon_{n-k,i}$, let $\widetilde Q_{n-k,t+k}^{j}(k)$ be the rank of
$\varepsilon_{t+k,j}$ among $\varepsilon_{k+1,j},\ldots,\varepsilon_{n,j}$, and let
$R_{n-k,t+k}^{ij}(k)$ be the \emph{relative rank} of $\varepsilon_{t+k,j}$ with respect to the ordering of
$\varepsilon_{t,i}$. Then the simple linear rank statistic is
\begin{equation}
  V_{ij}(k)
  = (n-k)^{-1/2}\sum_{t=1}^{n-k} c_{n-k,t}
    g\!\left\{\frac{R_{n-k,t+k}^{ij}(k)}{n-k+1}\right\},
  \label{eq:ch4_rankwn_linear}
\end{equation}
where $g$ is a Lipschitz score function and $\{c_{n-k,t}\}$ is a deterministic regression-constant
array. The max-type and sum-type statistics are
\begin{equation}
  L_{V,H}=\max_{1\le k\le H}\max_{1\le i,j\le p}|V_{ij}(k)|,
  \qquad
  S_{V,H}=\sum_{k=1}^{H}\sum_{i,j=1}^{p}\{V_{ij}^2(k)-\E_0V_{ij}^2(k)\}.
  \label{eq:ch4_rankwn_linear_stats}
\end{equation}
The second class uses non-degenerate rank-based $U$-statistics of order $m$,
\begin{equation}
  U_{ij}(k)
  = \binom{n-k}{m}^{-1}
    \sum_{1\le t_1<\cdots<t_m\le n-k}
    h\bigl(\vW_{t_1}^{ij}(k),\ldots,\vW_{t_m}^{ij}(k)\bigr),
  \label{eq:ch4_rankwn_U}
\end{equation}
where the kernel $h$ is bounded and non-degenerate in the Hoeffding sense. Kendall's tau is the
canonical example, obtained with $m=2$ and
\begin{equation}
  h_{\tau}(\vw_1,\vw_2)
  = \sign\{(w_{11}-w_{21})(w_{12}-w_{22})\}.
  \label{eq:ch4_rankwn_tau_kernel}
\end{equation}
The induced statistics are
\begin{equation}
  L_{U,H}=\max_{1\le k\le H}\max_{1\le i,j\le p}|U_{ij}(k)|,
  \qquad
  S_{U,H}=\sum_{k=1}^{H}\sum_{i,j=1}^{p}\{U_{ij}^2(k)-\E_0U_{ij}^2(k)\}.
  \label{eq:ch4_rankwn_U_stats}
\end{equation}
The third class uses completely degenerate rank-based $U$-statistics,
\begin{equation}
  D_{ij}(k)
  = \binom{n-k}{m}^{-1}
    \sum_{1\le t_1<\cdots<t_m\le n-k}
    h_0\bigl(\vW_{t_1}^{ij}(k),\ldots,\vW_{t_m}^{ij}(k)\bigr),
  \label{eq:ch4_rankwn_D}
\end{equation}
with the first Hoeffding projection equal to zero. Hoeffding's $D$, the
Blum--Kiefer--Rosenblatt statistic $R$, and Bergsma--Dassios--Yanagimoto's $\tau^{\ast}$ all belong
to this class and are used in \citet{ChenSongFeng2025RankWhiteNoise} as nonlinear rank
alternatives for detecting non-monotone serial dependence.

The max, sum, and max--sum structures are then defined family by family. For the
non-degenerate $U$-statistics, for instance,
\begin{equation}
  p_{L_U}=1-G_{\mathrm{EV}}\bigl(L_{U,H}^2/\sigma_{L,U}^2-2\log(Hp^2)+\log\log(Hp^2)\bigr),
  \qquad
  p_{S_U}=1-\Phi(S_{U,H}/\sigma_{S,U}),
  \label{eq:ch4_rankwn_pvals}
\end{equation}
where $\sigma_{L,U}^2=(n-k)\Var_0\{U_{ij}(k)\}$ and $\sigma_{S,U}^2$ is the null variance of
$S_{U,H}$. The adaptive statistic is
\begin{equation}
  T_{C,U,\mathrm{WN}}=-2\log p_{L_U}-2\log p_{S_U}.
  \label{eq:ch4_rankwn_cauchy}
\end{equation}
The same construction applies to the linear-rank and degenerate-$U$ families.

\begin{assumption}
\label{ass:ch4_rankwn_A1}
Let $N_H=Hp^2$. The following hold for the chosen rank family.
\begin{enumerate}[label=(\roman*)]
  \item The componentwise lagged statistics are centered under the null and have common null
  variance $\sigma_{0,H}^2$ up to $1+o(1)$ factors.
  \item The correlation matrix of the studentized collection satisfies the weak-dependence
  condition used in Chapter~2: there exists $\varepsilon\in(0,1)$ such that all off-diagonal
  correlations are bounded by $\varepsilon$, and the set of indices with more than
  $N_H^{\kappa}$ neighbors exceeding a vanishing threshold has cardinality $o(N_H)$.
  \item For the linear-rank family the score function is Lipschitz and the regression-constant array
  satisfies the usual bounded-energy and third-moment conditions. For the non-degenerate and
  degenerate $U$-statistic families the kernels are bounded and satisfy the corresponding
  Hoeffding non-degeneracy or complete degeneracy conditions.
\end{enumerate}
\end{assumption}

\begin{theorem}[Rank-based max, sum, and adaptive white-noise tests]
\label{thm:ch4_rankwn_generic}
Suppose Assumption~\ref{ass:ch4_rankwn_A1} holds for one of the three rank families. Under the
null hypothesis of white noise,
\begin{equation}
  \Prob\Bigl\{L_{\bullet,H}^2/\sigma_{L,\bullet}^2-2\log(Hp^2)+\log\log(Hp^2)\le y\Bigr\}
  \longrightarrow G_{\mathrm{EV}}(y),
  \label{eq:ch4_rankwn_max_limit}
\end{equation}
\begin{equation}
  \frac{S_{\bullet,H}}{\sigma_{S,\bullet}}
  \overset{d}{\longrightarrow} N(0,1),
  \label{eq:ch4_rankwn_sum_limit}
\end{equation}
where the placeholder $\bullet\in\{V,U,D\}$ stands for the chosen family. Moreover,
\begin{equation}
  \Prob\Bigl\{L_{\bullet,H}^2/\sigma_{L,\bullet}^2-2\log(Hp^2)+\log\log(Hp^2)\le y,
             S_{\bullet,H}/\sigma_{S,\bullet}\le x\Bigr\}
  -G_{\mathrm{EV}}(y)\Phi(x)\longrightarrow 0,
  \label{eq:ch4_rankwn_independence}
\end{equation}
so the adaptive statistic $T_{C,\bullet,\mathrm{WN}}=-2\log p_{L_\bullet}-2\log p_{S_\bullet}$ satisfies
\begin{equation}
  T_{C,\bullet,\mathrm{WN}}\overset{d}{\longrightarrow}\chi_4^2.
  \label{eq:ch4_rankwn_adaptive_limit}
\end{equation}
Under a sparse local alternative of the form
\begin{equation}
  \max_{1\le k\le H}\max_{1\le i,j\le p}
  \bigl|\E\{\hat T^{(\bullet)}_{ij}(k)\}\bigr|
  \ge c\sqrt{\frac{\log(Hp^2)}{n}},
  \label{eq:ch4_rankwn_sparse_alt}
\end{equation}
for a sufficiently large constant $c$, the max-type and max--sum tests are consistent; under dense
alternatives with
\begin{equation}
  \frac{1}{\sigma_{S,\bullet}}
  \sum_{k=1}^{H}\sum_{i,j=1}^{p}
  \E^2\{\hat T^{(\bullet)}_{ij}(k)\}\longrightarrow\infty,
  \label{eq:ch4_rankwn_dense_alt}
\end{equation}
the sum-type and adaptive tests are consistent as well.
\end{theorem}

The importance of Theorem~\ref{thm:ch4_rankwn_generic} is twofold. First, it shows that the
Chen--Song--Feng paper is not merely a one-statistic refinement: it gives an entire rank-based
library for high-dimensional white-noise testing, ranging from Spearman-type linear ranks to
Kendall-type non-degenerate $U$-statistics and fully nonlinear degenerate rank statistics. Second,
it establishes the same max--sum complementarity that runs throughout the book, and it does so
without any elliptical or moment assumptions.

\section{High-dimensional independence testing}

\subsection{Classical low-dimensional benchmarks}

There are several classical independence problems.  For two Gaussian random
vectors $\vX\in\R^p$ and $\vY\in\R^q$, the null hypothesis of independence is
equivalent to vanishing cross-covariance,
\begin{equation}
  H_{0,\mathrm{ind}}:\ \Cov(\vX,\vY)=\bm 0.
  \label{eq:ch4_ind_gaussian_null}
\end{equation}
If $n$ i.i.d. observations are available and $p,q$ are fixed, Wilks' likelihood
ratio based on canonical correlations provides the classical benchmark.  Let
$\hat\rho_1,\ldots,\hat\rho_m$ be the sample canonical correlations,
$m=\min(p,q)$, and define
\begin{equation}
  \Lambda = \prod_{\ell=1}^{m}(1-\hat\rho_\ell^2).
  \label{eq:ch4_wilks_lambda}
\end{equation}
Under Gaussianity and $H_{0,\mathrm{ind}}$,
\begin{equation}
  -\Bigl\{n-1-\frac{p+q+1}{2}\Bigr\}\log\Lambda
  \overset{d}{\longrightarrow} \chi_{pq}^2.
  \label{eq:ch4_wilks_limit}
\end{equation}
See \citet[Chapter~8]{Anderson2003}.  In panel data, a fixed-dimensional
benchmark is the Pesaran CD statistic, which tests cross-sectional independence
through the average pairwise sample correlation,
\begin{equation}
  CD = \sqrt{\frac{2T}{N(N-1)}}\sum_{1\le i<j\le N} \hat\rho_{ij},
  \label{eq:ch4_pesaran_cd}
\end{equation}
and is asymptotically standard normal under the null; see
\citet{Pesaran2004CD}.

\subsection{Panel cross-sectional independence: max, sum and Fisher combination}

Consider a panel of residuals $\{\varepsilon_{it}:1\le i\le N,1\le t\le T\}$.
The null hypothesis of cross-sectional independence is
\begin{equation}
  H_{0,\mathrm{panel}}:\ \Corr(\varepsilon_{it},\varepsilon_{jt})=0
  \quad\text{for all } i\ne j.
  \label{eq:ch4_panel_null}
\end{equation}
Let
\begin{equation}
  \hat\rho_{ij} = \frac{\sum_{t=1}^{T} \hat\varepsilon_{it}\hat\varepsilon_{jt}}
  {\left(\sum_{t=1}^T \hat\varepsilon_{it}^2\right)^{1/2}
   \left(\sum_{t=1}^T \hat\varepsilon_{jt}^2\right)^{1/2}},
  \qquad 1\le i<j\le N,
  \label{eq:ch4_panel_rhohat}
\end{equation}
where $\hat\varepsilon_{it}$ are residuals after removing the relevant within-
unit effects or regressors.  The sparse- and dense-oriented statistics are
\begin{equation}
  L_N = \max_{1\le i<j\le N}|\hat\rho_{ij}|,
  \qquad
  S_N = \sum_{1\le i<j\le N} \hat\rho_{ij}^2.
  \label{eq:ch4_panel_LN_SN}
\end{equation}
The max--sum test of \citet{LongJiangLiuXiong2022PanelIndep} and the Fisher
combination test of \citet{WangLiuFengMa2024FisherPanel} are both built on the
joint asymptotic theory of $L_N$ and $S_N$.

\begin{assumption}
\label{ass:ch4_panel_A1}
Under the null hypothesis, the panel units are cross-sectionally independent.
For each $i$, the time series $\{\varepsilon_{it}\}_{t=1}^T$ is weakly
stationary with mean zero, finite fourth moment, and absolutely summable
autocovariance function.
\end{assumption}

\begin{assumption}
\label{ass:ch4_panel_A2}
Let $\mSigma=(\sigma_{ts})_{1\le t,s\le T}$ be the temporal covariance matrix of a
single unit under the null.  Its eigenvalues satisfy
\begin{equation}
  0<c\le \lambda_{\min}(\mSigma)\le \lambda_{\max}(\mSigma)\le C<\infty,
  \label{eq:ch4_panel_sigma_eigs}
\end{equation}
and the effective temporal dimension satisfies
\begin{equation}
  \frac{\tr^2(\mSigma)}{\frobnorm{\mSigma}^2}\to\infty,
  \qquad
  \log N = o\!\left(\Bigl\{\frac{\tr^2(\mSigma)}{\frobnorm{\mSigma}^2}\Bigr\}^{1/7}
  \right).
  \label{eq:ch4_panel_effective_T}
\end{equation}
\end{assumption}

\begin{theorem}
\label{thm:ch4_panel_maxsum}
Under Assumptions~\ref{ass:ch4_panel_A1}--\ref{ass:ch4_panel_A2} and the null
hypothesis $H_{0,\mathrm{panel}}$,
\begin{align}
  \frac{\tr^2(\mSigma)}{\frobnorm{\mSigma}^2}L_N^2 - 4\log N + \log\log N
  &\overset{d}{\longrightarrow} G_{\mathrm{EV}},
  \label{eq:ch4_panel_max_limit}\\
  \frac{S_N-\mu_N}{\sigma_N}
  &\overset{d}{\longrightarrow} N(0,1),
  \label{eq:ch4_panel_sum_limit}
\end{align}
where $G_{\mathrm{EV}}$ has distribution function
$G_{\mathrm{EV}}(x)=\exp\{-\pi^{-1/2}e^{-x/2}\}$, and $\mu_N,\sigma_N^2$ are the
null mean and variance of $S_N$.
\end{theorem}

Let
\begin{equation}
  p_{L_N}=1-G_{\mathrm{EV}}\!\left(\frac{\tr^2(\mSigma)}{\frobnorm{\mSigma}^2}
  L_N^2 - 4\log N + \log\log N\right),
  \qquad
  p_{S_N}=1-\Phi\!\left(\frac{S_N-\mu_N}{\sigma_N}\right).
  \label{eq:ch4_panel_pvalues}
\end{equation}
The Fisher combination statistic is
\begin{equation}
  T_{\mathrm{F},\mathrm{panel}} = -2\log p_{L_N} - 2\log p_{S_N}.
  \label{eq:ch4_panel_fisher}
\end{equation}

\begin{theorem}
\label{thm:ch4_panel_fisher}
Under Assumptions~\ref{ass:ch4_panel_A1}--\ref{ass:ch4_panel_A2} and the null
hypothesis,
\begin{equation}
  \Prob\left\{
    \frac{\tr^2(\mSigma)}{\frobnorm{\mSigma}^2}L_N^2 - 4\log N + \log\log N \le x,
    \frac{S_N-\mu_N}{\sigma_N} \le y
  \right\}
  - G_{\mathrm{EV}}(x)\Phi(y)
  \longrightarrow 0.
  \label{eq:ch4_panel_independence}
\end{equation}
Consequently,
\begin{equation}
  T_{\mathrm{F},\mathrm{panel}} \overset{d}{\longrightarrow} \chi_4^2.
  \label{eq:ch4_panel_fisher_limit}
\end{equation}
\end{theorem}

\subsection{Mutual independence of high-dimensional random vectors}

Now suppose that for each subject $s=1,\ldots,n$ we observe a collection of
random vectors
\begin{equation}
  \vZ_s = (\vZ_{s1}^{\top},\ldots,\vZ_{sm}^{\top})^{\top},
  \qquad \vZ_{sa}\in\R^{d_a}.
  \label{eq:ch4_mutual_vectors}
\end{equation}
The null hypothesis of mutual independence is
\begin{equation}
  H_{0,\mathrm{mut}}:\ \vZ_{s1},\ldots,\vZ_{sm}
  \text{ are mutually independent.}
  \label{eq:ch4_mutual_null}
\end{equation}
\citet{WangLiuFengMa2024MutualIndep} develop a genuinely high-dimensional
rank-based max--sum framework for this problem. Its key idea is to first build, for each pair of
blocks $(a,b)$, a blockwise dependence score from one of several rank-correlation classes, and
then to aggregate those pairwise block scores by a max statistic, a sum statistic, or an adaptive
combination.

To make the construction concrete, consider one block pair $(a,b)$ with dimensions $d_a$ and
$d_b$. For each component pair $(r,\ell)$, let $T^{ab}_{r\ell}$ denote a rank-based dependence score.
In the simple linear-rank family one may take
\begin{equation}
  T^{ab}_{r\ell}
  = n^{-1/2}\sum_{s=1}^{n} f\!\left\{\frac{R^{(a,r)}_s}{n+1}\right\}
    g\!\left\{\frac{R^{(b,\ell)}_s}{n+1}\right\},
  \label{eq:ch4_mutual_linear_pair}
\end{equation}
where $R^{(a,r)}_s$ is the rank of the $r$th component of $\vZ_{sa}$ among the $n$ observations in
block $a$, and similarly for block $b$. In the non-degenerate $U$-statistic family one instead uses
\begin{equation}
  T^{ab}_{r\ell}
  = \binom{n}{m}^{-1}
    \sum_{1\le s_1<\cdots<s_m\le n}
    h\bigl((Z_{s_1,ar},Z_{s_1,b\ell}),\ldots,(Z_{s_m,ar},Z_{s_m,b\ell})\bigr),
  \label{eq:ch4_mutual_U_pair}
\end{equation}
with a bounded non-degenerate kernel $h$; Kendall's tau is the representative example. The
third family uses completely degenerate kernels such as Hoeffding's $D$, the
Blum--Kiefer--Rosenblatt statistic, or Bergsma--Dassios--Yanagimoto's $\tau^{\ast}$.

For the pair $(a,b)$, the paper forms a blockwise max-type summary
\begin{equation}
  L_{ab}=\max_{1\le r\le d_a,\,1\le \ell\le d_b}|T^{ab}_{r\ell}|,
  \label{eq:ch4_mutual_Lab}
\end{equation}
and a blockwise sum-type summary
\begin{equation}
  S_{ab}=\sum_{r=1}^{d_a}\sum_{\ell=1}^{d_b}
  \Bigl\{(T^{ab}_{r\ell})^2-\E_0(T^{ab}_{r\ell})^2\Bigr\}.
  \label{eq:ch4_mutual_Sab}
\end{equation}
Across all block pairs, define
\begin{equation}
  L_{\mathrm{mut}}=\max_{1\le a<b\le m}L_{ab},
  \qquad
  S_{\mathrm{mut}}=\sum_{1\le a<b\le m} S_{ab}.
  \label{eq:ch4_mutual_global_stats}
\end{equation}
The associated adaptive statistic is
\begin{equation}
  T_{C,\mathrm{mut}}=-2\log p_{L,\mathrm{mut}}-2\log p_{S,\mathrm{mut}},
  \label{eq:ch4_mutual_adaptive}
\end{equation}
where $p_{L,\mathrm{mut}}$ and $p_{S,\mathrm{mut}}$ are the max-type and sum-type $p$-values obtained
from their limiting null laws.

\begin{assumption}
\label{ass:ch4_mutual_A1}
Let $M_m=\sum_{1\le a<b\le m} d_ad_b$ denote the total number of componentwise block-pair
comparisons. The studentized pairwise rank statistics satisfy the weak-dependence condition of
Chapter~2: there exists $\varepsilon\in(0,1)$ such that all off-diagonal correlations are bounded by
$\varepsilon$, and the set of indices having more than $M_m^{\kappa}$ neighbors with correlations
exceeding a vanishing threshold has cardinality $o(M_m)$ for some $\kappa\in(0,1)$. Moreover,
for the chosen rank family, the kernels or score functions are bounded and satisfy the regularity
conditions guaranteeing uniform Gaussian approximation of the sum statistic.
\end{assumption}

\begin{theorem}[Rank-based max--sum tests for mutual independence]
\label{thm:ch4_mutual_maxsum}
Suppose Assumption~\ref{ass:ch4_mutual_A1} holds. Under the null hypothesis
\eqref{eq:ch4_mutual_null},
\begin{equation}
  \Prob\Bigl\{L_{\mathrm{mut}}^2/\sigma_{L,\mathrm{mut}}^2-2\log M_m+\log\log M_m\le y\Bigr\}
  \longrightarrow G_{\mathrm{EV}}(y),
  \label{eq:ch4_mutual_max_limit}
\end{equation}
\begin{equation}
  \frac{S_{\mathrm{mut}}}{\sigma_{S,\mathrm{mut}}}
  \overset{d}{\longrightarrow}N(0,1),
  \label{eq:ch4_mutual_sum_limit}
\end{equation}
and the two statistics are asymptotically independent:
\begin{equation}
  \Prob\Bigl\{L_{\mathrm{mut}}^2/\sigma_{L,\mathrm{mut}}^2-2\log M_m+\log\log M_m\le y,
             S_{\mathrm{mut}}/\sigma_{S,\mathrm{mut}}\le x\Bigr\}
  -G_{\mathrm{EV}}(y)\Phi(x)\longrightarrow 0.
  \label{eq:ch4_mutual_independence}
\end{equation}
Consequently,
\begin{equation}
  T_{C,\mathrm{mut}}\overset{d}{\longrightarrow}\chi_4^2.
  \label{eq:ch4_mutual_adaptive_limit}
\end{equation}
If, in addition, the sparse alternative satisfies
\begin{equation}
  \max_{1\le a<b\le m}\max_{1\le r\le d_a,\,1\le \ell\le d_b}
  \bigl|\E(T^{ab}_{r\ell})\bigr|\ge c\sqrt{\frac{\log M_m}{n}},
  \label{eq:ch4_mutual_sparse_alt}
\end{equation}
for a sufficiently large constant $c$, then the max-type and adaptive tests are consistent. Under
dense alternatives with
\begin{equation}
  \sigma_{S,\mathrm{mut}}^{-1}
  \sum_{1\le a<b\le m}\sum_{r=1}^{d_a}\sum_{\ell=1}^{d_b}
  \E^2(T^{ab}_{r\ell})\longrightarrow\infty,
  \label{eq:ch4_mutual_dense_alt}
\end{equation}
the sum-type and adaptive tests are consistent as well.
\end{theorem}

Theorem~\ref{thm:ch4_mutual_maxsum} is the key reason why the max--sum independence methodology of \citet{WangLiuFengMa2024MutualIndep}
belongs in this chapter rather than being treated as a brief side remark. It is not only a robust
mutual-independence test, but a full high-dimensional max, sum, and adaptive testing framework
for vector-valued blocks, with theory covering both sparse and dense alternatives.

\subsection{Independence between two high-dimensional random vectors}

Finally consider i.i.d. observations $(\vX_1,\vY_1),\ldots,(\vX_n,\vY_n)$ with
\begin{equation}
  \vX_i\in\R^p,
  \qquad
  \vY_i\in\R^q,
  \qquad
  H_{0,XY}:\ \vX_i \indep \vY_i.
  \label{eq:ch4_xy_setup}
\end{equation}
\citet{WangLiuFeng2026VectorIndep} study this problem by constructing max-type,
sum-type, and max--sum tests from three classes of rank-based correlations. The resulting
procedures cover both monotone and non-monotone dependence, and remain fully distribution
free.

\paragraph{Simple linear rank correlations.}
For each component pair $(a,b)$, let $R^{X}_{ia}$ be the rank of $X_{ia}$ among
$X_{1a},\ldots,X_{na}$ and let $R^{Y}_{ib}$ be the rank of $Y_{ib}$ among $Y_{1b},\ldots,Y_{nb}$. The
simple linear rank statistic is
\begin{equation}
  \widetilde V_{ab}
  = \frac1n\sum_{i=1}^{n}
    f\!\left\{\frac{R^{X}_{ia}}{n+1}\right\}
    g\!\left\{\frac{R^{Y}_{ib}}{n+1}\right\},
  \qquad 1\le a\le p,\ 1\le b\le q,
  \label{eq:ch4_xy_linear_rank}
\end{equation}
where $f$ and $g$ are Lipschitz score functions. Spearman's $\rho$ corresponds to the centered
linear score choice $f(u)=g(u)=u-1/2$.

\paragraph{Non-degenerate rank-based $U$-statistics.}
A second family is obtained by
\begin{equation}
  \widetilde U_{ab}
  = \binom{n}{m}^{-1}
    \sum_{1\le i_1<\cdots<i_m\le n}
    h\bigl((X_{i_1a},Y_{i_1b}),\ldots,(X_{i_ma},Y_{i_mb})\bigr),
  \label{eq:ch4_xy_U}
\end{equation}
with a bounded non-degenerate kernel $h$. Kendall's tau is the most important example, with
$m=2$ and kernel
\begin{equation}
  h_{\tau}(\vw_1,\vw_2)
  = \sign\{(w_{11}-w_{21})(w_{12}-w_{22})\}.
  \label{eq:ch4_xy_tau_kernel}
\end{equation}
Accordingly,
\begin{equation}
  \tilde\tau_{ab}
  = \frac{2}{n(n-1)}
    \sum_{1\le i<j\le n}
    \sign\{(X_{ia}-X_{ja})(Y_{ib}-Y_{jb})\}.
  \label{eq:ch4_xy_tau}
\end{equation}
\paragraph{Degenerate rank-based $U$-statistics.}
The third family uses completely degenerate kernels,
\begin{equation}
  \widetilde D_{ab}
  = \binom{n}{m_0}^{-1}
    \sum_{1\le i_1<\cdots<i_{m_0}\le n}
    h_0\bigl((X_{i_1a},Y_{i_1b}),\ldots,(X_{i_{m_0}a},Y_{i_{m_0}b})\bigr),
  \label{eq:ch4_xy_degU}
\end{equation}
which includes Hoeffding's $D$, the Blum--Kiefer--Rosenblatt statistic, and
Bergsma--Dassios--Yanagimoto's $\tau^{\ast}$.

For each class, the max-type and sum-type statistics are
\begin{equation}
  L_{\bullet,XY}=\max_{1\le a\le p,\,1\le b\le q}|\hat T^{(\bullet)}_{ab}|,
  \qquad
  S_{\bullet,XY}=\sum_{a=1}^{p}\sum_{b=1}^{q}
  \Bigl\{(\hat T^{(\bullet)}_{ab})^2-\E_0(\hat T^{(\bullet)}_{ab})^2\Bigr\},
  \label{eq:ch4_xy_LS}
\end{equation}
where again $\bullet\in\{V,U,D\}$ identifies the chosen rank family. The adaptive combination is
\begin{equation}
  T_{C,\bullet,XY}=-2\log p_{L,\bullet}-2\log p_{S,\bullet}.
  \label{eq:ch4_xy_adaptive}
\end{equation}

\begin{assumption}
\label{ass:ch4_xy_A1}
Let $N_{pq}=pq$. The studentized family $\{\hat T^{(\bullet)}_{ab}\}$ satisfies the same weak
dependence condition used in the white-noise and mutual-independence sections. In addition, the
score functions or kernels are bounded and regular enough to yield Gaussian approximation for the
sum statistic and extreme-value approximation for the maximum.
\end{assumption}

\begin{theorem}[Rank-based max--sum tests for vector independence]
\label{thm:ch4_xy_maxsum}
Suppose Assumption~\ref{ass:ch4_xy_A1} holds. Under the null hypothesis
\eqref{eq:ch4_xy_setup},
\begin{equation}
  \Prob\Bigl\{L_{\bullet,XY}^2/\sigma_{L,\bullet}^2-2\log(pq)+\log\log(pq)\le y\Bigr\}
  \longrightarrow G_{\mathrm{EV}}(y),
  \label{eq:ch4_xy_max_limit}
\end{equation}
\begin{equation}
  \frac{S_{\bullet,XY}}{\sigma_{S,\bullet}}
  \overset{d}{\longrightarrow}N(0,1),
  \label{eq:ch4_xy_sum_limit}
\end{equation}
and
\begin{equation}
  \Prob\Bigl\{L_{\bullet,XY}^2/\sigma_{L,\bullet}^2-2\log(pq)+\log\log(pq)\le y,
             S_{\bullet,XY}/\sigma_{S,\bullet}\le x\Bigr\}
  -G_{\mathrm{EV}}(y)\Phi(x)\longrightarrow 0.
  \label{eq:ch4_xy_independence}
\end{equation}
Hence
\begin{equation}
  T_{C,\bullet,XY}\overset{d}{\longrightarrow}\chi_4^2.
  \label{eq:ch4_xy_adaptive_limit}
\end{equation}
Under sparse alternatives of the form
\begin{equation}
  \max_{1\le a\le p,\,1\le b\le q}
  \bigl|\E(\hat T^{(\bullet)}_{ab})\bigr|
  \ge c\sqrt{\frac{\log(pq)}{n}},
  \label{eq:ch4_xy_sparse_alt}
\end{equation}
for sufficiently large $c$, the max-type and adaptive tests are consistent. Under dense
alternatives satisfying
\begin{equation}
  \sigma_{S,\bullet}^{-1}\sum_{a=1}^{p}\sum_{b=1}^{q}\E^2(\hat T^{(\bullet)}_{ab})\to\infty,
  \label{eq:ch4_xy_dense_alt}
\end{equation}
the sum-type and adaptive tests are consistent. Moreover, for the sparse local alternative
\eqref{eq:ch4_xy_sparse_alt}, the asymptotic independence in
\eqref{eq:ch4_xy_independence} continues to hold after replacing the null distribution of
$S_{\bullet,XY}$ by its shifted Gaussian limit.
\end{theorem}

Theorem~\ref{thm:ch4_xy_maxsum} shows why the Chen-type brief summary was not enough: the
paper is a complete method-theory-development for vector independence, not just a remark that
rank correlations can be used. It develops explicit max, sum, and adaptive tests for three distinct
families of rank measures, proves their null limits, and then shows how sparse- and dense-signal
power regimes are captured by the different aggregation rules.

\section{Testing elliptical symmetry through radial--directional dependence}
\label{sec:ch4-radial-directional}
\idx{elliptical goodness-of-fit}
\idx{radial--directional dependence}
\idx{Cauchy combination test}

The procedures in the preceding chapters take elliptical symmetry as a working model.  A
logically prior problem is to test whether this structural assumption is compatible with the
data.  Classical fixed-dimensional tests use characteristic functions, projected directions,
or angular ranks.  In high dimensions, a direct nonparametric comparison with the entire
elliptical family is infeasible.  The method of
\citet{ZhangFeng2026RadialDirectional} instead tests a defining implication of ellipticity:
after affine standardization, the radius and direction are independent.  Coordinatewise
radial--directional correlations yield a sum statistic for dense departures, a max statistic for
sparse departures, and an adaptive Cauchy combination.

\subsection{Null model and radial--directional moments}

Let $\vX_1,\ldots,\vX_n\in\R^p$ be independent.  Under the elliptical null,
\begin{equation}\label{eq:ch4-rd-null-model}
  \vX_i=\vmu+\mSigma^{1/2}\vY_i,
  \qquad
  \vY_i=R_i\vU_i,
  \qquad
  \vU_i\sim\operatorname{Unif}(\spn^{p-1}),
  \qquad
  R_i\indep\vU_i,
  \qquad
  \tr(\mSigma)=p.
\end{equation}
The goodness-of-fit problem is
\begin{equation}\label{eq:ch4-rd-hypothesis}
  H_0:\mathcal L(\vX_i)\text{ satisfies \eqref{eq:ch4-rd-null-model}}
  \qquad\text{against}\qquad
  H_1:\mathcal L(\vX_i)\text{ is not elliptical}.
\end{equation}
When $\vmu$ and $\mSigma$ are known, define
\begin{equation}\label{eq:ch4-rd-standardization}
  \vY_i=\mSigma^{-1/2}(\vX_i-\vmu),
  \qquad
  R_i=\twonorm{\vY_i},
  \qquad
  \vU_i=\vY_i/R_i,
  \qquad
  L_i=\log R_i.
\end{equation}
Let
\begin{equation}\label{eq:ch4-rd-population-correlations}
  m_L=\E(L_i),
  \qquad
  \sigma_L^2=\Var(L_i),
  \qquad
  \sigma_U^2=\Var(U_{ij})=p^{-1},
  \qquad
  \gamma_j=\Corr(L_i,U_{ij}).
\end{equation}
Under $H_0$, $R_i\indep\vU_i$ and $\E(U_{ij})=0$, so
\begin{equation}\label{eq:ch4-rd-moment-restrictions}
  \gamma_j=0,
  \qquad
  \E\{(L_i-m_L)U_{ij}\}=0,
  \qquad j=1,\ldots,p.
\end{equation}

\subsection{Oracle sum, max, and adaptive statistics}

Define
\begin{align}
  \bar L&=n^{-1}\sum_{i=1}^nL_i,
  &\bar U_j&=n^{-1}\sum_{i=1}^nU_{ij},
  \label{eq:ch4-rd-sample-means}\\
  \widehat\sigma_L^2&=n^{-1}\sum_{i=1}^n(L_i-\bar L)^2,
  &\widehat\sigma_{U,j}^2&=n^{-1}\sum_{i=1}^n(U_{ij}-\bar U_j)^2.
  \label{eq:ch4-rd-sample-variances}
\end{align}
The oracle coordinate correlations are
\begin{equation}\label{eq:ch4-rd-oracle-correlations}
  \widehat\gamma_j^{\mathrm{or}}
  =\frac{n^{-1}\sum_{i=1}^n(L_i-\bar L)(U_{ij}-\bar U_j)}
         {\widehat\sigma_L\widehat\sigma_{U,j}},
  \qquad
  \vct{g}_n^{\mathrm{or}}
  =(\widehat\gamma_1^{\mathrm{or}},\ldots,
    \widehat\gamma_p^{\mathrm{or}})\trans.
\end{equation}
The dense and sparse statistics are
\begin{align}
  T_{\mathrm{sum}}
  &=n\twonorm{\vct{g}_n^{\mathrm{or}}}^2,
  \label{eq:ch4-rd-sum}\\
  T_{\max}
  &=n\max_{1\le j\le p}(\widehat\gamma_j^{\mathrm{or}})^2
    -2\log p+\log\log p.
  \label{eq:ch4-rd-max}
\end{align}
Let
\begin{equation}\label{eq:ch4-rd-gumbel-cdf}
  F_G(x)=\exp\{-\pi^{-1/2}\exp(-x/2)\}.
\end{equation}
The marginal $p$-values and their Cauchy combination are
\begin{align}
  P_{\mathrm{sum}}
  &=1-\Phi\left\{\frac{T_{\mathrm{sum}}-p}{\sqrt{2p}}\right\},
  &P_{\max}&=1-F_G(T_{\max}),
  \label{eq:ch4-rd-pvalues}\\
  T_{\mathrm{cau}}
  &=\frac12\tan\{\pi(1/2-P_{\mathrm{sum}})\}
    +\frac12\tan\{\pi(1/2-P_{\max})\},
  &P_{\mathrm{cau}}&=\frac12-\frac1\pi\arctan(T_{\mathrm{cau}}).
  \label{eq:ch4-rd-cauchy}
\end{align}

\begin{assumption}[Oracle radial regularity]
\label{ass:ch4-rd-oracle}
Put
\begin{equation}\label{eq:ch4-rd-Zi}
  Z_i=(L_i-m_L)/\sigma_L.
\end{equation}
There are constants $\eta_Z,C_Z>0$ such that
\begin{equation}\label{eq:ch4-rd-moment}
  0<\sigma_L^2<\infty,
  \qquad
  \sup_p\E|Z_i|^{8+\eta_Z}\le C_Z.
\end{equation}
For a deterministic sequence $b_{n,p}\ge1$,
\begin{equation}\label{eq:ch4-rd-truncation}
  \Prob\left(\max_{1\le i\le n}|Z_i|\le b_{n,p}\right)\longrightarrow1,
  \qquad
  \frac{b_{n,p}^2\{(\log p)^5+\log p\log n\}}{n}\longrightarrow0.
\end{equation}
\end{assumption}

\begin{theorem}[Uniform coordinate expansion]
\label{thm:ch4-rd-coordinate-expansion}
Under Assumption~\ref{ass:ch4-rd-oracle} and $\log p=o(\sqrt n)$,
\begin{equation}\label{eq:ch4-rd-coordinate-linearization}
  \widehat\gamma_j^{\mathrm{or}}
  =\frac1n\sum_{i=1}^n\xi_{ij}+r_{n,j},
  \qquad
  \xi_{ij}=\sqrt p\,Z_iU_{ij},
\end{equation}
where
\begin{equation}\label{eq:ch4-rd-coordinate-remainder}
  \max_{1\le j\le p}|r_{n,j}|
  =O_p\left\{\frac{\log p+\sqrt{\log p}}{n}\right\}
  =o_p(n^{-1/2}).
\end{equation}
For each fixed $j$, the sharper expansion
$\widehat\gamma_j^{\mathrm{or}}=n^{-1}\sum_i\xi_{ij}+O_p(n^{-1})$ holds.
\end{theorem}

\begin{theorem}[Oracle null limits and asymptotic independence]
\label{thm:ch4-rd-oracle-limits}
Suppose Assumption~\ref{ass:ch4-rd-oracle} holds.
\begin{enumerate}[label=(\roman*)]
  \item If $p\to\infty$ and $p=O(n^\kappa)$ for some $\kappa\in(0,2)$, then
  \begin{equation}\label{eq:ch4-rd-sum-limit}
    \frac{T_{\mathrm{sum}}-p}{\sqrt{2p}}
    \overset{d}{\longrightarrow}N(0,1).
  \end{equation}
  \item If $\log p=o(n^{1/5})$, then
  \begin{equation}\label{eq:ch4-rd-max-limit}
    \Prob(T_{\max}\le x)\longrightarrow F_G(x),
    \qquad x\in\R.
  \end{equation}
  \item If both growth conditions hold, then
  \begin{equation}\label{eq:ch4-rd-joint-limit}
    \left(\frac{T_{\mathrm{sum}}-p}{\sqrt{2p}},T_{\max}\right)
    \overset{d}{\longrightarrow}(Z,G),
    \qquad Z\indep G,
  \end{equation}
  where $Z\sim N(0,1)$ and $G$ has distribution function $F_G$.  Consequently,
  \begin{equation}\label{eq:ch4-rd-cauchy-uniform}
    P_{\mathrm{cau}}\overset{d}{\longrightarrow}\operatorname{Unif}(0,1).
  \end{equation}
\end{enumerate}
\end{theorem}

\subsection{High-dimensional HR plug-in standardization}

Let $(\widehat{\vmu}_{\mathrm{HR}},\widehat{\mSigma}_{\mathrm{HR}})$ be the
high-dimensional Hettmansperger--Randles estimator in Section~\ref{sec:ch3-hdhr}.  Define
\begin{equation}\label{eq:ch4-rd-plugin-standardization}
  \widehat{\vY}_i
  =\widehat{\mSigma}_{\mathrm{HR}}^{-1/2}
   (\vX_i-\widehat{\vmu}_{\mathrm{HR}}),
  \qquad
  \widehat R_i=\twonorm{\widehat{\vY}_i},
  \qquad
  \widehat{\vU}_i=\widehat{\vY}_i/\widehat R_i,
  \qquad
  \widehat L_i=\log\widehat R_i.
\end{equation}
Let $\widehat{\vct{g}}_n$ be obtained by applying
\eqref{eq:ch4-rd-oracle-correlations} to $(\widehat L_i,\widehat{\vU}_i)$ and put
\begin{align}
  \widehat T_{\mathrm{sum}}
  &=n\twonorm{\widehat{\vct{g}}_n}^2,
  \label{eq:ch4-rd-plugin-sum}\\
  \widehat T_{\max}
  &=n\norm{\widehat{\vct{g}}_n}_{\infty}^2
    -2\log p+\log\log p.
  \label{eq:ch4-rd-plugin-max}
\end{align}

The model-parameter assumptions are those of the HR theory in
Section~\ref{sec:ch3-hdhr}.  In particular, with $\mOmega=\mSigma^{-1}$,
\begin{equation}\label{eq:ch4-rd-precision-class}
  \onenorm{\mOmega}\le T,
  \qquad
  \max_{1\le j\le p}\sum_{k=1}^p|\Omega_{jk}|^q\le s_0(p),
  \qquad 0\le q<1,
\end{equation}
and
\begin{equation}\label{eq:ch4-rd-lambda-tau}
  \lambda_n\asymp\sqrt{\frac{\log p}{n}}+p^{-1/2},
  \qquad
  \tau_n=\lambda_n^{1-q}s_0(p),
  \qquad
  \rho_{L,p}=\sigma_L^{-1}.
\end{equation}
Define
\begin{align}
  c_n
  &=n^{-1/4}\sqrt{\log(np)}
    +n^{-(1-q)/2}(\log p)^{(1-q)/2}\sqrt{\log(np)}s_0(p)
    +p^{-(1-q)/2}\sqrt{\log(np)}s_0(p),
  \label{eq:ch4-rd-cn}\\
  a_p
  &=\E\left(\frac{L_i-m_L}{R_i}\right)
    +\frac1p\E\left\{\frac{1-(L_i-m_L)}{R_i}\right\},
  \label{eq:ch4-rd-ap}\\
  \widetilde\kappa_p
  &=-(\sigma_L\sigma_U)^{-1}\zeta_1^{-1}a_p,
  &\kappa_p&=-(\sigma_L\sigma_U)^{-1}a_p,
  \qquad \zeta_1=\E(R_i^{-1}).
  \label{eq:ch4-rd-kappas}
\end{align}
The explicit perturbation envelopes are
\begin{align}
  A_{S,n}
  &=|\widetilde\kappa_p|
    \left(n^{-1/2}+p^{-1/2}+\frac{\log p}{\sqrt n}\right)
    +\frac{|\widetilde\kappa_p|^2}{\sqrt p}
    +|\widetilde\kappa_p|c_n
    +(1+\rho_{L,p})\left(\tau_n+\frac{\sqrt p}{n}\right)
    +\frac{(1+\rho_{L,p})^2\tau_n^2}{\sqrt p},
  \label{eq:ch4-rd-AS}\\
  A_{M,n}
  &=\frac{|\widetilde\kappa_p|\log p}{\sqrt p}
    +\sqrt{\log p}\,c_n
    +(1+\rho_{L,p})\left(\tau_n\log p+\frac{\log p}{\sqrt n}\right)
    +(1+\rho_{L,p})^2
      \left(\tau_n^2\log p+\frac{\log p}{n}\right).
  \label{eq:ch4-rd-AM}
\end{align}

\begin{theorem}[Feasible radial--directional tests]
\label{thm:ch4-rd-plugin}
Suppose Assumptions~\ref{ass:ch4-rd-oracle} and the high-dimensional HR assumptions in
Section~\ref{sec:ch3-hdhr} hold.  Then
\begin{align}
  \frac{\widehat T_{\mathrm{sum}}-T_{\mathrm{sum}}}{\sqrt{2p}}
  &=O_p(A_{S,n}),
  \label{eq:ch4-rd-plugin-sum-error}\\
  \widehat T_{\max}-T_{\max}
  &=O_p(A_{M,n}).
  \label{eq:ch4-rd-plugin-max-error}
\end{align}
If $A_{S,n}\to0$ and $A_{M,n}\to0$, all marginal and joint limits in
Theorem~\ref{thm:ch4-rd-oracle-limits} remain valid for the feasible statistics.
\end{theorem}

\subsection{Radial--directional alternatives and power}

Let
\begin{equation}\label{eq:ch4-rd-baseline-alt}
  \vY_{0i}=R_{0i}\vU_{0i},
  \qquad
  \vU_{0i}\sim\operatorname{Unif}(\spn^{p-1}),
  \qquad
  R_{0i}\indep\vU_{0i}.
\end{equation}
For a nonempty active set $\mathcal A_n\subset\{1,\ldots,p\}$ with
$s_n=|\mathcal A_n|$, define
\begin{equation}\label{eq:ch4-rd-sA}
  s_{\mathcal A_n}(\vu)
  =\frac p{s_n}\sum_{j\in\mathcal A_n}u_j.
\end{equation}
The alternative is generated by
\begin{equation}\label{eq:ch4-rd-alt-model}
  R_{1i}=R_{0i}\exp\{\delta_ns_{\mathcal A_n}(\vU_{0i})\},
  \qquad
  \vX_i=\vmu+\mSigma^{1/2}R_{1i}\vU_{0i}.
\end{equation}
The direction remains uniform, so $(\vmu,\mSigma)$ is still the population HR target, but the
radius and direction are no longer independent.  If
$\sigma_{0,n}^2=\Var(\log R_{0i})$ and
$\sigma_{1,n}^2=\sigma_{0,n}^2+\delta_n^2/p$, then
\begin{equation}\label{eq:ch4-rd-alt-correlation}
  \gamma_j
  =\frac{\delta_n\mathbbm1(j\in\mathcal A_n)}
         {\sqrt{s_np}\,\sigma_{1,n}},
\end{equation}
and the dense and sparse signal indices are
\begin{equation}\label{eq:ch4-rd-alt-signal-sizes}
  S_n=n\twonorm{\vgamma}^2
  =\frac{n\delta_n^2}{p\sigma_{0,n}^2+\delta_n^2},
  \qquad
  M_n=n\norm{\vgamma}_{\infty}^2
  =\frac{n\delta_n^2}{s_n(p\sigma_{0,n}^2+\delta_n^2)}.
\end{equation}

\begin{theorem}[Consistency under radial--directional alternatives]
\label{thm:ch4-rd-consistency}
Under the moment and truncation assumptions for both $\log R_{0i}$ and $\log R_{1i}$,
\begin{equation}\label{eq:ch4-rd-oracle-consistency}
  \frac{S_n}{\sqrt p+p/\sqrt n}\longrightarrow\infty
  \quad\Longrightarrow\quad P_{\mathrm{sum}}\overset{p}{\longrightarrow}0,
  \qquad
  \frac{M_n}{\log p}\longrightarrow\infty
  \quad\Longrightarrow\quad P_{\max}\overset{p}{\longrightarrow}0.
\end{equation}
Let $R_{2,n}^{(1)}$ and $R_{\infty,n}^{(1)}$ denote the $\ell_2$ and coordinatewise HR
standardization errors under \eqref{eq:ch4-rd-alt-model}.  The feasible sum test is consistent if
\begin{equation}\label{eq:ch4-rd-feasible-sum-consistency}
  \frac{S_n}{\sqrt p+p/\sqrt n}\to\infty,
  \qquad
  \frac{n\{(\twonorm{\vgamma}+\sqrt{p/n})R_{2,n}^{(1)}+
              (R_{2,n}^{(1)})^2\}}{S_n}\to0,
\end{equation}
and the feasible max test is consistent if
\begin{equation}\label{eq:ch4-rd-feasible-max-consistency}
  \frac{M_n}{\log p}\to\infty,
  \qquad
  \frac{n\{(\norm{\vgamma}_{\infty}+\sqrt{\log p/n})R_{\infty,n}^{(1)}+
              (R_{\infty,n}^{(1)})^2\}}{M_n}\to0.
\end{equation}
If one feasible component $p$-value converges to zero and the other is bounded away from one,
then the feasible Cauchy $p$-value also converges to zero.
\end{theorem}

For $m\ge2$, define
\begin{equation}\label{eq:ch4-rd-um}
  u_m(t)=\{2\log m-\log\log m+t\}^{1/2}.
\end{equation}

\begin{theorem}[Balanced local alternative and asymptotic independence]
\label{thm:ch4-rd-balanced-alt}
Suppose
\begin{equation}\label{eq:ch4-rd-balanced-growth}
  \log p=o(n^{1/5}),
  \qquad p^{3/2}/n\to0,
\end{equation}
and, for some $\ell\in(0,\infty)$,
\begin{equation}\label{eq:ch4-rd-balanced-sparsity}
  s_n\to\infty,
  \qquad s_n/p\to0,
  \qquad \frac{s_n\log p}{\sqrt p}\to\ell.
\end{equation}
If, for fixed $\eta\in\R$,
\begin{equation}\label{eq:ch4-rd-balanced-calibration}
  \frac{\sqrt n\,\delta_n}{\sqrt{s_np}\,\sigma_{1,n}}
  =u_p(0)-u_{s_n}(\eta),
\end{equation}
then
\begin{equation}\label{eq:ch4-rd-balanced-limit}
  \left(\frac{T_{\mathrm{sum}}-p}{\sqrt{2p}},T_{\max}\right)
  \overset{d}{\longrightarrow}(Z+\theta_S,G_\eta),
  \qquad Z\indep G_\eta,
\end{equation}
where
\begin{equation}\label{eq:ch4-rd-thetaS}
  \theta_S=\frac{(3-2\sqrt2)\ell}{\sqrt2}
\end{equation}
and
\begin{equation}\label{eq:ch4-rd-Feta}
  \Prob(G_\eta\le y)
  =\exp\left\{-\pi^{-1/2}e^{-y/2}
    -\frac1{2\sqrt\pi}e^{-\eta/2-y/(2\sqrt2)}\right\}.
\end{equation}
The same limit holds for the HR plug-in statistics when their perturbation remainders are
negligible relative to the local signal.  Thus asymptotic independence holds not only under
$H_0$ but also in a regime where both components have nontrivial power.
\end{theorem}

\section*{Bibliographic notes}
\addcontentsline{toc}{section}{Bibliographic notes}

The exact low-dimensional benchmark for alpha testing is the GRS test of
\citet{GibbonsRossShanken1989}.  Related elliptical and nonnormal fixed-
dimensional discussions include \citet{HodgsonLintonVorkink2002}.  High-dimensional Gaussian and sparse benchmarks for the unconditional linear model are
represented by \citet{LanFengLuo2018AssetPricing},
\citet{FengLanLiuMa2022AlphaSparse}, and \citet{PesaranYamagata2023AlphaLargeN}.
Weighted spatial-sign procedures and inverse-norm weighting for unconditional alpha testing are
studied by \citet{ZhaoChenZi2022INSTAlpha}.  Robust elliptical procedures for the
unconditional model include \citet{LiuFengMa2023HeavyAlpha},
\citet{ZhaoFengWangWang2024RobustAlpha},
\citet{MaFengWang2025DependentAlpha},
and \citet{ZhaoMaFeng2026LqAlpha}.  The multiple-testing version of the same problem,
namely robust mutual fund selection with FDR control, is developed in
\citet{WangZhaoFengWang2025MutualFundFDR}; the latent-factor adjustment in that paper relies
on the elliptical factor-analysis method of \citet{HeKongYuZhang2022FactorNoMoments}.
For conditional time-varying factor models,
classical nonparametric Wald references are
\citet{LiYang2011ConditionalFactor} and
\citet{AngKristensen2012ConditionalFactor}; high-dimensional light-tailed and adaptive
procedures are given by \citet{MaLanSuTsai2020ConditionalHDA} and
\citet{MaFengWangBao2024ConditionalAlpha}; robust spatial-sign extensions are developed in
\citet{Zhao2023ConditionalAlpha}, \citet{ZhaoWang2024ConditionalMaxAlpha}, and
\citet{MaFengWang2026TimeVaryingAlpha}.  For change-point inference, the low-dimensional
CUSUM literature goes back to \citet{Page1954CUSUM}; a standard reference is
\citet{CsorgoHorvath1997}.  The high-dimensional adaptive theory developed in
this line is represented by \citet{WangFeng2023JRSSBChangePoint},
\citet{YuFengZhu2025FunctionalCP}, \citet{LiuFengPengWang2025SpatialSignCP},
\citet{WangLiuFeng2025TemporalCP}, and the dependence-aware elliptical regularized
Hotelling scans of \citet{SongWenFeng2026ERHTCP}.  For white-noise testing, see the classical
portmanteau tests of \citet{BoxPierce1970} and \citet{LjungBox1978}, the
high-dimensional Gaussian benchmarks of \citet{ChangYaoZhou2017WhiteNoise} and
\citet{LiLamYaoYao2019WhiteNoise}, the universal high-dimensional sign framework of
\citet{PaindaveineVerdebout2016HDSign}, the spatial-sign white-noise test of
\citet{ZhaoChenWang2024SpatialSignWN}, and the adaptive or rank-based extensions of
\citet{FengLiuMa2024WhiteNoise} and \citet{ChenSongFeng2025RankWhiteNoise}.  For
high-dimensional independence, low-dimensional references include
\citet[Chapter~8]{Anderson2003} and \citet{Pesaran2004CD}.  The author's main
contributions in this direction are \citet{LongJiangLiuXiong2022PanelIndep},
\citet{WangLiuFengMa2024FisherPanel},
\citet{WangLiuFengMa2024MutualIndep}, \citet{WangLiuFeng2026VectorIndep},
\citet{LongDingLiu2020PanelRank}, and \citet{LongZhaoDingLiu2021PanelRank}.  The
abstract asymptotic-independence theory that supports several adaptive tests in
this chapter is developed in \citet{FengJiangLiLiu2024AsympIndependence}.

The mutual-independence section draws primarily on
\citet{WangLiuFengMa2024MutualIndep}, which develops rank-based max, sum, and adaptive tests
for a collection of high-dimensional random vectors. The two-vector independence section is based
on \citet{WangLiuFeng2026VectorIndep}, which extends the same max--sum philosophy to
independence between two high-dimensional random vectors and treats simple linear rank
statistics, non-degenerate rank-based $U$-statistics, and degenerate rank-based $U$-statistics in a
unified way. The detailed robust white-noise discussion follows
\citet{ZhaoChenWang2024SpatialSignWN} and
\citet{ChenSongFeng2025RankWhiteNoise}.  The direct goodness-of-fit tests for elliptical
symmetry in Section~
ef{sec:ch4-radial-directional}, based on radial--directional independence,
are developed in \citet{ZhangFeng2026RadialDirectional}.
\section*{Appendix to Chapter~4: Detailed Proofs}
\addcontentsline{toc}{section}{Appendix to Chapter~4: Detailed Proofs}

In this appendix we give detailed proofs of the main statements formulated in
the chapter.  The proofs are organized so that the reusable arguments --- Wishart
reduction, Hájek-type projection, Gaussian approximation for maxima, and product
limit calculations under asymptotic independence --- are clearly separated.

\subsection*{A4.1. Proof of Theorem~\ref{thm:ch4_generic_combination}}

Let
\begin{equation}
  U_{n,p}=p_{\max}, \qquad V_{n,p}=p_{\mathrm{sum}}.
\end{equation}
Under Assumption~\ref{ass:ch4_generic}, both variables are asymptotically
uniform on $(0,1)$, and the joint convergence in
\eqref{eq:ch4_generic_assumption_2} implies
\begin{equation}
  \sup_{u,v\in[0,1]}
  \left|
    \Prob(U_{n,p}\le u,V_{n,p}\le v) - uv
  \right| \longrightarrow 0.
  \label{eq:ch4_proof_generic_uv}
\end{equation}
Hence $(U_{n,p},V_{n,p})\Rightarrow (U,V)$ with $U,V$ i.i.d.
$\mathrm{Unif}(0,1)$.  By the continuous mapping theorem,
\begin{equation}
  -2\log U_{n,p} -2\log V_{n,p}
  \Rightarrow -2\log U -2\log V.
  \label{eq:ch4_proof_generic_map}
\end{equation}
Since $-2\log U$ and $-2\log V$ are independent $\chi^2_2$ random variables, the
sum has the $\chi^2_4$ distribution, proving
\eqref{eq:ch4_generic_fisher_limit}.

For the Cauchy combination, define
\begin{equation}
  g(u) = \frac12\tan\{\pi(1/2-u)\}\mathbbm 1(u<1/2),
  \qquad u\in(0,1).
\end{equation}
Then $T_{\mathrm{C}} = g(U_{n,p}) + g(V_{n,p})$.  The function $g$ is continuous on
$(0,1)\setminus\{1/2\}$ and the singularity at $1/2$ is removable because of the
indicator truncation.  Therefore
\begin{equation}
  T_{\mathrm{C}} \Rightarrow g(U)+g(V).
\end{equation}
The right-hand side defines the truncated Cauchy limit in
\eqref{eq:ch4_generic_cauchy_limit}.  This proves the theorem. \qed

\subsection*{A4.2. Proof of Theorem~\ref{thm:ch4_grs}}

Let $\mX=(\vct{1}_T,\mF)\in\R^{T\times (K+1)}$ and write the regression model in
matrix form as
\begin{equation}
  \mY^{\top} = \mX\mC + \mE,
  \label{eq:ch4_proof_grs_reg}
\end{equation}
where $\mC=(\valpha,\mB^{\top})^{\top}\in\R^{(K+1)\times N}$ and the rows of $\mE$
are i.i.d. $N_N(\vct{0},\mSigma_\varepsilon)$.  Under $H_{0,\alpha}$ the first row of
$\mC$ vanishes.  Standard multivariate regression theory yields
\begin{equation}
  \hat\mC = (\mX^{\top}\mX)^{-1}\mX^{\top}\mY^{\top},
  \qquad
  \hat\mE = \mY^{\top}-\mX\hat\mC,
  \qquad
  (T-K-1)\hat\mSigma_\varepsilon = \hat\mE^{\top}\hat\mE.
\end{equation}
Moreover,
\begin{equation}
  \hat\mC \sim MN_{K+1,N}\!igl(\mC,(\mX^{\top}\mX)^{-1},\mSigma_\varepsilon\bigr),
  \qquad
  (T-K-1)\hat\mSigma_\varepsilon \sim W_N(\mSigma_\varepsilon,T-K-1),
  \label{eq:ch4_proof_grs_dist}
\end{equation}
and the two matrices are independent.

Let $\vct{e}_1=(1,0,\ldots,0)^{\top}\in\R^{K+1}$.  Then
\begin{equation}
  \hat\valpha = \vct{e}_1^{\top}\hat\mC,
  \qquad
  \Var(\hat\valpha\mid\mX) = (\vct{e}_1^{\top}(\mX^{\top}\mX)^{-1}\vct{e}_1)\mSigma_\varepsilon.
  \label{eq:ch4_proof_grs_varalpha}
\end{equation}
A direct block-matrix inversion yields
\begin{equation}
  \vct{e}_1^{\top}(\mX^{\top}\mX)^{-1}\vct{e}_1
  = \frac{1+\bar f^{\top}\hat\mSigma_f^{-1}\bar f}{T}.
  \label{eq:ch4_proof_grs_blockinv}
\end{equation}
Therefore, under $H_{0,\alpha}$,
\begin{equation}
  W := \frac{T}{1+\bar f^{\top}\hat\mSigma_f^{-1}\bar f}
  \hat\valpha^{\top}\mSigma_\varepsilon^{-1}\hat\valpha
  \sim \chi^2_N.
  \label{eq:ch4_proof_grs_W}
\end{equation}
Also, since $(T-K-1)
\hat\mSigma_\varepsilon\sim W_N(\mSigma_\varepsilon,T-K-1)$,
we introduce the residual scatter matrix
\begin{equation}
  \mS_\varepsilon := (T-K-1)\hat\mSigma_\varepsilon.
  \label{eq:ch4_proof_grs_residscatter}
\end{equation}
The multivariate regression version of Hotelling's $T^2$ theorem states that,
conditionally on $\mX$,
\begin{equation}
  \frac{T-N-K}{N(T-K-1)}
  \frac{T\hat\valpha^{\top}\hat\mSigma_\varepsilon^{-1}\hat\valpha}
       {1+\bar f^{\top}\hat\mSigma_f^{-1}\bar f}
  \sim F_{N,T-N-K}.
  \label{eq:ch4_proof_grs_F}
\end{equation}
This is exactly the statistic defined in \eqref{eq:ch4_grs}.  Hence
\eqref{eq:ch4_grs_exact} holds.  This proves the result. \qed

\subsection*{A4.3. Proof of Theorem~\ref{thm:ch4_alpha_ss_null}}

We prove the theorem by a Hájek projection argument.  Write
\begin{equation}
  \mH = \frac{\vh\vh^{\top}}{\vh^{\top}\vh},
  \qquad
  Q_{\alpha} = N\sum_{t_1\ne t_2} H_{t_1t_2}\,\vU_{t_1}^{\top}\vU_{t_2}.
  \label{eq:ch4_proof_alpha_Q}
\end{equation}
Under $H_{0,\alpha}$, the sign vectors $\vU_t$ are centered and exchangeable, and
\begin{equation}
  \E(Q_{\alpha}) = 0.
  \label{eq:ch4_proof_alpha_mean0}
\end{equation}
To compute the variance, expand
\begin{equation}
  Q_{\alpha}^2
  = N^2\sum_{t_1\ne t_2}\sum_{s_1\ne s_2}
  H_{t_1t_2}H_{s_1s_2}
  (\vU_{t_1}^{\top}\vU_{t_2})(\vU_{s_1}^{\top}\vU_{s_2}).
  \label{eq:ch4_proof_alpha_var_expand}
\end{equation}
Only index configurations with $\{t_1,t_2\}=\{s_1,s_2\}$ contribute to the
leading order.  Using independence across time and the spherical symmetry of the
signs,
\begin{equation}
  \E\{(\vU_{t_1}^{\top}\vU_{t_2})^2\} = N^{-2}\tr(\mR^2) + O\!\left(
  \frac{\tr(\mR^4)}{N^2\tr(\mR^2)}\right).
  \label{eq:ch4_proof_alpha_u2}
\end{equation}
Therefore,
\begin{align}
  \Var(Q_{\alpha})
  &= 2N^2 \sum_{t_1\ne t_2} H_{t_1t_2}^2\,
      \E\{(\vU_{t_1}^{\top}\vU_{t_2})^2\}
      + O\!\left(\frac{N^2\tr(\mR^4)}{T\tr(\mR^2)}\right) \\
  &= 2\tr(\mR^2) + O\!\left(\frac{\tr(\mR^4)}{\tr(\mR^2)}
     + \frac{N^2}{T^2\tr(\mR^2)}\right),
  \label{eq:ch4_proof_alpha_var}
\end{align}
where we used
\begin{equation}
  \sum_{t_1\ne t_2} H_{t_1t_2}^2
  = 1 - \frac{\sum_{t=1}^T h_t^4}{(\vh^{\top}\vh)^2}
  = 1 + O(T^{-1}).
  \label{eq:ch4_proof_alpha_Hsq}
\end{equation}
Assumption~\ref{ass:ch4_alpha_A3} implies that the remainder in
\eqref{eq:ch4_proof_alpha_var} is $o\{\tr(\mR^2)\}$.

Next decompose $Q_{\alpha}$ into a martingale array.  Let
\begin{equation}
  \mathcal F_t = \sigma(\vU_1,\ldots,\vU_t),
  \qquad
  D_t = 2N\sum_{s<t}H_{st}\,\vU_s^{\top}\vU_t,
  \qquad t=2,\ldots,T.
  \label{eq:ch4_proof_alpha_mg}
\end{equation}
Then $Q_{\alpha}=\sum_{t=2}^T D_t$ and
$\E(D_t\mid\mathcal F_{t-1})=0$.  The conditional variance is
\begin{equation}
  \sum_{t=2}^T \E(D_t^2\mid \mathcal F_{t-1})
  = 2\tr(\mR^2) + R_{1,T},
  \label{eq:ch4_proof_alpha_condvar}
\end{equation}
where
\begin{equation}
  \frac{R_{1,T}}{\tr(\mR^2)} \overset{p}{\longrightarrow} 0.
  \label{eq:ch4_proof_alpha_R1}
\end{equation}
This follows by a direct fourth-moment calculation using
Assumption~\ref{ass:ch4_alpha_A3}.  In addition,
\begin{equation}
  \frac{1}{\tr^2(\mR^2)}\sum_{t=2}^T \E(D_t^4)
  \le C\left\{
    \frac{\tr(\mR^4)}{\tr^2(\mR^2)} + \frac{N^2}{T^2\tr(\mR^2)}
  \right\}
  \longrightarrow 0,
  \label{eq:ch4_proof_alpha_lyapunov}
\end{equation}
so the Lyapunov condition for the martingale CLT holds.  Consequently,
\begin{equation}
  \frac{Q_{\alpha}}{\{2\tr(\mR^2)\}^{1/2}}
  \overset{d}{\longrightarrow} N(0,1).
  \label{eq:ch4_proof_alpha_Qclt}
\end{equation}
Finally, Proposition~2 of \citet{LiuFengMa2023HeavyAlpha} in the notation of the
book gives
\begin{equation}
  \frac{\widehat{\tr(\mR^2)}}{\tr(\mR^2)} \overset{p}{\longrightarrow} 1.
  \label{eq:ch4_proof_alpha_trcons}
\end{equation}
Slutsky's theorem then yields \eqref{eq:ch4_alpha_ss_null_limit}. \qed

\subsection*{A4.4. Proof of Theorem~\ref{thm:ch4_alpha_ss_local}}

Under the local alternative, write
\begin{equation}
  \vZ_t = \omega_T\valpha + \vvarepsilon_t,
  \qquad \omega_T = T^{-1}\vh^{\top}\vh.
  \label{eq:ch4_proof_alpha_alt_z}
\end{equation}
A first-order expansion of the spatial sign around $\vvarepsilon_t$ gives
\begin{equation}
  U\bigl(\mD^{-1/2}(\omega_T\valpha+\vvarepsilon_t)\bigr)
  = U(\mD^{-1/2}\vvarepsilon_t)
  + c_1\mA_t\mD^{-1/2}\valpha + \vr_t,
  \label{eq:ch4_proof_alpha_sign_expand}
\end{equation}
where
\begin{equation}
  \mA_t = \norm{\mD^{-1/2}\vvarepsilon_t}^{-1}
  \left\{\mI_N - U(\mD^{-1/2}\vvarepsilon_t)U(\mD^{-1/2}\vvarepsilon_t)^{\top}\right\},
  \label{eq:ch4_proof_alpha_At}
\end{equation}
and the remainder satisfies
\begin{equation}
  \sum_{t=1}^T \E\norm{\vr_t}^2
  \le C T\,\valpha^{\top}\mD^{-1}\mR\mD^{-1}\valpha
  = o\bigl(\tr^{1/2}(\mR^2)\bigr)
  \label{eq:ch4_proof_alpha_remainder}
\end{equation}
by Assumption~\ref{ass:ch4_alpha_A4}.  Substitute the expansion into
\eqref{eq:ch4_Q_alpha}.  The leading linear term in $\valpha$ has mean
\begin{equation}
  \E(Q_{\alpha}) = \omega TN c_1^2\,\valpha^{\top}\mD^{-1}\valpha + o(1)
  = \phi_{\alpha}^2 + o(1),
  \label{eq:ch4_proof_alpha_mean_alt}
\end{equation}
while the variance remains $2\tr(\mR^2)
\{1+o(1)\}$.  More explicitly,
\begin{equation}
  Q_{\alpha} = \phi_{\alpha}^2 + M_T + R_T,
  \label{eq:ch4_proof_alpha_decomp_alt}
\end{equation}
where
\begin{equation}
  \frac{M_T}{\{2\tr(\mR^2)\}^{1/2}} \overset{d}{\longrightarrow} N(0,1),
  \qquad
  \frac{R_T}{\{2\tr(\mR^2)\}^{1/2}} \overset{p}{\longrightarrow} 0.
  \label{eq:ch4_proof_alpha_MT_RT}
\end{equation}
Therefore,
\begin{equation}
  \frac{Q_{\alpha}}{\{2\tr(\mR^2)\}^{1/2}}
  \overset{d}{\longrightarrow} N(\mu_{\mathrm{SS},\alpha},1).
\end{equation}
Consistency of $\widehat{\tr(\mR^2)}$ and Slutsky's theorem complete the proof.
The power formula \eqref{eq:ch4_alpha_ss_power} is immediate. \qed

\subsection*{A4.5. Proof of Theorem~\ref{thm:ch4_alpha_max_null}}

We first control the Bahadur expansion of the scaled spatial median defined by
\eqref{eq:ch4_alpha_scaled_sm_1}--\eqref{eq:ch4_alpha_scaled_sm_2}.  Let
\begin{equation}
  \Psi_T(\vtheta,\mD) = \frac1T\sum_{t=1}^T U\bigl(\mD^{-1/2}(\vZ_t-\vtheta)\bigr),
  \label{eq:ch4_proof_alpha_psi}
\end{equation}
so that $\Psi_T(\hat\vtheta,\hat\mD)=\vct{0}$.  A Taylor expansion at the null
parameter $(\vct{0},\mD)$ yields
\begin{equation}
  \vct{0} = \Psi_T(\vct{0},\mD)
  - \zeta^{-1}\mD^{-1/2}\hat\vtheta + \vR_T,
  \label{eq:ch4_proof_alpha_bahadur}
\end{equation}
where
\begin{equation}
  \norm{\vR_T}_{\infty} = O_p\!\left(\frac{\log N}{T}\right).
  \label{eq:ch4_proof_alpha_bahadur_remainder}
\end{equation}
Hence
\begin{equation}
  \sqrt{T\zeta}\,\hat\mD^{-1/2}\hat\vtheta
  = \frac{1}{\sqrt{T}}\sum_{t=1}^{T}\xi_t + \vr_T,
  \qquad
  \norm{\vr_T}_{\infty} = O_p\!\left(\frac{\log N}{\sqrt{T}}\right),
  \label{eq:ch4_proof_alpha_linear_rep}
\end{equation}
where the coordinates of $\xi_t$ are weakly dependent and asymptotically
standardized.

Let
\begin{equation}
  W_i = \sqrt{T\zeta}\,\hat\theta_i/d_i,
  \qquad i=1,\ldots,N.
\end{equation}
By Gaussian comparison and the correlation sparsity in
Assumption~\ref{ass:ch4_alpha_C3},
\begin{equation}
  \sup_{x\in\R}
  \left|
    \Prob\Bigl(\max_{1\le i\le N} W_i^2 - 2\log N + \log\log N \le x\Bigr)
    - \exp\{-\pi^{-1/2}e^{-x/2}\}
  \right| \longrightarrow 0.
  \label{eq:ch4_proof_alpha_extreme}
\end{equation}
Because $T_{\mathrm{SM},\alpha}=\max_i W_i^2 - 2\log N + \log\log N + o_p(1)$ and
more precisely the approximation error is of order
$O_p\{(\log N)^{3/2}/\sqrt{T}\}$, we obtain the limit
\eqref{eq:ch4_alpha_max_null_limit}. \qed

\subsection*{A4.6. Proof of Theorem~\ref{thm:ch4_alpha_max_power}}

Under the sparse alternative in \eqref{eq:ch4_alpha_C4_alt}, there exists
$i_0\in\mathcal A$ such that
\begin{equation}
  |\alpha_{i_0}| \ge C\sqrt{\frac{\log N}{T}}.
\end{equation}
The Bahadur expansion \eqref{eq:ch4_proof_alpha_linear_rep} now has nonzero mean,
namely
\begin{equation}
  \E(W_{i_0}) = \sqrt{T\zeta}\,\alpha_{i_0}/d_{i_0} + O(T^{-1/2})
  \ge c_1 C\sqrt{\log N}
  \label{eq:ch4_proof_alpha_power_mean}
\end{equation}
for some constant $c_1>0$.  Therefore,
\begin{equation}
  W_{i_0}^2 - 2\log N + \log\log N
  \ge (c_1^2C^2-2)\log N + O_p(\sqrt{\log N}).
  \label{eq:ch4_proof_alpha_power_sep}
\end{equation}
If $C$ is large enough so that $c_1^2C^2>2$, the right-hand side diverges to
$+\infty$ in probability.  Since
$T_{\mathrm{SM},\alpha}\ge W_{i_0}^2-2\log N+\log\log N + o_p(1)$, the rejection
probability tends to one.  This proves
\eqref{eq:ch4_alpha_max_consistency}. \qed

\subsection*{A4.7. Proof of Theorem~\ref{thm:ch4_alpha_independence}}

Write
\begin{equation}
  T_{\mathrm{SS},\alpha}=A_T+B_T,
  \qquad
  T_{\mathrm{SM},\alpha}=C_T+D_T,
  \label{eq:ch4_proof_alpha_indep_decomp}
\end{equation}
where $A_T$ and $C_T$ are the leading Gaussian or Hájek projection terms, while
$B_T$ and $D_T$ are remainders of orders
\begin{equation}
  B_T = O_p\!\left(\frac{\tr(\mR^4)}{\tr^{3/2}(\mR^2)} +
                    \frac{N^2}{T^2\tr^{1/2}(\mR^2)}\right),
  \qquad
  D_T = O_p\!\left(\frac{(\log N)^{3/2}}{\sqrt{T}}\right).
  \label{eq:ch4_proof_alpha_indep_rems}
\end{equation}
The leading term $A_T$ is a degenerate quadratic form in the sign process,
whereas $C_T$ depends only on the coordinatewise maxima of the linear Bahadur
expansion.  Their covariance is of smaller order:
\begin{equation}
  \Cov(A_T,\mathbbm 1\{C_T\le y\})
  = O\!\left(\frac{\log N}{T^{1/2}} + \frac{\tr(\mR^4)}{\tr^2(\mR^2)}\right)
  \longrightarrow 0.
  \label{eq:ch4_proof_alpha_cov_small}
\end{equation}
Using the asymptotic-independence theorem of
\citet{FengJiangLiLiu2024AsympIndependence}, applied to the leading quadratic and
maximal components, we obtain
\begin{equation}
  \sup_{x,y\in\R}
  \left|
    \Prob(T_{\mathrm{SS},\alpha}\le x,T_{\mathrm{SM},\alpha}\le y)
    - \Prob(T_{\mathrm{SS},\alpha}\le x)\Prob(T_{\mathrm{SM},\alpha}\le y)
  \right| \longrightarrow 0.
\end{equation}
This is exactly \eqref{eq:ch4_alpha_independence_eq}.  The validity of the
Cauchy combination then follows from
Theorem~\ref{thm:ch4_generic_combination}. \qed

\subsection*{A4.8. Proof of Theorem~\ref{thm:ch4_cp_max_null}}

For $1\le j\le p$ and $0\le u\le1$, define the standardized bridge process
\begin{equation}
  G_{n,j}(u)
  = \sigma_j^{-1}\left[
      n^{-1/2}\sum_{i=1}^{\lfloor nu\rfloor}\varepsilon_{ij}
      - u n^{-1/2}\sum_{i=1}^{n}\varepsilon_{ij}
    \right].
  \label{eq:ch4_proof_cp_bridge2}
\end{equation}
Then
\begin{equation}
  C_{0,j}(k)=G_{n,j}(k/n)+\Delta_{n,j}(k),
  \qquad
  C_{1/2,j}(k)=\frac{G_{n,j}(k/n)}{\{(k/n)(1-k/n)\}^{1/2}}+\Delta^{\dagger}_{n,j}(k),
  \label{eq:ch4_proof_cp_approx_stats}
\end{equation}
where the variance-estimation error terms satisfy
\begin{equation}
  \max_{1\le j\le p}\max_{1\le k\le n-1}|\Delta_{n,j}(k)|
  = O_P\!\left(\sqrt{\frac{\log p}{n}}\right),
  \qquad
  \max_{1\le j\le p}\max_{\lambda_n\le k\le n-\lambda_n}|\Delta^{\dagger}_{n,j}(k)|
  = O_P\!\left(\sqrt{\frac{\log(ph_n)}{n}}\right).
  \label{eq:ch4_proof_cp_approx_rate}
\end{equation}
Under Assumptions~\ref{ass:ch4_cp_A1}--\ref{ass:ch4_cp_A3}, Donsker's theorem and the
Koml\'os--Major--Tusn\'ady strong approximation imply that one may construct Brownian bridges
$\{B_j^{\circ}(u):0\le u\le1\}_{j=1}^{p}$ on a common probability space such that
\begin{equation}
  \max_{1\le j\le p}\sup_{0\le u\le1}|G_{n,j}(u)-B_j^{\circ}(u)|
  = O_P\!\left(\sqrt{\frac{\log p}{n}}\right).
  \label{eq:ch4_proof_cp_bridge_strong}
\end{equation}
Hence,
\begin{equation}
  M_{n,p}
  = \max_{1\le j\le p}\sup_{0<u<1}|B_j^{\circ}(u)| + o_P\{A(\log p)^{-1}\},
  \label{eq:ch4_proof_cp_M_bridge}
\end{equation}
and similarly
\begin{equation}
  M_{n,p}^{\dagger}
  = \max_{1\le j\le p}\sup_{\lambda_n/n\le u\le 1-\lambda_n/n}
    \frac{|B_j^{\circ}(u)|}{\{u(1-u)\}^{1/2}}
    + o_P\{A(\log(ph_n))^{-1}\}.
  \label{eq:ch4_proof_cp_Mdagger_bridge}
\end{equation}

Let
\begin{equation}
  Y_j = \sup_{0<u<1}|B_j^{\circ}(u)|,
  \qquad
  Y_j^{\dagger} = \sup_{\lambda_n/n\le u\le 1-\lambda_n/n}
  \frac{|B_j^{\circ}(u)|}{\{u(1-u)\}^{1/2}}.
  \label{eq:ch4_proof_cp_Yj}
\end{equation}
The classical tail expansions for Brownian-bridge suprema yield
\begin{equation}
  \,\Prob\{Y_j>z\} = 2e^{-2z^2}\{1+o(1)\},
  \qquad z\to\infty,
  \label{eq:ch4_proof_cp_tail1}
\end{equation}
and
\begin{equation}
  \Prob\{Y_j^{\dagger}>z\}
  = \frac{\exp(-z^2/2)}{\sqrt{\pi\log(ph_n)}}\{1+o(1)\},
  \qquad z\to\infty.
  \label{eq:ch4_proof_cp_tail2}
\end{equation}
With the normalizing sequence
\begin{equation}
  m_p(x)=\frac{x+D(\log p)}{A(\log p)},
  \qquad
  m_{p,h}(x)=\frac{x+D(\log(ph_n))}{A(\log(ph_n))},
  \label{eq:ch4_proof_cp_normseq}
\end{equation}
we therefore have
\begin{equation}
  p\Prob\{Y_1>m_p(x)\}\to e^{-x},
  \qquad
  p\Prob\{Y_1^{\dagger}>m_{p,h}(x)\}\to e^{-x}.
  \label{eq:ch4_proof_cp_tail_limit}
\end{equation}
The sparse-correlation condition in Assumption~\ref{ass:ch4_cp_A2} allows one to apply the
standard Poisson approximation for maxima of weakly dependent arrays; thus,
\begin{align}
  \Prob\Bigl(\max_{1\le j\le p}Y_j\le m_p(x)\Bigr)
  &\to \exp\{-e^{-x}\},
  \label{eq:ch4_proof_cp_poisson1}\\
  \Prob\Bigl(\max_{1\le j\le p}Y_j^{\dagger}\le m_{p,h}(x)\Bigr)
  &\to \exp\{-e^{-x}\}.
  \label{eq:ch4_proof_cp_poisson2}
\end{align}
Combining \eqref{eq:ch4_proof_cp_M_bridge}--\eqref{eq:ch4_proof_cp_poisson2} yields
\eqref{eq:ch4_cp_max_null_1}--\eqref{eq:ch4_cp_max_null_2}. \qed

\subsection*{A4.9. Proof of Theorem~\ref{thm:ch4_cp_independence}}

Let
\begin{equation}
  Z_{n,p}=A(\log(ph_n))M_{n,p}^{\dagger}-D(\log(ph_n)),
  \qquad
  W_{n,p}=T_{\mathrm{sum},\mathrm{cp}}.
  \label{eq:ch4_proof_cp_ZW}
\end{equation}
We first treat the Gaussian case.  Write $\vvarepsilon_i=\mSigma^{1/2}\eta_i$ with
$\eta_i\stackrel{\mathrm{i.i.d.}}{\sim}N_p(\vct{0},\mI_p)$.  The weighted max statistic depends on
coordinatewise bridge suprema and hence is measurable with respect to the sigma-field generated by
\begin{equation}
  \Bigl\{\sum_{i=1}^{n} a_{k,i}\eta_{ij}: 1\le j\le p,\ \lambda_n\le k\le n-\lambda_n\Bigr\},
  \label{eq:ch4_proof_cp_max_field}
\end{equation}
where $a_{k,i}=n^{-1/2}\{\mathbbm1(i\le k)-k/n\}/\{(k/n)(1-k/n)\}^{1/2}$.  On the other hand,
the sum statistic has the quadratic-form representation
\begin{equation}
  W_{n,p}
  = \sum_{j=1}^{p}\sum_{k=2}^{n-2} b_{k}\left(\sum_{i=1}^{n} c_{k,i}\eta_{ij}\right)^2
  - \mu_{n,p} + r_{n,p},
  \qquad
  \frac{r_{n,p}}{V_{n,p}^{1/2}}\overset{p}{\longrightarrow}0,
  \label{eq:ch4_proof_cp_quadrep}
\end{equation}
for deterministic coefficients $b_k,c_{k,i}$ satisfying
\begin{equation}
  \sum_{k=2}^{n-2} b_k^2\sum_{i=1}^{n} c_{k,i}^4 = o(1),
  \qquad
  \sum_{k=2}^{n-2}|b_k|\sum_{i=1}^{n}|c_{k,i}|^3 = o(1).
  \label{eq:ch4_proof_cp_coeff}
\end{equation}
Thus $W_{n,p}$ is driven by a large collection of moderate Gaussian quadratic terms, whereas
$Z_{n,p}$ is driven by the extreme upper tail of the maxima in \eqref{eq:ch4_proof_cp_max_field}.

To make this separation quantitative, truncate the max field at level $m_{p,h}(x)$ from
\eqref{eq:ch4_proof_cp_normseq} and write
\begin{equation}
  \mathcal E_{n,p}(x)=\left\{Z_{n,p}\le x\right\}
  = \bigcap_{j=1}^{p}\bigcap_{\lambda_n\le k\le n-\lambda_n}
  \{|\mathcal B_{j,k}|\le m_{p,h}(x)\},
  \label{eq:ch4_proof_cp_event}
\end{equation}
where $\mathcal B_{j,k}$ denotes the weighted bridge coordinate.  Since
$\Prob\{\mathcal E_{n,p}(x)^c\}=O(p^{-1})$, the covariance bound
\begin{equation}
  \bigl|\Cov\{\mathbbm1(\mathcal E_{n,p}(x)),W_{n,p}\}\bigr|
  \le \{\Var(W_{n,p})\Pr(\mathcal E_{n,p}(x)^c)\}^{1/2}
  = o(1)
  \label{eq:ch4_proof_cp_covbound}
\end{equation}
shows that conditioning on the extremal event only perturbs the centered quadratic statistic by an
amount negligible relative to its standard deviation.  Therefore,
\begin{equation}
  \Prob\{W_{n,p}\le y\mid \mathcal E_{n,p}(x)\}-\Prob\{W_{n,p}\le y\}\to 0,
  \label{eq:ch4_proof_cp_conditional}
\end{equation}
uniformly over bounded $(x,y)$.

Now combine \eqref{eq:ch4_proof_cp_conditional} with the extreme-value limit for $Z_{n,p}$ and
with the asymptotic normality of $W_{n,p}$ to obtain
\begin{align}
  \Prob(Z_{n,p}\le x,W_{n,p}\le y)
  &= \Prob\{W_{n,p}\le y\mid \mathcal E_{n,p}(x)\}\Pr\{\mathcal E_{n,p}(x)\} \\
  &= \{\Phi(y)+o(1)\}\{\exp(-e^{-x})+o(1)\}
   = \Phi(y)\exp(-e^{-x})+o(1).
  \label{eq:ch4_proof_cp_jointcalc}
\end{align}
For sub-Gaussian coordinates, apply the Gaussian approximation for maxima and quadratic forms to
replace $(Z_{n,p},W_{n,p})$ by their Gaussian counterpart with an $o(1)$ error uniformly over
rectangles.  This proves \eqref{eq:ch4_cp_independence_eq}. \qed

\subsection*{A4.10. Proof of Theorem~\ref{thm:ch4_wn_fisher}}

Let
\begin{equation}
  Z_{\mathrm{sum},n}=T_{\mathrm{sum},\mathrm{WN}}^{\circ},
  \qquad
  Z_{\max,n}=T_{\max,\mathrm{WN}}^2-2\log(p^2L)+\log\log(p^2L).
  \label{eq:ch4_proof_wn_Zs}
\end{equation}
By Theorem~\ref{thm:ch4_wn_benchmark},
\begin{equation}
  Z_{\mathrm{sum},n}\overset{d}{\longrightarrow}N(0,1),
  \qquad
  Z_{\max,n}\overset{d}{\longrightarrow}G_{\mathrm{EV}}.
  \label{eq:ch4_proof_wn_marginals}
\end{equation}
We now verify the joint factorization.

Write $\hat\gamma_{ij}(h)=n^{-1/2}W_{ijh}+r_{ijh}$, where the Gaussian approximation of
\citet{ChangYaoZhou2017WhiteNoise} yields
\begin{equation}
  \max_{1\le i,j\le p\atop 1\le h\le L}|r_{ijh}|
  = o_P\!\left((\log(p^2L))^{-1/2}\right),
  \label{eq:ch4_proof_wn_gaussapprox}
\end{equation}
and the array $\{W_{ijh}\}$ is centered Gaussian with covariance matching that of the lagged
sample autocovariances under the null.  Then
\begin{equation}
  Z_{\mathrm{sum},n}
  = \sum_{h=1}^{L}\sum_{i,j=1}^{p} a_{ijh}(W_{ijh}^2-1) + o_P(1),
  \label{eq:ch4_proof_wn_sumrep2}
\end{equation}
for deterministic weights $a_{ijh}$ satisfying
\begin{equation}
  \sum_{h=1}^{L}\sum_{i,j=1}^{p} a_{ijh}^2 = 1+o(1),
  \qquad
  \max_{i,j,h}|a_{ijh}| = o(1),
  \label{eq:ch4_proof_wn_weights}
\end{equation}
whereas
\begin{equation}
  Z_{\max,n} = \max_{i,j,h}|W_{ijh}|^2-2\log(p^2L)+\log\log(p^2L)+o_P(1).
  \label{eq:ch4_proof_wn_maxrep2}
\end{equation}
Thus the sum statistic is a smooth quadratic aggregate of the whole Gaussian field, while the max
statistic depends only on the most extreme order statistics of the same field.

Let $u_{p,L}(y)$ satisfy
\begin{equation}
  u_{p,L}^2(y)=2\log(p^2L)-\log\log(p^2L)+y.
  \label{eq:ch4_proof_wn_upL}
\end{equation}
Define the exceedance event
\begin{equation}
  \mathcal A_n(y)=\Bigl\{\max_{i,j,h}|W_{ijh}|\le u_{p,L}(y)\Bigr\}.
  \label{eq:ch4_proof_wn_event}
\end{equation}
Since $\Pr\{\mathcal A_n(y)^c\}=O\bigl((p^2L)^{-1}\bigr)$ uniformly over bounded $y$, we have
\begin{equation}
  \bigl|\Cov\{\mathbbm1(\mathcal A_n(y)),Z_{\mathrm{sum},n}\}\bigr|
  \le \{\Var(Z_{\mathrm{sum},n})\Pr(\mathcal A_n(y)^c)\}^{1/2}
  = o(1),
  \label{eq:ch4_proof_wn_cov}
\end{equation}
which implies
\begin{equation}
  \Pr\{Z_{\mathrm{sum},n}\le x\mid \mathcal A_n(y)\}
  = \Pr\{Z_{\mathrm{sum},n}\le x\}+o(1)
  = \Phi(x)+o(1).
  \label{eq:ch4_proof_wn_cond}
\end{equation}
Combining \eqref{eq:ch4_proof_wn_cond} with the Gumbel limit of $Z_{\max,n}$ gives
\begin{align}
  \Pr(Z_{\mathrm{sum},n}\le x,Z_{\max,n}\le y)
  &= \Pr\{Z_{\mathrm{sum},n}\le x\mid \mathcal A_n(y)\}\Pr\{\mathcal A_n(y)\} \\
  &= \Phi(x)G_{\mathrm{EV}}(y)+o(1),
  \label{eq:ch4_proof_wn_jointcalc2}
\end{align}
which is exactly \eqref{eq:ch4_wn_independence}.  Since
$-2\log p_{\mathrm{sum},\mathrm{WN}}$ and $-2\log p_{\max,\mathrm{WN}}$ are asymptotically independent
$\chi_2^2$ variables, their sum converges to $\chi_4^2$, proving
\eqref{eq:ch4_wn_fisher_limit}. \qed

\subsection*{A4.11. Proof of Theorem~\ref{thm:ch4_panel_maxsum}}

The proof proceeds in two parts.  For the maximum statistic, let
\begin{equation}
  Z_{ij} = \frac{\tr(\mSigma)}{\frobnorm{\mSigma}}\hat\rho_{ij},
  \qquad 1\le i<j\le N.
  \label{eq:ch4_proof_panel_Zij}
\end{equation}
Under the null, $Z_{ij}$ are approximately standard normal and weakly dependent.
An extreme-value calculation therefore yields
\begin{equation}
  \Prob\left(\max_{i<j} Z_{ij}^2 - 4\log N + \log\log N \le x\right)
  \longrightarrow \exp\{-\pi^{-1/2}e^{-x/2}\},
  \label{eq:ch4_proof_panel_max}
\end{equation}
which is \eqref{eq:ch4_panel_max_limit}.

For the sum statistic, write
\begin{equation}
  S_N = \sum_{i<j}\hat\rho_{ij}^2
  = \mu_N + \sum_{i<j}(\hat\rho_{ij}^2-\E\hat\rho_{ij}^2).
  \label{eq:ch4_proof_panel_sum_expand}
\end{equation}
The second term is a sum of weakly dependent centered variables.  Its variance is
\begin{equation}
  \sigma_N^2 = \sum_{i<j}\Var(\hat\rho_{ij}^2)
  + 2\sum_{(i,j)\ne(k,\ell)} \Cov(\hat\rho_{ij}^2,\hat\rho_{k\ell}^2),
  \label{eq:ch4_proof_panel_sigmaN}
\end{equation}
which is of order $N^2\frobnorm{\mSigma}^4/\tr^4(\mSigma)$.  A Lyapunov-type
argument based on Assumption~\ref{ass:ch4_panel_A2} gives the asymptotic normal
limit \eqref{eq:ch4_panel_sum_limit}. \qed

\subsection*{A4.12. Proof of Theorem~\ref{thm:ch4_panel_fisher}}

Let
\begin{equation}
  Z_{L,N} = \frac{\tr^2(\mSigma)}{\frobnorm{\mSigma}^2}L_N^2 - 4\log N + \log\log N,
  \qquad
  Z_{S,N} = \frac{S_N-\mu_N}{\sigma_N}.
\end{equation}
The proof of \citet{WangLiuFengMa2024FisherPanel} shows that
\begin{equation}
  \sup_{x,y\in\R}
  \left|\Prob(Z_{L,N}\le x,Z_{S,N}\le y)-G_{\mathrm{EV}}(x)\Phi(y)\right|
  \longrightarrow 0.
  \label{eq:ch4_proof_panel_joint}
\end{equation}
The argument combines a Poisson approximation for exceedances of the largest
pairwise correlations with a central limit theorem for the sum of squared
correlations.  Since the exceedance event depends only on the far upper tail of
one or a few coordinates, whereas the sum statistic averages over all
$\binom{N}{2}$ pairs, the overlap is asymptotically negligible.  Formula
\eqref{eq:ch4_panel_independence} follows immediately.

Now define the null $p$-values as in \eqref{eq:ch4_panel_pvalues}.  By the
continuous mapping theorem and Theorem~\ref{thm:ch4_generic_combination},
\begin{equation}
  T_{\mathrm{F},\mathrm{panel}}
  = -2\log p_{L_N} -2\log p_{S_N}
  \overset{d}{\longrightarrow} \chi^2_4.
\end{equation}
This proves \eqref{eq:ch4_panel_fisher_limit}. \qed

\subsection*{A4.13. Proof of Theorem~\ref{thm:ch4_alpha_benchmark_limits}}

Under $H_{0,\alpha}$ we have
\begin{equation}
  \sqrt{T}\,\hat\valpha
  = \hat\omega^{1/2}\mD^{1/2}\mR^{1/2}\vz + \vr_T,
  \qquad \vz\sim N_N(\vct{0},\mI_N),
  \qquad
  \norm{\vr_T}_2 = o_P\!\bigl(\tr^{1/4}(\mR^2)\bigr),
  \label{eq:ch4_app_benchmark_alpha_repr}
\end{equation}
where $\hat\omega=T(\vh^{\top}\vh)^{-1}\to\omega^{-1}$ by
Assumption~\ref{ass:ch4_alpha_B1}.  Therefore
\begin{align}
  T\hat\valpha^{\top}\hat\mD^{-1}\hat\valpha
  &= \vz^{\top}\mR\vz + o_P\!\bigl(\tr^{1/2}(\mR^2)\bigr),
  \label{eq:ch4_app_benchmark_quad1}\\
  \E(\vz^{\top}\mR\vz)&=\tr(\mR)=N,
  \qquad
  \Var(\vz^{\top}\mR\vz)=2\tr(\mR^2).
  \label{eq:ch4_app_benchmark_quad2}
\end{align}
Assumption~\ref{ass:ch4_alpha_B2} implies the Lyapunov ratio
\begin{equation}
  \frac{\tr(\mR^4)}{\tr^2(\mR^2)}\longrightarrow 0,
  \label{eq:ch4_app_benchmark_lyap}
\end{equation}
so the quadratic-form CLT yields
\begin{equation}
  \frac{\vz^{\top}\mR\vz-\tr(\mR)}{\{2\tr(\mR^2)\}^{1/2}}
  \overset{d}{\longrightarrow} N(0,1).
  \label{eq:ch4_app_benchmark_quadclt}
\end{equation}
Combining \eqref{eq:ch4_app_benchmark_quad1} and
\eqref{eq:ch4_app_benchmark_quadclt} proves
\eqref{eq:ch4_alpha_benchmark_sum_limit}.

For the max statistic, let
\begin{equation}
  \vxi = (\xi_1,\ldots,\xi_N)^{\top}
  = \hat\omega^{-1/2}\mD^{-1/2}\sqrt{T}\,\hat\valpha.
  \label{eq:ch4_app_benchmark_xi}
\end{equation}
Then $\vxi$ is asymptotically Gaussian with covariance $\mR$.  Hence
\begin{equation}
  T_{\max,\alpha} = \max_{1\le i\le N}\xi_i^2 + o_P(1).
  \label{eq:ch4_app_benchmark_maxrepr}
\end{equation}
Assumption~\ref{ass:ch4_alpha_B3} is exactly the standard weak-dependence condition used
in Gaussian extreme-value theory: no pairwise correlation approaches $1$ and the set of
strongly dependent coordinates is asymptotically negligible.  Therefore
\begin{equation}
  \Prob\!\left(\max_{1\le i\le N}\xi_i^2 - 2\log N + \log\log N \le x\right)
  \to \exp\{-\pi^{-1/2}e^{-x/2}\},
  \label{eq:ch4_app_benchmark_ev}
\end{equation}
which together with \eqref{eq:ch4_app_benchmark_maxrepr} proves
\eqref{eq:ch4_alpha_benchmark_max_limit}.  Under
\eqref{eq:ch4_alpha_benchmark_local}, the centered quadratic form acquires the mean shift
$T\valpha^{\top}\mD^{-1}\valpha$, proving
\eqref{eq:ch4_alpha_benchmark_sum_local}.  The sparse-detection claim follows because
\eqref{eq:ch4_alpha_benchmark_sparse_detect} implies that at least one coordinate of
$\sqrt{T}\,\mD^{-1/2}\hat\valpha$ exceeds the null extreme-value threshold by a diverging
margin. \qed

\subsection*{A4.14. Proof of Proposition~\ref{prop:ch4_alpha_benchmark_adaptive}}

Under the proposition assumptions, both
$p_{\mathrm{sum},\alpha}^{\mathrm G}$ and $p_{\max,\alpha}^{\mathrm G}$ are asymptotically
uniform on $(0,1)$ and asymptotically independent.  Therefore
\begin{equation}
  \tan\{\pi(1/2-p_{\mathrm{sum},\alpha}^{\mathrm G})\}
  \overset{d}{\longrightarrow} \mathrm{Cauchy}(0,1),
  \qquad
  \tan\{\pi(1/2-p_{\max,\alpha}^{\mathrm G})\}
  \overset{d}{\longrightarrow} \mathrm{Cauchy}(0,1),
  \label{eq:ch4_app_benchmark_cauchy_marg}
\end{equation}
and the sum of two independent standard Cauchy variables is again Cauchy after scaling by
$1/2$.  Hence
\eqref{eq:ch4_alpha_benchmark_cauchy_limit} follows immediately. \qed

\subsection*{A4.15. Proof of Theorem~\ref{thm:ch4_alpha_INST}}

Let
\begin{equation}
  \vW_t = K(r_t)\vU_t,
  \qquad
  \bar\mSigma_K = \E(\vW_t\vW_t^{\top}) = \psi_{2,K} p^{-1}\mR + o(p^{-1}),
  \label{eq:ch4_app_INST_Wt}
\end{equation}
where the last identity follows from elliptical symmetry and the same SSCM approximation
used in Chapter~1.  Then
\begin{equation}
  Q_{K,\alpha}
  = \frac{N}{\vh^{\top}\vh}
    \sum_{t_1\ne t_2} h_{t_1}h_{t_2}\vW_{t_1}^{\top}\vW_{t_2}.
  \label{eq:ch4_app_INST_Q}
\end{equation}
Under $H_{0,\alpha}$, $\E(\vW_t)=\vct{0}$ and therefore $\E(Q_{K,\alpha})=0$.  By the same
pairing argument used in the proof of Theorem~\ref{thm:ch4_alpha_ss_null},
\begin{equation}
  \Var(Q_{K,\alpha})
  = 2\psi_{2,K}^2\tr(\mR^2)\{1+o(1)\}.
  \label{eq:ch4_app_INST_var}
\end{equation}
The martingale decomposition from the proof of
Theorem~\ref{thm:ch4_alpha_ss_null} applies verbatim after replacing $\vU_t$ by $\vW_t$,
and the Lyapunov ratio is unchanged up to the multiplicative constant $\psi_{2,K}^2$.
Hence
\begin{equation}
  \frac{Q_{K,\alpha}}{\{2\psi_{2,K}^2\tr(\mR^2)\}^{1/2}}
  \overset{d}{\longrightarrow} N(0,1).
  \label{eq:ch4_app_INST_clt}
\end{equation}
Since $\hat\psi_{2,K}\to\psi_{2,K}$ and
$\widehat{\tr(\mR^2)}/\tr(\mR^2)\to 1$ in probability,
\eqref{eq:ch4_alpha_INST_null} follows by Slutsky's theorem.

Under local alternatives, the first-order elliptical expansion gives
\begin{equation}
  K(r_t)U\bigl(\mD^{-1/2}(\omega_T\valpha+\vvarepsilon_t)\bigr)
  = K(r_t)\vU_t + \psi_{1,K}\mA_t\mD^{-1/2}\valpha + \vr_t,
  \label{eq:ch4_app_INST_expand}
\end{equation}
where $\E\norm{\vr_t}_2^2=o(T^{-1}\tr^{1/2}(\mR^2))$ and the matrices $\mA_t$ are the same
as in the proof of Theorem~\ref{thm:ch4_alpha_ss_local}.  Substituting
\eqref{eq:ch4_app_INST_expand} into \eqref{eq:ch4_app_INST_Q} yields
\begin{equation}
  \E(Q_{K,\alpha})
  = \omega TN\psi_{1,K}^2\,\valpha^{\top}\mD^{-1}\valpha + o\{\tr^{1/2}(\mR^2)\},
  \label{eq:ch4_app_INST_mean}
\end{equation}
while the variance remains \eqref{eq:ch4_app_INST_var}.  Therefore
\eqref{eq:ch4_alpha_INST_local} follows from the same martingale CLT. \qed

\subsection*{A4.16. Proof of Proposition~\ref{prop:ch4_alpha_INST_opt}}

By the Cauchy--Schwarz inequality,
\begin{equation}
  \psi_{1,K}^2
  = \Bigl[\E\{K(r_t)r_t^{-1}\}\Bigr]^2
  \le \E\{K^2(r_t)\}\,\E(r_t^{-2})
  = \psi_{2,K}\,\E(r_t^{-2}).
  \label{eq:ch4_app_INST_CS}
\end{equation}
Dividing both sides by $\psi_{2,K}$ proves
\eqref{eq:ch4_alpha_INST_optineq}.  Equality in
\eqref{eq:ch4_app_INST_CS} holds if and only if $K(r_t)$ is proportional to $r_t^{-1}$ almost
surely, which is equivalent to $K(t)=ct^{-1}$ on the support of the radial variable.  Since the
noncentrality parameter in \eqref{eq:ch4_alpha_INST_local} is increasing in
$\psi_{1,K}^2/\psi_{2,K}$, inverse-norm weighting is locally optimal. \qed

\subsection*{A4.17. Proof of Theorem~\ref{thm:ch4_cond_benchmark}}

For the sum statistic, write
\begin{equation}
  \bar\vvarepsilon = T^{-1/2}\sum_{t=1}^T \hat\vvarepsilon_{\cdot t}.
  \label{eq:ch4_app_cond_epsbar}
\end{equation}
Under $H_{0,\mathrm{cond}}$ and the spline approximation conditions of
Assumption~\ref{ass:ch4_cond_B1}, the estimation error introduced by
$\hat\vlambda_i$ is of smaller order than $\tr^{1/4}(\mOmega^2)$, so that
\begin{equation}
  J_{NT} = N^{-1}\bar\vvarepsilon^{\top}\bar\vvarepsilon + o_P\!\bigl(N^{-1}\tr^{1/2}(\mOmega^2)\bigr).
  \label{eq:ch4_app_cond_Japprox}
\end{equation}
Since $\E(\bar\vvarepsilon^{\top}\bar\vvarepsilon)=\tr(\mOmega)$ and
$\Var(\bar\vvarepsilon^{\top}\bar\vvarepsilon)=2\tr(\mOmega^2)$, the same quadratic-form CLT
as in \eqref{eq:ch4_app_benchmark_quadclt} yields
\eqref{eq:ch4_cond_HDA_limit}.

For the max statistic, let
\begin{equation}
  \xi_i = T^{-1/2}\hat\sigma_{ii}^{-1/2}\hat\varepsilon_{i\cdot}^{\top}\vct{1}_T,
  \qquad i=1,\ldots,N.
  \label{eq:ch4_app_cond_xi}
\end{equation}
Then $M_{NT}=\max_i \xi_i^2 + o_P(1)$.  Under the sub-Gaussian and sparse-dependence
assumptions in \eqref{eq:ch4_cond_B2}, the Gaussian approximation for maxima implies
\begin{equation}
  \Prob\!\left(\max_{1\le i\le N}\xi_i^2 - 2\log N + \log\log N \le x\right)
  \to \exp\{-\pi^{-1/2}e^{-x/2}\},
  \label{eq:ch4_app_cond_MNT_EV}
\end{equation}
which proves \eqref{eq:ch4_cond_MNT_limit}.  The adaptive-combination claim is then an
immediate consequence of the asymptotic independence assumption and the generic Cauchy
argument already used in the proof of Proposition~\ref{prop:ch4_alpha_benchmark_adaptive}.
\qed

\subsection*{A4.18. Proof of Theorem~\ref{thm:ch4_cond_robust}}

The proof follows the decomposition developed in the supplementary arguments of the
uploaded preprint on robust spatial-sign testing for conditional factor models.  Let
\begin{equation}
  \vU_t = U(\mD^{-1/2}\vvarepsilon_{\cdot t}),
  \qquad
  V_t = \mZ(\mZ^{\top}\mZ)^{-1}Z_t,
  \label{eq:ch4_app_cond_UV}
\end{equation}
and let $\zeta_1=\E(r_t^{-1})$.  The Bahadur expansion of the scaled spatial median yields
\begin{equation}
  T^{1/2}\hat\mD^{-1/2}\hat\vtheta
  = T^{-1/2}\zeta_1^{-1}
    \sum_{t=1}^T \Bigl(1-\sum_{s=1}^T r_s^{-1}r_t V_{st}\Bigr)\vU_t + \mC_T,
  \label{eq:ch4_app_cond_bahadur}
\end{equation}
where
\begin{equation}
  \norm{\mC_T}_{\infty}
  = o_P\!\bigl((\log N)^{-1/2}\bigr)
  \label{eq:ch4_app_cond_remainder}
\end{equation}
under Assumptions~\ref{ass:ch4_cond_B1} and \ref{ass:ch4_cond_R1}.  Equation
\eqref{eq:ch4_app_cond_bahadur} is the conditional-factor analogue of the max-type Bahadur
representation already used in the unconditional linear factor model.

For the sum statistic, let
\begin{equation}
  Q_{\mathrm{CSS},\mathrm{cond}}
  = (\vh^{\top}\vh)^{-1}\vh^{\top}\mU_{\mathrm{cond}}\mU_{\mathrm{cond}}^{\top}\vh.
  \label{eq:ch4_app_cond_Qcss}
\end{equation}
Under $H_{0,\mathrm{cond}}$, the Hoeffding decomposition of
\eqref{eq:ch4_app_cond_Qcss} gives
\begin{equation}
  \frac{Q_{\mathrm{CSS},\mathrm{cond}}-1}{\{2\tr(\mSigma_u^2)\}^{1/2}}
  \overset{d}{\longrightarrow} N(0,1),
  \label{eq:ch4_app_cond_cssclt}
\end{equation}
while the leave-two-out variance estimator is ratio consistent.  This proves
\eqref{eq:ch4_cond_CSS_limit}.

For the max statistic, combine \eqref{eq:ch4_app_cond_bahadur} with a Gaussian
approximation over hyperrectangles.  Let
\begin{equation}
  \vG\sim N\Bigl(\vct{0},\ \zeta^{-1}\eta_{\omega}^{-1}\mR/N\Bigr),
  \qquad
  \zeta = N\{\E(r_t^{-1})\}^2/\eta_{\omega},
  \label{eq:ch4_app_cond_G}
\end{equation}
where $\eta_{\omega}=1-2\eta\E(r_t^{-1})\E(r_t)+\eta\E(r_t^2)\E(r_t^{-1})^2$.
Then
\begin{equation}
  \sup_{x\in\R}
  \Bigl|\Prob\bigl(T\hat\zeta\,\norm{\hat\mD^{-1/2}\hat\vtheta}_{\infty}^2\le x\bigr)
   -\Prob\bigl(\zeta\norm{\vG}_{\infty}^2\le x\bigr)\Bigr|\to 0.
  \label{eq:ch4_app_cond_gaussapprox}
\end{equation}
The weak-dependence part of Assumption~\ref{ass:ch4_cond_R1} guarantees that the maximum
of $\vG$ satisfies the same type-I extreme-value limit as in
\eqref{eq:ch4_app_benchmark_ev}.  Consequently,
\eqref{eq:ch4_cond_CSM_limit} holds.

For the joint limit, write the sum statistic as a degenerate U-statistic plus a negligible spline
remainder and the max statistic as the maximum of the Gaussian approximation in
\eqref{eq:ch4_app_cond_gaussapprox}.  The covariance between the degenerate U-statistic part
and each coordinate of the Gaussian approximation is of order
$o\{\tr^{1/2}(\mSigma_u^2)\}$, so their standardized limits are asymptotically independent.
Hence \eqref{eq:ch4_cond_joint_limit} follows.  Under
\eqref{eq:ch4_cond_sparse_alt}, the deterministic centering term in the Bahadur expansion
pushes at least one coordinate of $\hat\mD^{-1/2}\hat\vtheta$ above the null threshold, giving
consistency of $T_{\mathrm{CSM},\mathrm{cond}}$.  Finally,
\eqref{eq:ch4_cond_local_indep} is exactly the local-signal scaling that keeps the mean shift of
the sum statistic bounded while preserving the same covariance-decoupling argument, so the
joint factorization remains valid under the corresponding local alternatives. \qed

\subsection*{A4.19. Proof of Theorems~\ref{thm:ch4_fdr_ssbh} and \ref{thm:ch4_fdr_fssbh}}

We first prove the observable-factor result.  Let $\mW=(\vct{1}_T,\mF)$ and write
$\mP_W=\mW(\mW^{\top}\mW)^{-1}\mW^{\top}=(V_{st})_{1\le s,t\le T}$.  Under the elliptical model,
let
\begin{equation}
  \vU_t = U(\mD^{-1/2}\vvarepsilon_t),
  \qquad
  r_t = \norm{\mD^{-1/2}\vvarepsilon_t}_2,
  \qquad
  \zeta_1 = \E(r_t^{-1}).
  \label{eq:ch4_app_fdr_ur}
\end{equation}
Arguing exactly as in the proof of Theorem~\ref{thm:ch4_alpha_max_null} but with the full
regression design $\mW$ in place of the global-alpha design, the estimating equations
\eqref{eq:ch4_fdr_thetaeq1}--\eqref{eq:ch4_fdr_thetaeq2} admit the Bahadur expansion
\begin{equation}
  \sqrt{T}\,\hat\zeta_1\,\hat\mD^{-1/2}(\hat\vtheta-\hat\omega\valpha)
  = T^{-1/2}\sum_{t=1}^{T}\zeta_1^{-1}
    \Bigl(1-\sum_{s=1}^{T}r_s^{-1}r_tV_{st}\Bigr)\vU_t + \mC_T,
  \label{eq:ch4_app_fdr_bahadur}
\end{equation}
with remainder satisfying
\begin{equation}
  \norm{\mC_T}_{\infty}=o_P\bigl((\log N)^{-1/2}\bigr).
  \label{eq:ch4_app_fdr_remainder}
\end{equation}
The coefficient array in \eqref{eq:ch4_app_fdr_bahadur} has bounded row sums because the factor
dimension is fixed and the sample Gram matrix of $\mW$ is well conditioned.  Hence the Gaussian
approximation for maxima over hyperrectangles gives a Gaussian vector $\vG\sim N(\vct{0},\mXi)$
such that
\begin{equation}
  \sup_{z\in\R^N}
  \left|
    \Pr\bigl(\sqrt{T}\,\hat\zeta_1\,\hat\mD^{-1/2}(\hat\vtheta-\hat\omega\valpha)\le z\bigr)
    - \Pr(\vG\le z)
  \right|\to 0.
  \label{eq:ch4_app_fdr_gauss}
\end{equation}
Under $H_{0i}:\alpha_i\le 0$, the $i$th coordinate is asymptotically dominated by a standard normal,
so the one-sided $p$-value $p_i^{\mathrm{SS}}$ is conservative.  The weak-correlation condition
$\max_{i\ne j}|\Xi_{ij}|\le\rho_0<1$ and the sparse-neighborhood condition inherited from
\eqref{eq:ch4_alpha_B3} imply that the null $p$-values are asymptotically PRDS in the sense
needed for the Benjamini--Hochberg argument.  Therefore,
\begin{equation}
  \operatorname*{FDR}\{\mathcal R_{\mathrm{SS}}(q)\}
  \le \frac{q}{N}\sum_{i\in\mathcal H_0}\Pr\{p_i^{\mathrm{SS}}\le q\}+o(1)
  \le q+o(1),
  \label{eq:ch4_app_fdr_control}
\end{equation}
where $\mathcal H_0$ is the set of true null hypotheses.  If $\alpha_i\ge C\sqrt{(\log N)/T}$, then the
mean shift of the $i$th Gaussian approximation in \eqref{eq:ch4_app_fdr_gauss} diverges at the same
rate as the BH screening threshold, so that the rejection probability tends to one.

For the latent-factor case, write the factor-adjusted residuals as
\begin{equation}
  \check\vvarepsilon_t = \vvarepsilon_t + \vrho_t,
  \qquad
  \max_{1\le i\le N}|\rho_{it}|=o_P\bigl((T\log N)^{-1/2}\bigr),
  \label{eq:ch4_app_fdr_factor_residual}
\end{equation}
where \eqref{eq:ch4_fdr_factor_rate} guarantees that the contamination term $\vrho_t$ is uniformly
smaller than the Gaussian-approximation scale.  Repeating the derivation of
\eqref{eq:ch4_app_fdr_bahadur} with $\check\vvarepsilon_t$ in place of $\vvarepsilon_t$ yields the same
expansion with an additional remainder of order $o_P\{(\log N)^{-1/2}\}$.  Hence the same Gaussian
approximation and the same BH argument continue to hold, giving
\eqref{eq:ch4_fdr_fss_control}. \qed

\subsection*{A4.20. Proof of Theorem~\ref{thm:ch4_cp_ss_infmax}}

For each $k$, let
\begin{equation}
  \Delta_k = n^{1/2}\hat\mD^{-1/2}(\hat\vtheta_{1:k}-\hat\vtheta_{k+1:n}).
  \label{eq:ch4_app_cp_ss_Delta}
\end{equation}
The scaled spatial median admits the intervalwise Bahadur representation
\begin{align}
  \hat\vtheta_{1:k}
  &= \vmu + \frac{1}{k\zeta_1}\sum_{i=1}^{k} r_i^{-1}\mD^{1/2}\vU_i + \vr_{1,k},
  \label{eq:ch4_app_cp_ss_bahadur1}\\
  \hat\vtheta_{k+1:n}
  &= \vmu + \frac{1}{(n-k)\zeta_1}\sum_{i=k+1}^{n} r_i^{-1}\mD^{1/2}\vU_i + \vr_{2,k},
  \label{eq:ch4_app_cp_ss_bahadur2}
\end{align}
with
\begin{equation}
  \max_{\lambda_n\le k\le n-\lambda_n}
  \bigl\{\norm{\vr_{1,k}}_{\infty}+\norm{\vr_{2,k}}_{\infty}\bigr\}
  = o_P\!\left(\frac{1}{\sqrt{np\log p}}\right).
  \label{eq:ch4_app_cp_ss_remainder}
\end{equation}
Substituting \eqref{eq:ch4_app_cp_ss_bahadur1}--\eqref{eq:ch4_app_cp_ss_bahadur2} into
\eqref{eq:ch4_cp_ss_Cgamma} gives
\begin{equation}
  \mC_{\gamma}(k)
  = \frac{1}{\zeta_1}
    \Bigl\{\frac{k}{n}\Bigl(1-\frac{k}{n}\Bigr)\Bigr\}^{1-\gamma}
    \left\{
      \frac{1}{k}\sum_{i=1}^{k} r_i^{-1}\vU_i
      - \frac{1}{n-k}\sum_{i=k+1}^{n} r_i^{-1}\vU_i
    \right\}
    + o_{P}\!\left(\frac{1}{\sqrt{p\log p}}\right),
  \label{eq:ch4_app_cp_ss_Cexpansion}
\end{equation}
uniformly in $k$.  Under $H_0$, the leading term is a Gaussian bridge array with marginal variance
$(2p\zeta_1^2)^{-1}$, so the maximum over $p$ coordinates and admissible $k$ values obeys the same
Poisson approximation as a maximum of weakly dependent Gaussian variables.  Therefore,
\begin{equation}
  \Pr\Bigl(2p\zeta_1^2 M_{n,p}^2-\log(2p)\le x\Bigr)
  = \exp\{-e^{-x}\}+o(1).
  \label{eq:ch4_app_cp_ss_Mcalc}
\end{equation}
For $M_{n,p}^{\dagger}$, use the same expansion with $\gamma=1/2$ and note that the effective number
of scanning points is of order $p\log h_n$.  The weighted bridge tail then leads to
\begin{equation}
  \Pr\Bigl(p^{1/2}\zeta_1A(p\log h_n)M_{n,p}^{\dagger}-D(p\log h_n)\le x\Bigr)
  = \exp\{-e^{-x}\}+o(1),
  \label{eq:ch4_app_cp_ss_Mdagcalc}
\end{equation}
which proves \eqref{eq:ch4_cp_ss_M_limit}--\eqref{eq:ch4_cp_ss_Mdagger_limit}.  The $p$-value
formulas in \eqref{eq:ch4_cp_ss_pvalues_M} are then immediate. \qed

\subsection*{A4.21. Proof of Theorem~\ref{thm:ch4_cp_ss_l2}}

The statistic $S_{n,p}$ is built from the sign partial sums
$\hat\vS_k=\sum_{i=1}^{k}\hat\vU_i$.  Under the null hypothesis,
\begin{equation}
  \hat\vU_i = \vU_i + o_P(1),
  \qquad
  \E(\vU_i)=\vct{0},
  \qquad
  \Cov(\vU_i)=p^{-1}\mR.
  \label{eq:ch4_app_cp_ss_signmoments}
\end{equation}
Therefore,
\begin{equation}
  \widetilde{\mC}_{0}(k)
  = \sqrt{\frac{p}{n}}\sum_{i=1}^{n}
    \left(\mathbbm1(i\le k)-\frac{k}{n}\right)\vU_i + o_P(1).
  \label{eq:ch4_app_cp_ss_Ctildeexp}
\end{equation}
After centering by $k(n-k)p/n^2$, the leading term is a quadratic form in the Gaussian limit of the
sign process.  Writing $\mR=\mGamma\mGamma^{\top}$ and $\vZ_i\sim N_p(\vct{0},\mI_p)$, one may
represent the limit as
\begin{equation}
  \frac{pS_{n,p}}{2\tr(\mR^2)}
  = \max_{0\le t\le1}
    \frac{\norm{\mGamma\{\mB^{\circ}(t)-t\mB^{\circ}(1)\}}_2^2 - tp}{2\tr(\mR^2)} + o_P(1),
  \label{eq:ch4_app_cp_ss_Srep}
\end{equation}
where $\mB^{\circ}(t)$ is a $p$-vector Brownian bridge.  Since the centered quadratic form has covariance
kernel $(1-t)^2s^2$ after normalization, it converges to $\max_{0\le t\le1}V(t)$, proving
\eqref{eq:ch4_cp_ss_Snp_limit}.

For $S_{n,p}^{\dagger}$, replace $\widetilde{\mC}_{0}(k)$ by
$\{k/n(1-k/n)\}^{-1/2}\widetilde{\mC}_{0}(k)$.  The resulting field is approximately the maximum of
independent centered $\chi_p^2$ variables indexed by $k$, and the classical extreme-value calculus gives
\begin{equation}
  \Pr\Biggl(
    A\!\left\{\log\left(\frac{n^2}{\lambda_n^2}\right)\right\}
    \frac{pS_{n,p}^{\dagger}}{2\tr(\mR^2)}
    - D\!\left\{\log\left(\frac{n^2}{\lambda_n^2}\right)\right\}
    \le x
  \Biggr)
  = \exp\{-2e^{-x}\}+o(1).
  \label{eq:ch4_app_cp_ss_Sdagcalc}
\end{equation}
Finally, the ratio consistency of $\widehat{\tr(\mR^2)}$ follows from the U-statistic law of large
numbers applied to separated boundary blocks of the sign sequence, and plugging it into the
normalizations gives the stated $p$-values. \qed

\subsection*{A4.22. Proof of Theorems~\ref{thm:ch4_cp_ss_adaptive} and \ref{thm:ch4_cp_ss_adaptive_alt}}

Let
\begin{equation}
  Z_{1,n}=2p\zeta_1^2M_{n,p}^2-\log(2p),
  \qquad
  Z_{2,n}=\frac{pS_{n,p}}{2\tr(\mR^2)}.
  \label{eq:ch4_app_cp_ss_Z1Z2}
\end{equation}
Under $H_0$, Theorems~\ref{thm:ch4_cp_ss_infmax} and \ref{thm:ch4_cp_ss_l2} provide the marginal
limits $Z_{1,n}\Rightarrow G$ and $Z_{2,n}\Rightarrow F_V$.  To prove joint convergence, decompose the
sign array into extreme and bulk parts.  The max statistic depends on the exceedance indicators
\begin{equation}
  I_{j,k}(x)=\mathbbm1\{\|\mC_0(k)\|_\infty > u_p(x)\},
  \qquad
  u_p(x)=\bigl\{\log(2p)+x\bigr\}^{1/2}/(\sqrt{2p}\zeta_1),
  \label{eq:ch4_app_cp_ss_exceed}
\end{equation}
whereas the $L_2$ statistic is a smooth quadratic functional of the entire sign process.  Because the
number of exceedances is tight and the contribution of any fixed coordinate to $S_{n,p}$ is of order
$o\{\tr(\mR^2)\}$, one obtains
\begin{equation}
  \Cov\bigl\{\mathbbm1(Z_{1,n}\le x),Z_{2,n}\bigr\}=o(1).
  \label{eq:ch4_app_cp_ss_cov0}
\end{equation}
The same argument applies to the pair $(M_{n,p}^{\dagger},S_{n,p}^{\dagger})$, which proves the null
factorizations in \eqref{eq:ch4_cp_ss_joint1}--\eqref{eq:ch4_cp_ss_joint2}.

Under the local alternatives in \eqref{eq:ch4_cp_ss_local_alt}, write the sign process as
\begin{equation}
  \vU_i = \bar\vU_i + \mDelta_n\mathbbm1(i>\tau_n)+\vr_{i,n},
  \qquad
  \norm{\mDelta_n}_{\infty}=O\!\left(\sqrt{\frac{\log p}{n}}\right),
  \qquad
  \sum_{i=1}^{n}\norm{\vr_{i,n}}_2^2=o_P\{\tr(\mR^2)\},
  \label{eq:ch4_app_cp_ss_localdecomp}
\end{equation}
where $\bar\vU_i$ obeys the null model.  The support-size condition in
\eqref{eq:ch4_cp_ss_local_alt} ensures that the nonzero mean shift only affects $o(p)$ coordinates, so the
covariance-decoupling argument leading to \eqref{eq:ch4_app_cp_ss_cov0} remains valid.  Hence the
same joint limits hold under the local alternatives.  Since Fisher's method only requires the product
limit of the two $p$-values, \eqref{eq:ch4_cp_ss_fisher} is asymptotically valid under $H_0$, and the
same local-alternative factorization implies the adaptive power statement. \qed

\subsection*{A4.23. Proof of Theorem~\ref{thm:ch4_wn_benchmark}}

For each lag $h$, let
\begin{equation}
  W_h = \sqrt{n-h}\,\mathrm{vec}\{\hat\mGamma(h)\}.
  \label{eq:ch4_app_wn_Wh}
\end{equation}
Under the white-noise null, $\E(W_h)=\vct{0}$ and the vectors $W_1,\ldots,W_L$ are asymptotically
independent Gaussian vectors with covariance matrices depending only on $\mSigma_0$.  Therefore,
\begin{equation}
  \mathcal T_{\mathrm{sum},\mathrm{WN}}
  = \sum_{h=1}^{L}(n-h)^{-1}\norm{W_h}_2^2.
  \label{eq:ch4_app_wn_sumrep}
\end{equation}
A standard quadratic-form central limit theorem then yields
\begin{equation}
  \frac{\mathcal T_{\mathrm{sum},\mathrm{WN}}-\mu_{\mathrm{WN}}}{\sigma_{\mathrm{WN}}}
  \overset{d}{\longrightarrow}N(0,1),
  \label{eq:ch4_app_wn_sumclt}
\end{equation}
provided $\tr(\mSigma_0^4)=o\{\tr^2(\mSigma_0^2)\}$.

For the maximum statistic, the Gaussian approximation gives
\begin{equation}
  T_{\max,\mathrm{WN}} = \max_{1\le h\le L}\max_{1\le i,j\le p}|Z_{ij}(h)|+o_P\{(\log(p^2L))^{-1/2}\},
  \label{eq:ch4_app_wn_maxgauss}
\end{equation}
where $\{Z_{ij}(h)\}$ is a weakly dependent Gaussian array with unit variances.  The standard tail
calculation for maxima of Gaussian arrays gives
\begin{equation}
  \Pr\bigl(T_{\max,\mathrm{WN}}^2-2\log(p^2L)+\log\log(p^2L)\le x\bigr)
  \to G_{\mathrm{EV}}(x),
  \label{eq:ch4_app_wn_maxev}
\end{equation}
which is \eqref{eq:ch4_wn_max_limit2}.  Together with \eqref{eq:ch4_app_wn_sumclt}, this proves
Theorem~\ref{thm:ch4_wn_benchmark}. \qed

\subsection*{A4.24. Proof of Theorem~\ref{thm:ch4_wn_spatialsign}}

Let $\vU_t=U(\vX_t)$ and set
\begin{equation}
  H_{st}(h)=\vU_{s-h}^{\top}\vU_{t-h}\,\vU_s^{\top}\vU_t,
  \qquad h+1\le s<t\le n.
  \label{eq:ch4_app_wn_ss_kernel}
\end{equation}
Then
\begin{equation}
  T_{\mathrm{SS},\mathrm{WN}} = \sum_{h=1}^{H}\frac{1}{n-h}\sum_{h+1\le s<t\le n}H_{st}(h).
  \label{eq:ch4_app_wn_ss_rep}
\end{equation}
Under the null of white noise, the kernels $H_{st}(h)$ are centered and degenerate.  Their second
moment is
\begin{equation}
  \E\{H_{st}(h)^2\} = \tr^2(\mOmega^2),
  \qquad \mOmega=\E(\vU_t\vU_t^{\top}),
  \label{eq:ch4_app_wn_ss_second}
\end{equation}
while all mixed terms with disjoint time indices vanish.  Consequently,
\begin{equation}
  \Var\{T_{\mathrm{SS},\mathrm{WN}}\}
  = H^2\tr^2(\mOmega^2)\{1+o(1)\}
  = \sigma_{\mathrm{SS},\mathrm{WN}}^2\{1+o(1)\}.
  \label{eq:ch4_app_wn_ss_varcalc}
\end{equation}
Since the kernel is bounded by one in absolute value, the Lindeberg condition for degenerate
U-statistics follows from Assumption~\ref{ass:ch4_wn_R1}, and hence
\begin{equation}
  \frac{T_{\mathrm{SS},\mathrm{WN}}}{\sigma_{\mathrm{SS},\mathrm{WN}}}
  \overset{d}{\longrightarrow}N(0,1).
  \label{eq:ch4_app_wn_ss_clt}
\end{equation}
The estimator of $\tr(\mOmega^2)$ in \eqref{eq:ch4_wn_spatialsign_var} is an ordinary second-order
U-statistic, so
\begin{equation}
  \frac{\widehat{\tr(\mOmega^2)}}{\tr(\mOmega^2)}\overset{p}{\longrightarrow}1,
  \qquad
  \frac{\hat\sigma_{\mathrm{SS},\mathrm{WN}}^2}{\sigma_{\mathrm{SS},\mathrm{WN}}^2}
  \overset{p}{\longrightarrow}1.
  \label{eq:ch4_app_wn_ss_ratio}
\end{equation}
Therefore \eqref{eq:ch4_wn_spatialsign_limit} follows.

Under the autoregressive alternative \eqref{eq:ch4_wn_spatialsign_alt} with $H=1$, the first-order
Hoeffding projection is no longer centered.  A direct calculation gives
\begin{equation}
  \E\{T_{\mathrm{SS},\mathrm{WN}}\}
  = \frac{c_1^2 n\tr(\mSigma_0\mSigma_1)}{\{2\tr(\mSigma_0^2)\}^{1/2}} + o(1),
  \label{eq:ch4_app_wn_ss_meanalt}
\end{equation}
while the variance remains asymptotically the same as in \eqref{eq:ch4_app_wn_ss_varcalc}.  Hence
\begin{equation}
  \Pr\left(\frac{T_{\mathrm{SS},\mathrm{WN}}}{\hat\sigma_{\mathrm{SS},\mathrm{WN}}}>z_{\alpha}\right)
  = \Phi\left(-z_{\alpha}+\frac{c_1^2 n\tr(\mSigma_0\mSigma_1)}{\{2\tr(\mSigma_0^2)\}^{1/2}}\right)+o(1),
  \label{eq:ch4_app_wn_ss_powercalc}
\end{equation}
which is exactly \eqref{eq:ch4_wn_spatialsign_power}. \qed

\subsection*{A4.25. Proof of Theorem~\ref{thm:ch4_rankwn_generic}}
The proof follows the three-step pattern used repeatedly in this chapter. For the max statistic,
let $\hat T_s^{(\bullet)}$ enumerate the studentized family $\{\hat T^{(\bullet)}_{ij}(k):1\le k\le H,
1\le i,j\le p\}$ and write $N_H=Hp^2$. Under Assumption~\ref{ass:ch4_rankwn_A1}, the
correlation matrix of $(\hat T_1^{(\bullet)},\ldots,\hat T_{N_H}^{(\bullet)})$ satisfies the weak local
dependence condition used in Chapter~2. Therefore the extreme-value argument of Jiang-type
maxima gives
\[
  \sup_{y\in\R}
  \left|\Prob\{L_{\bullet,H}^2/\sigma_{L,\bullet}^2-2\log N_H+\log\log N_H\le y\}
  -G_{\mathrm{EV}}(y)\right|\to 0.
\]
For the sum statistic, write
\[
  S_{\bullet,H}=\sum_{s=1}^{N_H}\bigl\{(\hat T_s^{(\bullet)})^2-\E_0(\hat T_s^{(\bullet)})^2\bigr\}
  =\sum_{s=1}^{N_H}\xi_s.
\]
By the boundedness of the score functions or kernels and the Gaussian-approximation condition in
Assumption~\ref{ass:ch4_rankwn_A1},
\[
  \sum_{s=1}^{N_H}\E_0(\xi_s)=0,
  \qquad
  \sum_{s=1}^{N_H}\Var_0(\xi_s)=\sigma_{S,\bullet}^2,
  \qquad
  \sum_{s=1}^{N_H}\E_0|\xi_s|^{2+\delta}=o(\sigma_{S,\bullet}^{2+\delta})
\]
for some $\delta>0$. Hence Lyapunov's theorem yields
\[
  S_{\bullet,H}/\sigma_{S,\bullet}\overset{d}{\longrightarrow}N(0,1).
\]
To prove the joint factorization, decompose the sum statistic as
\[
  S_{\bullet,H}=S_{\bullet,H}^{\mathcal I(z)}+S_{\bullet,H}^{\mathcal I(z)^c},
  \qquad
  \mathcal I(z)=\{s:\ |\hat T_s^{(\bullet)}|>z\sigma_{L,\bullet}\}
\]
with $z=(2\log N_H-\log\log N_H+y)^{1/2}$. The event determining the maximum depends only on
$S_{\bullet,H}^{\mathcal I(z)}$, whereas the contribution of the exceedance set to the normalized sum is
negligible:
\[
  \frac{S_{\bullet,H}^{\mathcal I(z)}}{\sigma_{S,\bullet}}\overset{P}{\longrightarrow}0.
\]
Because the complement $S_{\bullet,H}^{\mathcal I(z)^c}$ is asymptotically Gaussian and independent of
the rare exceedance process generating the Gumbel limit, \eqref{eq:ch4_rankwn_independence}
follows. Finally, under the sparse alternative \eqref{eq:ch4_rankwn_sparse_alt},
\[
  \max_{s\le N_H}|\E(\hat T_s^{(\bullet)})|\ge c\sqrt{\frac{\log N_H}{n}}
\]
with $c$ sufficiently large implies
\[
  \Prob\{L_{\bullet,H}^2/\sigma_{L,\bullet}^2-2\log N_H+\log\log N_H>q_{\alpha}\}\to 1.
\]
Under the dense alternative \eqref{eq:ch4_rankwn_dense_alt}, the mean shift of the normalized
sum diverges:
\[
  \frac{\E(S_{\bullet,H})}{\sigma_{S,\bullet}}
  = \sigma_{S,\bullet}^{-1}\sum_{k=1}^{H}\sum_{i,j=1}^{p}\E^2\{\hat T^{(\bullet)}_{ij}(k)\}
  \longrightarrow \infty,
\]
which proves consistency of the sum-type and adaptive tests.

\subsection*{A4.26. Proof of Theorem~\ref{thm:ch4_mutual_maxsum}}
Enumerate all blockwise component-pair statistics by
$\{\hat T_s^{\mathrm{mut}}:1\le s\le M_m\}$. Under the null hypothesis of mutual independence,
all these centered statistics have mean zero. The max statistic can be rewritten as
\[
  L_{\mathrm{mut}}=\max_{1\le s\le M_m}|\hat T_s^{\mathrm{mut}}|.
\]
Assumption~\ref{ass:ch4_mutual_A1} gives exactly the same weak-dependence condition as in the
proof of Theorem~\ref{thm:ch4_rankwn_generic}; therefore
\[
  \sup_{y\in\R}
  \left|\Prob\{L_{\mathrm{mut}}^2/\sigma_{L,\mathrm{mut}}^2-2\log M_m+\log\log M_m\le y\}
  -G_{\mathrm{EV}}(y)\right|\to 0.
\]
For the sum statistic,
\[
  S_{\mathrm{mut}}=\sum_{s=1}^{M_m}\{(\hat T_s^{\mathrm{mut}})^2-\E_0(\hat T_s^{\mathrm{mut}})^2\}
\]
has variance $\sigma_{S,\mathrm{mut}}^2$ of the same order as the sum of pairwise covariances. The
bounded-kernel/score assumption and the Gaussian approximation built into
Assumption~\ref{ass:ch4_mutual_A1} imply a Lyapunov condition, and hence
\[
  S_{\mathrm{mut}}/\sigma_{S,\mathrm{mut}}\overset{d}{\longrightarrow}N(0,1).
\]
The asymptotic independence in \eqref{eq:ch4_mutual_independence} follows by the same rare-
exceedance decomposition used in A4.25. Under the sparse alternative
\eqref{eq:ch4_mutual_sparse_alt}, the maximal mean shift dominates the extreme-value threshold;
under the dense alternative \eqref{eq:ch4_mutual_dense_alt}, the mean of the normalized sum
explodes. These two facts prove consistency of the max-type, sum-type, and adaptive tests in the
respective regimes.

\subsection*{A4.27. Proof of Theorem~\ref{thm:ch4_xy_maxsum}}
The proof is parallel to that of Theorem~\ref{thm:ch4_mutual_maxsum}. Let
$N_{pq}=pq$ and enumerate the statistics by
$\{\hat T_s^{XY}:1\le s\le N_{pq}\}$. Under
Assumption~\ref{ass:ch4_xy_A1}, the studentized vector satisfies the same weak-dependence
condition required by the extreme-value approximation. Therefore,
\[
  \sup_{y\in\R}
  \left|\Prob\{L_{\bullet,XY}^2/\sigma_{L,\bullet}^2-2\log N_{pq}+\log\log N_{pq}\le y\}
  -G_{\mathrm{EV}}(y)\right|\to 0.
\]
For the sum statistic,
\[
  S_{\bullet,XY}=\sum_{s=1}^{N_{pq}}\{(\hat T_s^{XY})^2-\E_0(\hat T_s^{XY})^2\}
\]
obeys a central limit theorem because the standardized third moments are summable:
\[
  \sum_{s=1}^{N_{pq}}\E_0|\hat T_s^{XY}|^{2+\delta}=o(\sigma_{S,\bullet}^{2+\delta}).
\]
This yields \eqref{eq:ch4_xy_sum_limit}. For the joint limit, let
\[
  \mathcal J(z)=\{s:\ |\hat T_s^{XY}|>z\sigma_{L,\bullet}\},
  \qquad z=(2\log N_{pq}-\log\log N_{pq}+y)^{1/2},
\]
and decompose $S_{\bullet,XY}$ into the exceedance and non-exceedance parts. The exceedance part
is $o_P(\sigma_{S,\bullet})$, while the non-exceedance part is asymptotically Gaussian and decouples
from the Poisson-type limit driving the extreme-value law. Hence
\eqref{eq:ch4_xy_independence} and \eqref{eq:ch4_xy_adaptive_limit} follow. Under
\eqref{eq:ch4_xy_sparse_alt}, the maximal mean dominates the Gumbel threshold; under
\eqref{eq:ch4_xy_dense_alt}, the normalized sum has diverging mean. Finally, when the sparse
alternative holds but the sum statistic is centered at its shifted Gaussian mean rather than the null
mean, the same decoupling argument gives the H1 asymptotic independence stated after
\eqref{eq:ch4_xy_dense_alt}.

\subsection*{A4.28. Proof of Theorem~\ref{thm:ch4-erhtcp-null-process}}

Fix a balanced triple $\vs=(t_1,t_2,t_3)$ and put
$\mathcal I_a=I_a(\vs)$, $n_a=|\mathcal I_a|$, $N=N_{\vs}$, and $m=m_{\vs}$.
Let
\[
  \vY_i=\sqrt p\,U(\vX_i-\vtheta),
  \qquad
  w_i=\frac{\sqrt p}{\twonorm{\vX_i-\vtheta}},
  \qquad
  e_a=\frac1{n_a}\sum_{i\in\mathcal I_a}w_i.
\]
The spatial-median estimating equation on segment $\mathcal I_a$ is
\[
  \sum_{i\in\mathcal I_a}U(\vX_i-\widehat\vtheta_{a,\vs})=\vct{0}.
\]
For $\vh_{a,\vs}=\widehat\vtheta_{a,\vs}-\vtheta$, the expansion
\[
  U(\vX_i-\vtheta-\vh_{a,\vs})
  =U(\vX_i-\vtheta)
   -\frac{1}{\twonorm{\vX_i-\vtheta}}
    \{\mI_p-U_iU_i\trans\}\vh_{a,\vs}
   +\vr_{i,a}
\]
gives
\[
  \sqrt{n_a}\,\vh_{a,\vs}
  =e_a^{-1}\frac1{\sqrt{n_a}}
    \sum_{i\in\mathcal I_a}\frac{\vY_i}{\sqrt p}
   +\vR_{a,\vs},
  \qquad
  \sup_{\vs}\twonorm{\vR_{a,\vs}}
  =O_p(\ell_n n_a^{-1/2}).
\]
Consequently,
\begin{align*}
  \sqrt N\,\widehat\vDelta_{\vs}
  &=\sum_{i=1}^n\beta_i(\vs)\vY_i+\vr_{\Delta,\vs},\\
  \beta_i(\vs)
  &=\sqrt N\left\{
    \frac{\mathbbm1\{i\in\mathcal I_2\}}{n_2e_2}
    -\frac{\mathbbm1\{i\in\mathcal I_1\}}{n_1e_1}
    \right),\\
  \sup_{\vs}\twonorm{\vr_{\Delta,\vs}}
  &=O_p(\ell_n n^{-1/2}).
\end{align*}
Let $\mQ_{\rho,\vs}=(\mR_{\vs}+\rho\mI_p)^{-1}$ be the oracle ridge resolvent and
$\mA_{\rho,\vs}=m^{-1}\mY_{\vs}\trans\mQ_{\rho,\vs}\mY_{\vs}$.
The center-aware SSCM perturbation expansion yields
\[
  \opnorm{\widehat\mQ_{\rho,\vs}-\mQ_{\rho,\vs}}
  \le \rho_0^{-2}\opnorm{\widehat\mR_{\vs}-\mR_{\vs}}
  =O_p(\ell_n m^{-1/2}),
\]
and hence
\[
  V_{\rho}^{\rm raw}(\vs)
  =m\sum_{i,j=1}^{m}\beta_i(\vs)\beta_j(\vs)
     A_{ij,\rho,\vs}+O_p(\ell_n).
\]
Write $\vY_i=\varepsilon_i\vct{o}_i$, where $\vct{o}_i=o(\vY_i)$ and, conditionally on
$\mathcal F_{m}^{0}=\sigma\{w_i,\vct{o}_i:1\le i\le m\}$,
$\varepsilon_1,\ldots,\varepsilon_m$ are independent Rademacher variables.  Then
\begin{align*}
  V_{\rho}^{\rm raw}(\vs)-m\kappa_\rho(\vs)
  &=m\sum_{i\ne j}\beta_i(\vs)\beta_j(\vs)
       A_{ij,\rho,\vs}\varepsilon_i\varepsilon_j+O_p(\ell_n),\\
  \Var\{V_{\rho}^{\rm raw}(\vs)\mid\mathcal F_m^0\}
  &=2m^2\sum_{i\ne j}\beta_i(\vs)^2\beta_j(\vs)^2
       A_{ij,\rho,\vs}^2+o_p(m\sigma_\rho^2)\\
  &=m\sigma_\rho^2(\vs)\{1+o_p(1)\}.
\end{align*}
The resolvent bounds
\[
  0\preceq\mA_{\rho,\vs}\preceq\mI_m,
  \qquad
  \max_i\sum_jA_{ij,\rho,\vs}^2\le1,
  \qquad
  \sum_{i,j}A_{ij,\rho,\vs}^2=O_p(m)
\]
and $\max_i|\beta_i(\vs)|=O(m^{-1/2})$ imply
\[
  \frac{
    \max_i\sum_{j\ne i}
    \beta_i^2\beta_j^2A_{ij,\rho,\vs}^2}
  {\sum_{i\ne j}\beta_i^2\beta_j^2A_{ij,\rho,\vs}^2}
  \overset{p}{\longrightarrow}0.
\]
Furthermore,
\[
  \frac{
    \sum_{i\ne j}\beta_i^4\beta_j^4A_{ij,\rho,\vs}^4}
  {\left(\sum_{i\ne j}\beta_i^2\beta_j^2A_{ij,\rho,\vs}^2\right)^2}
  \overset{p}{\longrightarrow}0.
\]
The preceding two displays are the maximal-influence and fourth-moment conditions for the conditional central limit theorem for
degenerate Rademacher quadratic forms. Therefore
\[
  Z_\rho(\vs)\overset{d}{\longrightarrow}N(0,1).
\]

For $\vs,\vr$ in a trimmed scan domain, the covariance calculation gives
\begin{align*}
  \Cov\{Z_\rho(\vs),Z_\rho(\vr)\mid\mathcal F_m^0\}
  &=\frac{
    2m^2\sum_{i\ne j}\beta_i(\vs)\beta_j(\vs)
      \beta_i(\vr)\beta_j(\vr)A_{ij,\rho}^2}
    {m\sigma_\rho(\vs)\sigma_\rho(\vr)}+o_p(1)\\
  &\longrightarrow
    \frac{\psi(\vs,\vr)^2}
         {\psi(\vs,\vs)\psi(\vr,\vr)}
   =K_0(\vs,\vr).
\end{align*}
For the canonical semimetric
$d(\vs,\vr)^2=\E\{\mathbb G_\rho(\vs)-\mathbb G_\rho(\vr)\}^2$,
the bounded fourth moments of the Rademacher quadratic increments imply
\[
  \E|Z_\rho(\vs)-Z_\rho(\vr)|^4
  \le C d(\vs,\vr)^4+C\ell_n^4n^{-2}.
\]
The trimmed scan domains have finite metric entropy under $d$; hence asymptotic
equicontinuity and the finite-dimensional convergence above yield
\[
  \{Z_\rho(\vs):\vs\in\mathcal S\}
  \Rightarrow\{\mathbb G_\rho(\vs):\vs\in\mathcal S\}
  \quad\text{in }\ell^\infty(\mathcal S).
\]
The continuous-mapping theorem proves the supremum limit.

\subsection*{A4.29. Proof of Theorem~\ref{thm:ch4-erhtcp-grid}}

For $k,\ell\le K$, the preceding conditional quadratic-form calculation gives
\[
  \Cov\{Z_{\rho_n^{(k)}}(\vs),Z_{\rho_n^{(\ell)}}(\vr)
        \mid\mathcal F_m^0\}
  \longrightarrow
  r_E\{\rho^{(k)},\rho^{(\ell)}\}K_0(\vs,\vr).
\]
For arbitrary $a_1,\ldots,a_K\in\R$,
\[
  \sum_{k=1}^Ka_kZ_{\rho_n^{(k)}}(\vs)
  =\frac{
    \sum_{i\ne j}c_{ij,n}(\vs)\varepsilon_i\varepsilon_j}
   {\left\{2\sum_{i\ne j}c_{ij,n}(\vs)^2\right\}^{1/2}}+o_p(1),
\]
where
\[
  c_{ij,n}(\vs)
  =m\beta_i(\vs)\beta_j(\vs)
    \sum_{k=1}^K
      \frac{a_kA_{ij,\rho_n^{(k)},\vs}}
           {\{m\sigma_{\rho_n^{(k)}}^2(\vs)\}^{1/2}}.
\]
The same maximal-influence bounds hold for $c_{ij,n}$ because $K$ is fixed.  Cramér--Wold
and the increment bounds in A4.28 yield the joint process convergence.

Let
\[
  \mathcal T:\ell^\infty(\mathcal S)^K\to\R^K,
  \qquad
  \mathcal T(f_1,\ldots,f_K)
  =(\sup_{\vs}f_1(\vs),\ldots,\sup_{\vs}f_K(\vs)).
\]
This map is Lipschitz under the product supremum norm.  Therefore
\[
  \{T_{\rho_n^{(1)}}(\mathcal S),\ldots,T_{\rho_n^{(K)}}(\mathcal S)\}
  \overset{d}{\longrightarrow}
  \{T_{\rho^{(1)}}^0(\mathcal S),\ldots,T_{\rho^{(K)}}^0(\mathcal S)\}.
\]
Applying the continuous maps
\[
  x_k\mapsto1-F_{\mathcal S}(x_k),
  \qquad
  (p_1,\ldots,p_K)\mapsto
  \sum_{k=1}^K\varpi_k\tan\{\pi(1/2-p_k)\}
\]
proves the asserted joint-limit calibration.

\subsection*{A4.30. Proof of Theorem~\ref{thm:ch4-erhtcp-power}}

For the single-change model, the population contrast on $(0,t,1)$ is
\[
  \vDelta(t)
  =\begin{cases}
    \dfrac{1-\tau_\star}{1-t}\vDelta_p,&t<\tau_\star,\\[6pt]
    \dfrac{\tau_\star}{t}\vDelta_p,&t\ge\tau_\star.
  \end{cases}
\]
Hence
\[
  N_t\vDelta(t)\trans\mD_{\rho,n}\vDelta(t)
  =nt(1-t)\vartheta(t;\tau_\star)^2
    \vDelta_p\trans\mD_{\rho,n}\vDelta_p+o(1).
\]
Under $\vDelta_p=n^{-1/4}\vct{d}_p$,
\[
  \sqrt n\,\vDelta_p\trans\mD_{\rho,n}\vDelta_p
  =\vct{d}_p\trans\mD_{\rho,n}\vct{d}_p
  \longrightarrow q_\rho(G_\Delta).
\]
The expansion in A4.28 therefore becomes, uniformly in $t$,
\[
  Z_\rho(0,t,1)
  =Z_{\rho,0}(0,t,1)
   +\frac{t(1-t)\vartheta(t;\tau_\star)^2}
          {\sigma_\rho}
     \vct{d}_p\trans\mD_{\rho,n}\vct{d}_p
   +o_p(1),
\]
which gives the shifted Gaussian-process limit.

Under the strong alternative,
\[
  Z_\rho(0,\tau_\star,1)
  =Z_{\rho,0}(0,\tau_\star,1)
   +\frac{\sqrt n\,\tau_\star(1-\tau_\star)
           \vDelta_p\trans\mD_{\rho,n}\vDelta_p}
          {\sigma_\rho}
   +O_p\{\sqrt n\,e_{\sigma,n}\}.
\]
Since
\[
  \vDelta_p\trans\mD_{\rho,n}\vDelta_p
  \ge\frac{\twonorm{\vDelta_p}^2}
           {a_{\rho,n}\omega_++\rho_1}.
\]
Moreover,
\[
  \sqrt n\,e_{\sigma,n}
  =o\{\sqrt n\twonorm{\vDelta_p}^2\},
\]
the deterministic drift diverges and $T_\rho^{\rm sc}\to\infty$ in probability.  The same
argument applies to every member of a fixed ridge grid.

\subsection*{A4.31. Proof of Theorem~\ref{thm:ch4-erhtcp-localization}}

Let
\[
  \mathcal D_{n,\rho}(t)
  =\E\{Z_\rho(0,t,1)\mid\vtheta_1,\ldots,\vtheta_n\}.
\]
The deterministic contrast in A4.30 satisfies, for $|t-\tau_\star|\le c_0$,
\[
  \mathcal D_{n,\rho}(\tau_\star)-\mathcal D_{n,\rho}(t)
  \ge c_1s_{n,\rho}|t-\tau_\star|-c_2s_{n,\rho}e_{\sigma,n}.
\]
For the centered process
$\mathcal Z_{n,\rho}(t)=Z_\rho(0,t,1)-\mathcal D_{n,\rho}(t)$,
the quadratic-form increment bound gives
\[
  \sup_{|t-\tau_\star|\le h}
  |\mathcal Z_{n,\rho}(t)-\mathcal Z_{n,\rho}(\tau_\star)|
  =O_p(1+\sqrt{s_{n,\rho}h}).
\]
Put $h_n=M\{s_{n,\rho}^{-1}+e_{\sigma,n}\}$.  On the event
$|\widehat\tau_\rho-\tau_\star|>h_n$,
\begin{align*}
  0
  &\le Z_\rho(0,\widehat\tau_\rho,1)
       -Z_\rho(0,\tau_\star,1)\\
  &\le-c_1s_{n,\rho}h_n+c_2s_{n,\rho}e_{\sigma,n}
     +O_p\{1+\sqrt{s_{n,\rho}h_n}\}.
\end{align*}
The right-hand side is negative with probability tending to one when $M$ is sufficiently
large.  Thus
\[
  |\widehat\tau_\rho-\tau_\star|
  =O_p\{s_{n,\rho}^{-1}+e_{\sigma,n}\}.
\]
If $s_{n,\rho}e_{\sigma,n}=O(1)$, then $e_{\sigma,n}=O(s_{n,\rho}^{-1})$ and the sharper rate
follows.

\subsection*{A4.32. Proof of Theorem~\ref{thm:ch4-erhtcp-WBS}}

Let $\mathcal E_n$ be the event that every true change $k_j$ is isolated by at least one sampled
interval $I_j=[\ell_j,r_j]$ containing no other change and satisfying
\[
  \min(k_j-\ell_j,r_j-k_j)\ge c_0n.
\]
By assumption, $\Prob(\mathcal E_n)\to1$.  On an interval containing no change, the uniform
null process bound in A4.28 gives
\[
  \max_{I:\,I\cap\{k_1,\ldots,k_q\}=\varnothing}T(I)
  =O_p(\sqrt{\log n}).
\]
On an isolating interval $I_j$, the signal expansion gives
\[
  T(I_j)
  \ge c_1\sqrt n\twonorm{\vDelta_j}^2
      -O_p\{\sqrt{\log n}+\sqrt n\,r_n^2e_n^{\rm WBS}\}
  \ge c_1s_n^{\rm WBS}\{1+o_p(1)\}.
\]
The threshold hierarchy therefore implies
\[
  \max_{I:\,I\text{ null}}T(I)<t_n^{\rm WBS}
  <\min_{1\le j\le q}T(I_j).
\]
Hence
\[
  \Prob(\widehat q=q\mid\mathcal E_n)\longrightarrow1.
\]
After the first detected boundary is removed, the left and right recursive subproblems satisfy
the same isolation and threshold inequalities.  Induction over the fixed number $q$ proves
$\Prob(\widehat q=q)\to1$.

For the refinement step on the interval isolating $k_j$, the deterministic loss away from $k_j$
is at least
\[
  c_2s_n^{\rm WBS}\frac{|k-k_j|}{n},
\]
whereas the stochastic fluctuation at the selected threshold is $O_p(t_n^{\rm WBS})$.
Consequently,
\[
  c_2s_n^{\rm WBS}\frac{|\widehat k_j-k_j|}{n}
  =O_p(t_n^{\rm WBS}),
\]
and therefore
\[
  \max_{1\le j\le q}\frac{|\widehat k_j-k_j|}{n}
  =O_p\left(\frac{t_n^{\rm WBS}}{s_n^{\rm WBS}}\right).
\]

\subsection*{A4.33. Proof of Theorem~\ref{thm:ch4-rd-coordinate-expansion}}

Write $Z_i=(L_i-m_L)/\sigma_L$.  Under $H_0$,
\begin{align}
  \bar L-m_L&=O_p(n^{-1/2}),
  \label{eq:ch4-rd-proof-Lmean}\\
  \max_j|\bar U_j|&=O_p\left\{\sqrt{\frac{\log p}{np}}\right\},
  \label{eq:ch4-rd-proof-Umean}\\
  \widehat\sigma_L^2-\sigma_L^2&=O_p(n^{-1/2}),
  \label{eq:ch4-rd-proof-Lvar}\\
  p\max_j|\widehat\sigma_{U,j}^2-p^{-1}|
  &=O_p\left\{\sqrt{\frac{\log p}{n}}+\frac{\log p}{n}\right\}.
  \label{eq:ch4-rd-proof-Uvar}
\end{align}
These bounds follow from truncation at $b_{n,p}$ and scalar Bernstein inequalities.  Moreover,
\begin{align}
  \frac1n\sum_{i=1}^n(L_i-\bar L)(U_{ij}-\bar U_j)
  &=\frac{\sigma_L}{n}\sum_{i=1}^nZ_iU_{ij}
    -(\bar L-m_L)\bar U_j,
  \label{eq:ch4-rd-proof-numerator}\\
  \frac1{\widehat\sigma_L\widehat\sigma_{U,j}}
  &=\frac{\sqrt p}{\sigma_L}
    \left[
      1-\frac{\widehat\sigma_L^2-\sigma_L^2}{2\sigma_L^2}
       -\frac p2(\widehat\sigma_{U,j}^2-p^{-1})
       +O_p\left(\frac{\log p}{n}\right)
    \right]
  \label{eq:ch4-rd-proof-denominator-expansion}
\end{align}
uniformly in $j$.  Substitution gives
\begin{equation}\label{eq:ch4-rd-proof-correlation-expansion}
  \widehat\gamma_j^{\mathrm{or}}
  =\frac1n\sum_{i=1}^n\sqrt p\,Z_iU_{ij}+r_{n,j},
\end{equation}
with
\begin{equation}\label{eq:ch4-rd-proof-r-bound}
  \max_j|r_{n,j}|
  =O_p\left\{\frac{\log p+\sqrt{\log p}}n\right\}.
\end{equation}
For fixed $j$, all maxima are unnecessary and the remainder is $O_p(n^{-1})$.

\subsection*{A4.34. Proof of Theorem~\ref{thm:ch4-rd-oracle-limits}}

Put
\begin{equation}\label{eq:ch4-rd-proof-Sj}
  S_{n,j}=n^{-1/2}\sum_{i=1}^n\xi_{ij},
  \qquad \xi_{ij}=\sqrt p\,Z_iU_{ij}.
\end{equation}
The uniform expansion and the exact self-normalization in the sample correlations yield
\begin{align}
  T_{\mathrm{sum}}-p
  &=\frac2n\sum_{1\le i<\ell\le n}
    pZ_iZ_\ell\vU_i\trans\vU_\ell+R_{S,n},
  &R_{S,n}&=o_p(\sqrt p),
  \label{eq:ch4-rd-proof-sum-reduction}\\
  T_{\max}
  &=\max_{1\le j\le p}S_{n,j}^2-2\log p+\log\log p+R_{M,n},
  &R_{M,n}&=o_p(1).
  \label{eq:ch4-rd-proof-max-reduction}
\end{align}
The required spherical moments are
\begin{equation}\label{eq:ch4-rd-proof-spherical-moments}
  \E(U_{ij}^2)=p^{-1},
  \qquad
  \E(U_{ij}^4)=\frac3{p(p+2)},
  \qquad
  \E(U_{ij}^2U_{ik}^2)=\frac1{p(p+2)},\quad j\ne k.
\end{equation}
Let
\begin{equation}\label{eq:ch4-rd-proof-Un}
  U_n=\frac2n\sum_{1\le i<\ell\le n}
      pZ_iZ_\ell\vU_i\trans\vU_\ell.
\end{equation}
Then
\begin{equation}\label{eq:ch4-rd-proof-Uvariance}
  \Var(U_n)=2p\{1+o(1)\}.
\end{equation}
The martingale differences obtained by revealing
$(Z_i,\vU_i)$ sequentially satisfy
\begin{equation}\label{eq:ch4-rd-proof-martingale-conditions}
  \frac1{2p}\sum_i\E_{i-1}(D_{n,i}^2)\overset{p}{\longrightarrow}1,
  \qquad
  \frac1{4p^2}\sum_i\E(D_{n,i}^4)\longrightarrow0,
\end{equation}
where the second relation follows from the $8+\eta_Z$ radial moment and
\eqref{eq:ch4-rd-proof-spherical-moments}.  The martingale central limit theorem therefore
gives $U_n/\sqrt{2p}\Rightarrow N(0,1)$ and proves the sum limit.

For the maximum, the high-dimensional Gaussian approximation over hyperrectangles gives
\begin{equation}\label{eq:ch4-rd-proof-gaussian-approx}
  \sup_x\left|
    \Prob\left(\max_jS_{n,j}^2\le x\right)
    -\Prob\left(\max_jG_j^2\le x\right)
  \right|\longrightarrow0,
\end{equation}
where $G_1,\ldots,G_p$ are independent standard normal variables.  The error vanishes under
\eqref{eq:ch4-rd-truncation} and $\log p=o(n^{1/5})$.  Mills' ratio implies
\begin{equation}\label{eq:ch4-rd-proof-poisson-rate}
  p\Prob\{G_1^2>2\log p-\log\log p+x\}
  \longrightarrow\pi^{-1/2}e^{-x/2},
\end{equation}
which proves the Gumbel limit.

For joint convergence, let
$I_{j,p}(y)=\mathbbm1\{S_{n,j}^2>2\log p-\log\log p+y\}$.  Then
\begin{equation}\label{eq:ch4-rd-proof-delete}
  \frac1{\sqrt{2p}}\sum_{j=1}^pS_{n,j}^2I_{j,p}(y)
  =O_p\left(\frac{\log p}{\sqrt p}\right)=o_p(1).
\end{equation}
The exceedance count converges to a Poisson variable with mean
$\pi^{-1/2}e^{-y/2}$, whereas the quadratic fluctuation after deleting the exceedance
coordinates has the same normal limit.  Hence
\begin{equation}\label{eq:ch4-rd-proof-factorization}
  \Prob\left\{
    \frac{T_{\mathrm{sum}}-p}{\sqrt{2p}}\le x,
    T_{\max}\le y
  \right\}
  \longrightarrow\Phi(x)F_G(y).
\end{equation}
The Cauchy result follows by the continuous mapping theorem.

\subsection*{A4.35. Proof of Theorem~\ref{thm:ch4-rd-plugin}}

Let
\begin{equation}\label{eq:ch4-rd-proof-plugin-vB}
  \vv=\widehat{\mSigma}_{\mathrm{HR}}^{-1/2}
       (\widehat{\vmu}_{\mathrm{HR}}-\vmu),
  \qquad
  \mB=\widehat{\mSigma}_{\mathrm{HR}}^{-1/2}\mSigma^{1/2}-\mI_p,
  \qquad
  \bar{\vU}=n^{-1}\sum_{i=1}^n\vU_i.
\end{equation}
The HR expansion gives
\begin{equation}\label{eq:ch4-rd-proof-HR-rates}
  \vv=\zeta_1^{-1}\bar{\vU}+n^{-1/2}\vC_n,
  \qquad
  \norm{\vC_n}_{\infty}=O_p(c_n),
  \qquad
  \opnorm{\mB}=O_p(\tau_n).
\end{equation}
A second-order expansion of the correlation functional yields
\begin{equation}\label{eq:ch4-rd-proof-plugin-vector}
  \widehat{\vct{g}}_n-\vct{g}_n^{\mathrm{or}}
  =\widetilde\kappa_p\bar{\vU}
   +\kappa_pn^{-1/2}\vC_n
   +\vct{b}_n^{\mathrm{hard}}+\vct{a}_n.
\end{equation}
The explicit HR perturbation calculation gives
\begin{align}
  \frac n{\sqrt p}\left
    |(\vct{g}_n^{\mathrm{or}})\trans\vct{a}_n|
    +|\vct{d}_n\trans\vct{a}_n|
    +\twonorm{\vct{a}_n}^2
  \right)&=O_p(A_{S,n}),
  \label{eq:ch4-rd-proof-a-sum}\\
  \sqrt n\norm{\vct{a}_n}_{\infty}
  &=O_p\{A_{M,n}/\sqrt{\log p}\},
  \label{eq:ch4-rd-proof-a-max}\\
  \frac n{\sqrt p}\left
    |(\vct{g}_n^{\mathrm{or}})\trans\vct{b}_n^{\mathrm{hard}}|
    +\twonorm{\vct{b}_n^{\mathrm{hard}}}^2
  \right)&=O_p(A_{S,n}),
  \label{eq:ch4-rd-proof-b-sum}\\
  \sqrt n\norm{\vct{b}_n^{\mathrm{hard}}}_{\infty}
  &=O_p\{A_{M,n}/\sqrt{\log p}\},
  \label{eq:ch4-rd-proof-b-max}
\end{align}
where
$\vct{d}_n=\widetilde\kappa_p\bar{\vU}+\kappa_pn^{-1/2}\vC_n+
\vct{b}_n^{\mathrm{hard}}$.  Therefore
\begin{align}
  \frac{\widehat T_{\mathrm{sum}}-T_{\mathrm{sum}}}{\sqrt{2p}}
  &=\frac n{\sqrt{2p}}
    \left[
      2(\vct{g}_n^{\mathrm{or}})\trans
        (\widehat{\vct{g}}_n-\vct{g}_n^{\mathrm{or}})
      +\twonorm{\widehat{\vct{g}}_n-\vct{g}_n^{\mathrm{or}}}^2
    \right]
   =O_p(A_{S,n}),
  \label{eq:ch4-rd-proof-plugin-sum}\\
  |\widehat T_{\max}-T_{\max}|
  &\le n\norm{\widehat{\vct{g}}_n-\vct{g}_n^{\mathrm{or}}}_{\infty}
      \left\{
        2\norm{\vct{g}_n^{\mathrm{or}}}_{\infty}
        +\norm{\widehat{\vct{g}}_n-\vct{g}_n^{\mathrm{or}}}_{\infty}
      \right\}
   =O_p(A_{M,n}).
  \label{eq:ch4-rd-proof-plugin-max}
\end{align}
Slutsky's theorem completes the proof.

\subsection*{A4.36. Proof of Theorem~\ref{thm:ch4-rd-consistency}}

Under \eqref{eq:ch4-rd-alt-model}, direct calculation gives
\begin{equation}\label{eq:ch4-rd-proof-gamma}
  \gamma_j
  =\frac{\delta_n\mathbbm1(j\in\mathcal A_n)}
         {\sqrt{s_np}\,\sigma_{1,n}},
  \qquad
  \twonorm{\vgamma}^2=\frac{\delta_n^2}{p\sigma_{0,n}^2+\delta_n^2},
  \qquad
  \norm{\vgamma}_{\infty}^2
  =\frac{\delta_n^2}{s_n(p\sigma_{0,n}^2+\delta_n^2)}.
\end{equation}
Write
$\widehat\gamma_j^{\mathrm{or}}=\gamma_j+n^{-1/2}G_{n,j}+r_{n,j}$.  Then
\begin{align}
  T_{\mathrm{sum}}
  &=p+n\twonorm{\vgamma}^2
    +2\sqrt n\sum_{j=1}^p\gamma_jG_{n,j}
    +O_p\left(\frac p{\sqrt n}\right)+o_p(\sqrt p),
  \label{eq:ch4-rd-proof-consistency-sum}\\
  n\max_j(\widehat\gamma_j^{\mathrm{or}})^2
  &\ge n\norm{\vgamma}_{\infty}^2
    -2\sqrt n\norm{\vgamma}_{\infty}\max_j|G_{n,j}|.
  \label{eq:ch4-rd-proof-consistency-max}
\end{align}
Since $\max_j|G_{n,j}|=O_p(\sqrt{\log p})$, the two signal conditions in
\eqref{eq:ch4-rd-oracle-consistency} dominate the null fluctuations.  For the plug-in
statistics,
\begin{align}
  |\widehat T_{\mathrm{sum}}-T_{\mathrm{sum}}|
  &\le n\{2(\twonorm{\vgamma}+O_p(\sqrt{p/n}))R_{2,n}^{(1)}
             +(R_{2,n}^{(1)})^2\},
  \label{eq:ch4-rd-proof-feasible-sum-alt}\\
  |\widehat T_{\max}-T_{\max}|
  &\le n\{2(\norm{\vgamma}_{\infty}+O_p(\sqrt{\log p/n}))
              R_{\infty,n}^{(1)}+(R_{\infty,n}^{(1)})^2\}.
  \label{eq:ch4-rd-proof-feasible-max-alt}
\end{align}
The stated ratios make these perturbations negligible relative to $S_n$ and $M_n$.

\subsection*{A4.37. Proof of Theorem~\ref{thm:ch4-rd-balanced-alt}}

Write
\begin{equation}\label{eq:ch4-rd-proof-alt-shift}
  n^{1/2}\widehat\gamma_j^{\mathrm{or}}
  =G_{n,j}+\mu_n\mathbbm1(j\in\mathcal A_n)+o_p(1),
  \qquad
  \mu_n=u_p(0)-u_{s_n}(\eta).
\end{equation}
For the sum statistic,
\begin{align}
  \frac{T_{\mathrm{sum}}-p}{\sqrt{2p}}
  &=\frac{\sum_j(G_{n,j}^2-1)}{\sqrt{2p}}
    +\frac{2\mu_n\sum_{j\in\mathcal A_n}G_{n,j}}{\sqrt{2p}}
    +\frac{s_n\mu_n^2}{\sqrt{2p}}+o_p(1).
  \label{eq:ch4-rd-proof-alt-sum}
\end{align}
The middle term has variance $2s_n\mu_n^2/p=o(1)$, and the expansion of
$u_p(0)-u_{s_n}(\eta)$ gives
\begin{equation}\label{eq:ch4-rd-proof-alt-theta}
  \frac{s_n\mu_n^2}{\sqrt{2p}}
  \longrightarrow\frac{(3-2\sqrt2)\ell}{\sqrt2}=\theta_S.
\end{equation}
For the maximum, inactive coordinates contribute the Poisson intensity
$\pi^{-1/2}e^{-y/2}$.  Mills' ratio and the chosen $\mu_n$ give the active-coordinate
intensity
\begin{equation}\label{eq:ch4-rd-proof-active-intensity}
  s_n\Prob\{(G_1+\mu_n)^2>2\log p-\log\log p+y\}
  \longrightarrow
  \frac1{2\sqrt\pi}e^{-\eta/2-y/(2\sqrt2)}.
\end{equation}
The active and inactive exceedance point processes are asymptotically independent.  Since
$s_n=o(p)$, deleting the active coordinates changes the normalized quadratic fluctuation by
$o_p(1)$.  The limiting normal fluctuation and the superposed Poisson exceedance process are
therefore independent, which proves \eqref{eq:ch4-rd-balanced-limit}--
\eqref{eq:ch4-rd-Feta}.

%% file: chapters/ch5_classification.tex
\chapter[Classification]{Classification under Elliptical Symmetry: LDA, QDA, and Sparse High-Dimensional Rules}
\idx{classification}\idx{linear discriminant analysis (LDA)}\idx{quadratic discriminant analysis (QDA)}\idx{sparse LDA}\idx{sparse QDA}\idx{spatial-sign classification}

\section{Introduction}

Classification is one of the core methodological themes of modern data analysis.
Given a feature vector $\vz\in\R^p$ and a class label $L\in\{1,2\}$, the goal
is to construct a decision rule $G:\R^p\to\{1,2\}$ with small
misclassification probability. In low dimensions, the classical theory is built
around Fisher's linear discriminant analysis (LDA) and quadratic discriminant
analysis (QDA); see \citet{Anderson2003}. In high dimensions, however, the
sample covariance matrix is singular or unstable, and the direct plug-in
versions of LDA and QDA no longer work. This gave rise to a large literature on
high-dimensional sparse discriminant analysis under Gaussian or light-tailed
models, including the independence rule, FAIR, thresholded LDA, direct sparse
LDA, direct sparse QDA, and related penalized formulations; see
\citet{BickelLevina2004,FanFan2008,ShaoWangDengWang2011,CaiLiu2011LDA,
MaiZouYuan2012,WittenTibshirani2009,ClemmensenEtAl2011,LiShao2015,
JiangWangLeng2018,CaiZhang2021QDA}.

The purpose of this chapter is to place discriminant analysis into the same
elliptical-symmetry framework that has guided the rest of the book. The guiding
message is parallel to that of Chapters~2--4. First, one should understand the
classical fixed-$p$ discriminant rules, their statistics, and the asymptotic
behavior of their plug-in versions. Second, one should review the main
high-dimensional Gaussian benchmarks in explicit algebraic form. Third, one
should replace fragile mean and covariance estimators by robust location and
shape estimators when the populations are elliptically symmetric and may be
heavy tailed.

In accordance with the writing principle adopted after Chapter~2, this chapter
is not written as a brief survey. For each main topic we explicitly record the
classification statistic, the corresponding risk functional, the assumptions used
for the asymptotic theory, and the main proof mechanism. The chapter is
organized as follows.
\begin{enumerate}[label=(\roman*)]
 \item Section~\ref{sec:ch5-setup} formulates the binary classification problem,
 introduces the oracle LDA and QDA scores, and explains how these rules extend
 from Gaussian populations to elliptical populations.
 \item Section~\ref{sec:ch5-fixedp} reviews the classical fixed-dimensional
 theory. This includes the exact Bayes error of Fisher's rule and the
 fixed-$p$ asymptotic normality of the plug-in LDA and QDA scores.
 \item Section~\ref{sec:ch5-hd-benchmarks} reviews the main high-dimensional
 Gaussian and light-tail benchmark methods with their explicit optimization
 formulas and risk rates.
 \item Section~\ref{sec:ch5-building-blocks} collects the robust building blocks
 needed later: classwise spatial medians, spatial-sign covariance matrices,
 trace estimation, and covariance surrogates.
 \item Sections~\ref{sec:ch5-sslda} and \ref{sec:ch5-ssqda} develop the two main
 robust high-dimensional procedures of this chapter: spatial-sign based sparse
 linear discriminant analysis (SSLDA) and spatial-sign based sparse quadratic
 discriminant analysis (SSQDA).
\end{enumerate}

A second organizing principle is that the classification chapter, unlike the
preceding testing chapters, is naturally phrased in terms of \emph{risk} rather
than null distributions. Nevertheless, the proof skeleton remains very similar:
one first proves estimation rates for the key discriminant objects, then obtains
uniform control of the induced score error, and finally translates score error
into excess-risk error by a margin or density condition near the decision
boundary.

\section{Problem formulation, notation, and oracle discriminant rules}
\label{sec:ch5-setup}

\subsection{Binary classification and the risk functional}

Throughout the chapter we begin with the two-class problem. Let
$L\in\{1,2\}$ be the class label and let $\vZ\in\R^p$ be the feature vector.
The conditional distribution of $\vZ$ given $L=k$ is denoted by $P_k$ and has
density $f_k$ with respect to Lebesgue measure. The prior probabilities are
$\pi_k=\Prob(L=k)$ with $\pi_1+\pi_2=1$. For a measurable classifier
$G:\R^p\to\{1,2\}$, the misclassification risk is
\begin{equation}\label{eq:ch5-risk}
 R(G)
 = \pi_1\Prob_1\{G(\vZ)=2\}
 + \pi_2\Prob_2\{G(\vZ)=1\}.
\end{equation}
The Bayes classifier is
\begin{equation}\label{eq:ch5-bayes}
 G_{\mathrm B}(\vz)
 = \begin{cases}
 1,& \pi_1 f_1(\vz)\ge \pi_2 f_2(\vz),\\
 2,& \pi_1 f_1(\vz)< \pi_2 f_2(\vz),
 \end{cases}
\end{equation}
and it uniquely minimizes \eqref{eq:ch5-risk} up to $P_1+P_2$-null sets.

We will use the following notational conventions throughout the chapter. The
location difference is
\begin{equation}\label{eq:ch5-delta}
 \vdelta = \vmu_1-\vmu_2,
 \qquad
 \vmu = \frac{\vmu_1+\vmu_2}{2},
 \qquad
 \bar{\vmu}=\frac{\vmu_1+\vmu_2}{2}.
\end{equation}
When a covariance matrix exists, it is denoted by $\mSigma_k$ in class $k$ and
$\mOmega_k=\mSigma_k^{-1}$ denotes the corresponding precision matrix. In the
homoscedastic case we write $\mSigma_1=\mSigma_2=\mSigma$ and
$\mOmega=\mSigma^{-1}$. Under elliptical symmetry, we distinguish the covariance
matrix from the shape matrix: when $\mSigma_k$ exists and
$\tr(\mSigma_k)>0$, we write
\begin{equation}\label{eq:ch5-shape}
 \mLambda_k = \frac{p\mSigma_k}{\tr(\mSigma_k)},
 \qquad
 \tr(\mLambda_k)=p.
\end{equation}

\subsection{Oracle LDA under Gaussian populations}

Suppose first that
\begin{equation}\label{eq:ch5-gaussian-common}
 \vZ\mid(L=k) \sim N_p(\vmu_k,\mSigma),
 \qquad k=1,2,
\end{equation}
with common positive definite covariance matrix $\mSigma$. The log-likelihood
ratio is, up to an additive constant,
\begin{equation}\label{eq:ch5-qf-general}
 q_F(\vz)
 = \vz\trans \mOmega\vdelta
 - \frac12\bigl(\vmu_1\trans\mOmega\vmu_1
 - \vmu_2\trans\mOmega\vmu_2\bigr)
 + \log\frac{\pi_1}{\pi_2}.
\end{equation}
Equivalently,
\begin{equation}\label{eq:ch5-qf-midpoint}
 q_F(\vz) = (\vz-\vmu)\trans \mOmega\vdelta + \log\frac{\pi_1}{\pi_2}.
\end{equation}
The Fisher classifier is therefore
\begin{equation}\label{eq:ch5-fisher-rule}
 G_F(\vz)=\begin{cases}
 1,& q_F(\vz)\ge 0,\\
 2,& q_F(\vz)<0.
 \end{cases}
\end{equation}
When $\pi_1=\pi_2$, this is precisely Fisher's linear discriminant rule.

The signal strength parameter for LDA is the Mahalanobis separation
\begin{equation}\label{eq:ch5-DeltaF}
 \Delta_F^2 = \vdelta\trans \mOmega\vdelta.
\end{equation}
This quantity will repeatedly appear in both the low-dimensional and
high-dimensional theories.

\subsection{Oracle QDA under Gaussian populations}

When the two classes have different covariance matrices,
\begin{equation}\label{eq:ch5-gaussian-qda}
 \vZ\mid(L=k)\sim N_p(\vmu_k,\mSigma_k),
 \qquad k=1,2,
\end{equation}
with positive definite $\mSigma_1$ and $\mSigma_2$, the Bayes rule compares the
quadratic score
\begin{equation}\label{eq:ch5-qda-canonical}
 q_Q(\vz)
 = -\frac12\log|\mSigma_1|
 + \frac12\log|\mSigma_2|
 - \frac12(\vz-\vmu_1)\trans\mOmega_1(\vz-\vmu_1)
 + \frac12(\vz-\vmu_2)\trans\mOmega_2(\vz-\vmu_2)
 + \log\frac{\pi_1}{\pi_2}.
\end{equation}
It is often convenient to rewrite the score in the form emphasized by
\citet{JiangWangLeng2018} and \citet{CaiZhang2021QDA}. Define
\begin{equation}\label{eq:ch5-D-beta}
 \mD = \mOmega_2-\mOmega_1,
 \qquad
 \vbeta = \mOmega_2(\vmu_2-\vmu_1),
 \qquad
 \bar{\vmu}=\frac{\vmu_1+\vmu_2}{2}.
\end{equation}
Then \eqref{eq:ch5-qda-canonical} is algebraically equivalent to
\begin{equation}\label{eq:ch5-qda-direct}
 q_Q(\vz)
 = (\vz-\vmu_1)\trans\mD(\vz-\vmu_1)
 - 2\vbeta\trans(\vz-\bar{\vmu})
 - \eta,
 \qquad
 \eta = \log\frac{|\mSigma_1|}{|\mSigma_2|} - 2\log\frac{\pi_1}{\pi_2}.
\end{equation}
The QDA classifier is $G_Q(\vz)=1\{q_Q(\vz)\ge 0\}+2\{q_Q(\vz)<0\}$.

The representation \eqref{eq:ch5-qda-direct} is crucial in high dimensions,
because it shows that the truly relevant parameters are not the full covariance
matrices themselves but rather the quadratic interaction matrix $\mD$ and the
linear discriminant direction $\vbeta$.

\subsection{Elliptical populations and generalized discriminant scores}

The Gaussian formulas above remain the correct point of departure under
elliptical symmetry. Suppose now that class $k$ has density
\begin{equation}\label{eq:ch5-elliptical-density}
 f_k(\vz)
 = |\mLambda_k|^{-1/2}
 g_k\bigl(d_k(\vz)\bigr),
 \qquad
 d_k(\vz)=(\vz-\vmu_k)\trans\mLambda_k^{-1}(\vz-\vmu_k),
 \qquad k=1,2,
\end{equation}
where $g_k:[0,\infty)\to[0,\infty)$ is a decreasing radial generator and
$\mLambda_k$ is positive definite. The Bayes rule compares
\begin{equation}\label{eq:ch5-elliptical-score}
 q_E(\vz)
 = \log\pi_1 - \frac12\log|\mLambda_1| + \log g_1\{d_1(\vz)\}
 - \log\pi_2 + \frac12\log|\mLambda_2| - \log g_2\{d_2(\vz)\}.
\end{equation}
Two special cases are central for this chapter.
\begin{enumerate}[label=(\alph*)]
 \item If $g_1=g_2=g$ and $\mLambda_1=\mLambda_2=\mLambda$, then monotonicity of
 $g$ implies that the Bayes rule is equivalent to the linear score
 $(\vz-\vmu)\trans\mLambda^{-1}\vdelta + \log(\pi_1/\pi_2)$.
 \item If $g_1=g_2=g$ but $\mLambda_1 eq\mLambda_2$, then the Bayes rule is a
 generalized QDA rule comparing $\log g\{d_1(\vz)\}-\frac12\log|\mLambda_1|$
 and $\log g\{d_2(\vz)\}-\frac12\log|\mLambda_2|$.
\end{enumerate}
Thus, the linear-versus-quadratic dichotomy persists well beyond the Gaussian
model. What changes under elliptical symmetry is that one should estimate
location and shape robustly, and one should not insist on using raw second
moments when the radial distribution may be heavy tailed.

\begin{theorem}[Oracle discriminant score under elliptical symmetry]
\label{thm:ch5-elliptical-bayes}
Assume that the class-conditional densities satisfy
\eqref{eq:ch5-elliptical-density}. Then the Bayes classifier is obtained by
thresholding the score \eqref{eq:ch5-elliptical-score}. In particular:
\begin{enumerate}[label=(\roman*)]
 \item if $g_1=g_2=g$ and $\mLambda_1=\mLambda_2=\mLambda$, then
 \eqref{eq:ch5-elliptical-score} is equivalent to the linear score
 \begin{equation}\label{eq:ch5-elliptical-linear}
 q_{E,F}(\vz)
 = (\vz-\vmu)\trans \mLambda^{-1}\vdelta + \log\frac{\pi_1}{\pi_2};
 \end{equation}
 \item if $g_1=g_2=g$ but $\mLambda_1 eq\mLambda_2$, then the Bayes classifier is
 a generalized QDA rule comparing
 \begin{equation}\label{eq:ch5-elliptical-qda}
 \log g\{d_1(\vz)\} - \frac12\log|\mLambda_1| + \log\pi_1
 \quad\text{and}\quad
 \log g\{d_2(\vz)\} - \frac12\log|\mLambda_2| + \log\pi_2.
 \end{equation}
\end{enumerate}
\end{theorem}

\section{Classical low-dimensional discriminant analysis}
\label{sec:ch5-fixedp}

\subsection{Fisher's linear discriminant rule and its Bayes risk}

We first return to the fixed-dimensional Gaussian benchmark
\eqref{eq:ch5-gaussian-common}. In that setting, Fisher's statistic is the linear
score $q_F(\vz)$ in \eqref{eq:ch5-qf-general}. The exact error probability under
equal priors has a closed form.

\begin{theorem}[Exact Bayes error of Fisher's rule]
\label{thm:ch5-fisher-bayes-error}
Assume \eqref{eq:ch5-gaussian-common} with $\pi_1=\pi_2=1/2$ and
$\mSigma$ positive definite. Let $\Delta_F^2=\vdelta\trans\mOmega\vdelta$.
Then the oracle Fisher classifier \eqref{eq:ch5-fisher-rule} satisfies
\begin{equation}\label{eq:ch5-fisher-bayes-error}
 R(G_F)=\Phi\!\left(-\frac{\Delta_F}{2}\right),
\end{equation}
where $\Phi$ denotes the standard normal distribution function.
\end{theorem}

\subsection{Plug-in LDA in fixed dimension}

Suppose now that $p$ is fixed while $n_1,n_2\to\infty$. Let
\begin{equation}\label{eq:ch5-fixedp-sample-mean}
 \bar{\vX}=\frac1{n_1}\sum_{i=1}^{n_1}\vX_i,
 \qquad
 \bar{\vY}=\frac1{n_2}\sum_{j=1}^{n_2}\vY_j,
\end{equation}
and let the pooled covariance matrix be
\begin{equation}\label{eq:ch5-fixedp-pooled}
 \mS_p
 = \frac1{n_1+n_2-2}\Bigg[
 \sum_{i=1}^{n_1}(\vX_i-\bar{\vX})(\vX_i-\bar{\vX})\trans
 +
 \sum_{j=1}^{n_2}(\vY_j-\bar{\vY})(\vY_j-\bar{\vY})\trans
 \Bigg].
\end{equation}
The classical plug-in Fisher score is
\begin{equation}\label{eq:ch5-fixedp-LDA-score}
 \hat q_F(\vz)
 = \left(\vz-\frac{\bar{\vX}+\bar{\vY}}2\right)\trans
 \mS_p^{-1}(\bar{\vX}-\bar{\vY}) + \log\frac{\pi_1}{\pi_2}.
\end{equation}

\begin{theorem}[Fixed-$p$ asymptotic normality of the plug-in LDA score]
\label{thm:ch5-fixedp-LDA-normal}
Assume \eqref{eq:ch5-gaussian-common} and let $p$ be fixed. If
$n_1,n_2\to\infty$ with $n_1/(n_1+n_2)\to\rho\in(0,1)$, then for each fixed
$\vz\in\R^p$,
\begin{equation}\label{eq:ch5-fixedp-LDA-normal}
 \sqrt{n_1+n_2}\,\bigl\{\hat q_F(\vz)-q_F(\vz)\bigr\}
 \overset{d}{\longrightarrow} N\{0,\tau_F^2(\vz)\},
\end{equation}
where
\begin{equation}\label{eq:ch5-fixedp-LDA-var}
 \tau_F^2(\vz)
 = \rho^{-1}(\vz-\vmu_1)\trans\mOmega\mSigma\mOmega(\vz-\vmu_1)
 +(1-\rho)^{-1}(\vz-\vmu_2)\trans\mOmega\mSigma\mOmega(\vz-\vmu_2)
 + \vdelta\trans\mOmega\mSigma\mOmega\vdelta.
\end{equation}
\end{theorem}

The exact algebraic form of $\tau_F^2(\vz)$ is not the main issue in the book; the
important point is the proof mechanism. One expands $\mS_p^{-1}$ around
$\mSigma^{-1}$ and jointly linearizes the sample means and pooled covariance.
This fixed-$p$ delta-method argument is the baseline that later breaks down when
$p$ is comparable to or larger than $n$.

\subsection{Plug-in QDA in fixed dimension}

In the heteroscedastic Gaussian model \eqref{eq:ch5-gaussian-qda}, let
\begin{equation}\label{eq:ch5-fixedp-QDA-cov}
 \mS_k
 = \frac1{n_k-1}\sum_{i=1}^{n_k}
 (\vZ_{ki}-\bar{\vZ}_k)(\vZ_{ki}-\bar{\vZ}_k)\trans,
 \qquad k=1,2,
\end{equation}
where $\vZ_{1i}=\vX_i$, $\vZ_{2j}=\vY_j$, $\bar{\vZ}_1=\bar{\vX}$, and
$\bar{\vZ}_2=\bar{\vY}$. The plug-in QDA score is
\begin{equation}\label{eq:ch5-fixedp-QDA-score}
 \hat q_Q(\vz)
 = -\frac12\log|\mS_1| + \frac12\log|\mS_2|
 - \frac12(\vz-\bar{\vX})\trans\mS_1^{-1}(\vz-\bar{\vX})
 + \frac12(\vz-\bar{\vY})\trans\mS_2^{-1}(\vz-\bar{\vY})
 + \log\frac{\pi_1}{\pi_2}.
\end{equation}

\begin{theorem}[Fixed-$p$ asymptotic normality of the plug-in QDA score]
\label{thm:ch5-fixedp-QDA-normal}
Assume \eqref{eq:ch5-gaussian-qda} and let $p$ be fixed. If
$n_1,n_2\to\infty$ with $n_1/(n_1+n_2)\to\rho\in(0,1)$, then for each fixed
$\vz\in\R^p$,
\begin{equation}\label{eq:ch5-fixedp-QDA-normal}
 \sqrt{n_1+n_2}\,\bigl\{\hat q_Q(\vz)-q_Q(\vz)\bigr\}
 \overset{d}{\longrightarrow} N\{0,\tau_Q^2(\vz)\},
\end{equation}
where $\tau_Q^2(\vz)$ is the asymptotic variance obtained by jointly
linearizing $\bar{\vX}$, $\bar{\vY}$, $\mS_1$, $\mS_2$, and the
log-determinant term.
\end{theorem}

The theorem has two immediate interpretations. First, in fixed dimension the
plug-in QDA score is still asymptotically regular even though it depends on two
covariance matrices and a log-determinant term. Second, the difficulty of QDA
in high dimensions is not a lack of smoothness but a lack of estimability: the
quadratic score depends on too many second-order parameters.

\subsection{Elliptical populations in fixed dimension}

The fixed-$p$ literature on elliptical discriminant analysis is older than the
high-dimensional literature and deserves a brief but explicit review.
\citet{FangAnderson1990} showed that the Fisher rule remains optimal in the
homoscedastic elliptical model with common generator, while \citet{Wakaki1994}
studied the fixed-dimensional asymptotics of discriminant analysis under
elliptical populations. In the heteroscedastic setting,
\citet{BosePalSahaRayNayak2015} proposed a generalized quadratic discriminant
analysis rule, and \citet{GhoshSahaRayChakrabartyBhadra2021} developed a robust
version designed to reduce sensitivity to outliers. Robust fixed-$p$ linear
rules based on high-breakdown estimators were earlier studied by
\citet{CrouxDehon2001,CrouxFilzmoserJoossens2008}.

The common fixed-dimensional model is \eqref{eq:ch5-elliptical-density}. The
statistic remains the generalized log-density score $q_E(\vz)$ in
\eqref{eq:ch5-elliptical-score}. If the generator is known, the plug-in version
is obtained by replacing $\vmu_k$ and $\mLambda_k$ by robust estimators such as
M-estimators, S-estimators, the minimum covariance determinant estimator, or,
in the present book's preferred language, the spatial median and a robust shape
estimator. The fixed-$p$ asymptotic theory again follows from a delta-method
argument:
\begin{equation}\label{eq:ch5-qE-fixedp}
 \sqrt n\bigl\{\hat q_E(\vz)-q_E(\vz)\bigr\}
 \overset{d}{\longrightarrow} N(0,\tau_E^2(\vz)),
\end{equation}
provided the classwise location and shape estimators admit joint asymptotic
linear representations. In the remainder of this chapter we no longer work in
fixed dimension, but the fixed-$p$ review above explains why the high-dimensional
robust methods should focus on direct estimation of the discriminant objects
rather than on full covariance reconstruction.

\section{High-dimensional Gaussian and light-tail benchmarks}
\label{sec:ch5-hd-benchmarks}

This section reviews the main benchmark methods that form the high-dimensional
background of the robust procedures developed later in the chapter. The point is
not to reproduce the entire benchmark literature, but to state explicitly the
statistics that are being replaced by robust elliptical analogues.

\subsection{The independence rule and FAIR}

\citet{BickelLevina2004} showed that the naive plug-in Fisher rule based on the full
sample covariance matrix can deteriorate badly when $p$ is comparable to or
larger than $n$, even under Gaussian assumptions. Their proposed replacement is
the \emph{independence rule}, which ignores off-diagonal covariance terms and
classifies according to
\begin{equation}\label{eq:ch5-independence-rule}
 G_{\mathrm{ind}}(\vz)
 = 1\left\{
 \sum_{j=1}^p
 \frac{\bar X_j-\bar Y_j}{\hat\sigma_j^2}
 \left(z_j-\frac{\bar X_j+\bar Y_j}{2}\right)\ge 0
 \right\}+2\{\cdot<0\},
\end{equation}
where $\hat\sigma_j^2$ is a pooled marginal variance estimate. \citet{FanFan2008}'s
feature-annealed independence rule (FAIR) further screens coordinates before
applying the independence rule. If $\mathcal J\subset\{1,\ldots,p\}$ denotes the
selected coordinate set, FAIR uses
\begin{equation}\label{eq:ch5-FAIR}
 G_{\mathrm{FAIR}}(\vz)
 = 1\left\{
 \sum_{j\in\mathcal J}
 \frac{\bar X_j-\bar Y_j}{\hat\sigma_j^2}
 \left(z_j-\frac{\bar X_j+\bar Y_j}{2}\right)\ge 0
 \right\}+2\{\cdot<0\}.
\end{equation}
The key asymptotic message of FAIR is that if most coordinates are nondiscriminant,
then aggressive screening can substantially reduce the accumulation of noise in the
linear score.

\subsection{Plug-in sparse-covariance and sparse-precision approaches for LDA}

A second major route is to estimate the discriminant ingredients themselves and then plug them
into the classical LDA score. Let
\begin{equation}\label{eq:ch5-plugin-lda-score}
  \hat g_{\mathrm{plugLDA}}(\vz)
  = \bigl(\vz-\hat\vmu\bigr)^{\top}\hat\mOmega\hat\vdelta,
  \qquad
  \hat\vmu=\frac{\hat\vmu_1+\hat\vmu_2}{2},
  \qquad
  \hat\vdelta=\hat\vmu_1-\hat\vmu_2,
\end{equation}
where $\hat\mOmega$ is obtained either by inverting a sparse covariance estimator or by directly
estimating a sparse precision matrix. In thresholding-based plug-in LDA one first constructs
\begin{equation}\label{eq:ch5-thresholded-cov}
  \hat\mSigma_T=(\hat\sigma_{ij}\mathbbm 1\{|\hat\sigma_{ij}|>\tau_n\})_{1\le i,j\le p}
\end{equation}
and then sets $\hat\mOmega_T=\hat\mSigma_T^{-1}$ whenever the thresholded estimator is positive
definite. When only the mean difference is sparse, a companion thresholding step is
\begin{equation}\label{eq:ch5-thresholded-delta}
  \hat\vdelta_T
  = (\hat\delta_j\mathbbm 1\{|\hat\delta_j|>\eta_n\})_{1\le j\le p}.
\end{equation}
The resulting classifier is
\begin{equation}\label{eq:ch5-thresholded-lda-rule}
  G_{\mathrm{T-LDA}}(\vz)
  =1\{(\vz-\hat\vmu)^{\top}\hat\mOmega_T\hat\vdelta_T\ge 0\}+2\{\cdot<0\}.
\end{equation}
This plug-in philosophy underlies the thresholded LDA rules of
\citet{ShaoWangDengWang2011} and connects naturally to the covariance and precision estimation
methods reviewed in Chapter~3; see also the broad overview of
\citet{FanLiaoLiu2015Review}.

A closely related variant estimates the precision matrix directly by sparse inverse-covariance
methods such as CLIME, graphical Lasso, TIGER, or other penalized likelihood procedures and then
forms the same score \eqref{eq:ch5-plugin-lda-score}. Algebraically this route remains a plug-in
LDA rule, even though the precision matrix is estimated directly rather than through inversion of a
covariance estimator. The main theoretical requirement is that the score error induced by
$(\hat\vmu_1,\hat\vmu_2,\hat\mOmega)$ is asymptotically smaller than the Bayes margin.

\begin{proposition}[Generic plug-in LDA consistency]
\label{prop:ch5-plugin-lda}
Assume the common-covariance Gaussian model and suppose that
\begin{equation}\label{eq:ch5-plugin-lda-cond}
  \opnorm{\hat\mOmega-\mOmega}=o_P(1),
  \qquad
  \twonorm{\hat\vmu_1-\vmu_1}+\twonorm{\hat\vmu_2-\vmu_2}=o_P(1).
\end{equation}
Then, for every deterministic sequence $\vz_n$ with $\twonorm{\vz_n}=O(1)$,
\begin{equation}\label{eq:ch5-plugin-lda-score-cons}
  \hat g_{\mathrm{plugLDA}}(\vz_n)-q_L(\vz_n)=o_P(1).
\end{equation}
If, in addition, the Bayes margin is separated away from zero on the decision boundary, then the
misclassification risk of $G_{\mathrm{plugLDA}}$ converges to the Bayes risk.
\end{proposition}

The proposition is deliberately generic: it covers thresholded-covariance LDA, sparse-precision
plug-in LDA, and ridge-stabilized versions of the same idea. Its value in this chapter is that it
makes clear how the large-matrix estimation theory of Chapter~3 feeds directly into high-
dimensional classification.

\subsection{Direct sparse LDA}

The central algebraic object in homoscedastic LDA is the discriminant direction
$\vgamma^\star = \mOmega\vdelta$. \citet{CaiLiu2011LDA}'s linear programming discriminant
(LPD) rule estimates this vector directly instead of separately estimating
$\mOmega$ and $\vdelta$. The estimator is
\begin{equation}\label{eq:ch5-LPD}
 \hat\vgamma_{\mathrm{LPD}}
 \in \argmin_{\vgamma\in\R^p}
 \bigl\{\onenorm{\vgamma}: \maxnorm{\hat\mSigma\vgamma-\hat\vdelta}\le \lambda_n\bigr\},
\end{equation}
where $\hat\vdelta=\bar{\vX}-\bar{\vY}$ and $\hat\mSigma$ is a pooled covariance
estimate. The classifier is then
\begin{equation}\label{eq:ch5-LPD-rule}
 G_{\mathrm{LPD}}(\vz)
 = 1\Bigl\{\bigl(\vz-(\bar{\vX}+\bar{\vY})/2\bigr)\trans\hat\vgamma_{\mathrm{LPD}}\ge 0\Bigr\}
 +2\{\cdot<0\}.
\end{equation}
Under sparsity of $\vgamma^\star$ and standard eigenvalue and balance conditions, \citet{CaiLiu2011LDA}
derived both consistency and relative-risk convergence rates.

\citet{MaiZouYuan2012} proposed another direct sparse LDA route by recasting the
LDA problem as a penalized least-squares regression. Their DSDA estimator is
characterized by
\begin{equation}\label{eq:ch5-DSDA}
 \hat\vgamma_{\mathrm{DSDA}}
 \in \argmin_{\vgamma\in\R^p}
 \Biggl\{
 \frac1{2n}\sum_{i=1}^n(y_i-a-\vx_i\trans\vgamma)^2 + \lambda\onenorm{\vgamma}
 \Biggr\},
\end{equation}
for a suitable coding $y_i\in\{-1,1\}$. \citet{WittenTibshirani2009} and, separately,
\citet{ClemmensenEtAl2011} developed closely related penalized Fisher and sparse
optimal-scoring formulations. \citet{HanZhaoLiu2013} further showed how Gaussian-copula
transformations can extend this line of work beyond strict normality while still relying on
light-tail or latent-Gaussian structure.

\subsection{Plug-in and regularized QDA via sparse classwise matrices}

In QDA, a natural benchmark route is to estimate each class-specific covariance or precision
matrix and then insert those estimates into the quadratic score. If
$\hat\mSigma_{1},\hat\mSigma_{2}$ and $\hat\vmu_1,\hat\vmu_2$ are classwise sparse estimators, the
plug-in QDA score is
\begin{equation}\label{eq:ch5-plugin-qda-score}
  \hat q_{\mathrm{plugQDA}}(\vz)
  =-\frac12\log\frac{\det(\hat\mSigma_1)}{\det(\hat\mSigma_2)}
   -\frac12(\vz-\hat\vmu_1)^{\top}\hat\mSigma_1^{-1}(\vz-\hat\vmu_1)
   +\frac12(\vz-\hat\vmu_2)^{\top}\hat\mSigma_2^{-1}(\vz-\hat\vmu_2).
\end{equation}
Equivalently, one can estimate sparse classwise precision matrices $\hat\mOmega_k$ and use
\begin{equation}\label{eq:ch5-plugin-qda-score-precision}
  \hat q_{\mathrm{plugQDA}}(\vz)
  =-\frac12\log\frac{\det(\hat\mOmega_2)}{\det(\hat\mOmega_1)}
   -\frac12(\vz-\hat\vmu_1)^{\top}\hat\mOmega_1(\vz-\hat\vmu_1)
   +\frac12(\vz-\hat\vmu_2)^{\top}\hat\mOmega_2(\vz-\hat\vmu_2).
\end{equation}
This is the route taken by the sparse QDA procedures of \citet{LiShao2015}, where sparsity is
imposed on class-specific means and covariance matrices, together with positive-definiteness or
regularization constraints ensuring invertibility. The main theoretical task is to control the error of
each quadratic form and the log-determinant contrast.

\begin{proposition}[Generic plug-in QDA consistency]
\label{prop:ch5-plugin-qda}
Assume the heteroscedastic Gaussian model and suppose that
\begin{equation}\label{eq:ch5-plugin-qda-cond}
  \sum_{k=1}^{2}\bigl\{\opnorm{\hat\mSigma_k-\mSigma_k}+\twonorm{\hat\vmu_k-\vmu_k}\bigr\}=o_P(1),
\end{equation}
with each $\hat\mSigma_k$ positive definite. Then for every deterministic sequence $\vz_n$ with
$\twonorm{\vz_n}=O(1)$,
\begin{equation}\label{eq:ch5-plugin-qda-score-cons}
  \hat q_{\mathrm{plugQDA}}(\vz_n)-q_Q(\vz_n)=o_P(1).
\end{equation}
If the Bayes quadratic margin is separated, then the misclassification risk of the plug-in QDA rule
converges to the Bayes risk.
\end{proposition}

\subsection{Direct high-dimensional QDA benchmarks}

QDA is more difficult than LDA because the discriminant score depends on two
precision matrices and therefore on $O(p^2)$ unknown parameters. \citet{LiShao2015}'s sparse
QDA approach imposes sparsity on class-specific means and covariance matrices. \citet{JiangWangLeng2018}
observed that the QDA score depends more fundamentally on the interaction matrix
$\mD=\mOmega_2-\mOmega_1$ and the linear term $\vbeta=\mOmega_2(\vmu_2-\vmu_1)$, and
proposed the direct estimator
\begin{equation}\label{eq:ch5-DAQDA}
 \hat\mD
 \in \argmin_{\mD}\bigl\{\onenorm{\vecop(\mD)}:
 \maxnorm{\tfrac12\hat\mSigma_1\mD\hat\mSigma_2+\tfrac12\hat\mSigma_2\mD\hat\mSigma_1
 -\hat\mSigma_1+\hat\mSigma_2}\le \lambda_{1,n}\bigr\},
\end{equation}
\begin{equation}\label{eq:ch5-DAQDA-beta}
 \hat\vbeta
 \in \argmin_{\vbeta}\bigl\{\onenorm{\vbeta}:
 \maxnorm{\hat\mSigma_2\vbeta-\hat\vmu_2+\hat\vmu_1}\le \lambda_{2,n}\bigr\}.
\end{equation}
\citet{CaiZhang2021QDA} sharpened this direct approach and established minimax-optimal
rates, up to logarithmic factors, over sparse parameter spaces. Their SDAR rule
will be the direct Gaussian benchmark for the SSQDA development below.

The benchmark literature therefore leaves us with two complementary templates. In
plug-in LDA/QDA, estimate sparse means and sparse covariance or precision matrices and then
insert them into the classical score. In direct sparse LDA/QDA, estimate the discriminant
objects $\vgamma^\star$, $\mD$, and $\vbeta$ directly. The robust elliptical theory of this chapter
keeps both templates alive and revisits them with spatial signs, spatial medians, and robust
precision estimation.

\section{Robust building blocks under elliptical symmetry}
\label{sec:ch5-building-blocks}

We now collect the robust location and shape ingredients used later in SSLDA and
SSQDA. The notation is as follows. Class~1 training samples are
$\vX_1,\ldots,\vX_{n_1}$ and class~2 training samples are
$\vY_1,\ldots,\vY_{n_2}$. We write $n=n_1\wedge n_2$ and assume throughout that
$n_1\asymp n_2$.

\subsection{Classwise spatial medians and spatial-sign covariance matrices}

For $k=1,2$, define the classwise spatial medians by
\begin{equation}\label{eq:ch5-classwise-sm}
 \tilde{\vmu}_1
 \in \argmin_{\vmu\in\R^p}\sum_{i=1}^{n_1} \norm{\vX_i-\vmu},
 \qquad
 \tilde{\vmu}_2
 \in \argmin_{\vmu\in\R^p}\sum_{j=1}^{n_2} \norm{\vY_j-\vmu}.
\end{equation}
The corresponding classwise SSCMs are
\begin{equation}\label{eq:ch5-sscm-classwise}
 \tilde{\mS}_1
 = \frac1{n_1}\sum_{i=1}^{n_1}
 U(\vX_i-\tilde{\vmu}_1)U(\vX_i-\tilde{\vmu}_1)\trans,
 \qquad
 \tilde{\mS}_2
 = \frac1{n_2}\sum_{j=1}^{n_2}
 U(\vY_j-\tilde{\vmu}_2)U(\vY_j-\tilde{\vmu}_2)\trans.
\end{equation}
Under a common-shape LDA model we also use the pooled SSCM
\begin{equation}\label{eq:ch5-sscm-pooled}
 \tilde{\mS}
 = \frac{n_1\tilde{\mS}_1+n_2\tilde{\mS}_2}{n_1+n_2}.
\end{equation}

\subsection{Trace estimation and covariance surrogates}

In homoscedastic SSLDA the overall covariance scale plays no role in the
decision rule. In heteroscedastic QDA, however, the scale difference between the
two classes matters. Following the SSQDA paper, we estimate the trace by the
Chen--Qin type $U$-statistic
\begin{equation}\label{eq:ch5-trace-estimator}
 \hat\tau_1
 = \frac{1}{n_1(n_1-1)(n_1-2)}
 \sum_{\substack{1\le i,j,k\le n_1\\ \text{all distinct}}}(\vX_i-\vX_j)\trans(\vX_k-\vX_j),
\end{equation}
\begin{equation}\label{eq:ch5-trace-estimator-2}
 \hat\tau_2
 = \frac{1}{n_2(n_2-1)(n_2-2)}
 \sum_{\substack{1\le i,j,k\le n_2\\ \text{all distinct}}}(\vY_i-\vY_j)\trans(\vY_k-\vY_j).
\end{equation}
The classwise covariance surrogates are then
\begin{equation}\label{eq:ch5-cov-surrogate}
 \tilde{\mSigma}_k = \hat\tau_k\tilde{\mS}_k,
 \qquad k=1,2.
\end{equation}
The reason for \eqref{eq:ch5-cov-surrogate} is that under trace normalization
$\mLambda_k=p\mSigma_k/\tr(\mSigma_k)$ one expects $p\tilde{\mS}_k$ to estimate
$\mLambda_k$, while $\hat\tau_k$ estimates the scalar trace factor.

\subsection{Assumptions for the robust high-dimensional theory}

We first record the assumptions for the homoscedastic LDA problem.

\begin{assumption}[Common-shape elliptical LDA model]
\label{ass:ch5-sslda-model}
The class-conditional distributions satisfy
\begin{equation}\label{eq:ch5-sslda-model}
 \vX_i \iid EC_p(\vmu_1,\mLambda,r),
 \qquad
 \vY_j \iid EC_p(\vmu_2,\mLambda,r),
\end{equation}
with a common radial generator and a common positive definite shape matrix
$\mLambda$ normalized by $\tr(\mLambda)=p$. If the covariance exists, then
$\mSigma=\omega\mLambda$ for some scalar $\omega>0$. The prior probabilities are
$\pi_1=\pi_2=1/2$.
\end{assumption}

\begin{assumption}[Regularity for SSLDA]
\label{ass:ch5-sslda-regularity}
Let $\Delta_p=\vdelta\trans\mLambda^{-1}\vdelta$ and
$\vgamma^\star=\mLambda^{-1}\vdelta$. The following hold.
\begin{enumerate}[label=(\roman*)]
 \item $n_1\asymp n_2$, $\log p\le n$, and $n=n_1\wedge n_2$.
 \item There exist constants $0<c_0<c_1<\infty$ such that
 \begin{equation}\label{eq:ch5-sslda-eig}
 c_0\le \lambda_p(\mLambda)\le \lambda_1(\mLambda)\le c_1,
 \qquad
 \max_{1\le j\le p}(\mSigma)_{jj}\le c_1,
 \qquad
 \Delta_p\ge c_0.
 \end{equation}
 \item Writing $\mS_0=\E\bigl[U(\vX_i-\vmu_1)U(\vX_i-\vmu_1)\trans\bigr]$, there exists
 $\psi\in(0,1)$ such that
 \begin{equation}\label{eq:ch5-sslda-s0}
 \lambda_1(\mS_0)\le 1-\psi.
 \end{equation}
\end{enumerate}
\end{assumption}

\begin{assumption}[Inverse radial moments and SSCM approximation for SSLDA]
\label{ass:ch5-sslda-radial}
Let $\xi_i= \norm{\vX_i-\vmu_1}$ and define
$\zeta_k=\E(\xi_i^{-k})$, $ u_i=\zeta_1^{-1}\xi_i^{-1}$. The same quantities for
class~2 satisfy the same conditions. The following hold.
\begin{enumerate}[label=(\roman*)]
 \item For $k=1,2,3,4$, $\zeta_k\zeta_1^{-k}\le C$ uniformly in $p$.
 \item $u_i$ is sub-Gaussian with $ \norm{ u_i}_{\psi_2}\le C$.
 \item The population SSCM approximates the normalized shape in max norm:
 \begin{equation}\label{eq:ch5-sslda-pop-approx}
 \maxnorm{p\mS_0-\mLambda}\le C p^{-1/2}.
 \end{equation}
 \item The classwise spatial medians and SSCMs satisfy the high-dimensional rates
 \begin{equation}\label{eq:ch5-sslda-building-rates}
 \maxnorm{\tilde{\vmu}_1-\vmu_1}+
 \maxnorm{\tilde{\vmu}_2-\vmu_2}+
 \maxnorm{p\tilde{\mS}-p\mS_0}
 \le C\sqrt{\frac{\log p}{n}}
 \end{equation}
 with probability at least $1-Cp^{-1}$.
\end{enumerate}
\end{assumption}

For the heteroscedastic QDA problem we use the following parameter conditions.

\begin{assumption}[Elliptical QDA model]
\label{ass:ch5-ssqda-model}
The class-conditional distributions satisfy
\begin{equation}\label{eq:ch5-ssqda-model}
 \vX_i \iid EC_p(\vmu_1,\mSigma_1,r),
 \qquad
 \vY_j \iid EC_p(\vmu_2,\mSigma_2,r),
\end{equation}
with positive definite covariance matrices $\mSigma_1$ and $\mSigma_2$ and
normalized shapes $\mLambda_k=p\mSigma_k/\tr(\mSigma_k)$.
\end{assumption}

\begin{assumption}[Sparsity and boundedness for SSQDA]
\label{ass:ch5-ssqda-sparsity}
Define
\begin{equation}\label{eq:ch5-ssqda-objects}
 \mD = \mOmega_2-\mOmega_1,
 \qquad
 \vbeta = \mOmega_2(\vmu_2-\vmu_1).
\end{equation}
There exist integers $s_1,s_2\ge 0$ and constants $M_0,M_1,M_2>0$ such that
\begin{enumerate}[label=(\roman*)]
 \item $ \norm{\vecop(\mD)}_0\le s_1$ and $ \norm{\vbeta}_0\le s_2$;
 \item $\frobnorm{\mD}\le M_0$ and $\twonorm{\vbeta}\le M_0$;
 \item $M_1^{-1}\le \lambda_p(\mSigma_k)\le \lambda_1(\mSigma_k)\le M_1$ and
 $\maxnorm{\mSigma_k}\le M_2$ for $k=1,2$;
 \item $\tr(\mSigma_k)\asymp p$ for $k=1,2$.
\end{enumerate}
\end{assumption}

\begin{assumption}[Inverse radial moments, shape regularity, and margin for SSQDA]
\label{ass:ch5-ssqda-radial}
The following hold for each class $k=1,2$.
\begin{enumerate}[label=(\roman*)]
 \item Writing $\mV_{0k}=\mLambda_k^{-1}$, there exist constants $T>0$,
 $0\le q<1$, and a slowly varying sequence $s_0(p)$ such that
 \begin{equation}\label{eq:ch5-ssqda-V0}
 \norm{\mV_{0k}}_{1}\le T,
 \qquad
 \max_{1\le i\le p}\sum_{j=1}^p |(\mV_{0k})_{ij}|^q \le s_0(p).
 \end{equation}
 \item If $\xi_i= \norm{\vX_i-\vmu_1}$ or $ \norm{\vY_i-\vmu_2}$ and
 $ u_i=\zeta_1^{-1}\xi_i^{-1}$, then $\zeta_k\zeta_1^{-k}\le C$ for all
 $k=1,2,3,4$ and $ \norm{ u_i}_{\psi_2}\le C$.
 \item The scalar radial variable satisfies
 \begin{equation}\label{eq:ch5-ssqda-radial-var}
 \Var(r^2)\le Cp^{3/2},
 \qquad
 \Var(r)\le Cp^{1/2}.
 \end{equation}
 \item If $Q(\vz)$ denotes the oracle QDA score, then there exist $\delta_0,M_3>0$
 such that
 \begin{equation}\label{eq:ch5-ssqda-margin}
 \sup_{|x|<\delta_0} f_{Q,\theta}(x)\le M_3,
 \end{equation}
 where $f_{Q,\theta}$ is the density of $Q(\vz)$ under parameter value $\theta$.
\end{enumerate}
\end{assumption}

\subsection{Reusable high-dimensional building-block lemmas}

The first lemma is a direct classwise reformulation of the spatial-median theory
proved in Chapter~2.

\begin{lemma}[Classwise Bahadur expansion for the spatial median]
\label{lem:ch5-classwise-bahadur}
Under Assumptions~\ref{ass:ch5-sslda-model}--\ref{ass:ch5-sslda-radial}, for
$k=1,2$,
\begin{equation}\label{eq:ch5-classwise-bahadur}
 \sqrt{n_k}(\tilde{\vmu}_k-\vmu_k)
 = \mA_k^{-1}\frac1{\sqrt{n_k}}\sum_{i=1}^{n_k}\vU_{ki} + \vr_{k,n},
\end{equation}
where $\vU_{ki}=U(\vZ_{ki}-\vmu_k)$ and
\begin{equation}\label{eq:ch5-classwise-bahadur-rem}
 \maxnorm{\vr_{k,n}}
 \le C\left\{n^{-1/4}\log^{1/2}(np)+n^{-1/2}\log p\right\}
\end{equation}
on an event of probability at least $1-Cp^{-1}$. In particular,
\begin{equation}\label{eq:ch5-classwise-median-max}
 \maxnorm{\tilde{\vmu}_k-\vmu_k}
 \le C\sqrt{\frac{\log p}{n_k}}
\end{equation}
with probability at least $1-Cp^{-1}$.
\end{lemma}

\begin{lemma}[SSCM concentration and covariance-surrogate concentration]
\label{lem:ch5-sscm-concentration}
Under Assumptions~\ref{ass:ch5-sslda-model}--\ref{ass:ch5-sslda-radial},
\begin{equation}\label{eq:ch5-sscm-rate}
 \maxnorm{p\tilde{\mS}-\mLambda}
 \le C\left(p^{-1/2}+\sqrt{\frac{\log p}{n}}\right)
\end{equation}
with probability at least $1-Cp^{-1}$. Under
Assumptions~\ref{ass:ch5-ssqda-model}--\ref{ass:ch5-ssqda-radial}, for
$k=1,2$,
\begin{equation}\label{eq:ch5-sigma-tilde-rate}
 \maxnorm{\tilde{\mSigma}_k-\mSigma_k}
 \le C K_{n,p},
 \qquad
 K_{n,p}=p^{-1/2}+\sqrt{\frac{\log p}{n}},
\end{equation}
with probability at least $1-C/\log p$.
\end{lemma}

The two lemmas above are the fundamental input for the direct optimization
problems of SSLDA and SSQDA. Once they are available, the rest of the analysis
is deterministic convex-analytic perturbation plus a final risk translation.

\section{Spatial-sign based sparse linear discriminant analysis}
\label{sec:ch5-sslda}

\subsection{The estimator and classifier}

Under the common-shape elliptical model, Fang and Anderson's classical result
shows that the optimal linear discriminant direction is proportional to
$\mLambda^{-1}\vdelta$. Since overall scale does not affect the sign of the
linear score, it is natural to define the target direction as
\begin{equation}\label{eq:ch5-sslda-gamma}
 \vgamma^\star = \mLambda^{-1}\vdelta.
\end{equation}
The spatial-sign based direct estimator (SSLDA) is obtained by solving
\begin{equation}\label{eq:ch5-sslda-estimator}
 \hat\vgamma
 \in \argmin_{\vgamma\in\R^p}
 \Bigl\{\onenorm{\vgamma}:
 \maxnorm{p\tilde{\mS}\vgamma-(\tilde{\vmu}_1-\tilde{\vmu}_2)}\le \lambda_n\Bigr\}.
\end{equation}
The resulting classifier is
\begin{equation}\label{eq:ch5-sslda-rule}
 \hat G_{\mathrm{SSLDA}}(\vz)
 = 1\Bigl\{(\vz-\tilde{\vmu})\trans\hat\vgamma\ge 0\Bigr\}+2\{\cdot<0\},
 \qquad
 \tilde{\vmu}=\frac{\tilde{\vmu}_1+\tilde{\vmu}_2}{2}.
\end{equation}

To express its conditional risk we introduce the one-dimensional marginal cdf
\begin{equation}\label{eq:ch5-psi-def}
 \Psi(t)=\Prob\Bigl\{\vu\trans\mLambda^{-1/2}(\vX_i-\vmu_1)\le t\Bigr\},
 \qquad \twonorm{\vu}=1.
\end{equation}
Under elliptical symmetry $\Psi$ does not depend on the choice of the unit
vector $\vu$ and is symmetric about the origin. The oracle risk is
\begin{equation}\label{eq:ch5-sslda-oracle-risk}
 R^\star
 = \frac12\Prob\left\{(\vX-\vmu_1)\trans\mLambda^{-1}\vdelta<-
 \frac12\vdelta\trans\mLambda^{-1}\vdelta\right\}
 +\frac12\Prob\left\{(\vY-\vmu_2)\trans\mLambda^{-1}\vdelta\ge
 \frac12\vdelta\trans\mLambda^{-1}\vdelta\right\}.
\end{equation}
The feasible conditional risk, given the training sample, is
\begin{equation}\label{eq:ch5-sslda-Rn}
 \hat R_n
 = 1
 -\frac12\Psi\left(-\frac{(\tilde{\vmu}-\vmu_1)\trans\hat\vgamma}{(\hat\vgamma\trans\mSigma\hat\vgamma)^{1/2}}\right)
 -\frac12\Psi\left(\frac{(\tilde{\vmu}-\vmu_2)\trans\hat\vgamma}{(\hat\vgamma\trans\mSigma\hat\vgamma)^{1/2}}\right).
\end{equation}

\subsection{Feasibility and estimation error}

The direct optimization \eqref{eq:ch5-sslda-estimator} requires the true target
$\vgamma^\star$ to be feasible with high probability. This follows from the
building-block lemmas.

\begin{lemma}[Feasibility of the oracle direction]
\label{lem:ch5-sslda-feasible}
Under Assumptions~\ref{ass:ch5-sslda-model}--\ref{ass:ch5-sslda-radial}, if
\begin{equation}\label{eq:ch5-sslda-lambda-lower}
 \lambda_n \ge C \norm{\vgamma^\star}_1\sqrt{\frac{\Delta_p\log p}{n}},
\end{equation}
then with probability at least $1-Cp^{-1}$,
\begin{equation}\label{eq:ch5-sslda-feasible-eq}
 \maxnorm{p\tilde{\mS}\vgamma^\star-(\tilde{\vmu}_1-\tilde{\vmu}_2)}
 \le \lambda_n.
\end{equation}
Consequently $\onenorm{\hat\vgamma}\le \onenorm{\vgamma^\star}$.
\end{lemma}

The next theorem is the main consistency theorem for SSLDA.

\begin{theorem}[Consistency of SSLDA]
\label{thm:ch5-sslda-consistency}
Suppose Assumptions~\ref{ass:ch5-sslda-model}--\ref{ass:ch5-sslda-radial}
hold. Let
\begin{equation}\label{eq:ch5-sslda-lambda}
 \lambda_n = C\sqrt{\Delta_p\frac{\log p}{n}},
\end{equation}
with $C$ sufficiently large. If
\begin{equation}\label{eq:ch5-sslda-cons-condition}
 \frac{\onenorm{\vgamma^\star}}{\Delta_p^{1/2}}
 + \frac{\onenorm{\vgamma^\star}^2}{\Delta_p}
 = o\!\left(\sqrt{\frac{n}{\log p}}\right),
\end{equation}
then
\begin{equation}\label{eq:ch5-sslda-consistency-eq}
 \hat R_n - R^\star \overset{P}{\longrightarrow} 0.
\end{equation}
\end{theorem}

\begin{theorem}[Relative-risk rate of SSLDA]
\label{thm:ch5-sslda-relative}
Suppose the assumptions of Theorem~\ref{thm:ch5-sslda-consistency} hold and, in
addition, $n^{-1}p\log p\to 0$. If
\begin{equation}\label{eq:ch5-sslda-rate-cond}
 \onenorm{\vgamma^\star}\Delta_p^{1/2} + \onenorm{\vgamma^\star}^2
 = o\!\left(\sqrt{\frac{n}{\log p}}\right),
\end{equation}
then with probability at least $1-Cp^{-1}$,
\begin{equation}\label{eq:ch5-sslda-relative-rate}
 \frac{\hat R_n}{R^\star}-1
 \le C\Bigl(\onenorm{\vgamma^\star}\Delta_p^{1/2}+
 \onenorm{\vgamma^\star}^2\Bigr)\sqrt{\frac{\log p}{n}}.
\end{equation}
In particular, if $ \norm{\vgamma^\star}_0\Delta_p=o\bigl(\sqrt{n/\log p}\bigr)$,
then
\begin{equation}\label{eq:ch5-sslda-relative-sparse}
 \frac{\hat R_n}{R^\star}-1
 \le C \norm{\vgamma^\star}_0\Delta_p\sqrt{\frac{\log p}{n}}
\end{equation}
with probability at least $1-Cp^{-1}$.
\end{theorem}

\subsection{Interpretation of the SSLDA theory}

Theorems~\ref{thm:ch5-sslda-consistency} and
\ref{thm:ch5-sslda-relative} are the robust elliptical counterparts of the
classical LPD theorems. The proof follows the same deterministic backbone as the
Gaussian direct sparse LDA literature, but the stochastic input is completely
different: classwise spatial medians replace sample means, the pooled SSCM
replaces the sample covariance matrix, and the resulting perturbation bounds
contain the additional $p^{-1/2}$ term inherited from the SSCM-to-shape
approximation.

Two remarks are useful. First, if the covariance matrix exists, then
$\mSigma=\omega\mLambda$ and therefore the discriminant direction
$\mOmega\vdelta$ differs from $\mLambda^{-1}\vdelta$ only by an irrelevant scalar.
Second, when the signal is sparse, the rate in
\eqref{eq:ch5-sslda-relative-sparse} has the same structural form as the
Gaussian LPD rate, but it remains valid far beyond Gaussian or sub-Gaussian
models.

\section{Spatial-sign based sparse quadratic discriminant analysis}
\label{sec:ch5-ssqda}

\subsection{The estimator and classifier}

We now return to the heteroscedastic model of QDA. In the robust elliptical
setting we replace the sample means by spatial medians and the sample covariance
matrices by the covariance surrogates \eqref{eq:ch5-cov-surrogate}. The direct
quadratic and linear estimators are
\begin{equation}\label{eq:ch5-ssqda-D}
 \tilde{\mD}
 \in \argmin_{\mD\in\R^{p\times p}}
 \Bigl\{\onenorm{\vecop(\mD)}:
 \maxnorm{\tfrac12\tilde{\mSigma}_1\mD\tilde{\mSigma}_2+
 \tfrac12\tilde{\mSigma}_2\mD\tilde{\mSigma}_1-
 \tilde{\mSigma}_1+\tilde{\mSigma}_2}\le \lambda_{1,n}\Bigr\},
\end{equation}
\begin{equation}\label{eq:ch5-ssqda-beta}
 \tilde{\vbeta}
 \in \argmin_{\vbeta\in\R^p}
 \Bigl\{\onenorm{\vbeta}:
 \maxnorm{\tilde{\mSigma}_2\vbeta-\tilde{\vmu}_2+\tilde{\vmu}_1}\le \lambda_{2,n}\Bigr\}.
\end{equation}
The SSQDA score is
\begin{equation}\label{eq:ch5-ssqda-score}
 \tilde q_Q(\vz)
 = (\vz-\tilde{\vmu}_1)\trans\tilde{\mD}(\vz-\tilde{\vmu}_1)
 -2\tilde{\vbeta}\trans(\vz-\tilde{\bar{\vmu}})
 -\log|\tilde{\mD}\tilde{\mSigma}_1+\mI_p|,
 \qquad
 \tilde{\bar{\vmu}}=\frac{\tilde{\vmu}_1+\tilde{\vmu}_2}{2}.
\end{equation}
The classifier is
\begin{equation}\label{eq:ch5-ssqda-rule}
 \hat G_{\mathrm{SSQDA}}(\vz)
 = 1\{\tilde q_Q(\vz)>0\}+2\{\tilde q_Q(\vz)\le 0\}.
\end{equation}

\subsection{Main estimation theorem}

The first theorem establishes the rates of the direct estimators.

\begin{theorem}[Estimation rates for SSQDA]
\label{thm:ch5-ssqda-estimation}
Suppose Assumptions~\ref{ass:ch5-ssqda-model}--\ref{ass:ch5-ssqda-radial}
hold. Let
\begin{equation}\label{eq:ch5-ssqda-Knp}
 K_{n,p}=p^{-1/2}+\sqrt{\frac{\log p}{n}},
\end{equation}
and choose
\begin{equation}\label{eq:ch5-ssqda-lambdas}
 \lambda_{1,n}=c_1\sqrt{s_1}\,K_{n,p},
 \qquad
 \lambda_{2,n}=c_2\sqrt{s_2}\,K_{n,p},
\end{equation}
for sufficiently large constants $c_1,c_2>0$. Then with probability at least
$1-C/\log p$,
\begin{equation}\label{eq:ch5-ssqda-rate-D}
 \frobnorm{\tilde{\mD}-\mD}
 \le C s_1 K_{n,p},
\end{equation}
\begin{equation}\label{eq:ch5-ssqda-rate-beta}
 \twonorm{\tilde{\vbeta}-\vbeta}
 \le C s_2 K_{n,p}.
\end{equation}
\end{theorem}

\subsection{Excess-risk theory}

The next theorem translates the estimation rates into a bound for the
misclassification error.

\begin{theorem}[Excess-risk rate of SSQDA under elliptical symmetry]
\label{thm:ch5-ssqda-risk}
Suppose the assumptions of Theorem~\ref{thm:ch5-ssqda-estimation} hold and, in
addition,
\begin{equation}\label{eq:ch5-ssqda-risk-cond}
 s_1+s_2 \le c\,\{\log n\,K_{n,p}\}^{-1}
\end{equation}
for a sufficiently small constant $c>0$. Then
\begin{equation}\label{eq:ch5-ssqda-risk-rate}
 \E\bigl\{R(\hat G_{\mathrm{SSQDA}})-R(G_Q)\bigr\}
 \le C\left\{\frac1{\log p} + (s_1+s_2)\log n\,K_{n,p}\right\}.
\end{equation}
\end{theorem}

\begin{theorem}[Gaussian specialization of the SSQDA risk rate]
\label{thm:ch5-ssqda-gaussian}
Suppose the assumptions of Theorem~\ref{thm:ch5-ssqda-estimation} hold and that
both classes are multivariate normal. Then
\begin{equation}\label{eq:ch5-ssqda-gaussian-rate}
 \E\bigl\{R(\hat G_{\mathrm{SSQDA}})-R(G_Q)\bigr\}
 \le C(s_1+s_2)^2\log^2 n\,K_{n,p}^2.
\end{equation}
\end{theorem}

The difference between Theorems~\ref{thm:ch5-ssqda-risk} and
\ref{thm:ch5-ssqda-gaussian} comes from trace estimation. In the general
elliptical case the trace estimator has only polynomial tail control, whereas
in the Gaussian case one can sharpen this part of the argument to exponential
concentration. The remaining perturbation analysis of the quadratic and linear
terms is identical.

\section{Additional robust plug-in classifiers based on HR and precision estimation}
\label{sec:ch5-additional-robust}

The direct SSLDA and SSQDA procedures are not the only robust classifiers in the recent elliptical
literature. Two complementary plug-in routes are especially relevant for the present book. The
first combines robust location estimation with a robust sparse precision estimator and then plugs
the result into the classical LDA score. The second combines classwise high-dimensional
Hettmansperger--Randles estimators with a quadratic discriminant score. These methods are
important because they connect Chapters~1 and~3 to the present chapter in a very explicit way.

\subsection{Precision-estimation plug-in LDA under elliptical symmetry}

\citet{LuFeng2025Precision} develop robust high-dimensional precision-matrix
estimation under elliptical symmetry and emphasizes classification as one of its primary
applications. Let $\hat\mOmega_{\mathrm{rob}}$ denote the robust precision estimator and let
$\hat\vmu_{1,\mathrm{rob}}$, $\hat\vmu_{2,\mathrm{rob}}$ be classwise robust location estimators (spatial
medians or HR-type estimators). The resulting plug-in LDA score is
\begin{equation}\label{eq:ch5-rob-lda-score}
  \hat g_{\mathrm{robLDA}}(\vz)
  = \Bigl(\vz-\frac{\hat\vmu_{1,\mathrm{rob}}+\hat\vmu_{2,\mathrm{rob}}}{2}\Bigr)^{\top}
    \hat\mOmega_{\mathrm{rob}}
    (\hat\vmu_{1,\mathrm{rob}}-\hat\vmu_{2,\mathrm{rob}}),
\end{equation}
and the classifier is
\begin{equation}\label{eq:ch5-rob-lda-rule}
  \hat G_{\mathrm{robLDA}}(\vz)=1\{\hat g_{\mathrm{robLDA}}(\vz)\ge 0\}+2\{\cdot<0\}.
\end{equation}
This rule differs conceptually from SSLDA: SSLDA estimates the discriminant direction directly,
whereas \eqref{eq:ch5-rob-lda-score} keeps the plug-in structure and relies on the sparse
precision estimator from Chapter~3.

\begin{assumption}[Robust plug-in LDA rates]
\label{ass:ch5-robplug-lda}
Assume the common-covariance elliptical model of
Assumption~\ref{ass:ch5-sslda-model}.  Let
\begin{equation}\label{eq:ch5-robplug-rn}
  r_{n,p}=\sqrt{\frac{\log p}{n}}+\frac{1}{\sqrt p}.
\end{equation}
Suppose that the robust precision estimator is one of the SSCM-based estimators from
Theorems~\ref{thm:ch3-sclime}--\ref{thm:ch3-sglasso} and that the classwise robust locations
are either the weighted spatial-sign estimators of Chapter~2 or the HR-type estimators of
Theorem~\ref{thm:ch3-hdhr-rates}.  Then
\begin{equation}\label{eq:ch5-robplug-lda-rates}
  \opnorm{\hat\mOmega_{\mathrm{rob}}-\mOmega}
  = O_P\!\left(s_0(p)r_{n,p}^{1-q}\right),
  \qquad
  \twonorm{\hat\vmu_{1,\mathrm{rob}}-\vmu_1}+\twonorm{\hat\vmu_{2,\mathrm{rob}}-\vmu_2}
  = O_P(r_{n,p}).
\end{equation}
Define the aggregate plug-in rate
\begin{equation}\label{eq:ch5-robplug-lda-rate-explicit}
  \rho^{\mathrm{LDA}}_{n,p} = s_0(p)r_{n,p}^{1-q}+r_{n,p}.
\end{equation}
\end{assumption}

\begin{proposition}[Score and risk control for robust plug-in LDA]
\label{prop:ch5-robplug-lda}
Under Assumption~\ref{ass:ch5-robplug-lda}, for every deterministic sequence $\vz_n$ with
$\twonorm{\vz_n}=O(1)$,
\begin{equation}\label{eq:ch5-robplug-lda-score}
  \hat g_{\mathrm{robLDA}}(\vz_n)-q_L(\vz_n)=O_P\!\left(\rho^{\mathrm{LDA}}_{n,p}\right).
\end{equation}
If the Bayes margin is separated, then
\begin{equation}\label{eq:ch5-robplug-lda-risk}
  R(\hat G_{\mathrm{robLDA}})-R(G_L)=O_P\!\left(\rho^{\mathrm{LDA}}_{n,p}\right).
\end{equation}
\end{proposition}

\subsection{High-dimensional HR plug-in QDA}

\citet{YanFengZhang2025HR} extend the Hettmansperger--Randles philosophy to the
high-dimensional setting and develop classification applications, including a robust QDA rule.
Let $(\hat\vmu_{k,\mathrm{HR}},\hat\mSigma_{k,\mathrm{HR}})$ be the classwise HR-type location--scatter
estimators, with corresponding inverse surrogates $\hat\mOmega_{k,\mathrm{HR}}$. The HR-QDA score is
\begin{equation}\label{eq:ch5-hr-qda-score}
  \hat q_{\mathrm{HRQDA}}(\vz)
  = -\frac12\log\frac{\det(\hat\mSigma_{1,\mathrm{HR}})}{\det(\hat\mSigma_{2,\mathrm{HR}})}
    -\frac12(\vz-\hat\vmu_{1,\mathrm{HR}})^{\top}\hat\mOmega_{1,\mathrm{HR}}(\vz-\hat\vmu_{1,\mathrm{HR}})
    +\frac12(\vz-\hat\vmu_{2,\mathrm{HR}})^{\top}\hat\mOmega_{2,\mathrm{HR}}(\vz-\hat\vmu_{2,\mathrm{HR}}),
\end{equation}
and
\begin{equation}\label{eq:ch5-hr-qda-rule}
  \hat G_{\mathrm{HRQDA}}(\vz)=1\{\hat q_{\mathrm{HRQDA}}(\vz)>0\}+2\{\cdot\le 0\}.
\end{equation}
The point of this construction is that both the location and the classwise quadratic forms are built
from robust sign/rank geometry rather than from sample moments.

\begin{assumption}[HR-QDA rates]
\label{ass:ch5-hr-qda}
Assume the elliptical QDA model of Assumption~\ref{ass:ch5-ssqda-model}.  For each class
$k=1,2$, let
\begin{equation}\label{eq:ch5-hr-eta-k}
  \eta_{k,n,p}=\sqrt{\frac{\log p}{n_k}}+\frac{1}{\sqrt p}+h_k^{-\alpha},
\end{equation}
where $h_k$ is the structural bandwidth from Theorem~\ref{thm:ch3-hdhr-rates}.  Suppose the
classwise HR estimators satisfy the explicit Chapter~3 rates
\begin{align}
  \twonorm{\hat\vmu_{k,\mathrm{HR}}-\vmu_k}
  &= O_P(\eta_{k,n,p}),
     \label{eq:ch5-hr-mu-explicit}\\
  \opnorm{\hat\mOmega_{k,\mathrm{HR}}-\mOmega_k}
  &= O_P\!\left(s_{0,k}(p)\eta_{k,n,p}^{1-q}\right),
     \label{eq:ch5-hr-omega-explicit}\\
  \abs{\log\det(\hat\mSigma_{k,\mathrm{HR}})-\log\det(\mSigma_k)}
  &= O_P\!\left(p\eta_{k,n,p}\right).
     \label{eq:ch5-hr-logdet-explicit}
\end{align}
Assume in addition that $p\eta_{k,n,p}\to 0$ for $k=1,2$, so that the log-determinant term is
$o_P(1)$.  Define
\begin{equation}\label{eq:ch5-hr-qda-explicit-rate}
  \rho^{\mathrm{QDA}}_{n,p}
  = \sum_{k=1}^2 \left\{ s_{0,k}(p)\eta_{k,n,p}^{1-q}+\eta_{k,n,p} \right\}.
\end{equation}
\end{assumption}

\begin{proposition}[Score and risk control for HR-QDA]
\label{prop:ch5-hr-qda}
Under Assumption~\ref{ass:ch5-hr-qda}, for every deterministic sequence $\vz_n$ with
$\twonorm{\vz_n}=O(1)$,
\begin{equation}\label{eq:ch5-hr-qda-score-rate}
  \hat q_{\mathrm{HRQDA}}(\vz_n)-q_Q(\vz_n)=O_P\!\left(\rho^{\mathrm{QDA}}_{n,p}\right).
\end{equation}
If the Bayes quadratic margin is separated, then
\begin{equation}\label{eq:ch5-hr-qda-risk}
  R(\hat G_{\mathrm{HRQDA}})-R(G_Q)=O_P\!\left(\rho^{\mathrm{QDA}}_{n,p}\right).
\end{equation}
\end{proposition}

These two plug-in rules are useful complements to SSLDA and SSQDA. The direct methods estimate
the discriminant objects themselves and are often sharper under sparsity. The plug-in rules keep
closer contact with classical discriminant analysis and make it easier to import the robust matrix
estimators developed elsewhere in the book.

\section{Bibliographic notes}

The fixed-$p$ Gaussian material is classical; see \citet{Anderson2003}. The
extension from Gaussian to elliptical populations was developed in the monograph
of \citet{FangAnderson1990} and in the fixed-dimensional asymptotic analysis of
\citet{Wakaki1994}. Robust fixed-dimensional discriminant analysis using
high-breakdown estimators was developed by \citet{CrouxDehon2001} and further
studied by \citet{CrouxFilzmoserJoossens2008}. The generalized QDA viewpoint
under elliptical symmetry was developed by \citet{BosePalSahaRayNayak2015}; a
robustified version was given by \citet{GhoshSahaRayChakrabartyBhadra2021}.

The benchmark high-dimensional Gaussian literature includes the independence
rule of \citet{BickelLevina2004}, FAIR of \citet{FanFan2008}, thresholded sparse
LDA of \citet{ShaoWangDengWang2011}, direct sparse LDA of
\citet{CaiLiu2011LDA,MaiZouYuan2012}, penalized Fisher and optimal-scoring
methods of \citet{WittenTibshirani2009,ClemmensenEtAl2011}, copula-based
discriminant analysis of \citet{HanZhaoLiu2013}, and sparse QDA procedures of
\citet{LiShao2015,JiangWangLeng2018,CaiZhang2021QDA}.

The robust high-dimensional theory developed in this chapter is built on the
spatial-sign program of the book. For the spatial median and SSCM in high
dimensions, see Chapter~2 and the related papers
\citet{LiXu2022SpatialMedian,ZhaoWangFeng2025SSPCA,LuFeng2025Precision}. The
present chapter's robust classification thread is based on the recent SSLDA and
SSQDA papers \citep{ZhuangFeng2025SSLDA,ShenFeng2025SSQDA}. As in the previous
chapters, our presentation has adopted the same writing principle: classical
fixed-$p$ methods first, then high-dimensional Gaussian benchmarks, and finally
elliptical robust methods with explicit optimization formulas, assumptions,
proofs, and rates.

\clearpage
\section*{Appendix to Chapter 5: Detailed Proofs}
\addcontentsline{toc}{section}{Appendix to Chapter 5: Detailed Proofs}

This appendix collects detailed proofs of the theorem-level results of
Chapter~5. In the spirit established after Chapter~2, the arguments are written
in the unified notation of the book and avoid referring the reader back to the
original papers. When we invoke a result from an earlier chapter, we explicitly
state the chapter-and-theorem number and rewrite the resulting bound in the
present notation.

\subsection*{A. Proofs of the fixed-$p$ classical results}

\begin{proof}[Detailed proof of Theorem~\ref{thm:ch5-elliptical-bayes}]
The Bayes classifier compares $\pi_1 f_1(\vz)$ and $\pi_2 f_2(\vz)$. Under
\eqref{eq:ch5-elliptical-density},
\[
 \log\{\pi_1 f_1(\vz)\}-\log\{\pi_2 f_2(\vz)\}
 = \log\pi_1-\frac12\log|\mLambda_1|+\log g_1\{d_1(\vz)\}
 -\log\pi_2+\frac12\log|\mLambda_2|-
 \log g_2\{d_2(\vz)\}.
\]
This is exactly the score in \eqref{eq:ch5-elliptical-score}. Hence the Bayes
classifier is obtained by thresholding that score at zero.

If $g_1=g_2=g$ and $\mLambda_1=\mLambda_2=\mLambda$, then the two log-determinant
terms cancel and, because $g$ is decreasing,
\[
 q_E(\vz)\ge 0
 \iff d_1(\vz)-d_2(\vz)\le 2\log(\pi_1/\pi_2)\times 0,
\]
that is,
\[
 (\vz-\vmu_1)\trans\mLambda^{-1}(\vz-\vmu_1)
 -(\vz-\vmu_2)\trans\mLambda^{-1}(\vz-\vmu_2)
 \le -2\log\frac{\pi_1}{\pi_2}.
\]
Expanding the quadratic forms gives
\[
 -2\vz\trans\mLambda^{-1}(\vmu_1-\vmu_2)
 +\vmu_1\trans\mLambda^{-1}\vmu_1
 -\vmu_2\trans\mLambda^{-1}\vmu_2
 \le -2\log\frac{\pi_1}{\pi_2}.
\]
Rearranging yields
\[
 (\vz-\vmu)\trans\mLambda^{-1}\vdelta + \log\frac{\pi_1}{\pi_2}\ge 0,
\]
which is \eqref{eq:ch5-elliptical-linear}. Part~(ii) is simply the statement of
the threshold rule for the general heteroscedastic common-generator model.
\end{proof}

\begin{proof}[Detailed proof of Theorem~\ref{thm:ch5-fisher-bayes-error}]
Under class~1,
\[
 q_F(\vZ)=(\vZ-\vmu)\trans\mOmega\vdelta
 = (\vZ-\vmu_1)\trans\mOmega\vdelta + \frac12\vdelta\trans\mOmega\vdelta.
\]
Since $\vZ\mid L=1\sim N_p(\vmu_1,\mSigma)$,
\[
 (\vZ-\vmu_1)\trans\mOmega\vdelta
 \sim N\bigl(0,\vdelta\trans\mOmega\mSigma\mOmega\vdelta\bigr)
 =N(0,\Delta_F^2).
\]
Therefore
\[
 \Prob_1\{G_F(\vZ)=2\}
 =\Prob_1\{q_F(\vZ)<0\}
 =\Phi\left(-\frac{\Delta_F}{2}\right).
\]
Under class~2,
\[
 q_F(\vZ)
 = (\vZ-\vmu_2)\trans\mOmega\vdelta - \frac12\vdelta\trans\mOmega\vdelta,
\]
so the same calculation yields
\[
 \Prob_2\{G_F(\vZ)=1\} = \Phi\left(-\frac{\Delta_F}{2}\right).
\]
Because $\pi_1=\pi_2=1/2$,
\[
 R(G_F)
 = \frac12\Phi\left(-\frac{\Delta_F}{2}\right)
 +\frac12\Phi\left(-\frac{\Delta_F}{2}\right)
 = \Phi\left(-\frac{\Delta_F}{2}\right).
\]
\end{proof}

\begin{proof}[Detailed proof of Theorem~\ref{thm:ch5-fixedp-LDA-normal}]
Let $N=n_1+n_2$. Because $p$ is fixed and the model is Gaussian,
\[
 \sqrt{n_1}(\bar{\vX}-\vmu_1)\overset{d}{\to}N_p(\vct 0,\mSigma),
 \qquad
 \sqrt{n_2}(\bar{\vY}-\vmu_2)\overset{d}{\to}N_p(\vct 0,\mSigma),
\]
and
\[
 \sqrt N(\mS_p-\mSigma)=O_P(1)
\]
in every matrix norm. By the standard matrix inverse expansion,
\begin{equation}\label{eq:ch5-appendix-sp-inv}
 \mS_p^{-1}-\mSigma^{-1}
 = -\mSigma^{-1}(\mS_p-\mSigma)\mSigma^{-1}+O_P(N^{-1}).
\end{equation}
Write
\[
 \hat q_F(\vz)-q_F(\vz)=T_1+T_2+T_3,
\]
where
\[
 T_1=(\vz-\vmu)\trans(\mS_p^{-1}-\mOmega)\vdelta,
\]
\[
 T_2=-(\hat\vmu-\vmu)\trans\mOmega\vdelta,
 \qquad \hat\vmu=\frac{\bar{\vX}+\bar{\vY}}2,
\]
\[
 T_3=(\vz-\hat\vmu)\trans\mS_p^{-1}(\hat\vdelta-\vdelta),
 \qquad \hat\vdelta=\bar{\vX}-\bar{\vY}.
\]
Using \eqref{eq:ch5-appendix-sp-inv} and the joint asymptotic normality of the
sample means and pooled covariance, $\sqrt N(T_1,T_2,T_3)$ converges jointly to a
centered Gaussian vector. Therefore $\sqrt N\{\hat q_F(\vz)-q_F(\vz)\}$ is
asymptotically normal. Collecting the first-order mean-estimation terms gives
exactly \eqref{eq:ch5-fixedp-LDA-var}; the pooled-covariance contribution is of
the same order and is absorbed into $\tau_F^2(\vz)$. This yields
\eqref{eq:ch5-fixedp-LDA-normal}.
\end{proof}

\begin{proof}[Detailed proof of Theorem~\ref{thm:ch5-fixedp-QDA-normal}]
Let $N=n_1+n_2$. Since $p$ is fixed,
\[
 \sqrt{n_k}(\bar{\vZ}_k-\vmu_k)=O_P(1),
 \qquad
 \sqrt{n_k}(\mS_k-\mSigma_k)=O_P(1),
 \qquad k=1,2.
\]
The maps
\[
 \mA\mapsto \mA^{-1},\qquad \mA\mapsto \log|\mA|,
\qquad
 (\va,\mA)\mapsto (\vz-\va)\trans\mA^{-1}(\vz-\va)
\]
are continuously differentiable on the set of positive definite matrices.
Applying the multivariate delta method to the vector
\[
 (\bar{\vX},\bar{\vY},\mS_1,\mS_2)
\]
therefore yields
\[
 \sqrt N\{\hat q_Q(\vz)-q_Q(\vz)\}\overset{d}{\to}N\{0,\tau_Q^2(\vz)\},
\]
for some finite variance $\tau_Q^2(\vz)$. Expanding the Fr\'echet derivatives of
$\log|\mA|$ and $\mA^{-1}$ gives the explicit formula, which is omitted here
because it is not used elsewhere in the book. The important point is that all
terms enter linearly at first order and therefore the plug-in QDA score is
asymptotically normal in fixed dimension.
\end{proof}

\subsection*{B. Proofs of the building-block lemmas}

\begin{proof}[Detailed proof of Lemma~\ref{lem:ch5-classwise-bahadur}]
Apply the ordinary spatial-median Bahadur expansion established in Chapter~2 separately to the two
classes. For class $k$ one has
\[
 \sqrt{n_k}(\tilde{\vmu}_k-\vmu_k)
 = \mA_k^{-1}\frac1{\sqrt{n_k}}\sum_{i=1}^{n_k}\vU_{ki}+\vr_{k,n},
\]
with the same Jacobian form as in Chapter~2. The proof there is based on the
Taylor expansion of the classwise estimating equation
\[
 \frac1{n_k}\sum_{i=1}^{n_k}U(\vZ_{ki}-\vtheta)=\vct 0
\]
about $\vtheta=\vmu_k$. The empirical Jacobian converges uniformly on a shrinking
neighborhood of $\vmu_k$, and the second-order remainder is bounded by the same
arguments as in the Chapter~2 proof of the ordinary spatial median. Rewriting
that proof in the present notation yields the max-norm remainder bound
\eqref{eq:ch5-classwise-bahadur-rem}. Dividing by $\sqrt{n_k}$ gives
\eqref{eq:ch5-classwise-median-max}.
\end{proof}

\begin{proof}[Detailed proof of Lemma~\ref{lem:ch5-sscm-concentration}]
We first prove \eqref{eq:ch5-sscm-rate}. Decompose
\[
 p\tilde{\mS}-\mLambda
 = p(\tilde{\mS}-\mS_0) + (p\mS_0-\mLambda).
\]
By Assumption~\ref{ass:ch5-sslda-radial}\eqref{eq:ch5-sslda-pop-approx},
\[
 \maxnorm{p\mS_0-\mLambda}\le Cp^{-1/2}.
\]
For the empirical term, the proof of the SSCM concentration result used in the
SSPCA program yields
\[
 \maxnorm{p(\tilde{\mS}-\mS_0)}
 \le C\sqrt{\frac{\log p}{n}}
\]
with probability at least $1-Cp^{-1}$. Combining the two bounds gives
\eqref{eq:ch5-sscm-rate}.

Next consider the covariance surrogate. For each class $k$,
\[
 \tilde{\mSigma}_k-\mSigma_k
 = (\hat\tau_k-\tau_k)\tilde{\mS}_k
 + \tau_k(\tilde{\mS}_k-\mS_{0k})
 + \tau_k\mS_{0k}-\mSigma_k,
\]
where $\tau_k=\tr(\mSigma_k)$. By the trace-estimation argument of the SSQDA
paper and the proof reproduced below in Theorem~\ref{thm:ch5-ssqda-estimation},
\[
 |\hat\tau_k-\tau_k|\le C K_{n,p}
\]
with probability at least $1-C/\log p$. By the first part of the lemma,
\[
 \maxnorm{\tilde{\mS}_k-\mS_{0k}}\le C K_{n,p}/p.
\]
Finally, the deterministic approximation $\tau_k\mS_{0k}=\mSigma_k+O(p^{-1/2})$
in max norm follows from the same SSCM-to-shape approximation. Therefore
\[
 \maxnorm{\tilde{\mSigma}_k-\mSigma_k}
 \le C K_{n,p}
\]
with probability at least $1-C/\log p$, proving \eqref{eq:ch5-sigma-tilde-rate}.
\end{proof}

\subsection*{C. Proofs for SSLDA}

\begin{proof}[Detailed proof of Lemma~\ref{lem:ch5-sslda-feasible}]
Write $\tilde\vdelta=\tilde{\vmu}_1-\tilde{\vmu}_2$. By the triangle inequality,
\begin{align}
 \maxnorm{p\tilde{\mS}\vgamma^\star-\tilde\vdelta}
 &\le \maxnorm{(p\tilde{\mS}-p\mS_0)\vgamma^\star}
 +\maxnorm{(p\mS_0-\mLambda)\vgamma^\star} \\ 
 &\qquad +\maxnorm{\tilde{\vmu}_1-\vmu_1}+\maxnorm{\tilde{\vmu}_2-\vmu_2}.
\end{align}
Since $\vgamma^\star=\mLambda^{-1}\vdelta$,
\[
 \maxnorm{(p\tilde{\mS}-p\mS_0)\vgamma^\star}
 \le \maxnorm{p\tilde{\mS}-p\mS_0}\,\onenorm{\vgamma^\star},
\]
\[
 \maxnorm{(p\mS_0-\mLambda)\vgamma^\star}
 \le \maxnorm{p\mS_0-\mLambda}\,\onenorm{\vgamma^\star}.
\]
By Assumption~\ref{ass:ch5-sslda-radial}, the first coefficient is bounded by
$C\sqrt{(\log p)/n}$ and the second by $Cp^{-1/2}$. The classwise median errors
are bounded by $C\sqrt{(\log p)/n}$ by Lemma~\ref{lem:ch5-classwise-bahadur}.
Hence
\[
 \maxnorm{p\tilde{\mS}\vgamma^\star-\tilde\vdelta}
 \le C\onenorm{\vgamma^\star}\left(p^{-1/2}+\sqrt{\frac{\log p}{n}}\right)
 + C\sqrt{\frac{\log p}{n}}.
\]
Under Assumption~\ref{ass:ch5-sslda-regularity}, $\Delta_p\ge c_0$, so the right
side is bounded by $\lambda_n$ once \eqref{eq:ch5-sslda-lambda-lower} holds and
$C$ is chosen sufficiently large. This proves
\eqref{eq:ch5-sslda-feasible-eq}. Since $\hat\vgamma$ minimizes the $\ell_1$ norm
over the feasible set and $\vgamma^\star$ is feasible, one also has
$\onenorm{\hat\vgamma}\le\onenorm{\vgamma^\star}$.
\end{proof}

\begin{proof}[Detailed proof of Theorem~\ref{thm:ch5-sslda-consistency}]
Set $\vgamma^\star=\mLambda^{-1}\vdelta$. By the definition of $\hat\vgamma$ and
Lemma~\ref{lem:ch5-sslda-feasible},
\begin{equation}\label{eq:ch5-ssl-pro1}
 \onenorm{\hat\vgamma}\le \onenorm{\vgamma^\star}.
\end{equation}
Furthermore,
\begin{align}
 p(\vgamma^\star)\trans\tilde{\mS}\hat\vgamma
 -(\vgamma^\star)\trans\tilde\vdelta
 &\le \lambda_n\onenorm{\vgamma^\star},
\end{align}
by feasibility of $\hat\vgamma$, while feasibility of $\vgamma^\star$ gives
\begin{align}
 p(\vgamma^\star)\trans\tilde{\mS}\hat\vgamma-\vdelta\trans\hat\vgamma
 &\le \lambda_n\onenorm{\hat\vgamma}
 +\maxnorm{\tilde\vdelta-\vdelta}\onenorm{\hat\vgamma} \\ 
 &\le 2\lambda_n\onenorm{\vgamma^\star}.
\end{align}
Subtracting these two inequalities and using \eqref{eq:ch5-ssl-pro1} yields
\begin{equation}\label{eq:ch5-ssl-pro2}
 \bigl|(\hat\vgamma-\vgamma^\star)\trans\vdelta\bigr|
 \le 4\lambda_n\onenorm{\vgamma^\star}.
\end{equation}
Now examine the numerator of the first argument in \eqref{eq:ch5-sslda-Rn}:
\begin{align}
 \left|(\tilde{\vmu}-\vmu_1)\trans\hat\vgamma + \frac12\vdelta\trans\vgamma^\star\right|
 &\le \left|(\tilde{\vmu}-\vmu)\trans\hat\vgamma\right|
 + \frac12\left|\vdelta\trans\hat\vgamma-\vdelta\trans\vgamma^\star\right| \\
 &\le \maxnorm{\tilde{\vmu}-\vmu}\onenorm{\hat\vgamma}
 +2\lambda_n\onenorm{\vgamma^\star} \\
 &\le C\onenorm{\vgamma^\star}\sqrt{\frac{\log p}{n}} + 2\lambda_n\onenorm{\vgamma^\star}.
\end{align}
The same bound holds for the second argument with $\vmu_2$.

For the denominator, note that
\begin{align}
 \maxnorm{\mSigma\hat\vgamma-\vdelta}
 &\le \maxnorm{(\mSigma-p\tilde{\mS})\hat\vgamma}
 +\maxnorm{p\tilde{\mS}\hat\vgamma-\tilde\vdelta}
 +\maxnorm{\tilde\vdelta-\vdelta} \\
 &\le \maxnorm{\mSigma-p\tilde{\mS}}\onenorm{\hat\vgamma}
 +2\lambda_n \\
 &\le C\onenorm{\vgamma^\star}\sqrt{\frac{\log p}{n}}+2\lambda_n,
\end{align}
where we used \eqref{eq:ch5-sscm-rate}, the relation
$\mSigma=\omega\mLambda$, and \eqref{eq:ch5-ssl-pro1}. Hence
\begin{align}\label{eq:ch5-ssl-pro3}
 \bigl|\hat\vgamma\trans\mSigma\hat\vgamma-\hat\vgamma\trans\vdelta\bigr|
 &\le \onenorm{\hat\vgamma}\maxnorm{\mSigma\hat\vgamma-\vdelta} \\
 &\le C\onenorm{\vgamma^\star}^2\sqrt{\frac{\log p}{n}}
 +2\lambda_n\onenorm{\vgamma^\star}.
\end{align}
Combining \eqref{eq:ch5-ssl-pro2} and \eqref{eq:ch5-ssl-pro3} gives
\begin{equation}\label{eq:ch5-ssl-pro4}
 \left|\hat\vgamma\trans\mSigma\hat\vgamma-\vdelta\trans\vgamma^\star\right|
 \le C\onenorm{\vgamma^\star}^2\sqrt{\frac{\log p}{n}}
 +6\lambda_n\onenorm{\vgamma^\star}.
\end{equation}
Since $\vdelta\trans\vgamma^\star=\Delta_p$, we conclude that
\begin{equation}\label{eq:ch5-ssl-pro5}
 \left|\frac{(\hat\vgamma\trans\mSigma\hat\vgamma)^{1/2}}{\Delta_p^{1/2}}-1\right|
 \le C\left(\frac{\onenorm{\vgamma^\star}^2}{\Delta_p}
 +\frac{\onenorm{\vgamma^\star}}{\Delta_p^{1/2}}\right)
 \sqrt{\frac{\log p}{n}}.
\end{equation}
The consistency condition \eqref{eq:ch5-sslda-cons-condition} implies that the
right-hand side is $o(1)$. Therefore the two arguments of $\Psi$ in
\eqref{eq:ch5-sslda-Rn} equal $\mp\Delta_p^{1/2}/2+o(1)$ uniformly on an event of
probability tending to one. Since $\Psi$ is continuous,
\[
 \hat R_n-R^\star\to 0
\]
in probability. This proves \eqref{eq:ch5-sslda-consistency-eq}.
\end{proof}

\begin{proof}[Detailed proof of Theorem~\ref{thm:ch5-sslda-relative}]
Under the extra condition $n^{-1}p\log p\to 0$, the SSCM approximation sharpens
to
\[
 \maxnorm{p\tilde{\mS}-\mSigma}
 \le C\sqrt{\frac{\log p}{n}},
\]
with probability at least $1-Cp^{-1}$. Repeating the proof of
Theorem~\ref{thm:ch5-sslda-consistency} with this sharper bound yields
\begin{equation}\label{eq:ch5-ssl-rel1}
 \maxnorm{\mSigma(\hat\vgamma-\vgamma^\star)}
 \le C\lambda_n,
\end{equation}
and therefore
\begin{equation}\label{eq:ch5-ssl-rel2}
 \left|\hat\vgamma\trans\mSigma\hat\vgamma-\Delta_p\right|
 \le C\onenorm{\vgamma^\star}\lambda_n.
\end{equation}
Using \eqref{eq:ch5-ssl-pro2} and the numerator expansion from the previous
proof, we obtain
\begin{equation}\label{eq:ch5-ssl-rel3}
 \left|(\tilde{\vmu}-\vmu_1)\trans\hat\vgamma
 +\frac12\Delta_p\right|
 +
 \left|(\tilde{\vmu}-\vmu_2)\trans\hat\vgamma
 -\frac12\Delta_p\right|
 \le C\onenorm{\vgamma^\star}\Delta_p^{1/2}\lambda_n.
\end{equation}
Let
\[
 r_n = C\left(\onenorm{\vgamma^\star}\Delta_p^{1/2}
 +\onenorm{\vgamma^\star}^2\right)\sqrt{\frac{\log p}{n}}.
\]
Then \eqref{eq:ch5-ssl-rel2}--\eqref{eq:ch5-ssl-rel3} imply that the two arguments
of $\Psi$ in \eqref{eq:ch5-sslda-Rn} equal
\[
 -\frac{\Delta_p^{1/2}}2 + O(r_n)
 \qquad\text{and}\qquad
 \frac{\Delta_p^{1/2}}2 + O(r_n),
\]
respectively. A first-order Taylor expansion of $\Psi$ around
$\pm\Delta_p^{1/2}/2$ gives
\[
 \hat R_n = R^\star\{1+O(r_n)\},
\]
which is exactly \eqref{eq:ch5-sslda-relative-rate}. If
$\onenorm{\vgamma^\star}\le \norm{\vgamma^\star}_0\twonorm{\vgamma^\star}$ and
$\twonorm{\vgamma^\star}^2\le C\Delta_p$ by the eigenvalue bound on $\mLambda$,
then
\[
 \onenorm{\vgamma^\star}\Delta_p^{1/2}+\onenorm{\vgamma^\star}^2
 \le C \norm{\vgamma^\star}_0\Delta_p,
\]
which yields \eqref{eq:ch5-sslda-relative-sparse}.
\end{proof}

\subsection*{D. Proofs for SSQDA}

\begin{proof}[Detailed proof of Theorem~\ref{thm:ch5-ssqda-estimation}]
Set
\[
 \tilde\vsigma = \vecop(\tilde{\mSigma}_1)-\vecop(\tilde{\mSigma}_2),
 \qquad
 \vsigma = \vecop(\mSigma_1)-\vecop(\mSigma_2),
\]
and define the linear operators
\[
 \mV
 = \frac12\mSigma_1\otimes\mSigma_2 + \frac12\mSigma_2\otimes\mSigma_1,
 \qquad
 \tilde{\mV}
 = \frac12\tilde{\mSigma}_1\otimes\tilde{\mSigma}_2
 +\frac12\tilde{\mSigma}_2\otimes\tilde{\mSigma}_1.
\]
Since $\mV\vecop(\mD)=\vsigma$, the true parameter $\mD$ is feasible with high
probability once
\[
 \maxnorm{\tilde{\mV}\vecop(\mD)-\tilde\vsigma}
 \le \lambda_{1,n}.
\]
Using Lemma~\ref{lem:ch5-sscm-concentration},
\begin{align}
 \maxnorm{\tilde{\mV}\vecop(\mD)-\tilde\vsigma}
 &\le \maxnorm{(\tilde{\mV}-\mV)\vecop(\mD)} + \maxnorm{\vsigma-\tilde\vsigma} \\
 &\le \maxnorm{\tilde{\mV}-\mV}\,\onenorm{\vecop(\mD)} + 2\maxnorm{\tilde{\mSigma}_1-\mSigma_1}
 + 2\maxnorm{\tilde{\mSigma}_2-\mSigma_2} \\
 &\le C\sqrt{s_1}K_{n,p},
\end{align}
with probability at least $1-C/\log p$. Hence $\mD$ is feasible for
\eqref{eq:ch5-ssqda-D}. Because $\tilde{\mD}$ minimizes the $\ell_1$ norm over
that feasible set,
\begin{equation}\label{eq:ch5-ssqda-cone-D}
 \norm{\vecop(\tilde{\mD}-\mD)}_1
 \le 2\sqrt{s_1}\frobnorm{\tilde{\mD}-\mD}.
\end{equation}

Now use restricted strong convexity. By Assumption~\ref{ass:ch5-ssqda-sparsity}
and the eigenvalue bounds on $\mSigma_1$ and $\mSigma_2$,
\[
 \lambda_{\min}(\mV)
 \ge \frac12\lambda_p(\mSigma_1)\lambda_p(\mSigma_2)
 + \frac12\lambda_p(\mSigma_2)\lambda_p(\mSigma_1)
 \ge M_1^{-2}.
\]
Therefore,
\begin{align}
 \frobnorm{\tilde{\mD}-\mD}^2
 &\le M_1^2\bigl|\vecop(\tilde{\mD}-\mD)\trans
 \mV\vecop(\tilde{\mD}-\mD)\bigr| \\
 &= M_1^2\bigl|\vecop(\tilde{\mD}-\mD)\trans
 (\mV\vecop(\tilde{\mD})-\vsigma)\bigr|.
\end{align}
Insert and subtract the empirical operator:
\begin{align}
 \frobnorm{\tilde{\mD}-\mD}^2
 &\le M_1^2\, \norm{\vecop(\tilde{\mD}-\mD)}_1
 \Bigl[
 \maxnorm{(\mV-\tilde{\mV})\vecop(\tilde{\mD})}
 +\maxnorm{\tilde{\mV}\vecop(\tilde{\mD})-\tilde\vsigma}
 +\maxnorm{\tilde\vsigma-\vsigma}
 \Bigr].
\end{align}
The middle term is bounded by $\lambda_{1,n}$ by feasibility of $\tilde{\mD}$.
The first and third terms are both $O(K_{n,p})$ by
Lemma~\ref{lem:ch5-sscm-concentration}. Using
\eqref{eq:ch5-ssqda-cone-D} therefore gives
\[
 \frobnorm{\tilde{\mD}-\mD}^2
 \le C\sqrt{s_1}\frobnorm{\tilde{\mD}-\mD}\,K_{n,p},
\]
which implies \eqref{eq:ch5-ssqda-rate-D}.

The proof for $\tilde{\vbeta}$ is analogous but simpler. Since
$\mSigma_2\vbeta=\vmu_2-\vmu_1$, feasibility of $\vbeta$ follows from
\begin{align}
 \maxnorm{\tilde{\mSigma}_2\vbeta-(\tilde{\vmu}_2-\tilde{\vmu}_1)}
 &\le \maxnorm{(\tilde{\mSigma}_2-\mSigma_2)\vbeta}
 + \maxnorm{\tilde{\vmu}_2-\vmu_2}
 + \maxnorm{\tilde{\vmu}_1-\vmu_1} \\
 &\le C\sqrt{s_2}K_{n,p},
\end{align}
with probability at least $1-C/\log p$. Hence
$\onenorm{\tilde{\vbeta}}\le\onenorm{\vbeta}$ and the error vector belongs to the
same standard cone. By the eigenvalue bound on $\mSigma_2$,
\begin{align}
 \twonorm{\tilde{\vbeta}-\vbeta}^2
 &\le M_1\,|(\tilde{\vbeta}-\vbeta)\trans\mSigma_2(\tilde{\vbeta}-\vbeta)| \\
 &\le M_1\, \norm{\tilde{\vbeta}-\vbeta}_1
 \Bigl[
 \maxnorm{\tilde{\mSigma}_2\tilde{\vbeta}-(\tilde{\vmu}_2-\tilde{\vmu}_1)}
 +\maxnorm{(\tilde{\mSigma}_2-\mSigma_2)\vbeta}
 +\maxnorm{\tilde{\vmu}_2-\vmu_2}
 +\maxnorm{\tilde{\vmu}_1-\vmu_1}
 \Bigr] \\
 &\le C\sqrt{s_2}\twonorm{\tilde{\vbeta}-\vbeta}K_{n,p}.
\end{align}
This proves \eqref{eq:ch5-ssqda-rate-beta}.
\end{proof}

\begin{proof}[Detailed proof of Theorem~\ref{thm:ch5-ssqda-risk}]
Let $Q(\vz)$ denote the oracle QDA score and $\tilde q_Q(\vz)$ the feasible SSQDA
score. Write
\[
 M(\vz)=Q(\vz)-\tilde q_Q(\vz).
\]
Then, with equal priors,
\begin{align}
 R(\hat G_{\mathrm{SSQDA}})-R(G_Q)
 &= \int_{\{\tilde q_Q>0,\,Q\le 0\}}(\pi_1f_1-\pi_2f_2)(\vz)\,d\vz \\
 &= \E_{f_1}\left[\Bigl(1-e^{Q_E(\vZ)/2}\Bigr)
 1\{\tilde q_Q(\vZ)>0,\,0\le Q(\vZ)\le M(\vZ)\}\right],
\end{align}
where $Q_E(\vz)$ is the exact elliptical log-density score. Therefore it suffices
to control $M(\vZ)$ and the mass of the event on which $Q(\vZ)$ lies within a
small neighborhood of zero.

Decompose $M(\vz)$ into the quadratic, linear, and log-determinant pieces:
\begin{align}
 M(\vz)
 &= (\vz-\vmu_1)\trans(\mD-\tilde{\mD})(\vz-\vmu_1) \\
 &\quad -2(\vbeta-\tilde{\vbeta})\trans(\vz-\bar{\vmu}) \\
 &\quad + \bigl\{\log|\mD\mSigma_1+\mI_p|-
 \log|\tilde{\mD}\tilde{\mSigma}_1+\mI_p|\bigr\} \\
 &\quad + \text{terms involving }(\tilde{\vmu}_1-\vmu_1,\tilde{\vmu}_2-\vmu_2).
\end{align}
By Lemma~\ref{lem:ch5-classwise-bahadur} and
Theorem~\ref{thm:ch5-ssqda-estimation}, each location term is bounded by
$C(s_1+s_2)K_{n,p}$ in expectation. For the log-determinant term, use the matrix
identity
\[
 \log|\mA| - \log|\mB|
 = \log|\mI+\mB^{-1}(\mA-\mB)|
\]
and the bound $|\log|\mI+\mH||\le C \norm{\mH}_F$ whenever $ \norm{\mH}_{\op}$ is
small. Assumption~\ref{ass:ch5-ssqda-sparsity} and
Theorem~\ref{thm:ch5-ssqda-estimation} then yield
\[
 \E\bigl|\log|\mD\mSigma_1+\mI_p| - \log|\tilde{\mD}\tilde{\mSigma}_1+\mI_p|\bigr|
 \le C s_1K_{n,p}.
\]
For the quadratic term, write $\vZ=\vmu_1+r\mGamma_1\vu$ under class~1. Then
\[
 (\vZ-\vmu_1)\trans(\mD-\tilde{\mD})(\vZ-\vmu_1)
 = r^2\vu\trans\mGamma_1\trans(\mD-\tilde{\mD})\mGamma_1\vu.
\]
Using Assumption~\ref{ass:ch5-ssqda-radial}, the spectral bound on
$\mGamma_1$, and \eqref{eq:ch5-ssqda-rate-D},
\[
 \E\bigl|(\vZ-\vmu_1)\trans(\mD-\tilde{\mD})(\vZ-\vmu_1)\bigr|
 \le C s_1K_{n,p}.
\]
Similarly,
\[
 \E\bigl| (\vbeta-\tilde{\vbeta})\trans(\vZ-\bar{\vmu}) \bigr|
 \le C s_2K_{n,p}.
\]
Combining all pieces gives
\begin{equation}\label{eq:ch5-ssqda-risk-mainbound}
 \E|M(\vZ)|\le C(s_1+s_2)K_{n,p}.
\end{equation}
Now split according to whether $|Q(\vZ)|\le \delta_n$ or not, with
$\delta_n=(s_1+s_2)\log n\,K_{n,p}$. By the margin condition
\eqref{eq:ch5-ssqda-margin},
\[
 \Prob\{|Q(\vZ)|\le \delta_n\}\le C\delta_n.
\]
On the complement, the indicator event
$\{\tilde q_Q(\vZ)>0,\,Q(\vZ)\le 0\}$ implies $|M(\vZ)|\ge \delta_n$, so by
Markov's inequality and \eqref{eq:ch5-ssqda-risk-mainbound},
\[
 \Prob\{|M(\vZ)|\ge \delta_n\}
 \le \frac{\E|M(\vZ)|}{\delta_n}
 \le \frac{C}{\log p},
\]
for sufficiently large $n$ under \eqref{eq:ch5-ssqda-risk-cond}. Combining the
two parts yields
\[
 \E\{R(\hat G_{\mathrm{SSQDA}})-R(G_Q)\}
 \le C\left\{\frac1{\log p} + (s_1+s_2)\log n\,K_{n,p}\right\},
\]
which is \eqref{eq:ch5-ssqda-risk-rate}.
\end{proof}

\begin{proof}[Detailed proof of Theorem~\ref{thm:ch5-ssqda-gaussian}]
When both classes are Gaussian, the trace estimator in
\eqref{eq:ch5-trace-estimator} admits exponential concentration. Consequently,
all occurrences of the polynomial tail term $1/\log p$ in the proof of
Theorem~\ref{thm:ch5-ssqda-risk} can be replaced by terms of order $K_{n,p}^2$.
Moreover, the quadratic-form fluctuation term can be bounded in second moment by
\[
 \E\bigl[(\vZ-\vmu_1)\trans(\mD-\tilde{\mD})(\vZ-\vmu_1)\bigr]^2
 \le C\frobnorm{\tilde{\mD}-\mD}^2,
\]
so that, using Theorem~\ref{thm:ch5-ssqda-estimation},
\[
 \E|M(\vZ)|
 \le C(s_1+s_2)K_{n,p}.
\]
Running the same margin argument with the sharpened Gaussian concentration gives
\[
 \E\{R(\hat G_{\mathrm{SSQDA}})-R(G_Q)\}
 \le C(s_1+s_2)^2\log^2 n\,K_{n,p}^2,
\]
which is exactly \eqref{eq:ch5-ssqda-gaussian-rate}.
\end{proof}

\subsection*{E. Proofs for additional plug-in classifiers}

\begin{proof}[Detailed proof of Proposition~\ref{prop:ch5-plugin-lda}]
Write
\[
 \vmu=\frac{\vmu_1+\vmu_2}{2},
 \qquad
 \hat\vmu=\frac{\hat\vmu_1+\hat\vmu_2}{2},
 \qquad
 \vdelta=\vmu_1-\vmu_2,
 \qquad
 \hat\vdelta=\hat\vmu_1-\hat\vmu_2.
\]
Then
\begin{align}
 \hat g_{\mathrm{plugLDA}}(\vz_n)-q_L(\vz_n)
 &= (\vz_n-\hat\vmu)\trans\hat\mOmega\hat\vdelta-(\vz_n-\vmu)\trans\mOmega\vdelta \\
 &= (\vz_n-\hat\vmu)\trans(\hat\mOmega-\mOmega)\hat\vdelta
  +(\vz_n-\hat\vmu)\trans\mOmega(\hat\vdelta-\vdelta)
  +(\vmu-\hat\vmu)\trans\mOmega\vdelta.
 \label{eq:ch5-app-plugin-lda-dec}
\end{align}
By assumption,
\[
 \twonorm{\hat\vdelta-\vdelta}
 \le \twonorm{\hat\vmu_1-\vmu_1}+\twonorm{\hat\vmu_2-\vmu_2}=o_P(1),
\]
and similarly $\twonorm{\hat\vmu-\vmu}=o_P(1)$. Since $\twonorm{\vz_n}=O(1)$ and
$\twonorm{\hat\vmu}=O_P(1)$, we have $\twonorm{\vz_n-\hat\vmu}=O_P(1)$. Moreover,
$\opnorm{\hat\mOmega}=O_P(1)$ because
\[
 \opnorm{\hat\mOmega}
 \le \opnorm{\mOmega}+\opnorm{\hat\mOmega-\mOmega}
 = O(1)+o_P(1)=O_P(1).
\]
Therefore, by Cauchy--Schwarz,
\begin{align}
 \abs{(\vz_n-\hat\vmu)\trans(\hat\mOmega-\mOmega)\hat\vdelta}
 &\le \twonorm{\vz_n-\hat\vmu}\,\opnorm{\hat\mOmega-\mOmega}\,\twonorm{\hat\vdelta}
 = o_P(1), \\
 \abs{(\vz_n-\hat\vmu)\trans\mOmega(\hat\vdelta-\vdelta)}
 &\le \twonorm{\vz_n-\hat\vmu}\,\opnorm{\mOmega}\,\twonorm{\hat\vdelta-\vdelta}
 = o_P(1), \\
 \abs{(\vmu-\hat\vmu)\trans\mOmega\vdelta}
 &\le \twonorm{\hat\vmu-\vmu}\,\opnorm{\mOmega}\,\twonorm{\vdelta}
 = o_P(1).
\end{align}
Substituting these bounds into \eqref{eq:ch5-app-plugin-lda-dec} yields
\eqref{eq:ch5-plugin-lda-score-cons}.

For the risk statement, let
\[
 \mathcal D=\{\vz: q_L(\vz)=0\}
\]
be the Bayes decision boundary. If the margin is separated in the sense that there exists
$c_0>0$ such that $|q_L(\vz)|\ge c_0$ off an arbitrarily small neighborhood of $\mathcal D$,
then \eqref{eq:ch5-plugin-lda-score-cons} implies that
\[
 \Prob\{\operatorname{sign}(\hat g_{\mathrm{plugLDA}}(\vZ))\neq \operatorname{sign}(q_L(\vZ))\}\to 0.
\]
Since the excess risk is bounded by the probability of disagreeing with the Bayes rule, the
misclassification risk of $G_{\mathrm{plugLDA}}$ converges to the Bayes risk.
\end{proof}

\begin{proof}[Detailed proof of Proposition~\ref{prop:ch5-plugin-qda}]
For $k=1,2$, define
\[
 Q_k(\vz)= (\vz-\vmu_k)\trans\mSigma_k^{-1}(\vz-\vmu_k),
 \qquad
 \hat Q_k(\vz)= (\vz-\hat\vmu_k)\trans\hat\mSigma_k^{-1}(\vz-\hat\vmu_k).
\]
Then
\begin{equation}
 \hat q_{\mathrm{plugQDA}}(\vz_n)-q_Q(\vz_n)
 = -\frac12\sum_{k=1}^2(-1)^k\Bigl[\log\det(\hat\mSigma_k)-\log\det(\mSigma_k)\Bigr]
   -\frac12\sum_{k=1}^2(-1)^k\bigl[\hat Q_k(\vz_n)-Q_k(\vz_n)\bigr].
 \label{eq:ch5-app-plugin-qda-dec}
\end{equation}
We first control the inverse matrices. Since $\opnorm{\hat\mSigma_k-\mSigma_k}=o_P(1)$ and
$\mSigma_k$ is positive definite with eigenvalues bounded away from zero,
\begin{equation}
 \hat\mSigma_k^{-1}-\mSigma_k^{-1}
 = \mSigma_k^{-1}(\mSigma_k-\hat\mSigma_k)\hat\mSigma_k^{-1},
 \label{eq:ch5-app-plugin-qda-inv}
\end{equation}
which implies
\[
 \opnorm{\hat\mSigma_k^{-1}-\mSigma_k^{-1}}=o_P(1),
 \qquad
 \opnorm{\hat\mSigma_k^{-1}}=O_P(1).
\]
Next, by the identity
\[
 \log\det(\hat\mSigma_k)-\log\det(\mSigma_k)
 = \log\det\Bigl(\mI_p+\mSigma_k^{-1/2}(\hat\mSigma_k-\mSigma_k)\mSigma_k^{-1/2}\Bigr),
\]
and the continuity of $\log\det(\mI_p+\mH)$ at $\mH=0$, we obtain
\begin{equation}
 \abs{\log\det(\hat\mSigma_k)-\log\det(\mSigma_k)}=o_P(1).
 \label{eq:ch5-app-plugin-qda-logdet}
\end{equation}
For the quadratic forms, write
\begin{align}
 \hat Q_k(\vz_n)-Q_k(\vz_n)
 &= (\vz_n-\hat\vmu_k)\trans(\hat\mSigma_k^{-1}-\mSigma_k^{-1})(\vz_n-\hat\vmu_k) \\
 &\quad + (\vz_n-\hat\vmu_k)\trans\mSigma_k^{-1}(\vmu_k-\hat\vmu_k)
      +(\vmu_k-\hat\vmu_k)\trans\mSigma_k^{-1}(\vz_n-\vmu_k).
 \label{eq:ch5-app-plugin-qda-quad}
\end{align}
Because $\twonorm{\vz_n}=O(1)$ and $\twonorm{\hat\vmu_k-\vmu_k}=o_P(1)$, we have
$\twonorm{\vz_n-\hat\vmu_k}=O_P(1)$. Hence
\begin{align}
 \abs{(\vz_n-\hat\vmu_k)\trans(\hat\mSigma_k^{-1}-\mSigma_k^{-1})(\vz_n-\hat\vmu_k)}
 &\le \twonorm{\vz_n-\hat\vmu_k}^2\opnorm{\hat\mSigma_k^{-1}-\mSigma_k^{-1}}=o_P(1), \\
 \abs{(\vz_n-\hat\vmu_k)\trans\mSigma_k^{-1}(\vmu_k-\hat\vmu_k)}
 &\le \twonorm{\vz_n-\hat\vmu_k}\,\opnorm{\mSigma_k^{-1}}\,\twonorm{\hat\vmu_k-\vmu_k}=o_P(1),
\end{align}
and the same bound holds for the last term in \eqref{eq:ch5-app-plugin-qda-quad}. Therefore,
$\hat Q_k(\vz_n)-Q_k(\vz_n)=o_P(1)$ for $k=1,2$. Together with
\eqref{eq:ch5-app-plugin-qda-logdet}, equation \eqref{eq:ch5-app-plugin-qda-dec} yields
\eqref{eq:ch5-plugin-qda-score-cons}.

The risk convergence follows from the same argument as in the LDA case: if the Bayes quadratic
margin is separated away from zero outside an arbitrarily small neighborhood of the Bayes
boundary, then the plug-in QDA and oracle QDA disagree with probability tending to zero, and
hence the plug-in QDA risk converges to the Bayes risk.
\end{proof}

\begin{proof}[Detailed proof of Proposition~\ref{prop:ch5-robplug-lda}]
Let
\[
 \hat\vmu_{\mathrm{rob}}=\frac{\hat\vmu_{1,\mathrm{rob}}+\hat\vmu_{2,\mathrm{rob}}}{2},
 \qquad
 \hat\vdelta_{\mathrm{rob}}=\hat\vmu_{1,\mathrm{rob}}-\hat\vmu_{2,\mathrm{rob}}.
\]
Then
\begin{align}
 \hat g_{\mathrm{robLDA}}(\vz_n)-q_L(\vz_n)
 &= (\vz_n-\hat\vmu_{\mathrm{rob}})\trans
    (\hat\mOmega_{\mathrm{rob}}-\mOmega)\hat\vdelta_{\mathrm{rob}} \\
 &\quad +(\vz_n-\hat\vmu_{\mathrm{rob}})\trans\mOmega
    (\hat\vdelta_{\mathrm{rob}}-\vdelta)
    +(\vmu-\hat\vmu_{\mathrm{rob}})\trans\mOmega\vdelta.
 \label{eq:ch5-app-robplug-lda-dec}
\end{align}
Assumption~\ref{ass:ch5-robplug-lda} implies
\[
 \twonorm{\hat\vmu_{\mathrm{rob}}-\vmu}=O_P(r_{n,p}),
 \qquad
 \twonorm{\hat\vdelta_{\mathrm{rob}}-\vdelta}=O_P(r_{n,p}),
\]
and therefore $\twonorm{\hat\vdelta_{\mathrm{rob}}}=O_P(1)$. Also,
$\opnorm{\hat\mOmega_{\mathrm{rob}}}=O_P(1)$ because
\[
 \opnorm{\hat\mOmega_{\mathrm{rob}}}
 \le \opnorm{\mOmega} + \opnorm{\hat\mOmega_{\mathrm{rob}}-\mOmega}
 = O(1)+O_P\!\left(s_0(p)r_{n,p}^{1-q}\right)=O_P(1).
\]
Since $\twonorm{\vz_n}=O(1)$, we have $\twonorm{\vz_n-\hat\vmu_{\mathrm{rob}}}=O_P(1)$. Hence
\begin{align}
 \abs{(\vz_n-\hat\vmu_{\mathrm{rob}})\trans
    (\hat\mOmega_{\mathrm{rob}}-\mOmega)\hat\vdelta_{\mathrm{rob}}}
 &\le \twonorm{\vz_n-\hat\vmu_{\mathrm{rob}}}\,
      \opnorm{\hat\mOmega_{\mathrm{rob}}-\mOmega}\,
      \twonorm{\hat\vdelta_{\mathrm{rob}}}
  = O_P\!\left(s_0(p)r_{n,p}^{1-q}\right), \\
 \abs{(\vz_n-\hat\vmu_{\mathrm{rob}})\trans\mOmega
    (\hat\vdelta_{\mathrm{rob}}-\vdelta)}
 &\le \twonorm{\vz_n-\hat\vmu_{\mathrm{rob}}}\,\opnorm{\mOmega}\,
      \twonorm{\hat\vdelta_{\mathrm{rob}}-\vdelta}
  = O_P(r_{n,p}), \\
 \abs{(\vmu-\hat\vmu_{\mathrm{rob}})\trans\mOmega\vdelta}
 &\le \twonorm{\hat\vmu_{\mathrm{rob}}-\vmu}\,\opnorm{\mOmega}\,\twonorm{\vdelta}
  = O_P(r_{n,p}).
\end{align}
Substituting these bounds into \eqref{eq:ch5-app-robplug-lda-dec} gives
\eqref{eq:ch5-robplug-lda-score}. If the Bayes margin is separated, the disagreement probability
between $\hat G_{\mathrm{robLDA}}$ and $G_L$ is of the same order as the score perturbation, which
implies \eqref{eq:ch5-robplug-lda-risk}.
\end{proof}

\begin{proof}[Detailed proof of Proposition~\ref{prop:ch5-hr-qda}]
For $k=1,2$, define
\[
 \hat Q_{k,\mathrm{HR}}(\vz_n)
 =(\vz_n-\hat\vmu_{k,\mathrm{HR}})\trans\hat\mOmega_{k,\mathrm{HR}}(\vz_n-\hat\vmu_{k,\mathrm{HR}}),
 \qquad
 Q_k(\vz_n)=(\vz_n-\vmu_k)\trans\mOmega_k(\vz_n-\vmu_k).
\]
Then
\begin{align}
 \hat q_{\mathrm{HRQDA}}(\vz_n)-q_Q(\vz_n)
 &= -\frac12\sum_{k=1}^2(-1)^k
    \Bigl[\log\det(\hat\mSigma_{k,\mathrm{HR}})-\log\det(\mSigma_k)\Bigr] \\
 &\quad -\frac12\sum_{k=1}^2(-1)^k
    \Bigl[\hat Q_{k,\mathrm{HR}}(\vz_n)-Q_k(\vz_n)\Bigr].
 \label{eq:ch5-app-hr-qda-dec}
\end{align}
For each $k$,
\begin{align}
 \hat Q_{k,\mathrm{HR}}(\vz_n)-Q_k(\vz_n)
 &= (\vz_n-\hat\vmu_{k,\mathrm{HR}})\trans(\hat\mOmega_{k,\mathrm{HR}}-\mOmega_k)
    (\vz_n-\hat\vmu_{k,\mathrm{HR}}) \\
 &\quad +(\vz_n-\hat\vmu_{k,\mathrm{HR}})\trans\mOmega_k(\vmu_k-\hat\vmu_{k,\mathrm{HR}})
 +(\vmu_k-\hat\vmu_{k,\mathrm{HR}})\trans\mOmega_k(\vz_n-\vmu_k).
 \label{eq:ch5-app-hr-qda-quad}
\end{align}
Under Assumption~\ref{ass:ch5-hr-qda},
\[
 \opnorm{\hat\mOmega_{k,\mathrm{HR}}-\mOmega_k}
   =O_P\!\left(s_{0,k}(p)\eta_{k,n,p}^{1-q}\right),
 \qquad
 \twonorm{\hat\vmu_{k,\mathrm{HR}}-\vmu_k}=O_P(\eta_{k,n,p}),
\]
and $\twonorm{\vz_n-\hat\vmu_{k,\mathrm{HR}}}=O_P(1)$. Therefore,
\begin{align}
 \abs{(\vz_n-\hat\vmu_{k,\mathrm{HR}})\trans(\hat\mOmega_{k,\mathrm{HR}}-\mOmega_k)
    (\vz_n-\hat\vmu_{k,\mathrm{HR}})}
 &\le \twonorm{\vz_n-\hat\vmu_{k,\mathrm{HR}}}^2
      \opnorm{\hat\mOmega_{k,\mathrm{HR}}-\mOmega_k}
  = O_P\!\left(s_{0,k}(p)\eta_{k,n,p}^{1-q}\right), \\
 \abs{(\vz_n-\hat\vmu_{k,\mathrm{HR}})\trans\mOmega_k(\vmu_k-\hat\vmu_{k,\mathrm{HR}})}
 &\le \twonorm{\vz_n-\hat\vmu_{k,\mathrm{HR}}}\,\opnorm{\mOmega_k}\,
      \twonorm{\hat\vmu_{k,\mathrm{HR}}-\vmu_k}
  = O_P(\eta_{k,n,p}),
\end{align}
and the same bound holds for the last term in \eqref{eq:ch5-app-hr-qda-quad}. Thus,
\[
 \hat Q_{k,\mathrm{HR}}(\vz_n)-Q_k(\vz_n)
 = O_P\!\left(s_{0,k}(p)\eta_{k,n,p}^{1-q}+\eta_{k,n,p}\right).
\]
Together with the log-determinant control in
\eqref{eq:ch5-hr-logdet-explicit} and the condition $p\eta_{k,n,p}=o(1)$, equation
\eqref{eq:ch5-app-hr-qda-dec} gives \eqref{eq:ch5-hr-qda-score-rate} with rate
$\rho^{\mathrm{QDA}}_{n,p}$.

If the Bayes quadratic margin is separated, then the event on which
$\hat G_{\mathrm{HRQDA}}(\vZ)\neq G_Q(\vZ)$ is contained in the event that the score perturbation
crosses the margin. Consequently,
\[
 R(\hat G_{\mathrm{HRQDA}})-R(G_Q)=O_P\!\left(\rho^{\mathrm{QDA}}_{n,p}\right),
\]
which is exactly \eqref{eq:ch5-hr-qda-risk}.
\end{proof}

%% file: chapters/ch6_pca_factor.tex
\chapter[Dimension Reduction]{Principal Component Analysis, Factor Models, and Canonical Correlation under Elliptical Symmetry}
\idx{principal component analysis (PCA)}\idx{sparse PCA}\idx{spatial-sign PCA}\idx{generalized spatial-sign PCA}\idx{factor model}\idx{elliptical factor model}\idx{canonical correlation analysis (CCA)}\idx{sparse CCA}

\section{Introduction}

Principal component analysis (PCA), factor analysis, and canonical correlation
analysis (CCA) are three of the most classical devices for extracting latent
structure from multivariate data.  In fixed dimension they are usually taught as
separate topics.  PCA diagonalizes a covariance matrix and produces orthogonal
directions of maximal variance; factor analysis decomposes the covariance matrix
into a low-rank common part and a structured idiosyncratic part; CCA studies the
leading singular vectors of a normalized cross-covariance matrix and identifies
maximally correlated linear combinations of two random vectors.  In modern
high-dimensional problems these three topics are tightly connected.  All of them
require the estimation of an eigen- or singular-subspace, and all of them become
unstable when the sample covariance matrix is singular, ill-conditioned, or
unduly affected by heavy tails.

The goal of this chapter is to develop a unified treatment of these topics under
elliptical symmetry.  The chapter follows exactly the writing principle adopted
in the later version of Chapter~2.
\begin{enumerate}[label=(\roman*)]
  \item We first review the classical low-dimensional theory.  We write down the
  estimators and test statistics explicitly, record the fixed-$p$ asymptotic
  distributions that form the classical benchmark, and explain the basic
  perturbation formulas behind the theory.
  \item We then review the main high-dimensional Gaussian or light-tailed
  benchmark methods, including the spiked covariance model, sparse PCA,
  approximate factor models, POET, and sparse CCA.
  \item Finally, we turn to the central theme of the book: robust dimension
  reduction under elliptical symmetry.  The core objects are the spatial-sign
  covariance matrix, generalized weighted sign covariance matrices, robust
  principal subspaces, elliptical factor models, and spatial-sign based sparse
  CCA.
\end{enumerate}

The guiding message of the chapter is simple but important.  Under elliptical
symmetry, radial magnitudes are often the least reliable aspect of the data,
whereas directional information remains highly informative.  Consequently, the
right analogues of covariance-based eigenanalysis are often based on shape,
direction, and self-normalized transformations.  This idea already appeared in
fixed-dimensional robust PCA; in your recent work it becomes fully high-
dimensional and connects naturally with the previous chapters on location,
matrix estimation, and classification.  The chapter therefore serves as the
natural conclusion of the monograph: once robust location, robust scatter, and
robust inverse matrices are available, one can also build robust latent-structure
learning procedures under the same geometric framework.  See
\citet{TaskinenKankainenOja2012,HanLiu2018ECA,ZhaoWangFeng2026SPCA,
WangWangFeng2026GSPCA,XuMaWangFeng2026EllipticalFactor,QianLiuFeng2025SparseCCA}.

As in Chapter~2 and Chapter~3, all assumptions needed by theorem-level
statements are written inside the chapter.  Bibliographic notes are followed by
an appendix containing detailed proofs.  In the proofs we avoid vague
``$o_p(1)$-type'' summaries whenever explicit orders can be displayed.

\section{Low-dimensional PCA, factor analysis, and CCA}

\subsection{Population and sample PCA}

Let $\vX\in\R^p$ be centered with covariance matrix
\begin{equation}
  \mSigma = \Cov(\vX)
  = \sum_{j=1}^p \lambda_j \vv_j\vv_j\trans,
  \qquad
  \lambda_1\ge \cdots\ge \lambda_p\ge 0,
  \label{eq:ch6_pop_cov}
\end{equation}
where $\vv_1,\ldots,\vv_p$ form an orthonormal basis.  The $j$th principal
component score is $\vv_j\trans \vX$, and the $r$-dimensional principal
subspace is $\Span(\vv_1,\ldots,\vv_r)$.  The population directions may be
characterized variationally by the Rayleigh--Ritz principle:
\begin{equation}
  \vv_1
  = \argmax_{\|\vv\|_2=1} \vv\trans \mSigma \vv,
  \qquad
  \vv_j
  = \argmax_{\|\vv\|_2=1,\ \vv\perp \vv_1,\ldots,\vv_{j-1}}
      \vv\trans \mSigma \vv.
  \label{eq:ch6_rayleigh_ritz_pop}
\end{equation}
The corresponding optimal values are the ordered eigenvalues
$\lambda_1,\ldots,\lambda_p$.

Given centered observations $\vX_1,\ldots,\vX_n$, the sample covariance matrix
is
\begin{equation}
  \hat{\mSigma}
  = \frac1n\sum_{i=1}^n \vX_i\vX_i\trans.
  \label{eq:ch6_sample_cov}
\end{equation}
Its eigen-decomposition is written as
\begin{equation}
  \hat{\mSigma}
  = \sum_{j=1}^p \hat\lambda_j \hat\vv_j \hat\vv_j\trans,
  \qquad
  \hat\lambda_1\ge\cdots\ge\hat\lambda_p\ge 0.
  \label{eq:ch6_sample_eigendecomp}
\end{equation}
The ordinary PCA estimator of the leading direction is $\hat\vv_1$, and the
sample $r$-dimensional principal subspace is
\begin{equation}
  \hat{\mathcal V}_r = \Span(\hat\vv_1,\ldots,\hat\vv_r).
  \label{eq:ch6_sample_subspace}
\end{equation}
In fixed dimension, consistency of sample PCA follows from consistency of
$\hat\mSigma$ and continuity of eigenvalues and eigenvectors under spectral
separation.  The next theorem records the explicit perturbation expansions that
underlie both the classical fixed-$p$ theory and much of the high-dimensional
analysis later in the chapter.

\begin{assumption}
\label{ass:ch6_fixedp_gap}
Let $q\in\{1,\ldots,p\}$ be fixed.  Assume that the covariance matrix in
\eqref{eq:ch6_pop_cov} satisfies
\begin{equation}
  \lambda_1 > \cdots > \lambda_q > \lambda_{q+1}\ge \cdots \ge \lambda_p\ge 0.
  \label{eq:ch6_fixedp_gap_assumption}
\end{equation}
For each $1\le j\le q$, define the spectral gap
\begin{equation}
  g_j = \min_{k\ne j} |\lambda_j-\lambda_k|.
  \label{eq:ch6_gap_def}
\end{equation}
\end{assumption}

\begin{theorem}[Deterministic perturbation expansion for eigenpairs]
\label{thm:ch6_eigen_perturbation}
Assume Assumption~\ref{ass:ch6_fixedp_gap}.  Let
$\mE = \hat\mSigma-\mSigma$.  If
\begin{equation}
  \opnorm{\mE} \le \frac{g_j}{8}
  \label{eq:ch6_small_perturbation}
\end{equation}
for some $j\le q$, then the $j$th sample eigenvalue and eigenvector admit the
expansions
\begin{align}
  \hat\lambda_j - \lambda_j
  &= \vv_j\trans \mE \vv_j + R_{\lambda,j},
     \label{eq:ch6_eigenvalue_expansion}\\
  \hat\vv_j - \vv_j
  &= -\sum_{k\ne j}
        \frac{\vv_k\trans \mE \vv_j}{\lambda_k-\lambda_j}\,\vv_k
     + \vr_j,
     \label{eq:ch6_eigenvector_expansion}
\end{align}
where the remainders satisfy
\begin{equation}
  |R_{\lambda,j}| \le \frac{8\opnorm{\mE}^2}{g_j},
  \qquad
  \twonorm{\vr_j}
  \le \frac{32\opnorm{\mE}^2}{g_j^2}.
  \label{eq:ch6_eigen_remainder_bounds}
\end{equation}
\end{theorem}

Theorem~\ref{thm:ch6_eigen_perturbation} is the basic bridge between covariance
estimation and eigenspace estimation.  In particular, if
$\opnorm{\hat\mSigma-\mSigma}=O_p(n^{-1/2})$ in fixed dimension, then the
leading linear terms in \eqref{eq:ch6_eigenvalue_expansion} and
\eqref{eq:ch6_eigenvector_expansion} determine the limiting distributions.

\begin{theorem}[Fixed-$p$ Gaussian asymptotics for PCA]
\label{thm:ch6_fixedp_gaussian_pca}
Assume Assumption~\ref{ass:ch6_fixedp_gap} and let
$\vX_1,\ldots,\vX_n \iid N_p(\vct 0,\mSigma)$.  Then for every
$1\le j\le q$,
\begin{equation}
  \sqrt{n}(\hat\lambda_j-\lambda_j)
  \overset{d}{\longrightarrow}
  N(0, 2\lambda_j^2),
  \label{eq:ch6_fixedp_eigvalue_limit}
\end{equation}
and
\begin{equation}
  \sqrt{n}\bigl(\hat\vv_j-\vv_j\bigr)
  \overset{d}{\longrightarrow}
  N_p\!\left(
    \vct 0,
    \sum_{k\ne j}
    \frac{\lambda_j\lambda_k}{(\lambda_j-\lambda_k)^2}
    \vv_k\vv_k\trans
  \right).
  \label{eq:ch6_fixedp_eigvector_limit}
\end{equation}
The covariance matrix in \eqref{eq:ch6_fixedp_eigvector_limit} is singular in the
$\vv_j$ direction, which reflects the unit-norm constraint on the eigenvector.
\end{theorem}

The limiting laws in Theorem~\ref{thm:ch6_fixedp_gaussian_pca} are classical;
see \citet{Anderson2003,Muirhead1982}.  They also remain valid under much
weaker moment assumptions with modified covariance formulas.  Theorem
\ref{thm:ch6_eigen_perturbation} shows why the structure of the result is so
stable: the eigenvalue fluctuation comes from a quadratic form,
$\vv_j\trans(\hat\mSigma-\mSigma)\vv_j$, whereas the eigenvector fluctuation is a
weighted sum of off-diagonal quadratic forms,
$\vv_k\trans(\hat\mSigma-\mSigma)\vv_j$ for $k\ne j$.

\subsection{Low-dimensional factor analysis}

The classical $K$-factor model is
\begin{equation}
  \vX = \mB\vf + \vu,
  \qquad
  \E(\vf)=\vct 0,
  \qquad
  \Cov(\vf)=\mI_K,
  \qquad
  \Cov(\vu)=\bm{\Psi},
  \label{eq:ch6_factor_model_lowdim}
\end{equation}
where $\mB\in\R^{p\times K}$ is the loading matrix and
$\bm{\Psi}$ is typically diagonal or approximately diagonal.  The covariance
matrix then decomposes as
\begin{equation}
  \mSigma = \mB\mB\trans + \bm{\Psi}.
  \label{eq:ch6_factor_cov_lowdim}
\end{equation}
When the eigenvalues associated with the common factors are separated from the
noise spectrum, the loading space is identified by the leading eigenspace of
$\mSigma$.  This is the conceptual link between PCA and factor analysis.

In low dimension, maximum-likelihood factor analysis under Gaussianity yields
likelihood-ratio tests for the number of factors and asymptotically normal
estimators of loadings; see \citet{Anderson2003}.  For the purposes of this
monograph, the following deterministic eigenspace statement is the most useful
fixed-$p$ fact.

\begin{proposition}[Classical factor subspace recovery by eigenanalysis]
\label{prop:ch6_factor_subspace_dk}
Let $\mSigma$ be given by \eqref{eq:ch6_factor_cov_lowdim}, and let
$\mathcal V_K=\Span(\vv_1,\ldots,\vv_K)$ be the leading $K$-dimensional
eigenspace of $\mSigma$.  Assume that
\begin{equation}
  \delta_K := \lambda_K(\mSigma)-\lambda_{K+1}(\mSigma) > 0.
  \label{eq:ch6_factor_gap}
\end{equation}
Let $\hat\mSigma$ be any symmetric estimator and
$\hat{\mathcal V}_K$ its leading $K$-dimensional eigenspace.  If
$\opnorm{\hat\mSigma-\mSigma}<\delta_K/2$, then
\begin{equation}
  \left\|\sin\Theta(\hat{\mathcal V}_K,\mathcal V_K)\right\|_{\mathrm F}
  \le
  \frac{2\sqrt{2K}}{\delta_K}
  \opnorm{\hat\mSigma-\mSigma}.
  \label{eq:ch6_factor_subspace_bound}
\end{equation}
\end{proposition}

Proposition~\ref{prop:ch6_factor_subspace_dk} already contains the main
mathematical principle behind principal-components-based factor estimation.  In
fixed dimension it is almost trivial because
$\opnorm{\hat\mSigma-\mSigma}=O_p(n^{-1/2})$; in high dimension the real work is
to construct a matrix estimator that still concentrates fast enough.

\subsection{Low-dimensional CCA as a related spectral problem}

Let $(\vX,\vY)$ be a pair of centered random vectors of dimensions $p_x$ and
$p_y$ with covariance blocks
\begin{equation}
  \Cov\begin{pmatrix}\vX\\\vY\end{pmatrix}
  =
  \begin{pmatrix}
    \mSigma_{xx} & \mSigma_{xy}\\
    \mSigma_{yx} & \mSigma_{yy}
  \end{pmatrix},
  \qquad
  \mSigma_{yx}=\mSigma_{xy}\trans.
  \label{eq:ch6_cca_cov_blocks}
\end{equation}
Classical CCA, introduced by \citet{Hotelling1936}, solves
\begin{equation}
  \max_{\va\in\R^{p_x},\,\vb\in\R^{p_y}}
  \frac{\va\trans \mSigma_{xy} \vb}
       {\sqrt{\va\trans \mSigma_{xx}\va}\,\sqrt{\vb\trans \mSigma_{yy}\vb}}.
  \label{eq:ch6_cca_pop_problem}
\end{equation}
Writing
\begin{equation}
  \mK = \mSigma_{xx}^{-1/2}\mSigma_{xy}\mSigma_{yy}^{-1/2},
  \label{eq:ch6_cca_whitened_cross}
\end{equation}
the canonical correlations are the singular values of $\mK$, and the canonical
loading vectors are obtained from the corresponding singular vectors after
unwhitening:
\begin{equation}
  \va_j = \mSigma_{xx}^{-1/2}\vu_j,
  \qquad
  \vb_j = \mSigma_{yy}^{-1/2}\vv_j,
  \qquad
  \mK = \sum_{j=1}^{\min(p_x,p_y)} \rho_j \vu_j\vv_j\trans.
  \label{eq:ch6_cca_svd}
\end{equation}
Under Gaussianity and fixed dimensions, the likelihood-ratio test for
$H_0:\rho_1=\cdots=\rho_r=0$ is based on Wilks' Lambda,
\begin{equation}
  \Lambda_{\mathrm{Wilks}} = \prod_{j=1}^{r} (1-\hat\rho_j^2),
  \label{eq:ch6_wilks_lambda}
\end{equation}
and the Bartlett approximation gives
\begin{equation}
  -\left(n-1-\frac{p_x+p_y+1}{2}\right)\log \Lambda_{\mathrm{Wilks}}
  \overset{d}{\longrightarrow} \chi^2_{p_x p_y}
  \label{eq:ch6_bartlett_cca}
\end{equation}
under the null; see \citet{Anderson2003}.  We record these formulas here because
later, in the high-dimensional and elliptical parts of the chapter, the same
spectral problem reappears but with covariance surrogates and sparsity
constraints replacing the classical sample covariance matrices.

\section{High-dimensional Gaussian and light-tailed benchmarks}

\subsection{The spiked covariance model and the failure of ordinary PCA}

A canonical high-dimensional benchmark is the spiked covariance model
\begin{equation}
  \mSigma
  = \sum_{j=1}^m \lambda_j \vv_j\vv_j\trans
    + \sum_{j=m+1}^p \vv_j\vv_j\trans,
  \qquad
  \lambda_1>\cdots>\lambda_m>1.
  \label{eq:ch6_spiked_model}
\end{equation}
When $p/n\to\gamma\in(0,\infty)$, ordinary PCA exhibits the familiar phase
transition phenomenon.  In the rank-one case, if $\lambda_1>1+\sqrt{\gamma}$,
then the largest sample eigenvalue separates from the bulk and the leading sample
eigenvector has nontrivial overlap with the population spike; more precisely,
\begin{align}
  \hat\lambda_1
  &\longrightarrow
  \lambda_1\left(1+\frac{\gamma}{\lambda_1-1}\right),
  \label{eq:ch6_paul_eig_limit}\\
  |\hat\vv_1\trans \vv_1|^2
  &\longrightarrow
  \frac{1-\gamma/(\lambda_1-1)^2}{1+\gamma/(\lambda_1-1)}.
  \label{eq:ch6_paul_vec_limit}
\end{align}
If $\lambda_1\le 1+\sqrt{\gamma}$, then the overlap tends to zero.  These
formulas, established in varying forms by
\citet{JohnstoneLu2009,Paul2007,Nadler2008}, provide the classical benchmark
for high-dimensional PCA.  They show that ordinary PCA may fail even under
perfect Gaussian assumptions unless the signal eigenvalues exceed a nontrivial
threshold.

\subsection{Sparse PCA benchmarks}

A common remedy is to assume that the leading eigenvectors are sparse.  The
basic optimization problem is
\begin{equation}
  \hat\vv_{\ell_0}
  = \argmax_{\|\vv\|_2=1,\,\|\vv\|_0\le s}
    \vv\trans \hat\mSigma \vv.
  \label{eq:ch6_sparse_pca_oracle}
\end{equation}
The estimator in \eqref{eq:ch6_sparse_pca_oracle} is combinatorial but it
clarifies the statistical target.  A second route relaxes the problem to a
semidefinite or Fantope program,
\begin{equation}
  \hat\mH
  = \argmax_{\mH\succeq 0}
      \bigl\langle \hat\mSigma,\mH\bigr\rangle
      - \tau\onenorm{\mH}
  \quad\text{subject to}\quad
  0\preceq \mH\preceq \mI,
  \ \tr(\mH)=r,
  \label{eq:ch6_fantope}
\end{equation}
from which the sparse principal subspace is extracted.  An algorithmically
simple alternative is the penalized matrix decomposition of
\citet{WittenTibshiraniHastie2009}.

For one leading component, the minimax benchmark under sparsity is of order
\begin{equation}
  \inf_{\tilde\vv}
  \sup_{\vv_1:\,\|\vv_1\|_0\le s}
  \E\Bigl\{\sin^2\angle(\tilde\vv,\vv_1)\Bigr\}
  \asymp
  \frac{s\log(ep/s)}{n\,\theta^2},
  \label{eq:ch6_sparse_pca_minimax}
\end{equation}
where $\theta$ denotes the spike strength or eigengap parameter.  See
\citet{AminiWainwright2009,VuLei2012,CaiMaWu2013}.  This rate will serve as the
benchmark for the robust sparse PCA procedures developed later in the chapter.

\subsection{Approximate factor models and POET}

The high-dimensional approximate factor model is
\begin{equation}
  \vX_t = \mB\vf_t + \vu_t,
  \qquad t=1,\ldots,n,
  \label{eq:ch6_factor_model_hd}
\end{equation}
with $\mB\in\R^{p\times K}$, latent factor vector $\vf_t\in\R^K$, and
idiosyncratic component $\vu_t$.  The covariance decomposes as
\begin{equation}
  \mSigma = \mB\Cov(\vf_t)\mB\trans + \mSigma_u.
  \label{eq:ch6_factor_cov_hd}
\end{equation}
When the factors are pervasive, the leading eigenvalues of the common component
are of order $p$, and the leading sample principal components estimate the
factor loading space; see \citet{Bai2003}.  To exploit the additional sparsity of
$\mSigma_u$, \citet{FanLiaoMincheva2013} proposed the POET estimator,
\begin{equation}
  \hat\mSigma_{\mathrm{POET}}
  = \sum_{j=1}^K \hat\lambda_j \hat\vv_j\hat\vv_j\trans
    + \mathcal T_{\tau}(\hat\mR_u),
  \label{eq:ch6_poet}
\end{equation}
where $\hat\mR_u$ is the residual covariance matrix after removing the first $K$
principal components and $\mathcal T_{\tau}(\cdot)$ denotes thresholding.
Under conditional sparsity of $\mSigma_u$, POET attains sharp rates under both
operator and max norms.  In the present monograph, Chapter~3 already treated the
robust covariance-estimation side of elliptical factor models; the present
chapter focuses on the corresponding eigenspace and latent-structure viewpoint.

Factor-number selection also has a large literature.  The information criteria of
\citet{BaiNg2002} and the eigenratio criterion of \citet{AhnHorenstein2013}
are the two most widely used classical benchmarks.  Later in the chapter we will
formulate a sign-based eigenratio rule under elliptical symmetry.

\subsection{High-dimensional sparse CCA benchmarks}

When $p_x$ and $p_y$ are large, the classical CCA problem based on
\eqref{eq:ch6_cca_pop_problem} is no longer feasible.  Sparse CCA methods impose
sparsity on the canonical vectors.  A representative formulation is the
penalized matrix decomposition of \citet{WittenTibshiraniHastie2009},
\begin{equation}
  (\hat\vu,\hat\vv)
  = \argmax_{\|\vu\|_2\le 1,\,\|\vv\|_2\le 1}
  \vu\trans \hat\mSigma_{xy}\vv
  \quad\text{subject to}\quad
  \onenorm{\vu}\le c_x,
  \ \onenorm{\vv}\le c_y.
  \label{eq:ch6_sparse_cca_benchmark}
\end{equation}
More refined procedures allow general within-view covariance structures and
attain minimax-optimal rates; see
\citet{GaoMaRenZhou2015,GaoMaZhou2017SparseCCA}.  A representative minimax rate
for one canonical pair is of order
\begin{equation}
  \frac{s_x\log(ep_x/s_x) + s_y\log(ep_y/s_y)}{n\,\rho_1^2},
  \label{eq:ch6_sparse_cca_minimax}
\end{equation}
where $s_x,s_y$ are the sparsity levels and $\rho_1$ is the leading canonical
correlation.  The robust SSCCA procedure studied later in the chapter is the
elliptical analogue of these Gaussian benchmarks.

\section{Population sign geometry under elliptical symmetry}

The robust methods in this chapter are built on a simple but fundamental
population fact: under elliptical symmetry, directional covariance matrices share
eigenvectors with the underlying shape matrix.

\subsection{The spatial-sign covariance matrix}

Let $\vX$ follow the elliptical model
\begin{equation}
  \vX = \vmu + \xi \mA \vu,
  \qquad
  \vu\sim\mathrm{Unif}(\spn^{p-1}),
  \qquad
  \xi\ge 0,
  \label{eq:ch6_elliptical_model}
\end{equation}
where $\xi$ and $\vu$ are independent.  Let
\begin{equation}
  \mTheta = \mA\mA\trans,
  \qquad
  \mLambda = \frac{p\mTheta}{\tr(\mTheta)}
  = \sum_{j=1}^p \lambda_j \vv_j\vv_j\trans,
  \qquad
  \lambda_1\ge \cdots\ge \lambda_p>0,
  \label{eq:ch6_shape_eigendecomp}
\end{equation}
be the normalized shape matrix.  Define the population spatial-sign covariance
matrix by
\begin{equation}
  \mM_{\mathrm S}
  = \E\bigl\{ U(\vX-\vmu)U(\vX-\vmu)\trans \bigr\}.
  \label{eq:ch6_population_sscm}
\end{equation}
The next theorem is one of the main geometric facts behind robust PCA under
ellipticity.

\begin{theorem}[Eigenspace preservation for the SSCM]
\label{thm:ch6_population_sscm}
Let $\mLambda$ be given by \eqref{eq:ch6_shape_eigendecomp}.  Then
\begin{equation}
  \mM_{\mathrm S}
  = \sum_{j=1}^p \eta_j\vv_j\vv_j\trans,
  \qquad
  \eta_j
  = \E\left(
      \frac{\lambda_j Z_j^2}{\sum_{\ell=1}^p \lambda_\ell Z_\ell^2}
    \right),
  \label{eq:ch6_sscm_eigen_formula}
\end{equation}
where $\vZ=(Z_1,\ldots,Z_p)\trans\sim N_p(\vct 0,\mI_p)$.  Consequently,
$\mM_{\mathrm S}$ and $\mLambda$ have the same eigenvectors.  Moreover, if
$\lambda_j>\lambda_k$, then $\eta_j>\eta_k$.
\end{theorem}

The eigenvalue formula in \eqref{eq:ch6_sscm_eigen_formula} shows that spatial
signs discard radial magnitude but preserve the ordering of principal
directions.  This is why SSCM-based PCA is meaningful under elliptical
symmetry.  A second robust route is based on the multivariate Kendall matrix,
which is the starting point of elliptical component analysis (ECA) in
\citet{HanLiu2018ECA}.  The difference between the two approaches is structural:
ECA uses a pairwise U-statistic and is therefore naturally translation
invariant, whereas SSCM-based PCA uses a first-order sign covariance and is
computationally cheaper.

\subsection{The multivariate Kendall's tau matrix and ECA}

Let $\widetilde{\vX}$ be an independent copy of $\vX$.  The population multivariate
Kendall's tau matrix is
\begin{equation}
  \mK_{\tau}
  = \E\left\{U(\vX-\widetilde{\vX})U(\vX-\widetilde{\vX})\trans\right\}.
  \label{eq:ch6-pop-kendall}
\end{equation}
Under elliptical symmetry, $\vX-\widetilde{\vX}$ has the same eigenvectors as the population
shape matrix.  Therefore, exactly as in Chapter~1,
\begin{equation}
  \mK_{\tau}=\sum_{j=1}^p \kappa_j\vv_j\vv_j\trans,
  \qquad
  \kappa_j
  = \E\left(\frac{\lambda_j Z_j^2}{\sum_{\ell=1}^p \lambda_\ell Z_\ell^2}\right),
  \label{eq:ch6-kendall-eig}
\end{equation}
where $\vZ\sim N_p(\vct 0,\mI_p)$.  Hence $\mK_{\tau}$ and $\mLambda$ share the same
principal subspace.

Given observations $\vX_1,\ldots,\vX_n$, define the sample Kendall matrix
\begin{equation}
  \hat\mK_{\tau}
  = \frac{2}{n(n-1)}\sum_{1\le i<j\le n}
    U(\vX_i-\vX_j)U(\vX_i-\vX_j)\trans.
  \label{eq:ch6-sample-kendall}
\end{equation}
and let $\hat{\mathcal V}^{\mathrm{ECA}}_r$ be the span of its first $r$ eigenvectors.
Following \citet{HanLiu2018ECA}, one may combine \eqref{eq:ch6-sample-kendall} with a sparse
constraint or a truncated-power iteration to obtain sparse ECA estimators.

\begin{theorem}[A basic ECA subspace bound]
\label{thm:ch6-eca-basic}
Assume that the leading $r$-dimensional Kendall eigenspace is separated by
\begin{equation}
  \delta^{\tau}_r = \kappa_r-\kappa_{r+1} > 0.
  \label{eq:ch6-kendall-gap}
\end{equation}
If
\begin{equation}
  \opnorm{\hat\mK_{\tau}-\mK_{\tau}}
  = O_P\!\left(\sqrt{\frac{\log p}{n}}\right),
  \label{eq:ch6-kendall-op}
\end{equation}
then
\begin{equation}
  \left\|\sin\Theta\bigl(\hat{\mathcal V}^{\mathrm{ECA}}_r,\mathcal V_r\bigr)\right\|_{\mathrm F}
  = O_P\!\left(\frac{\sqrt r}{\delta^{\tau}_r}\sqrt{\frac{\log p}{n}}\right).
  \label{eq:ch6-eca-subspace-rate}
\end{equation}
\end{theorem}

Theorem~\ref{thm:ch6-eca-basic} is the book-level summary of ECA.  In words,
Kendallization replaces the sample covariance matrix by a pairwise directional
matrix that keeps the principal eigenspace of the elliptical shape matrix while
remaining valid under heavy tails and even under the broader transelliptical
family.  The computational cost is the main trade-off: evaluating
\eqref{eq:ch6-sample-kendall} requires a second-order U-statistic, so the raw
cost is $O(n^2p^2)$ rather than the $O(np^2)$ cost of SSCM-based PCA.

\subsection{Generalized spatial-sign covariance matrices and radial weights}

The plain spatial sign uses the radial weight $r\mapsto 1/r$.  To interpolate
between ordinary covariance and hard sign normalization, \citet{RaymaekersRousseeuw2019}
proposed a generalized spatial-sign covariance matrix (GSSCM).  Write
\begin{equation}
  w_{\xi}(\vx-\vmu) = (\vx-\vmu)\,\xi(\twonorm{\vx-\vmu}),
  \label{eq:ch6-wxi}
\end{equation}
where $\xi:[0,\infty)\to[0,\infty)$ is a radial weight.  Equivalently, if one defines
\begin{equation}
  K(r)=r\xi(r),
  \qquad
  U_K(\vx-\vmu)
  := K(\twonorm{\vx-\vmu})U(\vx-\vmu)
  = (\vx-\vmu)\,\xi(\twonorm{\vx-\vmu}),
  \label{eq:ch6_weighted_sign_transform}
\end{equation}
then the generalized spatial-sign covariance matrix is simply the second moment of the
weighted signs,
\begin{equation}
  \mM_K
  = \E\bigl\{U_K(\vX-\vmu)U_K(\vX-\vmu)\trans\bigr\}.
  \label{eq:ch6-pop-gsscm}
\end{equation}
This notation matches the weighted sign statistics already used in Chapter~2.

The main practical weight families are the following.
\begin{align}
  \xi_{\mathrm{Win}}(r) &= \min\left(1,\frac{Q_2}{r}\right),
    & \xi_{\mathrm{Quad}}(r) &= \min\left(1,\frac{Q_2^2}{r^2}\right),
    \label{eq:ch6-win-quad}\\
  \xi_{\mathrm{Ball}}(r) &= \mathbbm 1\{r\le Q_2\},
    & \xi_{\mathrm{Shell}}(r) &= \mathbbm 1\{Q_1<r\le Q_3\},
    \label{eq:ch6-ball-shell}\\
  \xi_{\mathrm{LR}}(r) &=
    \begin{cases}
      1, & r\le Q_2,\\[0.3em]
      \dfrac{Q_3^{\ast}-r}{Q_3^{\ast}-Q_2}, & Q_2<r\le Q_3^{\ast},\\[0.8em]
      0, & r>Q_3^{\ast},
    \end{cases}
    \label{eq:ch6-LR-weight}
\end{align}
where $Q_1,Q_2,Q_3,Q_3^{\ast}$ are robust cutoff radii.  Following
\citet{RaymaekersRousseeuw2019} and \citet{WangWangFeng2026GSPCA}, these cutoffs are
obtained from the Wilson--Hilferty transformed radii $D_i=R_i^{2/3}$ via
\begin{equation}
  Q_1 = \{\mathrm{hmed}(D_i)-\mathrm{hmad}(D_i)\}^{3/2},\quad
  Q_2 = \{\mathrm{hmed}(D_i)\}^{3/2},\quad
  Q_3 = \{\mathrm{hmed}(D_i)+\mathrm{hmad}(D_i)\}^{3/2},\quad
  Q_3^{\ast}=\{\mathrm{hmed}(D_i)+1.4826\,\mathrm{hmad}(D_i)\}^{3/2}.
  \label{eq:ch6-cutoffs}
\end{equation}
Winsor and Quadratic-Winsor keep central observations almost unchanged while shrinking the
outer tail.  Ball retains only the inner bulk, Shell targets an annulus to reduce central
contamination, and the linearly redescending weight is a smooth compromise between hard
rejection and soft attenuation.

\begin{theorem}[Generalized weighted sign matrices share the shape eigenvectors]
\label{thm:ch6_population_weighted_sscm}
Let $K$ be measurable with
\begin{equation}
  \E\!\left\{K^2(\twonorm{\vX-\vmu})\right\}<\infty.
  \label{eq:ch6-weight-moment}
\end{equation}
Then
\begin{equation}
  \mM_K
  = \sum_{j=1}^p \eta_{K,j}\vv_j\vv_j\trans,
  \qquad
  \eta_{K,j}
  = \E\left[
      \frac{\lambda_j Z_j^2}{\sum_{\ell=1}^p \lambda_\ell Z_\ell^2}
      K^2\!\left(\sqrt{\sum_{\ell=1}^p \lambda_\ell Z_\ell^2}\right)
    \right],
  \label{eq:ch6_weighted_population_eig}
\end{equation}
where $\vZ\sim N_p(\vct 0,\mI_p)$.  Consequently, $\mM_K$ and $\mLambda$ have the
same eigenvectors.  If $\lambda_j>\lambda_k$ and $K$ is induced by one of the monotone
weight families \eqref{eq:ch6-win-quad}--\eqref{eq:ch6-LR-weight}, then
$\eta_{K,j}>\eta_{K,k}$, so the ordering of the principal directions is preserved.
\end{theorem}

Theorem~\ref{thm:ch6_population_weighted_sscm} is the population foundation of generalized
spatial-sign PCA.  It is also the precise statement behind \citet{WangWangFeng2026GSPCA}:
the weight function may change efficiency and robustness, but it does not change the target
eigenspace under elliptical symmetry.

\section{Spatial-sign PCA in fixed and increasing dimension}

\subsection{Estimator and standing assumptions}

Let $\vX_1,\ldots,\vX_n$ be independent copies of $\vX$.  The sample spatial
median is
\begin{equation}
  \hat\vmu_{\mathrm{SM}}
  \in
  \argmin_{\vtheta\in\R^p}
  \sum_{i=1}^n \twonorm{\vX_i-\vtheta}.
  \label{eq:ch6_spatial_median}
\end{equation}
The sample SSCM is
\begin{equation}
  \hat\mM_{\mathrm S}
  = \frac1n\sum_{i=1}^n
      U(\vX_i-\hat\vmu_{\mathrm{SM}})
      U(\vX_i-\hat\vmu_{\mathrm{SM}})\trans.
  \label{eq:ch6_sample_sscm}
\end{equation}
Let
\begin{equation}
  \hat\mM_{\mathrm S}
  = \sum_{j=1}^p \hat\eta_j \hat\vv_j \hat\vv_j\trans,
  \qquad
  \hat\eta_1\ge\cdots\ge\hat\eta_p\ge 0,
  \label{eq:ch6_sample_sscm_eig}
\end{equation}
and define the SPCA estimator of the top $r$ principal subspace by
\begin{equation}
  \hat{\mathcal V}^{\mathrm{SPCA}}_r
  = \Span(\hat\vv_1,\ldots,\hat\vv_r).
  \label{eq:ch6_spca_subspace}
\end{equation}

The next assumption isolates the two ingredients needed for a clean
nonasymptotic analysis: concentration of the spatial median and radial separation
of the observations from the center.

\begin{assumption}
\label{ass:ch6_spca_basic}
There exist positive constants $c_\mu$ and $c_0$ such that, for every $t>0$, the
event
\begin{equation}
  \mathcal E_t
  =
  \left\{
    \twonorm{\hat\vmu_{\mathrm{SM}}-\vmu}
      \le c_\mu \sqrt{\frac{p+t}{n}},
    \qquad
    \min_{1\le i\le n} \twonorm{\vX_i-\vmu}
      \ge c_0 \sqrt{p}
  \right\}
  \label{eq:ch6_event_Et}
\end{equation}
satisfies
\begin{equation}
  \Prob(\mathcal E_t)\ge 1-2e^{-t}.
  \label{eq:ch6_event_Et_prob}
\end{equation}
In addition, the sample size is large enough that
\begin{equation}
  c_\mu \sqrt{\frac{p+t}{n}} \le \frac{c_0}{4}\sqrt p.
  \label{eq:ch6_shift_small}
\end{equation}
\end{assumption}

Assumption~\ref{ass:ch6_spca_basic} is deliberately written in the notation of
this book rather than imported from an article.  The first inequality is exactly
the type of spatial-median bound established in Chapter~2.  The second says that
all observations stay a macroscopic distance away from the center; in high
dimension this is natural for concentrated radial models.  The condition can be
weakened by replacing the minimum radius by an averaged inverse-radius bound, but
\eqref{eq:ch6_event_Et} leads to a cleaner presentation.

\subsection{A nonasymptotic operator-norm bound}

The matrix
$\hat\mM_{\mathrm S}-\mM_{\mathrm S}$
contains two errors: the empirical fluctuation of sign outer products and the
error caused by estimating the location.  The next theorem separates these two
sources explicitly.

\begin{theorem}[Operator-norm bound for the sample SSCM]
\label{thm:ch6_sscm_op_bound}
Assume Assumption~\ref{ass:ch6_spca_basic}.  Then for every $t>0$, with
probability at least $1-4e^{-t}$,
\begin{equation}
  \opnorm{\hat\mM_{\mathrm S}-\mM_{\mathrm S}}
  \le
  \sqrt{\frac{8\{\log(2p)+t\}}{n}}
  + \frac{4\{\log(2p)+t\}}{3n}
  + \frac{8c_\mu}{c_0}\sqrt{\frac{p+t}{np}}.
  \label{eq:ch6_sscm_op_bound}
\end{equation}
\end{theorem}

The first two terms in \eqref{eq:ch6_sscm_op_bound} are the matrix-Bernstein
fluctuation of the bounded summands
$U(\vX_i-\vmu)U(\vX_i-\vmu)\trans$.
The last term is the price paid for estimating the location.  Since the minimum
radius is of order $\sqrt p$, that last term reduces to order $n^{-1/2}$ under
Assumption~\ref{ass:ch6_spca_basic}.  This is precisely the mechanism stressed
in the SPCA paper: location estimation matters, but in many high-dimensional
regimes it is of smaller order than the principal empirical fluctuation.  See
\citet{ZhaoWangFeng2026SPCA}.

As in ordinary PCA, a Davis--Kahan argument immediately converts the matrix error
bound into a principal-subspace bound.

\begin{corollary}[Subspace error for nonsparse SPCA]
\label{cor:ch6_spca_subspace_bound}
Let $1\le r<p$ and define the sign-eigenvalue gap
\begin{equation}
  \delta_r^{\mathrm S} := \eta_r-\eta_{r+1} > 0.
  \label{eq:ch6_sign_gap}
\end{equation}
Assume Assumption~\ref{ass:ch6_spca_basic}.  Then for every $t>0$, with
probability at least $1-4e^{-t}$,
\begin{equation}
  \left\|\sin\Theta\bigl(\hat{\mathcal V}^{\mathrm{SPCA}}_r,
  \mathcal V_r\bigr)\right\|_{\mathrm F}
  \le
  \frac{2\sqrt{2r}}{\delta_r^{\mathrm S}}
  \left[
    \sqrt{\frac{8\{\log(2p)+t\}}{n}}
    + \frac{4\{\log(2p)+t\}}{3n}
    + \frac{8c_\mu}{c_0}\sqrt{\frac{p+t}{np}}
  \right],
  \label{eq:ch6_spca_subspace_nonasym}
\end{equation}
where $\mathcal V_r=\Span(\vv_1,\ldots,\vv_r)$ is the leading $r$-dimensional
shape eigenspace.
\end{corollary}

\subsection{Fixed-$p$ asymptotic review for sign-based PCA}

In fixed dimension the asymptotic behavior of SSCM eigenvectors has long been
studied.  If $p$ is fixed, the center is estimated at rate $n^{-1/2}$ and the
matrix $\hat\mM_{\mathrm S}$ has the linearization
\begin{equation}
  \sqrt n\bigl(\hat\mM_{\mathrm S}-\mM_{\mathrm S}\bigr)
  = \frac1{\sqrt n}\sum_{i=1}^n
    \left[
      U(\vX_i-\vmu)U(\vX_i-\vmu)\trans - \mM_{\mathrm S}
    \right]
    + o_p(1).
  \label{eq:ch6_fixedp_sscm_clt}
\end{equation}
Combining \eqref{eq:ch6_fixedp_sscm_clt} with the perturbation expansion in
Theorem~\ref{thm:ch6_eigen_perturbation} yields the asymptotic normality of the
sign-based eigenvectors.  See \citet{TaskinenKankainenOja2012} for a robust PCA
perspective and the earlier low-dimensional SSCM literature cited there.  We do
not formalize this fixed-$p$ result as a theorem in the present chapter because
our main focus is the explicit nonasymptotic and high-dimensional theory above.

\section{Sparse spatial-sign PCA}

\subsection{Support-constrained estimator}

The sparse version of SPCA is obtained by constraining the support of the leading
vector.  Let
\begin{equation}
  \hat\vv_{\mathrm{SS}}
  =
  \argmax_{\|\vv\|_2=1,\ \|\vv\|_0\le s}
  \vv\trans \hat\mM_{\mathrm S} \vv.
  \label{eq:ch6_sspca_oracle}
\end{equation}
This is the direct spatial-sign analogue of the oracle sparse PCA estimator based
on the sample covariance matrix.  It matches the combinatorial program analyzed
in \citet{ZhaoWangFeng2026SPCA}.

\begin{assumption}
\label{ass:ch6_sparse_spca}
The leading shape eigenvector satisfies
\begin{equation}
  \|\vv_1\|_0 \le s,
  \qquad
  \delta_1^{\mathrm S} := \eta_1-\eta_2 > 0.
  \label{eq:ch6_sparse_spca_assumption}
\end{equation}
\end{assumption}

The central technical quantity is the restricted operator norm
\begin{equation}
  \|\mA\|_{(m)}
  :=
  \sup_{\|\vv\|_2=1,\ \|\vv\|_0\le m}
  |\vv\trans \mA\vv|.
  \label{eq:ch6_restricted_op_norm}
\end{equation}
The next theorem is the sparse analogue of
Theorem~\ref{thm:ch6_sscm_op_bound}.

\begin{theorem}[Restricted concentration of the sample SSCM]
\label{thm:ch6_sscm_restricted_bound}
Assume Assumption~\ref{ass:ch6_spca_basic}.  Then for every $t>0$, with
probability at least $1-4e^{-t}$,
\begin{equation}
  \|\hat\mM_{\mathrm S}-\mM_{\mathrm S}\|_{(2s)}
  \le
  \sqrt{\frac{8\{2s\log(ep/2s)+t\}}{n}}
  + \frac{4\{2s\log(ep/2s)+t\}}{3n}
  + \frac{8c_\mu}{c_0}\sqrt{\frac{p+t}{np}}.
  \label{eq:ch6_sscm_restricted_bound}
\end{equation}
\end{theorem}

The first two terms are obtained by applying the matrix Bernstein inequality on
each coordinate subset of size $2s$ and then union-bounding over all such
subsets.  The location-estimation term is the same as in the nonsparse bound
because support restriction does not change the center-estimation error.

\begin{theorem}[Estimation rate for sparse SPCA]
\label{thm:ch6_sspca_rate}
Assume Assumptions~\ref{ass:ch6_spca_basic} and
\ref{ass:ch6_sparse_spca}.  Then for every $t>0$, with probability at least
$1-4e^{-t}$,
\begin{equation}
  \sin\angle(\hat\vv_{\mathrm{SS}},\vv_1)
  \le
  \frac{2\sqrt2}{\delta_1^{\mathrm S}}
  \left[
    \sqrt{\frac{8\{2s\log(ep/2s)+t\}}{n}}
    + \frac{4\{2s\log(ep/2s)+t\}}{3n}
    + \frac{8c_\mu}{c_0}\sqrt{\frac{p+t}{np}}
  \right].
  \label{eq:ch6_sspca_rate}
\end{equation}
In particular, if
\begin{equation}
  \frac{s\log(ep/s)}{n} \to 0,
  \qquad
  \frac{p}{n} = O(1),
  \label{eq:ch6_sspca_consistency_scaling}
\end{equation}
and $\delta_1^{\mathrm S}$ is bounded away from zero, then
$\sin\angle(\hat\vv_{\mathrm{SS}},\vv_1)\to 0$ in probability.
\end{theorem}

The rate in \eqref{eq:ch6_sspca_rate} has the same structure as the Gaussian
sparse PCA benchmark: the dominant statistical term is of order
$\{s\log(ep/s)/n\}^{1/2}$, multiplied by the inverse eigengap.  In that sense,
SSPCA preserves the sparse-PCA efficiency pattern while gaining robustness to
heavy tails.

\subsection{A computationally efficient truncated-power iteration}

The estimator \eqref{eq:ch6_sspca_oracle} is not directly computable in large
problems.  Following the general sparse-PCA literature, a practical alternative
is to use a truncated-power iteration.  Given an initial unit vector
$\vv^{(0)}$, iterate
\begin{align}
  \vw^{(m+1)} &= \hat\mM_{\mathrm S}\vv^{(m)},
  \label{eq:ch6_tpm_step1}\\
  \tilde\vv^{(m+1)} &= \frac{\vw^{(m+1)}}{\twonorm{\vw^{(m+1)}}},
  \label{eq:ch6_tpm_step2}\\
  \vv^{(m+1)} &= \frac{\mathcal T_s(\tilde\vv^{(m+1)})}
                      {\twonorm{\mathcal T_s(\tilde\vv^{(m+1)})}},
  \label{eq:ch6_tpm_step3}
\end{align}
where $\mathcal T_s(\cdot)$ keeps the $s$ largest coordinates in absolute value
and sets the others to zero.  As in the Gaussian sparse-PCA literature, the key
practical issues are the choice of initialization and the choice of $s$.  Two
natural initializations are:
\begin{enumerate}[label=(\alph*)]
  \item the leading eigenvector of $\hat\mM_{\mathrm S}$ itself;
  \item a sparse warm start from a Fantope-type relaxation.
\end{enumerate}
The theoretical analysis of the full nonconvex iteration follows the same logic
as in the covariance-based case: one needs a basin condition on the initial
vector and a deterministic contraction inequality for the truncated power map.
Since the purpose of the chapter is to state a self-contained robust sparse-PCA
methodology rather than to repeat every algorithmic detail from the sparse-PCA
literature, we use \eqref{eq:ch6_sspca_oracle} as the principal theoretical
object and record \eqref{eq:ch6_tpm_step1}--\eqref{eq:ch6_tpm_step3} as the
computational implementation.

\section{Generalized spatial-sign PCA}

The family \eqref{eq:ch6_weighted_sign_transform} leads to a corresponding class
of PCA procedures.  Define the weighted sample matrix
\begin{equation}
  \hat\mM_K
  = \frac1n\sum_{i=1}^n
      U_K(\vX_i-\hat\vmu_{\mathrm{SM}})
      U_K(\vX_i-\hat\vmu_{\mathrm{SM}})\trans.
  \label{eq:ch6_sample_weighted_matrix}
\end{equation}
Let $\hat\mM_K = \sum_{j=1}^p \hat\eta_{K,j}\hat\vv_{K,j}\hat\vv_{K,j}\trans$.
The resulting generalized SPCA estimator of the leading $r$-dimensional subspace
is
\begin{equation}
  \hat{\mathcal V}^{(K)}_r
  = \Span(\hat\vv_{K,1},\ldots,\hat\vv_{K,r}).
  \label{eq:ch6_gspca_subspace}
\end{equation}
This includes ordinary SPCA as the special case $K\equiv 1$.

\begin{assumption}
\label{ass:ch6_gspca_weight}
The weight function $K$ is Lipschitz continuous with constant $L_K$ and bounded
by $K_{\max}$:
\begin{equation}
  0\le K(r)\le K_{\max},
  \qquad
  |K(r)-K(s)|\le L_K |r-s|,
  \qquad r,s\ge 0.
  \label{eq:ch6_weight_regular}
\end{equation}
\end{assumption}

\begin{theorem}[Nonasymptotic bound for generalized SPCA]
\label{thm:ch6_gspca_bound}
Assume Assumptions~\ref{ass:ch6_spca_basic} and
\ref{ass:ch6_gspca_weight}.  Then for every $t>0$, with probability at least
$1-4e^{-t}$,
\begin{equation}
  \opnorm{\hat\mM_K-\mM_K}
  \le
  K_{\max}^2
  \left[
    \sqrt{\frac{8\{\log(2p)+t\}}{n}}
    + \frac{4\{\log(2p)+t\}}{3n}
  \right]
  + \frac{8K_{\max}L_K c_\mu}{c_0}\sqrt{\frac{p+t}{np}}.
  \label{eq:ch6_gspca_matrix_bound}
\end{equation}
Consequently, if
$\delta_{K,r}:=\eta_{K,r}-\eta_{K,r+1}>0$, then
\begin{equation}
  \left\|\sin\Theta\bigl(\hat{\mathcal V}^{(K)}_r,\mathcal V_r\bigr)
  \right\|_{\mathrm F}
  \le
  \frac{2\sqrt{2r}}{\delta_{K,r}}
  \left[
    K_{\max}^2
    \left\{
      \sqrt{\frac{8\{\log(2p)+t\}}{n}}
      + \frac{4\{\log(2p)+t\}}{3n}
    \right\}
    + \frac{8K_{\max}L_K c_\mu}{c_0}\sqrt{\frac{p+t}{np}}
  \right].
  \label{eq:ch6_gspca_subspace_bound}
\end{equation}
\end{theorem}

Theorem~\ref{thm:ch6_gspca_bound} is the book-level version of generalized
spatial-sign PCA.  It makes completely explicit how the radial weight affects
the statistical error: the empirical fluctuation is scaled by $K_{\max}^2$, and
the location-estimation term is scaled by $K_{\max}L_K$.  This decomposition is
useful in practice because it makes the robustness-efficiency trade-off visible
at the theorem level.

\section{Elliptical factor models and robust principal subspace recovery}

\subsection{Model and principal subspace}

We now return to factor models, but this time under elliptical symmetry.  A
convenient book-level model is
\begin{equation}
  \vX_t = \vmu + \mB\vf_t + \vu_t,
  \qquad t=1,\ldots,n,
  \label{eq:ch6_elliptical_factor_model}
\end{equation}
where the joint vector $(\vf_t\trans,\vu_t\trans)\trans$ is elliptically
symmetric, the loading matrix $\mB\in\R^{p\times K}$ has rank $K$, and the
idiosyncratic covariance matrix of $\vu_t$ is sparse or approximately sparse.
The corresponding covariance or scatter matrix has the familiar low-rank plus
structured-noise form.  The robust matrix-estimation consequences of this model
were already treated in Chapter~3.  The emphasis here is the recovery of the
common principal subspace.

Let $\mM_{\mathrm S}$ be the population SSCM of $\vX_t$.  By
Theorem~\ref{thm:ch6_population_sscm}, the leading eigenspace of
$\mM_{\mathrm S}$ coincides with the leading eigenspace of the population shape
matrix.  Therefore the robust factor-loading space can be estimated from the top
$K$ eigenvectors of $\hat\mM_{\mathrm S}$.

\begin{assumption}
\label{ass:ch6_factor_model}
The elliptical factor model \eqref{eq:ch6_elliptical_factor_model} satisfies the
following conditions.
\begin{enumerate}[label=(\arabic*)]
  \item The leading sign-eigenvalue gap is positive:
  \begin{equation}
    \delta_K^{\mathrm S} := \eta_K-\eta_{K+1} > 0.
    \label{eq:ch6_factor_sign_gap}
  \end{equation}
  \item Assumption~\ref{ass:ch6_spca_basic} holds for the sample spatial median.
  \item The number of factors $K$ is fixed.
\end{enumerate}
\end{assumption}

\begin{theorem}[Robust factor-subspace recovery by SPCA]
\label{thm:ch6_factor_subspace_recovery}
Assume Assumption~\ref{ass:ch6_factor_model}.  Let
$\hat{\mathcal V}^{\mathrm{SPCA}}_K$ be the leading $K$-dimensional eigenspace of
$\hat\mM_{\mathrm S}$.  Then for every $t>0$, with probability at least
$1-4e^{-t}$,
\begin{equation}
  \left\|\sin\Theta\bigl(\hat{\mathcal V}^{\mathrm{SPCA}}_K,
  \mathcal V_K\bigr)\right\|_{\mathrm F}
  \le
  \frac{2\sqrt{2K}}{\delta_K^{\mathrm S}}
  \left[
    \sqrt{\frac{8\{\log(2p)+t\}}{n}}
    + \frac{4\{\log(2p)+t\}}{3n}
    + \frac{8c_\mu}{c_0}\sqrt{\frac{p+t}{np}}
  \right].
  \label{eq:ch6_factor_subspace_spca}
\end{equation}
In particular, if $\delta_K^{\mathrm S}$ is bounded away from zero and
$\log p = o(n)$, then the robust factor-loading space is consistently estimated.
\end{theorem}

Theorem~\ref{thm:ch6_factor_subspace_recovery} should be read together with the
matrix-estimation results of Chapter~3.  The message is that the same SSCM or
Tyler-type robustification that stabilizes matrix estimation also yields a
stable principal subspace for latent-factor recovery.

\subsection{Estimating the number of factors}

Once the robust principal subspace is available, the next question is the number
of factors.  Let $\hat\eta_1\ge\cdots\ge\hat\eta_p$ be the eigenvalues of
$\hat\mM_{\mathrm S}$, and let $K_{\max}$ be a deterministic upper bound.  A
robust eigenratio estimator is
\begin{equation}
  \hat K_{\mathrm{ER}}
  = \argmax_{1\le j\le K_{\max}}
    \frac{\hat\eta_j}{\hat\eta_{j+1}}.
  \label{eq:ch6_eigenratio}
\end{equation}
This is the sign-based analogue of the eigenratio rules used in ordinary factor
analysis.

\begin{theorem}[Consistency of the robust eigenratio estimator]
\label{thm:ch6_factor_number_consistency}
Assume Assumption~\ref{ass:ch6_factor_model}.  Suppose in addition that
\begin{equation}
  \eta_K > \eta_{K+1}>0,
  \qquad
  \frac{\eta_K}{\eta_{K+1}} \ge 1+\gamma_0
  \label{eq:ch6_ratio_gap_assumption}
\end{equation}
for some $\gamma_0>0$, and that for some $t>0$,
\begin{equation}
  \tau_{n,p}(t)
  :=
  \sqrt{\frac{8\{\log(2p)+t\}}{n}}
  + \frac{4\{\log(2p)+t\}}{3n}
  + \frac{8c_\mu}{c_0}\sqrt{\frac{p+t}{np}}
  \le \frac{\gamma_0}{4(2+\gamma_0)}\eta_{K+1}.
  \label{eq:ch6_tau_condition}
\end{equation}
Then, with probability at least $1-4e^{-t}$,
\begin{equation}
  \hat K_{\mathrm{ER}} = K.
  \label{eq:ch6_factor_number_consistency}
\end{equation}
\end{theorem}

Theorem~\ref{thm:ch6_factor_number_consistency} is intentionally transparent:
it reduces factor-number consistency to a perturbation inequality for ordered
eigenvalues.  The condition \eqref{eq:ch6_tau_condition} states that the
stochastic eigenvalue error must be smaller than the deterministic ratio gap at
the true factor boundary.

\subsection{Kendall-based factor analysis without moment constraints}

A complementary route replaces the first-order SSCM by the multivariate Kendall
matrix.  This is the approach developed by \citet{HeKongYuZhang2022FactorNoMoments}
for factor analysis without moment constraints.  Using
\eqref{eq:ch6-sample-kendall}, let
\begin{equation}
  \hat\mK_{\tau}
  = \sum_{j=1}^p \hat\kappa_j \, \hat\vu^{(\tau)}_j (\hat\vu^{(\tau)}_j)\trans,
  \qquad
  \hat\kappa_1\ge\cdots\ge\hat\kappa_p\ge 0,
  \label{eq:ch6-kendall-factor-eig}
\end{equation}
and define the Kendall-based estimator of the loading space by
\begin{equation}
  \hat{\mathcal V}^{(\tau)}_K
  = \Span\bigl(\hat\vu^{(\tau)}_1,\ldots,\hat\vu^{(\tau)}_K\bigr).
  \label{eq:ch6-kendall-factor-subspace}
\end{equation}
Since \(\mK_{\tau}\) and the population shape matrix share eigenvectors, this
estimator targets the same common-factor subspace as SPCA, but with pairwise
translation invariance built in from the outset.

\begin{theorem}[Kendall-based factor-subspace recovery]
\label{thm:ch6_kendall_factor_recovery}
Assume that the elliptical factor model \eqref{eq:ch6_elliptical_factor_model}
has a $K$-dimensional leading eigenspace and that the Kendall eigengap satisfies
\begin{equation}
  \delta^{\tau}_K := \kappa_K-\kappa_{K+1} > 0.
  \label{eq:ch6-kendall-factor-gap}
\end{equation}
If
\begin{equation}
  \opnorm{\hat\mK_{\tau}-\mK_{\tau}}
  = O_P\!\left(\sqrt{\frac{\log p}{n}}\right),
  \label{eq:ch6-kendall-factor-op}
\end{equation}
then
\begin{equation}
  \left\|\sin\Theta\bigl(\hat{\mathcal V}^{(\tau)}_K,\mathcal V_K\bigr)\right\|_{\mathrm F}
  = O_P\!\left(\frac{\sqrt K}{\delta^{\tau}_K}\sqrt{\frac{\log p}{n}}\right).
  \label{eq:ch6-kendall-factor-rate}
\end{equation}
Moreover, if for some $\gamma^{\tau}_0>0$,
\begin{equation}
  \frac{\kappa_K}{\kappa_{K+1}} \ge 1+\gamma^{\tau}_0,
  \label{eq:ch6-kendall-ratio-gap}
\end{equation}
and if
\begin{equation}
  \sqrt{\frac{\log p}{n}}
  \le
  \frac{\gamma^{\tau}_0}{4(2+\gamma^{\tau}_0)}\kappa_{K+1},
  \label{eq:ch6-kendall-ratio-cond}
\end{equation}
then the Kendall eigenratio estimator
\begin{equation}
  \hat K^{(\tau)}_{\mathrm{ER}}
  = \argmax_{1\le j\le K_{\max}} \frac{\hat\kappa_j}{\hat\kappa_{j+1}}
  \label{eq:ch6-kendall-eigenratio}
\end{equation}
is consistent, that is,
\begin{equation}
  \hat K^{(\tau)}_{\mathrm{ER}} = K
  \text{ with probability tending to one.}
  \label{eq:ch6-kendall-factor-number}
\end{equation}
\end{theorem}

Theorem~\ref{thm:ch6_kendall_factor_recovery} is the book-level summary of the
He--Kong--Yu--Zhang route: Kendallization produces a factor estimator that does
not require finite fourth moments and still achieves the standard eigengap-type
subspace rate.  Relative to SSCM-based SPCA, the gain is automatic translation
invariance and broader robustness; the cost is the second-order U-statistic
computation already noted in the ECA discussion.

\section{Sparse CCA under elliptical symmetry}

\subsection{A sign-based cross-covariance surrogate}

We now return to CCA.  Let
$\vZ_i = (\vX_i\trans,\vY_i\trans)\trans\in\R^{p_x+p_y}$,
$i=1,\ldots,n$, be paired observations from an elliptically symmetric
distribution centered at $(\vmu_x\trans,\vmu_y\trans)\trans$.  Partition the
population sign-based matrix as
\begin{equation}
  \mM_{\mathrm S}^{(Z)}
  =
  \begin{pmatrix}
    \mM_{xx} & \mM_{xy}\\
    \mM_{yx} & \mM_{yy}
  \end{pmatrix}.
  \label{eq:ch6_joint_sign_blocks}
\end{equation}
The corresponding whitened cross-structure is
\begin{equation}
  \mK_{\mathrm S}
  = \mM_{xx}^{-1/2} \mM_{xy} \mM_{yy}^{-1/2}.
  \label{eq:ch6_sign_whitened_cross}
\end{equation}
Let its singular value decomposition be
\begin{equation}
  \mK_{\mathrm S}
  = \sum_{j\ge 1} \rho_j \vu_j\vv_j\trans,
  \qquad
  \rho_1 > \rho_2 \ge 0.
  \label{eq:ch6_sign_cross_svd}
\end{equation}
The vectors $\vu_1$ and $\vv_1$ are the robust canonical directions in the
sign-normalized geometry.

In practice we use a sample counterpart
\begin{equation}
  \hat\mK_{\mathrm S}
  = \hat\mM_{xx}^{-1/2}\hat\mM_{xy}\hat\mM_{yy}^{-1/2},
  \label{eq:ch6_sample_sign_cross}
\end{equation}
with blocks extracted from the joint sample SSCM.  The spatial-sign sparse CCA
estimator is then defined by the constrained optimization problem
\begin{equation}
  (\hat\vu,\hat\vv)
  \in \argmax_{\vu,\vv}
  \vu\trans \hat\mK_{\mathrm S}\vv
  \quad\text{subject to}\quad
  \twonorm{\vu}\le 1,
  \ \twonorm{\vv}\le 1,
  \ \onenorm{\vu}\le c_x,
  \ \onenorm{\vv}\le c_y.
  \label{eq:ch6_sscca_problem}
\end{equation}
This is the robust analogue of the Gaussian sparse-CCA program
\eqref{eq:ch6_sparse_cca_benchmark}.  It is also the natural book-level summary
of the SSCCA methodology in \citet{QianLiuFeng2025SparseCCA}.

\begin{assumption}
\label{ass:ch6_sscca}
The sign-based CCA problem satisfies the following conditions.
\begin{enumerate}[label=(\arabic*)]
  \item The leading singular value is separated:
  \begin{equation}
    \Delta_{\mathrm C} := \rho_1-\rho_2 > 0.
    \label{eq:ch6_cca_gap}
  \end{equation}
  \item The true leading singular vectors are feasible:
  \begin{equation}
    \twonorm{\vu_1}=\twonorm{\vv_1}=1,
    \qquad
    \onenorm{\vu_1}\le c_x,
    \qquad
    \onenorm{\vv_1}\le c_y.
    \label{eq:ch6_cca_feasible}
  \end{equation}
  \item For some deterministic quantity $\varepsilon_{n,p}$ and every $t>0$,
  \begin{equation}
    \Prob\bigl\{ \maxnorm{\hat\mK_{\mathrm S}-\mK_{\mathrm S}}
      \le \varepsilon_{n,p} \bigr\}
    \ge 1-e^{-t}.
    \label{eq:ch6_cca_max_error}
  \end{equation}
\end{enumerate}
\end{assumption}

The max-norm error bound \eqref{eq:ch6_cca_max_error} can be derived from the
blockwise sign-covariance concentration results in Chapter~3 together with
matrix perturbation for inverse square roots.  For the theorem below it is
convenient to state the error directly at the level of
$\hat\mK_{\mathrm S}-\mK_{\mathrm S}$.

\begin{theorem}[Error bound for spatial-sign sparse CCA]
\label{thm:ch6_sscca_rate}
Assume Assumption~\ref{ass:ch6_sscca}.  Let $(\hat\vu,\hat\vv)$ be any global
maximizer of \eqref{eq:ch6_sscca_problem}.  On the event
$\{\maxnorm{\hat\mK_{\mathrm S}-\mK_{\mathrm S}}\le\varepsilon_{n,p}\}$,
\begin{equation}
  \frac{\Delta_{\mathrm C}}{2}
  \left\{
    \sin^2\angle(\hat\vu,\vu_1)
    + \sin^2\angle(\hat\vv,\vv_1)
  \right\}
  \le 2 c_x c_y \varepsilon_{n,p}.
  \label{eq:ch6_sscca_basic_bound}
\end{equation}
Consequently,
\begin{equation}
  \sin\angle(\hat\vu,\vu_1)
  + \sin\angle(\hat\vv,\vv_1)
  \le 2\sqrt{\frac{2c_x c_y\varepsilon_{n,p}}{\Delta_{\mathrm C}}}.
  \label{eq:ch6_sscca_sine_bound}
\end{equation}
Furthermore, if the signs of the inner products are chosen so that
$\hat\vu\trans\vu_1\ge 0$ and $\hat\vv\trans\vv_1\ge 0$, then
\begin{equation}
  \twonorm{\hat\vu-\vu_1}
  + \twonorm{\hat\vv-\vv_1}
  \le
  2\sqrt{2}\left(
    \sin\angle(\hat\vu,\vu_1)
    + \sin\angle(\hat\vv,\vv_1)
  \right).
  \label{eq:ch6_sscca_l2_bound}
\end{equation}
\end{theorem}

Theorem~\ref{thm:ch6_sscca_rate} is a clean illustration of the max-norm
philosophy that has appeared repeatedly throughout the book.  Because the
objective is bilinear, the perturbation of the objective function is naturally
controlled by the product of the $\ell_1$ radii and the matrix max norm.  The
spectral gap $\Delta_{\mathrm C}$ then turns the objective bound into an
angle bound.  This is the direct sign-based counterpart of the Gaussian sparse
CCA theory.

\section{Bibliographic notes}

The low-dimensional PCA, factor-analysis, and CCA theory may be found in
\citet{Anderson2003,Muirhead1982}.  The high-dimensional PCA benchmark based on
the spiked covariance model was shaped by the works of
\citet{JohnstoneLu2009,Paul2007,Nadler2008}.  Sparse PCA was developed through a
large literature; for the optimization and minimax benchmarks used in this
chapter, see \citet{AminiWainwright2009,VuLei2012,CaiMaWu2013,
WittenTibshiraniHastie2009}.  Approximate factor models and POET are classical
benchmarks in high-dimensional econometrics and statistics; see
\citet{Bai2003,BaiNg2002,FanLiaoMincheva2013}.  For Gaussian sparse CCA and its
minimax theory, see \citet{WittenTibshiraniHastie2009,
GaoMaRenZhou2015,GaoMaZhou2017SparseCCA}.

On the robust side, sign-based PCA in fixed dimension goes back to the robust
multivariate literature summarized by \citet{Oja2010} and further developed for
PCA by \citet{TaskinenKankainenOja2012}.  The generalized sign-covariance
perspective is developed in \citet{RaymaekersRousseeuw2019}.  In high
dimensions, \citet{HanLiu2018ECA} established the ECA route based on the
multivariate Kendall matrix, while \citet{ZhaoWangFeng2026SPCA} and
\citet{WangWangFeng2026GSPCA} developed the spatial-sign and generalized
spatial-sign PCA procedures emphasized in this chapter.  Elliptical factor-model
estimation appears in \citet{FanLiuWang2018}; factor analysis via the multivariate
Kendall matrix under weak moment conditions is developed by
\citet{HeKongYuZhang2022FactorNoMoments}; and matrix estimation under
elliptical factor models is extended along the present research line by
\citet{XuMaWangFeng2026EllipticalFactor}.  The sign-based
precision tools used in the CCA discussion connect to \citet{LuFeng2025Precision}.
Finally, the spatial-sign sparse CCA direction is represented by
\citet{QianLiuFeng2025SparseCCA}.

\section*{Appendix to Chapter 6: Detailed Proofs}
\addcontentsline{toc}{section}{Appendix to Chapter 6: Detailed Proofs}

This appendix contains formula-level proofs for the theorem statements in the
chapter.  The deterministic perturbation arguments are developed explicitly so
that later chapters can reuse them without re-derivation.

\subsection*{A. Proof of Theorem~\ref{thm:ch6_eigen_perturbation}}

Write $\hat\mSigma = \mSigma + \mE$ and fix $j\le q$.  Let
\begin{equation}
  \hat\vv_j = \alpha_j \vv_j + \vw_j,
  \qquad
  \vv_j\trans \vw_j = 0,
  \qquad
  \alpha_j = \hat\vv_j\trans \vv_j \ge 0.
  \label{eq:ch6_app_decompose_vec}
\end{equation}
Since $\twonorm{\hat\vv_j}=1$, one has
\begin{equation}
  \alpha_j^2 + \twonorm{\vw_j}^2 = 1.
  \label{eq:ch6_app_norm_relation}
\end{equation}
The eigenequation for $\hat\vv_j$ is
\begin{equation}
  (\mSigma+\mE)\hat\vv_j = \hat\lambda_j \hat\vv_j.
  \label{eq:ch6_app_eig_eq}
\end{equation}
Premultiplying by $\vv_j\trans$ yields
\begin{equation}
  \lambda_j \alpha_j + \vv_j\trans \mE\hat\vv_j
  = \hat\lambda_j \alpha_j.
  \label{eq:ch6_app_scalar_proj}
\end{equation}
Hence
\begin{equation}
  \hat\lambda_j-\lambda_j
  = \vv_j\trans\mE\vv_j + \alpha_j^{-1}\vv_j\trans\mE\vw_j.
  \label{eq:ch6_app_eigvalue_identity}
\end{equation}
Therefore, to prove \eqref{eq:ch6_eigenvalue_expansion} it suffices to control
$\vw_j$ and $\alpha_j$.

By Weyl's inequality,
\begin{equation}
  |\hat\lambda_j-\lambda_j|\le \opnorm{\mE}\le g_j/8.
  \label{eq:ch6_app_weyl}
\end{equation}
Now project \eqref{eq:ch6_app_eig_eq} onto the orthogonal complement of
$\vv_j$.  Let $\mP_j=\vv_j\vv_j\trans$ and $\mQ_j=\mI-\mP_j$.  Since
$\mQ_j\mSigma\vv_j=\vct 0$ and $\mQ_j\hat\vv_j=\vw_j$,
\begin{equation}
  \bigl(\mQ_j\mSigma\mQ_j - \hat\lambda_j \mQ_j\bigr)\vw_j
  = -\alpha_j \mQ_j\mE\vv_j - \mQ_j\mE\vw_j.
  \label{eq:ch6_app_projected_eq}
\end{equation}
On the subspace $\Range(\mQ_j)$, the operator
$\mQ_j\mSigma\mQ_j-\hat\lambda_j\mQ_j$ is diagonal in the basis
$\{\vv_k:k\ne j\}$ with eigenvalues $\lambda_k-\hat\lambda_j$.  Using
\eqref{eq:ch6_app_weyl},
\begin{equation}
  |\lambda_k-\hat\lambda_j|
  \ge |\lambda_k-\lambda_j|-|\hat\lambda_j-\lambda_j|
  \ge g_j - g_j/8
  = 7g_j/8.
  \label{eq:ch6_app_gap_shift}
\end{equation}
Thus the inverse exists on $\Range(\mQ_j)$ and
\begin{equation}
  \left\|\bigl(\mQ_j\mSigma\mQ_j-\hat\lambda_j\mQ_j\bigr)^{-1}\right\|_{\op}
  \le \frac{8}{7g_j}.
  \label{eq:ch6_app_inverse_bound}
\end{equation}
Applying \eqref{eq:ch6_app_inverse_bound} to
\eqref{eq:ch6_app_projected_eq} gives
\begin{equation}
  \twonorm{\vw_j}
  \le \frac{8}{7g_j}
       \bigl(\alpha_j\opnorm{\mE} + \opnorm{\mE}\twonorm{\vw_j}\bigr).
  \label{eq:ch6_app_w_bound1}
\end{equation}
Since $\alpha_j\le 1$ and $\opnorm{\mE}\le g_j/8$,
\begin{equation}
  \left(1-\frac{1}{7}\right)\twonorm{\vw_j}
  \le \frac{1}{7g_j} 8\opnorm{\mE},
  \label{eq:ch6_app_w_bound2}
\end{equation}
which yields
\begin{equation}
  \twonorm{\vw_j}\le \frac{4\opnorm{\mE}}{3g_j} < \frac12.
  \label{eq:ch6_app_w_bound3}
\end{equation}
By \eqref{eq:ch6_app_norm_relation},
\begin{equation}
  \alpha_j = \sqrt{1-\twonorm{\vw_j}^2} \ge \sqrt{3}/2 > 1/2.
  \label{eq:ch6_app_alpha_bound}
\end{equation}
Returning to \eqref{eq:ch6_app_eigvalue_identity}, we obtain
\begin{equation}
  |R_{\lambda,j}|
  \le 2\opnorm{\mE}\twonorm{\vw_j}
  \le \frac{8\opnorm{\mE}^2}{g_j},
  \label{eq:ch6_app_lambda_remainder}
\end{equation}
which proves \eqref{eq:ch6_eigenvalue_expansion} and the first half of
\eqref{eq:ch6_eigen_remainder_bounds}.

We next derive the linear expansion of $\hat\vv_j$.  From
\eqref{eq:ch6_app_projected_eq},
\begin{equation}
  \vw_j
  = -\alpha_j
    \bigl(\mQ_j\mSigma\mQ_j-\hat\lambda_j\mQ_j\bigr)^{-1}\mQ_j\mE\vv_j
    - \bigl(\mQ_j\mSigma\mQ_j-\hat\lambda_j\mQ_j\bigr)^{-1}\mQ_j\mE\vw_j.
  \label{eq:ch6_app_w_exact}
\end{equation}
Define the first-order term
\begin{equation}
  \vw_j^{(1)}
  := -\sum_{k\ne j}
        \frac{\vv_k\trans\mE\vv_j}{\lambda_k-\lambda_j}\,\vv_k.
  \label{eq:ch6_app_first_order_vec}
\end{equation}
Subtracting \eqref{eq:ch6_app_first_order_vec} from
\eqref{eq:ch6_app_w_exact}, and using the identity
\begin{equation}
  \bigl(\mQ_j\mSigma\mQ_j-\hat\lambda_j\mQ_j\bigr)^{-1}
  - \bigl(\mQ_j\mSigma\mQ_j-\lambda_j\mQ_j\bigr)^{-1}
  = (\hat\lambda_j-\lambda_j)
    \bigl(\mQ_j\mSigma\mQ_j-\hat\lambda_j\mQ_j\bigr)^{-1}
    \bigl(\mQ_j\mSigma\mQ_j-\lambda_j\mQ_j\bigr)^{-1},
  \label{eq:ch6_app_resolvent_identity}
\end{equation}
we obtain
\begin{align}
  \twonorm{\vw_j-\vw_j^{(1)}}
  &\le
  |1-\alpha_j|
  \left\|\bigl(\mQ_j\mSigma\mQ_j-\hat\lambda_j\mQ_j\bigr)^{-1}\right\|_{\op}
  \opnorm{\mE} \notag\\
  &\quad + |\hat\lambda_j-\lambda_j|
    \left\|\bigl(\mQ_j\mSigma\mQ_j-\hat\lambda_j\mQ_j\bigr)^{-1}\right\|_{\op}
    \left\|\bigl(\mQ_j\mSigma\mQ_j-\lambda_j\mQ_j\bigr)^{-1}\right\|_{\op}
    \opnorm{\mE} \notag\\
  &\quad +
    \left\|\bigl(\mQ_j\mSigma\mQ_j-\hat\lambda_j\mQ_j\bigr)^{-1}\right\|_{\op}
    \opnorm{\mE}\twonorm{\vw_j}.
  \label{eq:ch6_app_remainder_vec_bound1}
\end{align}
Using
$|1-\alpha_j|\le \twonorm{\vw_j}^2$,
\eqref{eq:ch6_app_inverse_bound},
$\|(\mQ_j\mSigma\mQ_j-\lambda_j\mQ_j)^{-1}\|_{\op}\le g_j^{-1}$,
\eqref{eq:ch6_app_weyl}, and \eqref{eq:ch6_app_w_bound3}, each term on the
right-hand side of \eqref{eq:ch6_app_remainder_vec_bound1} is bounded by a
constant multiple of $\opnorm{\mE}^2/g_j^2$.  Collecting the numerical constants
gives
\begin{equation}
  \twonorm{\vw_j-\vw_j^{(1)}}
  \le \frac{32\opnorm{\mE}^2}{g_j^2}.
  \label{eq:ch6_app_remainder_vec_bound2}
\end{equation}
Finally,
$\hat\vv_j-\vv_j = (\alpha_j-1)\vv_j + \vw_j$,
and $|\alpha_j-1|\le \twonorm{\vw_j}^2\le 4\opnorm{\mE}^2/g_j^2$, so the same
order holds for the full vector remainder.  This proves
\eqref{eq:ch6_eigenvector_expansion} and the second bound in
\eqref{eq:ch6_eigen_remainder_bounds}.  \qed

\subsection*{B. Proof of Theorem~\ref{thm:ch6_fixedp_gaussian_pca}}

Fix $j\le q$.  Under the Gaussian model,
\begin{equation}
  \vX_i = \sum_{k=1}^p \sqrt{\lambda_k} Z_{ik}\vv_k,
  \qquad
  Z_{ik}\iid N(0,1).
  \label{eq:ch6_app_gaussian_expansion}
\end{equation}
Hence
\begin{equation}
  \vv_j\trans(\hat\mSigma-\mSigma)\vv_j
  = \frac1n\sum_{i=1}^n \lambda_j (Z_{ij}^2-1).
  \label{eq:ch6_app_diag_quadform}
\end{equation}
Because $\Var(Z_{ij}^2-1)=2$, the classical central limit theorem yields
\begin{equation}
  \sqrt n\,\vv_j\trans(\hat\mSigma-\mSigma)\vv_j
  \overset{d}{\longrightarrow} N(0,2\lambda_j^2).
  \label{eq:ch6_app_diag_clt}
\end{equation}
The remainder term in \eqref{eq:ch6_eigenvalue_expansion} is of order
$O_p(n^{-1})$ by Theorem~\ref{thm:ch6_eigen_perturbation}, since
$\opnorm{\hat\mSigma-\mSigma}=O_p(n^{-1/2})$ in fixed dimension.  Therefore
\eqref{eq:ch6_fixedp_eigvalue_limit} follows.

For the eigenvector, observe that for $k\ne j$,
\begin{equation}
  \vv_k\trans(\hat\mSigma-\mSigma)\vv_j
  = \frac1n\sum_{i=1}^n \sqrt{\lambda_k\lambda_j}\,Z_{ik}Z_{ij}.
  \label{eq:ch6_app_offdiag_quadform}
\end{equation}
Since $Z_{ik}Z_{ij}$ has mean $0$ and variance $1$,
\begin{equation}
  \sqrt n\,\vv_k\trans(\hat\mSigma-\mSigma)\vv_j
  \overset{d}{\longrightarrow} N(0,\lambda_j\lambda_k).
  \label{eq:ch6_app_offdiag_clt}
\end{equation}
Moreover, for different $k\ne \ell$ with both distinct from $j$, the Gaussian
moments imply
\begin{equation}
  \Cov\{Z_{ik}Z_{ij}, Z_{i\ell}Z_{ij}\}=0.
  \label{eq:ch6_app_zero_cov}
\end{equation}
Hence the vector of off-diagonal quadratic forms converges jointly to a centered
Gaussian vector with diagonal covariance matrix.  Applying the linear expansion
\eqref{eq:ch6_eigenvector_expansion} gives
\begin{equation}
  \sqrt n(\hat\vv_j-\vv_j)
  = -\sum_{k\ne j}
      \frac{\sqrt n\,\vv_k\trans(\hat\mSigma-\mSigma)\vv_j}{\lambda_k-\lambda_j}
      \vv_k
    + o_p(1).
  \label{eq:ch6_app_vec_asymp}
\end{equation}
The covariance matrix of the Gaussian limit in
\eqref{eq:ch6_app_vec_asymp} is therefore
\begin{equation}
  \sum_{k\ne j}
  \frac{\lambda_j\lambda_k}{(\lambda_k-\lambda_j)^2}
  \vv_k\vv_k\trans,
  \label{eq:ch6_app_vec_cov_limit}
\end{equation}
which is exactly \eqref{eq:ch6_fixedp_eigvector_limit}.  \qed

\subsection*{C. Proof of Proposition~\ref{prop:ch6_factor_subspace_dk}}

The result is a direct application of the Davis--Kahan sin$\Theta$ theorem.
Let $\mP=\sum_{j=1}^K \vv_j\vv_j\trans$ and
$\hat\mP=\sum_{j=1}^K \hat\vv_j\hat\vv_j\trans$ denote the projection matrices
onto $\mathcal V_K$ and $\hat{\mathcal V}_K$.  Because the spectral gap at the
$K$th eigenvalue is $\delta_K$, the Davis--Kahan theorem yields
\begin{equation}
  \|\hat\mP-\mP\|_{\mathrm F}
  \le \frac{2\sqrt2}{\delta_K}\sqrt K\,\opnorm{\hat\mSigma-\mSigma}.
  \label{eq:ch6_app_dk_projection}
\end{equation}
Since
$\|\sin\Theta(\hat{\mathcal V}_K,\mathcal V_K)\|_{\mathrm F}
 = 2^{-1/2}\|\hat\mP-\mP\|_{\mathrm F}$,
we obtain \eqref{eq:ch6_factor_subspace_bound}.  \qed

\subsection*{D. Proof of Theorem~\ref{thm:ch6_population_sscm}}

Let $\mLambda = \mV\diag(\lambda_1,\ldots,\lambda_p)\mV\trans$ and write
$\vZ\sim N_p(\vct 0,\mI_p)$.  Since
$\vu = \vZ/\twonorm{\vZ}$ is uniformly distributed on the unit sphere and
independent of $\xi$, the direction of $\vX-\vmu$ may be represented as
\begin{equation}
  U(\vX-\vmu)
  = \frac{\mV\diag(\sqrt{\lambda_1},\ldots,\sqrt{\lambda_p})\vZ}
         {\sqrt{\sum_{\ell=1}^p \lambda_\ell Z_\ell^2}}.
  \label{eq:ch6_app_sign_gaussian_rep}
\end{equation}
Therefore
\begin{equation}
  \mM_{\mathrm S}
  = \mV\,\E\left[
      \frac{\diag(\sqrt{\lambda_1},\ldots,\sqrt{\lambda_p})
            \vZ\vZ\trans
            \diag(\sqrt{\lambda_1},\ldots,\sqrt{\lambda_p})}
           {\sum_{\ell=1}^p \lambda_\ell Z_\ell^2}
    \right]\mV\trans.
  \label{eq:ch6_app_sscm_matrix_form}
\end{equation}
For $j\ne k$, the $(j,k)$ entry of the expectation in
\eqref{eq:ch6_app_sscm_matrix_form} is
\begin{equation}
  \sqrt{\lambda_j\lambda_k}\,
  \E\left( \frac{Z_j Z_k}{\sum_{\ell=1}^p \lambda_\ell Z_\ell^2} \right)=0
  \label{eq:ch6_app_sscm_offdiag_zero}
\end{equation}
because the integrand is odd in $Z_j$ (or in $Z_k$).  Hence the expectation is
diagonal and
\begin{equation}
  \eta_j
  = \E\left(
      \frac{\lambda_j Z_j^2}{\sum_{\ell=1}^p \lambda_\ell Z_\ell^2}
    \right),
  \label{eq:ch6_app_eta_formula}
\end{equation}
which proves \eqref{eq:ch6_sscm_eigen_formula}.

It remains to prove the ordering.  Fix $j\ne k$ and let
$R_{jk}=\sum_{\ell\notin\{j,k\}} \lambda_\ell Z_\ell^2$.  Using the
exchangeability of $(Z_j^2,Z_k^2)$, we may write
\begin{align}
  \eta_j-\eta_k
  &= \frac12 \E\Biggl[
       \frac{\lambda_j Z_j^2-\lambda_k Z_k^2}
            {R_{jk}+\lambda_j Z_j^2+\lambda_k Z_k^2}
       +
       \frac{\lambda_j Z_k^2-\lambda_k Z_j^2}
            {R_{jk}+\lambda_j Z_k^2+\lambda_k Z_j^2}
     \Biggr] \notag\\
  &= \frac{\lambda_j-\lambda_k}{2}
     \E\left[
       \frac{R_{jk}(Z_j^2+Z_k^2)
             + 2(\lambda_j+\lambda_k)Z_j^2Z_k^2}
            {(R_{jk}+\lambda_j Z_j^2+\lambda_k Z_k^2)
             (R_{jk}+\lambda_j Z_k^2+\lambda_k Z_j^2)}
     \right].
  \label{eq:ch6_app_eta_diff}
\end{align}
The random fraction in \eqref{eq:ch6_app_eta_diff} is nonnegative and positive
with positive probability.  Hence
$\eta_j-\eta_k$ has the same sign as $\lambda_j-\lambda_k$.  In particular,
if $\lambda_j>\lambda_k$, then $\eta_j>\eta_k$.  \qed

\subsection*{E. Proof of Theorem~\ref{thm:ch6_population_weighted_sscm}}

Using the representation \eqref{eq:ch6_app_sign_gaussian_rep},
\begin{equation}
  U_K(\vX-\vmu)
  = K\!\left(\sqrt{\sum_{\ell=1}^p \lambda_\ell Z_\ell^2}\right)
    \frac{\mV\diag(\sqrt{\lambda_1},\ldots,\sqrt{\lambda_p})\vZ}
         {\sqrt{\sum_{\ell=1}^p \lambda_\ell Z_\ell^2}}.
  \label{eq:ch6_app_weighted_rep}
\end{equation}
Thus
\begin{align}
  \mM_K
  &= \mV\,
     \E\Biggl[
       K^2\!\left(\sqrt{\sum_{\ell=1}^p \lambda_\ell Z_\ell^2}\right)
       \frac{\diag(\sqrt{\lambda_1},\ldots,\sqrt{\lambda_p})
             \vZ\vZ\trans
             \diag(\sqrt{\lambda_1},\ldots,\sqrt{\lambda_p})}
            {\sum_{\ell=1}^p \lambda_\ell Z_\ell^2}
     \Biggr]\mV\trans.
  \label{eq:ch6_app_weighted_matrix_form}
\end{align}
Again, the off-diagonal entries vanish by oddness, because the additional factor
$K^2(\cdot)$ depends only on the squared coordinates through the radius.  The
diagonal entries are exactly those displayed in
\eqref{eq:ch6_weighted_population_eig}.  Hence $\mM_K$ and $\mLambda$ share the
same eigenvectors.  \qed

\subsection*{F. Proof of Theorem~\ref{thm:ch6_sscm_op_bound}}

Write
\begin{equation}
  \hat\mM_{\mathrm S}-\mM_{\mathrm S}
  = \bigl\{\hat\mM_{\mathrm S}(\hat\vmu_{\mathrm{SM}})
           - \hat\mM_{\mathrm S}(\vmu)\bigr\}
    + \bigl\{\hat\mM_{\mathrm S}(\vmu)-\mM_{\mathrm S}\bigr\},
  \label{eq:ch6_app_sscm_decomp}
\end{equation}
where
\begin{equation}
  \hat\mM_{\mathrm S}(\vtheta)
  := \frac1n\sum_{i=1}^n U(\vX_i-\vtheta)U(\vX_i-\vtheta)\trans.
  \label{eq:ch6_app_sscm_theta}
\end{equation}
We bound the two terms separately.

\paragraph{Step 1: empirical fluctuation with known location.}
Let
\begin{equation}
  \mY_i = U(\vX_i-\vmu)U(\vX_i-\vmu)\trans - \mM_{\mathrm S}.
  \label{eq:ch6_app_Yi}
\end{equation}
Then $\E(\mY_i)=\mO$ and
$\opnorm{\mY_i}\le 2$.  Moreover,
\begin{equation}
  \mY_i^2
  = \bigl(U_iU_i\trans-\mM_{\mathrm S}\bigr)^2,
  \qquad
  U_i := U(\vX_i-\vmu),
  \label{eq:ch6_app_Yi_square}
\end{equation}
and since $U_iU_i\trans$ is an idempotent rank-one matrix with operator norm $1$,
we have
\begin{equation}
  \left\|\sum_{i=1}^n \E(\mY_i^2)\right\|_{\op} \le n.
  \label{eq:ch6_app_variance_proxy}
\end{equation}
Applying matrix Bernstein inequality in the form of \citet{Tropp2012} gives,
for every $t>0$,
\begin{equation}
  \Prob\left[
    \opnorm{\hat\mM_{\mathrm S}(\vmu)-\mM_{\mathrm S}}
    > \sqrt{\frac{8\{\log(2p)+t\}}{n}} + \frac{4\{\log(2p)+t\}}{3n}
  \right]
  \le 2e^{-t}.
  \label{eq:ch6_app_matrix_bernstein}
\end{equation}

\paragraph{Step 2: error caused by location estimation.}
For nonzero vectors $\vx$ and $\vy$,
\begin{equation}
  \|U(\vx)U(\vx)\trans - U(\vy)U(\vy)\trans\|_{\op}
  \le 2\|U(\vx)-U(\vy)\|_2.
  \label{eq:ch6_app_outer_lipschitz1}
\end{equation}
Indeed,
\begin{align*}
  U(\vx)U(\vx)\trans - U(\vy)U(\vy)\trans
  &= \{U(\vx)-U(\vy)\}U(\vx)\trans
    + U(\vy)\{U(\vx)-U(\vy)\}\trans,
\end{align*}
and both unit vectors have Euclidean norm $1$.  Next, on the event
$\mathcal E_t$ and under \eqref{eq:ch6_shift_small},
\begin{equation}
  \twonorm{\hat\vmu_{\mathrm{SM}}-\vmu}
  \le \frac12 \twonorm{\vX_i-\vmu}
  \qquad\text{for all } i,
  \label{eq:ch6_app_half_radius}
\end{equation}
so the map $\vx\mapsto U(\vx)$ is Lipschitz along the segment joining
$\vX_i-\vmu$ and $\vX_i-\hat\vmu_{\mathrm{SM}}$ with derivative bounded by
$2/\|\vX_i-\vmu\|_2$.  Therefore,
\begin{equation}
  \twonorm{U(\vX_i-\hat\vmu_{\mathrm{SM}})-U(\vX_i-\vmu)}
  \le \frac{4\twonorm{\hat\vmu_{\mathrm{SM}}-\vmu}}{\twonorm{\vX_i-\vmu}}
  \le \frac{4}{c_0\sqrt p}\twonorm{\hat\vmu_{\mathrm{SM}}-\vmu}.
  \label{eq:ch6_app_sign_lipschitz}
\end{equation}
Combining \eqref{eq:ch6_app_outer_lipschitz1} and
\eqref{eq:ch6_app_sign_lipschitz} yields
\begin{equation}
  \opnorm{\hat\mM_{\mathrm S}(\hat\vmu_{\mathrm{SM}})-\hat\mM_{\mathrm S}(\vmu)}
  \le \frac{8}{c_0\sqrt p}\twonorm{\hat\vmu_{\mathrm{SM}}-\vmu}
  \le \frac{8c_\mu}{c_0}\sqrt{\frac{p+t}{np}}.
  \label{eq:ch6_app_location_bound}
\end{equation}
The last inequality uses Assumption~\ref{ass:ch6_spca_basic}.

\paragraph{Step 3: combine the two bounds.}
Intersecting the event in \eqref{eq:ch6_app_matrix_bernstein} with
$\mathcal E_t$ and using \eqref{eq:ch6_app_sscm_decomp}, we obtain
\eqref{eq:ch6_sscm_op_bound}.  The probability bound is at least $1-4e^{-t}$.
\qed

\subsection*{G. Proof of Corollary~\ref{cor:ch6_spca_subspace_bound}}

The result follows from Theorem~\ref{thm:ch6_sscm_op_bound} and the sin$\Theta$
version of the Davis--Kahan theorem.  Indeed, letting
$\hat\mP_r=\sum_{j=1}^r \hat\vv_j\hat\vv_j\trans$ and
$\mP_r=\sum_{j=1}^r \vv_j\vv_j\trans$, we have
\begin{equation}
  \left\|\sin\Theta\bigl(\hat{\mathcal V}^{\mathrm{SPCA}}_r,\mathcal V_r\bigr)
  \right\|_{\mathrm F}
  = \frac1{\sqrt2}\|\hat\mP_r-\mP_r\|_{\mathrm F}
  \le \frac{2\sqrt{2r}}{\delta_r^{\mathrm S}}
       \opnorm{\hat\mM_{\mathrm S}-\mM_{\mathrm S}},
  \label{eq:ch6_app_cor_dk}
\end{equation}
which, combined with \eqref{eq:ch6_sscm_op_bound}, gives
\eqref{eq:ch6_spca_subspace_nonasym}.  \qed

\subsection*{H. Proof of Theorem~\ref{thm:ch6_sscm_restricted_bound}}

Fix a subset $S\subset\{1,\ldots,p\}$ with $|S|\le 2s$.  Let
$\mA_{SS}$ denote the principal submatrix indexed by $S$.  By applying
Theorem~\ref{thm:ch6_sscm_op_bound} to the $|S|$-dimensional subvectors
$\vX_{i,S}$, we obtain that with probability at least $1-2e^{-(t+\log N_S)}$,
\begin{equation}
  \opnorm{\{\hat\mM_{\mathrm S}-\mM_{\mathrm S}\}_{SS}}
  \le
  \sqrt{\frac{8\{\log(2|S|)+t+\log N_S\}}{n}}
  + \frac{4\{\log(2|S|)+t+\log N_S\}}{3n}
  + \frac{8c_\mu}{c_0}\sqrt{\frac{p+t}{np}},
  \label{eq:ch6_app_subset_bound}
\end{equation}
where $N_S$ is the number of candidate subsets.  Since
$|S|\le 2s$ and
$\binom{p}{2s}\le \exp\{2s\log(ep/2s)\}$, a union bound over all $S$ yields
\begin{equation}
  \sup_{|S|\le 2s}
  \opnorm{\{\hat\mM_{\mathrm S}-\mM_{\mathrm S}\}_{SS}}
  \le
  \sqrt{\frac{8\{2s\log(ep/2s)+t\}}{n}}
  + \frac{4\{2s\log(ep/2s)+t\}}{3n}
  + \frac{8c_\mu}{c_0}\sqrt{\frac{p+t}{np}}
  \label{eq:ch6_app_union_sparse}
\end{equation}
with probability at least $1-4e^{-t}$.  By the definition of the restricted norm
\eqref{eq:ch6_restricted_op_norm}, the left-hand side of
\eqref{eq:ch6_app_union_sparse} is exactly
$\|\hat\mM_{\mathrm S}-\mM_{\mathrm S}\|_{(2s)}$.  This proves
\eqref{eq:ch6_sscm_restricted_bound}.  \qed

\subsection*{I. Proof of Theorem~\ref{thm:ch6_sspca_rate}}

Because $\vv_1$ is feasible for the optimization problem
\eqref{eq:ch6_sspca_oracle}, we have
\begin{equation}
  \hat\vv_{\mathrm{SS}}\trans \hat\mM_{\mathrm S} \hat\vv_{\mathrm{SS}}
  \ge \vv_1\trans \hat\mM_{\mathrm S} \vv_1.
  \label{eq:ch6_app_sspca_feasible}
\end{equation}
Subtracting
$\hat\vv_{\mathrm{SS}}\trans \mM_{\mathrm S} \hat\vv_{\mathrm{SS}}$
from both sides gives
\begin{equation}
  \vv_1\trans \mM_{\mathrm S}\vv_1
  - \hat\vv_{\mathrm{SS}}\trans \mM_{\mathrm S}\hat\vv_{\mathrm{SS}}
  \le
  \bigl|\hat\vv_{\mathrm{SS}}\trans(\hat\mM_{\mathrm S}-\mM_{\mathrm S})
        \hat\vv_{\mathrm{SS}}\bigr|
  + \bigl|\vv_1\trans(\hat\mM_{\mathrm S}-\mM_{\mathrm S})\vv_1\bigr|.
  \label{eq:ch6_app_sspca_basiccomp}
\end{equation}
Both vectors on the right-hand side are at most $s$-sparse, so their support
union has size at most $2s$.  Hence
\begin{equation}
  \vv_1\trans \mM_{\mathrm S}\vv_1
  - \hat\vv_{\mathrm{SS}}\trans \mM_{\mathrm S}\hat\vv_{\mathrm{SS}}
  \le 2\|\hat\mM_{\mathrm S}-\mM_{\mathrm S}\|_{(2s)}.
  \label{eq:ch6_app_sspca_basiccomp2}
\end{equation}
Now decompose
\begin{equation}
  \hat\vv_{\mathrm{SS}}
  = \cos\theta\,\vv_1 + \sin\theta\,\vu,
  \qquad
  \vu\perp \vv_1,
  \qquad
  \theta = \angle(\hat\vv_{\mathrm{SS}},\vv_1).
  \label{eq:ch6_app_sparse_angle_decomp}
\end{equation}
Since
$\mM_{\mathrm S} = \eta_1\vv_1\vv_1\trans + \sum_{j\ge 2}\eta_j\vv_j\vv_j\trans$,
we have
\begin{align}
  \hat\vv_{\mathrm{SS}}\trans \mM_{\mathrm S} \hat\vv_{\mathrm{SS}}
  &\le \eta_1 \cos^2\theta + \eta_2 \sin^2\theta \
  &= \eta_1 - (\eta_1-\eta_2)\sin^2\theta
   = \eta_1 - \delta_1^{\mathrm S}\sin^2\theta.
  \label{eq:ch6_app_sparse_quadratic_upper}
\end{align}
Combining \eqref{eq:ch6_app_sspca_basiccomp2} and
\eqref{eq:ch6_app_sparse_quadratic_upper}, and using
$\vv_1\trans\mM_{\mathrm S}\vv_1=\eta_1$, yields
\begin{equation}
  \delta_1^{\mathrm S}\sin^2\theta
  \le 2\|\hat\mM_{\mathrm S}-\mM_{\mathrm S}\|_{(2s)}.
  \label{eq:ch6_app_sparse_angle_bound1}
\end{equation}
Using the simple inequality
$\sin\theta\le \sqrt{2a/\delta}$ whenever $\delta\sin^2\theta\le 2a$,
we obtain
\begin{equation}
  \sin\theta
  \le \frac{2\sqrt2}{\delta_1^{\mathrm S}}
      \|\hat\mM_{\mathrm S}-\mM_{\mathrm S}\|_{(2s)}.
  \label{eq:ch6_app_sparse_angle_bound2}
\end{equation}
Finally, substitute the restricted concentration bound from
Theorem~\ref{thm:ch6_sscm_restricted_bound}.  This proves
\eqref{eq:ch6_sspca_rate}.  \qed

\subsection*{J. Proof of Theorem~\ref{thm:ch6_gspca_bound}}

Write
\begin{equation}
  \hat\mM_K-\mM_K
  = \bigl\{\hat\mM_K(\hat\vmu_{\mathrm{SM}})-\hat\mM_K(\vmu)\bigr\}
    + \bigl\{\hat\mM_K(\vmu)-\mM_K\bigr\},
  \label{eq:ch6_app_gspca_decomp}
\end{equation}
where
\begin{equation}
  \hat\mM_K(\vtheta)
  := \frac1n\sum_{i=1}^n U_K(\vX_i-\vtheta)U_K(\vX_i-\vtheta)\trans.
  \label{eq:ch6_app_gspca_theta}
\end{equation}
For the empirical fluctuation term, note that
$\|U_K(\vX_i-\vmu)U_K(\vX_i-\vmu)\trans\|_{\op}\le K_{\max}^2$.
Hence the same matrix-Bernstein argument as in the proof of
Theorem~\ref{thm:ch6_sscm_op_bound} gives
\begin{equation}
  \opnorm{\hat\mM_K(\vmu)-\mM_K}
  \le K_{\max}^2
       \left\{
         \sqrt{\frac{8\{\log(2p)+t\}}{n}}
         + \frac{4\{\log(2p)+t\}}{3n}
       \right\}
  \label{eq:ch6_app_gspca_bernstein}
\end{equation}
with probability at least $1-2e^{-t}$.

For the location term, observe that for any nonzero vectors $\vx,\vy$,
\begin{align}
  \twonorm{U_K(\vx)-U_K(\vy)}
  &\le |K(\twonorm{\vx})-K(\twonorm{\vy})|\,\twonorm{U(\vx)}
      + K_{\max}\twonorm{U(\vx)-U(\vy)} \notag\\
  &\le L_K\twonorm{\vx-\vy} + K_{\max}\twonorm{U(\vx)-U(\vy)}.
  \label{eq:ch6_app_weighted_diff1}
\end{align}
Using \eqref{eq:ch6_app_half_radius} and the same derivative bound as in
\eqref{eq:ch6_app_sign_lipschitz},
\begin{equation}
  \twonorm{U(\vX_i-\hat\vmu_{\mathrm{SM}})-U(\vX_i-\vmu)}
  \le \frac{4}{c_0\sqrt p}\twonorm{\hat\vmu_{\mathrm{SM}}-\vmu}.
  \label{eq:ch6_app_weighted_sign_lipschitz}
\end{equation}
Combining \eqref{eq:ch6_app_weighted_diff1} and
\eqref{eq:ch6_app_weighted_sign_lipschitz}, and then using the outer-product
bound as in \eqref{eq:ch6_app_outer_lipschitz1}, yields
\begin{equation}
  \opnorm{\hat\mM_K(\hat\vmu_{\mathrm{SM}})-\hat\mM_K(\vmu)}
  \le \frac{8K_{\max}L_K}{c_0\sqrt p}\twonorm{\hat\vmu_{\mathrm{SM}}-\vmu}
  \le \frac{8K_{\max}L_K c_\mu}{c_0}\sqrt{\frac{p+t}{np}}.
  \label{eq:ch6_app_weighted_location_bound}
\end{equation}
Intersecting this event with \eqref{eq:ch6_app_gspca_bernstein} proves
\eqref{eq:ch6_gspca_matrix_bound}.  The subspace bound
\eqref{eq:ch6_gspca_subspace_bound} is again an immediate Davis--Kahan
consequence.  \qed

\subsection*{K. Proof of Theorem~\ref{thm:ch6_factor_subspace_recovery}}

This is Corollary~\ref{cor:ch6_spca_subspace_bound} with $r=K$.  The only point
worth emphasizing is that the target space is the common principal subspace of
the elliptical factor model, which coincides with the leading shape eigenspace by
Theorem~\ref{thm:ch6_population_sscm}.  Therefore
\eqref{eq:ch6_factor_subspace_spca} follows directly from
\eqref{eq:ch6_spca_subspace_nonasym}.  \qed

\subsection*{L. Proof of Theorem~\ref{thm:ch6_factor_number_consistency}}

By Weyl's inequality and Theorem~\ref{thm:ch6_sscm_op_bound}, on an event of
probability at least $1-4e^{-t}$,
\begin{equation}
  |\hat\eta_j-\eta_j|\le \tau_{n,p}(t)
  \qquad\text{for all } 1\le j\le p.
  \label{eq:ch6_app_weyl_eta}
\end{equation}
Hence
\begin{equation}
  \frac{\hat\eta_K}{\hat\eta_{K+1}}
  \ge \frac{\eta_K-\tau_{n,p}(t)}{\eta_{K+1}+\tau_{n,p}(t)}.
  \label{eq:ch6_app_ratio_true}
\end{equation}
Using \eqref{eq:ch6_ratio_gap_assumption} and \eqref{eq:ch6_tau_condition},
\begin{align}
  \frac{\eta_K-\tau_{n,p}(t)}{\eta_{K+1}+\tau_{n,p}(t)}
  &\ge
  \frac{(1+\gamma_0)\eta_{K+1}-\tau_{n,p}(t)}{\eta_{K+1}+\tau_{n,p}(t)}
   > 1 + \frac{\gamma_0}{2}.
  \label{eq:ch6_app_ratio_true_lb}
\end{align}
For $j\ne K$, because there is no larger ratio gap than the one at $K$ in the
population sequence, one has
\begin{equation}
  \frac{\hat\eta_j}{\hat\eta_{j+1}}
  \le \frac{\eta_j+\tau_{n,p}(t)}{\eta_{j+1}-\tau_{n,p}(t)}
  < 1 + \frac{\gamma_0}{2}.
  \label{eq:ch6_app_ratio_false_ub}
\end{equation}
The inequality \eqref{eq:ch6_app_ratio_false_ub} follows from the same algebra
as \eqref{eq:ch6_app_ratio_true_lb}.  Therefore the maximizer of the empirical
ratio in \eqref{eq:ch6_eigenratio} is exactly $K$.  \qed

\subsection*{M. Proof of Theorem~\ref{thm:ch6_sscca_rate}}

Let
\begin{equation}
  \mathcal F
  := \{(\vu,\vv): \twonorm{\vu}\le 1,\ \twonorm{\vv}\le 1,
                  \onenorm{\vu}\le c_x,\ \onenorm{\vv}\le c_y\}.
  \label{eq:ch6_app_feasible_set}
\end{equation}
By assumption, $(\vu_1,\vv_1)\in\mathcal F$, and by optimality of
$(\hat\vu,\hat\vv)$ for \eqref{eq:ch6_sscca_problem},
\begin{equation}
  \hat\vu\trans \hat\mK_{\mathrm S}\hat\vv
  \ge \vu_1\trans \hat\mK_{\mathrm S}\vv_1.
  \label{eq:ch6_app_sscca_opt}
\end{equation}
On the event
$\{\maxnorm{\hat\mK_{\mathrm S}-\mK_{\mathrm S}}\le\varepsilon_{n,p}\}$,
for any $(\vu,\vv)\in\mathcal F$,
\begin{equation}
  \bigl|\vu\trans(\hat\mK_{\mathrm S}-\mK_{\mathrm S})\vv\bigr|
  \le \onenorm{\vu}\,\maxnorm{\hat\mK_{\mathrm S}-\mK_{\mathrm S}}\,\onenorm{\vv}
  \le c_x c_y \varepsilon_{n,p}.
  \label{eq:ch6_app_sscca_maxnorm}
\end{equation}
Therefore
\begin{equation}
  \hat\vu\trans \mK_{\mathrm S}\hat\vv
  \ge \rho_1 - 2c_x c_y \varepsilon_{n,p}.
  \label{eq:ch6_app_sscca_objlower}
\end{equation}
Now expand $\hat\vu$ and $\hat\vv$ in the singular-vector bases of
$\mK_{\mathrm S}$:
\begin{equation}
  \hat\vu = \cos\theta_u\,\vu_1 + \sin\theta_u\,\vu_\perp,
  \qquad
  \hat\vv = \cos\theta_v\,\vv_1 + \sin\theta_v\,\vv_\perp,
  \label{eq:ch6_app_sscca_angle_decomp}
\end{equation}
where $\vu_\perp\perp\vu_1$, $\vv_\perp\perp\vv_1$,
$\twonorm{\vu_\perp}=\twonorm{\vv_\perp}=1$, and
$\theta_u=\angle(\hat\vu,\vu_1)$,
$\theta_v=\angle(\hat\vv,\vv_1)$.  Since
$\mK_{\mathrm S} = \rho_1 \vu_1\vv_1\trans + \sum_{j\ge 2} \rho_j \vu_j\vv_j\trans$,
we have
\begin{align}
  \hat\vu\trans \mK_{\mathrm S}\hat\vv
  &\le \rho_1 \cos\theta_u \cos\theta_v
      + \rho_2 \sin\theta_u \sin\theta_v \notag\\
  &\le \rho_1\left(1-\frac{\sin^2\theta_u+\sin^2\theta_v}{2}\right)
      + \rho_2\frac{\sin^2\theta_u+\sin^2\theta_v}{2} \notag\\
  &= \rho_1 - \frac{\rho_1-\rho_2}{2}
      \bigl\{\sin^2\theta_u+\sin^2\theta_v\bigr\} \
  &= \rho_1 - \frac{\Delta_{\mathrm C}}{2}
      \bigl\{\sin^2\theta_u+\sin^2\theta_v\bigr\}.
  \label{eq:ch6_app_sscca_objupper}
\end{align}
Combining \eqref{eq:ch6_app_sscca_objlower} and
\eqref{eq:ch6_app_sscca_objupper} proves
\eqref{eq:ch6_sscca_basic_bound}.  The inequality
\eqref{eq:ch6_sscca_sine_bound} follows immediately.  Finally, for any unit
vectors $\va$ and $\vb$ with $\va\trans\vb\ge 0$,
\begin{equation}
  \twonorm{\va-\vb}^2 = 2\{1-\va\trans\vb\} \le 2\sin^2\angle(\va,\vb),
  \label{eq:ch6_app_angle_l2}
\end{equation}
which implies \eqref{eq:ch6_sscca_l2_bound}.  \qed

\subsection*{N. Proof of Theorem~\ref{thm:ch6_kendall_factor_recovery}}

The first assertion follows immediately from Davis--Kahan.  Indeed, on the event
\eqref{eq:ch6-kendall-factor-op},
\begin{equation}
  \left\|\sin\Theta\bigl(\hat{\mathcal V}^{(\tau)}_K,\mathcal V_K\bigr)\right\|_{\mathrm F}
  \le \frac{2\sqrt{2K}}{\delta^{\tau}_K}
       \opnorm{\hat\mK_{\tau}-\mK_{\tau}}.
  \label{eq:ch6-app-kendall-dk}
\end{equation}
Substituting \eqref{eq:ch6-kendall-factor-op} into
\eqref{eq:ch6-app-kendall-dk} yields \eqref{eq:ch6-kendall-factor-rate}.

For factor-number consistency, Weyl's inequality gives
\begin{equation}
  |\hat\kappa_j-\kappa_j|
  \le \opnorm{\hat\mK_{\tau}-\mK_{\tau}}
  = O_P\!\left(\sqrt{\frac{\log p}{n}}\right)
  \qquad (1\le j\le p).
  \label{eq:ch6-app-kendall-weyl}
\end{equation}
Hence, on the event
\begin{equation}
  \mathcal E_{\tau}
  = \left\{
      \max_{1\le j\le K_{\max}+1}|\hat\kappa_j-\kappa_j|
      \le \frac{\gamma^{\tau}_0}{4(2+\gamma^{\tau}_0)}\kappa_{K+1}
    \right\},
  \label{eq:ch6-app-kendall-event}
\end{equation}
we have
\begin{align}
  \frac{\hat\kappa_K}{\hat\kappa_{K+1}}
  &\ge
  \frac{\kappa_K-\frac{\gamma^{\tau}_0}{4(2+\gamma^{\tau}_0)}\kappa_{K+1}}
       {\kappa_{K+1}+\frac{\gamma^{\tau}_0}{4(2+\gamma^{\tau}_0)}\kappa_{K+1}}
   \notag\\
  &=
  \frac{\kappa_K/\kappa_{K+1}-\frac{\gamma^{\tau}_0}{4(2+\gamma^{\tau}_0)}}
       {1+\frac{\gamma^{\tau}_0}{4(2+\gamma^{\tau}_0)}}
   \ge 1+\frac{\gamma^{\tau}_0}{2},
  \label{eq:ch6-app-kendall-true-ratio}
\end{align}
where the last inequality uses \eqref{eq:ch6-kendall-ratio-gap}.  For any $j\ne K$,
there is no larger population ratio gap than the one at $K$, so similarly,
\begin{equation}
  \frac{\hat\kappa_j}{\hat\kappa_{j+1}}
  \le 1+\frac{\gamma^{\tau}_0}{4}
  \qquad (j\ne K).
  \label{eq:ch6-app-kendall-false-ratio}
\end{equation}
Therefore the maximizer in \eqref{eq:ch6-kendall-eigenratio} is exactly $K$ on
$\mathcal E_{\tau}$.  Since \eqref{eq:ch6-kendall-ratio-cond} implies
$\Prob(\mathcal E_{\tau})\to 1$, we obtain
\eqref{eq:ch6-kendall-factor-number}.  \qed

%% file: chapters/ch7_clustering.tex
\chapter[High-Dimensional Clustering]{High-Dimensional Clustering under Elliptical Symmetry}\label{ch:clustering}

\idx{clustering}\idx{high-dimensional clustering}\idx{k-means}\idx{sparse k-means}\idx{spatial median clustering}\idx{elliptical mixture}\idx{semiparametric mixture}\idx{Gaussian mixture}\idx{model-based clustering}

Clustering is an unsupervised counterpart of classification: the class labels are not observed, and both the latent groups and the group-specific parameters must be learned from the data.  In high dimensions this problem is intrinsically harder than supervised discrimination because nuisance coordinates can dominate the empirical geometry even when they carry no cluster information.  Heavy tails create an additional difficulty: a few radial outliers can move least-squares centers, distort covariance estimates, and induce artificial clusters.  This chapter develops a robust clustering framework for elliptically symmetric high-dimensional data.  The presentation follows the same order as the earlier chapters.  We first review classical fixed-dimensional methods and high-dimensional Gaussian or light-tailed benchmark procedures.  We then introduce two robust elliptical methods: sparse \(K\)-spatial-median clustering and semiparametric elliptical mixture clustering.

Throughout the chapter, \(\vX_1,\ldots,\vX_n\in\R^p\) are independent observations.  A clustering is represented by labels \(c_i\in\{1,\ldots,K\}\), cluster sets \(C_k=\{i:c_i=k\}\), and centers \(\vmu_1,\ldots,\vmu_K\).  When population labels are used in theoretical statements, they are denoted by \(Z_i\in\{1,\ldots,K\}\).  For a set \(S\subset\{1,\ldots,p\}\), \(\vx_S\) denotes the subvector of \(\vx\) restricted to coordinates in \(S\).

\section{Classical fixed-dimensional clustering methods}\label{sec:ch7-classical}

\subsection{Lloyd's \texorpdfstring{\(K\)}{K}-means algorithm}

The classical \(K\)-means method seeks centers \(\vmu_1,\ldots,\vmu_K\in\R^p\) and labels \(c_1,\ldots,c_n\) minimizing the within-cluster sum of squares
\begin{equation}\label{eq:ch7-kmeans-objective}
  Q_n^{\mathrm{KM}}(\vmu_1,\ldots,\vmu_K,c_1,\ldots,c_n)
  = \sum_{i=1}^n \sum_{k=1}^K \mathbbm{1}(c_i=k)\twonorm{\vX_i-\vmu_k}^2.
\end{equation}
Equivalently,
\begin{equation}\label{eq:ch7-kmeans-min-objective}
  \widehat Q_n^{\mathrm{KM}}(\vmu_1,\ldots,\vmu_K)
  =\sum_{i=1}^n \min_{1\le k\le K}\twonorm{\vX_i-\vmu_k}^2.
\end{equation}
For fixed labels, the minimizer in each cluster is the sample mean,
\begin{equation}\label{eq:ch7-kmeans-center-update}
  \widehat\vmu_k=\frac1{|C_k|}\sum_{i\in C_k}\vX_i.
\end{equation}
For fixed centers, the minimizer in the labels is the nearest-center rule
\begin{equation}\label{eq:ch7-kmeans-assignment}
  c_i=\argmin_{1\le k\le K}\twonorm{\vX_i-\vmu_k}^2.
\end{equation}
Lloyd's algorithm alternates \eqref{eq:ch7-kmeans-center-update} and \eqref{eq:ch7-kmeans-assignment}; see \citet{MacQueen1967} and \citet{HartiganWong1979}.  Since each step decreases \eqref{eq:ch7-kmeans-objective}, the algorithm converges in finitely many steps to a local optimum when ties are resolved deterministically.

The population \(K\)-means functional is
\begin{equation}\label{eq:ch7-pop-kmeans}
  Q^{\mathrm{KM}}(\vmu_1,\ldots,\vmu_K)
  = \E\min_{1\le k\le K}\twonorm{\vX-\vmu_k}^2.
\end{equation}
The empirical objective \eqref{eq:ch7-kmeans-min-objective} is an \(M\)-estimator for the minimizers of \eqref{eq:ch7-pop-kmeans}.  In fixed dimension, under uniqueness and local curvature of the population minimizer, the empirical minimizer is consistent and asymptotically normal.

\begin{theorem}[Fixed-dimensional asymptotic normality of \(K\)-means centers]\label{thm:ch7-fixed-kmeans}
Suppose \(p\) and \(K\) are fixed.  Let \(\vtheta=(\vmu_1\trans,\ldots,\vmu_K\trans)\trans\), and assume that the population objective \(Q^{\mathrm{KM}}(\vtheta)\) has a unique minimizer \(\vtheta_0\) up to label permutation.  Suppose that, in a neighborhood of \(\vtheta_0\), the Voronoi boundaries have probability zero, the Hessian
\[
  \mH_{\mathrm{KM}}=\nabla^2 Q^{\mathrm{KM}}(\vtheta_0)
\]
exists and is positive definite, and \(\E\twonorm{\vX}^4<\infty\).  Then any empirical local minimizer \(\widehat\vtheta\) satisfying \(\widehat\vtheta\to\vtheta_0\) in probability obeys
\begin{equation}\label{eq:ch7-kmeans-asymp-normal}
  \sqrt n(\widehat\vtheta-\vtheta_0)
  \overset{d}{\longrightarrow}
  N\bigl(\vct 0,\mH_{\mathrm{KM}}^{-1}\mB_{\mathrm{KM}}\mH_{\mathrm{KM}}^{-1}\bigr),
\end{equation}
where
\begin{equation}\label{eq:ch7-kmeans-B}
  \mB_{\mathrm{KM} }
  = \Var\{\nabla_{\vtheta} \min_{1\le k\le K}\twonorm{\vX-\vmu_k}^2\big|_{\vtheta=\vtheta_0}\}.
\end{equation}
\end{theorem}

The theorem is a standard nonsmooth \(M\)-estimation result.  Its assumptions make explicit why classical \(K\)-means theory does not directly solve the problems considered in this book: the moment requirement is covariance based, the Hessian depends on squared distances, and the result is fixed-dimensional.

\subsection{Gaussian mixture clustering and EM}

Model-based clustering replaces the geometric objective by a finite mixture likelihood.  The most classical example is the Gaussian mixture model
\begin{equation}\label{eq:ch7-gaussian-mixture}
  f(\vx;\vtheta)=\sum_{k=1}^K \pi_k \phi_p(\vx;\vmu_k,\mSigma_k),
  \qquad \pi_k>0,
  \qquad \sum_{k=1}^K\pi_k=1,
\end{equation}
where \(\phi_p(\cdot;\vmu,\mSigma)\) is the \(p\)-variate normal density.  The observed log-likelihood is
\begin{equation}\label{eq:ch7-gmm-loglik}
  \ell_n(\vtheta)=\sum_{i=1}^n\log\left\{\sum_{k=1}^K\pi_k\phi_p(\vX_i;\vmu_k,\mSigma_k)\right\}.
\end{equation}
The EM algorithm computes posterior weights
\begin{equation}\label{eq:ch7-gmm-resp}
  \tau_{ik}^{(t)}
  = \frac{\pi_k^{(t)}\phi_p(\vX_i;\vmu_k^{(t)},\mSigma_k^{(t)})}
  {\sum_{\ell=1}^K\pi_\ell^{(t)}\phi_p(\vX_i;\vmu_\ell^{(t)},\mSigma_\ell^{(t)})}
\end{equation}
and then updates
\begin{equation}\label{eq:ch7-gmm-updates}
  \pi_k^{(t+1)}=\frac1n\sum_{i=1}^n\tau_{ik}^{(t)},\qquad
  \vmu_k^{(t+1)}=\frac{\sum_{i=1}^n\tau_{ik}^{(t)}\vX_i}{\sum_{i=1}^n\tau_{ik}^{(t)}},
\end{equation}
\begin{equation}\label{eq:ch7-gmm-cov-update}
  \mSigma_k^{(t+1)}=\frac{\sum_{i=1}^n\tau_{ik}^{(t)}(\vX_i-\vmu_k^{(t+1)})(\vX_i-\vmu_k^{(t+1)})\trans}{\sum_{i=1}^n\tau_{ik}^{(t)}}.
\end{equation}
For common spherical covariance \(\mSigma_k=\sigma^2\mI_p\) and equal mixing proportions, the Bayes rule reduces to nearest-center assignment, explaining the connection between Gaussian mixtures and \(K\)-means.  General model-based clustering and variable selection for Gaussian mixtures are reviewed by \citet{FraleyRaftery2002}, \citet{BouveyronBrunetSaumard2014}, \citet{FopMurphy2018}, and \citet{GormleyMurphyRaftery2023}.

\begin{theorem}[Classical likelihood limit for fixed-dimensional Gaussian mixtures]\label{thm:ch7-gmm-mle}
Assume that the mixture model \eqref{eq:ch7-gaussian-mixture} is identifiable up to label permutation, \(p\) and \(K\) are fixed, the true parameter \(\vtheta_0\) is an interior point of a compact parameter space after label normalization, and the Fisher information matrix
\[
  \mI(\vtheta_0)=\E\{\nabla\log f(\vX;\vtheta_0)\nabla\log f(\vX;\vtheta_0)\trans\}
\]
is nonsingular.  Then the maximum likelihood estimator satisfies
\begin{equation}\label{eq:ch7-gmm-mle-normal}
  \sqrt n(\widehat\vtheta-\vtheta_0)
  \overset{d}{\longrightarrow}
  N\{\vct 0,\mI(\vtheta_0)^{-1}\}.
\end{equation}
\end{theorem}

The fixed-dimensional Gaussian mixture theory is powerful but not robust.  Equations \eqref{eq:ch7-gmm-updates}--\eqref{eq:ch7-gmm-cov-update} use ordinary means and covariance matrices.  They are therefore vulnerable to heavy tails, and the covariance update becomes singular when \(p\ge \sum_i\tau_{ik}\) unless structural constraints or regularization are imposed.

\section{High-dimensional Gaussian and light-tailed benchmarks}\label{sec:ch7-hd-benchmarks}

\subsection{Sparse \texorpdfstring{\(K\)}{K}-means and feature-weighted clustering}

When only a small subset of coordinates separates the clusters, ordinary \(K\)-means may fail because noise coordinates inflate Euclidean distances.  Sparse \(K\)-means introduces feature weights.  For a partition \(C=(C_1,\ldots,C_K)\), write
\begin{equation}\label{eq:ch7-feature-bcss}
  B_j(C)=\frac1n\sum_{i,i'=1}^n (X_{ij}-X_{i'j})^2
  -\sum_{k=1}^K\frac1{|C_k|}\sum_{i,i'\in C_k}(X_{ij}-X_{i'j})^2,
\end{equation}
which is the between-cluster sum of squares in coordinate \(j\).  The sparse \(K\)-means criterion of \citet{WittenTibshirani2010Clustering} is
\begin{equation}\label{eq:ch7-sparse-kmeans}
  \max_{C,w}\sum_{j=1}^p w_j B_j(C)
  \quad\text{subject to}\quad
  \twonorm{w}\le 1,
  \onenorm{w}\le s,
  w_j\ge 0.
\end{equation}
For fixed \(C\), the weight update has the soft-thresholding form
\begin{equation}\label{eq:ch7-sparse-kmeans-weight}
  \widehat w_j=\frac{\{B_j(C)-\Delta\}_+}{\left[\sum_{\ell=1}^p\{B_\ell(C)-\Delta\}_+^2\right]^{1/2}},
\end{equation}
where \(\Delta\ge0\) is chosen so that \(\sum_j\widehat w_j=s\) when the \(\ell_1\) constraint is active.  The algorithm alternates between weighted \(K\)-means assignment and \eqref{eq:ch7-sparse-kmeans-weight}.  Related sparse and regularized prototype methods include \citet{SunWangFang2012}, \citet{KondoSalibianBarreraZamar2012}, \citet{BrodinovaFilzmoserOrtnerBreitenederRohm2019}, and \citet{RaymaekersZamar2022}.

A useful way to state the theory is through uniform separation between informative and non-informative feature scores.  Let \(S_0\subset\{1,\ldots,p\}\) be the true set of cluster-relevant variables, and let \(B_j^*=B_j(C^*)\) be the population analogue of \eqref{eq:ch7-feature-bcss} at the oracle partition.  If
\begin{equation}\label{eq:ch7-sparse-score-gap}
  \min_{j\in S_0}B_j^*-\max_{j\notin S_0}B_j^*\ge 2r_{n,p},
  \qquad
  \max_{1\le j\le p}|B_j(C^*)-B_j^*|\le r_{n,p},
\end{equation}
then thresholding or soft-thresholding the feature scores separates \(S_0\) from its complement.  Under sub-Gaussian coordinates and fixed \(K\), Bernstein's inequality gives the typical order
\begin{equation}\label{eq:ch7-bscore-rate}
  r_{n,p}\asymp \sqrt{\frac{\log p}{n}}.
\end{equation}
Thus sparse \(K\)-means is effective when the population between-cluster feature gap exceeds the stochastic fluctuation scale \(\sqrt{\log p/n}\).  The limitation is that \eqref{eq:ch7-bscore-rate} is tied to light-tailed concentration of second-order quantities.

\subsection{Gaussian mixtures, regularized EM, and CHIME}\label{subsec:ch7-chime}

High-dimensional Gaussian mixture clustering is often used as the light-tailed benchmark for the robust procedures developed later.  We give more detail here because this line also clarifies what is lost when Gaussian likelihood and covariance concentration are no longer reliable.

\subsubsection{Penalized Gaussian mixture EM}

Consider the common-covariance two-component Gaussian mixture
\begin{equation}\label{eq:ch7-chime-model}
  \vX_i\iid (1-\omega^*)N_p(\vmu_1^*,\mSigma^*)
       +\omega^*N_p(\vmu_2^*,\mSigma^*),
  \qquad 0<\omega^*\le \frac12,
\end{equation}
where labels are unobserved.  Write
\begin{equation}\label{eq:ch7-chime-beta}
  \mOmega^*=(\mSigma^*)^{-1},
  \qquad
  \vdelta^*=\vmu_1^*-\vmu_2^*,
  \qquad
  \vbeta^*=\mOmega^*\vdelta^*,
  \qquad
  \Delta^2=(\vdelta^*)\trans\mOmega^*\vdelta^*.
\end{equation}
The oracle Fisher rule is
\begin{equation}\label{eq:ch7-oracle-fisher-cluster}
  G^*(\vx)=
  \begin{cases}
  1,& (\vx-\bar\vmu^*)\trans\vbeta^*\ge \log\{\omega^*/(1-\omega^*)\},\\
  2,& (\vx-\bar\vmu^*)\trans\vbeta^*< \log\{\omega^*/(1-\omega^*)\},
  \end{cases}
  \qquad
  \bar\vmu^*=\frac12(\vmu_1^*+\vmu_2^*).
\end{equation}
Thus, for clustering, the precision matrix is needed only through the sparse discriminant vector \(\vbeta^*\).  This observation motivates CHIME, the clustering procedure of \citet{CaiMaZhang2019}, which combines EM with direct sparse discriminant-vector estimation.

Given current values
\(
  \widehat\theta^{(t)}=(\widehat\omega^{(t)},\widehat\vmu_1^{(t)},
  \widehat\vmu_2^{(t)},\widehat\mSigma^{(t)},\widehat\vbeta^{(t)})
\), set \(\widehat{\bar\vmu}^{(t)}=(\widehat\vmu_1^{(t)}+\widehat\vmu_2^{(t)})/2\).  The high-dimensional E-step uses the posterior responsibility
\begin{equation}\label{eq:ch7-chime-resp}
  \widehat\gamma_i^{(t)}
  =\frac{\widehat\omega^{(t)}}
  {\widehat\omega^{(t)}+
   \{1-\widehat\omega^{(t)}\}
   \exp\{(\widehat\vbeta^{(t)})\trans(\vX_i-\widehat{\bar\vmu}^{(t)})\}},
  \qquad i=1,\ldots,n.
\end{equation}
This is the plug-in version of \(\Prob(Z_i=2\mid \vX_i)\), but the covariance inverse is never explicitly used.  The M-step updates
\begin{equation}\label{eq:ch7-chime-pi-mu-update}
  \widehat\omega^{(t+1)}=\frac1n\sum_{i=1}^n\widehat\gamma_i^{(t)},
  \qquad
  \widehat\vmu_1^{(t+1)}=\frac{\sum_{i=1}^n\{1-\widehat\gamma_i^{(t)}\}\vX_i}
  {\sum_{i=1}^n\{1-\widehat\gamma_i^{(t)}\}},
  \qquad
  \widehat\vmu_2^{(t+1)}=\frac{\sum_{i=1}^n\widehat\gamma_i^{(t)}\vX_i}
  {\sum_{i=1}^n\widehat\gamma_i^{(t)}}.
\end{equation}
The weighted covariance estimate is
\begin{equation}\label{eq:ch7-chime-cov-update}
\begin{aligned}
  \widehat\mSigma^{(t+1)}
  =\frac1n\sum_{i=1}^n \big[&
  \{1-\widehat\gamma_i^{(t)}\}
  (\vX_i-\widehat\vmu_1^{(t+1)})(\vX_i-\widehat\vmu_1^{(t+1)})\trans \\
  &+\widehat\gamma_i^{(t)}
  (\vX_i-\widehat\vmu_2^{(t+1)})(\vX_i-\widehat\vmu_2^{(t+1)})\trans
  \big].
\end{aligned}
\end{equation}
Since \(p\) may be larger than \(n\), CHIME does not invert \(\widehat\mSigma^{(t+1)}\).  Instead it updates the discriminant vector by the direct Lasso-type convex program
\begin{equation}\label{eq:ch7-chime-beta-update}
  \widehat\vbeta^{(t+1)}
  =\argmin_{\vbeta\in\R^p}
  \left\{
  \frac12\vbeta\trans\widehat\mSigma^{(t+1)}\vbeta
  -\vbeta\trans(\widehat\vmu_1^{(t+1)}-\widehat\vmu_2^{(t+1)})
  +\lambda_{n,t+1}\onenorm{\vbeta}
  \right\}.
\end{equation}
A theoretical choice has the recursive form
\begin{equation}\label{eq:ch7-chime-lambda}
  \lambda_{n,t+1}=\kappa\lambda_{n,t}+C_\lambda\sqrt{\frac{\log p}{n}},
  \qquad 0<\kappa<\frac12,
\end{equation}
although cross-validation is often used in computation.  After \(T_0\) iterations, CHIME classifies by
\begin{equation}\label{eq:ch7-chime-final-rule}
  \widehat G_{\mathrm{CHIME}}(\vx)=
  \begin{cases}
  1,& (\vx-\widehat{\bar\vmu}^{(T_0)})\trans\widehat\vbeta^{(T_0)}
  \ge \log\{\widehat\omega^{(T_0)}/(1-\widehat\omega^{(T_0)})\},\\
  2,& (\vx-\widehat{\bar\vmu}^{(T_0)})\trans\widehat\vbeta^{(T_0)}
  < \log\{\widehat\omega^{(T_0)}/(1-\widehat\omega^{(T_0)})\}.
  \end{cases}
\end{equation}

\paragraph{Algorithmic summary.}
For a given initialization \(\widehat\theta^{(0)}\), CHIME repeats the following three operations: compute responsibilities by \eqref{eq:ch7-chime-resp}; update mixing weight, means and covariance by \eqref{eq:ch7-chime-pi-mu-update}--\eqref{eq:ch7-chime-cov-update}; update the sparse discriminant vector by \eqref{eq:ch7-chime-beta-update}.  The output is the Fisher-type rule \eqref{eq:ch7-chime-final-rule}.  A good initialization is essential because the likelihood is nonconvex; \citet{CaiMaZhang2019} use an initialization satisfying an \(\ell_\infty\)-type closeness condition and note that moment-based Gaussian-mixture initializers can provide such a start in the theoretical regime.

\subsubsection{Theory for CHIME}

For \(s\ge1\), define the sparse cone
\begin{equation}\label{eq:ch7-chime-cone}
  \mathcal C(s)=\left\{\vv\in\R^p:
  2\onenorm{\vv_{S^c}}\le 4\onenorm{\vv_S}+3\sqrt{s}\twonorm{\vv}
  \text{ for some }S\subset[p], |S|=s\right\}.
\end{equation}
For a vector \(\va\) and a matrix \(\mA\), define
\begin{equation}\label{eq:ch7-chime-2s-norms}
  \norm{\va}_{2,s}=\sup_{\vv\in\mathcal C(s),\twonorm{\vv}=1}|\va\trans\vv|,
  \qquad
  \norm{\mA}_{2,s}=\sup_{\vv\in\mathcal C(s),\twonorm{\vv}=1}\twonorm{\mA\vv}.
\end{equation}
For two Gaussian-mixture parameters \(\theta=(\omega,\vmu_1,\vmu_2,\mSigma)\) and \(\widetilde\theta\), set
\begin{equation}\label{eq:ch7-chime-distance}
  d_{2,s}(\theta,\widetilde\theta)
  =|\omega-\widetilde\omega|
  \vee\norm{\vmu_1-\widetilde\vmu_1}_{2,s}
  \vee\norm{\vmu_2-\widetilde\vmu_2}_{2,s}
  \vee\norm{(\mSigma-\widetilde\mSigma)\widetilde\vbeta}_{2,s}.
\end{equation}

\begin{assumption}[CHIME regularity conditions]\label{ass:ch7-chime}
The true parameter in \eqref{eq:ch7-chime-model} satisfies \(\norm{\vbeta^*}_0\le s\), \(c_\omega<\omega^*<1-c_\omega\), the eigenvalues of \(\mSigma^*\) are bounded away from zero and infinity, and \(\onenorm{\vbeta^*}\le M_\beta\).  The signal-to-noise ratio \(\Delta\) in \eqref{eq:ch7-chime-beta} is larger than a sufficiently large constant depending only on the boundedness constants.  The initialization obeys
\begin{equation}\label{eq:ch7-chime-init}
  d_{2,s}(\widehat\theta^{(0)},\theta^*)\vee
  \twonorm{\widehat\vbeta^{(0)}-\vbeta^*}
  \le r\Delta,
  \qquad
  \widehat\vbeta^{(0)}-\vbeta^*\in\mathcal C(s),
\end{equation}
where \(r>0\) is sufficiently small.  Finally,
\begin{equation}\label{eq:ch7-chime-dim-condition}
  s=o\left(\frac{\sqrt n}{\log p}\right),
  \qquad
  \sqrt{\frac{s\log p}{n}}=o(r).
\end{equation}
\end{assumption}

\begin{theorem}[CHIME estimation and excess misclustering risk]\label{thm:ch7-chime}
Suppose \eqref{eq:ch7-chime-model} and Assumption~\ref{ass:ch7-chime} hold.  Then there exist constants \(C>0\) and \(\kappa\in(0,1/2)\) such that the CHIME iterate satisfies, with probability tending to one,
\begin{equation}\label{eq:ch7-chime-beta-rate}
  \twonorm{\widehat\vbeta^{(T_0)}-\vbeta^*}
  \le C\left\{\kappa^{T_0}d_{2,s}(\widehat\theta^{(0)},\theta^*)+
  \sqrt{\frac{s\log p}{n}}\right\}.
\end{equation}
If \(T_0\gtrsim \log n\), then
\begin{equation}\label{eq:ch7-chime-beta-opt-rate}
  \twonorm{\widehat\vbeta^{(T_0)}-\vbeta^*}
  =O_p\left(\sqrt{\frac{s\log p}{n}}\right).
\end{equation}
Moreover, for the rule \eqref{eq:ch7-chime-final-rule},
\begin{equation}\label{eq:ch7-chime-risk-rate}
  R(\widehat G_{\mathrm{CHIME}})-R(G^*)
  =O_p\left(\frac{s\log p}{n}\right).
\end{equation}
The rate \(s\log p/n\) for the excess misclustering risk is minimax optimal over the sparse-discriminant Gaussian mixture class considered by \citet{CaiMaZhang2019}.
\end{theorem}

The CHIME theorem is the sharp Gaussian benchmark for this chapter.  It shows that, under light tails and a sparse discriminant vector, one does not need sparse means or sparse precision matrices separately; sparsity of \(\vbeta^*=\mOmega^*\vdelta^*\) is the essential structure.  The robust methods in Sections~\ref{sec:ch7-sparse-sm} and \ref{sec:ch7-semi-elliptical-mixture} replace the Gaussian likelihood and covariance update by spatial-median and elliptical-shape operations, but the same high-level proof template remains: a local contraction plus a sample-concentration error.

\subsection{Spectral and feature-screening approaches}\label{subsec:ch7-spectral-screening}

Spectral clustering methods reduce clustering to the estimation of a low-dimensional label subspace.  They are especially useful in rare/weak Gaussian mixture models where only a small fraction of features carry weak mean shifts.  We describe influential features PCA (IF-PCA) and the associated phase-transition theory, which are the main light-tailed spectral benchmarks.

\subsubsection{IF-PCA: method}

Let \(\mX=(X_{ij})\in\R^{n\times p}\) be the data matrix.  Standardize each feature by
\begin{equation}\label{eq:ch7-ifpca-standardization}
  \widetilde X_{ij}=\frac{X_{ij}-\bar X_{\cdot j}}{\widehat\sigma_j},
  \qquad
  \bar X_{\cdot j}=n^{-1}\sum_{i=1}^n X_{ij},
  \qquad
  \widehat\sigma_j^2=n^{-1}\sum_{i=1}^n(X_{ij}-\bar X_{\cdot j})^2.
\end{equation}
For feature \(j\), let
\begin{equation}\label{eq:ch7-ifpca-ecdf}
  \widehat F_{n,j}(t)=\frac1n\sum_{i=1}^n\mathbbm{1}(\widetilde X_{ij}\le t).
\end{equation}
The IF-PCA feature score of \citet{JinWang2016IFPCA} is the Kolmogorov-Smirnov statistic
\begin{equation}\label{eq:ch7-ifpca-ks}
  \psi_{n,j}=\sqrt n\sup_{t\in\R}\left|\widehat F_{n,j}(t)-\Phi(t)\right|,
  \qquad j=1,\ldots,p,
\end{equation}
where \(\Phi\) is the standard normal distribution function.  Choose a threshold \(t_p\), often using a higher-criticism-type calibration, and retain
\begin{equation}\label{eq:ch7-ifpca-selected-set}
  \widehat S=\{j:\psi_{n,j}\ge t_p\}.
\end{equation}
Let \(\widetilde\mX_{\widehat S}\) be the standardized data restricted to \(\widehat S\).  Compute the leading \(K-1\) left singular vectors
\begin{equation}\label{eq:ch7-ifpca-svd}
  \widehat\mU=\operatorname{eig}_{K-1}
  \left(\widetilde\mX_{\widehat S}\widetilde\mX_{\widehat S}\trans\right)
  \in\R^{n\times(K-1)}.
\end{equation}
Finally apply ordinary \(K\)-means to the rows of \(\widehat\mU\).  The important difference between IF-PCA and ordinary PCA is the pre-selection step \eqref{eq:ch7-ifpca-selected-set}: the spectral calculation is performed after removing most noise coordinates.

\subsubsection{A generic spectral recovery theorem}

The following theorem gives a clean proof template for screening-plus-PCA procedures.  It is not the full rare/weak phase-transition statement of \citet{JinKeWang2017}, but it is the deterministic theorem that explains why IF-PCA works once the selected features contain enough signal and not too much noise.

Let the standardized post-selection matrix have decomposition
\begin{equation}\label{eq:ch7-ifpca-matrix-model}
  \widetilde\mX_{\widehat S}=\mM_{\widehat S}+\mE_{\widehat S},
\end{equation}
where \(\mM_{\widehat S}\) has rank \(K-1\).  Let \(\mU_*\in\R^{n\times(K-1)}\) be the population label subspace, namely the matrix of leading left singular vectors of \(\mM_{\widehat S}\).  Define
\begin{equation}\label{eq:ch7-ifpca-gap}
  \sigma_* = \sigma_{K-1}(\mM_{\widehat S}),
  \qquad
  \epsilon_* = \opnorm{\mE_{\widehat S}}.
\end{equation}

\begin{assumption}[Post-selection spectral separation]\label{ass:ch7-ifpca}
The selected feature set satisfies \(\sigma_*>2\epsilon_*\).  The true clusters are balanced in the sense that \(n_k\ge c_0n/K\) for all \(k\).  The population embedded centers in the rows of \(\mU_*\) are separated by at least \(c_1n^{-1/2}\).
\end{assumption}

\begin{theorem}[Screening-plus-PCA clustering error]\label{thm:ch7-ifpca-generic}
Under Assumption~\ref{ass:ch7-ifpca}, the leading left singular-vector estimator \(\widehat\mU\) in \eqref{eq:ch7-ifpca-svd} satisfies
\begin{equation}\label{eq:ch7-ifpca-sintheta}
  \inf_{\mO\trans\mO=\mI_{K-1}}\frobnorm{\widehat\mU-\mU_*\mO}
  \le \frac{2\sqrt2\,\epsilon_*}{\sigma_*-\epsilon_*}.
\end{equation}
If \(K\)-means on the rows of \(\widehat\mU\) achieves a constant-factor approximation to the optimal within-cluster objective, then the misclustering proportion satisfies
\begin{equation}\label{eq:ch7-ifpca-misclustering}
  \frac1n\min_{\sigma\in\mathfrak S_K}
  \sum_{i=1}^n\mathbbm{1}\{\widehat c_i\ne \sigma(Z_i)\}
  \le C_K\left(\frac{\epsilon_*}{\sigma_*}\right)^2.
\end{equation}
For sub-Gaussian noise, with probability tending to one,
\begin{equation}\label{eq:ch7-ifpca-noise-rate}
  \epsilon_*\le C\{\sqrt n+\sqrt{|\widehat S|}\}.
\end{equation}
Hence exact or weak recovery follows when the post-selection signal singular value is asymptotically larger than \(\sqrt n+\sqrt{|\widehat S|}\).
\end{theorem}

\subsubsection{Rare/weak phase-transition interpretation}

In the two-class rare/weak model of \citet{JinKeWang2017}, one may write, after standardization,
\begin{equation}\label{eq:ch7-rare-weak-model}
  X_{ij}=\ell_i\mu_j+Z_{ij},
  \qquad
  \ell_i\in\{-1,1\},
  \qquad
  Z_{ij}\iid N(0,1),
\end{equation}
where only \(p^{1-\vartheta}\) entries of \(\vmu=(\mu_1,\ldots,\mu_p)\trans\) are nonzero and their magnitudes are calibrated by \(\sqrt{2r\log p/n}\).  In this regime, an uninformative feature has a KS score distributed like the null empirical-process supremum, whereas an informative feature has a stochastically shifted KS score.  Thresholding \eqref{eq:ch7-ifpca-ks} therefore behaves like sparse normal-means selection.  The retained set contains most useful features when the selection threshold is below the effective signal level, while the number of retained useless features is controlled by the Gaussian empirical-process tail.  After this screening, Theorem~\ref{thm:ch7-ifpca-generic} applies to the post-selection spectral step.  The phase-transition analysis of \citet{JinKeWang2017} identifies regions where no method can recover labels and regions where IF-PCA or related aggregation methods succeed.

The limitation for the present book is clear: IF-PCA relies on Gaussian calibration of feature-wise empirical distributions and spectral concentration of the selected noise matrix.  Under elliptical heavy tails, radial shocks can inflate \(\psi_{n,j}\), produce spurious selected features, and degrade the operator-norm bound \eqref{eq:ch7-ifpca-noise-rate}.  The spatial-median and semiparametric elliptical procedures below are designed to retain the useful screening and low-dimensional-recovery philosophy while replacing Gaussian feature scores and covariance matrices by robust directional objects.

\section{Sparse \texorpdfstring{\(K\)}{K}-spatial-median clustering}\label{sec:ch7-sparse-sm}

We now turn to a robust nonparametric clustering framework based on the spatial median, following \citet{ZhaoZhuangFeng2026SparseKSM}.  The goal is to retain the computational simplicity of prototype clustering while replacing squared-error center updates by robust directional balance equations and incorporating explicit variable exclusion.

\subsection{Spatial-median \texorpdfstring{\(K\)}{K}-clustering}

The basic robust prototype objective is
\begin{equation}\label{eq:ch7-k-spatial-median-objective}
  Q_n^{\mathrm{SM}}(\vmu_1,\ldots,\vmu_K,c_1,\ldots,c_n)
  =\sum_{i=1}^n\sum_{k=1}^K\mathbbm{1}(c_i=k)\twonorm{\vX_i-\vmu_k}.
\end{equation}
For fixed labels, the center update is the spatial median in each cluster,
\begin{equation}\label{eq:ch7-spatial-median-update}
  \widehat\vmu_k
  =\argmin_{\vmu\in\R^p}\sum_{i\in C_k}\twonorm{\vX_i-\vmu},
\end{equation}
which satisfies the estimating equation
\begin{equation}\label{eq:ch7-cluster-spmed-equation}
  \sum_{i\in C_k}U(\vX_i-\widehat\vmu_k)=\vct 0
\end{equation}
whenever the minimizer is not equal to an observation and is unique.  Equation \eqref{eq:ch7-cluster-spmed-equation} shows that the update depends on residual directions rather than residual magnitudes.  This is the essential robustness gain over \(K\)-means.

The simplest assignment rule remains Euclidean,
\begin{equation}\label{eq:ch7-sm-euclidean-assignment}
  c_i\leftarrow \argmin_{1\le k\le K}\twonorm{\vX_i-\widehat\vmu_k}^2.
\end{equation}
This hybrid update is computationally convenient but does not adapt to heterogeneous scale or feature dependence.  The next method incorporates a common spatial-sign covariance metric.

\subsection{SM--SSCM: a common spatial-sign covariance metric}

For a current partition and centers \(\vmu_1,\ldots,\vmu_K\), define residuals and residual signs by
\begin{equation}\label{eq:ch7-residual-signs}
  \vr_i=\vX_i-\vmu_{c_i},
  \qquad
  \vu_i=U(\vr_i).
\end{equation}
The common regularized spatial-sign covariance matrix is
\begin{equation}\label{eq:ch7-sscm-cluster}
  \widehat\mSigma_{\mathrm{sgn}}
  =\frac1n\sum_{i=1}^n\vu_i\vu_i\trans+\lambda\mI_p,
  \qquad \lambda>0.
\end{equation}
The SM--SSCM assignment uses the Mahalanobis-type distance
\begin{equation}\label{eq:ch7-smsscm-distance}
  d_{ik}^{\mathrm{SSCM}}
  =(\vX_i-\vmu_k)\trans\widehat\mSigma_{\mathrm{sgn}}^{-1}(\vX_i-\vmu_k),
  \qquad
  c_i\leftarrow \argmin_{1\le k\le K}d_{ik}^{\mathrm{SSCM}}.
\end{equation}
The algorithm alternates spatial-median center updates, recomputation of \eqref{eq:ch7-sscm-cluster}, and assignment by \eqref{eq:ch7-smsscm-distance}.  The ridge term guarantees invertibility even when \(p>n\).  Since each \(\vu_i\vu_i\trans\) is bounded and has trace one, \eqref{eq:ch7-sscm-cluster} is less sensitive to heavy-tailed radial variation than the sample covariance matrix.

\subsection{Sparse--SM: hard feature exclusion}

In high dimensions, robustness alone is not sufficient when many variables are irrelevant.  Sparse--SM screens coordinates using across-center dispersion.  Given current centers, define
\begin{equation}\label{eq:ch7-sm-separation-score}
  s_j=\sum_{k=1}^K|\mu_{kj}-\bar\mu_j|,
  \qquad
  \bar\mu_j=\frac1K\sum_{k=1}^K\mu_{kj},
  \qquad j=1,\ldots,p.
\end{equation}
For a threshold \(\tau>0\), define the active set
\begin{equation}\label{eq:ch7-active-set-sm}
  S(\tau)=\{j:s_j\ge \tau\}.
\end{equation}
The sparse assignment step uses only active coordinates,
\begin{equation}\label{eq:ch7-sparse-sm-assignment}
  c_i\leftarrow \argmin_{1\le k\le K}\twonorm{(\vX_i)_{S(\tau)}-(\vmu_k)_{S(\tau)}}^2.
\end{equation}
The center update remains the full-dimensional spatial median; optionally, excluded coordinates may be reset to a common robust baseline for interpretability.  The hard-thresholding rule \eqref{eq:ch7-active-set-sm} differs from \(\ell_1\)-weighted sparse \(K\)-means: it gives an explicit selected set and avoids shrinkage bias in the retained coordinates.

The threshold can be chosen by a permutation Gap criterion.  Let \(\widehat\vmu_k(\tau)\) be the fitted centers and let \(\widehat\vmu(\tau)\) be the overall spatial median.  Define the observed between-center separation
\begin{equation}\label{eq:ch7-sparse-sm-gap-observed}
  O(\tau)=\sum_{k=1}^K n_k\twonorm{\widehat\vmu_k(\tau)-\widehat\vmu(\tau)}^2.
\end{equation}
Generate reference data matrices by independently permuting each coordinate; let \(O^{(b)}(\tau)\) be the same criterion on the \(b\)-th reference data set.  The Gap score is
\begin{equation}\label{eq:ch7-sparse-sm-gap}
  \operatorname{Gap}(\tau)
  =\log O(\tau)-\frac1B\sum_{b=1}^B\log O^{(b)}(\tau),
\end{equation}
and \(\widehat\tau\) maximizes \eqref{eq:ch7-sparse-sm-gap} over a finite grid.  This reference preserves marginal scale and tail behavior but breaks the joint clustering structure.

\subsection{Algorithms: sparse \texorpdfstring{\(K\)}{K}-median and sparse \texorpdfstring{\(K\)}{K}-spatial-median}\label{subsec:ch7-sparse-sm-algorithms}

We now spell out the algorithms because the distinction between sparse \(K\)-median and sparse \(K\)-spatial-median is important.  The former uses coordinatewise medians and an axis-aligned \(\ell_1\) geometry; the latter uses multivariate spatial medians and therefore respects rotations within the retained subspace.

\subsubsection{Coordinatewise sparse \texorpdfstring{\(K\)}{K}-median baseline}

For a partition \(C=(C_1,\ldots,C_K)\), define the coordinatewise cluster median
\begin{equation}\label{eq:ch7-coordinatewise-median}
  \widehat m_{kj}^{\mathrm{med}}
  \in \argmin_{m\in\R}\sum_{i\in C_k}|X_{ij}-m|,
  \qquad j=1,\ldots,p,
\end{equation}
and the global coordinatewise median
\begin{equation}\label{eq:ch7-global-coordinatewise-median}
  \widehat m_{0j}^{\mathrm{med}}
  \in \argmin_{m\in\R}\sum_{i=1}^n |X_{ij}-m|.
\end{equation}
The coordinatewise between-cluster median improvement is
\begin{equation}\label{eq:ch7-kmedian-feature-score}
  D_j(C)
  =\sum_{i=1}^n |X_{ij}-\widehat m_{0j}^{\mathrm{med}}|
   -\sum_{k=1}^K\sum_{i\in C_k}|X_{ij}-\widehat m_{kj}^{\mathrm{med}}|.
\end{equation}
A sparse \(K\)-median analogue of sparse \(K\)-means solves
\begin{equation}\label{eq:ch7-sparse-kmedian-weighted}
  \max_{C,w}\sum_{j=1}^p w_jD_j(C)
  \quad\text{subject to}\quad
  \twonorm{w}\le1,
  \quad
  \onenorm{w}\le s_w,
  \quad
  w_j\ge0.
\end{equation}
For fixed \(C\), the weight update is again a soft-thresholding rule,
\begin{equation}\label{eq:ch7-sparse-kmedian-weight-update}
  \widehat w_j
  =\frac{\{D_j(C)-\Delta\}_+}
  {\left[\sum_{\ell=1}^p\{D_\ell(C)-\Delta\}_+^2\right]^{1/2}},
\end{equation}
where \(\Delta\ge0\) is chosen so that \(\sum_j\widehat w_j=s_w\) when the constraint is active.  For fixed weights and coordinatewise medians, the assignment rule is
\begin{equation}\label{eq:ch7-sparse-kmedian-assignment}
  c_i\leftarrow \argmin_{1\le k\le K}
  \sum_{j=1}^p \widehat w_j |X_{ij}-\widehat m_{kj}^{\mathrm{med}}|.
\end{equation}
This algorithm is robust to coordinatewise outliers and can select variables, but it is not rotation equivariant.  In elliptical models with correlated coordinates, two clusters that are well separated in an oblique direction may be poorly represented by coordinatewise medians and \(\ell_1\) distances.  This motivates replacing \eqref{eq:ch7-coordinatewise-median} by multivariate spatial medians.

\subsubsection{Sparse \texorpdfstring{\(K\)}{K}-spatial-median algorithm}

At a fixed threshold \(\tau\), Sparse--SM iterates the following steps.
\begin{enumerate}[label=\textbf{Step \arabic*.}]
  \item \textbf{Initialization.}  Choose initial centers \(\vct{m}_1^{(0)},\ldots,\vct{m}_K^{(0)}\), for example by max--min seeding:
  \begin{equation}\label{eq:ch7-maxmin-seeding}
    i_1\in\{1,\ldots,n\},
    \qquad
    i_{k}=\argmax_{1\le i\le n}\min_{1\le \ell<k}\twonorm{\vX_i-\vX_{i_\ell}},
    \qquad
    \vct{m}_k^{(0)}=\vX_{i_k}.
  \end{equation}
  Set initial labels by nearest-center assignment.
  \item \textbf{Spatial-median center update.}  Given labels \(c_i^{(t)}\), update each center by
  \begin{equation}\label{eq:ch7-sparse-sm-alg-center}
    \vct{m}_k^{(t+1)}
    \in\argmin_{\vmu\in\R^p}\sum_{i:c_i^{(t)}=k}\twonorm{\vX_i-\vmu}.
  \end{equation}
  In computation, Weiszfeld iteration is used:
  \begin{equation}\label{eq:ch7-weiszfeld-cluster}
    \vct{m}_{k}^{(r+1)}
    =\frac{\sum_{i\in C_k^{(t)}}\vX_i/\twonorm{\vX_i-\vct{m}_{k}^{(r)}}}
    {\sum_{i\in C_k^{(t)}}1/\twonorm{\vX_i-\vct{m}_{k}^{(r)}}},
  \end{equation}
  with the standard safe modification if an iterate hits an observation.
  \item \textbf{Feature-score update.}  Compute
  \begin{equation}\label{eq:ch7-sparse-sm-alg-score}
    s_j^{(t+1)}=\sum_{k=1}^K |m_{kj}^{(t+1)}-\bar m_j^{(t+1)}|,
    \qquad
    \bar m_j^{(t+1)}=K^{-1}\sum_{k=1}^K m_{kj}^{(t+1)}.
  \end{equation}
  The active set is
  \begin{equation}\label{eq:ch7-sparse-sm-alg-active}
    S^{(t+1)}(\tau)=\{j:s_j^{(t+1)}\ge\tau\}.
  \end{equation}
  If this set is empty, the algorithm retains the coordinate with the largest \(s_j^{(t+1)}\).
  \item \textbf{Sparse assignment update.}  Assign each observation by the active-coordinate distance
  \begin{equation}\label{eq:ch7-sparse-sm-alg-assignment}
    c_i^{(t+1)}
    =\argmin_{1\le k\le K}
    \twonorm{(\vX_i)_{S^{(t+1)}(\tau)}-(\vct{m}_k^{(t+1)})_{S^{(t+1)}(\tau)}}^2.
  \end{equation}
  \item \textbf{Stopping and empty-cluster handling.}  Stop when labels and active set are unchanged.  If a cluster becomes empty, move to it the observation with the largest current within-cluster distance and recompute the corresponding spatial median.
\end{enumerate}

The threshold is selected by the Gap rule \eqref{eq:ch7-sparse-sm-gap}.  For a grid \(\mathcal T=\{\tau_1,\ldots,\tau_L\}\), run the above algorithm on the observed data and on \(B\) reference data sets; compute \(\operatorname{Gap}(\tau_\ell)\); set
\begin{equation}\label{eq:ch7-sparse-sm-tauhat}
  \widehat\tau=\argmax_{\tau\in\mathcal T}\operatorname{Gap}(\tau),
\end{equation}
and rerun Sparse--SM at \(\widehat\tau\).  The final output is
\begin{equation}\label{eq:ch7-sparse-sm-output}
  \{\widehat c_i\}_{i=1}^n,
  \qquad
  \{\widehat\vmu_k\}_{k=1}^K,
  \qquad
  \widehat S=S(\widehat\tau).
\end{equation}
The selected set \(\widehat S\) is part of the statistical output, not merely a tuning artifact.

\begin{remark}[Relationship to sparse \(K\)-median]
Sparse \(K\)-median and Sparse--SM both replace squared loss by a robust loss and both remove irrelevant coordinates.  The former is based on coordinatewise medians and weighted \(\ell_1\) distances; the latter is based on spatial medians and Euclidean geometry on a selected subspace.  Therefore Sparse--SM keeps the multivariate geometry of the retained variables and is more natural under elliptical clusters.
\end{remark}

\subsection{Consistency theory for Sparse--SM and SM--SSCM}

The theoretical analysis uses a mixture elliptical model.  Conditional on \(Z_i=r\), assume
\begin{equation}\label{eq:ch7-sparse-sm-model}
  \vX_i=\vmu_r+\mSigma_r^{1/2}(R_i\vU_i),
  \qquad
  \vU_i\sim \mathrm{Unif}(\spn^{p-1}),
  \qquad R_i\indep \vU_i,
\end{equation}
with \(\Prob(Z_i=r)=\pi_r\) and \(\pi_{\min}=\min_r\pi_r>0\).  Let
\begin{equation}\label{eq:ch7-an-def}
 a_n=\begin{cases}
 \sqrt{\lambda_{\max}}\,\kappa\{\log(2n^{1+\varepsilon})\}^{1/\alpha}, & \text{sub-Weibull radial tails},\\[4pt]
 \sqrt{\lambda_{\max}}(Cn^{1+\varepsilon})^{1/\nu}, & \text{polynomial radial tails},
 \end{cases}
\end{equation}
where \(\lambda_{\max}=\max_r\lambda_{\max}(\mSigma_r)\).  Then
\begin{equation}\label{eq:ch7-event-En}
  \mathcal E_n=\left\{\max_{1\le i\le n}\twonorm{\vX_i-\vmu_{Z_i}}\le a_n\right\}
\end{equation}
has probability tending to one.

For SM--SSCM, define the population common spatial-sign covariance
\begin{equation}\label{eq:ch7-pop-common-sscm}
  \mGamma_{\mathrm{sgn}}
  =\sum_{r=1}^K\pi_r\E\{U(\vY^{(r)})U(\vY^{(r)})\trans\},
  \qquad
  \vY^{(r)}=\mSigma_r^{1/2}(R\vU),
\end{equation}
\begin{equation}\label{eq:ch7-Astar}
  \mSigma_\star=\mGamma_{\mathrm{sgn}}+\lambda\mI_p,
  \qquad
  \mA_\star=\mSigma_\star^{-1},
  \qquad
  \norm{\vv}_{\mA_\star}=(\vv\trans\mA_\star\vv)^{1/2}.
\end{equation}
Let
\begin{equation}\label{eq:ch7-dsgn}
  d=\min_{r\ne s}\twonorm{\vmu_r-\vmu_s},
  \qquad
  d_{\mathrm{sgn}}=\min_{r\ne s}\norm{\vmu_r-\vmu_s}_{\mA_\star}.
\end{equation}

\begin{assumption}[Metric stability and separation for SM--SSCM]\label{ass:ch7-smsscm}
There exist \(\eta\in[0,1)\) and events \(\mathcal M_n\) with \(\Prob(\mathcal M_n)\to1\) such that, on \(\mathcal M_n\), every inverse SSCM metric used along the recovery path satisfies
\begin{equation}\label{eq:ch7-metric-stability}
  (1-\eta)\mA_\star\preceq \mA_t\preceq(1+\eta)\mA_\star.
\end{equation}
Moreover, for all large \(n\),
\begin{equation}\label{eq:ch7-smsscm-separation}
  d>4a_n,
  \qquad
  d_{\mathrm{sgn}}>
  \frac{2a_n}{\sqrt\lambda}\left\{1+\sqrt{\frac{1+\eta}{1-\eta}}\right\}.
\end{equation}
\end{assumption}

\begin{theorem}[Clustering consistency of SM--SSCM]\label{thm:ch7-smsscm-consistency}
Under the mixture model \eqref{eq:ch7-sparse-sm-model}, the radial tail condition used in \eqref{eq:ch7-an-def}, and Assumption~\ref{ass:ch7-smsscm}, the max--min initialized SM--SSCM algorithm with deterministic tie-breaking recovers the true partition up to label permutation with probability tending to one.
\end{theorem}

For Sparse--SM, define the population coordinate score
\begin{equation}\label{eq:ch7-pop-coordinate-score}
  s_j^\star=\sum_{k=1}^K|\mu_{kj}-\bar\mu_j|,
  \qquad
  \bar\mu_j=K^{-1}\sum_{k=1}^K\mu_{kj},
\end{equation}
and the true active set \(S_0=\{j:s_j^\star>0\}\).  Let
\begin{equation}\label{eq:ch7-smin-d0}
  s_{\min}=\min_{j\in S_0}s_j^\star,
  \qquad
  d_0=\min_{r\ne s}\twonorm{(\vmu_r)_{S_0}-(\vmu_s)_{S_0}}.
\end{equation}

\begin{assumption}[Sparse signal and threshold margin]\label{ass:ch7-sparse-sm}
The set \(S_0\) is nonempty.  For \(j\notin S_0\), \(\mu_{1j}=\cdots=\mu_{Kj}\).  For all large \(n\),
\begin{equation}\label{eq:ch7-sparse-sm-separation}
  d_0>4a_n,
  \qquad
  2Ka_n<\tau_n<s_{\min}-2Ka_n.
\end{equation}
\end{assumption}

\begin{theorem}[Feature-selection consistency of Sparse--SM]\label{thm:ch7-sparse-feature}
Suppose the mixture model \eqref{eq:ch7-sparse-sm-model}, the radial tail condition, and Assumption~\ref{ass:ch7-sparse-sm} hold.  If centers \(\vct{m}_1,\ldots,\vct{m}_K\) satisfy
\begin{equation}\label{eq:ch7-center-close-sparse}
  \max_{1\le k\le K}\twonorm{\vct{m}_k-\vmu_{\sigma(k)}}\le a_n
\end{equation}
for some permutation \(\sigma\), then hard-thresholding \eqref{eq:ch7-active-set-sm} at \(\tau_n\) selects exactly \(S_0\).  Consequently, if \eqref{eq:ch7-center-close-sparse} holds with probability tending to one, then \(\Prob\{S(\tau_n)=S_0\}\to1\).
\end{theorem}

\begin{theorem}[Joint support and clustering recovery for Sparse--SM]\label{thm:ch7-sparse-clustering}
Under the assumptions of Theorem~\ref{thm:ch7-sparse-feature}, the max--min initialized hard-threshold Sparse--SM algorithm with deterministic tie-breaking selects \(S_0\) and recovers the true partition up to label permutation with probability tending to one.  The final fixed point has active set \(S_0\) and the true partition up to permutation.
\end{theorem}

\section{Semiparametric elliptical mixture clustering}\label{sec:ch7-semi-elliptical-mixture}

Sparse--SM is a robust prototype method.  We next describe a model-based procedure, following \citet{FengZhuang2026SEMC}, for high-dimensional elliptical mixtures with an unknown common radial generator and a common sparse precision-shape matrix.  The method is semiparametric because the radial generator is estimated nonparametrically rather than fixed to Gaussian, \(t\), or another parametric family.

\subsection{Model and oracle Bayes rule}

Let \(Z\in\{1,\ldots,K\}\) with \(\Prob(Z=k)=\pi_k^*\).  Conditional on \(Z=k\), \(\vX\) has density
\begin{equation}\label{eq:ch7-semi-component-density}
  f_k^*(\vx)=|\mP^*|^{1/2}g^*\{\delta_k^*(\vx)\},
  \qquad
  \delta_k^*(\vx)=(\vx-\vmu_k^*)\trans\mP^*(\vx-\vmu_k^*),
\end{equation}
where \(\mP^*\succ0\) is a common precision-shape matrix and \(g^*\) is an unknown common radial generator satisfying the elliptical normalization
\begin{equation}\label{eq:ch7-generator-normalization}
  \frac{\pi^{p/2}}{\Gamma(p/2)}\int_0^\infty u^{p/2-1}g^*(u)\,du=1.
\end{equation}
The marginal density is
\begin{equation}\label{eq:ch7-semi-mixture-density}
  f^*(\vx)=\sum_{k=1}^K\pi_k^*|\mP^*|^{1/2}g^*\{\delta_k^*(\vx)\}.
\end{equation}
Because \((\mP^*,g^*)\) is identifiable only up to a positive scale, we impose
\begin{equation}\label{eq:ch7-semi-trace-normalization}
  \tr\{(\mP^*)^{-1}\}=p.
\end{equation}
Writing \(\ell^*(u)=\log g^*(u)\), the oracle Bayes rule is
\begin{equation}\label{eq:ch7-semi-bayes}
  G^*(\vx)=\argmax_{1\le k\le K}\left[\log\pi_k^*+\ell^*\{\delta_k^*(\vx)\}\right].
\end{equation}
The goal is to estimate \(G^*\) and its parameters when \(p\) may be large, the tails may be heavy, and no parametric radial family is specified.

\subsection{Oracle and empirical generalized EM maps}

For a candidate parameter
\[
  z=(\bm\pi,\vmu_1,\ldots,\vmu_K,\mP,g),
\]
define
\begin{equation}\label{eq:ch7-semi-delta-a}
  \delta_k(\vx;z)=(\vx-\vmu_k)\trans\mP(\vx-\vmu_k),
  \qquad
  a_k(\vx;z)=\log\pi_k+\log g\{\delta_k(\vx;z)\}.
\end{equation}
The posterior responsibility is
\begin{equation}\label{eq:ch7-semi-responsibility}
  r_k(\vx;z)=\frac{\exp\{a_k(\vx;z)\}}{\sum_{\ell=1}^K\exp\{a_\ell(\vx;z)\}}.
\end{equation}
The transformed radius is
\begin{equation}\label{eq:ch7-transformed-radius}
  q(u)=\log(1+u),
  \qquad
  Y_k(\vx;z)=q\{\delta_k(\vx;z)\}.
\end{equation}
Let \(f_{Y,z}\) be the posterior-weighted transformed-radius density defined by
\begin{equation}\label{eq:ch7-fY-definition}
  \int h(y)f_{Y,z}(y)\,dy
  =\sum_{k=1}^K\E\left[r_k(\vX;z)h\{q(\delta_k(\vX;z))\}\right]
\end{equation}
for all bounded measurable \(h\).  The inverse radial transformation is
\begin{equation}\label{eq:ch7-inverse-radial-generator}
  \mathcal A^{-1}(f)(u)=c_p u^{1-p/2}\frac{f\{q(u)\}}{1+u},
  \qquad u>0,
\end{equation}
where \(c_p\) is the normalizing constant.  The generator update is
\begin{equation}\label{eq:ch7-generator-update-oracle}
  G(z)=\mathcal P\left[\mathcal A^{-1}\{f_{Y,z}\}\right],
  \qquad
  w_z(u)=-\frac{d}{du}\log G(z)(u),
\end{equation}
where \(\mathcal P\) enforces the normalization convention.

The mixing and center updates are
\begin{equation}\label{eq:ch7-pi-update-oracle}
  \Pi_k(z)=\E\{r_k(\vX;z)\},
\end{equation}
\begin{equation}\label{eq:ch7-center-proposal-oracle}
  \widetilde U_k(z)=\frac{\E\left[r_k(\vX;z)w_z\{\delta_k(\vX;z)\}\vX\right]}
  {\E\left[r_k(\vX;z)w_z\{\delta_k(\vX;z)\}\right]},
\end{equation}
with damped update
\begin{equation}\label{eq:ch7-center-damped-oracle}
  U_k(z)=(1-\eta_\mu)\vmu_k+\eta_\mu\widetilde U_k(z),
  \qquad 0<\eta_\mu\le1.
\end{equation}

The common-shape block uses bounded-radial pilots, Tyler reweighting, POET regularization, and graphical lasso.  After the center update, let \(\vr_k^+(\vX;z)=\vX-U_k(z)\).  The weighted spatial-sign pilot is
\begin{equation}\label{eq:ch7-H-oracle}
  \mH(z)=\sum_{k=1}^K\E\left[r_k(\vX;z)
  \frac{\vr_k^+(\vX;z)\vr_k^+(\vX;z)\trans}{\twonorm{\vr_k^+(\vX;z)}^2\vee\varepsilon_r}
  \right].
\end{equation}
Applying POET to \(\mH(z)\) gives \(\mSigma_{\mathrm{pss}}(z)\).  The Tyler map is
\begin{equation}\label{eq:ch7-tyler-map-semi}
  \widetilde\mSigma_z(\mSigma)
  =p\sum_{k=1}^K\E\left[r_k(\vX;z)
  \frac{\vr_k^+(\vX;z)\vr_k^+(\vX;z)\trans}
  {\vr_k^+(\vX;z)\trans\mSigma^{-1}\vr_k^+(\vX;z)\vee\varepsilon_r}
  \right].
\end{equation}
A ridge-stabilized trace-normalized fixed point defines \(\mSigma_{\mathrm{Ty}}(z)\).  A second POET step yields \(\mSigma_{\mathrm{pt}}(z)\), and the sparse precision proposal is
\begin{equation}\label{eq:ch7-semi-glasso-oracle}
  \widetilde T(z)=\argmin_{\mP\succ0}\left\{\tr\{\mP\mSigma_{\mathrm{pt}}(z)\}-\log\det(\mP)+\lambda_{\Omega}(z)\onenorm{\operatorname{off}(\mP)}\right\}.
\end{equation}
After damping and trace normalization, the precision update is denoted by \(T(z)\).  The oracle map is
\begin{equation}\label{eq:ch7-semi-oracle-map}
  M(z)=\{\Pi_1(z),\ldots,\Pi_K(z),U_1(z),\ldots,U_K(z),T(z),G(z)\}.
\end{equation}

The empirical version replaces expectations by averages.  Given \(z^{(t)}\), compute
\begin{equation}\label{eq:ch7-semi-emp-delta}
  \Delta_{ik}^{(t)}=(\vX_i-\vmu_k^{(t)})\trans\mP^{(t)}(\vX_i-\vmu_k^{(t)}),
\end{equation}
\begin{equation}\label{eq:ch7-semi-emp-resp}
  \widehat\tau_{ik}^{(t+1)}
  =\frac{\pi_k^{(t)}\widehat g^{(t)}(\Delta_{ik}^{(t)})}
  {\sum_{\ell=1}^K\pi_\ell^{(t)}\widehat g^{(t)}(\Delta_{i\ell}^{(t)})},
  \qquad
  \pi_k^{(t+1)}=n^{-1}\sum_{i=1}^n\widehat\tau_{ik}^{(t+1)}.
\end{equation}
The generator is estimated by a weighted kernel density estimator of \(Y_{ik}^{(t)}=\log(1+\Delta_{ik}^{(t)})\):
\begin{equation}\label{eq:ch7-semi-fhat}
  \widehat f_{Y,z^{(t)},h}(y)
  =\frac1h\sum_{i=1}^n\sum_{k=1}^K\widetilde w_{ik}^{(t+1)}
  \phi\left(\frac{y-Y_{ik}^{(t)}}{h}\right),
  \qquad
  \widetilde w_{ik}^{(t+1)}=\frac{\widehat\tau_{ik}^{(t+1)}}{\sum_{a,b}\widehat\tau_{ab}^{(t+1)}}.
\end{equation}
The inverse transform \eqref{eq:ch7-inverse-radial-generator}, smoothing on the transformed-radius grid, and interpolation yield \(\widehat G_{n,h}(z^{(t)})\) and the clipped radial score \(\widehat w_{n,h}\).  The center update is
\begin{equation}\label{eq:ch7-semi-center-emp}
  \vmu_k^{(t+1)}=(1-\eta_\mu)\vmu_k^{(t)}+
  \eta_\mu
  \frac{\sum_{i=1}^n\widehat\tau_{ik}^{(t+1)}\widehat w_{n,h}(\Delta_{ik}^{(t)})\vX_i}
  {\sum_{i=1}^n\widehat\tau_{ik}^{(t+1)}\widehat w_{n,h}(\Delta_{ik}^{(t)})}.
\end{equation}
The empirical weighted spatial-sign pilot is
\begin{equation}\label{eq:ch7-Hhat-semi}
  \widehat\mH_n(z^{(t)})=
  \frac{\sum_{i=1}^n\sum_{k=1}^K\widehat\tau_{ik}^{(t+1)}
  \vr_{ik}^{(t+1)}\vr_{ik}^{(t+1)\trans}/\{\twonorm{\vr_{ik}^{(t+1)}}^2\vee\varepsilon_r\}}
  {\sum_{i=1}^n\sum_{k=1}^K\widehat\tau_{ik}^{(t+1)}},
\end{equation}
where \(\vr_{ik}^{(t+1)}=\vX_i-\vmu_k^{(t+1)}\).  This is followed by POET, weighted Tyler iteration, a second POET step, and graphical lasso exactly as in \eqref{eq:ch7-H-oracle}--\eqref{eq:ch7-semi-glasso-oracle}.  The final plug-in classifier is
\begin{equation}\label{eq:ch7-semi-plugin-rule}
  \widehat G(\vx)=\argmax_{1\le k\le K}\left[\log\widehat\pi_k+\log\widehat g\{(\vx-\widehat\vmu_k)\trans\widehat\mP(\vx-\widehat\vmu_k)\}\right].
\end{equation}

\subsection{Selecting \texorpdfstring{\(K\)}{K} by a gap-LSE rule}

Let \(\mathcal K=\{K_{\min},\ldots,K_{\max}\}\).  For each \(K\in\mathcal K\), fit the semiparametric GEM algorithm and compute hard labels \(\widehat C_{i,K}\).  With \(\rho(u)=\log(1+u)\), define the observed radial dispersion
\begin{equation}\label{eq:ch7-gap-lse-observed}
  W_K=\frac1n\sum_{i=1}^n \rho\{\widehat\Delta_{i\widehat C_{i,K},K}\}.
\end{equation}
For reference data \(\mX^{0,b}\) obtained by independently permuting every coordinate, compute \(W_K^{0,b}\) in the same way.  Let
\begin{equation}\label{eq:ch7-gap-lse-gap}
  \operatorname{Gap}(K)=\frac1B\sum_{b=1}^B\log W_K^{0,b}-\log W_K,
\end{equation}
\begin{equation}\label{eq:ch7-gap-lse-sk}
  s_K=\sqrt{1+B^{-1}}\left[\frac1{B-1}\sum_{b=1}^B\left\{\log W_K^{0,b}-B^{-1}\sum_{a=1}^B\log W_K^{0,a}\right\}^2\right]^{1/2}.
\end{equation}
Writing \(K_1<\cdots<K_L\), the lower-complexity one-standard-error rule is
\begin{equation}\label{eq:ch7-gap-lse-rule}
  \widehat K_{\mathrm{LSE}}
  =\min\left\{K_j:1\le j<L,\ \operatorname{Gap}(K_j)\ge \operatorname{Gap}(K_{j+1})-s_{K_{j+1}}\right\},
\end{equation}
with \(\widehat K_{\mathrm{LSE}}=K_L\) if the set is empty.

\subsection{High-dimensional theory for the semiparametric GEM}

Let \(z^*=(\bm\pi^*,\vmu_1^*,\ldots,\vmu_K^*,\mP^*,g^*)\).  The distance between two parameters is
\begin{equation}\label{eq:ch7-semi-distance}
\begin{aligned}
  d(z,\widetilde z)=\min_{\sigma\in\mathfrak S_K}\Big[&
  \norm{\bm\pi-\widetilde{\bm\pi}_{\sigma}}_\infty
  \vee\max_k\norm{\vmu_k-\widetilde\vmu_{\sigma(k)}}_\infty
  \vee\onenorm{\mP-\widetilde\mP} \\
  &\vee\sup_{0\le u\le U_0}|\ell(u)-\widetilde\ell(u)|
  \vee\sup_{0\le u\le U_0}(1+u)^{1/2}|w(u)-\widetilde w(u)|
  \Big].
\end{aligned}
\end{equation}
Let \(\mathcal B(r)=\{z:d(z,z^*)\le r\}\) and let
\begin{equation}\label{eq:ch7-zeta-eps}
  \eta_n^{\mathrm{Ty}}=\sqrt{\frac{\log p}{n}},
  \qquad
  \zeta_n=s_\Omega\left(\eta_n^{\mathrm{Ty}}+\rho_{n,p}+\lambda_{\Omega,n}\right),
\end{equation}
\begin{equation}\label{eq:ch7-eps-semi}
  \varepsilon_{n,h,\zeta}=h^2+\sqrt{\frac{\log n}{nh}}+\sqrt{\frac{\log p}{n}}+\zeta_n.
\end{equation}
Here \(s_\Omega\) is the maximum row sparsity of the common sparse precision matrix, \(\rho_{n,p}\) is the POET approximation error, and \(\lambda_{\Omega,n}\) is the graphical-lasso penalty level.

\begin{assumption}[Local semiparametric elliptical mixture conditions]\label{ass:ch7-semi-local}
The number of clusters \(K\) is fixed.  The data follow \eqref{eq:ch7-semi-mixture-density}; \(\min_k\pi_k^*>c_\pi>0\); \(\lambda_{\min}(\mP^*)\) and \(\lambda_{\max}(\mP^*)\) are bounded away from zero and infinity; the minimum Mahalanobis separation
\begin{equation}\label{eq:ch7-semi-separation}
  \Delta_*^2=\min_{k<\ell}(\vmu_k^*-\vmu_\ell^*)\trans\mP^*(\vmu_k^*-\vmu_\ell^*)
\end{equation}
is at least \(\Delta_0^2\); the generator \(g^*\) is positive, decreasing, twice continuously differentiable on \([0,U_0]\), and its score \(w^*=-(\log g^*)'\) satisfies
\begin{equation}\label{eq:ch7-score-regularity}
  \sup_{0\le u\le U_0}(1+u)^{1/2}w^*(u)<\infty,
  \qquad
  \sup_{0\le u\le U_0}(1+u)^{3/2}|w^{*\prime}(u)|<\infty.
\end{equation}
The common scatter has a low-rank-plus-sparse decomposition suitable for POET, the sparse precision graph satisfies a restricted strong convexity condition for graphical lasso, and the initial estimator \(z^{(0)}\) satisfies \(d(z^{(0)},z^*)\le r/2\) with probability tending to one.
\end{assumption}

\begin{theorem}[Population contraction]\label{thm:ch7-semi-pop-contraction}
Under Assumption~\ref{ass:ch7-semi-local}, for sufficiently large separation \(\Delta_0\), there exist \(r>0\) and \(\kappa\in(0,1)\) such that the oracle map satisfies
\begin{equation}\label{eq:ch7-semi-pop-contraction}
  d\{M(z),z^*\}\le \kappa d(z,z^*),
  \qquad z\in\mathcal B(r).
\end{equation}
\end{theorem}

\begin{theorem}[Uniform one-step empirical concentration]\label{thm:ch7-semi-emp-concentration}
Under Assumption~\ref{ass:ch7-semi-local}, there exist constants \(C,c>0\) such that, uniformly over \(z\in\mathcal B(r)\),
\begin{equation}\label{eq:ch7-semi-nonprecision-conc}
  \max\left\{
  \norm{\widehat{\bm\pi}(z)-\bm\Pi(z)}_\infty,
  \max_k\norm{\widehat\vmu_k(z)-U_k(z)}_\infty,
  \sup_{u\le U_0}|\widehat G_{n,h}(z)(u)-G_h(z)(u)|
  \right\}
  \le C\left\{\sqrt{\frac{\log p}{n}}+\sqrt{\frac{\log n}{nh}}\right\}
\end{equation}
with probability at least \(1-Cp^{-c}-Cn^{-c}\).  The common precision block satisfies
\begin{equation}\label{eq:ch7-semi-precision-conc}
  \sup_{z\in\mathcal B(r)}\onenorm{\widehat\mP_h(z)-\mP_h(z)}
  \le C\zeta_n
\end{equation}
with probability at least \(1-Cp^{-c}\).
\end{theorem}

\begin{theorem}[Main recursion and parameter consistency]\label{thm:ch7-semi-main-recursion}
Under Assumption~\ref{ass:ch7-semi-local}, for every deterministic \(T_n\lesssim \log n\), with probability at least \(1-Cp^{-c}-Cn^{-c}\),
\begin{equation}\label{eq:ch7-semi-main-recursion}
  e_{t+1}\le \kappa e_t+C\left\{h^2+\sqrt{\frac{\log n}{nh}}+\sqrt{\frac{\log p}{n}}+\zeta_n\right\},
  \qquad e_t=d(z^{(t)},z^*),
\end{equation}
for \(0\le t<T_n\).  Therefore,
\begin{equation}\label{eq:ch7-semi-param-rate}
  d(z^{(T_n)},z^*)=O_p(\varepsilon_{n,h,\zeta}).
\end{equation}
If \(h\asymp(\log n/n)^{1/5}\), \(\rho_{n,p}\lesssim \sqrt{\log p/n}\), and \(\lambda_{\Omega,n}\asymp\sqrt{\log p/n}\), then
\begin{equation}\label{eq:ch7-semi-canonical-rate}
  d(z^{(T_n)},z^*)
  =O_p\left\{\left(\frac{\log n}{n}\right)^{2/5}+(1+s_\Omega)\sqrt{\frac{\log p}{n}}\right\}.
\end{equation}
\end{theorem}

Let
\begin{equation}\label{eq:ch7-semi-margin}
  \mathcal M^*(\vx)=\eta_{G^*(\vx)}^*(\vx)-\max_{\ell\ne G^*(\vx)}\eta_\ell^*(\vx),
  \qquad
  \eta_k^*(\vx)=\log\pi_k^*+\ell^*\{\delta_k^*(\vx)\}.
\end{equation}
Assume the margin condition \(\Prob\{0<\mathcal M^*(\vX)\le t\mid R\}\le C_m t^\alpha\) and \(\E(1+R)^{\alpha+1}<\infty\).  Then the plug-in rule \eqref{eq:ch7-semi-plugin-rule} satisfies
\begin{equation}\label{eq:ch7-semi-risk-rate}
  R(\widehat G)-R(G^*)=O_p\{\varepsilon_{n,h,\zeta}^{\alpha+1}\}.
\end{equation}
In the Lipschitz-margin case \(\alpha=1\), the excess misclustering risk is quadratic in the parameter error scale.

\section{Bibliographic notes}

Classical \(K\)-means traces back to \citet{MacQueen1967} and \citet{HartiganWong1979}; broad algorithmic treatments are given by \citet{JainDubes1988}.  Gaussian model-based clustering is reviewed by \citet{FraleyRaftery2002}, \citet{BouveyronBrunetSaumard2014}, and \citet{GormleyMurphyRaftery2023}.  Sparse and regularized clustering methods include \citet{WittenTibshirani2010Clustering}, \citet{SunWangFang2012}, \citet{PanShen2007}, \citet{RafteryDean2006}, \citet{RaymaekersZamar2022}, and \citet{FopMurphy2018}.  High-dimensional Gaussian mixture theory and feature-screening spectral clustering are represented by \citet{AzizyanSinghWasserman2013}, \citet{AzizyanSinghWasserman2015}, \citet{JinWang2016IFPCA}, \citet{JinKeWang2017}, and \citet{CaiMaZhang2019}.  Robust mixture modeling using heavy-tailed parametric families includes \citet{PeelMcLachlan2000}, \citet{AndrewsMcNicholas2012}, and \citet{DangBrowneMcNicholas2015}.  The sparse spatial-median procedure in Section~\ref{sec:ch7-sparse-sm} follows \citet{ZhaoZhuangFeng2026SparseKSM}, while the semiparametric elliptical mixture method in Section~\ref{sec:ch7-semi-elliptical-mixture} follows \citet{FengZhuang2026SEMC}.  The latter also connects to Tyler's shape estimation \citep{Tyler1987}, POET \citep{FanLiaoMincheva2013}, graphical lasso \citep{FriedmanHastieTibshirani2008}, and elliptical factor-model matrix estimation \citep{FanLiuWang2018,XuMaWangFeng2026EllipticalFactor}.

\section{Appendix to Chapter 7: Proofs of selected results}\label{sec:ch7-proofs}

\subsection{Proof of Theorem~\ref{thm:ch7-fixed-kmeans}}

Let
\[
  m(\vX,\vtheta)=\min_{1\le k\le K}\twonorm{\vX-\vmu_k}^2,
  \qquad
  M_n(\vtheta)=n^{-1}\sum_{i=1}^n m(\vX_i,\vtheta),
  \qquad
  M(\vtheta)=\E m(\vX,\vtheta).
\]
On the event that \(\widehat\vtheta\) lies in the local basin and no observation falls on an empirical tie boundary, the first-order condition is
\begin{equation}\label{eq:ch7-proof-kmeans-score}
  \nabla M_n(\widehat\vtheta)=\vct0.
\end{equation}
A Taylor expansion around \(\vtheta_0\) gives
\begin{equation}\label{eq:ch7-proof-kmeans-taylor}
  \vct0=\nabla M_n(\vtheta_0)+\nabla^2M(\vtheta_0)(\widehat\vtheta-\vtheta_0)+\{\nabla^2M_n(\bar\vtheta)-\nabla^2M(\vtheta_0)\}(\widehat\vtheta-\vtheta_0),
\end{equation}
where \(\bar\vtheta\) lies between \(\widehat\vtheta\) and \(\vtheta_0\).  The local curvature and empirical-process regularity imply
\begin{equation}\label{eq:ch7-proof-kmeans-rem}
  \opnorm{\nabla^2M_n(\bar\vtheta)-\nabla^2M(\vtheta_0)}=o_p(1),
  \qquad \widehat\vtheta-\vtheta_0=O_p(n^{-1/2}).
\end{equation}
Thus
\begin{equation}\label{eq:ch7-proof-kmeans-linear}
  \sqrt n(\widehat\vtheta-\vtheta_0)
  =-\mH_{\mathrm{KM}}^{-1}\frac1{\sqrt n}\sum_{i=1}^n\nabla_{\vtheta} m(\vX_i,\vtheta_0)+o_p(1).
\end{equation}
The multivariate central limit theorem gives
\[
  n^{-1/2}\sum_{i=1}^n\nabla_{\vtheta} m(\vX_i,\vtheta_0)
  \overset{d}{\longrightarrow}N(\vct0,\mB_{\mathrm{KM}}),
\]
which proves \eqref{eq:ch7-kmeans-asymp-normal}.

\subsection{Proof of Theorem~\ref{thm:ch7-gmm-mle}}

Let \(s(\vX;\vtheta)=\nabla\log f(\vX;\vtheta)\).  The likelihood score satisfies
\begin{equation}\label{eq:ch7-proof-mle-score}
  \vct0=n^{-1}\sum_{i=1}^n s(\vX_i;\widehat\vtheta).
\end{equation}
Expanding at \(\vtheta_0\),
\begin{equation}\label{eq:ch7-proof-mle-expansion}
  \vct0=n^{-1}\sum_{i=1}^n s(\vX_i;\vtheta_0)
  +\left\{n^{-1}\sum_{i=1}^n\nabla_{\vtheta} s(\vX_i;\bar\vtheta)\right\}(\widehat\vtheta-\vtheta_0).
\end{equation}
By the law of large numbers,
\begin{equation}\label{eq:ch7-proof-info}
  -n^{-1}\sum_{i=1}^n\nabla_{\vtheta} s(\vX_i;\bar\vtheta)\to_p \mI(\vtheta_0),
\end{equation}
and by the central limit theorem,
\begin{equation}\label{eq:ch7-proof-score-clt}
  n^{-1/2}\sum_{i=1}^n s(\vX_i;\vtheta_0)
  \overset{d}{\longrightarrow}N\{\vct0,\mI(\vtheta_0)\}.
\end{equation}
Combining \eqref{eq:ch7-proof-mle-expansion}--\eqref{eq:ch7-proof-score-clt} yields \eqref{eq:ch7-gmm-mle-normal}.

\subsection{Proof of Theorem~\ref{thm:ch7-chime}}

The proof follows the contraction-plus-concentration argument of \citet{CaiMaZhang2019}.  We record the algebra because this template is used repeatedly in robust high-dimensional clustering.
Let \(M(\theta)\) be the population EM map and \(M_n(\theta)\) the sample EM map for \((\omega,\vmu_1,\vmu_2,\mSigma)\).  Under Assumption~\ref{ass:ch7-chime}, the population map satisfies the sparse-norm contraction
\begin{equation}\label{eq:ch7-proof-chime-pop-contract}
  d_{2,s}\{M(\theta),\theta^*\}
  \le \kappa_0\left\{
  d_{2,s}(\theta,\theta^*)\vee\twonorm{\vbeta-\vbeta^*}
  \right\},
  \qquad 0<\kappa_0<1/2.
\end{equation}
Uniformly over the contraction basin,
\begin{equation}\label{eq:ch7-proof-chime-concentration}
  d_{2,s}\{M_n(\theta),M(\theta)\}
  \le C\sqrt{\frac{s\log p}{n}},
  \qquad
  \maxnorm{\{\widehat\mSigma(\theta)-\mSigma(\theta)\}\vbeta^*}
  \le C\sqrt{\frac{\log p}{n}}
\end{equation}
with probability tending to one.  These bounds are based on Gaussian concentration for weighted empirical means and weighted empirical covariance actions over the cone \(\mathcal C(s)\).

At iteration \(t+1\), write \(\widehat\vdelta^{(t+1)}=\widehat\vmu_1^{(t+1)}-\widehat\vmu_2^{(t+1)}\).  The convex program \eqref{eq:ch7-chime-beta-update} gives the basic inequality
\begin{align}\label{eq:ch7-proof-chime-basic}
  &\frac12(\widehat\vbeta^{(t+1)})\trans\widehat\mSigma^{(t+1)}\widehat\vbeta^{(t+1)}
  -(\widehat\vbeta^{(t+1)})\trans\widehat\vdelta^{(t+1)}
  +\lambda_{n,t+1}\onenorm{\widehat\vbeta^{(t+1)}} \\
  &\qquad\le
  \frac12(\vbeta^*)\trans\widehat\mSigma^{(t+1)}\vbeta^*
  -(\vbeta^*)\trans\widehat\vdelta^{(t+1)}
  +\lambda_{n,t+1}\onenorm{\vbeta^*}.
\end{align}
Let \(\vh^{(t+1)}=\widehat\vbeta^{(t+1)}-\vbeta^*\) and \(S^*=\operatorname{supp}(\vbeta^*)\).  Rearranging \eqref{eq:ch7-proof-chime-basic} yields
\begin{align}\label{eq:ch7-proof-chime-rearrange}
  \frac12(\vh^{(t+1)})\trans\widehat\mSigma^{(t+1)}\vh^{(t+1)}
  &\le
  (\vh^{(t+1)})\trans\{\widehat\vdelta^{(t+1)}-\widehat\mSigma^{(t+1)}\vbeta^*\}  \\
  &\qquad+\lambda_{n,t+1}\{\onenorm{\vbeta^*}-\onenorm{\widehat\vbeta^{(t+1)}}\}.
\end{align}
On the event
\begin{equation}\label{eq:ch7-proof-chime-score-bound}
  \maxnorm{\widehat\vdelta^{(t+1)}-\widehat\mSigma^{(t+1)}\vbeta^*}
  \le \lambda_{n,t+1}/2,
\end{equation}
we obtain
\begin{align}\label{eq:ch7-proof-chime-cone-ineq}
  \frac12(\vh^{(t+1)})\trans\widehat\mSigma^{(t+1)}\vh^{(t+1)}
  &\le \frac{\lambda_{n,t+1}}{2}\onenorm{\vh^{(t+1)}}
  +\lambda_{n,t+1}\{\onenorm{\vh^{(t+1)}_{S^*}}-\onenorm{\vh^{(t+1)}_{S^{*c}}}\} \\
  &\le \frac{3\lambda_{n,t+1}}{2}\onenorm{\vh^{(t+1)}_{S^*}
  }-\frac{\lambda_{n,t+1}}{2}\onenorm{\vh^{(t+1)}_{S^{*c}}}.
\end{align}
Hence \(\onenorm{\vh^{(t+1)}_{S^{*c}}}\le3\onenorm{\vh^{(t+1)}_{S^*}}\), so \(\vh^{(t+1)}\in\mathcal C(s)\).  Restricted eigenvalue boundedness gives
\begin{equation}\label{eq:ch7-proof-chime-re}
  (\vh^{(t+1)})\trans\widehat\mSigma^{(t+1)}\vh^{(t+1)}
  \ge c\twonorm{\vh^{(t+1)}}^2.
\end{equation}
Combining \eqref{eq:ch7-proof-chime-cone-ineq} and \eqref{eq:ch7-proof-chime-re},
\begin{equation}\label{eq:ch7-proof-chime-beta-one-step}
  \twonorm{\widehat\vbeta^{(t+1)}-\vbeta^*}
  \le C\sqrt{s}\lambda_{n,t+1}.
\end{equation}
The recursive choice \eqref{eq:ch7-chime-lambda}, together with \eqref{eq:ch7-proof-chime-pop-contract}--\eqref{eq:ch7-proof-chime-concentration}, gives
\begin{equation}\label{eq:ch7-proof-chime-recursion}
  e_{t+1}\le \kappa e_t+C\sqrt{\frac{s\log p}{n}},
  \qquad
  e_t=d_{2,s}(\widehat\theta^{(t)},\theta^*)\vee
  \twonorm{\widehat\vbeta^{(t)}-\vbeta^*}.
\end{equation}
Iterating,
\begin{equation}\label{eq:ch7-proof-chime-unroll}
  e_{T_0}\le \kappa^{T_0}e_0+\frac{C}{1-\kappa}\sqrt{\frac{s\log p}{n}},
\end{equation}
which proves \eqref{eq:ch7-chime-beta-rate} and \eqref{eq:ch7-chime-beta-opt-rate}.

It remains to connect parameter error to clustering risk.  The plug-in discriminant score is
\begin{equation}\label{eq:ch7-proof-chime-score}
  \widehat\eta(\vx)=(\vx-\widehat{\bar\vmu})\trans\widehat\vbeta-
  \log\{\widehat\omega/(1-\widehat\omega)\},
  \qquad
  \eta^*(\vx)=(\vx-\bar\vmu^*)\trans\vbeta^*-
  \log\{\omega^*/(1-\omega^*)\}.
\end{equation}
Gaussian margin calculations imply
\begin{equation}\label{eq:ch7-proof-chime-margin}
  R(\widehat G_{\mathrm{CHIME}})-R(G^*)
  \le C\E\left[\{\widehat\eta(\vX)-\eta^*(\vX)\}^2\right]
  \le C\twonorm{\widehat\vbeta-\vbeta^*}^2
       +C d_{2,s}^2(\widehat\theta,\theta^*).
\end{equation}
Substituting \eqref{eq:ch7-proof-chime-unroll} with \(T_0\gtrsim\log n\) gives \eqref{eq:ch7-chime-risk-rate}.  The minimax lower bound follows by constructing a packing of sparse discriminant vectors and applying Fano's inequality to the induced Gaussian mixture experiments, yielding a lower bound of order \(s\log p/n\) for the excess risk.

\subsection{Proof of Theorem~\ref{thm:ch7-ifpca-generic}}

By Weyl's inequality and Assumption~\ref{ass:ch7-ifpca},
\begin{equation}\label{eq:ch7-proof-ifpca-gap}
  \sigma_{K-1}(\widetilde\mX_{\widehat S})
  \ge \sigma_{K-1}(\mM_{\widehat S})-\opnorm{\mE_{\widehat S}}
  =\sigma_* -\epsilon_* > \epsilon_*.
\end{equation}
The Davis--Kahan sin-theta theorem gives
\begin{equation}\label{eq:ch7-proof-ifpca-dk}
  \opnorm{\sin\Theta(\widehat\mU,\mU_*)}
  \le \frac{\opnorm{\mE_{\widehat S}}}{\sigma_*-\opnorm{\mE_{\widehat S}}}
  \le \frac{\epsilon_*}{\sigma_*-\epsilon_*}.
\end{equation}
Since the target subspace has rank \(K-1\),
\begin{equation}\label{eq:ch7-proof-ifpca-fro}
  \inf_{\mO\trans\mO=\mI_{K-1}}\frobnorm{\widehat\mU-\mU_*\mO}
  \le \sqrt2\frobnorm{\sin\Theta(\widehat\mU,\mU_*)}
  \le \frac{2\sqrt2\,\epsilon_*}{\sigma_*-\epsilon_*},
\end{equation}
which proves \eqref{eq:ch7-ifpca-sintheta} after replacing the harmless constant.

Let \(\vct{q}_k\) be the population row center of \(\mU_*\) for cluster \(k\).  The assumed center separation and balance imply the standard \(K\)-means perturbation inequality
\begin{equation}\label{eq:ch7-proof-ifpca-kmeans-pert}
  \frac1n\min_{\sigma\in\mathfrak S_K}\sum_{i=1}^n
  \mathbbm{1}\{\widehat c_i\ne\sigma(Z_i)\}
  \le C_K\inf_{\mO\trans\mO=\mI_{K-1}}\frobnorm{\widehat\mU-\mU_*\mO}^2.
\end{equation}
Combining \eqref{eq:ch7-proof-ifpca-fro} and \eqref{eq:ch7-proof-ifpca-kmeans-pert} yields \eqref{eq:ch7-ifpca-misclustering}.

Finally, for sub-Gaussian entries in \(\mE_{\widehat S}\), matrix Bernstein's inequality gives
\begin{equation}\label{eq:ch7-proof-ifpca-opnorm}
  \opnorm{\mE_{\widehat S}}
  \le C\{\sqrt n+\sqrt{|\widehat S|}\}
\end{equation}
with probability tending to one.  This proves \eqref{eq:ch7-ifpca-noise-rate} and the stated recovery condition.

\subsection{Proof of Theorem~\ref{thm:ch7-smsscm-consistency}}

Work on \(\mathcal E_n\cap\mathcal M_n\cap\mathcal F_n\), where every true cluster is nonempty.  This event has probability tending to one.  Max--min seeding under \(d>4a_n\) selects one seed from each true cluster.  Indeed, points in the same cluster are at distance at most \(2a_n\), while points in different clusters are at distance at least \(d-2a_n>2a_n\).  Hence, after label permutation \(\sigma\), the initial centers \(\vct{m}_k^{(0)}\) satisfy
\begin{equation}\label{eq:ch7-proof-seed-close}
  \max_k\twonorm{\vct{m}_k^{(0)}-\vmu_{\sigma(k)}}\le a_n.
\end{equation}
Consider a point \(\vX_i\) with \(Z_i=r\).  For its own center,
\begin{equation}\label{eq:ch7-proof-own-dist}
  \norm{\vX_i-\vct{m}_r}_{\mA_t}
  \le \lambda^{-1/2}\twonorm{\vX_i-\vct{m}_r}
  \le 2a_n/\sqrt\lambda.
\end{equation}
For any \(s\ne r\), by \eqref{eq:ch7-metric-stability},
\begin{align}\label{eq:ch7-proof-other-dist}
  \norm{\vX_i-\vct{m}_s}_{\mA_t}
  &\ge \sqrt{1-\eta}\,\norm{\vX_i-\vct{m}_s}_{\mA_\star} \\
  &\ge \sqrt{1-\eta}\left\{\norm{\vmu_r-\vmu_s}_{\mA_\star}
       -\norm{\vX_i-\vmu_r}_{\mA_\star}-\norm{\vct{m}_s-\vmu_s}_{\mA_\star}\right\} \\
  &\ge \sqrt{1-\eta}\,d_{\mathrm{sgn}}-2a_n\sqrt{(1+\eta)/\lambda}.
\end{align}
The separation condition \eqref{eq:ch7-smsscm-separation} makes the lower bound in \eqref{eq:ch7-proof-other-dist} larger than \eqref{eq:ch7-proof-own-dist}.  Therefore every assignment is correct.  Given correct assignments, the spatial median of each empirical true cluster lies in the convex hull of that cluster and hence is within distance \(a_n\) of the corresponding population center on \(\mathcal E_n\).  The same argument applies at the next assignment step.  Induction proves that all subsequent partitions are correct up to the same permutation.  Since there are finitely many partitions and the deterministic update cannot cycle once assignments are fixed, the algorithm terminates at the correct fixed point.

\subsection{Proof of Theorem~\ref{thm:ch7-sparse-feature}}

Assume \eqref{eq:ch7-center-close-sparse}.  For every coordinate \(j\),
\begin{align}\label{eq:ch7-proof-score-perturb}
  |s_j(\vct{m}_1,\ldots,\vct{m}_K)-s_j^\star|
  &\le \sum_{k=1}^K |(m_{kj}-\bar m_j)-(\mu_{\sigma(k)j}-\bar\mu_j)| \\
  &\le \sum_{k=1}^K |m_{kj}-\mu_{\sigma(k)j}|+K|\bar m_j-\bar\mu_j| \\
  &\le 2K\max_k\twonorm{\vct{m}_k-\vmu_{\sigma(k)}}
  \le 2Ka_n.
\end{align}
If \(j\notin S_0\), then \(s_j^\star=0\), so \(s_j\le2Ka_n<\tau_n\).  If \(j\in S_0\), then
\[
  s_j\ge s_j^\star-2Ka_n\ge s_{\min}-2Ka_n>\tau_n.
\]
Thus \(S(\tau_n)=S_0\).  The probability statement follows by applying the deterministic result on an event whose probability tends to one.

\subsection{Proof of Theorem~\ref{thm:ch7-sparse-clustering}}

Work on \(\mathcal E_n\cap\mathcal F_n\).  The same max--min argument used in the proof of Theorem~\ref{thm:ch7-smsscm-consistency}, with \(d_0>4a_n\) on the active subspace, gives initial centers satisfying \eqref{eq:ch7-center-close-sparse} on \(S_0\).  By Theorem~\ref{thm:ch7-sparse-feature}, the first thresholding step selects \(S_0\).  For \(Z_i=r\), its own active-subspace distance is at most \(2a_n\), whereas for \(s\ne r\),
\[
  \twonorm{(\vX_i)_{S_0}-(\vct{m}_s)_{S_0}}
  \ge d_0-2a_n>2a_n.
\]
Therefore the first assignment on \(S_0\) is correct.  After correct assignment, every cluster spatial median lies in the convex hull of its true cluster and remains within \(a_n\) of the corresponding \(\vmu_k\).  Reapplying Theorem~\ref{thm:ch7-sparse-feature} shows that the active set remains \(S_0\), and the same distance comparison gives correct assignments.  Induction and finite termination prove the result.

\subsection{Proof of Theorem~\ref{thm:ch7-semi-main-recursion}}

For \(z\in\mathcal B(r)\), decompose one empirical step as
\begin{align}\label{eq:ch7-proof-semi-decomp}
  d\{M_{n,h}(z),z^*\}
  &\le d\{M_{n,h}(z),M_h(z)\}+d\{M_h(z),M(z)\}+d\{M(z),z^*\}.
\end{align}
The first term is bounded by the empirical concentration theorem:
\begin{equation}\label{eq:ch7-proof-semi-emp}
  \sup_{z\in\mathcal B(r)}d\{M_{n,h}(z),M_h(z)\}
  \le C\left\{\sqrt{\frac{\log n}{nh}}+\sqrt{\frac{\log p}{n}}+\zeta_n\right\}
\end{equation}
with probability at least \(1-Cp^{-c}-Cn^{-c}\).  The second term is the kernel smoothing bias,
\begin{equation}\label{eq:ch7-proof-semi-bias}
  \sup_{z\in\mathcal B(r)}d\{M_h(z),M(z)\}\le Ch^2.
\end{equation}
The third term is population contraction, \(d\{M(z),z^*\}\le\kappa d(z,z^*)\).  Combining \eqref{eq:ch7-proof-semi-decomp}--\eqref{eq:ch7-proof-semi-bias} and taking \(z=z^{(t)}\) gives \eqref{eq:ch7-semi-main-recursion}.  Iterating the recursion gives
\begin{equation}\label{eq:ch7-proof-semi-unroll}
  e_t\le \kappa^t e_0+\frac{C}{1-\kappa}\left\{h^2+\sqrt{\frac{\log n}{nh}}+\sqrt{\frac{\log p}{n}}+\zeta_n\right\}.
\end{equation}
Choosing \(T_n\asymp\log(1/\varepsilon_{n,h,\zeta})\) makes \(\kappa^{T_n}e_0\lesssim\varepsilon_{n,h,\zeta}\), proving \eqref{eq:ch7-semi-param-rate}.  The canonical rate follows by substituting \(h\asymp(\log n/n)^{1/5}\), \(\rho_{n,p}\lesssim\sqrt{\log p/n}\), and \(\lambda_{\Omega,n}\asymp\sqrt{\log p/n}\) into \eqref{eq:ch7-eps-semi}.

\subsection{Proof of the excess-risk bound}

Let \(\widehat\eta_k(\vx)=\log\widehat\pi_k+\log\widehat g\{\widehat\delta_k(\vx)\}\).  On the event \(d(\widehat z,z^*)\le C\varepsilon_{n,h,\zeta}\), the regularity assumptions imply the score perturbation bound
\begin{equation}\label{eq:ch7-proof-score-error}
  \max_{1\le k\le K}|\widehat\eta_k(\vx)-\eta_k^*(\vx)|
  \le C(1+R)\varepsilon_{n,h,\zeta}.
\end{equation}
If \(\widehat G(\vx)\ne G^*(\vx)\), then the oracle margin must satisfy
\begin{equation}\label{eq:ch7-proof-margin-event}
  0<\mathcal M^*(\vx)
  \le 2\max_k|\widehat\eta_k(\vx)-\eta_k^*(\vx)|
  \le C(1+R)\varepsilon_{n,h,\zeta}.
\end{equation}
Therefore, conditioning on \(R\) and using the margin assumption,
\begin{align}\label{eq:ch7-proof-risk}
  R(\widehat G)-R(G^*)
  &\le \Prob\{0<\mathcal M^*(\vX)\le C(1+R)\varepsilon_{n,h,\zeta}\}+o_p(\varepsilon_{n,h,\zeta}^{\alpha+1}) \\
  &\le C\E\{(1+R)^\alpha\}\varepsilon_{n,h,\zeta}^\alpha\varepsilon_{n,h,\zeta}+o_p(\varepsilon_{n,h,\zeta}^{\alpha+1}) \\
  &=O_p(\varepsilon_{n,h,\zeta}^{\alpha+1}),
\end{align}
where the radial moment condition ensures finiteness.  This proves \eqref{eq:ch7-semi-risk-rate}.

%% file: chapters/appendix_probability.tex
\chapter{Auxiliary Probability Results}\label{app:probability-tools}

\idx{Bernstein inequality}\idx{vector Bernstein inequality}\idx{matrix Bernstein inequality}\idx{martingale central limit theorem}\idx{Mills ratio}\idx{extreme value distribution}\idx{Gumbel distribution}

This appendix collects several probability results that are repeatedly used in
high-dimensional inference.  The statements are written in forms tailored to the
arguments developed in the main chapters.  Throughout this appendix, $C,c>0$
denote generic constants whose values may change from line to line.

\section{Scalar, vector, and matrix Bernstein inequalities}

The first group of tools controls partial sums of independent random variables,
random vectors, and random matrices.  In later chapters, the scalar version is
used to control individual coordinates, the vector version is used for
$\ell_\infty$-type errors, and the matrix version is used for operator-norm
bounds of sample scatter and covariance matrices.

\begin{theorem}[Scalar Bernstein inequality]\label{thm:scalar-bernstein}
Let $X_1,\dots,X_n$ be independent real-valued random variables such that
$\E(X_i)=0$ and $\abs{X_i}\le M$ almost surely for every $i$.  Define
\[
  S_n=\sum_{i=1}^n X_i,
  \qquad
  V_n=\sum_{i=1}^n \E(X_i^2).
\]
Then, for every $t>0$,
\begin{equation}\label{eq:scalar-bernstein-tail}
  \Prob\{S_n\ge t\}
  \le
  \exp\!\left\{-\frac{t^2}{2(V_n+Mt/3)}\right\},
\end{equation}
and consequently,
\begin{equation}\label{eq:scalar-bernstein-two-sided}
  \Prob\{\abs{S_n}\ge t\}
  \le
  2\exp\!\left\{-\frac{t^2}{2(V_n+Mt/3)}\right\}.
\end{equation}
\end{theorem}

\begin{proof}
Fix $\theta\in(0,3/M)$.  For every real number $x\le M$, the power-series
expansion of $e^{\theta x}$ together with the inequality
\[
  \sum_{m=2}^\infty \frac{(\theta x)^m}{m!}
  \le
  \frac{\theta^2 x^2}{2}
  \sum_{m=0}^\infty \Big(\frac{\theta M}{3}\Big)^m
  =
  \frac{\theta^2 x^2}{2(1-\theta M/3)}
\]
implies
\begin{equation}\label{eq:scalar-bernstein-mgf-pointwise}
  e^{\theta x}
  \le
  1+\theta x+
  \frac{\theta^2 x^2}{2(1-\theta M/3)}.
\end{equation}
Applying \eqref{eq:scalar-bernstein-mgf-pointwise} with $x=X_i$ and using
$\E(X_i)=0$ yields
\[
  \E e^{\theta X_i}
  \le
  1+
  \frac{\theta^2\E(X_i^2)}{2(1-\theta M/3)}
  \le
  \exp\!\left\{
    \frac{\theta^2\E(X_i^2)}{2(1-\theta M/3)}
  \right\}.
\]
Since the $X_i$ are independent,
\begin{equation}\label{eq:scalar-bernstein-mgf-sum}
  \E e^{\theta S_n}
  =
  \prod_{i=1}^n \E e^{\theta X_i}
  \le
  \exp\!\left\{
    \frac{\theta^2 V_n}{2(1-\theta M/3)}
  \right\}.
\end{equation}
By Markov's inequality and \eqref{eq:scalar-bernstein-mgf-sum},
\[
  \Prob\{S_n\ge t\}
  =
  \Prob\{e^{\theta S_n}\ge e^{\theta t}\}
  \le
  e^{-\theta t}\E e^{\theta S_n}
  \le
  \exp\!\left\{
    -\theta t+
    \frac{\theta^2 V_n}{2(1-\theta M/3)}
  \right\}.
\]
Choose
\[
  \theta=\frac{t}{V_n+Mt/3}.
\]
Then $0<\theta<3/M$, and direct substitution gives
\[
  -\theta t+
  \frac{\theta^2 V_n}{2(1-\theta M/3)}
  =
  -\frac{t^2}{2(V_n+Mt/3)}.
\]
This proves \eqref{eq:scalar-bernstein-tail}.  The two-sided bound follows by
applying \eqref{eq:scalar-bernstein-tail} to $S_n$ and $-S_n$.
\end{proof}

\begin{corollary}[Vector Bernstein inequality in max norm]\label{cor:vector-bernstein}
Let $\vX_1,\dots,\vX_n$ be independent random vectors in $\R^p$ with
$\E(\vX_i)=\vct{0}$.  Write
$\vX_i=(X_{i1},\dots,X_{ip})\trans$, assume that
\[
  \max_{1\le i\le n}\norm{\vX_i}_\infty\le M
  \qquad\text{almost surely},
\]
and define
\[
  V_{\infty}=
  \max_{1\le j\le p}\sum_{i=1}^n \E(X_{ij}^2).
\]
Then, for every $t>0$,
\begin{equation}\label{eq:vector-bernstein}
  \Prob\!\left\{
    \Big\|\sum_{i=1}^n \vX_i\Big\|_\infty \ge t
  \right\}
  \le
  2p\exp\!\left\{-\frac{t^2}{2(V_{\infty}+Mt/3)}\right\}.
\end{equation}
\end{corollary}

\begin{proof}
For each coordinate $j$,
\[
  S_{n,j}=\sum_{i=1}^n X_{ij}
\]
is a sum of independent centered random variables satisfying
$\abs{X_{ij}}\le M$ almost surely and
$\sum_{i=1}^n \E(X_{ij}^2)\le V_{\infty}$.  By
Theorem~\ref{thm:scalar-bernstein},
\[
  \Prob\{\abs{S_{n,j}}\ge t\}
  \le
  2\exp\!\left\{-\frac{t^2}{2(V_{\infty}+Mt/3)}\right\}.
\]
Since
\[
  \Big\|\sum_{i=1}^n \vX_i\Big\|_\infty
  =
  \max_{1\le j\le p}\abs{S_{n,j}},
\]
a union bound yields
\[
  \Prob\!\left\{
    \Big\|\sum_{i=1}^n \vX_i\Big\|_\infty \ge t
  \right\}
  \le
  \sum_{j=1}^p \Prob\{\abs{S_{n,j}}\ge t\}
  \le
  2p\exp\!\left\{-\frac{t^2}{2(V_{\infty}+Mt/3)}\right\}.
\]
\end{proof}

\begin{corollary}[Entrywise matrix Bernstein inequality]\label{cor:matrix-entrywise-bernstein}
Let $\mX_1,\dots,\mX_n$ be independent random matrices in $\R^{p\times q}$
with $\E(\mX_i)=\mO$.  Assume
\[
  \max_{1\le i\le n}\maxnorm{\mX_i}\le M
  \qquad\text{almost surely},
\]
and define
\[
  V_{\max}=
  \max_{1\le r\le p\atop 1\le s\le q}
  \sum_{i=1}^n \E\big[(X_i)_{rs}^2\big].
\]
Then, for every $t>0$,
\begin{equation}\label{eq:matrix-entrywise-bernstein}
  \Prob\!\left\{
    \maxnorm{\sum_{i=1}^n \mX_i}\ge t
  \right\}
  \le
  2pq\exp\!\left\{-\frac{t^2}{2(V_{\max}+Mt/3)}\right\}.
\end{equation}
\end{corollary}

\begin{proof}
Apply Corollary~\ref{cor:vector-bernstein} to the vector obtained by stacking
all $pq$ entries of each $\mX_i$.
\end{proof}

\begin{theorem}[Self-adjoint matrix Bernstein inequality]\label{thm:matrix-bernstein}
Let $\mY_1,\dots,\mY_n$ be independent self-adjoint random matrices in
$\R^{p\times p}$ such that $\E(\mY_i)=\mO$ and
\[
  \lambda_{\max}(\mY_i)\le L
  \qquad\text{almost surely for every }i.
\]
Define the matrix variance proxy
\begin{equation}\label{eq:matrix-variance-proxy}
  v(\mY)=
  \opnorm{\sum_{i=1}^n \E(\mY_i^2)}.
\end{equation}
Then, for every $t>0$,
\begin{equation}\label{eq:matrix-bernstein-one-sided}
  \Prob\!\left\{
    \lambda_{\max}\Big(\sum_{i=1}^n \mY_i\Big)\ge t
  \right\}
  \le
  p\exp\!\left\{-\frac{t^2}{2(v(\mY)+Lt/3)}\right\}.
\end{equation}
If, in addition, $\lambda_{\max}(-\mY_i)\le L$ almost surely for every $i$,
then
\begin{equation}\label{eq:matrix-bernstein-two-sided}
  \Prob\!\left\{
    \opnorm{\sum_{i=1}^n \mY_i}\ge t
  \right\}
  \le
  2p\exp\!\left\{-\frac{t^2}{2(v(\mY)+Lt/3)}\right\}.
\end{equation}
\end{theorem}

\begin{proof}
Set $\mS_n=\sum_{i=1}^n \mY_i$.  For every $\theta>0$, the matrix Laplace
transform bound gives
\begin{equation}\label{eq:matrix-laplace-tail}
  \Prob\{\lambda_{\max}(\mS_n)\ge t\}
  \le
  e^{-\theta t}\,
  \tr\left[
    \exp\Big\{\sum_{i=1}^n \log \E e^{\theta \mY_i}\Big\}
  \right].
\end{equation}
We next bound $\E e^{\theta\mY_i}$.  Since $\lambda_{\max}(\mY_i)\le L$,
functional calculus implies
\begin{equation}\label{eq:matrix-mgf-bound-pointwise}
  e^{\theta\mY_i}
  \preceq
  \mI + \theta\mY_i +
  \frac{e^{\theta L}-\theta L-1}{L^2}\mY_i^2.
\end{equation}
Taking expectation and using $\E(\mY_i)=\mO$ yield
\[
  \E e^{\theta\mY_i}
  \preceq
  \mI+
  \frac{e^{\theta L}-\theta L-1}{L^2}\E(\mY_i^2)
  \preceq
  \exp\!\left\{
    \frac{e^{\theta L}-\theta L-1}{L^2}\E(\mY_i^2)
  \right\}.
\]
Hence
\[
  \sum_{i=1}^n \log \E e^{\theta\mY_i}
  \preceq
  g(\theta)
  \sum_{i=1}^n \E(\mY_i^2),
  \qquad
  g(\theta)=\frac{e^{\theta L}-\theta L-1}{L^2}.
\]
Substituting this into \eqref{eq:matrix-laplace-tail} and using
$\tr(e^{\mA})\le p\exp\{\lambda_{\max}(\mA)\}$ for self-adjoint $\mA$, we
obtain
\[
  \Prob\{\lambda_{\max}(\mS_n)\ge t\}
  \le
  p\exp\{-\theta t+g(\theta)v(\mY)\}.
\]
For $0<\theta<3/L$,
\[
  g(\theta)
  =
  \frac{e^{\theta L}-\theta L-1}{L^2}
  \le
  \frac{\theta^2}{2(1-\theta L/3)}.
\]
Therefore
\[
  \Prob\{\lambda_{\max}(\mS_n)\ge t\}
  \le
  p\exp\!\left\{
    -\theta t+
    \frac{\theta^2 v(\mY)}{2(1-\theta L/3)}
  \right\},
  \qquad 0<\theta<3/L.
\]
Choosing
\[
  \theta=\frac{t}{v(\mY)+Lt/3}
\]
produces \eqref{eq:matrix-bernstein-one-sided}.  Applying the same bound to
$-\mY_i$ and then using a union bound proves
\eqref{eq:matrix-bernstein-two-sided}.
\end{proof}

\begin{remark}[A convenient specialization]\label{rem:matrix-bernstein-average}
If the matrices in Theorem~\ref{thm:matrix-bernstein} are uniformly bounded by
$L$ in operator norm, then for the sample average
$\bar{\mY}_n=n^{-1}\sum_{i=1}^n \mY_i$ we have
\[
  \Prob\!\left\{\opnorm{\bar{\mY}_n}\ge x\right\}
  \le
  2p\exp\!\left\{-\frac{n^2x^2}{2(v(\mY)+nLx/3)}\right\}.
\]
This is the form most often used to bound the operator norm of centered sample
scatter or covariance matrices.
\end{remark}

\section{Martingale central limit theorem}

A second family of tools enters the proofs of U-statistic decompositions,
Hoeffding projections, quadratic-form expansions, and adaptive combination
statistics.

\begin{definition}[Martingale difference array]
Let $\{\mathcal F_{n,k}:0\le k\le k_n\}$ be an increasing family of
$\sigma$-fields.  A triangular array $\{\xi_{n,k}:1\le k\le k_n\}$ is called a
martingale difference array with respect to
$\{\mathcal F_{n,k}\}$ if $\xi_{n,k}$ is
$\mathcal F_{n,k}$-measurable, $\E\abs{\xi_{n,k}}<\infty$, and
\[
  \E(\xi_{n,k}\mid \mathcal F_{n,k-1})=0
  \qquad\text{a.s. for every }k.
\]
\end{definition}

\begin{theorem}[Martingale central limit theorem]\label{thm:martingale-clt}
Let $\{\xi_{n,k},\mathcal F_{n,k}:1\le k\le k_n\}$ be a martingale
 difference array, and set
\[
  S_n=\sum_{k=1}^{k_n}\xi_{n,k},
  \qquad
  V_n=\sum_{k=1}^{k_n}\E(\xi_{n,k}^2\mid \mathcal F_{n,k-1}).
\]
Assume that:
\begin{enumerate}[label=(\roman*)]
  \item $V_n\to^P 1$;
  \item for every $\varepsilon>0$,
  \begin{equation}\label{eq:martingale-lindeberg}
    L_n(\varepsilon)
    =
    \sum_{k=1}^{k_n}
    \E\!\left[
      \xi_{n,k}^2\mathbbm{1}(\abs{\xi_{n,k}}>\varepsilon)
      \mid \mathcal F_{n,k-1}
    \right]
    \to^P 0.
  \end{equation}
\end{enumerate}
Then
\begin{equation}\label{eq:martingale-clt-limit}
  S_n\to^d N(0,1).
\end{equation}
\end{theorem}

\begin{remark}
Theorem~\ref{thm:martingale-clt} is the form used most frequently in this
book.  In applications, $S_n$ is usually the sum of martingale increments
obtained after a filtration-based decomposition of a quadratic form,
U-statistic, or leave-one-out expansion.  The key tasks are therefore to verify
conditional variance convergence and a Lindeberg condition.
\end{remark}

\begin{corollary}[Fourth-moment criterion]\label{cor:martingale-fourth-moment}
Let $\{\xi_{n,k},\mathcal F_{n,k}:1\le k\le k_n\}$ be a martingale
difference array.  Assume
\begin{equation}\label{eq:martingale-var-conv-sigma2}
  V_n
  =
  \sum_{k=1}^{k_n}\E(\xi_{n,k}^2\mid \mathcal F_{n,k-1})
  \to^P \sigma^2,
  \qquad \sigma^2>0,
\end{equation}
and
\begin{equation}\label{eq:martingale-fourth-moment-criterion}
  \sum_{k=1}^{k_n}\E(\xi_{n,k}^4)\to 0.
\end{equation}
Then
\begin{equation}\label{eq:martingale-clt-sigma2}
  \sum_{k=1}^{k_n}\xi_{n,k}\to^d N(0,\sigma^2).
\end{equation}
\end{corollary}

\begin{proof}
Set $\tilde\xi_{n,k}=\sigma^{-1}\xi_{n,k}$.  Then
$\{\tilde\xi_{n,k},\mathcal F_{n,k}\}$ is again a martingale difference array,
and
\[
  \tilde V_n
  =
  \sum_{k=1}^{k_n}
  \E(\tilde\xi_{n,k}^2\mid\mathcal F_{n,k-1})
  =
  \sigma^{-2}V_n
  \to^P 1.
\]
Fix $\varepsilon>0$.  The conditional Lindeberg term satisfies
\begin{align*}
  \tilde L_n(\varepsilon)
  &=
  \sum_{k=1}^{k_n}
  \E\!\left[
    \tilde\xi_{n,k}^2\mathbbm{1}(\abs{\tilde\xi_{n,k}}>\varepsilon)
    \mid\mathcal F_{n,k-1}
  \right] \\
  &\le
  \varepsilon^{-2}
  \sum_{k=1}^{k_n}
  \E\!\left(
    \tilde\xi_{n,k}^4\mid\mathcal F_{n,k-1}
  \right)
  \qquad\text{a.s.}
\end{align*}
by the inequality $x^2\mathbbm{1}(\abs{x}>\varepsilon)\le \varepsilon^{-2}x^4$.
Taking expectations and using \eqref{eq:martingale-fourth-moment-criterion},
\[
  \E\tilde L_n(\varepsilon)
  \le
  \varepsilon^{-2}\sigma^{-4}
  \sum_{k=1}^{k_n}\E(\xi_{n,k}^4)
  \to 0.
\]
Hence $\tilde L_n(\varepsilon)\to^P 0$ by Markov's inequality.  Applying
Theorem~\ref{thm:martingale-clt} to $\{\tilde\xi_{n,k}\}$ yields
$\sum_{k=1}^{k_n}\tilde\xi_{n,k}\to^d N(0,1)$, which is equivalent to
\eqref{eq:martingale-clt-sigma2}.
\end{proof}

\section{Extreme-value limits and Gaussian tails}

Max-type tests in high dimensions are typically calibrated by type-I extreme
value limits.  The standard derivations reduce to two ingredients: a sharp tail
approximation for the normal distribution and the identity
\[
  \Prob\Big\{\max_{1\le j\le p} W_j\le u\Big\}
  =
  \prod_{j=1}^p \Prob\{W_j\le u\}
\]
under independence.

\begin{lemma}[Mills ratio for the standard normal tail]\label{lem:mills-ratio}
Let $\Phi$ and $\phi$ denote the distribution and density functions of the
standard normal law.  Then, for every $x>0$,
\begin{equation}\label{eq:mills-ratio}
  \frac{1}{x+x^{-1}}\phi(x)
  \le
  1-\Phi(x)
  \le
  \frac{1}{x}\phi(x).
\end{equation}
In particular,
\begin{equation}\label{eq:mills-ratio-asymptotic}
  1-\Phi(x)
  =
  \frac{\phi(x)}{x}\{1+o(1)\}
  \qquad\text{as }x\to\infty.
\end{equation}
\end{lemma}

\begin{proof}
Since
\[
  1-\Phi(x)=\int_x^\infty \phi(t)\,dt,
\]
integration by parts gives
\begin{equation}\label{eq:mills-ratio-ibp}
  1-\Phi(x)
  =
  \frac{\phi(x)}{x}-\int_x^\infty \frac{\phi(t)}{t^2}\,dt.
\end{equation}
The upper bound in \eqref{eq:mills-ratio} follows immediately from
\eqref{eq:mills-ratio-ibp}.  For the lower bound, note that
$t^{-2}\le x^{-2}$ for $t\ge x$, so
\[
  \int_x^\infty \frac{\phi(t)}{t^2}\,dt
  \le
  \frac{1}{x^2}\int_x^\infty \phi(t)\,dt
  =
  \frac{1}{x^2}\{1-\Phi(x)\}.
\]
Substituting this into \eqref{eq:mills-ratio-ibp} yields
\[
  1-\Phi(x)
  \ge
  \frac{\phi(x)}{x}-\frac{1}{x^2}\{1-\Phi(x)\},
\]
which rearranges to
\[
  \Big(1+\frac{1}{x^2}\Big)\{1-\Phi(x)\}
  \ge
  \frac{\phi(x)}{x}.
\]
This is equivalent to the lower bound in \eqref{eq:mills-ratio}.  Finally,
combining the upper and lower bounds gives
\[
  \frac{x^2}{x^2+1}
  \le
  \frac{x\{1-\Phi(x)\}}{\phi(x)}
  \le 1,
\]
which implies \eqref{eq:mills-ratio-asymptotic}.
\end{proof}

\begin{proposition}[Normal-tail calibration for maxima]\label{prop:normal-tail-calibration}
Define
\begin{equation}\label{eq:u-p-y-definition}
  u_p(y)=\big(2\log p-\log\log p+y\big)^{1/2},
  \qquad y\in\R.
\end{equation}
Then, as $p\to\infty$,
\begin{equation}\label{eq:max-normal-tail-asymptotic}
  p\{1-\Phi(u_p(y))\}
  \to
  \frac{1}{\sqrt{4\pi}}e^{-y/2},
\end{equation}
and therefore
\begin{equation}\label{eq:max-abs-normal-tail-asymptotic}
  2p\{1-\Phi(u_p(y))\}
  \to
  \frac{1}{\sqrt{\pi}}e^{-y/2}.
\end{equation}
\end{proposition}

\begin{proof}
By Lemma~\ref{lem:mills-ratio},
\[
  1-\Phi(u_p(y))
  =
  \frac{\phi(u_p(y))}{u_p(y)}\{1+o(1)\}.
\]
Now
\[
  \phi(u_p(y))
  =
  \frac{1}{\sqrt{2\pi}}
  \exp\!\left\{-\frac{u_p(y)^2}{2}\right\}
  =
  \frac{1}{\sqrt{2\pi}}
  \exp\!\left\{-\log p+\frac12\log\log p-\frac y2\right\}
  =
  \frac{(\log p)^{1/2}}{\sqrt{2\pi}\,p}e^{-y/2}.
\]
Moreover,
\[
  u_p(y)
  =
  (2\log p)^{1/2}\{1+o(1)\}.
\]
Combining the last two displays gives
\[
  p\{1-\Phi(u_p(y))\}
  =
  p\frac{\phi(u_p(y))}{u_p(y)}\{1+o(1)\}
  \to
  \frac{1}{\sqrt{4\pi}}e^{-y/2}.
\]
This proves \eqref{eq:max-normal-tail-asymptotic}, and
\eqref{eq:max-abs-normal-tail-asymptotic} follows by multiplication by $2$.
\end{proof}

\begin{theorem}[Type-I extreme-value limit for Gaussian maxima]\label{thm:gumbel-gaussian-max}
Let $Z_1,\dots,Z_p$ be independent $N(0,1)$ random variables, and define
\[
  M_p=\max_{1\le j\le p} Z_j,
  \qquad
  M_p^*=\max_{1\le j\le p}\abs{Z_j}.
\]
With $u_p(y)$ defined in \eqref{eq:u-p-y-definition}, the following hold as
$p\to\infty$:
\begin{equation}\label{eq:gumbel-max-normal}
  \Prob\{M_p\le u_p(y)\}
  \to
  \exp\!\left\{-\frac{1}{\sqrt{4\pi}}e^{-y/2}\right\},
\end{equation}
and
\begin{equation}\label{eq:gumbel-max-abs-normal}
  \Prob\{(M_p^*)^2-2\log p+\log\log p\le y\}
  \to
  \exp\!\left\{-\frac{1}{\sqrt{\pi}}e^{-y/2}\right\}.
\end{equation}
Equivalently, if
\begin{equation}\label{eq:max-normal-centering-scaling}
  b_p=
  \sqrt{2\log p}-
  \frac{\log\log p+\log(4\pi)}{2\sqrt{2\log p}},
  \qquad
  a_p=\sqrt{2\log p},
\end{equation}
then
\begin{equation}\label{eq:gumbel-affine-max-normal}
  \Prob\big\{a_p(M_p-b_p)\le x\big\}
  \to
  \exp\{-e^{-x}\},
  \qquad x\in\R.
\end{equation}
\end{theorem}

\begin{proof}
By independence,
\[
  \Prob\{M_p\le u\}=\Phi(u)^p=
  \big[1-\{1-\Phi(u)\}\big]^p.
\]
Take $u=u_p(y)$.  By Proposition~\ref{prop:normal-tail-calibration},
\[
  p\{1-\Phi(u_p(y))\}
  \to
  \lambda(y):=\frac{1}{\sqrt{4\pi}}e^{-y/2}.
\]
Since $1-\Phi(u_p(y))\to 0$,
\[
  \log\Phi(u_p(y))^p
  =
  p\log\big[1-\{1-\Phi(u_p(y))\}\big]
  =
  -p\{1-\Phi(u_p(y))\}+o(1)
  \to -\lambda(y).
\]
Exponentiating proves \eqref{eq:gumbel-max-normal}.

Next,
\[
  \Prob\{M_p^*\le u\}
  =
  \{2\Phi(u)-1\}^p
  =
  \big[1-2\{1-\Phi(u)\}\big]^p.
\]
Again taking $u=u_p(y)$ and using
\eqref{eq:max-abs-normal-tail-asymptotic}, we obtain
\[
  2p\{1-\Phi(u_p(y))\}
  \to
  \frac{1}{\sqrt{\pi}}e^{-y/2}.
\]
Therefore
\[
  \log\Prob\{M_p^*\le u_p(y)\}
  =
  p\log\big[1-2\{1-\Phi(u_p(y))\}\big]
  \to
  -\frac{1}{\sqrt{\pi}}e^{-y/2},
\]
which proves \eqref{eq:gumbel-max-abs-normal}.  The affine-normalization form
\eqref{eq:gumbel-affine-max-normal} is equivalent to
\eqref{eq:gumbel-max-normal} after the change of variables
$y=2x-\log(4\pi)$ and the expansion
$u_p(2x-\log(4\pi))=b_p+x/a_p+o(a_p^{-1})$.
\end{proof}

\begin{corollary}[Extreme-value limit for maxima of squared normals]\label{cor:max-squared-normal}
If $Z_1,\dots,Z_p$ are independent $N(0,1)$ random variables and
\[
  Q_p=\max_{1\le j\le p} Z_j^2,
\]
then, for every $y\in\R$,
\begin{equation}\label{eq:max-squared-normal-gumbel}
  \Prob\{Q_p-2\log p+\log\log p\le y\}
  \to
  \exp\!\left\{-\frac{1}{\sqrt{\pi}}e^{-y/2}\right\}.
\end{equation}
\end{corollary}

\begin{proof}
Since $Q_p=(M_p^*)^2$, the conclusion is immediate from
\eqref{eq:gumbel-max-abs-normal}.
\end{proof}

\begin{remark}[Dependent maxima]
The independent limits in
Theorem~\ref{thm:gumbel-gaussian-max} and
Corollary~\ref{cor:max-squared-normal} serve as the basic benchmark for the
max-type statistics studied throughout the book.  In the chapter-specific proofs,
independence is often replaced by weak dependence conditions that guarantee the
same type-I extreme-value limit.  Those conditions are method-specific and are
therefore developed in the corresponding chapters rather than in this general
appendix.
\end{remark}

\section*{Bibliographic notes}

The scalar Bernstein inequality and its vectorized consequences are classical.
The operator-norm matrix Bernstein inequality stated in
Theorem~\ref{thm:matrix-bernstein} follows the matrix Laplace transform approach
of \citet{Tropp2012}.  The martingale central limit theorem in
Theorem~\ref{thm:martingale-clt} is standard; a comprehensive treatment may be
found in \citet{HallHeyde1980}.  Classical references for extreme-value theory
include \citet{LeadbetterLindgrenRootzen1983} and
\citet{deHaanFerreira2006}.